\documentclass{aa}  

\usepackage{graphicx}
\usepackage{txfonts}
\usepackage{enumerate}
\usepackage{paralist}
\bibpunct{(}{)}{;}{a}{{,}} 

\usepackage{wrapfig}

\usepackage[belowskip=-10pt,aboveskip=0pt]{caption}

\usepackage{mwe}

\RequirePackage[colorlinks,citecolor=blue,urlcolor=blue,pdfencoding=auto, psdextra]{hyperref}

\begin{document} 

\let\include\input

   \title{FR-type radio sources at 3 GHz VLA-COSMOS: \\ Relation to physical properties and large-scale environment }
   \titlerunning{FR-type radio sources at 3 GHz VLA-COSMOS}
   \subtitle{}
\let\cleardoublepage\clearpage

   \author{E. Vardoulaki\inst{1}\fnmsep\inst{2}\fnmsep\inst{3}\thanks{email: elenivard@gmail.com}\texorpdfstring{\href{https://orcid.org/0000-0002-4437-1773}{\protect\includegraphics{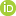}}}{}, 
  		E. F. Jim\'{e}nez Andrade\inst{4}\fnmsep\inst{3}\texorpdfstring{\href{https://orcid.org/0000-0002-2640-5917}{\protect\includegraphics{ORCID-iD_icon-16x16.png}}}{}, 
		I. Delvecchio\inst{5}\fnmsep\inst{6}\texorpdfstring{\href{https://orcid.org/0000-0001-8706-2252}{\protect\includegraphics{ORCID-iD_icon-16x16.png}}}{}, 
		V. Smol\v{c}i\'{c}\inst{7},
		E. Schinnerer\inst{8}\texorpdfstring{\href{https://orcid.org/0000-0002-3933-7677}{\protect\includegraphics{ORCID-iD_icon-16x16.png}}}{}, \\
		M. T. Sargent\inst{9}\texorpdfstring{\href{https://orcid.org/0000-0003-1033-9684}{\protect\includegraphics{ORCID-iD_icon-16x16.png}}}{},
		G. Gozaliasl\inst{10}\texorpdfstring{\href{https://orcid.org/0000-0002-0236-919X}{\protect\includegraphics{ORCID-iD_icon-16x16.png}}}{},
		A. Finoguenov\inst{10}\texorpdfstring{\href{https://orcid.org/0000-0002-4606-5403}{\protect\includegraphics{ORCID-iD_icon-16x16.png}}}{},
		M. Bondi\inst{11}\texorpdfstring{\href{https://orcid.org/0000-0002-9553-7999}{\protect\includegraphics{ORCID-iD_icon-16x16.png}}}{},
		G. Zamorani\inst{12}\texorpdfstring{\href{https://orcid.org/0000-0002-2318-301X}{\protect\includegraphics{ORCID-iD_icon-16x16.png}}}{},
		T. Badescu\inst{3},\\
		S. K. Leslie\inst{13}\texorpdfstring{\href{https://orcid.org/0000-0002-4826-8642}{\protect\includegraphics{ORCID-iD_icon-16x16.png}}}{},
		L. Ceraj\inst{7},
		K. Tisani\'{c}\inst{7}\texorpdfstring{\href{https://orcid.org/0000-0001-6382-4937}{\protect\includegraphics{ORCID-iD_icon-16x16.png}}}{},
		A. Karim\inst{3},
		B. Magnelli\inst{3},
		F. Bertoldi\inst{3},
		E. Romano-Diaz\inst{3}\texorpdfstring{\href{https://orcid.org/0000-0002-0071-3217}{\protect\includegraphics{ORCID-iD_icon-16x16.png}}}{},
		K. Harrington\inst{3}
		}

\authorrunning{Vardoulaki et al.}

   \institute{Th\"{u}ringer Landessternwarte, Sternwarte 5, 07778 Tautenburg, Germany
   \and
   Max-Planck-Institut f\"{u}r Radioastronomie, Auf dem H\"{u}gel 69, 53121 Bonn, Germany
   \and
   Argelander-Institut f\"{u}r Astronomie, Auf dem H\"{u}gel 71, D-53121 Bonn, Germany
\and
National Radio Astronomy Observatory, 520 Edgemont Road, Charlottesville, VA 22903, USA
\and 
 INAF - Osservatorio Astronomico di Brera, via Brera 28, I-20121, Milano, Italy \& via Bianchi 46, I-23807, Merate, Italy
 \and 
 CEA, Irfu, DAp, AIM, Universit\`e Paris-Saclay, Universit\`e de Paris, CNRS, F-91191 Gif-sur-Yvette, France
 \and
Department of Physics, Faculty of Science, University of Zagreb, Bijeni\v{c}ka cesta 32, 10000  Zagreb, Croatia
\and 
Max-Planck-Institut f\"{u}r Astronomie, K\"{o}nigstuhl 17, 69117, Heidelberg, Germany
\and
Astronomy Centre, Department of Physics and Astronomy, University of Sussex, Brighton, BN1 9QH, UK
\and
Department of Physics, University of Helsinki, P. O. Box 64, FI-00014 , Helsinki, Finland
\and
INAF - Istituto di Radioastronomia, Via Gobetti 101, 40129 Bologna, Italy 
\and
INAF-Osservatorio di Astrofisica e Scienza dello Spazio di Bologna, Via Piero Gobetti 93/3, I - 40129 Bologna, Italy
\and
Leiden Observatory, Leiden University, PO Box 9513, NL-2300 RA Leiden, the Netherlands
}

   \date{Received ; accepted }

 
  \abstract
   {Radio active galactic nuclei (AGN) are traditionally separated into two Fanaroff-Riley (FR) type classes, edge-brightened FRII sources or edge-darkened FRI sources. With the discovery of a plethora of radio AGN of different radio shapes, this dichotomy is becoming too simplistic in linking the radio structure to the physical properties of radio AGN, their hosts, and their environment. 
   }
   {We probe the physical properties and large-scale environment of radio AGN in the faintest FR population to date, and link them to their radio structure. We use the VLA-COSMOS Large Project at 3 GHz (3 GHz VLA-COSMOS), with a resolution and sensitivity of 0".75 and 2.3 $\mu$Jy/beam to explore the FR dichotomy down to $\mu$Jy levels. 
  }
   {We classified objects as FRIs, FRIIs, or hybrid FRI/FRII based on the surface-brightness distribution along their radio structure. Our control sample was the jet-less/compact radio AGN objects (COM AGN), which show excess radio emission at 3 GHz VLA-COSMOS exceeding what is coming from star-formation alone; this sample excludes FRs. The largest angular projected sizes of FR objects were measured by a machine-learning algorithm and also by hand, following a parametric approach to the FR classification. Eddington ratios were calculated using scaling relations from the X-rays, and we included the jet power by using radio luminosity as a probe. Furthermore, we investigated their host properties (star-formation ratio, stellar mass, morphology), and we explore their incidence within X-ray galaxy groups in COSMOS, and in the density fields and cosmic-web probes in COSMOS. 
}
   {Our sample is composed of 59 FRIIs, 32 FRI/FRIIs, 39 FRIs, and 1818 COM AGN at 0.03 $\le z \le$ 6. On average, FR objects have similar radio luminosities ($L_{\rm 3~GHz}\rm \sim 10^{23}~W~Hz^{-1}~sr^{-1}$), spanning a range of $\rm 10^{21-26}~W~Hz^{-1}~sr^{-1}$, and they lie at a median redshift of $z ~\sim ~1$. The median linear projected size of FRIIs is 106.6$^{238.2}_{36.9}$ kpc, larger than that of FRI/FRIIs and FRIs by a factor of 2-3. The COM AGN have sizes smaller than 30 kpc, with a median value of 1.7$^{4.7}_{1.5}$ kpc. The median Eddington ratio of FRIIs is 0.006$^{0.007}_{0.005}$, a factor of 2.5 less than in FRIs and a factor of 2 higher than in FRI/FRII. When the jet power is included, the median Eddington ratios of FRII and FRI/FRII increase by a factor of 12 and 15, respectively. FRs reside in their majority in massive quenched hosts ($M_{*}~> 10^{10.5} M_{\odot}$), with older episodes of star-formation linked to lower X-ray galaxy group temperatures, suggesting radio-mode AGN quenching. Regardless of their radio structure, FRs and COM AGN are found in all types and density environments (group or cluster, filaments, field).  
}
   {By relating the radio structure to radio luminosity, size, Eddington ratio, and large-scale environment, we find a broad distribution and overlap of FR and COM AGN populations. We discuss the need for a different classification scheme, that expands the classic FR classification by taking into consideration the physical properties of the objects rather than their projected radio structure which is frequency-, sensitivity- and resolution-dependent. This point is crucial in the advent of current and future all-sky radio surveys.
   }


   \maketitle
%

\section{Introduction}

Extragalactic radio sources associated with active galactic nuclei (AGN) have traditionally been classified based on the surface-brightness distribution along their radio structure, following the FR-type classification scheme of \cite{fr74}. Edge-brightened sources are deemed FRII and edge-darkened FRI. \cite{fr74} introduced this dichotomy, which was supported by the study of \cite{owenledlow94} and \cite{ledlowowen96}, described via the radio luminosity versus optical $R$-band luminosity diagram. In this diagram, FRIIs are more powerful at radio wavelengths than FRIs. The FR dichotomy was also supported by the study of \cite{Gopal-Krishna01} for redshifts $z~ <$ 0.5 and by \cite{vardoulaki10} at $z_{\rm med}~ \sim$ 1.25, with a few exceptions. It has further been suggested that FRI sources will eventually evolve into FRII \citep[e.g.][]{Gopal-Krishna88, kaiser07, turner15}, while other studies present different evolutionary paths \cite[e.g.][]{Kunert-Bajraszewska10}. The FR morphological dichotomy, is also believed to be a result of interaction of AGN jets with the environment \citep[e.g.][]{laing94, kaiser97} or due to mechanisms associated with jet production \citep[e.g.][]{meier01}. 

The linear projected sizes of FR-type objects in the sky can vary from sub-kpc/kpc to a few Mpc \citep[e.g.][]{blundell99b, Dabhade20}. In this wide distribution of sizes, FRIIs are traditionally larger than FRIs. \cite{Gopal-Krishna01} described FRIIs as more powerful, with powerful collimated jets, in contrast to FRIs. The reason for the different evolutionary picture in FRII and FRI jets according to \cite{Kunert-Bajraszewska10} is disruption of the FRI jets when going through the interstellar medium (ISM) of the host, resulting in loss of energy that prevents them from forming large FRII jets. 

FR objects are also categorised based on the properties of the black hole and on the efficiency of accretion onto the supermassive black hole (SMBH). FRIIs are in their majority high-excitation radio galaxies, while FRIs are mainly low-excitation radio galaxies \citep{kauffmann08, smolcic09, bestheckman12}. Thus FRIIs are thought to follow a model with efficient near-Eddington accretion onto the SMBH, which is described well by the unified AGN model \citep{heckmanbest14}. FRIs are related to inefficient sub-Eddington accretion onto the SMBH, an advection-dominated accretion flow (ADAF), giving rise to FRI-type jets \citep{heckmanbest14}. This division regarding accretion modes is supported by studies of bright-to-moderately faint ($\sim$ 100 mJy at 151 MHz) radio samples \cite[e.g.][]{mingo14,fernandes15}. When the intrinsically fainter radio universe is studied, this division starts to disappear: FRII and FRI sources exhibit sub-Eddington ratios and no clear division \citep[e.g.][]{lusso12}.

More recent studies of radio AGN have revealed a plethora of radio structures which deviate from a straight radio structure, introducing additional classifications, such as head-tail, one-sided, wide-/narrow-angle-tail, core-jet, core-lobe, twin-jet, fat-double, classic-double radio source, compact, jet-less, and even FR0 \citep[e.g.][]{baldi15, sadler16} and hybrid FRI/FRII \citep{Gopal-Krishna00, Gawronski06,  banfield15, kapinska17, harwood20}. Furthermore, when the faint radio universe is explored in enhanced sensitivity and resolution, radio sources are discovered that do not follow the FR-type classification. For instance, using a sample at $z~<$ 0.1, \cite{gendre13} reported that there is no dependence of the radio structure on radio luminosity but rather an overlap of populations (see their Fig. 8). \cite{mingo19} studied a sample of radio sources selected from the LOFAR Two-Metre Sky Survey (LoTSS-DR1), and showed these do not follow the classic FR dichotomy, with FRIIs observed up to three orders of magnitude fainter than the traditional FR break in radio luminosity and their hosts being fainter than expected.

From the literature, it is evident that the small-scale environment plays a role in shaping the radio structure of extended AGN. By small-scale environment we can either refer to the SMBH and feeding processes, or to the ISM. At the same time, the large-scale environment has also been shown to play a role, with more radio-luminous sources occupying massive and passive hosts \citep[e.g.][]{vardoulaki09, willott03, vardoulaki13} and preferring denser environments, such as galaxy clusters \citep[e.g.][]{magliocchetti18}. Previous studies have shown that FRIs prefer richer environments than FRIIs at low redshifts $z <$ 0.5 \citep[e.g.][at 408 MHz]{zirbel97}, at $z <$ 0.3 \citep[e.g.][at 1.4 GHz]{gendre13}, and at higher redshifts 1 $< z <$ 2 \citep[][at 1.4 GHz]{castignani14, chiaberge09}. More recent studies of 3C radio galaxies show the large-scale environments of FRI and FRII radio sources are similar \citep[e.g.][]{massaro20}. Recent reviews suggest that we should consider and study the AGN phenomenon as an interplay between small and large scales  through a self-regulated approach \citep{gaspari20}. It is clear that the AGN phenomenon and the different types of extended radio AGN, classified via the FR-type classification scheme, are not fully understood, as neither is the relation of AGN with their hosts and large-scale environment. AGN affect their hosts and environment through feedback mechanisms, which were introduced in models to constrain galaxy growth and to avoid having overly massive galaxies in the local universe \citep[e.g. the Illustris TNG simulation][]{weinberger18}. Feedback can be either positive, enhancing star formation, or negative, quenching star formation \citep[see][for a review]{fabian12}. The mechanisms in play involve radiative-mode and kinetic/jet-mode feedback, with the latter needed to explain quenching of star formation (SF) in massive galaxies as the maintenance mode of feedback \citep{fabian12}, and the suppression of cooling flow in massive cluster cores \citep{fabian03}. Recent studies \citep[e.g.][]{lacerda20} are showing that the main role of AGN in quenching is believed to be the removal and/or heating of the molecular gas instead of an additional suppression of star formation. Thus the role of the AGN is strongly linked to decreasing the molecular gas fraction of their host galaxies, leading to the quenching of star formation.

Some studies find the brightest radio AGN, thus FRIIs, to reside in massive hosts within clusters, while others find that FRIs should reside in denser environments. The picture of the relation of the radio structure, physical properties, and environment is therefore clearly still under debate. To better understand what affects the structure of radio sources, we need to carry out a systematic study of the radio structure and host/BH properties, and the large-scale environment of radio galaxies at both high resolution and sensitivity levels. For this purpose, we used the 3 GHz VLA-COSMOS Large Project \citep{smolcic17a} and extensive auxiliary data for the COSMOS\footnote{http://cosmos.astro.caltech.edu} field. These data cover a wide range of multi-wavelength properties and environmental probes that are essential to performing such a study.

With this paper we investigate the reason for the different radio structures associated with AGN, whether the FRI/FRII dichotomy is present at $\mu$Jy flux densities for median $z \sim$ 1, and the link of the radio structure to the physical properties of the sources and the large-scale environment. In Sec.~\ref{sec:sample} we present the sample. In Sec.~\ref{sec:analysis} we present the analysis related to the physical properties and environment of the sources in our sample as well as the results of this analysis: the linear projected size and radio luminosity in Sec.~\ref{sec:sizes}, the Eddington ratio in Sec.~\ref{sec:edd_ratios}, and hosts and large-scale environment in Sec.~\ref{sec:env}. In Sec.~\ref{sec:discuss} we discuss our findings and relate our results to the literature. Sec.~\ref{sec:conc} presents our conclusions. In Appendix~\ref{sec:measure_FR} we present a parametric approach to the FR classification, in Appendix~\ref{sec:notes_fr_obj} we provide notes on the objects and in Appendix~\ref{sec:auto_class} we present a semi-automatic method for measuring the largest projected angular size of a radio source. 

 Throughout this paper we use the convention for all spectral indices, $\alpha$, that the flux density $S_{\nu} \propto \nu^{-\alpha}$, where $\nu$ is the observing frequency. A low-density, $\Lambda$-dominated Universe in which $H_{0}=70~ {\rm km~s^{-1}Mpc^{-1}}$, $\Omega_{\rm M}=0.3$ and $\Omega_{\Lambda}=0.7$ is assumed throughout.

\section{Sample selection and radio classification}
\label{sec:sample}

Our sample is drawn from the VLA-COSMOS 3GHz Large Project \citep[][3 GHz VLA-COSMOS henceforth]{smolcic17a}, observed with the {\it Karl J. Jansky} Very Large Array (VLA) in S-band (centred at 3 GHz with a bandwidth of 2,048 MHz). The 3 GHz mosaic extends beyond the COSMOS field and covers 2.6 deg$^{2}$ at a resolution of 0.75 arcsec, while the median $rms$ in the 2 deg$^{2}$ of the COSMOS field is $\sim$ 2.3 $\mu$Jy/beam. Details of the observations and data reduction can be found in \cite{smolcic17a}. The source extraction was performed using the algorithm \textsc{blobcat} \citep{hales12}, which yielded $\sim$ 11,000 islands of radio emission or radio blobs. The final catalogue contains 10830 sources, 67 of which are multi-component sources, that is, they are composed of two or more radio blobs \citep[see][for detailed description]{vardoulaki19}, and the remainder are single-component sources \citep{smolcic17a}. 

\begin{figure}[!ht]
    \resizebox{\hsize}{!}
            {\includegraphics[trim={0cm 0cm 0cm 0.2cm},clip]{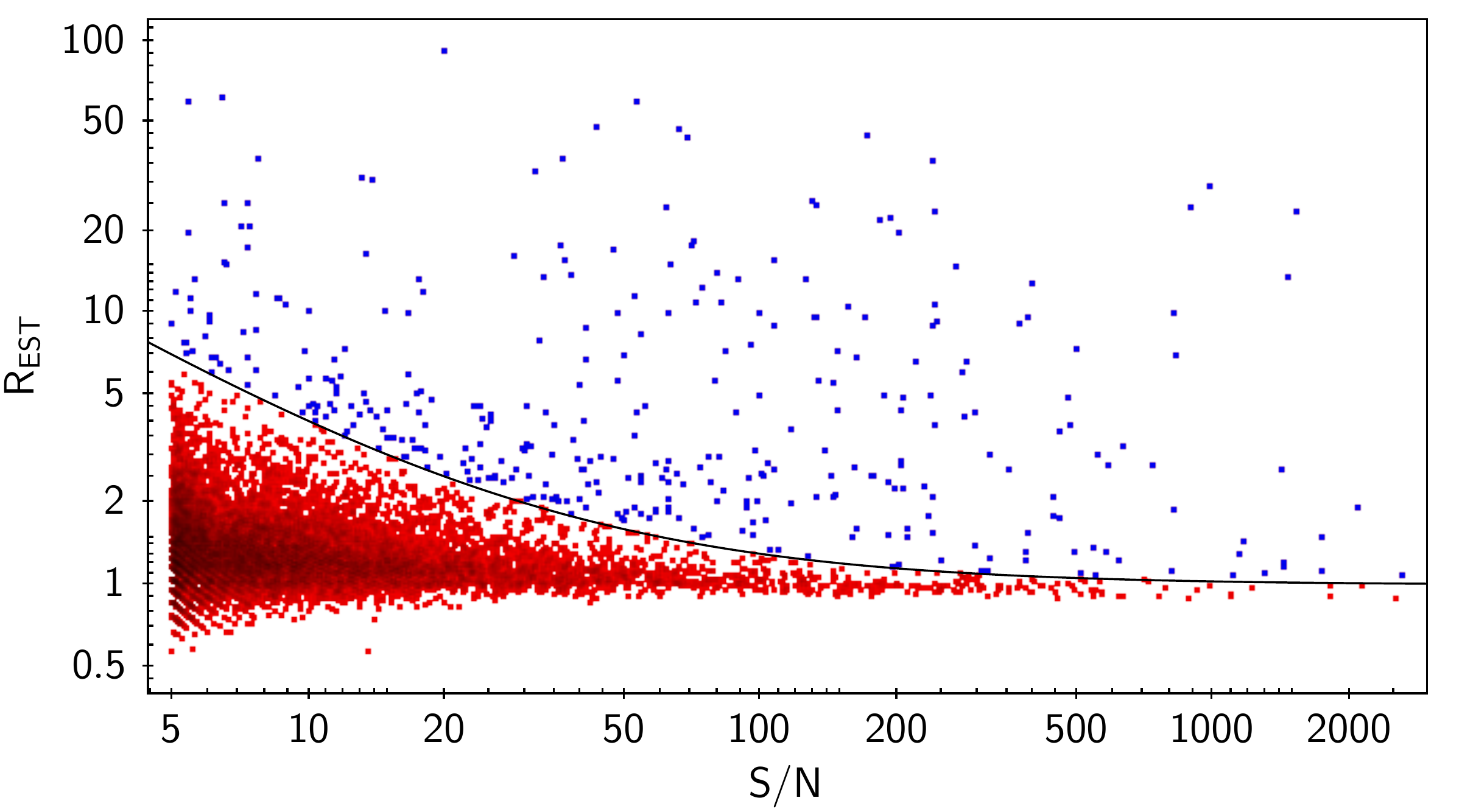}
 }
               \caption{R$_{\rm EST}$ parameter \citep[a unit-less size estimate from \textsc{blobcat};][]{hales12} versus signal-to-noise ratio for the 10899 entries in the \textsc{blobcat} catalogue identified from the 3 GHz mosaic of the COSMOS field. The solid black line represents the sources that obey the relation R$_{\rm EST} >$ 1+30/(S/N) and that were inspected visually. 
   }
              \label{fig:rest_snr}%
    \end{figure}

The aim of this paper is to study the FR-type radio sources in the 3 GHz VLA-COSMOS survey and explore the FR dichotomy. To identify which sources are extended amongst the 10830 radio sources in VLA-COSMOS, we used the \textsc{blobcat} size estimate parameter, R$_{\rm est}$, which provides an estimate of the size of each island/blob identified by the algorithm. The R$_{\rm est}$ parameter is not intended to be used for quantitative analysis but can be useful to identify blobs with a complex morphology. \cite{vardoulaki19}  provided a detailed description of this selection, which we briefly recall here. Based on the diagram in Fig.~\ref{fig:rest_snr}, we selected the sources that lie above the envelope given by the relationship R$_{\rm EST} >$ 1+30/(S/N). This envelope was chosen in order to include the most extended and brightest sources identified by \textsc{blobcat}. This selection yielded 351 blobs, which were visually inspected and matched to 350 sources. The matching procedure is described in detail in \cite{vardoulaki19}, which present the multi-components sources (composed of more than one radio blob) also studied in this paper. 

The matching was made with the aid of COSMOS auxiliary data in order to properly join the radio blobs to their parent source, when needed, and to the host galaxy. The host identification was made by visually crossmatching the radio core/body of galaxies with the optical/infrared stacked $YJHK_{S}$ image from the Ultra Deep Survey with the VISTA telescope \citep[Ultra-VISTA; see][and references therein]{laigle16, smolcic17b}, including regions observed at $z^{++}$ with an upgrade of the Subaru Suprime-Cam \cite[see][]{taniguchi07, smolcic17b, taniguchi15}. We also made use of the 1.4 GHz data \citep{schinnerer10} which, when available, served as confirmation of the matching procedure. Most of the multi-component sources (IDs from 10900-10966 in Table~\ref{table:data}) have a radio core, with the exception of sources 10915, 10926, and 10937 which show no radio core. Sources without an infrared host were matched on the basis that their radio emission resembled a radio AGN (sources 10908, 10922, 10924, 10932, and 10938). The single-component sources in our sample, all have associated infrared hosts, with the exception of sources 115 and 248.

\subsection{FR classification}
\label{sec:frclass}

\begin{figure}[!ht]
    \resizebox{\hsize}{!}
            {\includegraphics{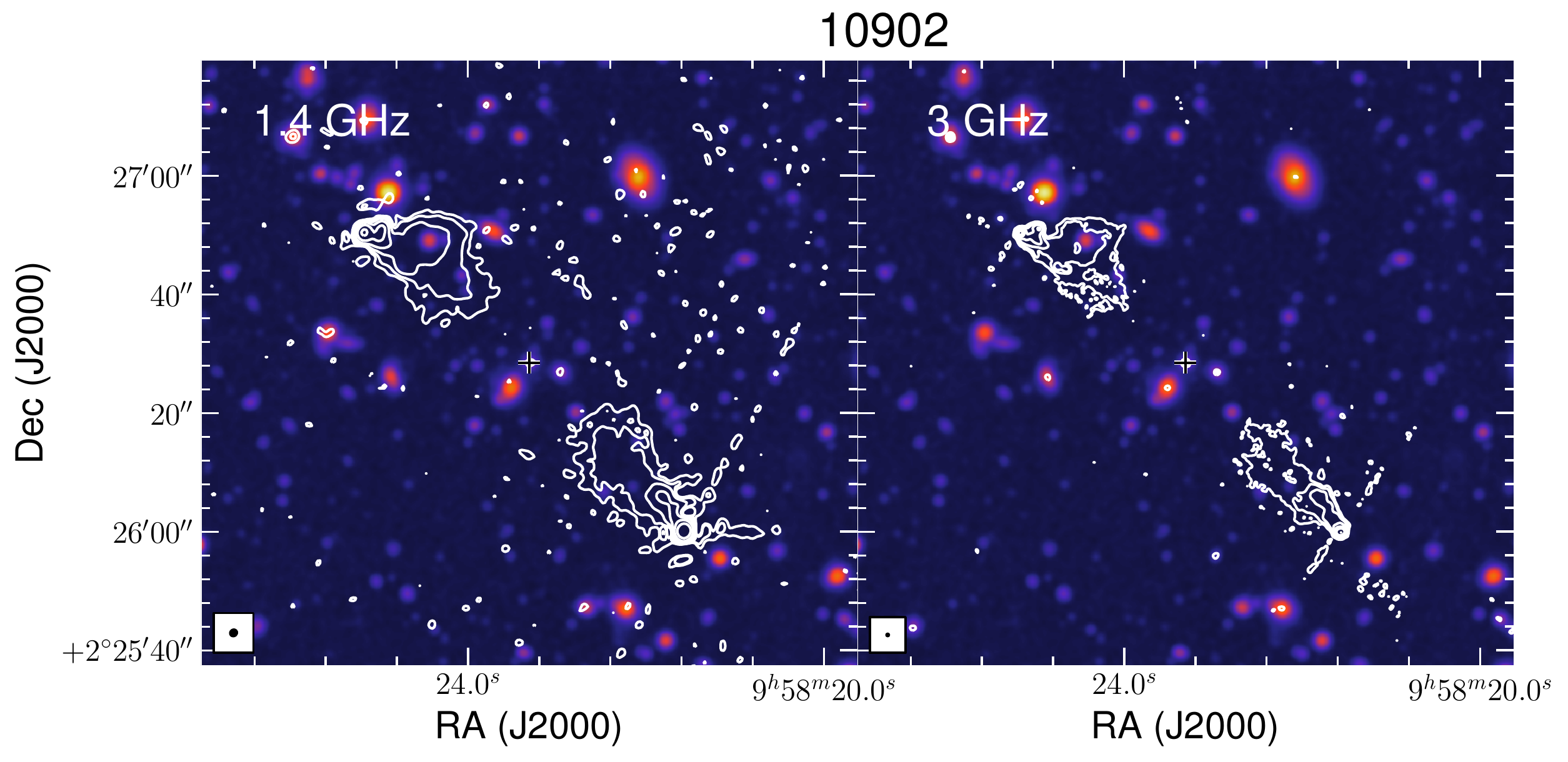}
            }
            \resizebox{\hsize}{!}
            {\includegraphics{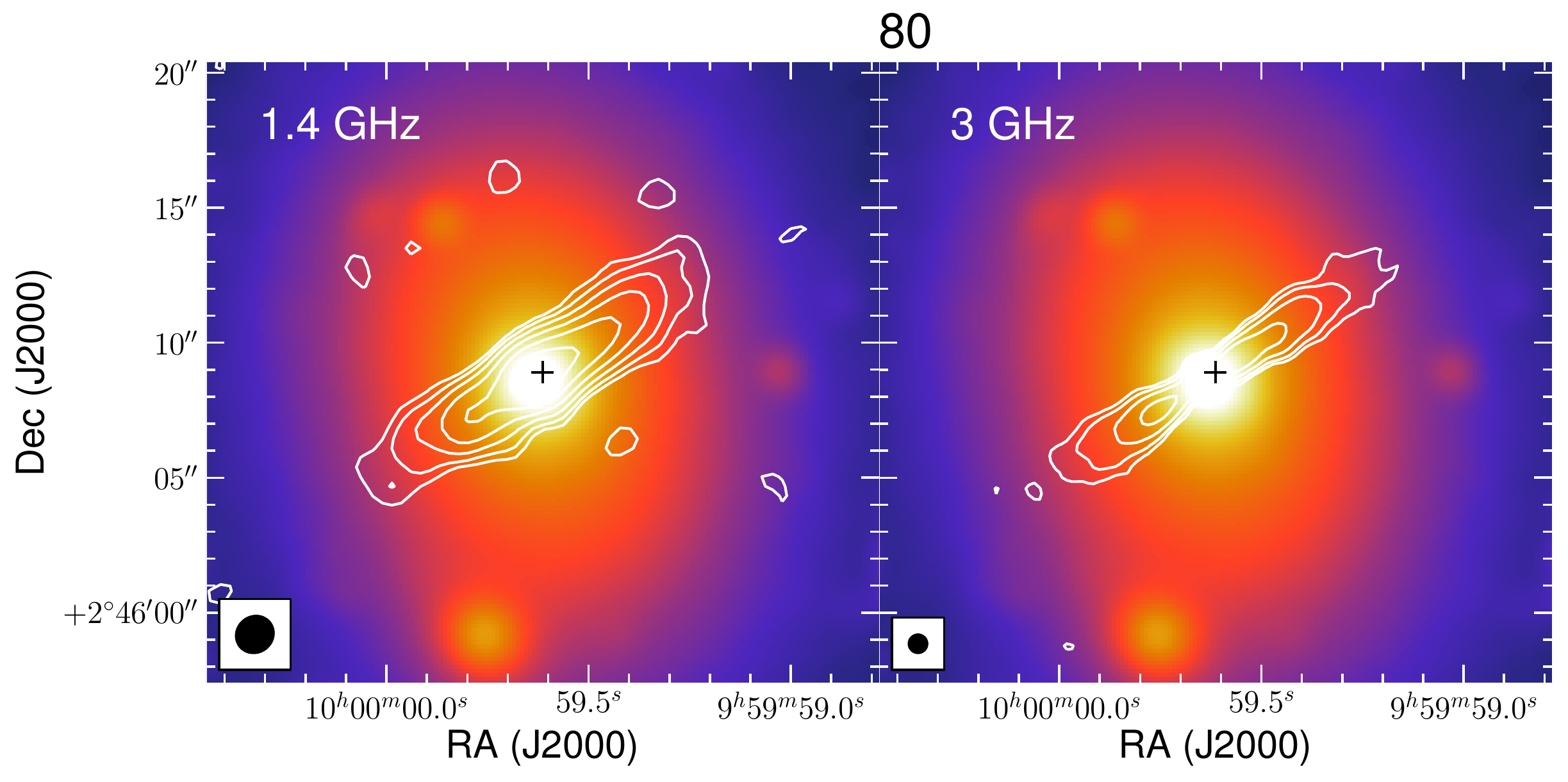}
            }
            \resizebox{\hsize}{!}
            {\includegraphics{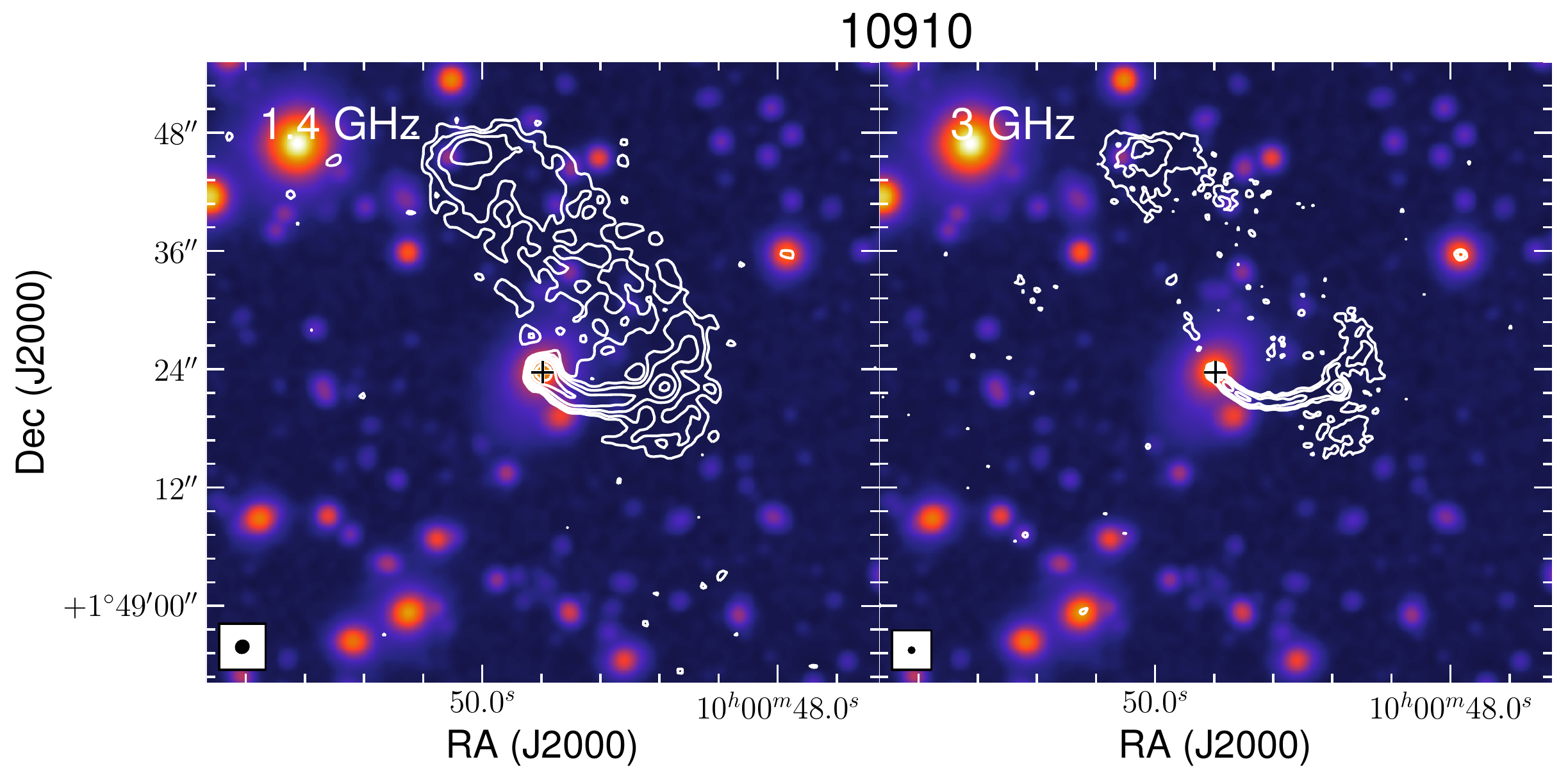}
            }
              \caption{Set of 1.4-GHz VLA (left) and 3 GHz VLA (right) stamps for examples of FRII ({\bf top}), FRI ({\bf middle}), and FRI/FRII ({\bf bottom}) objects, shown as white contours. These are overlaid on the Ultra-VISTA near-IR stacked image shown as a colour scale in arbitrary units. The beam size for the radio observations is shown at the bottom-left corner of the stamp: 1.4$\times$1.5 arcsec$^{2}$ for the 1.4 GHz and 0.75 arcsec FWHM for the 3 GHz maps. The contour levels are equally spaced on a log scale, with the lowest set at 3 $\sigma$ and the highest at the maximum peak flux-density of the radio structure. The remainder of the FR objects can be found in the Appendix in Fig.~\ref{fig:maps2}.
   }
              \label{fig:radmaps}%
    \end{figure}

We provide an FR-type classification based on the radio structure at 3 GHz, taking advantage of the COSMOS auxiliary data to identify the host galaxy, as described above. The 350 radio sources above the envelope in Fig.~\ref{fig:rest_snr} were classified by visual inspection in three stages. At stage one they were given to a team of non-experts on radio AGN with a set of guidelines. These guidelines (see Sec.~\ref{sec:stage1}) described how to visually separate the FRI, FRII, hybrid FRI/FRII, and non-FR-type radio sources based on the FR classification scheme \citep{fr74}. The second stage of FR classification was a revision of the results from the first stage, and a selection by two experts on FR-type radio sources of a sub-sample of 130 objects which exhibit jets and lobes\footnote{The 220 objects which were excluded from the FR sample did not show any signs of radio jets or lobes at 3 GHz. These can be a combination of star-forming galaxies (SFGs) and COM AGN. We selected the COM AGN from this sub-sample with the help of the radio excess flag \citep{delvecchio17} and used them in our analysis.}. These were taken to stage three, where we manually measured the distribution of flux-density along their structure based on the following criteria:
\begin{itemize}
      \item FRIIs or edge-brightened: the distance from the core to the brightest point along their structure is more than half of the total size of the source; these objects exhibit lobes (e.g. source 10902 in Fig.~\ref{fig:radmaps}). Single-lobed and one-side lobed objects were placed in this category.
      \item FRIs or edge-darkened: the distance from the core to the brightest point along their structure is less than half of the total size of the source; these objects exhibit jets (e.g. source 80 in Fig.~\ref{fig:radmaps}). Core single-jet and wide-angle tail objects were placed in this category.
      \item FRI/FRII: this is a hybrid {\bf class of objects} with one side being an FRII and the other an FRI (e.g. source 10910 in Fig.~\ref{fig:radmaps}).
\end{itemize}

This classification yielded 59 FRIIs, 32 FRI/FRIIs, and 39 FRIs. A detailed description of the FR classification is given in Appendix~\ref{sec:measure_FR}, and the final classification is presented in Table~\ref{table:measure_fr}; a question mark indicates an uncertain classification. In Table~\ref{table:data}, we report the 3 GHz FR classification and present the 1.4-GHz radio classification \citep{schinnerer10} for comparison. Notes on the objects are given in Appendix~\ref{sec:notes_fr_obj}. out of 130 radio sources in our sample, 46 are classified as FRI,  FRII, or wide-angle tail (WAT) based on the 1.4 GHz data. The remainder lack an FR-type classification. The 3 GHz sample includes 13 additional objects that are not identified at 1.4 GHz. These sources lie in masked areas or outside the coverage of the 1.4 GHz observations.

For the purposes of our analysis we compared the FR sample at 3 GHz to a control sample of radio AGN with compact or jet-less radio structure, that is, objects that do not exhibit lobes or jets in their radio structure but are associated with an AGN. These were selected on the basis of their radio excess \citep{delvecchio17}, that is, radio emission which exceeds that coming from star formation alone \citep[][see their Fig. 7-Top]{smolcic17b}. From these radio-excess objects we excluded all extended jet/lobed objects which were identified by cross-matching the FR sample to the radio-excess sample, and additionally performing visual inspection. The final sample of compact AGN (COM AGN henceforth) contained 1818 objects. 

The radio-excess selection is a rather conservative selection for radio AGN, which requires the ratio\footnote{No redshift-dependent threshold was applied.} of radio luminosity to the star-formation rate (SFR\footnote{Derived by fitting the spectral energy distribution \cite[SED;][]{delvecchio17, smolcic17b}.}; $L_{1.4\rm GHz}$/SFR$_{\rm IR}$) to be 3$\sigma$ the median value \citep{delvecchio17}. Objects with $L_{1.4\rm GHz}$/SFR$_{\rm IR}$ ratios higher than 3$\sigma$ are deemed radio-excess objects. As a result, we might be missing low-luminosity radio AGN with compact radio structure. These cannot be distinguished from star-forming radio sources via visual inspection as they show no clear signs of radio jets. Including these objects requires another approach and different AGN diagnostics than we used here. Thus our COM AGN sample is not complete, but misses low-luminosity compact radio AGN.

One important point we need to mention is that, if we had selected FR objects based on their radio excess, we would have missed $\sim$ 6\% of the FR objects in our sample, as they fall below the radio excess cut of \cite{delvecchio17}. This is shown in the bottom panel of Fig.~\ref{fig:n_z}, where FR objects without radio excess are marked as squares. They are randomly distributed in the $L-z$ plane of Fig.~\ref{fig:n_z}. The radio excess flag is given in Table~\ref{table:data}. Finally, we note that because the surface brightness decreases with redshift, we are only sensitive to bright sources as redshift increases, which will affect the number of FR sources we are able to detect at higher redshifts and thus limits our sample.

\subsection{Multi-wavelength data}
\label{sec:otherdata}

We made use of the 3 GHz VLA-COSMOS counterparts identified by \cite{smolcic17b}, who associated the 3 GHz radio sources with their hosts by using a multi-wavelength approach. Basic properties for the hosts are listed in Table~\ref{table:data} (redshift) and Table~\ref{table:hostprop} (SFR and stellar mass). The SFR and M$_{*}$ used in this analysis were calculated by \cite{delvecchio17} after fitting the multi-wavelength SED with \textsc{magphys} \citep{dacunha08} and the three-component SED-fitting code \textsc{sed3fit} by \cite{berta13}, which accounts for an additional AGN component. They used a Chabrier IMF. In brief, they exploited the optical to mid-infrared photometry from the COSMOS2015 catalogue \citep{laigle16}. To constrain the far-infrared part of the SED, they further included {\it Herschel} PACS \citep{lutz11} and SPIRE \citep{oliver12} data. For the higher redshift galaxies, they used a large dataset of sub-millimetre (sub-mm) photometry from JCMT/SCUBA-2, LABOCA, Bolocam, JCMT/AzTEC, MAMBO, ALMA, and PdBI \citep[see Sect. 2.2 in][ for references and discussion]{delvecchio17}. For classification purposes, they also used the Chandra-COSMOS \citep{elvis09, civano12} and COSMOS-Legacy \citep{civano16} X-ray catalogues. 

We further cross-correlated our FR and COM AGN samples with the most recent X-ray group catalogue for the COSMOS field, from \cite{gozaliasl19}, which is an updated version of the \cite{george11} X-ray group catalogue. By groups we refer to a set of galaxies with a common dark matter halo \citep{george11}. The \cite{gozaliasl19} catalogue includes 247 groups at 0.08 $\leq z <$ 1.53 from Chandra/XMM-Newton data, with halo masses {\bf $M_{ 200}=8\times 10^{12}-3\times 10^{14} M_{\odot}$} \citep[see][]{gozaliasl19}. To cross-correlate this with the FR and COM AGN objects in our sample, we used a search radius ($r_{200}$) within the virial radius of each group and the redshift of each object with $\Delta z = \pm 0.007\times(1+z_{\rm xgroup})$ in order to match the photometric redshift accuracy of our data \citep{laigle16}. Up to $z~ < 1.53$, we find that 24 out of 75 FRs (12 out of 43 FRIIs, 4 out of 23 FRI/FRIIs, 8 out of 32 FRIs) and 87 out of 963 COM AGN are associated with an X-ray group\footnote{These numbers correspond to the same field coverage between the X-ray groups and 3 GHz VLA-COSMOS. Of the 130 FRs, 44 lie above redshift z = 1.53.}.

\section{Analysis and results}
\label{sec:analysis}

\begin{figure}[!ht]
    \resizebox{\hsize}{!}
            {\includegraphics{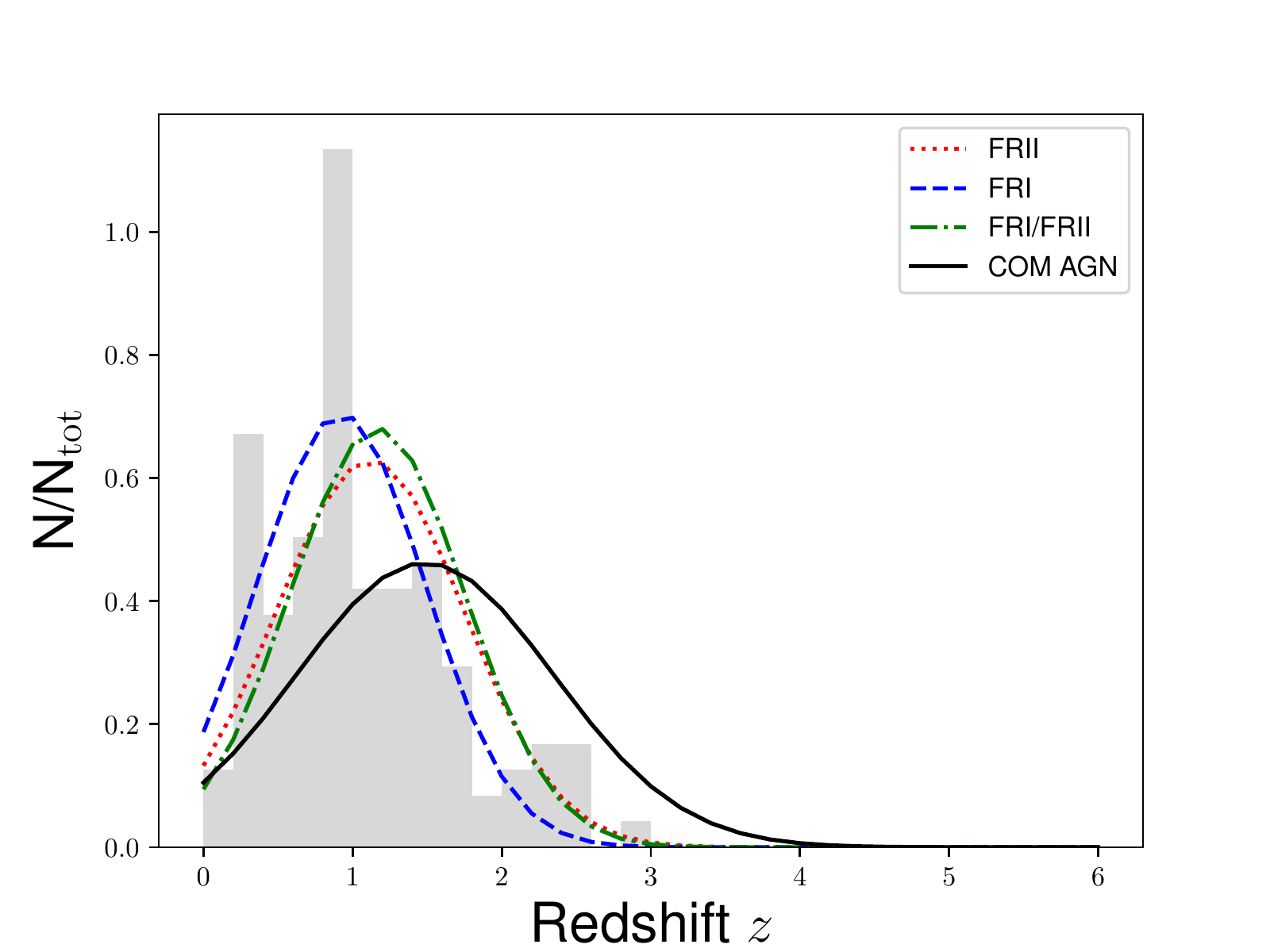}
 }
    \resizebox{\hsize}{!}
            {\includegraphics{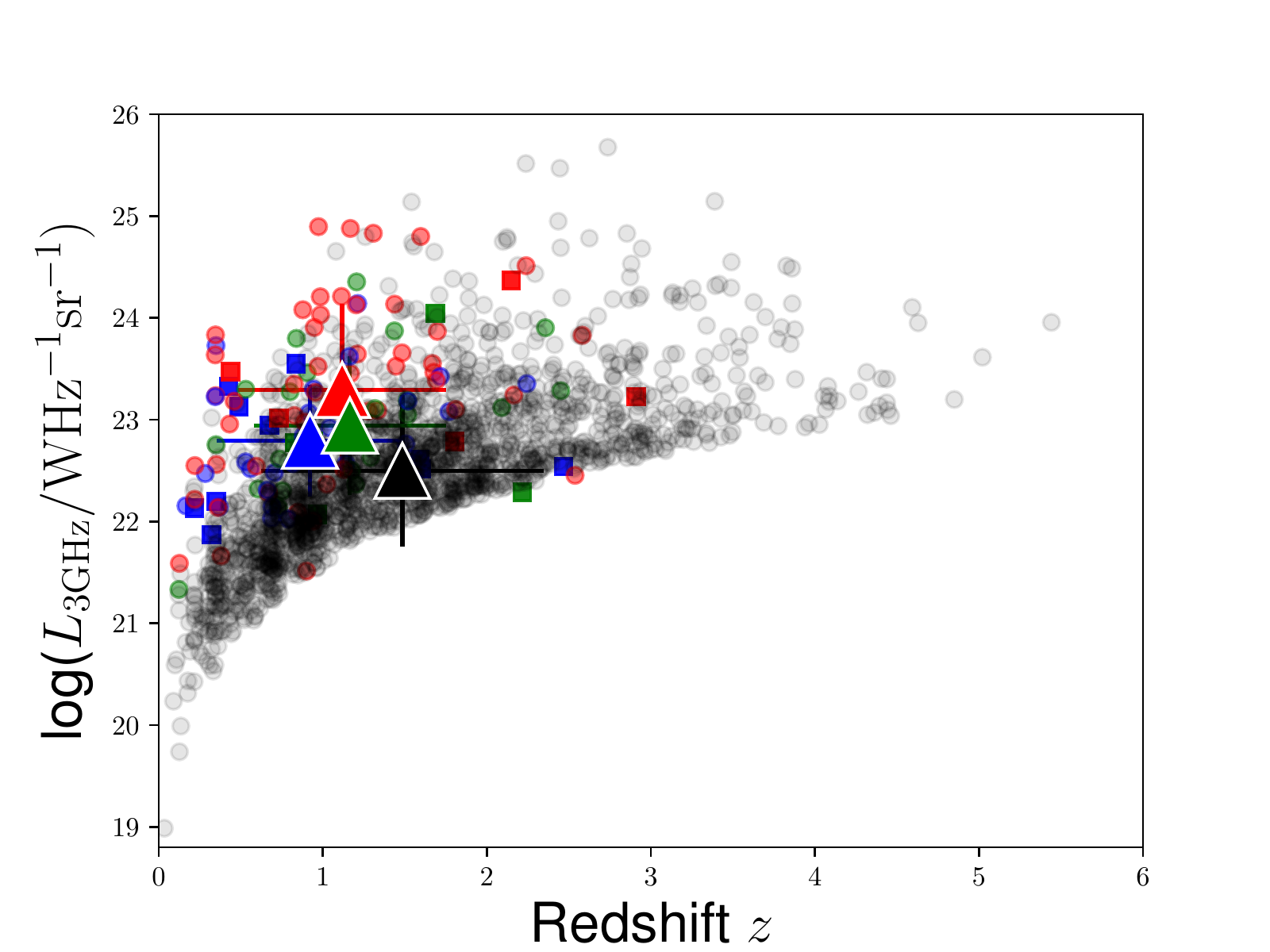}
 }
               \caption{{\bf Top}: Redshift distribution of the FR-type sources in our sample shown in grey. The Gaussian distributions show Gaussian fits to the redshift distribution of  different populations, colour-coded based on radio classification: FRIIs (red), FRIs (blue), and FRI/FRII (green). We also show the COM AGN control sample as a black solid line.
               {\bf Bottom}: Radio luminosity at 3 GHz vs. redshift for the FRIIs (red), FRIs (blue), FRI/FRII (green), and COM AGN (black) in our sample. The large triangles give the corresponding mean values for each population and the standard deviation. Squares are objects without radio excess.
   }
              \label{fig:n_z}%
    \end{figure}

Here we present the analysis and results of 130 FR-type radio sources from the 3 GHz VLA-COSMOS, which were classified visually  (59 FRIIs, 32 FRI/FRIIs, and 39 FRIs), and 1818 COM AGN. Most FR objects (119 out of 130) have a counterpart in the optical/infrared based on the \cite{smolcic17b} study on counterpart association of VLA-COSMOS detections. Likewise, the majority of the FR sources (119 out of 130; $\sim$ 91\%) have available redshifts, $\sim$ 57\% (68 out of 119) of which are spectroscopic and $\sim$ 43\% (51 out of 119) photometric. The control sample of compact COM AGN includes 1818 objects, all of which have counterpart association and redshifts, with a spectroscopic completeness of $\sim$ 32\% (575 out of 1818). The redshift distribution of the FR and COM AGN objects in our sample is presented in Fig.~\ref{fig:n_z}-Top, ranging between 0.03 $\le z \le$ 6, and the $L-z$ plane in Fig.~\ref{fig:n_z}-Bottom. The redshift distribution for the FR-type objects peaks around $z~ \sim$ 1, while for the COM AGN it peaks at slightly higher redshift\footnote{This difference in the mean redshift values between FRIIs or FRIs and COM AGN seems to be statistically significant. For example, a $Z$-test between FRI/FRII and COM AGN gives a {z}-score =  -3.35 and a {p}-value = 0.0008.} ($z~ \sim$ 1.5). Similarly, the FR-type objects have radio luminosities at 3 GHz of $L_{\rm 3~GHz} \sim~10^{23}\rm ~W~Hz^{-1}~sr^{-1}$\footnote{A steep radio spectral index of 0.8 is assumed in the calculation of radio luminosities at 3 GHz.}, while the COM AGN are on average fainter at $L_{3 \rm ~GHz} \sim\rm ~10^{22}~W~Hz^{-1}~sr^{-1}$ on average. At each redshift the $L_{3\rm~GHz}$ of FRs tend to be in the high $L_{3\rm~GHz}$ tail of the overall $L_{3\rm~GHz}$ distribution. Fig.~\ref{fig:n_z}-Bottom shows that we probe FR-type objects with radio luminosities higher than $L_{3\rm ~ GHz} \sim \rm ~10^{21}~W~Hz^{-1}~sr^{-1}$ and up to redshifts of $\sim 3$. This is expected due to the relation between surface brightness and redshift of $(1~+~z)^{-4}$, which means that the higher the redshift the more difficult it is to recover extended radio structures of low surface brightness. Table~\ref{tab:prop_pop} shows the median radio luminosity values for the different populations and their dispersion, indicating an overlap of distributions. These results show no clear dichotomy in radio luminosity at 3 GHz between FRIs and FRIIs, with median values and the 84 and 16 percentiles for FRIIs at log$_{10}( L_{3\rm~GHz} /\rm W~Hz^{-1}~sr^{-1}) = 23.30^{24.14}_{22.26}$ and for FRIs at log$_{10}(L_{3\rm~ GHz} /\rm W~Hz^{-1}~sr^{-1}) = 22.59^{23.32}_{22.15}$.

To investigate whether the differences between FRI- and FRII-type radio AGN presented in the classic FR classification scheme \citep{fr74} are inherent to their host galaxy/SMBH properties, or are acquired, for example, as a result of a denser environment, we compared the radio structure to physical properties of the radio sources and the large-scale environment. We refer to radio luminosity, size,
 and Eddington ratio as 'physical properties' and use host properties and kpc-/Mpc-scale surroundings as indicators of 'environment'. Average values from this analysis are given in Table~\ref{tab:prop_pop}.

\subsection{Linear projected sizes and radio luminosity}
\label{sec:sizes}

An important physical parameter for radio AGN is their linear projected size. This is not a straightforward parameter to measure, as most radio AGN with extended sizes are far from straight. They rather exhibit bends in their radio structure making the measurement more complex. Additionally, jet-less AGN require a different approach, as described below. The radio sizes of the objects in our sample were therefore measured with two different techniques. The FR-type objects were put through a semi-automatic machine learning code that  measures the largest angular size (LAS) of the sources in arcsec. The code is described in detail in Appendix~\ref{sec:auto_class}, and it provides accurate size estimates for $\sim$90\% of the FR sample. As a secondary check, we measured the largest angular sizes by hand, presented in Table~\ref{table:data}. As the machine-learning code does not provide robust size estimates for the full sample, we used the manual measurements in our analysis for consistency. A comparison of the estimated and manually measured sizes is given in Appendix~\ref{sec:auto_class} and Fig.~\ref{fig:las_comp}. We then converted the LAS quantity to linear projected size of the sources $D$ in kpc (Table~\ref{table:data}), by taking  the redshift of each object into account, resulting in sizes for 119 out of the 130 ($\sim$ 91\%) FRs in our sample. The COM AGN sizes were measured by a Gaussian fit using the publicly available code \textsc{pyBDSF} \citep{mohan15}. The size of the COM AGN used in our analysis is the intrinsic size, after deconvolution from the synthesised beam \citep{jimenez19}. Sizes for the COM AGN are given in the Appendix in Table~\ref{tab:com_D}. 

In Fig.~\ref{fig:L_D} we present the $L-D$ diagram for the sources in our sample. FR objects have on average similar radio luminosities at 3 GHz independent of FR type, and they also have similar luminosities to COM AGN. Their linear projected sizes, though, differ. FRs have sizes ranging from 10 kpc to 1 Mpc, forming the FR cloud, while COM AGN are smaller than $\sim$ 30 kpc at 3 GHz, forming the COM cloud in the $L-D$ diagram. Additionally, FRII-type objects are larger than FRI/FRII and FRI objects by a factor of $\sim$ 2 and $\sim$ 3 on average, respectively. Still, there is an overlap of the distributions and no clear dichotomy in FR type.

\begin{figure}[!ht]
  \resizebox{\hsize}{!}
 {\includegraphics{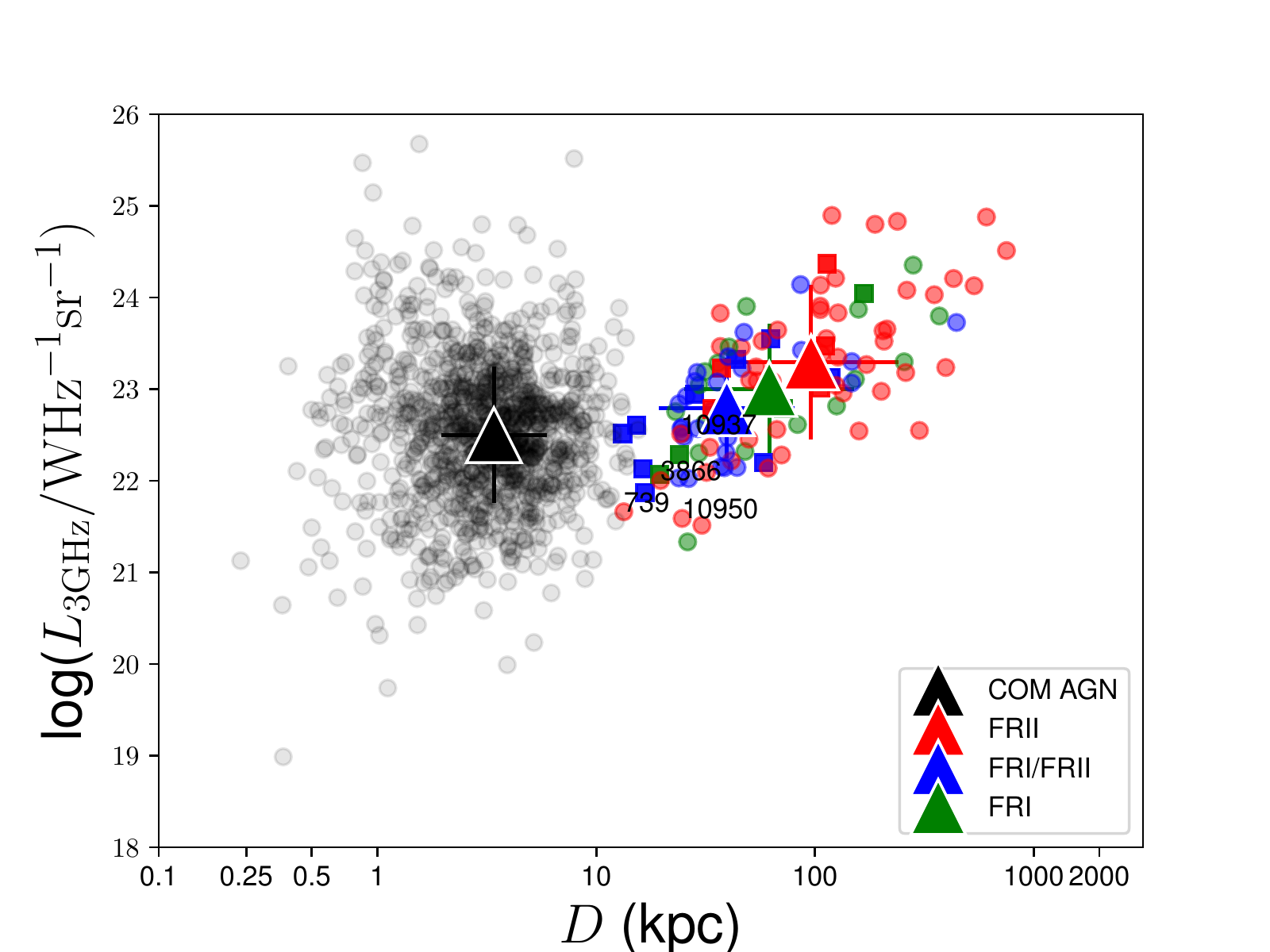}
             }
           
       \caption{Radio luminosity at 3 GHz vs. linear projected size $D$. Filled red circles denote FRII objects, filled blue circles denote FRI objects, filled green circles are FRI/FRII objects, and filled black circles denote compact AGN. FR objects without radio excess (see Sec.~\ref{sec:sample}) are shown as squares instead of circles. 
   }
              \label{fig:L_D}%
    \end{figure}

\subsection{Eddington ratios}
\label{sec:edd_ratios}

\begin{table*}[!ht]
\caption{Eddington ratios from radiative luminosity, not corrected for redshift dependence}             
\label{tab:eddrat_rad}      
\centering                          
\begin{tabular}{l l l l l l l l l}        
\hline\hline                 
radio   &            &   N  &      \multicolumn{3}{c}{Log$_{10}(\lambda_{\rm r})$}                &             \multicolumn{3}{c}{Log$_{10}(L_{\rm X}/L_{\rm radio})$}\\     
class            &           &        &     median   &   16\%   &  84\%  &   median  &   16\% &    84\% \\
\hline
FRII   &  detected  &  18   &    -2.20  &  -2.27 &    -2.11    &   1.76 &      1.72   &   1.81\\
          & stacked    &  30   &     -3.43 &   -3.55 &    -3.30   &   1.00 &       0.87  &   1.13 \\
\hline
FRI   &   detected  &  6    &   -1.82    &   -1.90   &  -1.74    &   2.64       &    2.58   &    2.71\\
         &   stacked   &  25   &$<$-3.52 &   $-$       &    $-$       &   $<$1.65 &    $-$       &    $-$    \\
\hline

FRI/FRII   &   detected  & 8     &   -2.43  &   -2.52   &  -2.26   &  1.88  &    1.83   &    1.92\\
          	&   stacked   &  12   &  -3.19  &   -3.37   &   -3.00 &   1.62 &      1.44 &      1.81\\
\hline
COM AGN  & detected  &  291     &    -1.96 &     -1.99 &     -1.94 &       3.47   &   3.45   &   3.49\\
                    & stacked    & 1386  &    -3.22 &     -3.35 &     -3.08 &       2.35 &     2.21 &     2.49\\
\hline
\end{tabular}
\tablefoot{Eddington ratios using radiative luminosity, $\lambda_{\rm r}$ = $L_{\rm rad}$ / $L_{\rm Edd}$ (see Sec.~\ref{sec:edd_ratios}). The 16th and 84th percentile values indicate uncertainties on the medians.}
\end{table*}

\begin{table*}[!ht]
\caption{Eddington ratios from radiative luminosity and $Q_{\rm jet}$, not corrected for redshift dependence}             
\label{tab:eddrat_qjet}      
\centering                          
\begin{tabular}{l l l l l l l l l}        
\hline\hline                 
radio   &            &   N  &      \multicolumn{3}{c}{Log$_{10}(\lambda_{\rm rk})$}                &             \multicolumn{3}{c}{Log$_{10}(L_{\rm X}/L_{\rm radio})$}\\    
 class           &           &        &     median   &   16\%   &  84\%  &   median  &   16\% &    84\% \\
\hline
FRII   &  detected  &  18   &    -1.13  &  -1.15 &    -1.10    &   1.76 &      1.71   &   1.81\\
          & stacked    &  30   &     -1.65 &   -1.65 &    -1.64   &   1.00 &       0.87  &   1.13 \\
\hline
FRI   &   detected  &  6    &   -1.50   &   -1.56   &  -1.43    &   2.64       &    2.58   &    2.71\\
         &   stacked   &  25   &$<$-2.21 &   $-$       &    $-$       &   $<$1.67 &    $-$       &    $-$    \\
\hline

FRI/FRII   &   detected  & 8     &   -1.32  &   -1.40  &  -1.28   &  1.88  &    1.83   &    1.93\\
          	&   stacked   &  12   &  -1.91  &   -1.92   &   -1.89 &   1.65 &     1.45 &      1.85\\
\hline
COM AGN  & detected  &  291     &   -1.79 &     -1.82 &     -1.77 &       3.47  &   3.45   &   3.49\\
                    & stacked    & 1386  &    -2.35 &     -2.37 &     -2.33 &       2.35 &    2.22 &     2.49\\
 \hline                                   
\end{tabular}
\tablefoot{Eddington ratios including radiative luminosity and kinetic energy $Q_{\rm jet}$, $\lambda_{\rm rk}$ = ($L_{\rm rad}$ + $Q_{\rm jet}$)/ $L_{\rm Edd}$ (see Sec.~\ref{sec:edd_ratios}). The 16th and 84th percentile values indicate uncertainties on the medians.}
\end{table*}

The Eddington ratio is a physical quantity that is directly related to how efficiently a black hole is accreting matter around it. It is the ratio of luminosity emitted by the source over the Eddington luminosity, that is the maximum luminosity an object can achieve when the gravitational pull and emitted radiation are balanced. We explore this quantity to  investigate how many of the objects in our sample are efficient or inefficient accreters, and to search for trends with radio structure. For the purposes of our study we calculated the Eddington ratios using the X-ray catalogue of \cite{marchesi16}. A cross-correlation with the Chandra COSMOS-Legacy Survey X-ray catalogue yields 19 FRIIs, 8 FRI/FRIIs, 6 FRIs, and 291 COM AGN with a secure X-ray detection. 

The Eddington ratio was calculated in two ways:
\begin{enumerate}
\item $\lambda_{\rm r}$ = $L_{\rm rad}$ / $L_{\rm Edd}$, i.e. the radiative luminosity over the Eddington luminosity. The intrinsic AGN X-ray luminosities $L_{\rm rad}$ were scaled to bolometric luminosities via a set of luminosity-dependent bolometric corrections by \cite{lusso12}. The Eddington luminosity was calculated using the standard conversion $M_{\rm BH}/M_{\rm bulge} \sim 0.002$ \citep{marconi03}. 
\item $\lambda_{\rm rk}$ = ($L_{\rm rad}$ + $Q_{\rm jet}$)/ $L_{\rm Edd}$, i.e. same as  $\lambda_{\rm r}$, but with the addition of the jet kinetic energy to the numerator. The kinetic energy was calculated from the radio luminosity at 1.4 GHz using the empirical relation by \cite{cavagnolo10}. The 1.4 GHz luminosity comes from 3 GHz fluxes, and was then converted to 1.4 GHz using a typical steep spectral index of 0.7 (if not detected at 1.4 GHz), or the observed 1.4-3 GHz slope (if detected at 1.4 GHz).
\end{enumerate}

We assumed $M_{\rm bulge}$ to be $M_{*}$, estimated from the fit to the SED \citep{delvecchio17}, which is a good approximation for massive quiescent galaxies such as those in our sample ($\gtrsim10^{11} M_{\odot}$; see Sec.~\ref{sec:env_host}). According to bulge-disk decomposition analysis \citep{dimauro18} the bulge mass fraction for star-forming galaxies typically is $>$ 70\% for $M_{*} > 10^{11} M_{\odot}$, so that our assumption is quite reasonable. X-ray detected COM AGN with X-ray luminosities higher than $10^{44}$ erg/s are only a small percentage ($\sim$ 4\%) of our COM AGN sample.

\begin{figure}[!ht]
  \resizebox{\hsize}{!}
 {\includegraphics{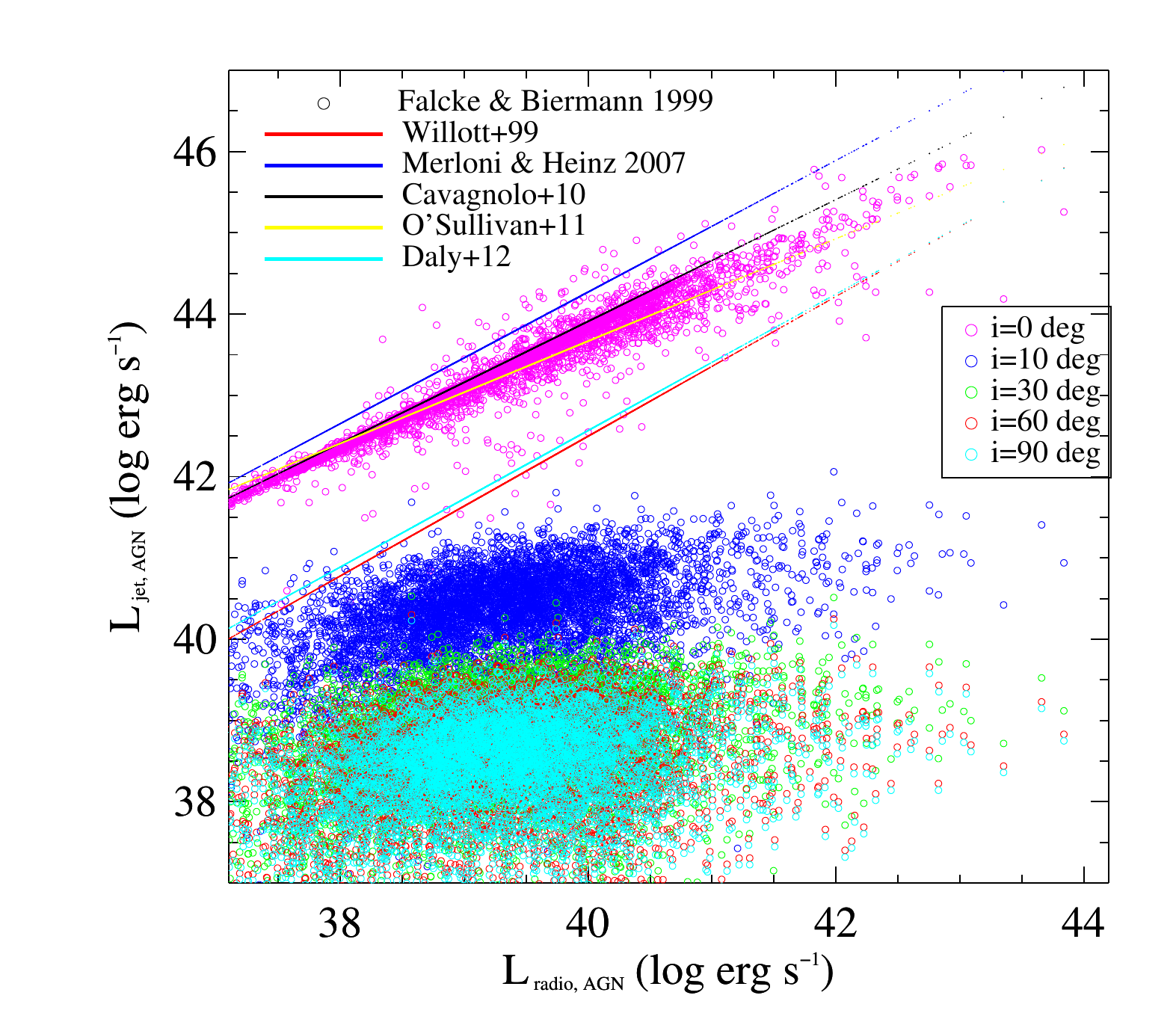}
            }
       \caption{Scaling relations between $L_{\rm jet}$ and radio luminosity at 1.4 GHz. The 3 GHz VLA-COSMOS data are plotted as coloured circles using the relations of \cite{falcke99} for different inclination angles of the jet, from 0 (face-on - towards the observer) to 90 (edge-on) degrees. Lines show scaling relations from the literature: the red line presents data from \cite{willott99}, the blue line data from \cite{merloni07}, the black line data from \cite{cavagnolo10}, the yellow line data from \cite{O'Sullivan11}, and the cyan line data from \cite{daly12}. \\
   }
              \label{fig:qjet_relations}%
    \end{figure}

In order to decide which scaling relation to use for the calculation of kinetic energy, or jet power $Q_{\rm jet}$, we compared several scaling relations between radio luminosity and jet power. This comparison is presented in Fig.~\ref{fig:qjet_relations}. We overplot the 3 GHz VLA-COSMOS data for different inclinations of the jet with respect to the observer. Objects with inclinations of 0 deg (face-on) follow the \cite{cavagnolo10} scaling relation well, therefore we took this as a conservative approach for the calculation of jet power. The radio jet power strongly depends on the viewing angle. The kinetic component of the Eddington ratio could be $>$ 100 times smaller, producing zero change in the Eddington ratio. For larger inclinations ($>$ 10 deg), none of the scaling relations shown represent the data.

For non-X-ray detected objects, we used a stacking approach to estimate a median Eddington ratio. We used the publicly available X-ray stacking tool {CSTACK}\footnote{http://lambic.astrosen.unam.mx/cstack/} developed by T. Miyaji. This tool provides stacked count rates and fluxes, as well as reliable uncertainties estimated from a bootstrapping procedure. Each bootstrap yields a mean stacked $L_{\rm X}$ (rest-frame 2-10 keV). After bootstrapping 500 times, we took the median of the resulting distribution in order to alleviate the effect of possible outliers. From the median $L_{\rm X}$, we subtracted the expected contribution arising from star formation\footnote{SFR is estimated from the fit to the IR+UV SED.}, given by the {\bf redshift-independent} $L_{\rm X}$-SFR relation derived by \cite{symeonidis14}, and we considered only the remaining X-ray emission (if any), which is likely attributable to the AGN. The X-ray emission was then corrected for nuclear obscuration, based on the hardness ratio \citep{xue10}, and by assuming an intrinsic power-law X-ray spectrum with a constant slope $\Gamma$ = 1.8 \cite[e.g.][]{tozzi06}. 

\begin{figure}[!ht]
   \centering
    \resizebox{\hsize}{!}
            {  
            \includegraphics[trim={2.5cm 0cm 2cm 0cm},clip]{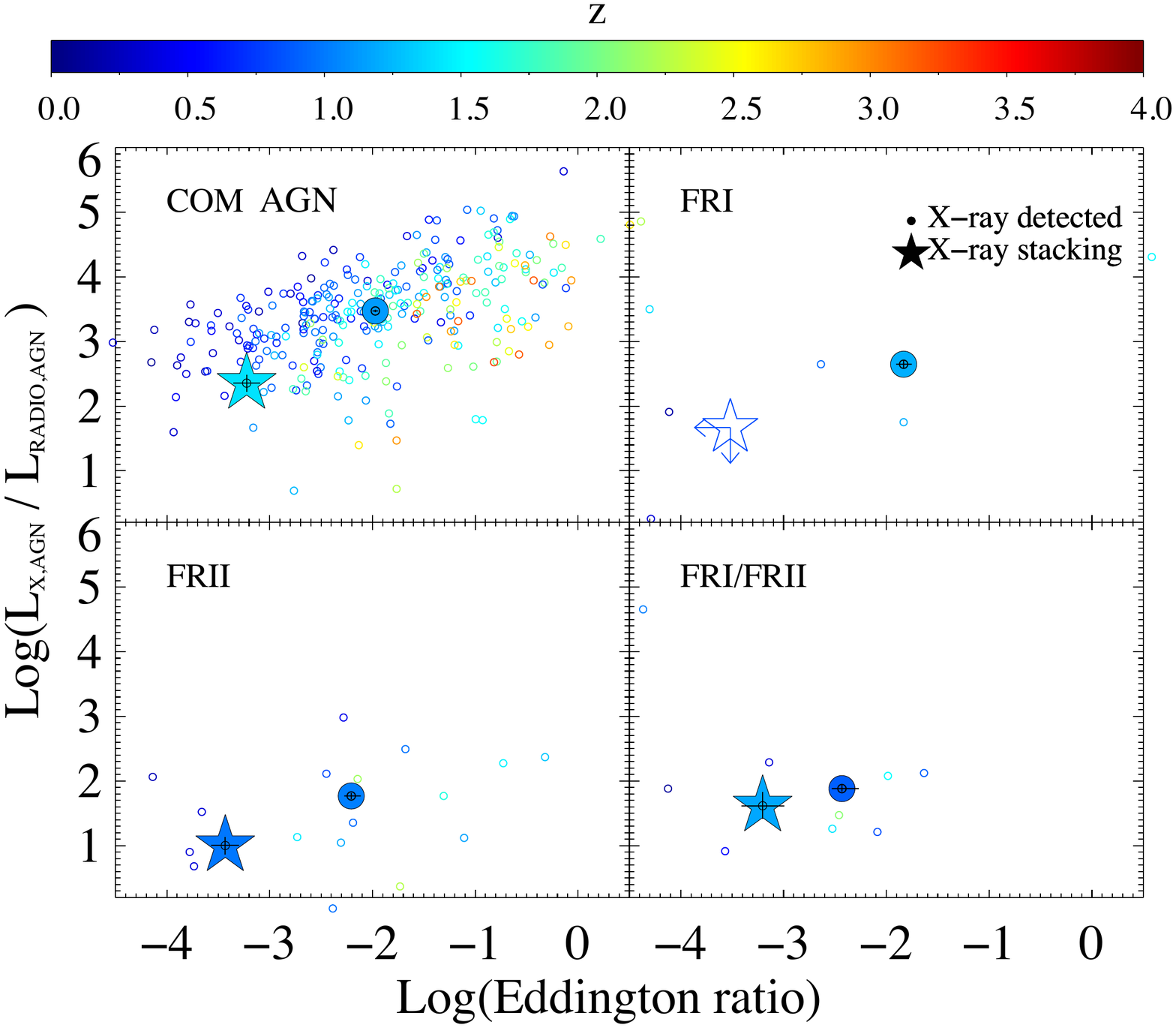}
          }

               \caption{The ratio of X-ray to radio luminosity at 1.4 GHz vs. radiative Eddington ratios for the FR and COM AGN objects. The {\bf top left} panel shows COM AGN, the {\bf top right} panel shows FRI, the {\bf bottom left} panel shows FRII, and the {\bf bottom right} panel shows FRI/FRII. Colours represent redshifts. Small circles are individual X-ray detections. Large circles show median values. Stars are the stacked values as listed in Table~\ref{tab:eddrat_rad}. The standard deviation is also plotted on the median and stacked values. Upper limits are given at 90\% level.   
   }
              \label{fig:classes_eddratio}%
    \end{figure}

\begin{figure}[!ht]
   \centering

  \resizebox{\hsize}{!}
            {
            
            \includegraphics[trim={2.5cm 0cm 2cm 0cm},clip]{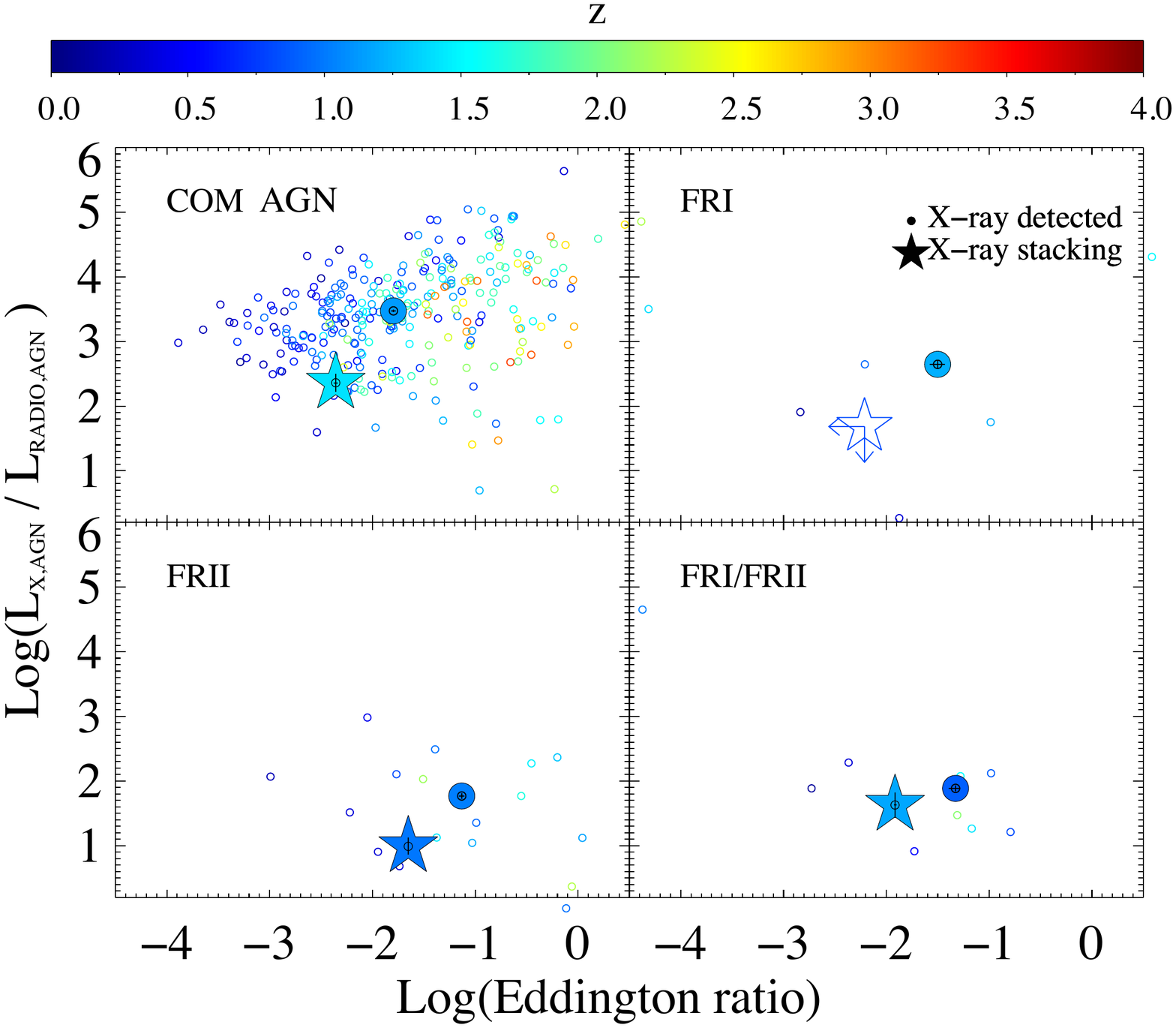}

      }

               \caption{The ratio of X-ray to radio luminosity at 1.4 GHz vs. radiative plus kinetic Eddington ratios for the FR and COM AGN objects. The {\bf top left} panel shows COM AGN, the {\bf top right} panel shows FRI, the {\bf bottom left} panel shows FRII, and the {\bf bottom right} panel shows FRI/FRII. Colours represent redshifts. Small circles are individual X-ray detections. Large circles show median values. Stars are the stacked values as listed in Table~\ref{tab:eddrat_qjet}. The standard deviation is also plotted on the median and stacked values. Upper limits are given at 90\% level.   
   }
              \label{fig:classes_eddratio_qjet}%
    \end{figure}

In Fig.~\ref{fig:classes_eddratio} we present the radiative Eddington ratio $\lambda_{\rm r}$ and in Fig.~\ref{fig:classes_eddratio_qjet} we present the radiative plus kinetic Eddington ratio $\lambda_{\rm rk}$ for the FRs and COM AGN with respect to the ratio of X-ray to radio luminosity.  
It is obvious that the addition of jet power boosts the Eddington ratio of FR sources, but only slightly increases the Eddington ratio of COM AGN.

The $L_{\rm}/L_{\rm radio}$ is higher for COM AGN because the average $L_{\rm radio}$ is lower than for the other classes (Fig.~\ref{fig:n_z}), even though the mean redshift is even higher. We plot this luminosity ratio in respect to the Eddington ratio in order compare how fast BHs are accreting relative to their mass against the predominant type of AGN feedback, radiative versus mechanical given by the $L_{\rm}/L_{\rm radio}$. In principle, we obtain information about which in form (mainly radiative or mainly mechanical) the feedback of the AGN is predominantly exerted as a function of BH accretion rate. Ideally, we should calculate the Eddington ratios from an independent tracer, not the X-rays, for instance, optical spectroscopy and the MgII emission line \citep[e.g.][]{mclure02}. The latter is not possible because optical spectra are available only for a small subset (A. Scultze priv. comm.). The target of this analysis is to compare different populations to each other, that is, the relative behaviour of FR classes and COM AGN, without overinterpreting the relation between $L_{\rm}/L_{\rm radio}$ and the Eddington ratio itself.

The stacked Eddington ratios of FRs also show a boost in their values with the addition of kinetic energy. This time, also COM AGN show a boost in the stacked $\lambda_{\rm rk}$ values. This is different from the measured values and in particular for COM AGN. The stacks allow us to reveal the fainter X-ray population that is not probed by the flux-limited X-ray data, and to lower the $L_{\rm X}/L_{\rm rad}$ ratio. The Eddington ratio is no longer dominated by the brightest X-ray sources, therefore the jet power has a stronger effect on the stacked Eddington ratio. Additionally, there appears to be a slight dichotomy in the stacked $\lambda_{\rm rk}$ values between FRIs and FRIIs, with FRIs having lower stacked Eddington ratios than FRIIs. Nevertheless, the FRI stacked values are upper limits. The boost given to the stacked $\lambda_{\rm rk}$ values is due to the radio luminosity, which is used to calculated the kinetic energy, and depends on redshift. The slight difference in the values of FRI and FRII objects can be attributed to the differences in radio luminosity between the populations, but it also depends on their redshift.

%

\begin{figure}[!ht]
    \resizebox{\hsize}{!}
            {\includegraphics[trim={0.5cm 0cm 0.5cm 0cm},clip]{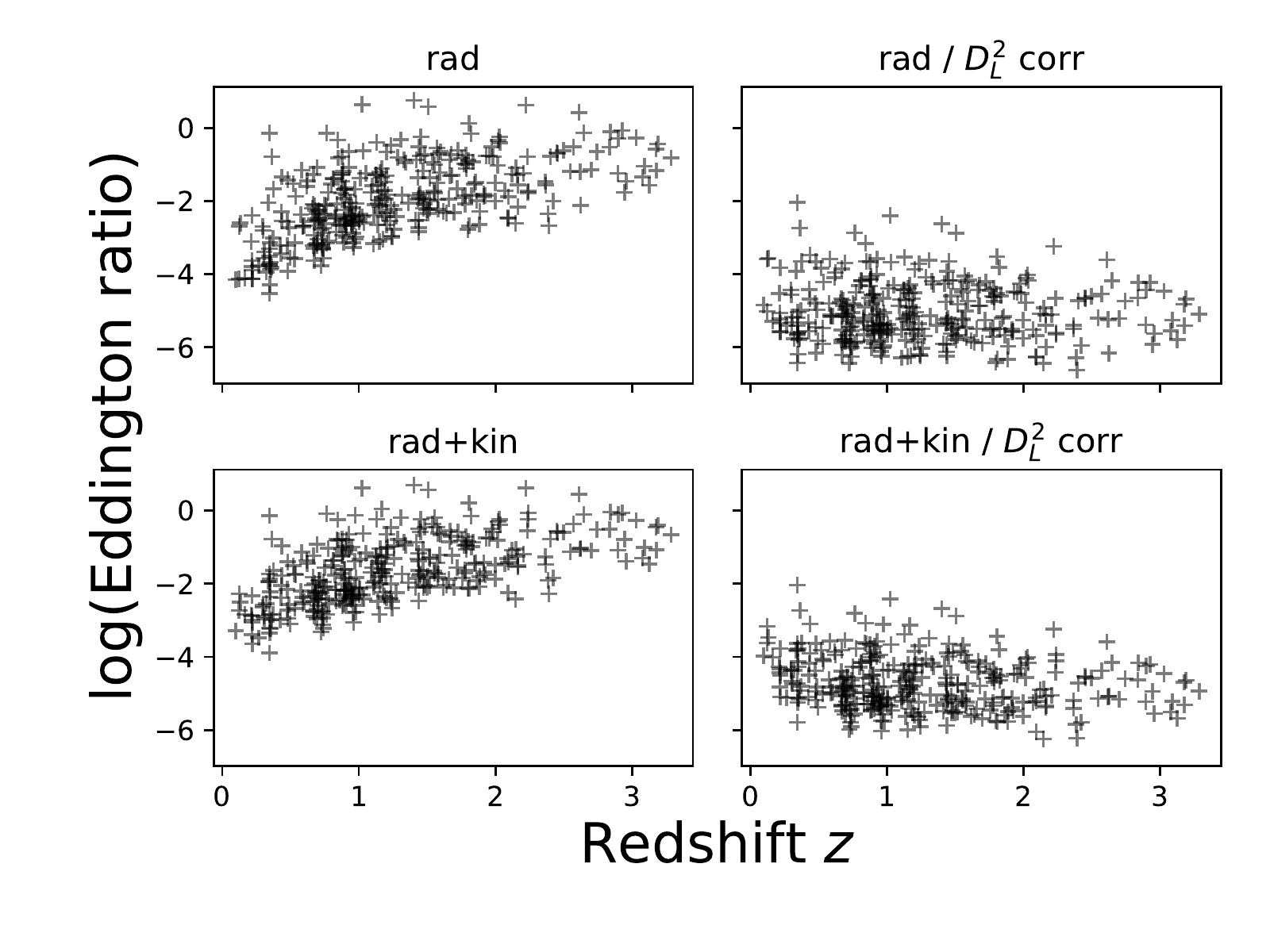}  
            }
               \caption{{\bf Left}: Redshift dependence of Eddington ratios $\lambda$ calculated from the X-rays. {\bf Right}: Eddington ratios corrected for redshift dependence by dividing with $D_{\rm L}^{2}$. The {\bf top} panels show the radiative Eddington ratios, and the {\bf bottom} panels present the radiative including kinetic energy contribution, as described in Sec.~\ref{sec:edd_ratios}.\\
   }
              \label{fig:eddrat_z}%
    \end{figure}

A redshift dependence of the Eddington ratio could affect the comparison of the FR populations to each other. This dependence enters the calculations through the X-ray data, which are flux limited. Another dependence on the Eddington ratio might be introduced by the stellar mass of the host used in the calculations. We investigated the relation between Eddington ratio and redshift and found an increase oin $\lambda$ with higher redshift (see Fig.~\ref{fig:eddrat_z}). We also investigated the relation between $M_{*}$ and redshift and found no dependence. We therefore assume that the relationship between $L_{\rm X}$ and redshift $z$ is the main driver for the increase of the Eddington ratio with redshift. To account\footnote{An alternative approach, in the case of large samples, would be to compare objects in the same redshift bin \citep[e.g.][]{fernandes15}.} for this dependence on redshift, we corrected the Eddington ratios by dividing with $D_{\rm L}^{2}$. For the purpose of comparing the FR and COM AGN populations to each other we will use the $D_{\rm L}^{2}$ corrected Eddington ratios throughout the paper, unless otherwise stated. We use $\lambda^{*}$ throughout for the $D_{\rm L}^{2}$ corrected Eddington ratio and $\lambda$ for the uncorrected. We caution not to take the $D_{\rm L}^{2}$ corrected Eddington ratios as absolute values, but rather as a means of comparison between the populations presented here. In Tables~\ref{tab:eddrat_rad}~and~\ref{tab:eddrat_qjet} we give the Eddington ratio values for the radiative and radiative plus kinetic calculations, respectively, before we applied the $D_{\rm L}^{2}$ correction. The uncorrected Eddington ratios are discussed in Sec.~\ref{sec:discuss}, where we compare them to literature results from previous radio studies. The Eddington ratio values for individual objects are presented in Fig.~\ref{tab:edd_qjet} in the Appendix. Finally, in Fig.~\ref{fig:hist_edd_ratio} we plot the histogram of Eddington ratios of X-ray detected sources split into different radio classes.

\begin{figure}[!ht]
  \resizebox{\hsize}{!}
 {\includegraphics[trim={1cm 0.8cm 0.5cm 0cm},clip]{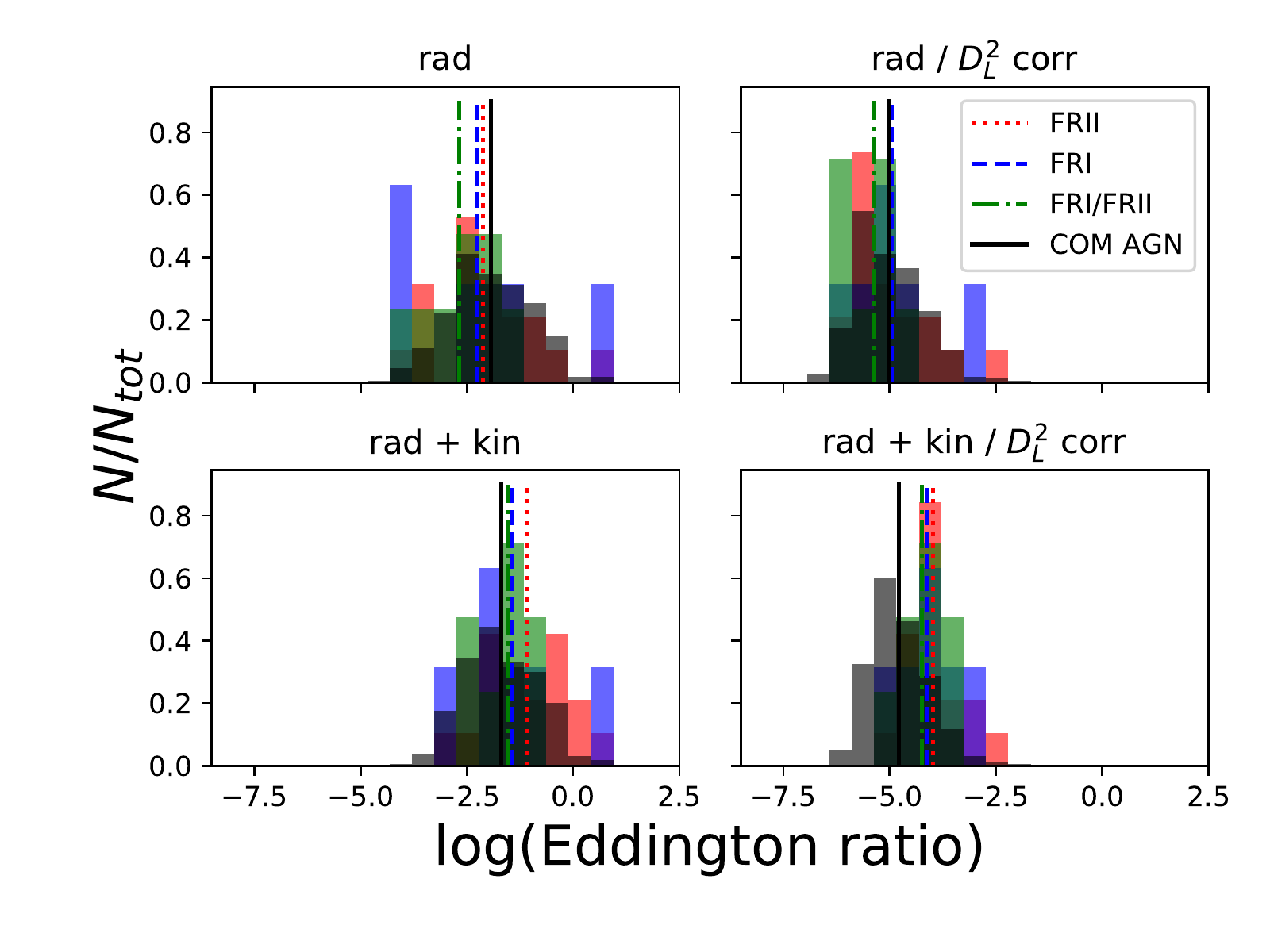}
            }
          
       \caption{Histogram of Eddington ratios of X-ray detected sources, colour-coded based on our classification scheme in FRII (red), FRI (blue), FRI/FRII (green) and COM AGN (black). The {\bf top} panels show the radiative Eddington ratio ({\bf left}) and the redshift-dependence corrected Eddington ratio ({\bf right}). The {\bf bottom} panels show the radiative and kinetic Eddington ratio ({\bf left}) and the redshift-dependence corrected Eddington ratio ({\bf right}). The median values of the Eddington ratio for each population are shown as lines: dotted red lines show FRIIs, dashed blue show FRIs, dashed-dotted green show FRI/FRIIs, and solid black lines show COM AGN.\\
   }
              \label{fig:hist_edd_ratio}%
    \end{figure}
\begin{figure}[!ht]
  \resizebox{\hsize}{!}
 {\includegraphics[trim={1cm 0.8cm 0.5cm 0cm},clip]{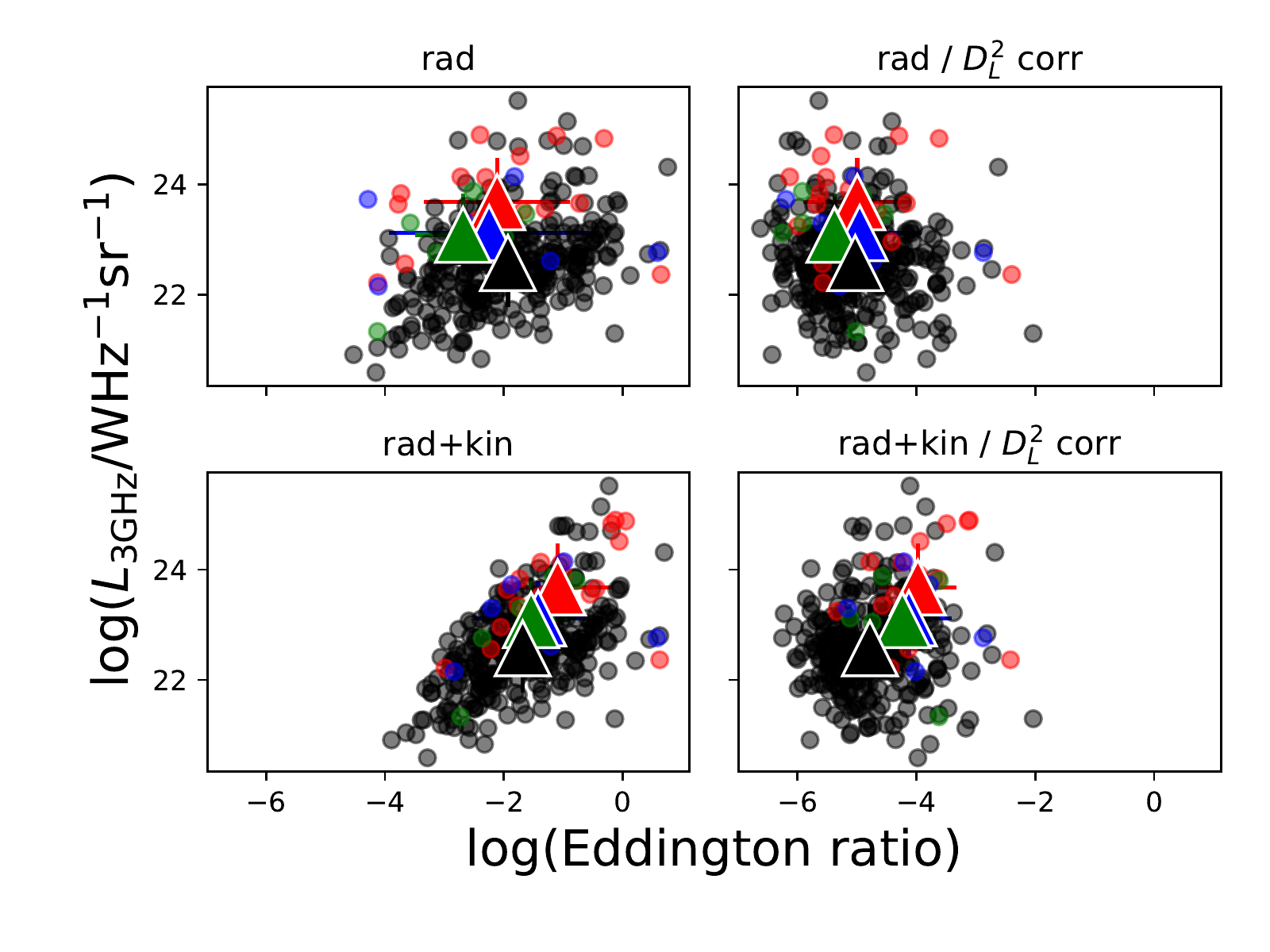}
            }
          
       \caption{Radio power at 3 GHz vs. Eddington ratio, radiative ({\bf top} panels) and kinetic ({\bf bottom} panels), for $D_{L}^{2}$ uncorrected ({\bf left} panels) and corrected values ({\bf right} panels).  FRII are shown in red, are shown in FRI blue, FRI/FRII are shown in green, and COM AGN are shown in black. Mean values and standard deviations are shown as large triangles, colour-coded based on the radio class. 
   }
              \label{fig:l3_eddrat}%
    \end{figure}

In Fig.~\ref{fig:l3_eddrat} we plot the radio luminosity at 3 GHz versus the Eddington ratio for corrected and uncorrected values. For the radiative case, FRIs, FRIIs, and FRI/FRIIs on average have similar $\lambda^{*}_{\rm r}$ values and a similar distribution of Eddington ratios. When the contribution from kinetic energy is added to the Eddington ratio, the mean values of $\lambda^{*}_{\rm rk}$ increase for all FR objects, and as a result of the difference in radio luminosity, there is an offset between the distributions. Still, no statistically significant dichotomy is found for the FR population when their Eddington ratios are considered (with and without the jet power).

For the COM AGN we note that the average $\lambda^{*}_{\rm r}$ and $\lambda^{*}_{\rm rk}$ values are similar and the kinetic energy does not contribute significantly to the Eddington ratio in this class of objects, because the radiative luminosity from the X-rays is the dominant contributor to the Eddington ratio. On average, the COM AGN are much brighter at X-rays than the FR-type objects (Figs~\ref{fig:classes_eddratio}~and~\ref{fig:classes_eddratio_qjet}), but the spread in the $L_{\rm X}/L_{\rm rad}$ ratio is large. Their $\lambda^{*}_{\rm r}$ Eddington ratios are on average similar to the FR objects within the error. The difference between FRs and COM AGN lies in the inclusion of jet power: FRs increase by  their jet power.

\subsection{Environmental probes}
\label{sec:env}

To address the large-scale environment of the objects in our sample, we used several environmental probes from kpc to Mpc scales. Below we describe the analysis and results in detail. We related the hosts (Sec.~\ref{sec:env_host}), X-ray groups (Sec.~\ref{sec:kpc_env}), and large-scale environment (Sec.~\ref{sec:densfields}) to the radio structure of the FRs and COM AGN in our sample.

\subsubsection{Host galaxies}
\label{sec:env_host}

\begin{figure*}[!ht]
    \resizebox{\hsize}{!}
            {\includegraphics[trim=0.2cm 0cm 1cm 0cm,clip=true]{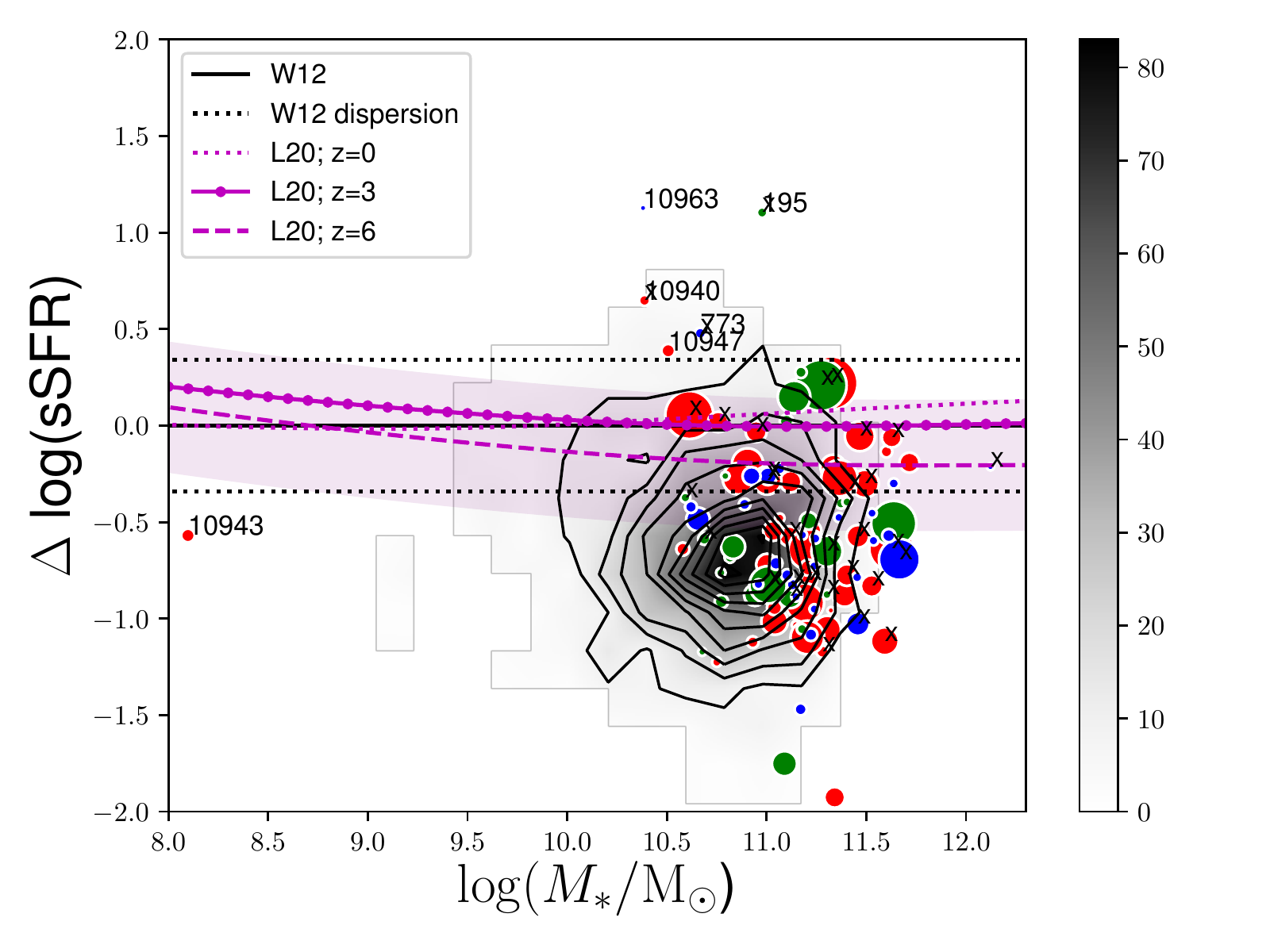}
            \includegraphics[trim=0.2cm 0cm 1cm 0cm,clip=true]{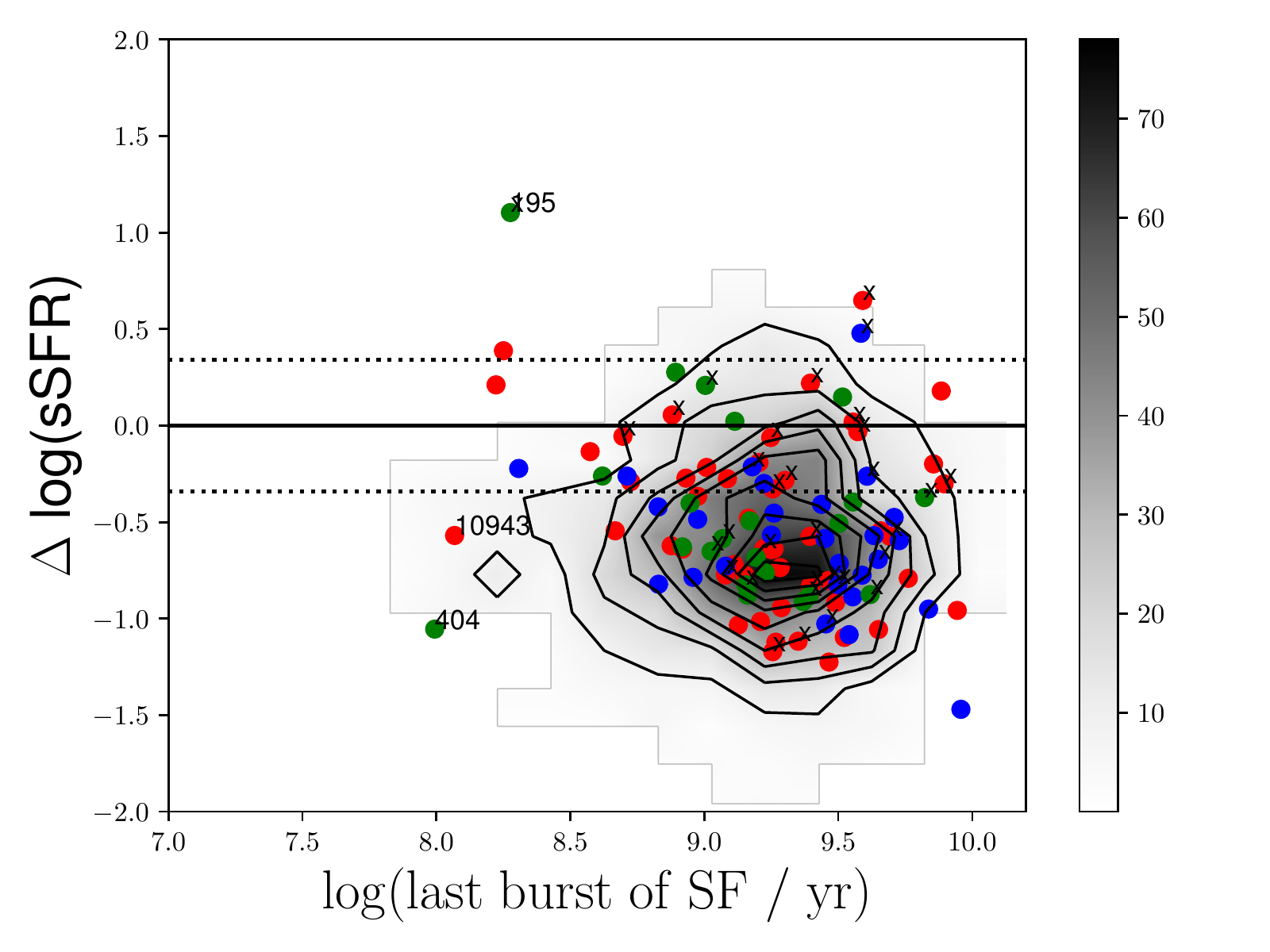}
 }
               \caption{{\bf Left}: $\Delta$sSFR, the difference between specific SFR and the specific SFR of objects in the MS, as a function of stellar mass for the four populations of radio AGN presented in this paper. For FR objects, the symbols are scaled based on their linear projected size, with larger symbols corresponding to larger objects. The jet-less COM AGN are shown in black as a density plot. The solid and dotted lines show the main sequence for star-forming galaxies and spread, based on \cite{whitaker12}. We also show in purple the \cite{leslie20} SFR-M* relation relative to the \cite{whitaker12} MS relation for $z$ = 0 (dotted purple line), $z$ = 3 (dotted-dashed purple line) and $z$ = 6 (dashed purple line) along with the dispersion. The diversion from the \cite{whitaker12} MS is insignificant for massive galaxies ($10^{10.5-11.5}$). {\bf Right}: $\Delta$sSFR, as on the left, vs. the star-formation history (SFH) of each object. SFHs are estimated from a fit to the SED as described in Sec.~\ref{sec:otherdata}. Median values are shown in Table~\ref{table:dssfr_mstar_sfh}. In both plots an x marks objects that have an X-ray detection based on the catalogue of \cite{marchesi16}.\\
   }
              \label{fig:ssfr_mstar}%
    \end{figure*}

In order to study what types of hosts the FRs and COM AGN of our sample inhabit, we plot the $\Delta$sSFR-$M_{*}$ diagram in Fig.~\ref{fig:ssfr_mstar}--Left. This shows the difference between the specific SFR (sSFR) of each object (SFR/$M_{*}$) and the specific SFR it would have at the main sequence (MS) for star-forming galaxies given its redshift (sSFR$_{\rm MS}$) versus the stellar mass; or the so-called "main-sequence offset $\Delta$(MS)". We also plot the main sequence for star forming galaxies \citep[see][]{whitaker12} as a solid black line. We chose the \cite{whitaker12} relation for this study to follow the study of \cite{delvecchio17}. More recent MS prescriptions include a bend at high $M_{*}$ \citep[e.g.][]{Speagle14, Schreiber15, lee15, scoville17, leslie20}, which would place the sources in our sample slightly closer to the MS, but still systematically below it. At $z~>~1$ we would have a negligible fraction of fully quiescent hosts \citep{Davidzon17}, because the $M_{*}$ function drops exponentially. To test whether our choice of MS biases our results, we overplot the \cite{leslie20} relation of SFR and $M_{*}$ in Fig.~\ref{fig:ssfr_mstar}--Left. The difference between the \cite{whitaker12} and \cite{leslie20} main sequences is insignificant up to $z$ = 3 and within the dispersion at $z$ = 6, in particular for massive galaxies ($\gtrsim 10^{11} M_{\odot}$). We therefore are confident that our choice of using the \cite{whitaker12} relation does not bias our results. 

Fig.~\ref{fig:ssfr_mstar}--Left shows that the majority of FR objects lie below the MS in the green valley and the red-and-dead region of the diagram. In Table~\ref{table:hostprop} we also present the optical morphology of these hosts from \cite{schinnerer10}. Not all FR objects in our sample reside in elliptical hosts. When the optical classification is applied, we find three elliptical hosts within the MS for star-forming galaxies (SFGs), and at the same time, 20 disk galaxies lie below the MS. Furthermore, we marked objects with their names for five cases that do not follow the general trend of the FR population. Objects 195, 773, 10940, 10947, and 10963 have hosts in the starburst (SB) region of the $\Delta$sSFR-$M_{*}$ diagram, above the MS for SFGs. Additionally, 10943 is an outlier at the low-mass end ($<10^{10.5}$ $M_{\odot}$) of the diagram. We visually inspected these outliers to avoid misidentifications. These objects have small and unusual shapes, but they either exhibit radio excess (195, 773, 10940, 10943, and 10947; Table~\ref{table:data}) and/or are classified as AGN based on their SED fit (773 and 10940; see Table~\ref{table:hostprop}). Object 10963 does not exhibit radio excess or an AGN SED, so that we remain cautious. We conclude that outliers of this type can exist in samples of radio AGN and may represent an early evolutionary stage, co-existence of starburst and AGN, or a starburst on its way to quenching.

Fig.~\ref{fig:ssfr_mstar} thus indicates that there is no observed dichotomy in the host properties of FR galaxies, rather we find a mix of distributions. Fig.~\ref{fig:ssfr_mstar} shows that the general trend of the FR populations suggests that objects move from the MS to the quiescent region through the green valley. Most of the COM AGN population follows the FR population trend in the $\Delta$sSFR-$M_{*}$ diagram, but with a higher fraction of sources at the SB and low stellar mass regions ($<10^{10.5}$ $M_{\odot}$). In Table~\ref{table:dssfr_mstar_sfh} we present the median properties for $\Delta$sSFR and $M_{*}$ for FRs and COMS AGN, as well as the results of a linear regression model fitted to the data. We find an anti-correlation between $\Delta$sSFR and $M_{*}$ for both FRs and COM AGN, indicating quenching of star formation. 

\begin{table*}[!ht]
\caption{Median and fitted values for $\Delta$sSFR-$M_{*}$ / SFH diagrams}             
\label{table:dssfr_mstar_sfh}      
\centering                          
\begin{tabular}{l   |  c c c  c | c    c  c  c  c c   }        
\hline\hline                 
   &  N & $< \Delta$sSFR $>$ &  $<M_{*}>$  & linear regression  & N & $< \Delta$sSFR $>$ & $<t_{\rm last ~burst}>$   \\    
 & & & & (intercept, slope) & & & \\
\hline
FRII & 52 & -0.57 (0.475) & 11.20 (0.532)  &                                     &  51 & -0.57 (0.438) & 9.25 (0.411)\\
FRI/FRII & 24 & -0.60 (0.564) & 11.09 (0.261) &                               &   22 & -0.54 (0.500) & 9.15 (0.413)\\
FRI & 29 & -0.58 (0.542) & 11.17 (0.360) &                                       & 27 & -0.58 (0.358) & 9.45 (0.377)\\
FR & 105 & -0.58 (0.516) & 11.17 (0.440) & 1.46, -0.18                   & 100 & -0.57 (0.437) & 9.25 (0.410) \\
COM AGN& 1800 & -0.60 (0.488) & 10.90 (0.558) & 0.27, -0.07      & 1738 &  -0.61 (0.481) & 9.24 (0.420) \\
\hline
\end{tabular}
\tablefoot{Median values derived for the FRs and COM AGN from Fig.~\ref{fig:ssfr_mstar}. Values in parenthesis are standard deviations. SFR, $M_{*}$ and $t_{\rm last ~burst}$ are estimated from the fit to the {\bf IR+UV} SED \citep{delvecchio17}. {\bf We also give values for the individual FR populations, for comparison.} Note: We do not provide the results of the linear regression between $\Delta$sSFR and $t_{\rm last ~burst}$ as the observed anti-correlation is artificially induced by the co-dependence between these quantities in the SED-fitting code.}
\end{table*}

Because FR objects display jets, we need to investigate the possibility that we witness radio-mode feedback from AGN in their hosts. For this purpose we plot the $\Delta$sSFR-SFH diagram shown in Fig.~\ref{fig:ssfr_mstar}--Right. We used the last burst of star formation to investigate the time of the last star-forming episode in each host. This quantity was estimated from the fit to the SED, as SFR and $M_{*}$ were \citep{delvecchio17}. We find that FR objects with more recent bursts of SF to still occupy the MS, while objects with later $t_{\rm last ~burst}$ are located below the MS for SFGs. This tells us that weaker star-forming galaxies have longer $t_{\rm last ~burst}$, which is also seen in the COM AGN sample. We note that this anti-correlation is expected from the co-dependance of sSFR and $t_{\rm last ~burst}$ in the SED-fitting code. Outliers with recent bursts below the MS  exist (e.g. 404 and 10943). We deduce from these plots that FR objects have quenched hosts because they systematically lie below the MS. We discuss this further in Sec.~\ref{sec:discuss}.

%

\subsubsection{Galaxy group environment: X-ray groups}
\label{sec:kpc_env}

\begin{figure}[!ht]
  \resizebox{\hsize}{!}
 {\includegraphics[angle=0]{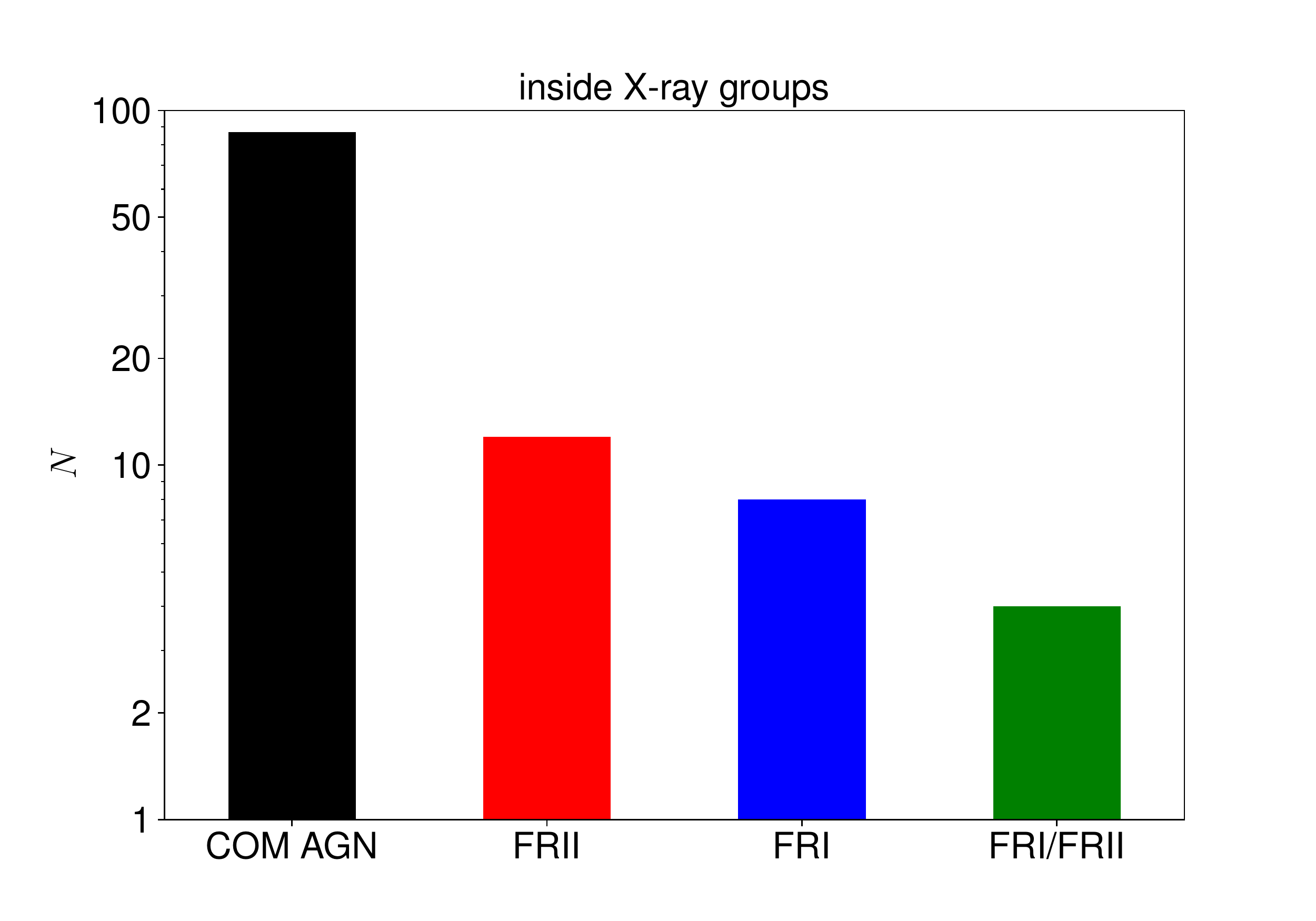}
}
  \resizebox{\hsize}{!}
 {\includegraphics[angle=0]{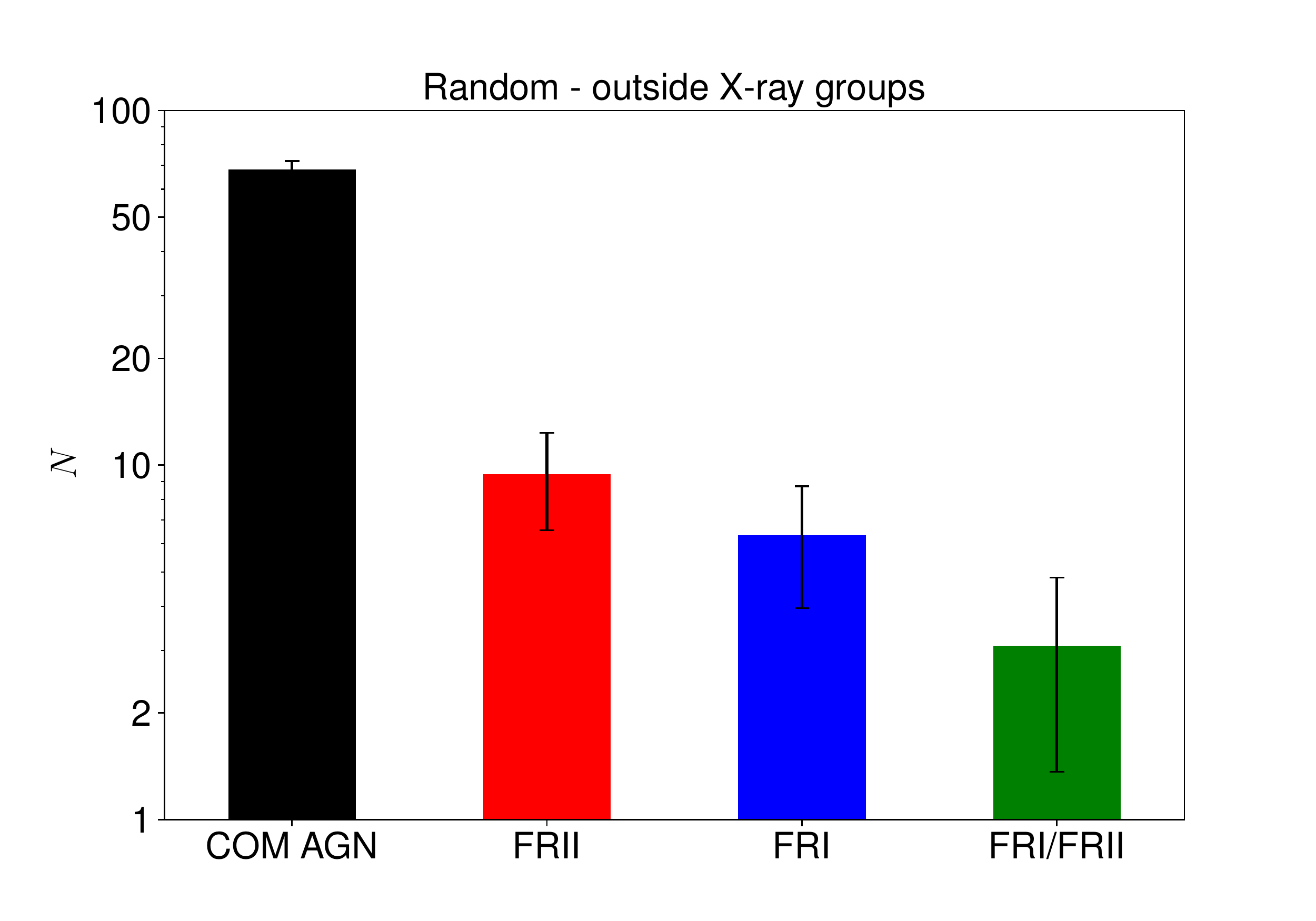}
            }
          
       \caption{Bar histogram of objects inside ({\bf top}) and outside ({\bf bottom}) the COSMOS X-ray groups from \cite{gozaliasl19} for 0.08 $\leq z <$ 1.53. The objects outside X-ray groups were selected by running a random number generator 1000 times with the same number of objects as the ones inside groups, and taking the mean and standard deviation. Red colour represents FRIIs, blue represents FRIs, green shows FRI/FRIIs, and black denotes COM AGN. The number of objects inside X-ray groups: 87 COM AGN, 12 FRII, 4 FRI/FRII, and 8 FRI. Outside the X-ray groups, the numbers are  68$\pm$3.9 COM AGN, 9$\pm$2.9 FRII, 3$\pm$1.7 FRI/FRII, and 6$\pm$2.3 FRI.\\
   }
   
              \label{fig:xrayinout}%
    \end{figure}

To probe whether our galaxies preferentially lie within galaxy groups, we cross-correlated their positions with the X-ray group catalogue of \cite{gozaliasl19}. Our goal was to investigate whether FR-type radio sources prefer one environment type over another. For example, do they tend to reside within group environment or in the field? For this purpose we compared objects that lay within the X-ray groups in COSMOS \citep{gozaliasl19} to the ones that lay outside X-ray groups. This is shown in Fig.~\ref{fig:xrayinout} where we present histograms for the FR and COM AGN objects within X-ray groups and in the field, for redshift 0.08 $\leq z <$ 1.53. Given that we have more objects outside the X-ray groups than inside in this redshift range, we randomly selected an equal number of objects for those outside as those inside the X-ray groups, to have an unbiased comparison. We randomly drew objects from the "outside a group" sub-sample from the four classes presented and compared the histograms in Fig.~\ref{fig:xrayinout}. Our results show that there is no preference for being inside X-ray groups or in the field for the radio AGN presented here. These results seem to be in contrast to the study of \cite{smolcic11}, who find that radio AGN from 1.4 GHz VLA-COSMOS preferentially lie within group environments. We suspect that this difference could be related to the different sensitivity and resolution of the 3 and 1.4 GHz surveys, with the former revealing smaller and fainter radio sources (927 radio AGN at 3 GHz outside the X-ray groups compared to 111 inside, up to redshift $z$ = 1.53). Another reason might be the different photometric catalogues used and the different methods applied. There is no strong trend regarding the different FR types either. We find slightly more FRIIs within X-ray groups than FRIs, and there are far fewer hybrids. These values are similar to the numbers of objects expected outside X-ray groups.

The objects lying outside X-ray groups could also belong to a group that has not been identified by X-ray observations as yet. Higher resolution and sensitivity X-ray observations could reveal more low-mass groups thath are not currently included in the \cite{gozaliasl19} catalogue. \cite{vardoulaki19} showed that by using the radio structure of jetted AGN as a probe, and how disturbed it is (i.e. bending caused by interaction with the large-scale environment) we can identify locations of possible X-ray groups which have not been identified by the \cite{gozaliasl19} X-ray observations of COSMOS.

\begin{figure}[!ht]
  \resizebox{\hsize}{!}
 {\includegraphics{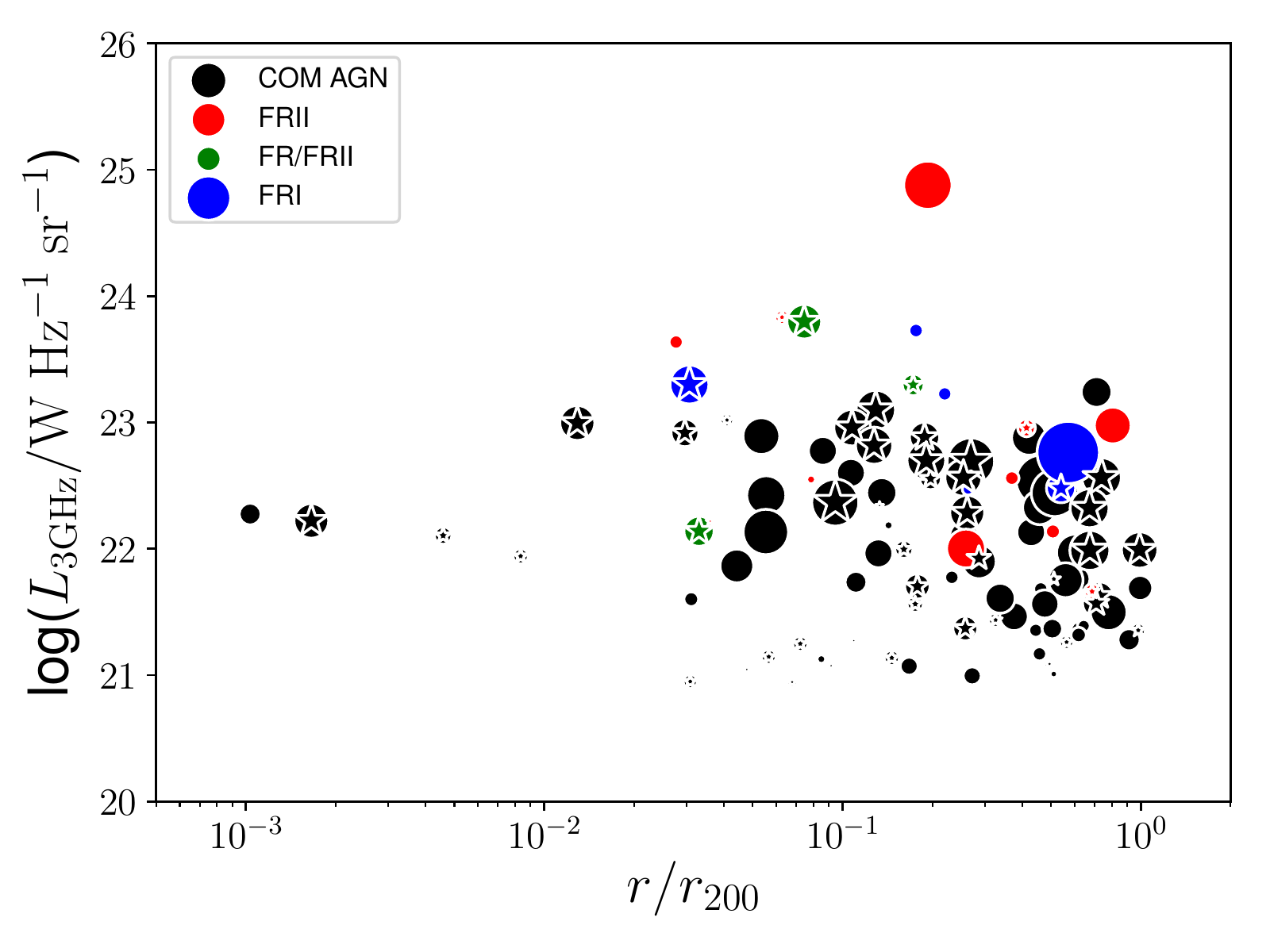}
            }

  \resizebox{\hsize}{!}
 {\includegraphics{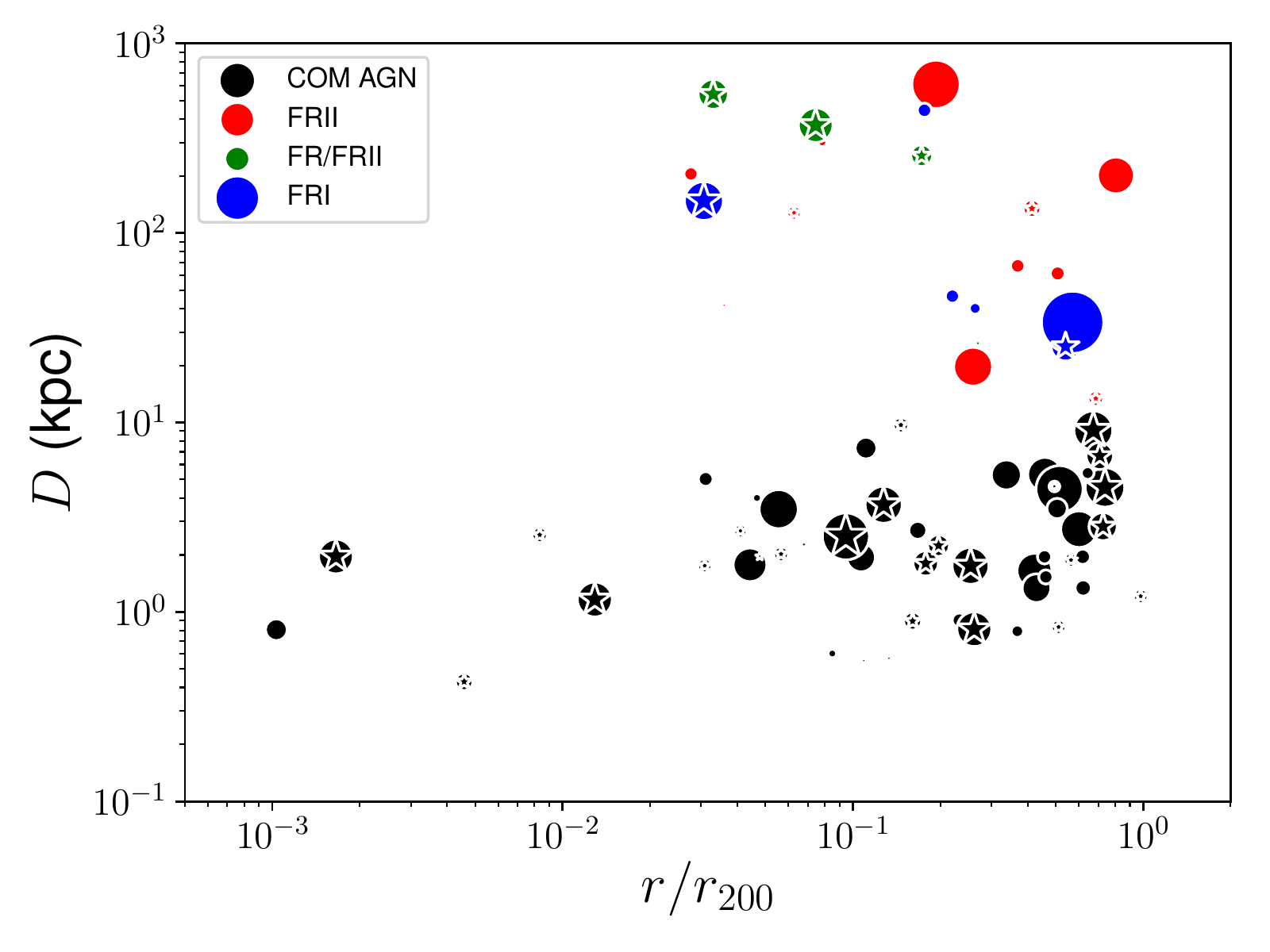}
            }
                      
       \caption{{\bf Top}: Radio luminosity at 3 GHz for the galaxies that lie within the X-ray groups in the COSMOS field \citep{gozaliasl19} vs. their distance from the group centre, normalised by the virial radius $r_{200}$, for 0.08 $\leq z <$ 1.53. 
       {\bf Bottom}: linear projected size $D$ as in Table~\ref{table:data} vs. distance from the group centre normalised to the virial radius for the same redshift bins as in the panel above. Red colour denotes FRIIs, blue is for FRIs, green denotes FRI/FRIIs, and black is for COM AGN. In both plots the symbol size is proportional to redshift (larger the symbol, larger the redshift). Stars highlight sources that are the brightest group galaxy.
   }
              \label{fig:xrayin}%
    \end{figure}

Additionally, we investigated whether FR-type objects have a preferred location in the X-ray groups they reside within and how does their location compared to jet-less AGN objects. This is shown in Fig.~\ref{fig:xrayin}, where we give the distance $r$ of the radio source from the group centre normalised to the virial radius $r_{200}$ of the X-ray group for redshifts 0.08 $\leq z <$ 1.53. We compare the $r/r_{200}$ ratio to the radio luminosity at 3 GHz and the linear projected size $D$. FR-type objects, independent of their type (FRII, FRI/FRII,  orFRI), can be found at any position within the virial radius of the X-ray group. This is also true for COM AGN. 

To investigate the preference in location within the X-ray groups for the FR and COM AGN populations, we first estimated the average number density of objects within the X-ray groups. For FRs this number is $<N_{\rm X}>$ = 1.55 $\pm$ 0.97, for COM AGN ,it is $<N_{\rm X}>$ = 2.06 $\pm$ 1.24, and for the whole X-ray group sample, it is $<N_{\rm X}>$ = 2.51 $\pm$ 1.37; the difference between the populations is not statistically significant. We then estimated their average distances from the X-ray group centres. FRs peak at $r/r_{200}$ = 0.26 $\pm$ 0.21 and COM AGN at $r/r_{200}$ = 0.31 $\pm$ 0.26. These results indicate that FRs and COM AGN reside close to the X-ray group centre on average. The latter is in line with the findings of \cite{smolcic11}, who used the 1.4 GHz VLA-COSMOS data. 

About half of the radio AGN in our sample, within X-ray groups, reside in brightest group galaxies (BGG), the most massive galaxy of the group. In particular $\sim$ 51\% (44 out of the 87) COM AGN and $\sim$ 46\% (11 out of 24) of FRs within X-ray groups are associated with a BGG. It has been shown by \cite{gozaliasl19} that BGGs are not always found in the centre of the X-ray group, suggesting the systems are not yet relaxed \citep{gozaliasl20}. The location of FRs within X-ray groups seems to be independent of physical properties such as radio luminosity and linear projected size of the source. Similarly for COM AGN, with the exception of a trend found between radio luminosity of COM AGN and distance from the X-ray group centre\footnote{We calculated the Pearson correlation coefficient between $L_{3~\rm~GHz}$ and $r/r_{200}$ in COM AGN and found an anti-correlation (slope = -0.19, P = 0.070).}, suggesting lower radio luminosities with increasing distance from the group centre. Otherwise, the brightest COM AGN tend to be closer to the X-ray group centre. Last, in Fig.~\ref{fig:xrayin} we also investigate the relation between redshift and $r/r_{200}$. Larger symbols represent objects further away. No trend becomes evident, however.

\begin{figure*}[!ht]
  \resizebox{\hsize}{!}
 {\includegraphics{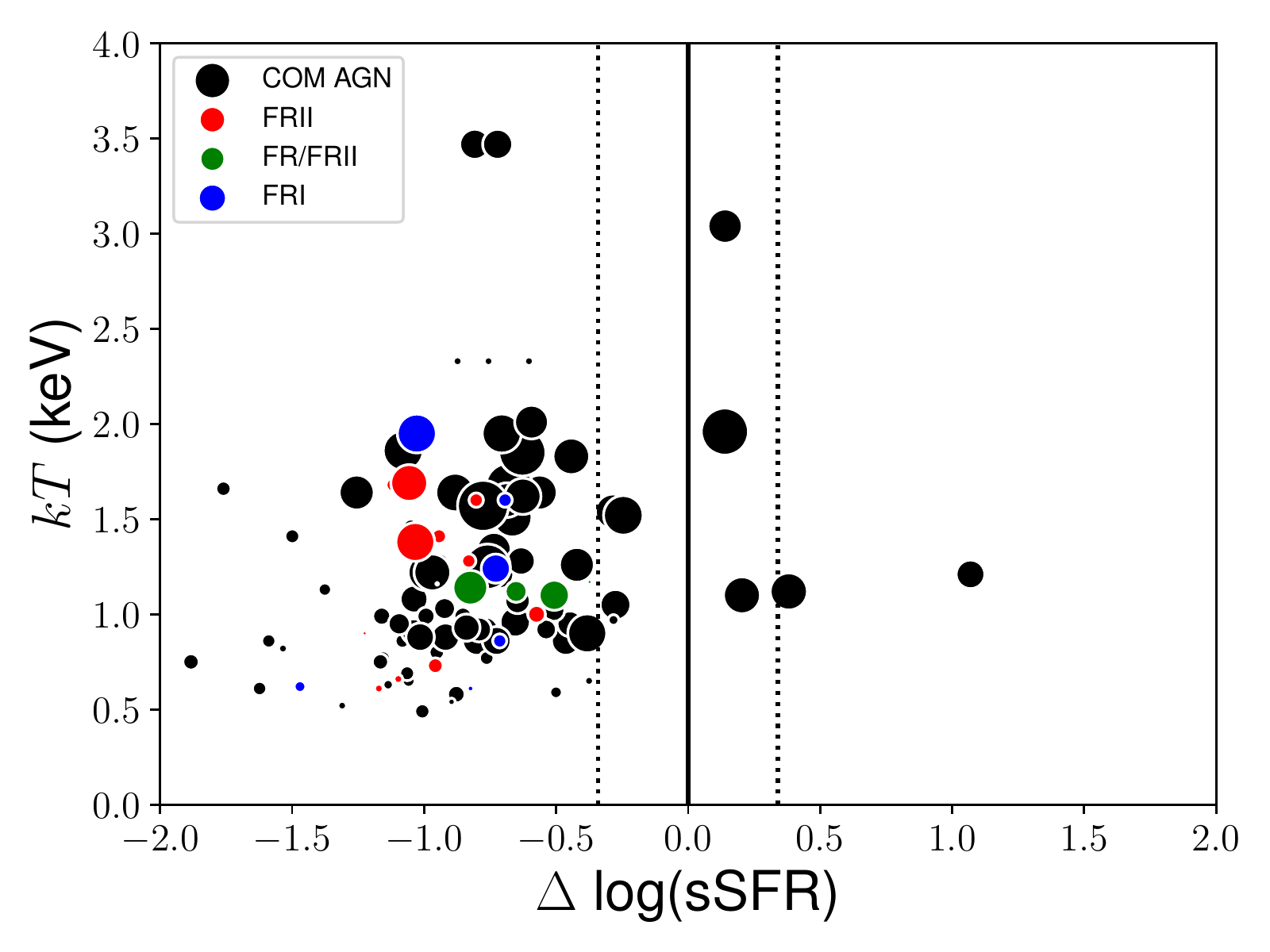}
           
 \includegraphics{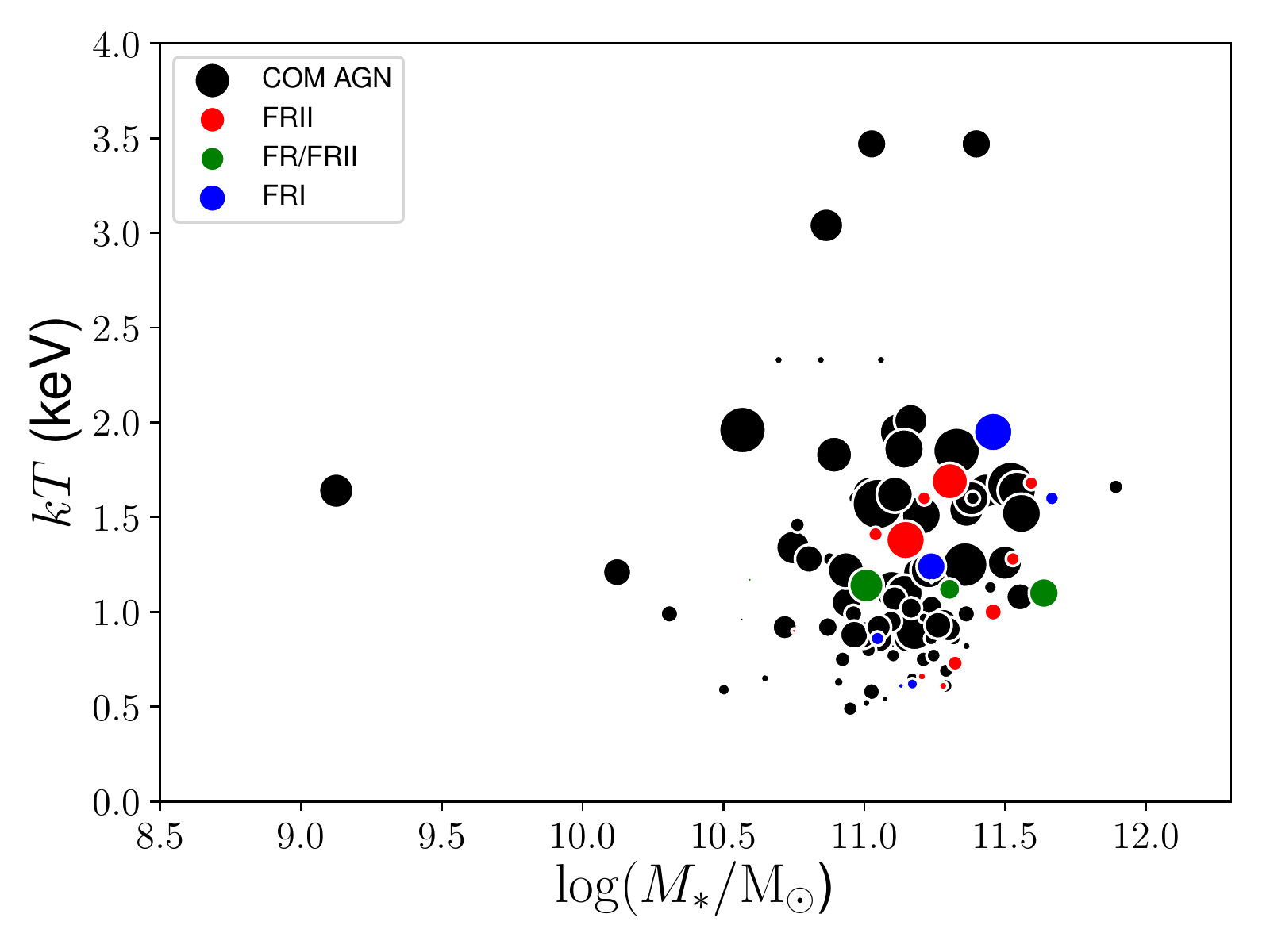}
            }
           
  \resizebox{\hsize}{!}
 {\includegraphics{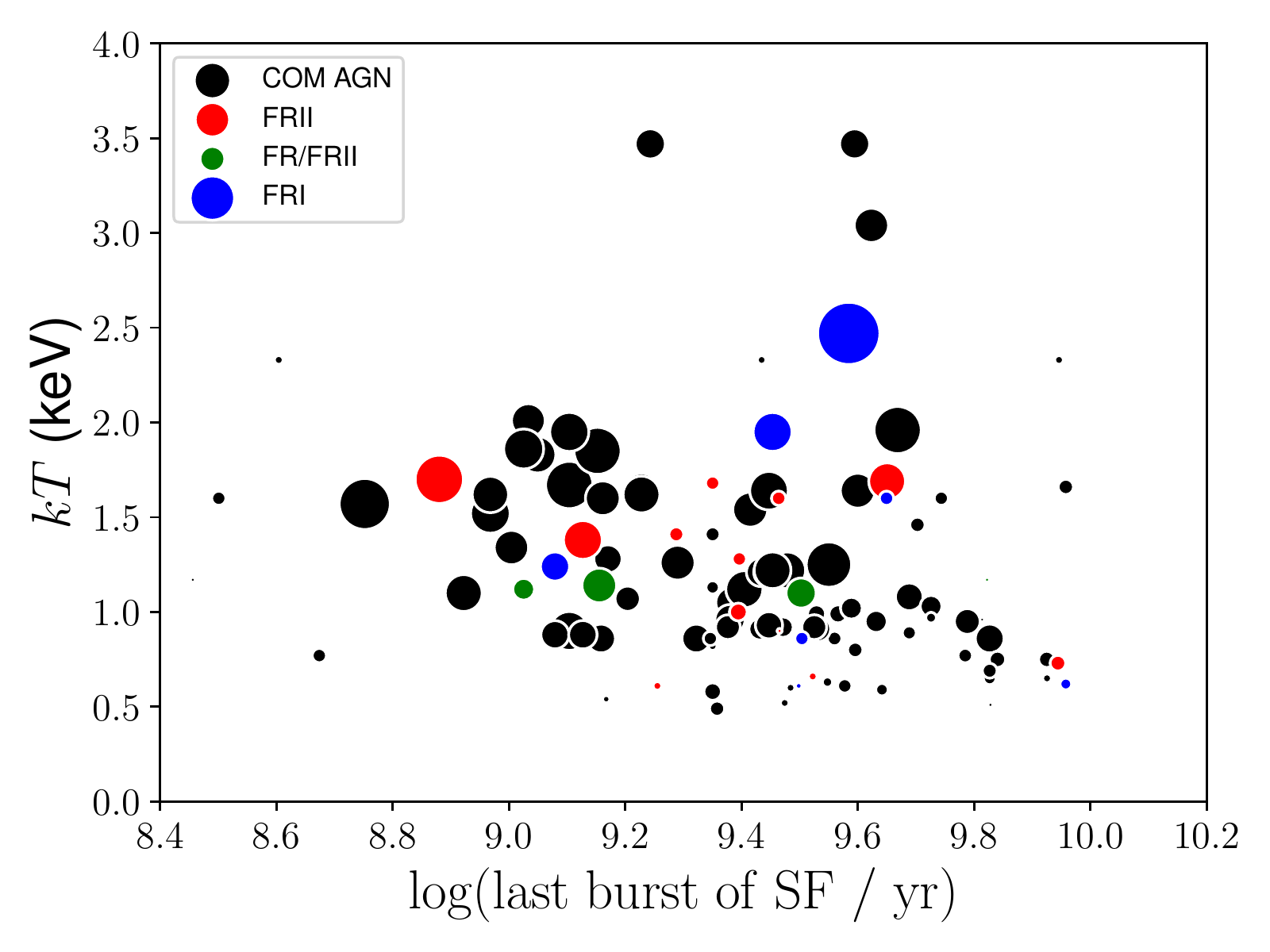}
            \includegraphics{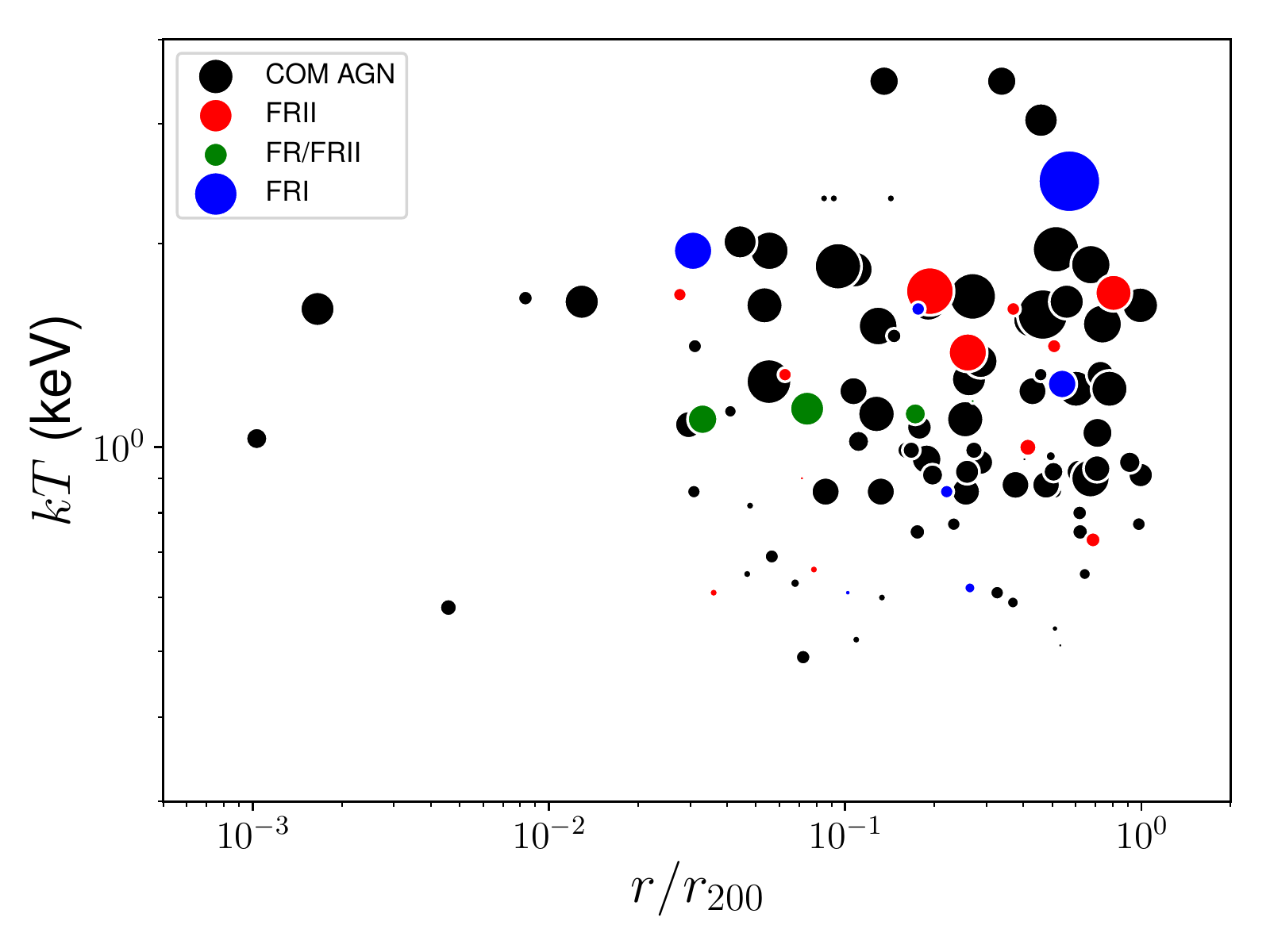}
            }        
            
       \caption{{\bf Top Left}: IGM temperature $kT$ of the X-ray group in keV vs. $\Delta$sSFR for the FRs and COM AGN that lie within the X-ray groups (0.08 $\leq z <$ 1.53) in the COSMOS field \citep{gozaliasl19}. The solid and dotted lines show the main sequence for star-forming galaxies and spread based on \cite{whitaker12}.
       {\bf Top Right}: IGM temperature $kT$ of the X-ray group in keV vs. $M_{*}$ of the FRs and COM AGN that lie within the X-ray groups. Colours: red for FRIIs, blue for FRIs, green for FRI/FRIIs and black for COM AGN. 
       {\bf Bottom Left}: IGM temperature in keV vs. last burst of star formation.  
       {\bf Bottom Right}: IGM temperature in keV vs. distance of FR or COM AGN from X-ray group centre, normalised to the virial radius $r_{200}$. Typical errors on the temperatures are 20\%. In all plots the symbol size is proportional to redshift (symbol size increases with redshift). Red symbols denote FRII, green symbols show FRI/FRII, blue symbols are for FRI, and black symbols denote COM AGN. The median IGM temperature corresponding to FRs and COM AGN is 1.16$\pm$0.46 and 1.04$\pm$0.59, respectively. \\
   }
              \label{fig:t_dssfr}%
    \end{figure*}

Finally, we investigated the hosts of FR and COM AGN within X-ray groups, plotting $\Delta$sSFR and $M_{*}$ versus the X-ray temperature of the group (Fig.~\ref{fig:t_dssfr}). Objects below the MS mainly lie in X-ray groups with temperatures between 0.5 and 2 keV, with some exceptions of COM AGN at higher temperatures. These FR objects lie in massive hosts, while there is a mild trend\footnote{We calculated the Pearson correlation coefficient between $kT$ and $z$ for FRs with $M_{*} > 10^{10.5}M_{\odot}$. We did not found a strong correlation (slope = 0.54, P = 0.009).} for quenched massive hosts at lower redshifts to be found in cooler X-ray groups than those at higher redshifts. Furthermore, the hosts with the oldest episode of SF are those in cooler X-ray groups, with some exceptions of COM AGN and FRIs. Both FRs and COM AGN moreover lie in similar IGM temperature X-ray groups on average, with median temperatures of 1.16$\pm$0.46 keV and 1.04$\pm$0.59 keV for FRs and COM AGN, respectively. There is no observed dichotomy in FRs regarding their X-ray group temperature. Finally, the location of the AGN within the X-ray group is not linked to the group temperature. \cite{dubois11} showed that the interaction between AGN energy released from jets and the ICM gas can result in the creation of cool-core clusters, assuming no metals are taken into account. Our results, showing a trend between quenched massive hosts and cooler X-ray groups might therefore support a scenario in which AGN radio-mode feedback is at play.

\subsubsection{Mpc-scale environments}
\label{sec:densfields}

To study the Mpc-scale environment we used 1) the density fields from \cite{scoville13} and 2) the large-scale environments from \cite{darvish15} and \cite{darvish17}, and we compared them to the radio structure of the FR and COM AGN objects in our sample. The density fields from \cite{scoville13} are given in redshift slices\footnote{Data are publicly available through at http://irsa.ipac.caltech.edu/data/COSMOS/ancillary/densities/} up to redshift of 3, thus any object in our sample above this redshift is excluded from the analysis. The total surface densities of galaxies per comoving Mpc$^{2}$, which were created using two techniques, adaptive smoothing and Voronoi tessellation, as described in \cite{scoville13}. The large-scale structure mapping made use of the $K$-band selected objects catalogue and photometric redshifts for COSMOS from the Ultra-VISTA survey, in addition to other COSMOS photometry \citep[see][]{Ilbert13}. \cite{darvish15} and \cite{darvish17} expanded the study of \cite{scoville13} to include specific types of environment, such as clusters, filaments, and the field, and also provided information on whether the galaxy is a central galaxy, or a satellite, or is isolated. 

\begin{figure}[!ht]
  \resizebox{\hsize}{!} 
 {\includegraphics[trim=0cm 0cm 0.8cm 0cm,clip=true]{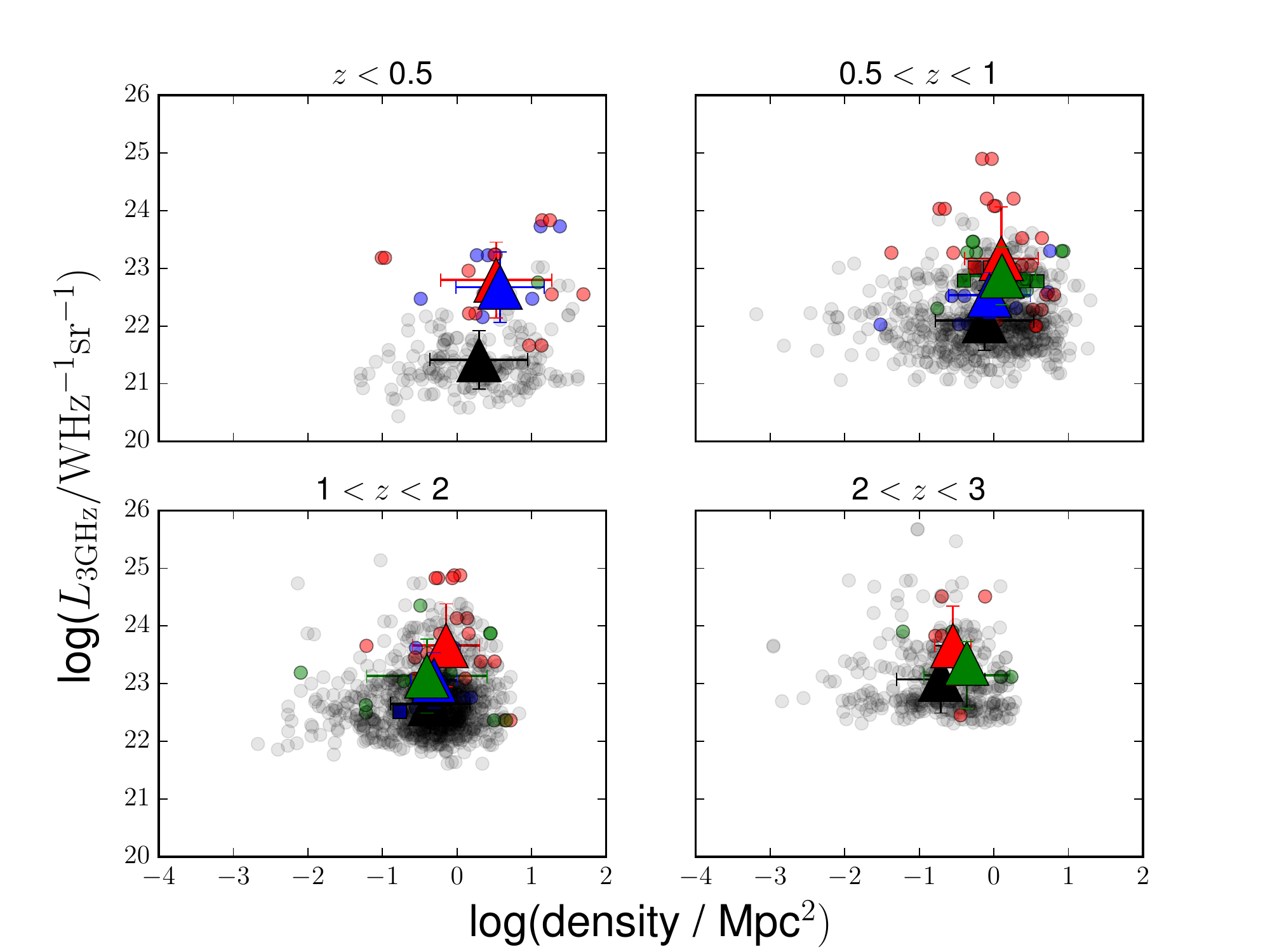}
            }
          
       \caption{Radio luminosity at 3 GHz vs. number density per Mpc$^{2}$ of galaxies in COSMOS, in four redshift bins. The density fields are from \cite{scoville13}. Symbols are the same as in Fig.~\ref{fig:L_D}. Circles are objects with radio excess, and squares are objects without radio excess \citep{delvecchio17}. The mean values are given by large triangles, along with their standard deviations for the FRIIs (red), FRIs (blue), FRI/FRIIs (green), and COM AGN (black). The bins are selected for a comparison with the literature.\\
   }
              \label{fig:dens_z}%
    \end{figure}

In Fig.~\ref{fig:dens_z} we present the density per Mpc$^{2}$ for the objects in our sample after cross-correlating with the density fields from \cite{scoville13}. There is a large scatter in the environments that COM AGN inhabit that is also seen for the FR-type objects. On average FRIs, FRIIs, FRI/FRIIs and COM AGN occupy similar density environments, as can be seen by their distributions. This suggests no preference in the environment between FR-type objects and no dichotomy in FR sources, which we discuss further in Sec.~\ref{sec:discuss}.  
Similar results are obtained when we use the over-densities from \cite{scoville13} instead of the number density/Mpc$^{2}$ of galaxies in the field.

\begin{figure}[!ht]
  \resizebox{\hsize}{!}
 {\includegraphics{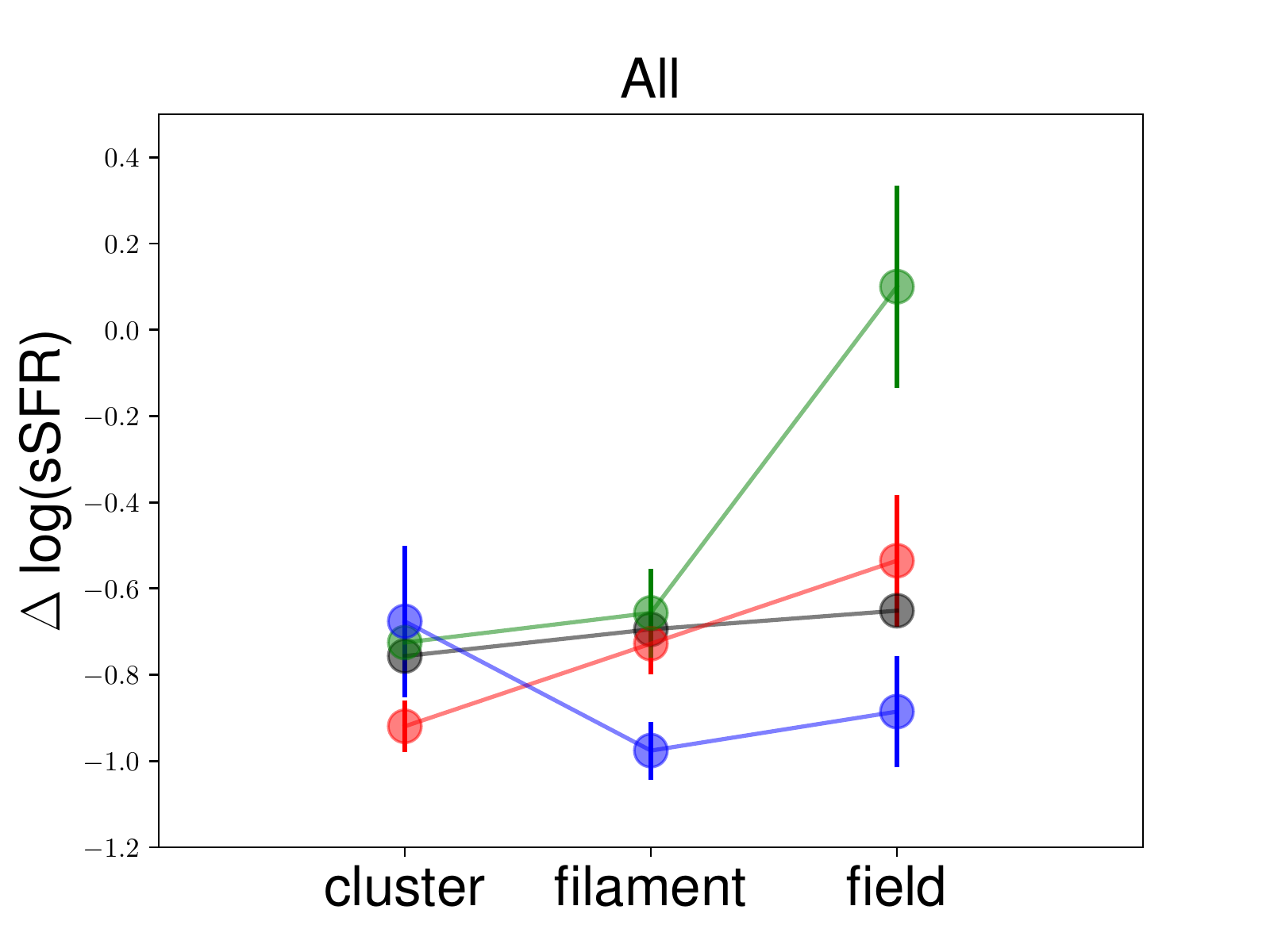}
 \includegraphics{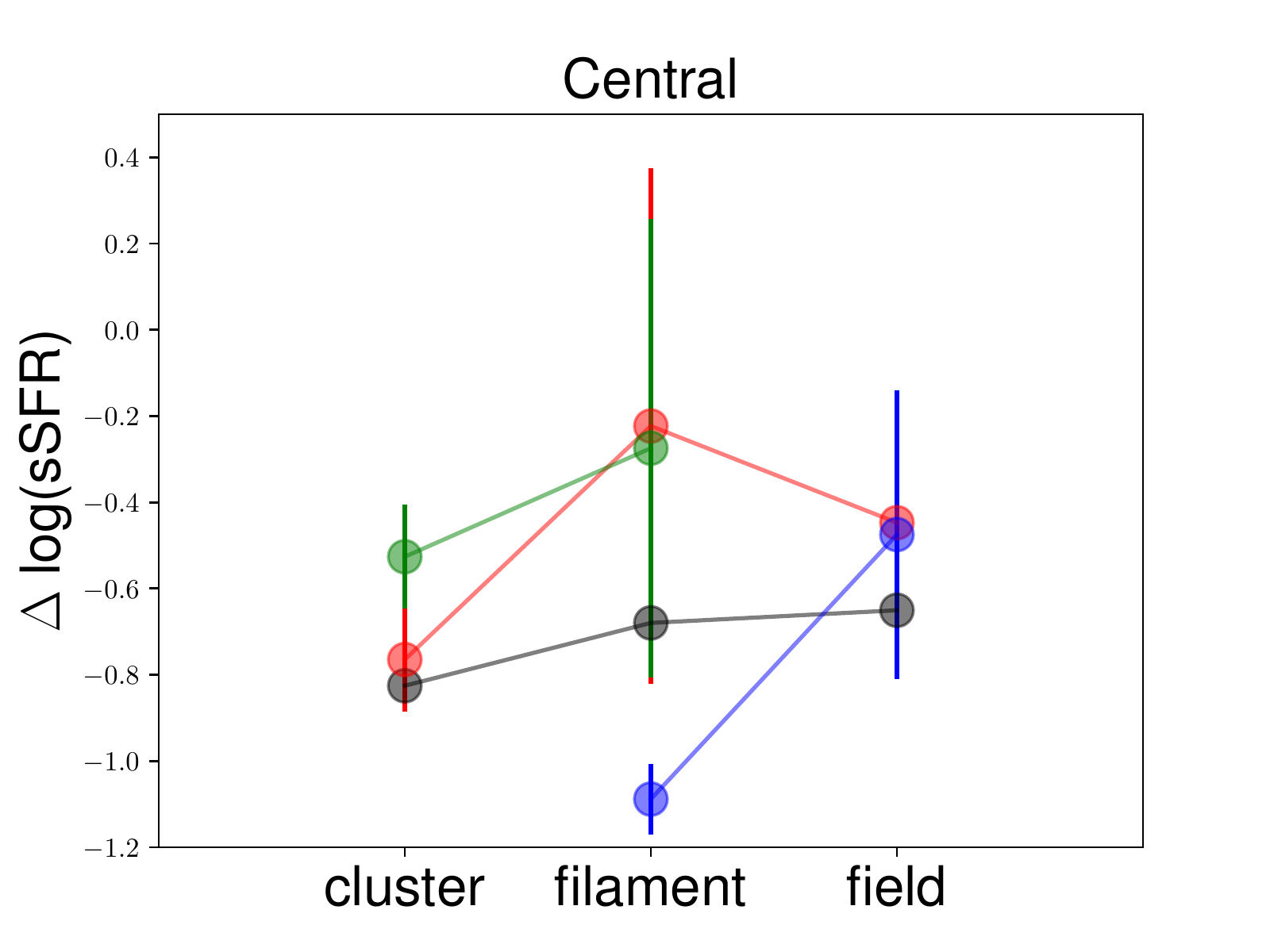}
            }
              \resizebox{\hsize}{!}
 {\includegraphics{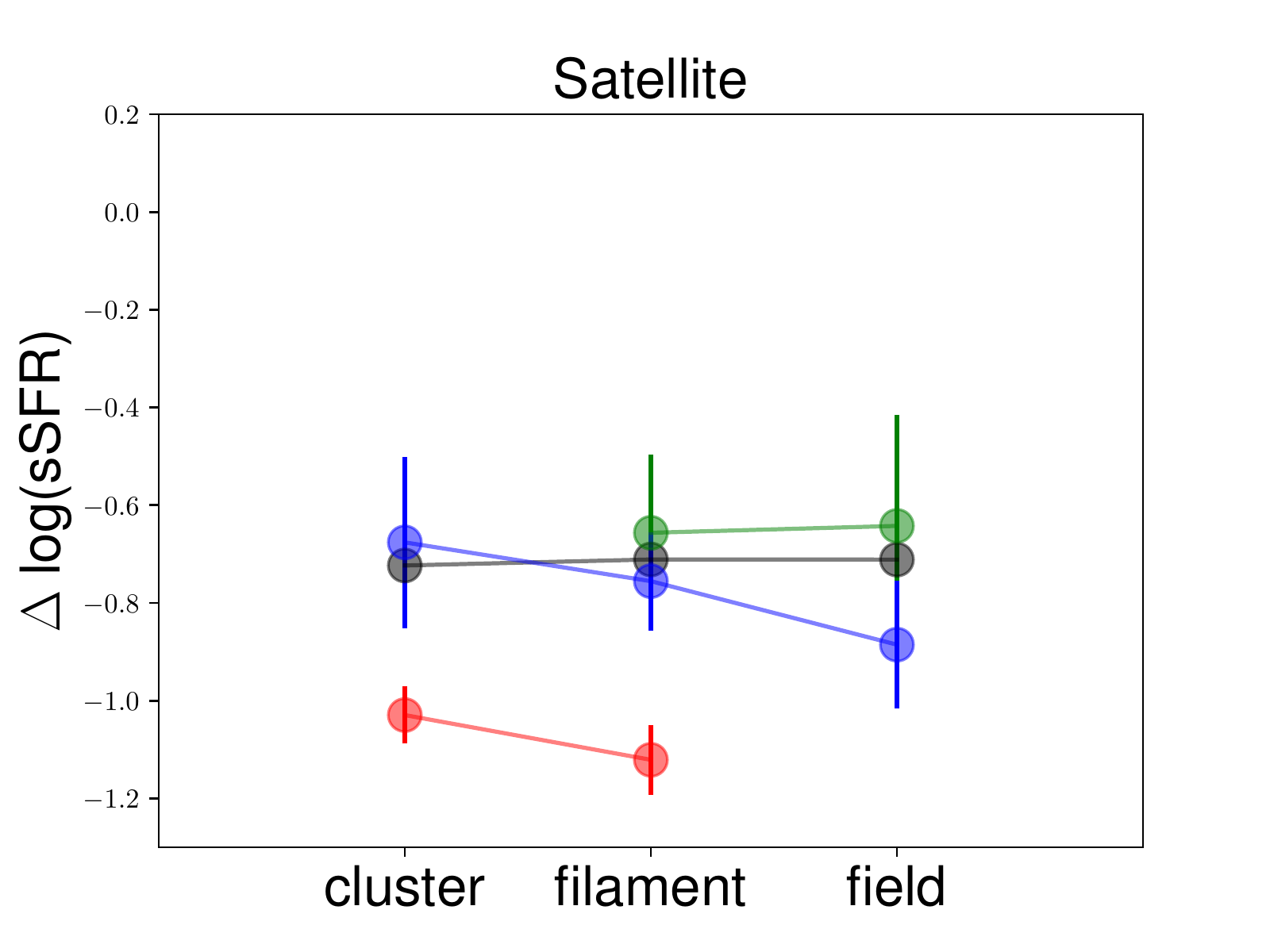}
 \includegraphics{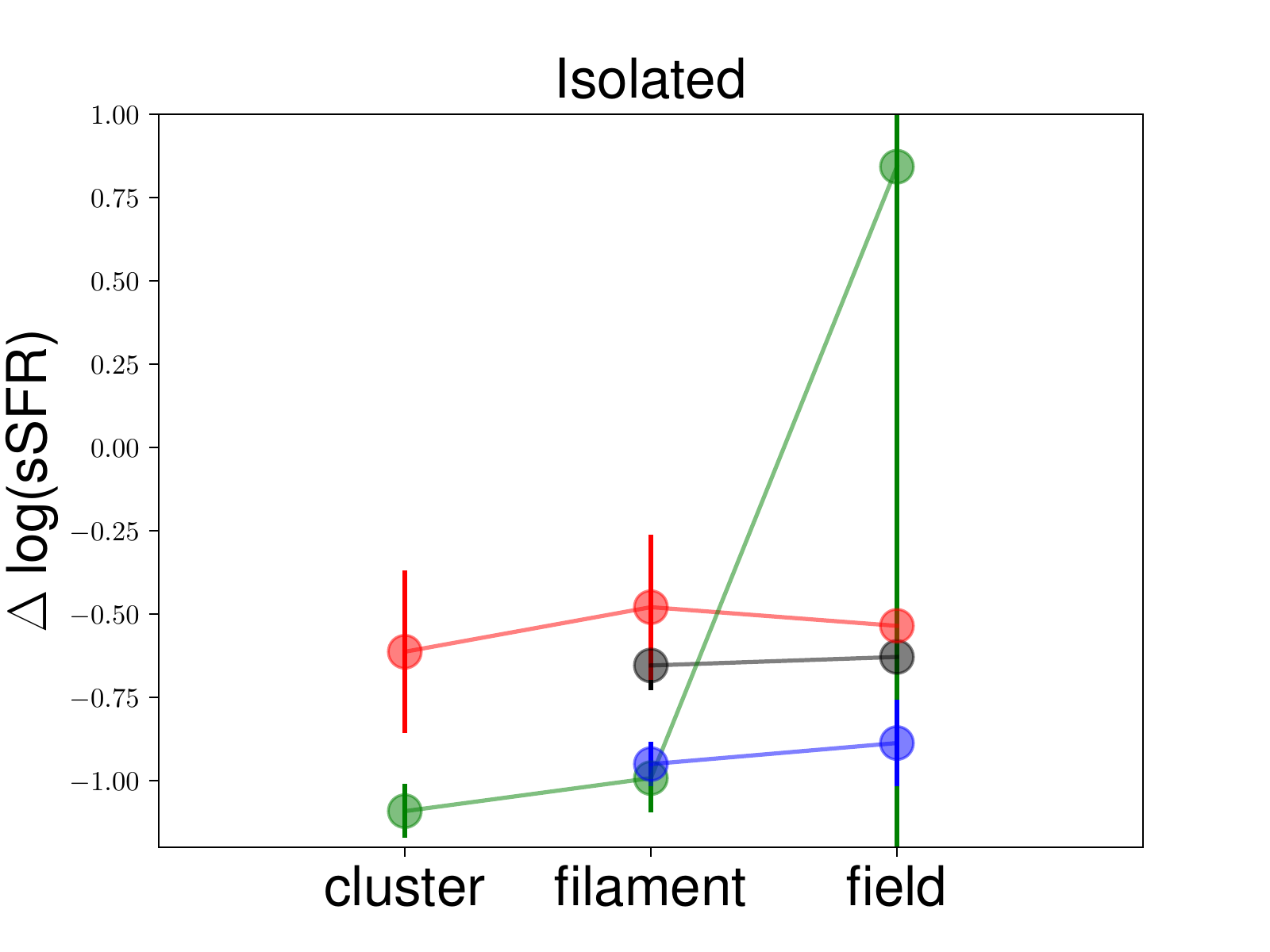}
            }
                 
       \caption{$\Delta$sSFR, the difference between specific SFR and the specific SFR of objects in the MS, for FRIIs (red), FRI/FRIIs (green), FRIs (blue), and COM AGN (black) with respect to environment (cluster, filament, and field) depending on galaxy type, with respect to environment as defined in \cite{darvish15}.
   }
              \label{fig:darvish_env}%
    \end{figure}

As a probe of the cosmic web, we used the study of \cite{darvish17} who separated the large-scale environment, the density fields in COSMOS, into cluster, filament, or field using a Hessian matrix. They further added a classification based on whether the galaxy was the most massive of a group (central), was within a group but not the most massive galaxy (satellite), and was not associated with a group (isolated) by applying a friends-of-friends algorithm. We cross-correlated their catalogue \citep{darvish17} with the FR and COM AGN samples. 
The results are presented in Table~\ref{tab:darvish_env} and Fig.~\ref{fig:darvish_env}. The match was made within a 30" radius and in a $\Delta z$ = 0.1 redshift slice. We found 
{\bf 22 FRIIs, 11 FRI/FRIIs, 15 FRIs, and 362 COM AGN.}

We note that all FRs that we cross-matched with the \cite{darvish17} environmental probes lie below the spread of the MS for SFGs. The largest difference is within a cluster environment for FR objects that lie in satellites galaxies. FRIs associated with satellite hosts are located closer to the MS for SFGs than the FRIIs and FRI/FRIIs in satellite hosts, which are embedded in the quiescent region, providing the only clear division in the current study between FR-type objects. The COM AGN show similar $\Delta$sSFR as FRIs that are associated with satellites inside clusters. 

\begin{figure}[!ht]
  \resizebox{\hsize}{!}
 {\includegraphics{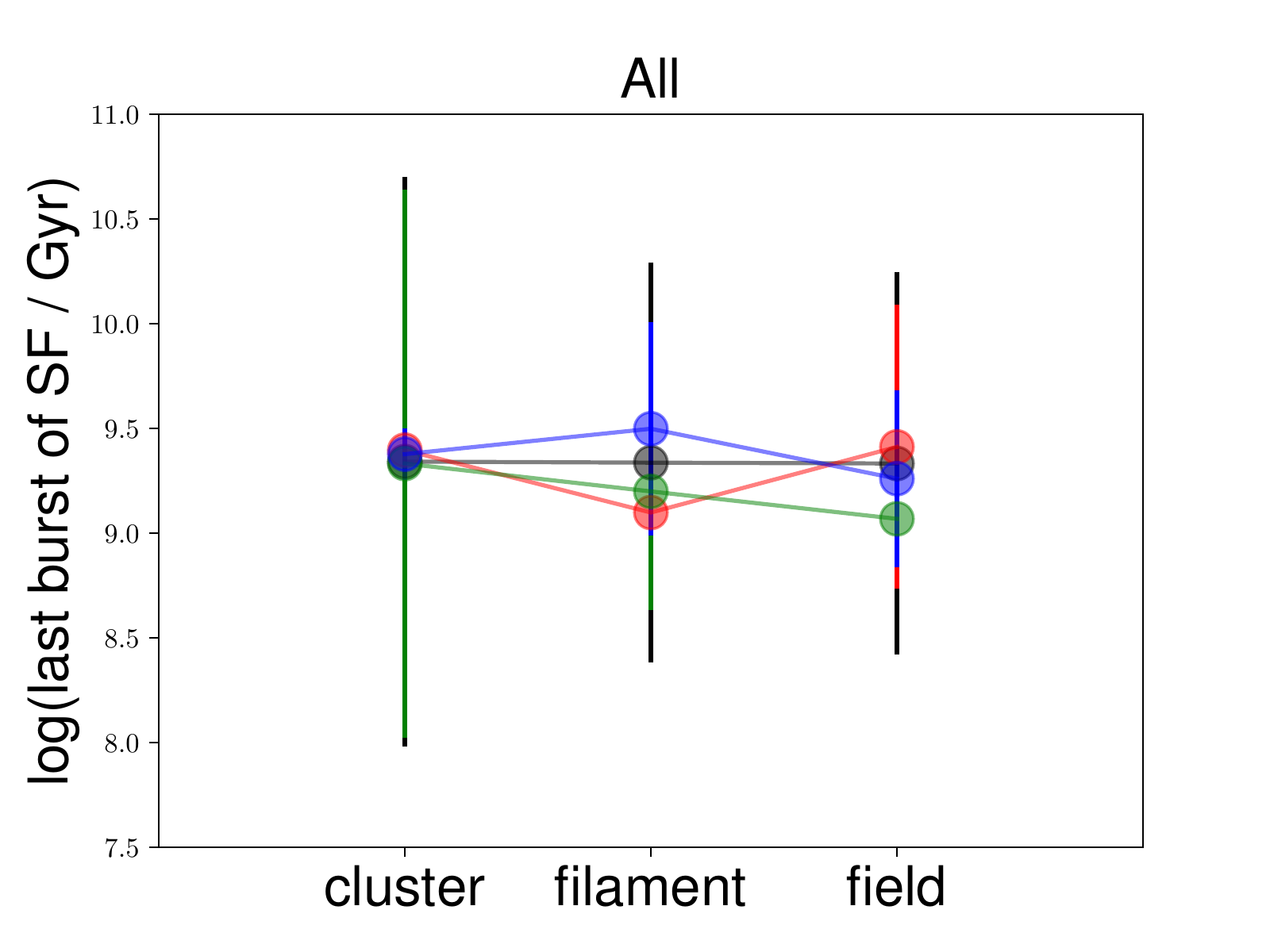}
 \includegraphics{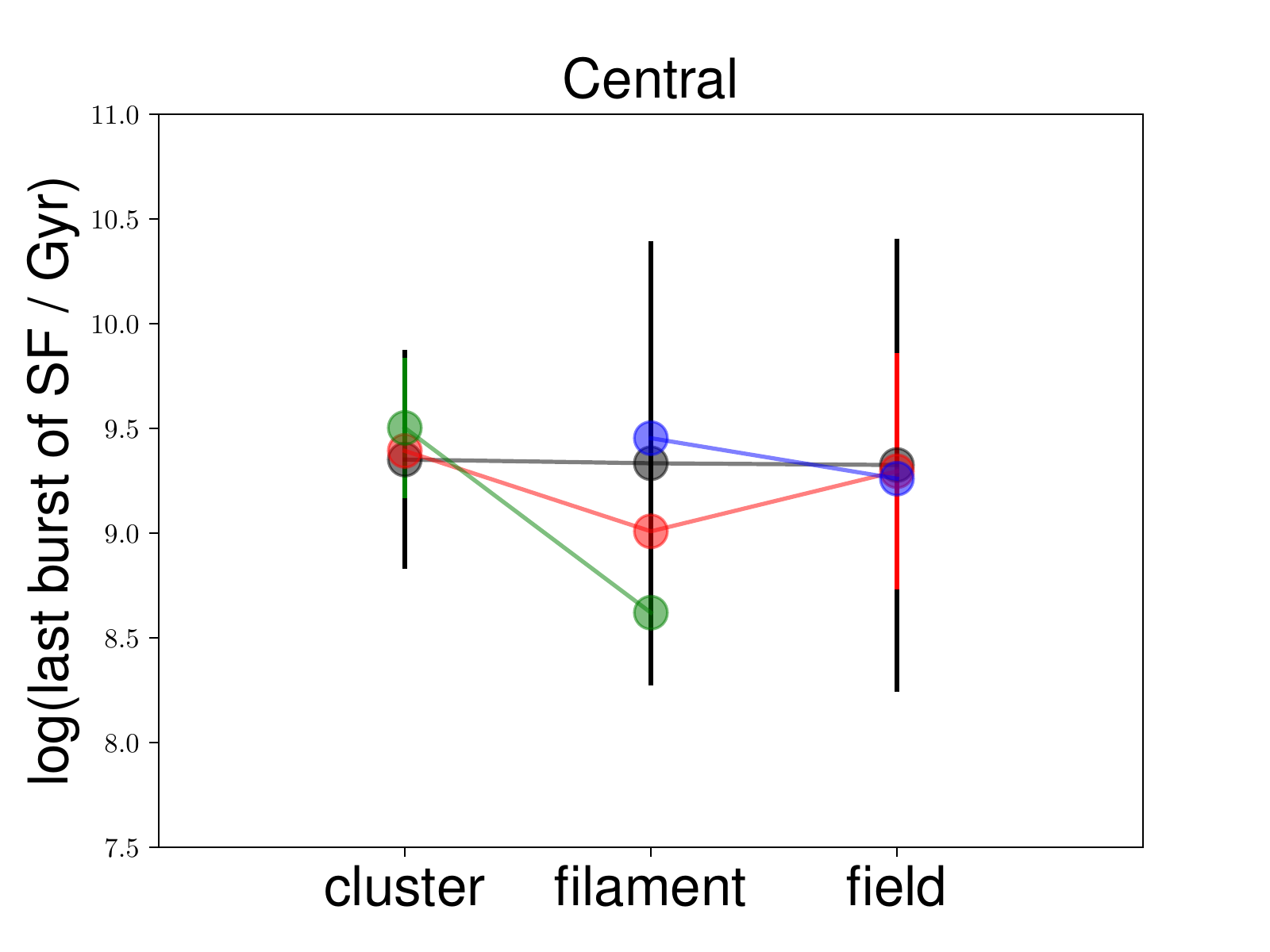}
            }
              \resizebox{\hsize}{!}
 {\includegraphics{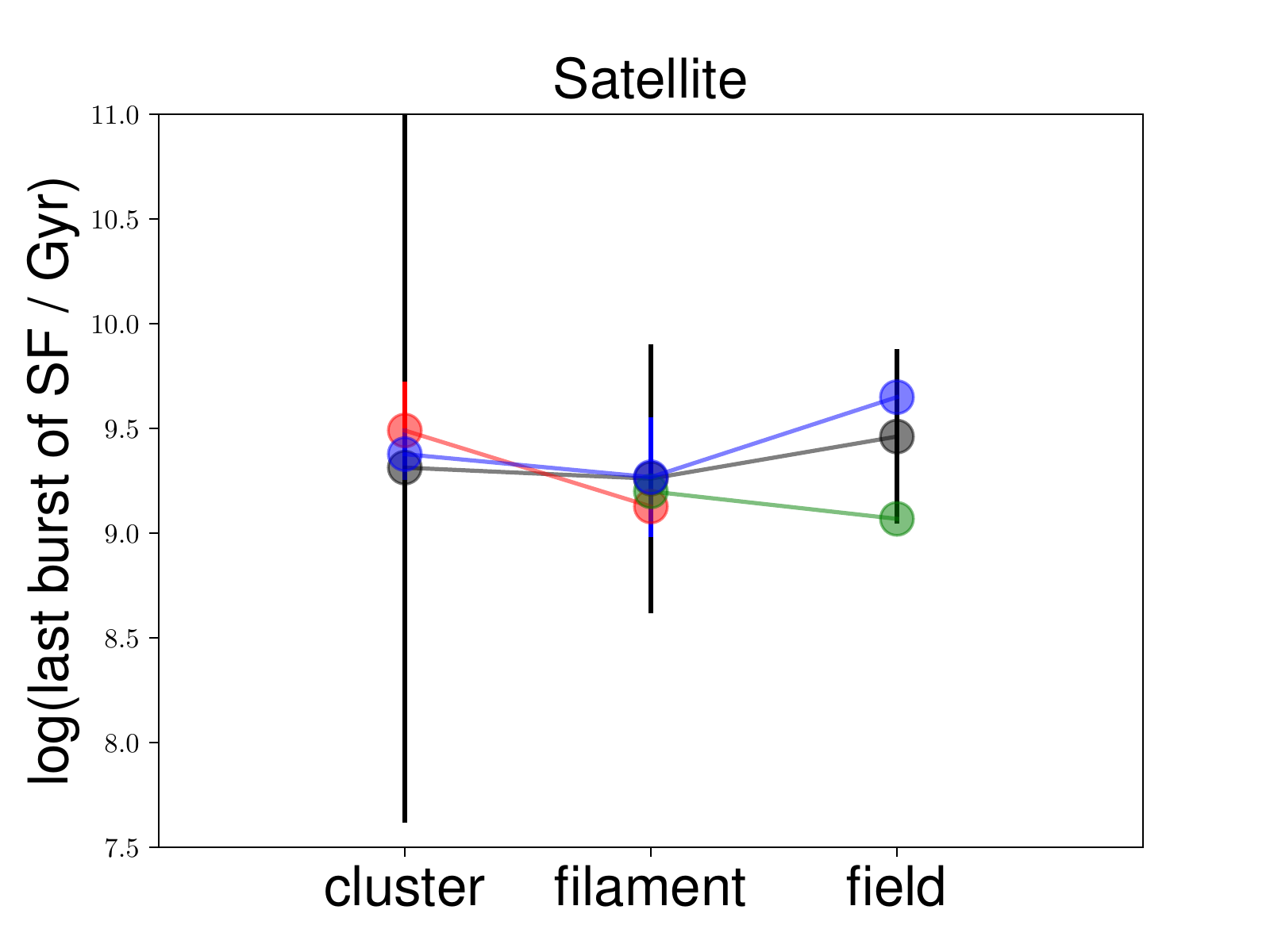}
 \includegraphics{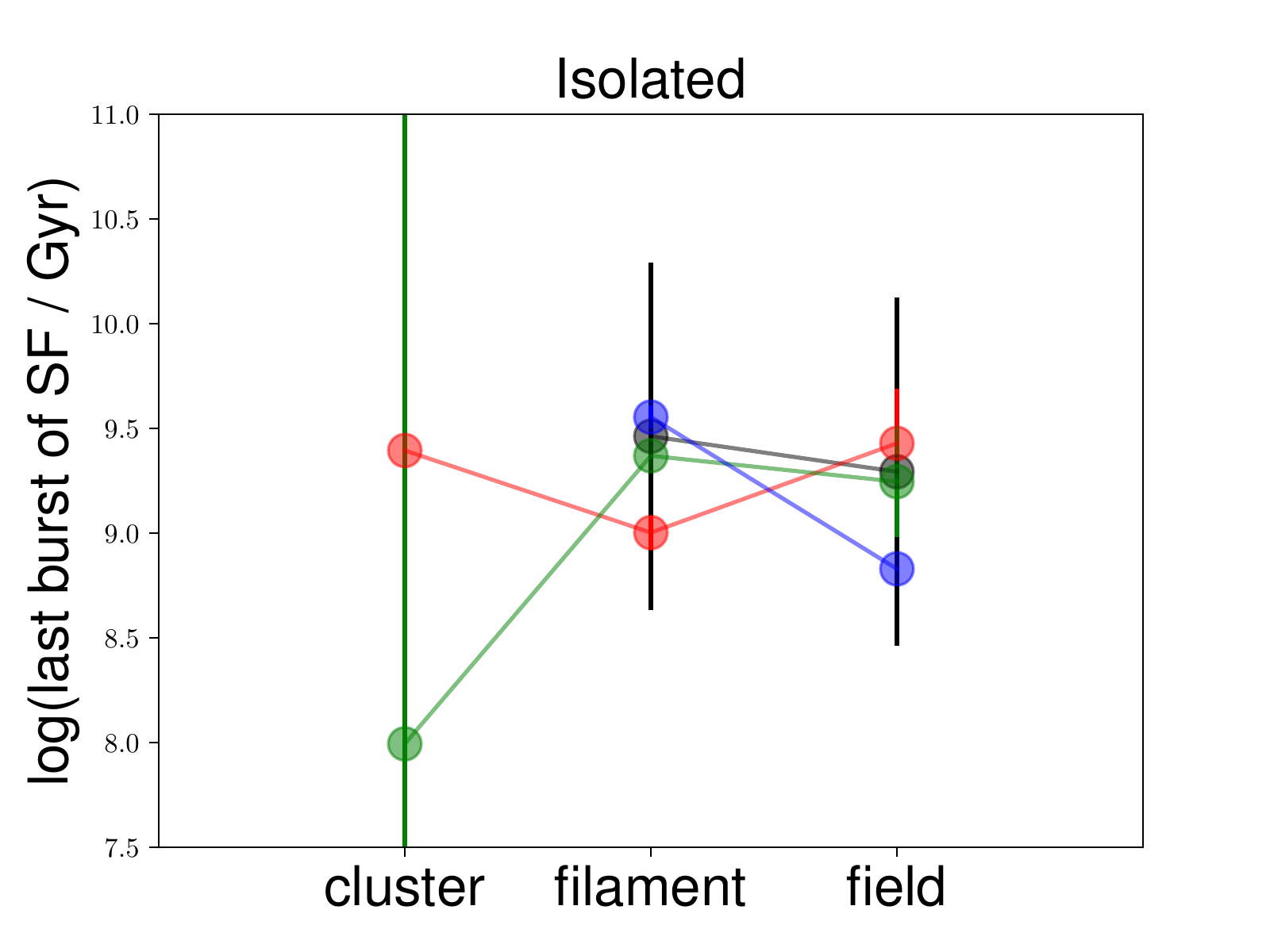}
            }
                 
       \caption{Last burst of star formation for FRIIs (red), FRI/FRIIs (green), FRIs (blue), and COM AGN (black) with respect to environment (cluster, filament, and field) depending on galaxy type, as defined in \cite{darvish15}.
   }
              \label{fig:darvish_env_tburst}%
    \end{figure}

We further investigated whether there are any links between FRs, the \cite{darvish17} environments\footnote{Darvish environments are large-scale filaments, and hence they cannot be characterised by a halo mass, but by a total mass or a mass overdensity.}, and the last burst of star formation. The purpose was to trace the effects of kinetic feedback from FRs in different environments. In Fig.~\ref{fig:darvish_env_tburst} we plot the FR objects, cross-correlated to these different environments versus, the last burst of star formation in their host. In combination with Fig.~\ref{fig:darvish_env}, we see that satellite hosts of FR objects within clusters have similar $t_{\rm last ~burst}$ but different $\Delta$sSFR values. 
The latter suggests differences in the quenching of SF: FRII quench their hosts more efficiently than FRIs and COM AGN. Because FRIIs are on average radio brighter than FRIs, this is an indication that radio-mode feedback from FRII that lie in satellite galaxies within clusters is the cause for star-formation quenching in their hosts. We note that the result is based on small number statistics, with 6 FRIIs and 3 FRIs in satellites within clusters. The green circle with high $\Delta$sSFR in Fig.~\ref{fig:darvish_env} for an isolated host is source 195, an outlier (see Fig.~\ref{fig:ssfr_mstar_sfg}).\\

%
\begin{table}[!ht]
\caption{FRs and COM AGN in different environments}             
\label{tab:darvish_env}      
\centering                          
\begin{tabular}{l l l l l l l l l }        
\hline\hline                 
 \multicolumn{1}{c}{Radio class}   &      \multicolumn{1}{c}{cluster}  &  \multicolumn{1}{c}{filament}   &  \multicolumn{1}{c}{field}                \\     
\hline
\hline 
FRII         & 10 (46\%)& 6 (27\%) & 6 (27\%) \\
FRI/FRII  & 4 (36\%) &  4 (36\%) & 3 (28\%) \\
FRI          & 4 (27\%) & 8 (53\%)& 3 (20\%)\\
COM AGN & 84 (23.2\%)& 132 (36.5\%)& 146 (40.3\%)\\
\hline
\end{tabular}
\tablefoot{Cross-correlation of the FR and COM AGN samples with \cite{darvish15} in respect to different environments (cluster, filament, field). In the parentheses we give the percentage over the total cross-matched number of objects with the same type (within the Darvish environments).}
\end{table}


\section{Discussion: Relating the FR structure to the physical properties and the large-scale environment}
\label{sec:discuss}

In the previous sections we presented an analysis on the properties of radio AGN from the 3 GHz VLA-COSMOS sample and related them to the radio structure. In Table~\ref{tab:prop_pop} we list the median derived properties, as well as the 16th and 84th percentiles, for the radio luminosity at 3 GHz, the redshift, the Eddington ratio (radiative and kinetic), and the number density of galaxies per Mpc$^{2}$. Below we discuss our results and compare them to literature, and to the semi-empirical simulation S$^{3}$-SEX \citep{wilman08} in order to understand what affects the FR structure and the physical mechanisms that drive the traditional FR dichotomy. \\

\begin{table*}[!ht]
\caption{Median derived properties of the FR and COM AGN objects}             
\label{tab:prop_pop}      
\centering                          
\begin{tabular}{l l l l l l l l l }        
\hline\hline                 
\multicolumn{1}{c}{radio}   &      \multicolumn{1}{c}{N}  &  \multicolumn{1}{c}{N}        & \multicolumn{4}{c}{median$_{16\%}^{84\%}$}            \\     
\multicolumn{1}{c}{class}   & \multicolumn{1}{c}{total}  &  \multicolumn{1}{c}{with $z$}  & \multicolumn{1}{c}{Log($L_{\rm 3GHz}$ /} & \multicolumn{1}{c}{$z$}   &   \multicolumn{1}{c}{$D$ (kpc)}  &  \multicolumn{1}{c}{$\lambda_{\rm r}$} &  \multicolumn{1}{c}{$\lambda_{\rm rk}$}  & \multicolumn{1}{c}{density/Mpc$^{2}$}     \\
 & & &  \multicolumn{1}{c}{$\rm W Hz^{-1} sr^{-1}$)}\\
\hline
\hline 
\\
FRII   &   59  & 56 & 23.30$_{22.26}^{24.14}$  &  0.97$_{0.42}^{1.69}$ &  106.6$_{36.9}^{238.2}$  & 0.006$_{0.005}^{0.007}$      &   0.074$_{0.069}^{0.078}$ & 0.90$_{0.05}^{3.73}$ \\ \\
\hline
\\
FRI/FRII & 32 & 27 & 22.81$_{22.35}^{23.51}$ &  0.96$_{0.73}^{1.65}$ &  47.8$_{23.9}^{255.8}$ &  0.003$_{0.002}^{0.005}$    &  0.046$_{0.039}^{0.052}$ & 0.95$_{0.39}^{3.49}$ \\ \\
\hline
\\
FRI   &   39   & 36 & 22.59$_{22.15}^{23.32}$ &   0.80$_{0.35}^{1.54}$ &  33.7$_{23.8}^{73.9}$ &  0.015$_{0.012}^{0.017}$      & 0.031$_{0.027}^{0.078}$ & 0.59$_{0.27}^{2.25}$\\ \\
\hline
\hline
\\
COM AGN  & 1818 & 1818 & 22.46$_{21.79}^{23.03}$ &  1.34$_{0.68}^{2.31}$ &  1.7$_{1.5}^{4.7}$  &  0.010$_{0.010}^{0.011}$   & 0.016$_{0.015}^{0.016}$ & 0.72$_{0.08}^{1.88}$\\ \\
\hline
\end{tabular}
\tablefoot{Median properties, with the 16th \& 84th percentiles, of the FR and COM AGN at 3 GHz VLA-COSMOS. We note that $\lambda_{\rm rk}$ values are consistent with face-on radio jets (low inclination with respect to the observer's line of sight), while the kinetic component could drop by $>$ 100 times at higher inclinations, converging to the $\lambda_{\rm r}$ values. Note: We have also calculated the median $D$ using the LAS from the semi-automatic ML method (for the secure measurements) presented in Appendix~\ref{sec:auto_class}, in combination with the by-hand measurements for non-secure measurements. The results are similar to the median $D$ derived with the by-hand only measured LAS: $D_{\rm FRII}$ = 105.6$^{212.9}_{30.4}$ kpc; $D_{\rm FRI/FRII}$ = 37.5$^{118.1}_{23.0}$ kpc; $D_{\rm FRI}$ = 29.0$^{60.4}_{16.6}$ kpc. The median $D$ values derived by using the combination of the ML and by-hand methods are up to $\sim$ 1 kpc smaller than the values calculated with the by-hand LAS measurements alone. This is related to the semi-automatic method underestimating the sizes in cases of diffuse radio emission. }
\end{table*}

\subsection{Radio luminosity and size}

In our study of 3 GHz VLA-COSMOS radio AGN we find that FRIIs are on average larger than FRIs and FRI/FRII by $\sim$ 69\% and $\sim$ 56\%, respectively, while there is an overlap in the distributions, as mentioned earlier. This is expected from the classic FR scheme wherein FRIIs are larger and brighter than FRIs \citep[e.g.][]{ledlowowen96}. However,  recent studies which address the FR dichotomy did not find a clear difference in size between FRIs and FRIIs \citep[e.g.][]{mingo19}. Several models of jet expansion approach the issue of how far an FR-type jet advances given a set of conditions \citep[e.g.][ and references within]{turner15}. Recently, \cite{Shen20} have shown that the radio power is the responsible driver for how extended the radio jet is. Still there is no clear picture on what affects the FR radio structure and when we should expect for an FRI- or and FRII-type source to form. 

The COM AGN at 3 GHz VLA-COSMOS have the smallest sizes, they lack jets, or their jets are not detectable. This is related to the resolution and sensitivity capabilities of our survey of 0".75 and 2.3 $\mu$Jy/beam. With the 3 GHz VLA-COSMOS we find FRI sources as small as 8 kpc (object 3065) and possible FRII sources of 13 kpc up to redshifts of $\sim$ 0.4 (object 739), or confirmed FRII sources of $\sim$ 24 kpc (object 10937 at $z$ = 1.128). 

When samples of radio AGN selected from radio surveys are studied, the biases introduced by the survey capabilities in detecting these objects should be taken into account. Our ability to detect FR-type radio AGN in surveys is limited by their surface brightness and redshift. We are only able to detect the brightest and youngest sources with increasing redshift, also known as the redshift-youth degeneracy \citep{blundell99}. Another possibility that makes it hard to detect high-redshift radio AGN can be related to their radio lobes being quenched by the cosmic microwave background radiation (CMB) \citep{Ghisellini15}. \cite{Ghisellini15} claimed that the parent population of high-redshift blazars, which are extended radio AGN, cannot be detected in current surveys due to the interaction of their lobes with the CMB, which quenched radio emission. These studies indicated that any trends with redshift are biased by our observing capabilities. Furthermore, the frequency of observation, although important in determining the actual size of the sources, is not as important for FRII sources that have bright hot spots with an intermediate spectral index ($\sim$ 0.5). On the contrary, it is very important for FRI, which have diffuse steep spectrum emission at their edges. This may explain the differences with samples of sources that are selected at low frequency, such as that of  \cite{mingo19}.

Fanaroff-Riley radio sources cover a wide range of values at 3 GHz of 10$^{21-26} \rm ~W~Hz~sr^{-1}$ in radio power, while COM AGN reach fainter values with radio powers of  10$^{19} \rm ~W~Hz~sr^{-1}$. This could be a surface brightness effect, as we discussed earlier, meaning we can only detect the lower luminosity radio emission if it is concentrated; if it is extended, its surface brightness might be too low. Furthermore, there is a large overlap in the distributions of FR in their radio luminosity. The traditional FR scheme \citep[e.g.][]{Gopal-Krishna01} in the local Universe ($z~ <$ 0.1) describes as FRIIs as brighter than FRIs, with a clear dichotomy in radio power. Our results do not show a clear dichotomy. \cite{vardoulaki10} have verified the FR dichotomy at redshifts $z_{\rm med}~ \sim$ 1.25 for an area of $\sim$ 5 deg$^{2}$ with depth 100mJy at 151 MHz, but with a small sample of 47 objects probing the FRI/FRII break at $L_{151 MHz} \sim 10^{25} \rm ~W~Hz^{-1}~sr^{-1}$. This translates into $L_{3 GHz} \sim 10^{24} \rm ~W~Hz^{-1}~sr^{-1}$ for $\alpha$ = 0.8, which is the high end of the radio powers we are probing at 3 GHz VLA-COSMOS (see Figs~\ref{fig:n_z}~\&~\ref{fig:L_D}). This suggests that the traditional FR dichotomy is based on populations which are much brighter and disappears when we probe much fainter populations of radio sources. Slight differences in radio power between FR classes remain, and further support the scenario according to which the difference might be due to accretion rate. We discuss this below.

\subsection{Accretion indicators: Eddington ratios}

Table~\ref{tab:prop_pop} shows that there is no clear dichotomy between FRIs and FRIIs and that the median values of Eddington ratios show a trend: FRIIs, which are slightly more radio bright on average, have lower $\lambda_{\rm r}$ values than FRIs, and FRI/FRIIs have the lowest ratios. This trend is linked to how X-ray bright these sources are on average, because the X-ray flux is used to calculate their radiative Eddington ratio. The median values suggest that FRIIs accrete matter onto their SMBHs less efficiently than FRIs, but the difference is not statistically significant. With the addition of the kinetic energy to the Eddington ratio, this picture changes. FRIIs and FRI/FRIIs get a boost (factor of 12 and 15, respectively) much more pronounced than FRIs (factor of 2). Since the only difference in the calculation of $\lambda_{\rm r}$ and $\lambda_{\rm rk}$ is the addition of jet power, using the radio luminosity as a proxy \citep[$P_{\rm jet} \propto P_{\rm radio}^{0.7}$;][]{cavagnolo10}, the difference in the boost between FRIs and FRIIs, of the order of 5-6 is expected; in Table~\ref{tab:prop_pop} we show that on average the difference in radio luminosity between FRIs and FRIIs is a factor of 5. COM AGN get only a slight boost at $\lambda_{\rm rk}$ (factor of 1.6); they have the lowest Eddington ratios on average amongst the radio AGN in our sample. 

\cite{lusso12} calculated Eddington ratios from the X-rays for the COSMOS type 1 and type 2 AGN, and found that these populations have sub-Eddington but still lower values than what is expected for highly accreting black holes. Our study verifies that the radio AGN population in COSMOS at 3 GHz is sub-Eddington. \cite{lusso12} furthermore showed that the average Eddington ratio increases with redshift for all types of AGN and black hole masses.  Our sample suffers from small number statistics in FRs, and we do not see an increasing trend with redshift. We only find a mild increase with redshift, more evident in the COM AGN sample, as shown in Figs~\ref{fig:classes_eddratio}~and~\ref{fig:classes_eddratio_qjet}.

For objects that are not detected at X-rays, we used stacking to obtain the median values of their Eddington ratios. These objects, although not detected at X-rays, are radio bright but have on-average very low Eddington ratios, of the order of $10^{-4}$. We investigate whether the Eddington ratios we are probing with X-ray using the bolometric corrections of \cite{lusso12} make physical sense. For Eddington ratio of the order of 0.0001 assuming $L_{\rm X} = 10^{42~(41)}~erg~s^{-1}$ for $z~ \sim 1$ \citep[i.e. below the detection limit at X-rays 2-10 keV][see their Fig. 7]{marchesi16}, bolometric correction of 2.2 and $L_{\rm Edd} = 10^{39.1}~(M_{\rm BH}/10^{8}~M_{\odot})~W$ \citep[e.g.][]{vardoulaki08}, we would expect to have $M_{\rm BH} \sim 10^{10~(9)}~M_{\odot}$. This is not surprising for radio-loud AGN jet-mode population, which is radiatively inefficient and has an advection dominated accretion flow (ADAF). Very high mass SMBHs are expected to be associated with these objects \citep[see Fig. 4 in][]{heckmanbest14}.

According to literature, FRIIs typically fall in the high-excitation class, with efficient accretion onto their SMBHs, and FRIs exhibit inefficient accretion with sub-Eddington values \citep{heckmanbest14}. \cite{kauffmann08} studied emission line radio AGN from the SDSS and found there is no dependence of radio power and accretion rate to black hole mass. This was also reported by the study of \cite{gendre13} of 206 radio galaxies below $z~ \leq$ 0.3 ($S_{\rm 1.4~GHz} \geq 1.5$), with no dependence between extended radio structure and accretion mode. Because we are calculating Eddington ratios using X-ray empirical relations and a different method, we cannot directly compare these studies to ours. We do find though that all radio AGN objects in our sample have sub-Eddington ratios, with values $<$ 1\% for $\lambda_{\rm r}$ on average, and that on average FRIIs accrete at similar rates to FRIs. We probe fainter populations of radio AGN than studied before, which are found to produce large (up to 1 Mpc) jets/lobes and are bright (up to $10^{25} \rm ~W~Hz^{-1}~sr^{-1}$ at 3 GHz) FR objects. 

When we keep radio power fixed, the Eddington ratio is not dependent on the radio structure in the radiative case (Fig.~\ref{fig:l3_eddrat}). In the radiative plus kinetic case, there is a slight dependence on radio power and radio structure, with a large overlap. Our results suggest there is no direct dependence of FR radio structure and FR radio power on the efficiency with which matter is accreted onto the SMBH, and for sub-Eddington accreters found in our sample there is a mixture of populations; the latter has also been shown in  \cite{gendre13}.

For the FR dichotomy, \cite{fernandes15} have shown, for a sample of $z~ \sim$ 1 radio sources (at the average redshift of the FR population in our sample) that the FR dichotomy is evident when the kinetic energy was included in the calculation of the Eddington ratio. Without the kinetic component, there is no dichotomy. Their sample was composed of much brighter samples of radio sources than ours, from the 3C, 6C, 7C, TOOT00 radio surveys with the faintest objects at $L_{\rm 151~MHz} \sim10^{25.4}~W~Hz^{-1}~sr^{-1}$. This translates to $L_{3 \rm ~GHz} \sim 10^{24}~W~Hz^{-1}~sr^{-1}$ for $\alpha$ = 0.7, again the region probing the brightest objects of our sample. We do not see such dichotomy in our sample related to accretion rate, but rather an overlap in their distributions. We conclude that the populations we probe do not present a clear dichotomy in FR radio structure and that the accretion mode does not dependent on the radio structure or on the radio luminosity.

\subsection{Large-scale environment}

We have explored several probes of the large-scale environment in which the FRs and COM AGN at 3 GHz VLA-COSMOS lie, from X-ray groups and density fields, as well as the type of large-scale environment and host. Table~\ref{tab:prop_pop} shows that the overall distribution of densities in Mpc-scale environments in FRs is similar, but the median values of FRIs within COSMOS show they lie in less dense environments than FRIIs and FRI/FRIIs. This is in contrast to the study of \cite{castignani14} of 32 FRIs at 1.4 GHz for 1 $< z <$ 2 at COSMOS. For comparison we inspect the panel of Fig.~\ref{fig:dens_z} that covers the same redshift range. We see that FRIs on average lie in less dense environments than FRIIs. We investigated further to understand the difference between the \cite{castignani14} and our results. In our study we find discrepancies in the FR classification between 1.4 and 3 GHz: 16 objects have a different classification. \cite{castignani14} consider as FRIs all objects below the FRI/FRII radio luminosity divide of $4 \times 10^{32} erg~s^{-1}~Hz^{-1}$. The classification method in \cite{castignani14} and our study is different, because we did not adopt a luminosity divide between the classes but rather followed the classic definition by \cite{fr74}; this is causes the discrepancy in our results.

Another important point is to explore whether the excitation mode is linked to the environment. \cite{gendre13} find a correlation, with high-excitation galaxies lying almost exclusively in low-density environments, while low-excitation galaxies are found in a wide range of environments. As we mentioned above, we cannot directly compare our results to the
literature results because different methods were applied to calculate the Eddington ratios. Nevertheless, we do not see this in our study (see Table~\ref{tab:prop_pop}), unless we only take the radiative Eddington ratios into account. The objects, FRIs in this case, with high Eddington ratios on average, live in less dense environments, but not exclusively; there is a wide range of environments. With the addition of kinetic energy, objects with high Eddington ratios, the FRIIs in our sample, lie in dense environments on average, but they are also found in a wide range of environments. 

For the location of the FR host within a group, our results show no preferred location within a group ($\sim$ 400 kpc - 1 Mpc; see Fig.~\ref{fig:xrayin}) with FR type. They can be associated with either the central or a satellite galaxy (see Fig.~\ref{fig:darvish_env}). Similarly, for the jet-less COM AGN, there is no preference for group environment or a host. 

%
   \begin{figure*}[!ht]
    \resizebox{\hsize}{!}
            {\includegraphics[trim=0cm 0cm 1cm 0cm,clip=true]{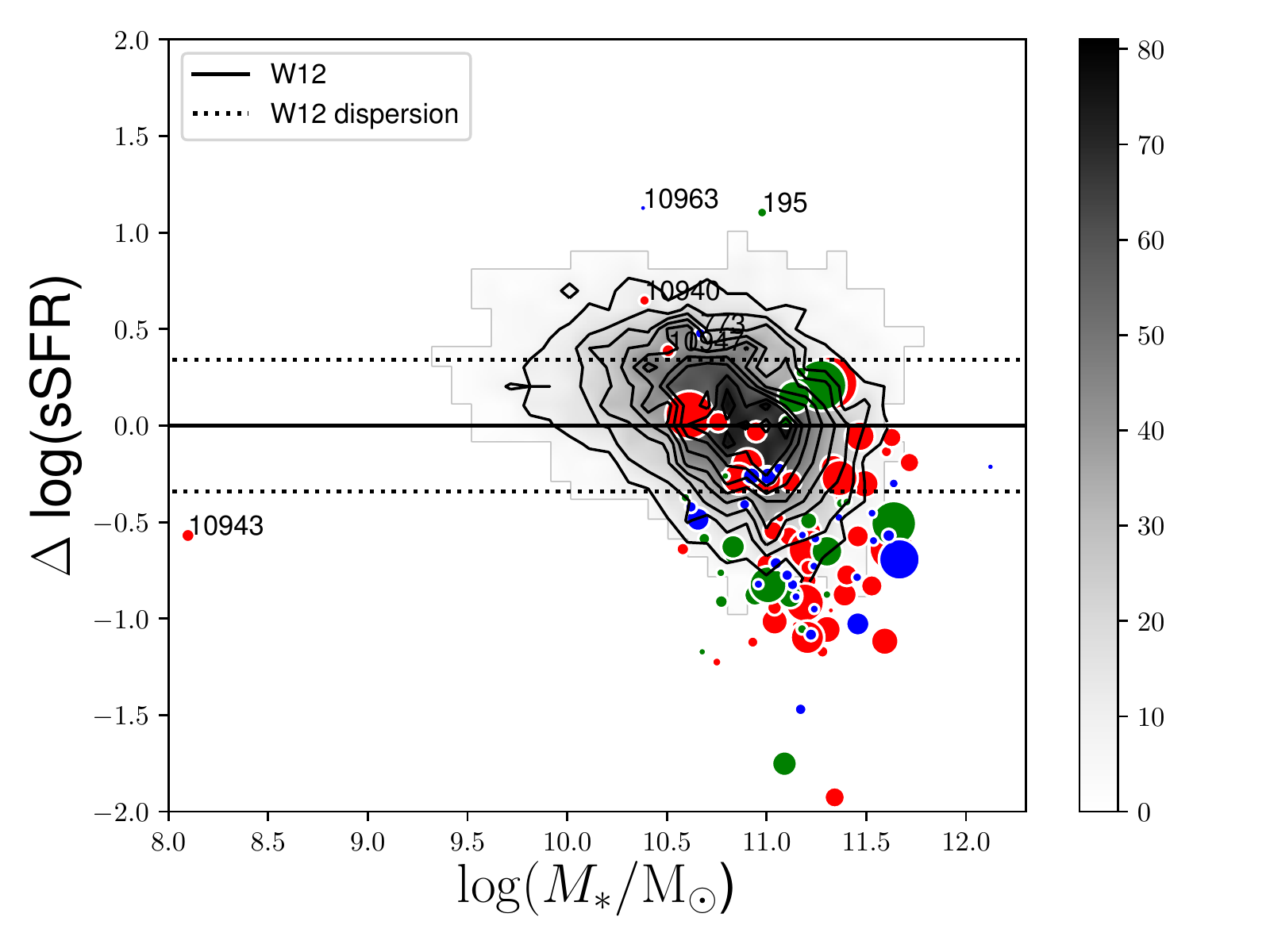}
            \includegraphics[trim=0cm 0cm 1cm 0cm,clip=true]{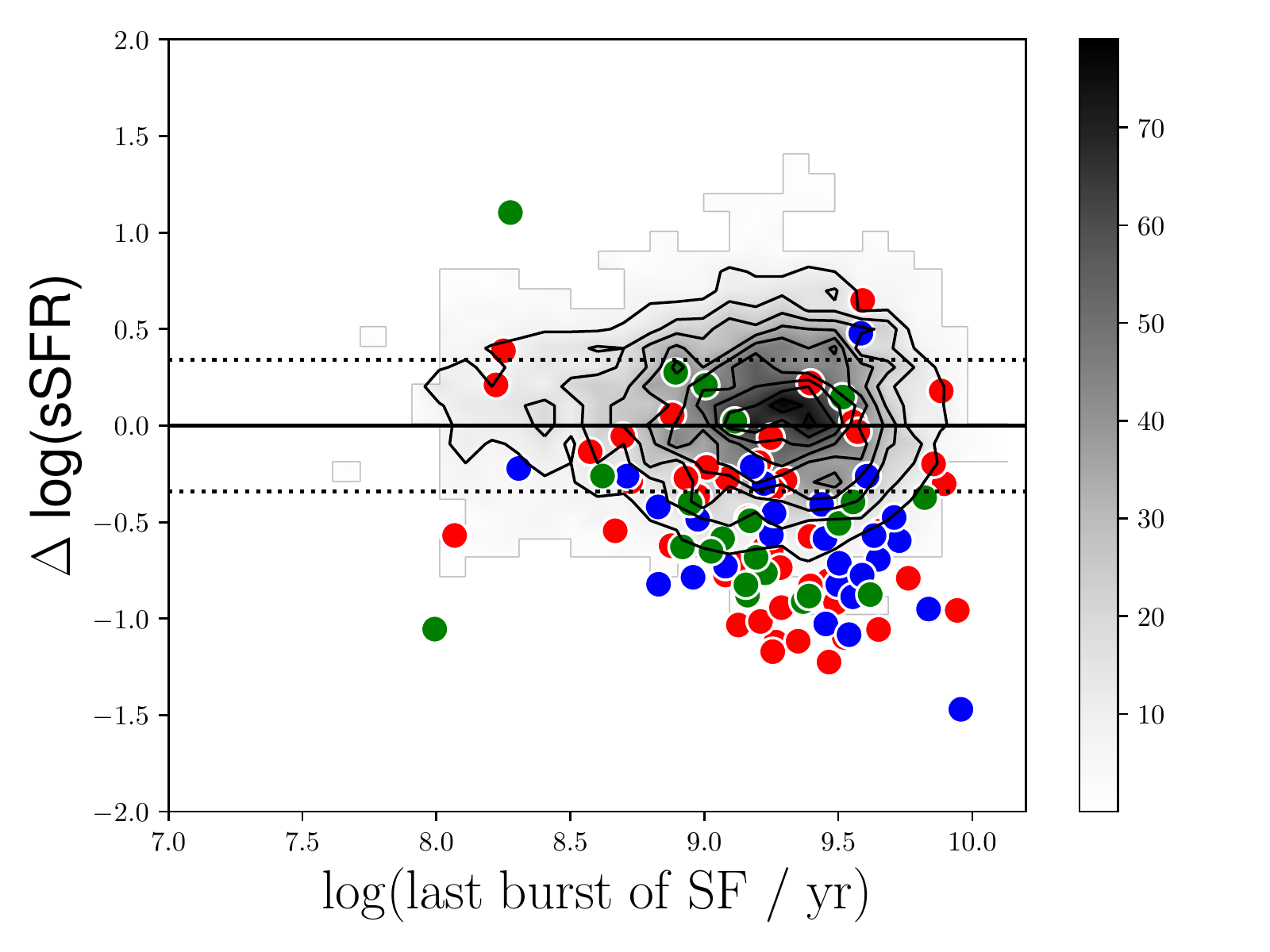}
 }
               \caption{{\bf Left}: $\Delta$sSFR, the difference between specific SFR and the specific SFR of objects in the MS, as a function of stellar mass for the FR objects, shown in colour, compared to the pure SFGs at 3 GHz \citep{smolcic17b}, shown in black as density plot. For FR objects, symbols are scaled based on their linear projected size. Larger symbols correspond to larger objects. FRIIs are shown in red, FRI/FRIIs in green and FRIs in blue. The solid and dotted lines show the main sequence for star forming galaxies and spread, based on \cite{whitaker12}. {\bf Right}: $\Delta$sSFR, as on the left, vs. the star-formation history (SFH) of each object. The SFHs are estimated from the fit to the SED as described in Sec.~\ref{sec:env_host}. We exclude COM AGN objects from these plots for clarity. COM AGN are shown in Fig.~\ref{fig:ssfr_mstar}. \\
   }
              \label{fig:ssfr_mstar_sfg}%
    \end{figure*}

\subsection{AGN quenching star formation}

Our results for the FR and COM AGN hosts, as presented in Fig.~\ref{fig:ssfr_mstar}, show that the AGN fraction is high in SFGs below the MS. This becomes more obvious when we compare the FR objects to the sample of SFGs in 3 GHz VLA-COSMOS survey in Fig.~\ref{fig:ssfr_mstar_sfg}. The latter include all radio sources at 3 GHz with confirmed hosts \citep{smolcic17b} except for radio AGN, that is the FRs and COM AGN. SFGs dominate the MS in Fig.~\ref{fig:ssfr_mstar_sfg}--Left, while FRs are mainly found in the green valley and red-and-dead region of the $\Delta$sSFR-M$_{*}$ diagram. In particular, 72 FRs lie below the MS compared to 751 SFGs, and 28 FRs within the MS compared to 4024 SFGs. We interpret the high fraction of FR objects below the MS as radio-mode feedback on the massive hosts ($> 10^{10.5} M_{\odot}$). We also see a continuation from the SFG cloud to the FR cloud below the MS for SFGs, indicating radio-mode feedback quenching SF. Radio-mode feedback, in contrast to radiative-mode feedback, is considered the maintenance mode of quenching in galaxies \citep{fabian12}, regulating star formation in massive galaxies by heating the galaxy halo and halting future rejuvenation of star formation caused by a fountain effect. 
We note that there is no dependence on the linear projected size of the FR objects and their location in the $\Delta$sSFR-M$_{*}$ diagram. 

COM AGN occupy the same region as FRs in the $\Delta$sSFR-M$_{*}$ diagram, but can be found also on less massive hosts ($\sim 10^{9.5-11.5} M_{\odot}$) below the MS when compared to FRs (see Fig.~\ref{fig:ssfr_mstar}). We note that most COM AGN cluster around $\sim 10^{10.8} M_{\odot}$, while FRs have more massive quenched hosts on average ($\sim 10^{11.3} M_{\odot}$). These results suggest that both FRs and COM AGN quench their hosts. As we mentioned before, the COM AGN sample might contain jetted sources which cannot be revealed with the current survey. Our findings are in line with the study of \cite{smolcic17c}, who reported radio-mode AGN feedback in the hosts of the 3 GHz VLA-COSMOS AGN at each cosmic epoch since $z~ \sim~5$. 

In Fig.~\ref{fig:ssfr_mstar} we marked sources observed at X-rays with Chandra \citep{marchesi16} with a cross. The idea was to investigate the presence of X-ray emission and the relation to galaxy quenching. About half of the FRs below the spread of the MS have X-ray identifications. This can suggest the presence the two mechanisms that quench star formation in their host, but to determine which mechanism is the responsible for permanently quenching star-formation we would need to know the duty cycle of these radio AGN and their lifetimes. We currently lack this information.

The radio AGN within X-ray groups that occupy massive hosts below the MS for SFGs (see Fig.~\ref{fig:t_dssfr}), reside in progressively cooler groups the older their episode of SF is.  
In other words, the objects that leave the MS for SFGs are the ones found in warmer X-ray groups which indicates that the IGM is heated by the AGN. When hosts have moved to the red-and-dead region of the $\Delta$sSFR-M$_{*}$ diagram, their group temperatures are cooler, suggesting a termination of the heating of the IGM through mechanical feedback.

\cite{darvish17} argued in their study of COSMOS large-scale environments that the role of the cosmic web environment is very important in controlling star formation in galaxy hosts, with satellite galaxies controlling the SF fraction in galaxies and with centrals controlling the overall SFR. Their sample included SFGs as well as quiescent galaxies up to $z~ \sim$ 1.2. They reported rapid quenching in most satellites as they transit through filaments from the field to clusters. As we have shown in Figs~\ref{fig:darvish_env}~and~\ref{fig:darvish_env_tburst}, we see indications for SF quenching in satellites of specific types of radio AGN, namely for FRIIs. The latter classes seem to quench their host much more efficiently than FRIs and COM AGN in a similar time from the last burst of star formation. FRIIs in our sample are slightly more powerful at radio  on average than FRIs at 3 GHz VLA-COSMOS survey. This could justify that the energy they release into their environment in the form of mechanical energy is higher than in FRIs, providing the necessary heating to the gas within the circum-galactic medium (CGM) to efficiently quench star formation in the host. Our finding that FRIIs could be more efficient in quenching their hosts is supported by simulations performed by \cite{perucho19}. They compared 3D hydrodynamical simulations to 2D axisymmetric simulations of relativistic outflows and to observational data from FRIIs to show that shock heating has a significant effect on feedback from AGN; in particular, FRII-type radio source are extremely fast and efficient in quenching their hosts.

\subsection{Comparison to the S$^{3}$-SEX semi-empirical simulation}

\begin{figure}[!ht]
    \resizebox{\hsize}{!}
            {\includegraphics{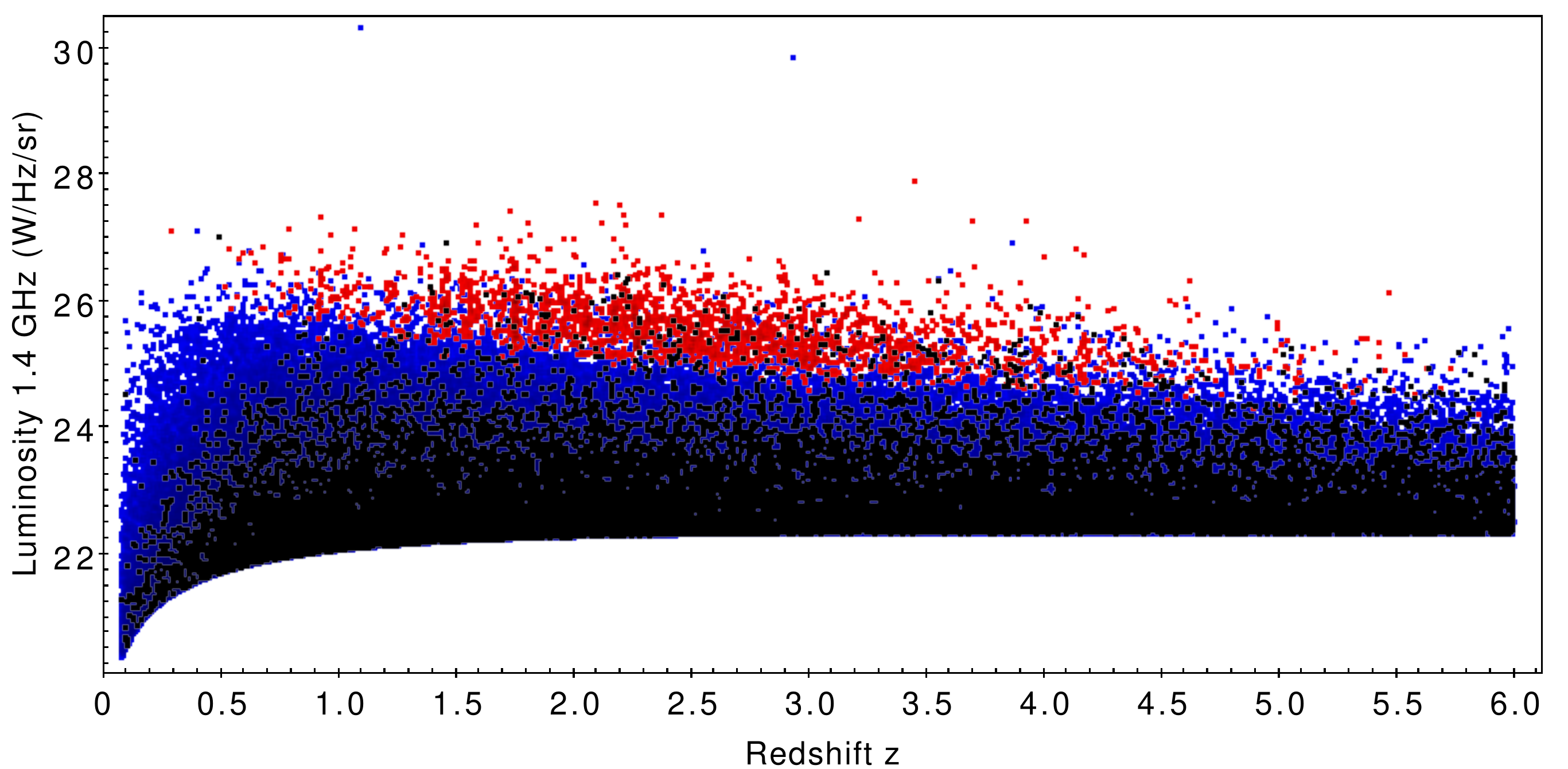}
 }
               \caption{Radio luminosity at 1.4 GHz vs. redshift from the S$^{3}$-SEX semi-empirical simulation \citep{wilman08} for a sky area of 20$\times$20 deg$^{2}$ centred at the central coordinates of the simulation, and for a flux density limit of 17 $\mu$Jy. Blue crosses are FRIs and red crosses are FRIIs, and GPS sources are plotted in black.
   }
              \label{fig:L14_z_s3sex}%
   \end{figure}

We compared our sample to the S$^{3}$-SEX semi-empirical simulation\footnote{http://s-cubed.physics.ox.ac.uk/s3\_sex} of \cite{wilman08}, a simulation of extragalactic radio continuum sources in a sky area of 20$\times$20 deg$^{2}$ out to a redshift of 20. The sources are drawn  from empirical data, and for the purposes of the simulation, extrapolated beyond the observational limits of the surveys. The simulation offers a radio AGN classification relevant to our study: FRI, FRII and gigahertz-peaked (GPS) sources, which are jet-less sources whose radio spectral energy distribution peaks at GHz frequencies. The 151 MHz luminosity function from \cite{willott01} was used to simulate these populations. We ran the online query for 1.4 GHz sources above flux densities\footnote{We converted the 10 $\mu$Jy flux-density limit of the 3 GHz VLA-COSMOS survey into 1.4 GHz using a standard steep radio spectral index of 0.7.} of 17 $\mu$Jy  up to redshift of 6. To avoid cosmic variance, we chose an area much larger than COSMOS and selected the full area covered by the simulation, which yielded 2,285,085 sources. Within this volume, there are 330,694 FRIs, 2,080 FRIIs, and 16,650 GPS sources. The result of the simulation is presented in Fig.~\ref{fig:L14_z_s3sex}. FRIs dominate the luminosity-redshift parameter space and reach the flux-density limit we selected. FRII radio sources are found above luminosities of $\sim$ 10$^{24} \rm ~W~Hz^{-1}~sr^{-1}$ at 1.4 GHz and at redshifts above $\sim$ 0.3. GPS sources are also widely distributed and reach down to the flux-density limit we chose.

To compare the 3 GHz FRs and COM AGN to the simulation, we scaled the S$^{3}$-SEX simulation down to the 2.6 deg$^{2}$ of the 3 GHz VLA-COSMOS survey and obtained $\sim$ 13139 sources in total with $\sim$ 1901 FRIs (8\%), $\sim$ 12 FRIIs (0.01\%), and $\sim$ 96 GPS (0.7\%) sources in the simulation. The radio spectral index $\alpha$ = 0.7 used to convert the flux-density limit of 3 GHz VLA-COSMOS into 1.4 GHz might affect the number of sources drawn by the simulation. In the 10830 sources within 3 GHz VLA-COSMOS, we have 39 FRIs (0.3\%), 32 FRI/FRII (0.2\%), 59 FRIIs (0.5\%), and 1818 COM AGN ($\sim$ 17\%). Despite the results of cosmic variance, because with COSMOS we only observe a small patch of the sky, there is a clear difference between the simulation and data, which is the number of FRII and FRI sources recovered. At 3 GHz VLA-COSMOS we recover five times more FRIIs than the S$^{3}$-SEX simulation for the same sky area and depth. This is mainly attributed to the combination of the high resolution and sensitivity of 3 GHz VLA-COSMOS survey, which can recover FRII-type radio sources at lower flux densities than probed before, and with lower surface brightness. This demonstrates that the traditional FR classification scheme of \cite{fr74} is surface brightness biased towards brighter and larger FRIIs. On the other hand the simulation predicts a factor of 50 more FRI sources than what we observe at the 3 GHz VLA-COSMOS survey. These are either not resolved by our survey and lie within the COM AGN population, or they are not observed because they are too faint to be detected at higher redshifts.

\begin{figure}[!ht]
    \resizebox{\hsize}{!}
            {\includegraphics{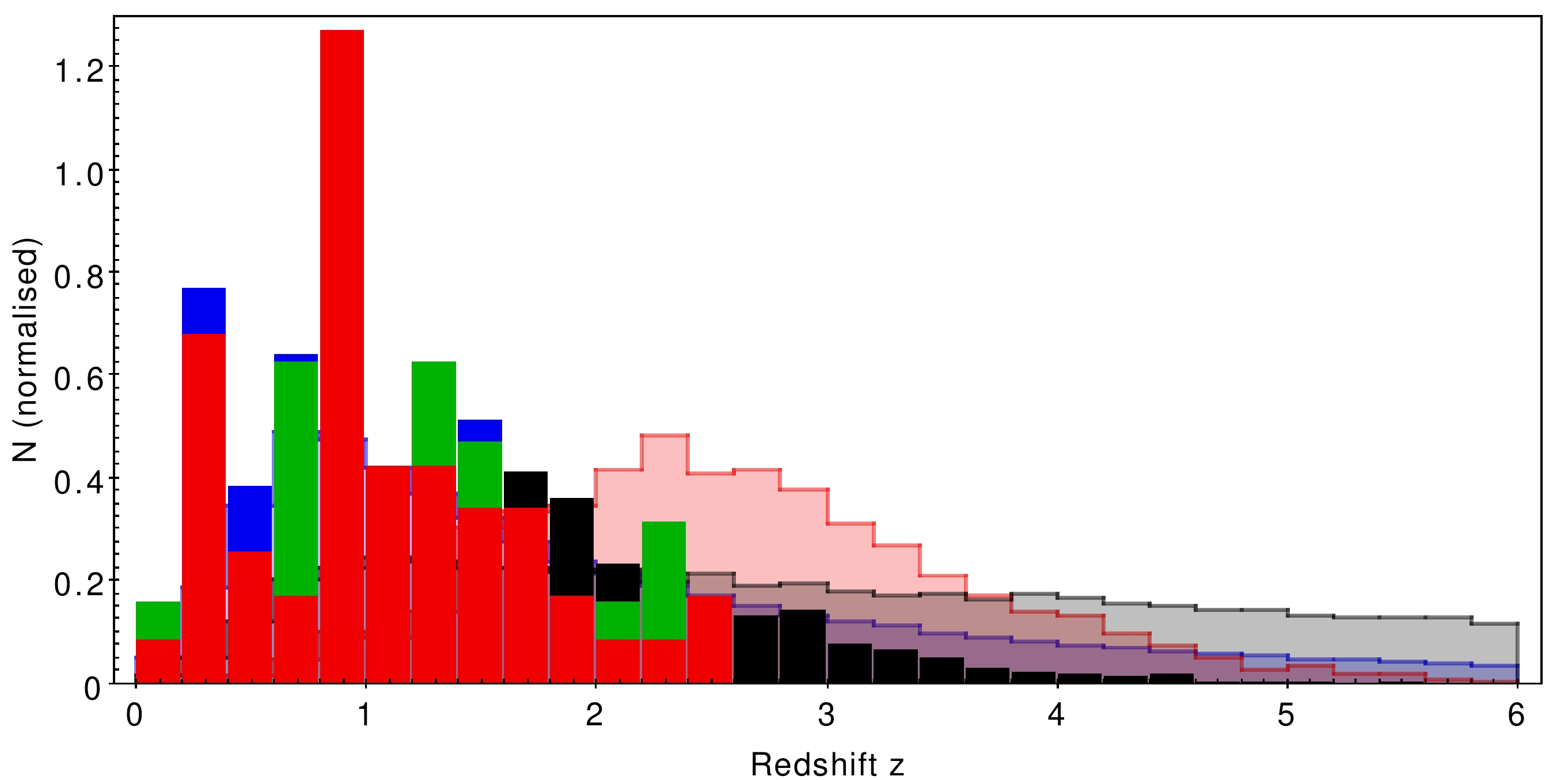}
 }
               \caption{Histogram of the redshift distribution between the radio AGN in the 3 GHz VLA-COSMOS sample and the S$^{3}$-SEX semi-empirical simulation \citep{wilman08} for a sky area of 20$\times$20 deg$^{2}$ centred at the central coordinates of the simulation, and for a flux density limit of 17 $\mu$Jy. The total area of the histogram bars is normalised to unity. Solid histograms correspond to 3 GHz data and semi-filled to the simulation. The bin size is 0.2. Red shows FRII, green shows FRI/FRII, blue shows FRI, and black shows COM AGN or GPS. 
               }
              \label{fig:hist_z_3ghz_s3sex}%
    \end{figure}

\begin{figure}[!ht]
    \resizebox{\hsize}{!}
            {\includegraphics{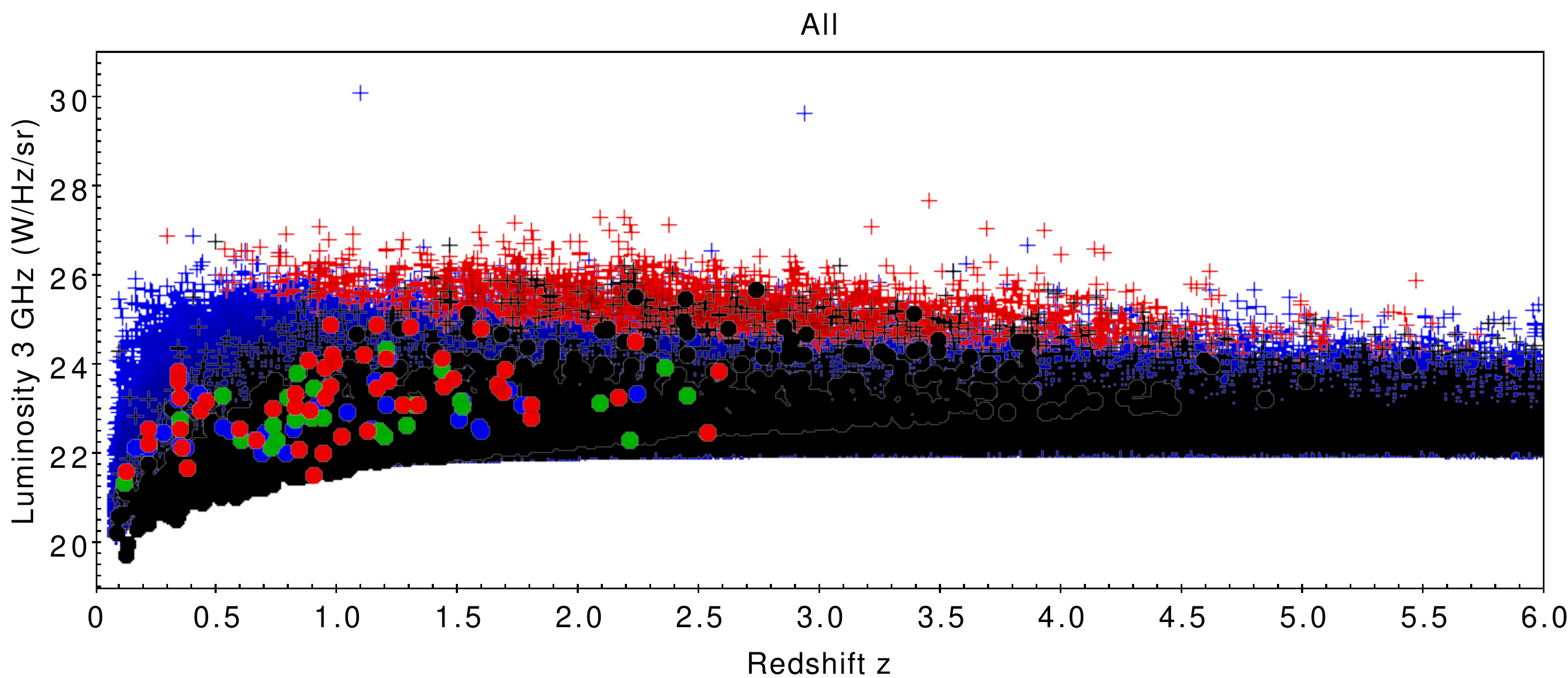}
            }
            \resizebox{\hsize}{!}
            {\includegraphics{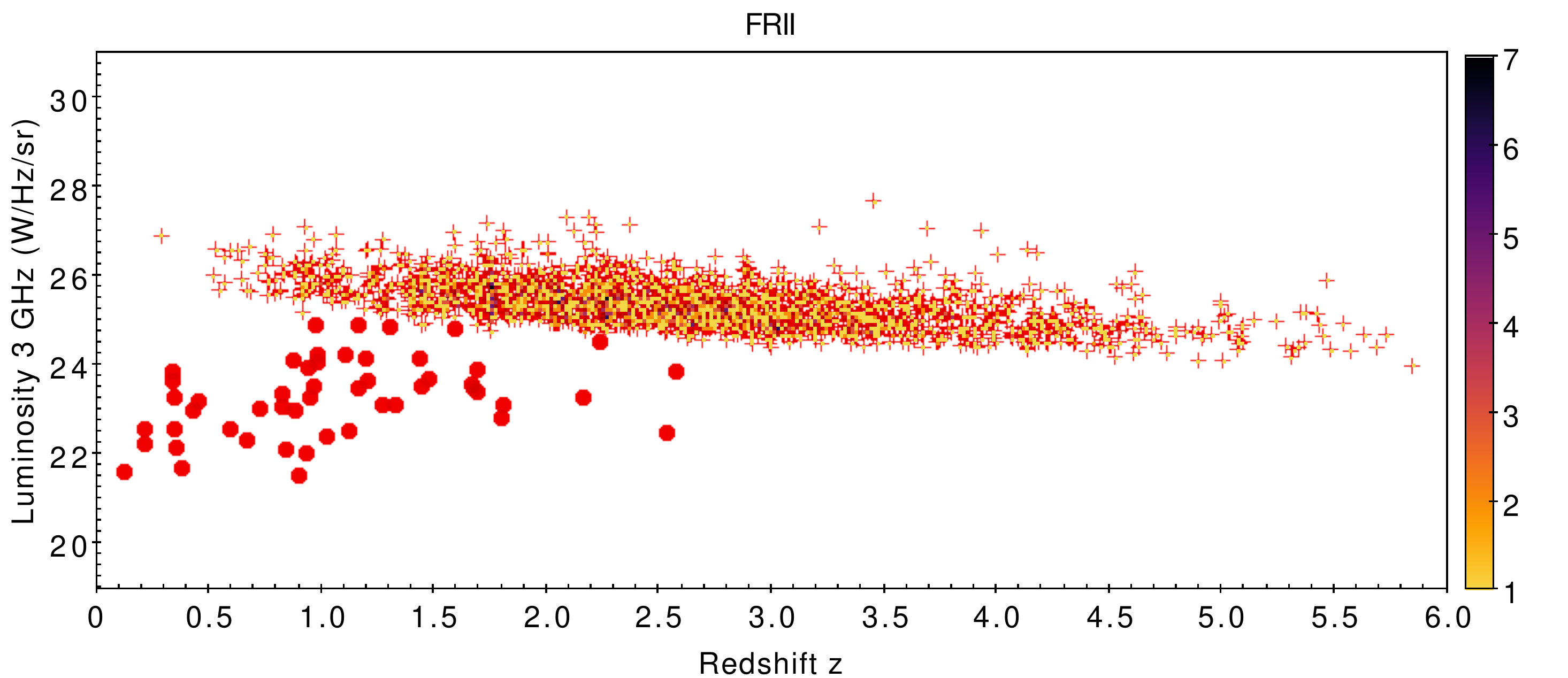} 
            }
            \resizebox{\hsize}{!}
 	{\includegraphics{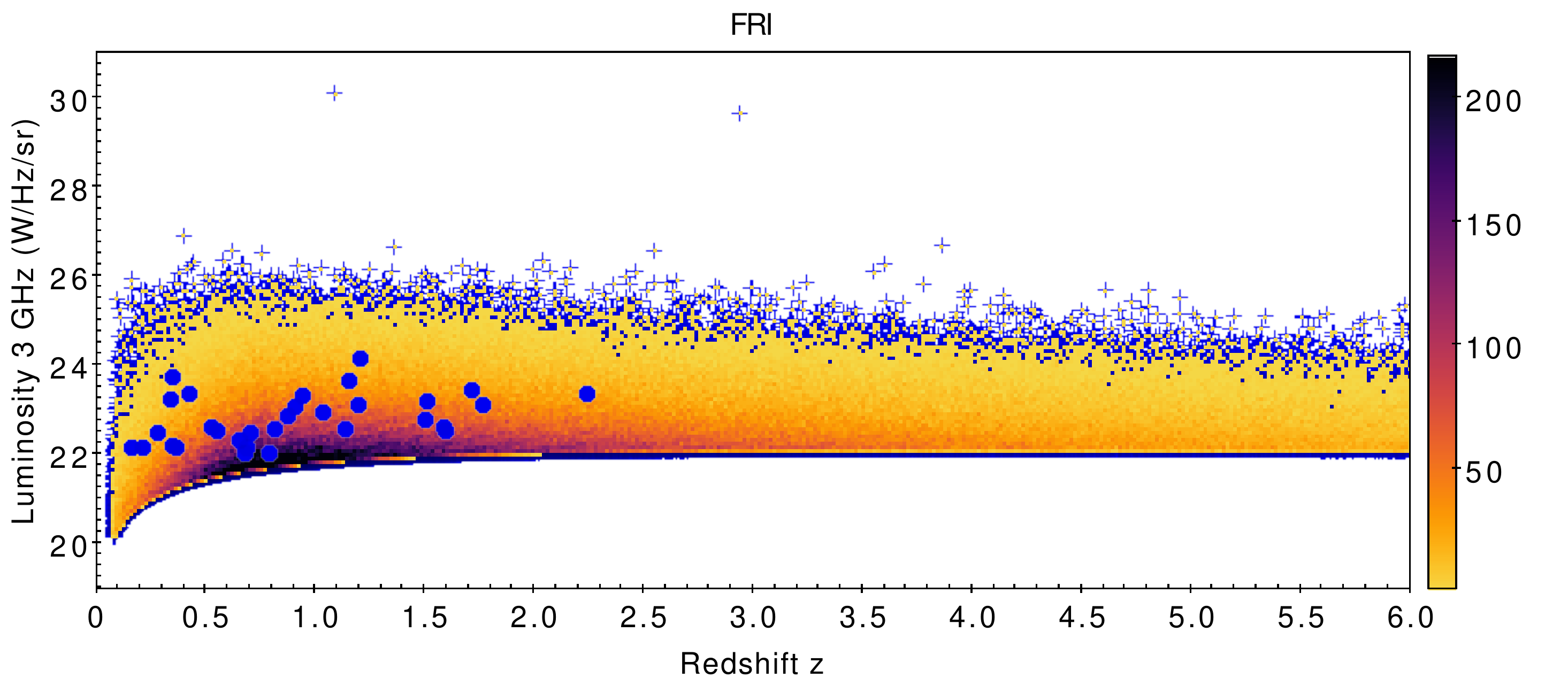}
            }
            \resizebox{\hsize}{!}
            {\includegraphics{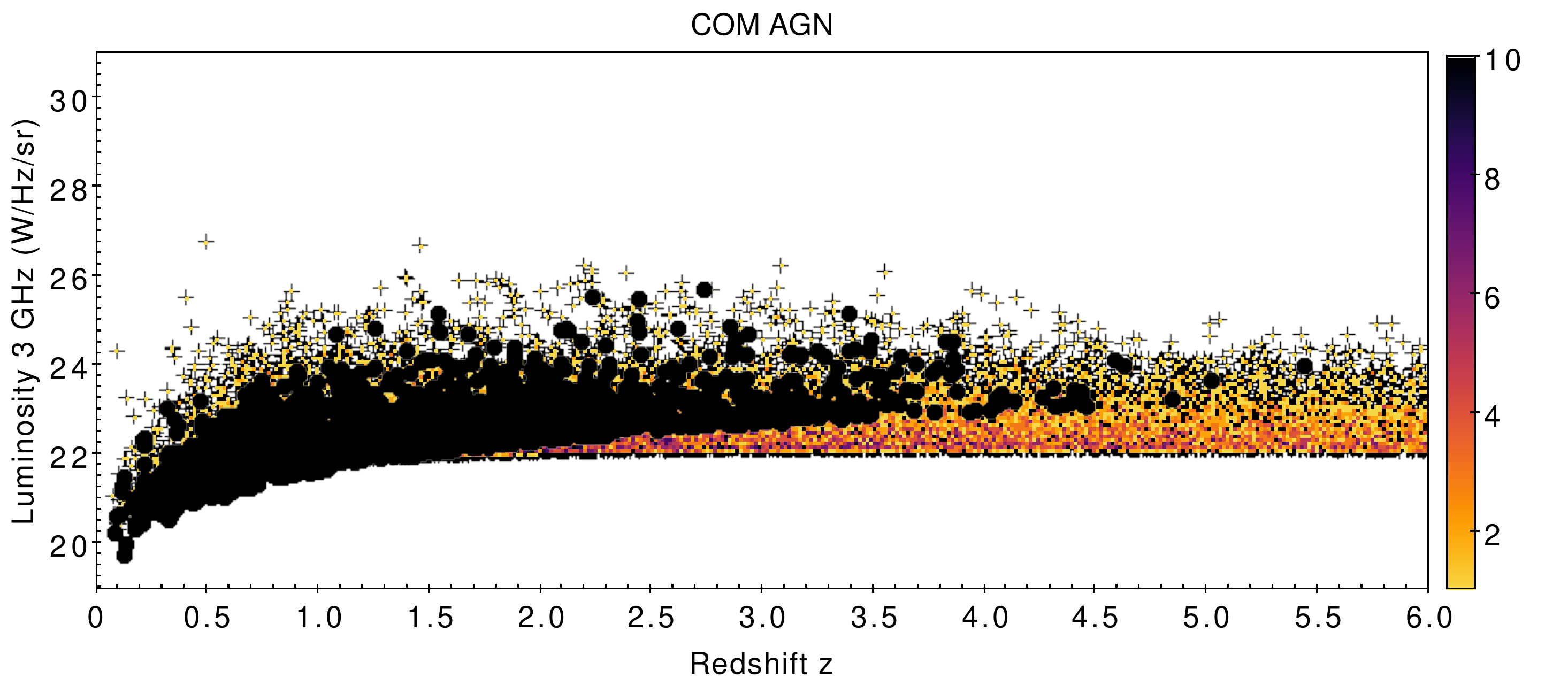} 
            }
               \caption{Radio luminosity at 3 GHz vs. redshift from the S$^{3}$-SEX semi-empirical simulation \citep{wilman08} for a sky area of 20$\times$20 deg$^{2}$ centred at the central coordinates of the simulation, and for a flux density limit of 17 $\mu$Jy. We use a typically steep radio spectral index to convert the 1.4 GHz radio fluxes provided by the simulation into 3 GHz. Blue crosses are FRIs and red crosses FRIIs, and GPS sources are shown in black. We overplot the 3 GHz VLA-COSMOS survey as solid circles with red (FRII), green (FRI/FRII), blue (FRI) and black (COM AGN). We also show panels for each class, for clarity. The bar at the right side gives the number density of sources in the simulation. The bottom panel shows GPS sources from the simulation and COM AGN from 3 GHz VLA-COSMOS survey, but a direct comparison cannot be made because COM AGN are not further classified as GPS.\\
               }
              \label{fig:L3_z_3ghz_s3sex}%
    \end{figure}

In Fig.~\ref{fig:hist_z_3ghz_s3sex} we present a normalised histogram of the redshift distribution for the S$^{3}$-SEX simulation, where we overplot the 3 GHz VLA-COSMOS FR and COM AGN objects. The difference in the redshift distribution of FRII sources is evident, with our survey revealing an FRII population at lower redshifts, that peaks around redshift of one; the simulation peaks above a redshift of two for FRIIs. Furthermore, FRIIs in our sample are fainter on average at 3 GHz than the FRIIs from the simulation\footnote{We converted the 1.4 GHz flux densities from the simulation into 3 GHz flux densities using $\alpha$ = 0.7. {Plotting the 1.4 GHz values versus redshift gives similar results.}}, as we show in Fig.~\ref{fig:L3_z_3ghz_s3sex}. In conclusion, the advantage of the 3 GHz VLA-COSMOS survey over past surveys and extrapolated data is that it recovers FRII-type radio AGN at lower redshifts and at lower flux densities than before. This is related to the surface brightness bias linked to the FR classification scheme, but it can also be due to the resolution. At higher redshifts, it is hard to disentangle the FRII radio structure. The smallest confirmed FRII at 3 GHz VLA-COSMOS is object 10937 ($D$ = 24.3 kpc, $z$ = 1.128; LAS = 2.96 arcsec). This is due to the capabilities of our survey: at $z$ = 1 (2) with a resolution of 0.75 arcsec we can resolve sources with sizes of 6 (6.2) kpc. Object 10937 is a double radio source whose lobes are $\sim$ 2 arcsec long separated by $\sim$ 1 arcsec, which at the redshift of the source would be $\sim$ 8 kpc (see Fig.~\ref{fig:maps2}).

Our sample and the simulation have similar redshift distributions for FRis up to $z~ \sim$ 2.5, where we do not detect any FRIs above this redshift. The radio powers of FRIs in our sample fall within the predicted values given by the simulation. For the COM AGN and the GPS sources of the simulation we cannot make a direct comparison because in our sample we do not further classify COM AGN as GPS or not. We present the plots, and we caution about the interpretation. 

Surface brightness and point source sensitivity could explain part of the discrepancy between the FRs in our sample and in the simulation of \cite{wilman08}, but not all of it. We conclude that the discrepancy highlights the point we make in this study: the way different researchers apply the FR classification scheme, depending on several properties (sensitivity, resolution, and frequency) as well as their own personal understanding of what a source looks like, introduce the bias in the FR classification scheme. Furthermore, the \cite{wilman08} simulation, although based on observational data, applies an idealised radio structure in order to model the FRI and FRII radio sources that enter the simulation, with homogeneous surface brightness and symmetric lobes (see their Sec. 2.5). Nature, though, is far from idealised, and as we see in this small sample from 3 GHz VLA-COSMOS, the sources that will be recovered by future deep and high-resolution surveys will need special care when compared to simulations and to other samples because of the biases we report in this study.

\subsection{What affects the FR structure?}
\label{sec:disc_what}

Classifying radio AGN based on the surface-brightness distribution along their radio structure is a very practical tool and is therefore widely adopted \citep[e.g.][]{vardoulaki08, mingo19}. It is not fully representative of what we see in nature, however. The classification depends on viewing angle, resolution, and sensitivity. Although the FR scheme helps in understanding the AGN energy output \citep[e.g.][]{croston18}, it is not yet clear what affects the FR structure. In this study of radio AGN in the 3 GHz VLA-COSMOS survey we adopted a parametric approach to classify sources to avoid introducing biasesion the classification. Our results for the 3 GHz VLA-COSMOS radio AGN population show no strong dependence of radio structure on physical properties (radio luminosity, size, and accretion rate) or large-scale environment. Furthermore, there is a mix of populations, and their distributions overlap. We do not see a clear dichotomy in FR type related to physical properties, as first reported by \cite{fr74}, rather a wide distribution of values. We note that the lack of correlation in FR radio structures and host properties/large-scale environment is not due to the uncertainties in the estimated properties, which are of the order of 0.1 dex \citep{delvecchio17, darvish17}. Comparison of jetted FR objects with jet-less COM AGN within the 3 GHz sample suggests that they are drawn from the same population (see Fig.~\ref{fig:ssfr_mstar}) when it comes to their host properties. They also lie in similar density environments. There are differences, however: COM AGN are more numerous, and fainter on average, they lie at higher redshifts, they have smaller sizes ($\sim$ 2 kpc), and have similar accretion rates only to FRIs when the kinetic energy is not taken into account. We note that this study does not take projection effects or redshift dependence into account, which could potentially change the FR classification because the surface brightness decreases with redshift as $(1~+~z)^{-4}$. This is related to the ability of recovering low surface brightness emission. Extensive simulations on this aspect have been reported in  \cite{liu19}, who presented a generic simulation. This is non-trivial to assess. Additionally, both Malmquist bias and resolution bias are at work.

We note that the high number of hybrid FRI/FRII radio sources in our sample is a natural consequence of our parametric classification scheme. We find 32 FRI/FRIIs ($\sim$ 25\% of the FRs in our sample), 9 (28\%) of which have an uncertain classification (Table~\ref{table:data}). Recent studies \citep[e.g.][]{banfield15, kapinska17, harwood20} have identified and studied the reasons for the existence of this hybrid class in brighter samples of radio galaxies, and attributed it to environmental effects and asymmetric density environments. Regarding the number of objects identified, \cite{Gopal-Krishna00} investigated the literature to find six hybrids, \cite{Gawronski06} report 21 hybrid candidates from the VLA Faint Images of the Radio Sky at Twenty-centimeters (FIRST) survey observed at 4.9 GHz where the sources have flux densities $S_{\rm 1.4 GHz} > 20~mJy$, much brighter than our sample. In our sample, we probed FRI/FRIIs below $S_{\rm 3 GHz}$ = 10 mJy, with only one (10904) at $S_{\rm 3 GHz}$ = 28 mJy. Results from Radio Galaxy Zoo \citep[e.g.][]{banfield15, kapinska17} revealed 25 candidate hybrids in the FIRST survey for redshifts up to 1 and brighter than 40 mJy at 1.4 GHz. Future surveys with their high resolution and sensitivity, in combination with studies probing the large-scale environment, will reveal more of these sources that were once considered to be rare \citep{Gopal-Krishna00, Gawronski06, kapinska17}. We furthermore explored how often we could miss faint outer lobes (or a faint extended envelope around bright peaks) that could be revealed in deeper imaging. We performed a back of the envelop calculation for FRI/FRIIs in our sample. We calculated the peak flux density over the faint envelope for the north lobe and dimmed it to the typical peak flux density of our sub-sample of FRI/FRIIs. For most sources we are able to see the lobe. 

In the 3 GHz VLA-COSMOS survey we probe radio AGN, extended or not, down to flux-densities of tens of $\mu$Jy. The lowest flux density we probe for an extended FR-type sources is object 3065 at 70 $\mu$Jy. This is achieved by the combination of the high resolution (0".75) and sensitivity (2.3 $\mu$Jy/beam) of the survey, which allows us to detect small FR sources down to low flux density limits, but it also affects the way they are classified based on the FR-type classification. When ew compare for the same object the 3 GHz to the 1.4 GHz VLA-COSMOS classifications, we find that $\sim$ 47\% (21 out of the 46 with classifications at 1.4 GHz) have a different classification, making the FR classification-scheme resolution and sensitivity dependent. In other words, there is a surface-brightness bias related to the FR classification scheme. This difference, where most FRII objects at 3 GHz are classified as FRI at 1.4 GHz, is probably due to the combination of the high sensitivity and resolution of the 3 GHz data. Additionally, the classification could depend on frequency and the effect of synchrotron ageing/losses, most probably are not so important in our study of comparing the 1.4 and 3 GHz FR classes, but would be more pronounced in a comparison of MHz and GHz frequencies. Thus trying to understand the relation of physical properties and environment with the FR radio structure is not trivial. 

Our results suggest that the FRI radio structure is not a result of the large-scale environment, but might be due to the effects of the environment within the host, that is on smaller scales. It has long been speculated that the ISM of the host can play a role in affecting the FR structure, causing the FR dichotomy. For example, \cite{bicknell95} attributed this difference to the different speeds the at which FR jets propagate through the ISM, which is supersonic for FRIIs or transonic and eventually subsonic for FRIs. \cite{meier13} explained why the FR dichotomy cannot originate very close to the black hole, on sub-pc scales, where the jets are produced, due to the existence of hybrid FR structures. Any asymmetries in the jet production would disappear within timescales of months for $M_{\rm BH} \sim 10^{9}$, while the differences are seen on timescales of tens of Myrs. They therefore attributed the FR dichotomy to kpc-scale or less, and they present a model where the recollimation shock caused the difference in FR structures, with the jets being reborn in the post-recollimation shock depending on the strength of the recollimation shock. When the recollimation shock, which occurs at pc scales, is strong enough, a supersonic or magnetosonic flow is obtained that results in an FRII jet. When it is not strong enough to dissipate much of the internal magnetic field, however, an FRI subsonic jet is produced. Another possibility is the type of ISM the jet encounters, being more or less dense. Recently, \cite{perucho20} computed a model in which FRI jets are disrupted as they encounter stars in the ISM of the host galaxy. According to this model, turbulence is one of the reasons that the jet loses energy and fails to develop into an FRII-type jet. With the currently available data for our sample we cannot verify this observationally, but it is a very interesting topic for future studies with high-resolution observations (e.g. ALMA) to probe the ISM of FR-type radio sources within COSMOS. With our current data for the FR radio AGN in COSMOS we cannot verify these studies. 

As we probe increasingly deeper into the radio universe, we will discover more of these discrepancies in the classic FR classification of \cite{fr74}. In this paper we highlighted the biases that are introduced by more sensitive observations and higher resolution. A study at similar sensitivity and resolution to ours but of a wider area, could better quantify the mentioned biases at full scale. Upcoming full sky radio surveys, in combination with multi-wavelength observations, will be able to do this. The FR classification scheme is a very practical tool, but future surveys, which will have to classify millions of sources, will run into the same issues we have demonstrated with our small sample. Our results can be a valuable addition to the development of future pipelines for classification purposes. If we wish to continue using the FR classification we should apply additional parameters to it, such as a radio power cut. Otherwise, we should start classifying extended radio AGN with a scheme that is more related to their physical properties. A prediction of how the source would be classified given different sensitivity, resolution, or frequency could also prove useful. Finally, as mentioned, the FR classification is applied to projected radio structures. In the future, and with the help of better algorithms, this might change to deprojected radio structures with a 3D representation.

\section{Conclusions}
\label{sec:conc}

We have investigated the connection of the radio structure in radio-selected AGN from the 3 GHz VLA-COSMOS survey \citep{smolcic17a} to their physical properties (radio power, size, and accretion rate) and large-scale environment (hosts, galaxy groups, and density fields). The purpose of this study was to address the complexity of connecting the radio structure to physical properties and to determine what drives the FR-type radio structure. We adopted a parametric classification and classified our sample into FRIIs, FRIs and FRI/FRIIs. We also included the jet-less COM AGN in our analysis. In summary, our results are:
   \begin{itemize}
   \item Within the 2.6 deg$^{2}$ of the COSMOS field at 3 GHz, we find 130 FR-type radio AGN (59 FRIIs, 32 FRI/FRIIs, and 39 FRIs) and 1818 COM AGN.
      \item We pushed the detection limit of FR-type radio sources to tens of $\mu$Jy at 3 GHz, which is deeper than past surveys. This is related to the combination of the high resolution (0".75) and sensitivity (2.3 mJy/beam) of 3 GHz VLA-COSMOS survey. The smallest FRII we detect is object 10937, with $D$ = 24.3 kpc at $z$ = 1.128, and the faintest FR object is 3065 at 70 $\mu$Jy. 
      \item Fanaroff-Riley objects can be detected up to redshifts of $z~ \sim$ 2.5, with a peak in their distribution around $z~ \sim$ 1, while COM AGN are detected up to $z~ \sim$ 6 with a peak around $z~ \sim$ 1.8.
        \item There is no clear dichotomy in the FR radio structure related to physical properties and large-scale environment. The traditional FR dichotomy is based on populations that are much brighter, and it disappears when we probe much fainter populations of radio sources ($L_{3~\rm~GHz} < 10^{25}~\rm W~Hz^{-1}~sr^{-1}$).
      \item The FR-type radio classification scheme is surface brightness biased. We find that $\sim$ 47\% (21 out of 46 with 1.4 GHz classifications) of the sources at 3 GHz have different FR classification from their counterparts at 1.4 GHz. 
      \item FRII objects are larger than FRI/FRII and FRI objects on average by a factor of 2 and 3, respectively, but there their distributions overlap. On average, FR objects have similar radio luminosities at 3 GHz. COM AGN are smaller in size than FRs and are located at larger distances on average, but they have similar radio luminosity at 3 GHz as FRs. 
      \item Kinetic energy or jet power boosts the Eddington ratio and thus the accretion power, as expected, but does not explain the classic FR dichotomy. The distributions of FRII, FRI/FRII and FRI objects overlap. 
       \item The radio AGN at 3 GHz VLA-COSMOS have sub-Eddington ratios. FRIIs accrete matter onto their BHs at similar rates as FRIs, and they produce large FRII objects with jets extending up to $\sim$ 1Mpc. 
      \item FRs and COM AGN are distributed randomly within the virial radius of their X-ray galaxy group, and their distributions peak around $r/r_{200} \sim$ 0.3 $\pm$ 0.2. Brighter COM AGN tend to lie closer to the X-ray group centre.
      \item FR objects mostly occupy massive hosts ($>10^{10.5} M_{\odot}$). At the same time there is an indication for radio-mode quenching of star formation in the hosts. FRIs in satellite hosts are less efficient in quenching SF than FRIIs. We also find five radio AGN with jets lying in the starburst region above the MS for SFGs.       
      \item Objects below the MS for SFGs, related to massive hosts, lie in cool X-ray groups with average IGM temperatures of $\sim$ 1 keV. Additionally, the older the episode of star formation the cooler the X-ray group in which FRs lie, suggesting quenching of the SF within X-ray groups by kinetic feedback. 
      \item In contrast to findings reported in literature, we do not find a connection between the density of the environment and the FR object type. We note these studies were not conucted at the same frequency, sensitivity and resolution as the 3 GHz VLA-COSMOS survey.
      \item The advantage of the 3 GHz VLA-COSMOS survey when compared to the S$^{3}$-SEX semi-empirical simulation is that it recovers FRII-type radio AGN at lower redshifts and at lower flux densities than expected. 
             \end{itemize}

The results of this study show that adopting the classic FR-type classification to categorise radio AGN should be used with care when different samples, selected at different depths and sensitivities, are compared. Care should also be taken when the classic FR classification is used in future radio surveys which will probe the radio universe with higher sensitivity and resolution. We have shown that the FR-type classification is surface-brightness biased, which makes it extremely difficult to investigate the reason for the FR dichotomy reported in \cite{fr74}. Although the FR-type classification is a useful tool for characterising extended radio sources associated with AGN, it can lead to inconsistent results when the caveats are not taken into account. A work-around for the caveats we mentioned is to expand the classic FR classification to be more representative of the bright and faint radio sky, given  current and future discoveries. An improved classification could include a stronger link to physical properties rather than the surface-brightness distribution, i.e. a radio power cut or/and an Eddington ratio cut, and orientation information that can be obtained by modelling the sources. Follow-up of our study should investigate the sub-kpc scale environments of radio AGN (e.g. with ALMA) to study the interaction of the jet and the ISM and to understand the role of small-scale environment to the radio structure.

\begin{acknowledgements}
      We would like to thank the anonymous referee for useful comments which significantly improved our manuscript. EV acknowledges funding from the DFG grant BE 1837/13-1 and from the VLBI group in MPIfR. ID is supported by the European Union's Horizon 2020 research and innovation program under the Marie Sk\l{}odowska-Curie grant agreement No 788679. Support for BM was provided by the DFG priority program 1573 "The physics of the interstellar medium". EV, EFJA, AK, BM and FB acknowledge support of the Collaborative Research Center 956, subproject A1 and C4, funded by the Deutsche Forschungsgemeinschaft (DFG). EV and EFJA would like to thank Nils Linz, Vishnu Balakrishnan for their assistance during the first phase of radio classification. This manuscript was submitted and reviewed during the covid-19 pandemic. 
\end{acknowledgements}

\bibliographystyle{aa} 

\clearpage

\appendix

\section{A parametric approach to FR classification}
\label{sec:measure_FR}

In this Section we describe the method we used to classify radio sources based on the FR classification scheme \citep[edge-brightened or edge-darkened radio sources,][]{fr74}. We followed a three-stage method, which involved visual inspection from non-experts (stage 1), visual inspection from experts (stage 2), and determination of FR type (stage 3).

\subsection{Stage 1: sample of 350 sources inspected by a team of non-experts on FR objects}
\label{sec:stage1}

The sample of 350 objects above the envelope in Fig.~\ref{fig:rest_snr} was given to seven investigators, non-experts in FR-type radio AGN in their majority, to classify the objects based on the following guidelines:
 \begin{itemize}
  	\item FRII: if objects are edge-brightened and exhibit lobes.
      	\item FRI: if objects are edge-darkened and exhibit jets.
     	\item FRI/FRII: if objects are FRI on one side and FRII on the other.
	\item radio source (RS): none of the above.
  \end{itemize}

This classification was then taken to stage 2, to be inspected by experts on FR-type objects, in order to avoid creating a sample of FR objects that is not representative of their nature.

\subsection{Stages 2 \& 3: determination of FR class}
\label{sec:stage2}

At stage 2 of the parametric approach, the 350 sources with preliminary classification were inspected by 2 experts on FR objects, and a sub-sample of 130 objects was taken to stage 3. 

Stage 3 involves manually determining which sources are edge-brightened and which edge-darkened. The reason we did not expand on our machine learning technique presented in Sec.~\ref{sec:auto_class} is the very peculiar nature of the majority of the objects, exhibiting bents, as well as the fact some of them are multi-component (composed of several radio blobs). And we leave the development of this machine learning code for FR classification as a future exercise.

In particular, we measured the length to the 3$\sigma$ contour, which is the maximum projected size of the jet/lobe, at both sides. We call the longer side DL and the shorter SL. We also measured the length to the brightest hotspot at each side of the core, where for the longer side we name it DL\_hs and for the shorter DS\_hs. We also measured the bent angle (BA in degrees), which is the angle that the two jets/lobes form in respect to each other. A fully symmetrical source will have BA = 180$^{\rm o}$, and if the jets/lobes are bent this angle will be smaller.

In case of one-sided jet/lobe we only measured the length to the brightest hotspot and compare it to the length of the 3$\sigma$ contour.
In case there are no obvious hotspots we report only the length to the 3$\sigma$ contour.
The measured values are listed in Table~\ref{table:measure_fr}, along with a note if the source is one-sided or not, and the resulting FR classification. 

FRII is a edge-brightened source, i.e. the distance to the hotspot is more than 0.5 times the distance to the maximum length of the source. If this is less than 0.5, then the source is edge-darkened or FRI.

To account for hybrid FRI/FRII objects, we measure both sides of the source i.e. those that have a lobe on one side and a jet on the other. Thus if DL$_{\rm hs} >$ 0.5$\times$DL then the larger side is FRII, otherwise it is FRI. Similarly with the sorter side, it's classified as FRII if DL$_{\rm hs} >$ 0.5$\times$DL, otherwise as FRI. 

\begin{figure}[!ht]
    \resizebox{\hsize}{!}
            {\includegraphics[trim={0cm 4cm 0cm 4cm},clip]{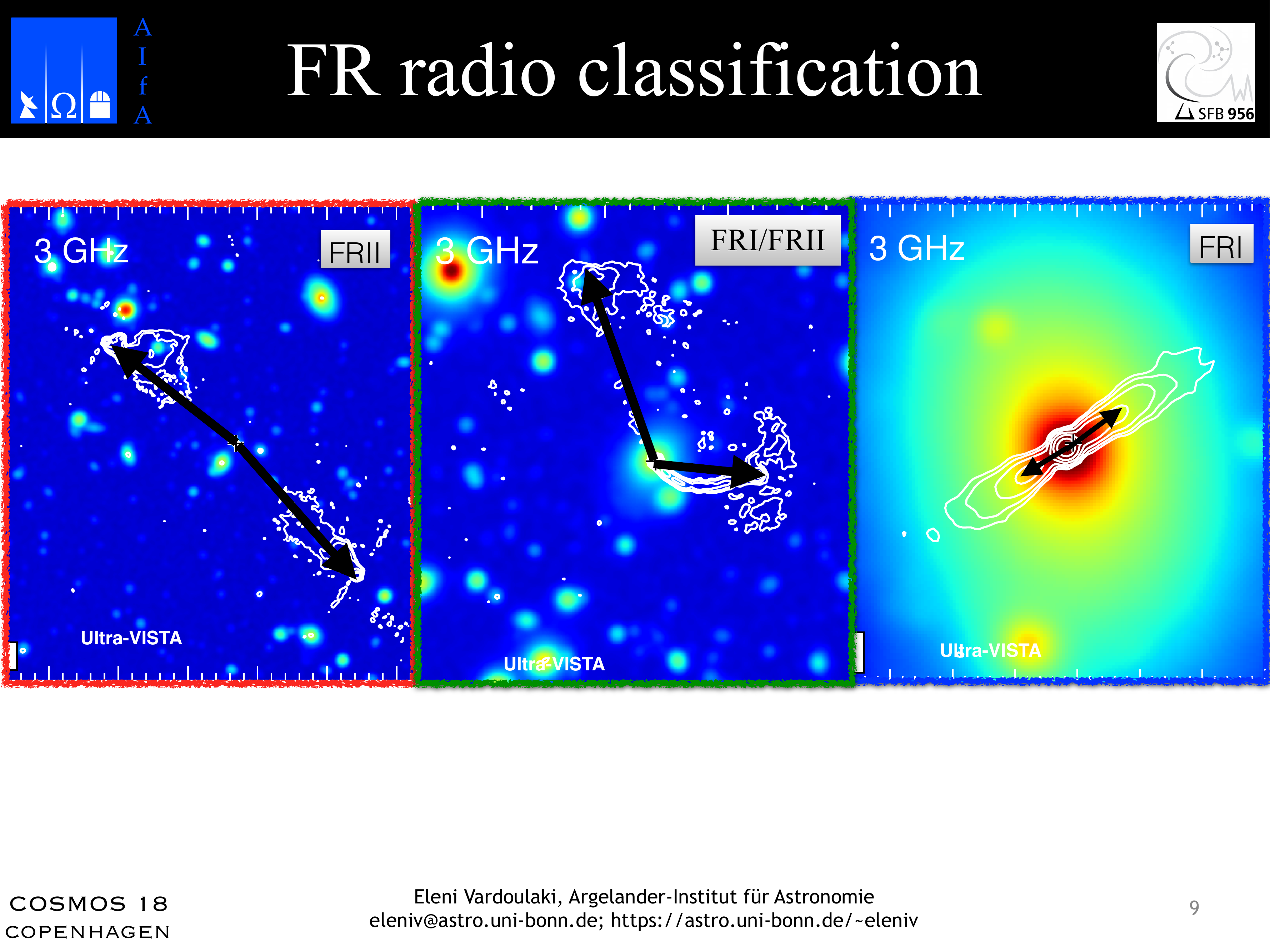}
            }
         
               \caption{Examples of FRII (left), FRI/FRII (middle), and FRI (right) radio sources. The arrows mark the point of high intensity along the radio structure. Contours are the 3 GHz data and the colour scale background is the Ultra-VISTA mosaic (see Fig.~\ref{fig:radmaps}). 
               }
              \label{fig:fr_marked}%
    \end{figure}

 In Fig.~\ref{fig:fr_marked} we give a visual example of stage 3 of the classification, where we classify objects in FRII, FRI/FRII or FRI based on the surface-brightness distribution along their radio structure.

\begin{table}
\caption{Results of parametric FR classification}             
\label{table:measure_fr}      
\centering          
\begin{tabular}{l c c c c c c l}     
\hline\hline       
                
\multicolumn{1}{c}{3 GHz}  & DL & DS & DL$\_{\rm hs}$ & DS$\_{\rm hs}$ & BA & FR \\ 
\hline                    
 \multicolumn{1}{c}{ID} &  \multicolumn{4}{c}{(arcsec)}  &  \multicolumn{1}{c}{(deg)} & class\\
 \hline                    
 \multicolumn{1}{c}{(1)} & \multicolumn{1}{c}{(2)}  & \multicolumn{1}{c}{(3)}& \multicolumn{1}{c}{(4)}& \multicolumn{1}{c}{(5)} & \multicolumn{1}{c}{(6)} & \multicolumn{1}{c}{(7)} \\
 \hline

26 & 5.2 & $-$ & 2.3 & $-$ &$-$ &FRI$^{\rm OS}$ \\
33 & 4.6 & 3.2 & 1.7 & 1.7 &         172 &FRI/FRII \\
38 & 3.4 & 1.9 & 1.1 & 1.4 &         132 &FRI/FRII? \\
44 &20.5 & 2.0 &11.4 & 1.3 &         170 &FRII \\
64 & 2.2 & 1.2 & 1.2 & $-$ &         180 &FRI/FRII? \\
80 & 8.6 & 7.6 & 1.5 & 2.1 &         176 &FRI \\
83 & 8.7 & 6.4 & 5.9 & 6.4 &         113 &FRII \\
89 & 2.5 & 2.2 & 1.5 & 1.3 &         174 &FRII \\
112 & 5.7 & 4.5 & 1.5 & 1.8 &         104 &FRI \\
115 & 5.2 & 4.6 & 2.7 & 2.2 &         164 &FRI/FRII \\
123 & 5.8 & 3.5 & 1.6 & 1.4 &          89 &FRI \\
137 & 4.8 & 4.3 & 2.6 & 3.1 &         170 &FRII \\
138 & 8.7 & 7.3 & 6.2 & 5.1 &         180 &FRII \\
145 &10.4 &10.2 & 6.8 & 7.6 &         111 &FRII \\
153 & 3.1 & 0.9 & 2.6 & $-$ &$-$ &FRI/FRII \\
160 & 3.5 & 3.5 & 2.3 & 2.0 &         180 &FRII \\
164 & 2.6 & 2.0 & $-$ & $-$ &         109 &FRI? \\
166 & 2.6 & 1.8 & 1.7 & 1.0 &         130 &FRII \\
177 &11.8 &11.6 & 8.4 & 5.4 &         133 &FRI/FRII \\
187 & 6.4 & 6.2 & 1.1 & 1.2 &         156 &FRI \\
195 & 2.5 & 1.2 & 1.7 & $-$ &$-$ &FRI/FRII \\
208 & 3.2 & 2.7 & 1.8 & 0.8 &         174 &FRI/FRII \\
213 & 2.0 & 1.9 & 0.8 & 0.9 &         180 &FRI? \\
233 & 4.5 & 3.4 & 1.8 & 1.9 &         180 &FRI/FRII? \\
236 & 2.6 & 1.2 & $-$ & $-$ &$-$ &FRI? \\
237 & 5.9 & $-$ & 3.2 & $-$ &$-$ &FRI/FRII$^{\rm OS}$ \\
247 &14.2 & 5.8 & 3.8 & 2.7 &         160 &FRI \\
248 & 4.3 & 4.1 & 3.2 & 2.3 &         178 &FRII \\
280 & 2.0 & 1.7 & $-$ & $-$ &         180 &FRI? \\
299 & 8.1 & 6.5 & 7.1 & 4.3 &         139 &FRII \\
307 & 1.9 & 0.8 & 1.2 & $-$ &$-$ &FRI/FRII? \\
311 & 2.4 & 2.2 & 1.7 & 1.0 &         175 &FRI/FRII \\
319 & 1.7 & 1.5 & 0.6 & 0.6 &         180 &FRI \\
327 & 4.7 & 3.6 & 3.8 & 2.0 &         160 &FRII \\
347 & 3.1 & 2.7 & $-$ & $-$ &         170 &FRI? \\
360 & 6.6 & 6.5 & 2.9 & 3.0 &          37 &FRI \\
386 & 3.8 & 1.2 & 1.3 & $-$ &         140 &FRI \\
404 & 4.2 & 2.2 & 2.3 & $-$ &         165 &FRI/FRII \\
433 & 8.3 & 6.4 & 2.6 & 2.6 &         180 &FRI \\
437 & 5.0 & 4.6 & 2.2 & 2.0 &         173 &FRI \\
503 & 3.4 & 0.8 & 2.3 & $-$ &$-$ &FRI/FRII \\
516 & 2.6 & 1.8 & $-$ & $-$ &         173 &FRI? \\
560 & 2.8 & 1.5 & 1.8 & $-$ &         130 &FRI/FRII \\
566 & 3.0 & 0.9 & 2.3 & $-$ &$-$ &FRI/FRII? \\
584 & 4.1 & 2.3 & 1.5 & $-$ &         167 &FRI \\
613 & 3.2 & 2.8 & 2.5 & 2.3 &         161 &FRII \\
619 & 3.1 & 1.0 & 1.5 & $-$ &$-$ &FRI \\

\hline                  
\end{tabular}
\tablefoot{Results from by-hand measurements of jet/lobe sizes to classify objects based on the FR classification scheme. {\bf Column 1}: 3 GHz ID; {\bf Column 2}: DL is the length of the larger side from the core to the 3$\sigma$ contour, or the maximum length of the jet/lobe. {\bf Column 3}: DS is the length of the shorter side from the core to the 3$\sigma$ contour, or the minimum length of the jet/lobe. {\bf Column 4}: DL$_{\rm hs}$ is the length of the larger side from the core to the brightest hot-spot of the jet/lobe. {\bf Column 5}: DS$_{\rm hs}$ is the length of the shorter side from the core to the brightest hot-spot of the jet/lobe. {\bf Column 6}: BA is the bent angle in degrees, which is the angle the jets/lobes form in respect to each other (180$^{\rm o}$ for symmetrical objects). {\bf Column 7}: FR class based on the previous measurements: FRI if DL$_{\rm hs} <$ 0.5$\times$DL and  DS$_{\rm hs} <$ 0.5$\times$DS (edge-darkened); FRII if DL$_{\rm hs} >$ 0.5$\times$DL and  DS$_{\rm hs} >$ 0.5$\times$DS (edge-brightened); FRI/FRII if one side is FRI-type and the other FRII-type; 'OS' denotes one-sided object; a '?' denotes uncertainty in the visual inspection.
}

\end{table}

\addtocounter{table}{-1}

\begin{table}
\caption{Results of parametric FR classification (continued)}             
\label{table:measure_fr}      
\centering          
\begin{tabular}{l c c c c c l}     
\hline\hline       
                
\multicolumn{1}{c}{3 GHz}  & DL & DS & DL$\_{\rm hs}$ & DS$\_{\rm hs}$ & BA & FR \\ 
\hline                    
 \multicolumn{1}{c}{ID} &  \multicolumn{4}{c}{(arcsec)}  &  \multicolumn{1}{c}{(deg)} & class\\
 \hline                    
 \multicolumn{1}{c}{(1)} & \multicolumn{1}{c}{(2)}  & \multicolumn{1}{c}{(3)}& \multicolumn{1}{c}{(4)}& \multicolumn{1}{c}{(5)} & \multicolumn{1}{c}{(6)} & \multicolumn{1}{c}{(7)} \\
 \hline
 629 & 6.9 & 6.9 & 1.5 & 3.7 &         169 &FRI/FRII \\
739 & 2.1 & 2.1 & 1.3 & 1.3 &         180 &FRII? \\
743 & 3.9 & 3.1 & 2.5 & 2.1 &         157 &FRII \\
746 & 6.9 & 4.3 & 2.5 & 2.5 &         141 &FRI/FRII \\
773 & 4.1 & 3.5 & 1.6 & 1.6 &         171 &FRI \\
798 & 2.2 & 2.0 & 0.7 & 0.8 &         180 &FRI \\
840 & 8.5 & 8.3 & 1.9 & 7.3 &         164 &FRI/FRII \\
936 & 3.2 & 3.1 & 1.8 & 1.9 &         174 &FRII \\
942 & 4.8 & 4.5 & 2.1 & 2.7 &         104 &FRI/FRII \\
976 & 4.9 & $-$ & $-$ & $-$ &$-$ &FRI$^{\rm OS}$ \\
1031 & 2.9 & 2.4 & 1.3 & 1.3 &         108 &FRI/FRII \\
1290 & 2.3 & 2.0 & $-$ & $-$ &         180 &FRI \\
1411 & 1.9 & 1.5 & 0.9 & 0.8 &         169 &FRI/FRII? \\
2251 & 4.8 & $-$ & $-$ & $-$ &$-$ &FRI$^{\rm OS}$ \\
2399 & 2.4 & 2.2 & 1.2 & 1.4 &         180 &FRII \\
2516 & 3.2 & 3.0 & 0.7 & 1.6 &         170 &FRI/FRII? \\
2660 & 2.4 & 2.0 & 1.4 & 1.2 &         175 &FRII \\
3065 & 3.9 & 1.6 & $-$ & $-$ &$-$ &FRI \\
3112 & 1.7 & 0.9 & 1.3 & 0.7 &         165 &FRII? \\
3528 & 2.0 & 1.4 & 1.4 & 0.7 &         166 &FRI/FRII \\
3866 & 2.3 & 1.4 & 1.7 & 1.4 &         176 &FRII? \\
7087 & 1.9 & 1.4 & 1.8 & 1.2 &         143 &FRII? \\
10900 &17.5 &13.0 &17.3 &16.0 &          96 &FRII \\
10901 &53.6 &51.9 &43.2 &39.0 &         176 &FRII \\
10902 &40.0 &38.5 &36.9 &35.3 &         170 &FRII \\
10903 & 9.0 & 8.1 & 3.6 & 2.2 &         123 &FRI \\
10904 &27.9 &21.7 &19.4 &10.0 &         177 &FRI/FRII \\
10905 &25.5 &24.0 &17.4 &16.8 &         175 &FRII \\
10906 & 9.9 & 9.0 & 6.0 & 4.7 &          94 &FRII \\
10907 & 6.6 & 4.2 & 5.6 & 3.9 &         148 &FRII \\
10908 & 5.9 & 4.9 & 5.4 & 3.8 &         179 &FRII \\
10909 & 8.2 & 7.3 & 6.7 & 5.5 &         167 &FRII \\
10910 &26.6 &24.0 &17.2 & 2.0 &         113 &FRI/FRII \\
10911 &22.2 & 9.7 &16.9 &15.1 &         176 &FRII \\
10912 &12.8 &11.7 & $-$ & $-$ &$-$ &FRI \\
10913 &66.5 &62.6 &30.8 &23.3 &         109 &FRI \\
10914 &13.3 &12.4 &10.0 & 1.5 &         163 &FRI/FRII \\
10915 & 5.5 & 4.0 & 3.9 & 2.0 &         158 &FRI/FRII \\
10916 &29.5 &25.5 &20.5 &19.0 &         175 &FRII \\
10917 & 4.5 & 1.6 & 4.0 & 3.5 &         166 &FRII \\
10918 &42.8 &14.4 &29.5 &10.1 &         139 &FRII \\
10919 &15.7 &14.7 &12.1 &10.5 &         179 &FRII \\
10920 & 9.4 & 8.4 & 9.1 & 4.8 &         169 &FRII \\
10921 &13.1 &11.0 & 9.6 & 2.2 &         168 &FRI/FRII \\
10922 &14.3 & 9.5 &11.3 & 8.8 &         179 &FRII \\
10923 &42.2 &40.8 &28.8 &23.2 &         166 &FRII \\
10924 &20.9 & $-$ &14.7 & $-$ &$-$ &FRI/FRII$^{\rm OS}$ \\
10925 &31.1 &28.4 &27.5 &25.6 &         173 &FRII \\
10926 & 4.4 & 3.0 & 2.2 & 1.0 &         152 &FRI \\
10927 &20.1 &19.0 & $-$ & $-$ &$-$ &FRI \\
10928 &24.5 &20.5 &14.5 &11.2 &         179 &FRII \\
10929 &10.9 &10.1 & $-$ & $-$ &$-$ &FRI \\
10930 & 6.3 & 2.7 & 4.8 & 3.3 &         169 &FRII \\
10931 &13.0 & 7.1 &11.3 & 8.1 &         111 &FRII \\
10932 & 4.8 & 3.8 & 2.4 & 1.8 &         164 &FRI \\
10933 &20.8 & 2.1 &16.7 & 5.3 &         162 &FRII \\
10934 & 6.8 & 1.7 & 5.3 & 2.5 &         175 &FRII \\
10935 &16.9 &13.0 &12.2 & 8.8 &         167 &FRII \\

\hline                  
\end{tabular}

\end{table}

\addtocounter{table}{-1}

\begin{table}
\caption{Results of parametric FR classification (continued)}             
\label{table:measure_fr}      
\centering          
\begin{tabular}{l c c c c c l}     
\hline\hline       
                
\multicolumn{1}{c}{3 GHz}  & DL & DS & DL$\_{\rm hs}$ & DS$\_{\rm hs}$ & BA & FR \\ 
\hline                    
 \multicolumn{1}{c}{ID} &  \multicolumn{4}{c}{(arcsec)}  &  \multicolumn{1}{c}{(deg)} & class\\
 \hline                    
 \multicolumn{1}{c}{(1)} & \multicolumn{1}{c}{(2)}  & \multicolumn{1}{c}{(3)}& \multicolumn{1}{c}{(4)}& \multicolumn{1}{c}{(5)} & \multicolumn{1}{c}{(6)} & \multicolumn{1}{c}{(7)} \\
 \hline
 10936 &35.7 &32.0 &21.8 & 1.4 &         177 &FRI/FRII \\
10937 & 2.4 & 1.5 & 2.2 & 1.3 &         170 &FRII \\
10938 & 3.1 & 2.4 & 1.9 & 1.4 &         163 &FRII \\
10939 & 4.8 & 3.6 & $-$ & $-$ &$-$ &FRI \\
 10940 & 3.1 & 1.9 & 2.1 & 1.1 &         180 &FRII \\
10941 & 3.9 & 2.8 & 1.7 & 0.7 &         167 &FRI \\
10942 &10.1 & 4.6 &10.1 & 5.5 &         179 &FRII \\
10943 & 4.3 & 3.6 & 3.4 & 2.4 &         171 &FRII \\
10945 & 6.2 & $-$ & 2.4 & $-$ &$-$ &FRI$^{\rm OS}$ \\
10947 & 3.9 & 3.2 & 3.2 & 2.5 &         165 &FRII \\
10948 &24.6 & 0.7 &24.3 & 0.7 &         179 &FRII \\
10949 &15.7 & 1.1 & 9.3 & 0.7 &         153 &FRII \\
10950 & 7.0 & 0.7 & 5.6 & 3.1 &         102 &FRII \\
10951 &10.7 & 2.5 & 6.2 & 5.0 &         175 &FRII \\
10952 &10.5 & 8.5 & 9.9 & 4.4 &         167 &FRII \\
10953 &17.6 & 2.0 &12.6 & 3.5 &         166 &FRII \\
10955 & 7.7 & 0.8 & $-$ & $-$ &$-$ &FRI \\
10956 &55.7 & 1.4 &39.5 & 1.9 &          57 &FRII \\
10957 &16.3 & $-$ & 6.3 & $-$ &$-$ &FRI$^{\rm OS}$ \\
10958 &15.0 & 1.4 & 4.9 & 1.0 &         101 &FRI/FRII \\
10959 &25.8 &22.8 &23.9 &21.1 &         173 &FRII \\
10962 & 8.8 & 7.5 & 8.6 & 6.2 &         175 &FRII \\
10963 & 3.0 & 1.6 & $-$ & $-$ &$-$ &FRI? \\
10964 & 3.3 & 1.7 & $-$ & $-$ &$-$ &FRI? \\
10966 & 7.3 & 5.4 & 7.3 & 4.6 &         155 &FRII \\
\hline                  
\end{tabular}

\end{table}

\section{Notes on the FR-type objects}
\label{sec:notes_fr_obj}

\noindent
{\bf 26}: This is classified as one-sided FRI source.\\
\noindent
{\bf 33}: Classified as a FRI/FRII radio source.\\ 
\noindent
{\bf 38}: A possible FRI/FRII radio source. This object lies in a masked region in the COSMOS2015 catalogue because of the presence of saturated or bright source in the optical-to-NIR bands, as shown in Fig.~1 of \cite{smolcic17b}. As a result it has not been matched with a host.\\
\noindent
{\bf 44}: A one-sided diffuse jet which is more evident at the VLA map. We classify it as one-sided FRI. This objects is a NAT and is in the same X-ray group as 10913.\\
\noindent
{\bf 64}: A possible FRI/FRII due to the elongated shape. Might be a young source and we might not be resolving the jets not even at our high resolution.\\
\noindent
{\bf 80}: A small very symmetric FRI source, with similar structure in both 1.4 and 3 GHz maps. It can also be classified as a twin-jet radio source. \\
\noindent
{\bf 83}: A small FRII source with a prominent bent in the structure on the south suggesting interaction with the intergalactic medium. \\
\noindent
{\bf 89}: A double source classified as FRII. \\
\noindent
{\bf 112}: The VLA-COSMOS map reveals a bent FRI source, or a wide-angle tail in particular, which is not evident in the lower resolution VLA map. This structure reveals strong interaction with the environment. \\
\noindent
{\bf 115}: This source is outside the coverage of the VLA-COSMOS survey and only observed at 3 GHz. It has clear core with joint jet/lobe-like structures, almost symmetric, and is classified as FRI.\\
\noindent
{\bf 123}: FRI radio source with particular structure, being more prominent on the south suggesting either diffuse emission or projection effect.\\
\noindent
{\bf 137}: FRI source which looks like a fat double radio source, i.e. diffuse lobes not separated from the rest of the source.\\
\noindent
{\bf 138}: FRII radio source since the lobes are prominent and the surface-brightness distribution along the structure is closer to the edges, i.e. the source is edge-brightened. \\
\noindent
{\bf 145}: FRII radio source, which suggests projections effects as it shows one jet-lobe structure on the south and a lobed structure on the north.\\
\noindent
{\bf 153}: Possible one-sided FRI source. The north radio blob might be associated with the underlying galaxy. The VLA radio classification (Eva Schinnerer priv. comm.) give a FRII, which we believe is a mis-classification.\\
\noindent
{\bf 160}: A double radio source, and in particular a fat-double which is thus classified as a FRI.\\
\noindent
{\bf 164}: Although the 1.4 GHz map shows a shapeless radio source, the 3 GHz map reveals a small bent radio source, which we classify as an FRI.\\
\noindent
{\bf 166}: We classify it as FRI cause of the twin-jet appearance at 3 GHz perpendicular to the host galaxy.\\
\noindent
{\bf 177}: A twin-jet FRI source with a bent at the south suggesting interaction with the environment. \\
\noindent
{\bf 187}: A fat-double FRI radio source with a slightly refined radio structure at 3 GHz than at 1.4 GHz.\\
\noindent
{\bf 195}: A possible FRI due to the jet-like feature on the north-west.\\
\noindent
{\bf 208}: A fat-double FRI from the 3 GHz map. \\
\noindent
{\bf 213}: The 3 GHz map reveals a fat-double structure thus we classify it as FRI.\\
\noindent
{\bf 233}: Elongated radio structure, possible FRI. \\
\noindent
{\bf 236}: We classify this object as FRI cause at 3 GHz the structure resembles a fat-double radio source. \\
\noindent
{\bf 237}: One-sided jet-like feature extending from the core. We classify it as FRI.\\
\noindent
{\bf 247}: A clear twin-jet FRI, where one jet appears longer than the other probably due to orientation effect.\\
\noindent
{\bf 248}: This radio source is outside the coverage of the VLA-COSMOS survey and the Ultra-VISTA coverage, thus we have no information on the host galaxy. It resembles a fat double and we classify it as FRI.\\
\noindent
{\bf 280}: Elongated radio structure at 3 GHz, classified as an FRI.\\
\noindent
{\bf 299}: A WAT source.\\
\noindent
{\bf 307}: A double radio source, with one of the components centred on the host galaxy. Classified as possible FRI.\\
\noindent
{\bf 311}: This object resembles a twin-jet FRI radio source. \\
\noindent
{\bf 319}: Looks like a young fat double FRI in the 3 GHz map.\\
\noindent
{\bf 327}: A fat double FRI radio source revealed in detail at 3 GHz.\\
\noindent
{\bf 347}: This source is not observed at 1.4 GHz nor with Ultra-VISTA, thus we don't have information on the host. We classify it as a possible FRII due to the double structure.\\
\noindent
{\bf 360}: This fat double FRI source shown interaction with the environment at the end of it's jets, which is particularly highlighted in the 3 GHz map.\\
\noindent
{\bf 386}: A hint of one-sided jet to the east. We classify it as possible FRI. \\
\noindent
{\bf 404}: This source displays a one-sided jet at the 3 GHz map, although there is no equivalent structure at the 1.4 GHz map. We classify it as FRI.\\
\noindent
{\bf 433}: A beautiful twin-jet FRI with very symmetric jets at 3 GHz. The 1.4 GHz map shows diffuse emission around the jets.\\
\noindent
{\bf 437}: A fat double radio source. We classify it as FRI.\\
\noindent
{\bf 503}: The 3 GHz map reveals a one-sided jet to the south-east. We classify it as FRI.\\
\noindent
{\bf 516}: This source shows small jets perpendicular to the host galaxy. We classify it as FRI.\\
\noindent
{\bf 560}: Classified as a possible FRI/FRII.\\
\noindent
{\bf 566}: Both the 1.4 and 3 GHz maps show a jet-like structure towards the south-west. We classify it as a possible FRI, unless it is related to star formation in the near-by galaxy. With our current resolution is hard to distinguish.\\
\noindent
{\bf 584}: A small twin-jet FRI in which the effects of orientation are evident, as the south-east jet is more elongated. The 1.4 GHz map show not signs of jet.\\
\noindent
{\bf 613}:
\noindent 
{\bf 619}: One-sided jet towards the south-east. We classify it as FRI. The blob on the north-west might not be associated with it, and it's too diffuse to be certain, plus it is not seen at the lowest resolution 1.4 GHz map.\\
\noindent
{\bf 629}: A very clear twin-jet FRI.\\
\noindent
{\bf 739}: A small and diffuse twin-jet FRI.\\
\noindent
{\bf 743}: A fat-double FRI with prominent core at 3 GHz. The core is not revealed at 1.4 GHz.\\
\noindent
{\bf 746}: A bent twin-jet FRI (WAT) in which the orientation effects are evident, as the south-east jet is larger than the north-west. The bent radio structure suggests interaction with the environment.\\
\noindent
{\bf 773} A small bent S-shaped FRI.\\
\noindent
{\bf 798}: The 3 GHz map reveals a fat-double FRI, not evident at 1.4 GHz.\\
\noindent
{\bf 840}: A twin-jet FRI with slightly bent radio structure suggesting interaction with the environment.\\
\noindent
{\bf 936}: A twin-jet FRI is revealed in the 3 GHz map. \\
\noindent
{\bf 942}: The 3 GHz map with it's high detail reveals a fat-double FRI radio source. If the galaxy at the north of the host emits synchrotron, this can contribute to the emission from the north lobe-like structure of the AGN and alter it's observed shape. Alternatively, the structure we are seeing might be due to orientation effects. A higher resolution map could help clear this situation and disentangle the emission from the AGN and the near-by galaxy.\\
\noindent
{\bf 976}: This object has no host. We classify it as one-sided FRI.\\
\noindent
{\bf 1031}: A possible FRII due to the prominent lobes, observed only at 3 GHz as it lies outside the 1.4 GHz survey coverage.\\
\noindent
{\bf 1290}: The 3 GHz map reveals a twin-jet FRI.\\
\noindent
{\bf 1411}: The 3 GHz map shows a double structure that resembles a fat-double FRI.\\
\noindent
{\bf 2251}: We classify it as one-sided FRI.\\
\noindent
{\bf 2399}: Small bent twin jets are seen in the 3 GHz map. We classify it as FRI. The object is not observed at 1.4 GHz as it falls outside the survey coverage.\\
\noindent
{\bf 2516}: 3 GHz map shows 2 radio components associated with the same optical/infrared counterpart; the core plus jet that are not detached and a detached lobe. The VLA structure differs in the sense the source is larger and resembles an FRI/FRII source with the SW lobe still attached. \\
\noindent
{\bf 2660}: The 1.4 and 3 GHz radio shapes are significantly different, with the latter revealing a fat-double FRI which is not seen at 1.4 GHz. \\
\noindent
{\bf JLVA 3065}: The 3 GHz map reveals a one-sided radio jet. We classify it as FRI.\\
\noindent
{\bf 3112}: In the 3 GHz map we see a fat-double radio structure. We classify the object as FRI.\\
\noindent
{\bf 3528}: We classify this object as FRI due to the fat-double radio structure.\\
\noindent
{\bf 3866}: We classify it as FRI at 3 GHz due to the fat-double radio structure.\\
\noindent
{\bf 7087}: This object is not observed at 1.4 GHz as it lies outside the survey coverage. At 3 GHz it resembles a fat-double FRI with a slight bent.\\
\noindent
{\bf 10900}: A bent FRII-type radio source, suggesting interaction with the environment. Emission at the east radio lobe is diffuse which is more pronounced at 3 GHz. \\
\noindent
{\bf 10901}: A symmetric FRII radio source with a prominent core. \\
\noindent
{\bf 10902}: A symmetric FRII radio source, where we can also see the core.  \\
\noindent
{\bf 10903}: Possible FRII radio source.\\
\noindent
{\bf 10904}: FRII radio source with projection effect on the south lobe. \\
\noindent
{\bf 10905}: Core and one-sided radio lobe. \\
\noindent
{\bf 10906}: At 1.4 GHz this is a fat-double FRI. At 3 GHz we classify it as FRII due to the separation of the components, and we also note the projection effect on the east lobe.\\
\noindent
{\bf 10907}: Possible FRII radio source due to the separation of the east lobe. \\
\noindent
{\bf 10908}: This object lies outside the VLA-COSMOS coverage and was not observed at 1.4 GHz. Furthermore there is no Ultra-VISTA coverage. The 3 GHz map reveals a FRII radio source.\\
\noindent
{\bf 10909}: A bent FRII radio source, probably a restarted source with a rotation in the projection of the lobes. The old AGN episode possibly gave the north-south lobes, and the new episode gives the north-west/south-east jet-like structures.\\
\noindent
{\bf 10910}: This is probably a relic FRII radio source with the south jet-lobe structure being stretched and bent due to interaction with the environment. The 1.4 GHz map reveals intense diffuse emission within the are covered by the source.\\
\noindent
{\bf 10911}: FRII radio source with the south lobe being rather diffuse.\\
\noindent
{\bf 10912}: One-sided lobe to the west of the core. We classify it as FRII. There is no other lobe evident in either of the maps. The small radio source at south-west is associated to a nearby galaxy.\\
\noindent
{\bf 10913}: The largest radio source in the sample in angular projected size, showing strong interaction with the environment. At 3 GHz the high resolution reveals the jet structure which is bent and the lobes dragged in the inter-galactic medium. We classify it as FRII. The one-sided radio source on the north-east, a NAT, is associated with another host, object 44 in our sample.\\
\noindent
{\bf 10914}: We classify this source as FRII. The peculiar radio structure is due to either projection effect or interaction with the environment. \\
\noindent
{\bf 10915}: A FRII radio source with peculiar lobes. No core is revealed at 3 GHz.\\
\noindent
{\bf 10916}: A symmetric FRII radio where we also see the core.\\
\noindent
{\bf 10917}: This object, although it is one-sided it resembles a fat-double. Thus we classify this as FRI. \\
\noindent
{\bf 10918}: A FRII with bent radio lobes, in particularly in the south where the lobe is bent perpendicular to the jet suggesting strong interaction with the environment. \\
\noindent
{\bf 10919}: A symmetric FRII radio source where we also see the core.\\
\noindent
{\bf 10920}: A symmetric FRII radio source where the east radio lobe is fainter than the west one. The core can also be seen in both maps.\\
\noindent
{\bf 10921}: A possible FRII due to the diffuse lobe-like structures at the east and west of the core.\\
\noindent
{\bf 10922}: This object is not observed at 1.4 GHz cause it lies outside the coverage. It also lacks a Ultra-VISTA detection. From the 3 GHz map we classify it as FRII, where we also see the core emission.\\
\noindent
{\bf 10923}: A symmetric FRII radio source with possible projection effect. The south-west lobe is joint with the jet from the core.\\
\noindent
{\bf 10924}: This object is not observed at 1.4 GHz as it lies outside the survey coverage, not with the Ultra-VISTA. At 3 GHz we observe an arched one-sided structure composed of 3 radio components, with a jet/lobe-like structure being the largest one. The radio position is marked at the brightest of the components. We classify it as possible FRII.\\
\noindent
{\bf 10925}: A symmetric FRII radio source with core emission observed at both maps.\\
\noindent
{\bf 10926}: The 1.4 GHz map shows a fat-double FRI, while the 3 GHz map reveals a FRII radio source. \\
\noindent
{\bf 10927}: We classify this as FRI. \\
\noindent
{\bf 10928}: FRII radio source with projection effect causing the lobes to have different structures.\\
\noindent
{\bf 10929}: At both maps we see a one-sided lobe north-east of the core. We thus classify it as FRII. There is no sign for another lobe on either of the maps.\\
\noindent
{\bf 10930}: A fat-double radio source classified as FRI. \\
\noindent
{\bf 10931}: Diffuse radio source and interaction with the environment in this WAT radio source. We classify it as FRI.\\
\noindent
{\bf 10932}: This source has not been observed at 1.4 GHz as it lies outside the coverage of the survey, and there is no Ultra-VISTA map either. It shows a possible FRII structure with core and two lobes on opposite directions. \\
\noindent
{\bf 10933}: We classify this source as FRI as it resembles a fat-double object and the lobe-like structures are closer to the core than the edges of the source. Furthermore we note that the 3 GHz map reveals some type of rotation in the emission that could be a projection effect, or a restarted AGN.\\
\noindent
{\bf 10934}: This object was classified as FRII by Schinnerer et al., but we believe it resembles a fat-double radio structure more so we classify it as FRI at 1.4 GHz. At 3 GHz the outer lobe-like structures get disattached from the rest of the body of the radio source forcing us to classify it as FRII.\\
\noindent
{\bf 10935}: FRII radio source at both 1.4 and 3 GHz, where the core is also observed.\\
\noindent
{\bf 10936}: FRII radio source with projection effect causing the different shape of the lobes.\\
\noindent
{\bf 10937}: This source is classified as FRI at 1.4 GHz cause it resembles a fat-double. At 3 GHz due to the higher resolution the lobes are separated, thus we classify it as FRII. The core is not observed. This is the smallest FRII in our sample.\\
\noindent
{\bf 10938}: This object has not been observed at 1.4 GHz as it lies outside the coverage of the survey, and there is no Ultra-VISTA map either. The 3 GHz map shows a FRII structure. We classify it as possible FRII due to lack of information regarding the host. \\
\noindent
{\bf 10939}: A peculiar one-sided radio structure, detached at 3 GHz from the core. We classify it as possible FRII.\\
\noindent
{\bf 10940}: We classify it as FRII. \\
\noindent
{\bf 10941}: We classify this as FRI.\\
\noindent
{\bf 10942}: We classify it as FRII. \\
\noindent
{\bf 10943}: We classify it as one-sided FRI. The source at north-east is probably associated with the nearby galaxy.\\
\noindent
{\bf 10945}: One-sided jet-like structure at 1.4 GHz, being refined as one-sided lobe structure at 3 GHz east of the core. There is no evidence for another lobe. We classify it as FRII.\\
\noindent
{\bf 10947}: At 1.4 GHz we have a FRI source, while at 3 GHz the source is composed of two radio components. We classify it as a possible FRII.\\
\noindent
{\bf 10948}: Diffuse radio jets. We classify it as FRII.\\
\noindent
{\bf 10949}: A bent WAT FRI radio source suggesting interaction with the environment.\\
\noindent
{\bf 10950}: The object is not observed at 1.4 GHz as it lies outside the survey coverage. It is a bent FRI radio source, otherwise a WAT. A small source in which the effects of interaction with the environment are evident. \\
\noindent
{\bf 10951}: Jet-like structures on opposite sides of the core, nevertheless they are very diffuse. We classify it as FRI at 3 GHz.\\
\noindent
{\bf 10952}: A bent twin-jet FRI, or a WAT radio source. The northern jet looks more diffuse at 3 GHz. The bent radio structure suggests interaction with the environment.\\
\noindent
{\bf 10953}: A twin-jet FRI with slight bent at the jet pointing towards the south-east. \\
\noindent
{\bf 10955}: One-sided lobe-like structure at 3 GHz with a bent towards the south. We classify it as possible FRII. \\
\noindent
{\bf 10956}: The most pronounced WAT FRI radio source of the sample, suggesting strong interaction with the environment. The 3 GHz map shows in detail the substructure along the jet.\\
\noindent
{\bf 10957}: One-sided jet-like structure at 3 GHz, while the 1.4 GHz map reveals diffuse emission around the jet. We classify it as FRI.\\
\noindent
{\bf 10958}: A bent twin-jet radio source, with diffuse emission towards the north-east jet. We classify it as FRI.\\
\noindent
{\bf 10959}: FRII radio source with visible core.\\
\noindent
{\bf 10962}: This source resembles a fat-double radio source with rather peculiar structure at 3 GHz. We classify it as FRI.\\
\noindent
{\bf 10963}: We classify it as possible FRI.\\
\noindent
{\bf 10964}: A double-like source at 3 GHz. The 1.4 GHz map it very different. We classify it as FRI. \\
\noindent
{\bf 10966}: We classify this as FRII since it is edge-brightened. There seems to be some interaction with the environment towards the end of the jet-lobe structure. Alternatively we would classify this source as a hybrid FRI/FRII source as the north-west jet-lobe resembles an FRII while the south-east structure resembles a bent FRI-type jet. It is a jet-bent source like 360.\\

\section{An approach for automatically measuring the largest
angular size (LAS) in VLA-COSMOS extended radio sources}
\label{sec:auto_class}

In the following lines we describe a Python script\footnote{Script developed by Eric F. Jim\'{e}nez-Andrade.} for automatically measuring the largest angular size (LAS) of extended radio sources identified in the VLA-COSMOS map at 3\,GHz in the COSMOS field. Thus, it aims to overcome very time-consuming routines to determine the LAS  ``by hand".  There are two main issues when trying to do so; first, a large diversity of complex morphologies together with large contamination of point like sources -- both real and spurious. Second, extended sources which split up into multi-components which are difficult to associate. To overcome those issues we propose to use advance image analysis algorithms in Python, namely: ``scikit-image skeleton"\footnote{\url{http://scikit-image.org/docs/dev/api/skimage.morphology.html}} and ``scikit-learn mean shift cluster''\footnote{\url{http://scikit-learn.org/stable/modules/generated/sklearn.cluster.MeanShift.html}}. In brief, the code works as follows.\\

{\bf I) Input, VLA-COSMOS cut-out}

We feed the script with a 160$\times$160 arcsec$^2$ cut-out from the VLA-COSMOS COSMOS 3\,GHz map centred at the position of the  radio source (as reported in the VLA-COSMOS COSMOS 3\,GHz catalogue). The cut-out size is chosen according to the largest  source in the sample, previously identified by eye.

{\bf II) Binary image}

We set up a threshold (N$\_$sigma) to identify ``islands'' of radio emission in the 160$\times$160 arcsec$^2$ cut-out. We use a conservative value of  5$\sigma$. The result is a binary image with pixel values equal 1 if the rms $>$  5$\sigma$ and 0 if rms $<$  5$\sigma$

{\bf III) Skeletonisation}
 
The "natural" step would be to associate these islands into sources, i.e. to explore clustering algorithms. Nevertheless, since we aim to associate extended radio emission into single sources we should first get rid of contaminant compact/point-like sources in the field.  Also, it should be noted that the main goal of the script is to determine the LAS of extended radio sources and, eventually, determine their FR class\footnote{Not presented here.} -- in the case of radio galaxies. Therefore, we also need to disentangle the pixels which belong to each lobe and, subsequently, determine the distance from the centre to the brightest ($d_b$) and farthest pixel ($d_f$); which will allow to apply the original  FR classification scheme \citep[FRII if $d_b/d_f>0.5$ and FRI if $d_b/d_f<0.5$][]f{r74}. The later task might be relatively simple for symmetric radio galaxies, however, the vast majority are asymmetric and complex extended radio galaxies. Thus, we need a comprehensive but simplistic representation of the source's shape.

We propose to use ``Skeletonisation algorithms'' to reduce binary objects  to 1 pixel wide representations. Skeletonisation is a process for reducing foreground regions in a binary image to a skeletal remnant, i.e. skeleton, that largely preserves the extent and connectivity of the original region while throwing away most of the original foreground pixels. This procedure allow us to measure the length  length of a shape by considering just the end points of the skeleton and finding the maximally separated pair of end points on the skeleton.

\begin{figure}[h]
	\begin{center}
		
		\includegraphics[width=9cm]{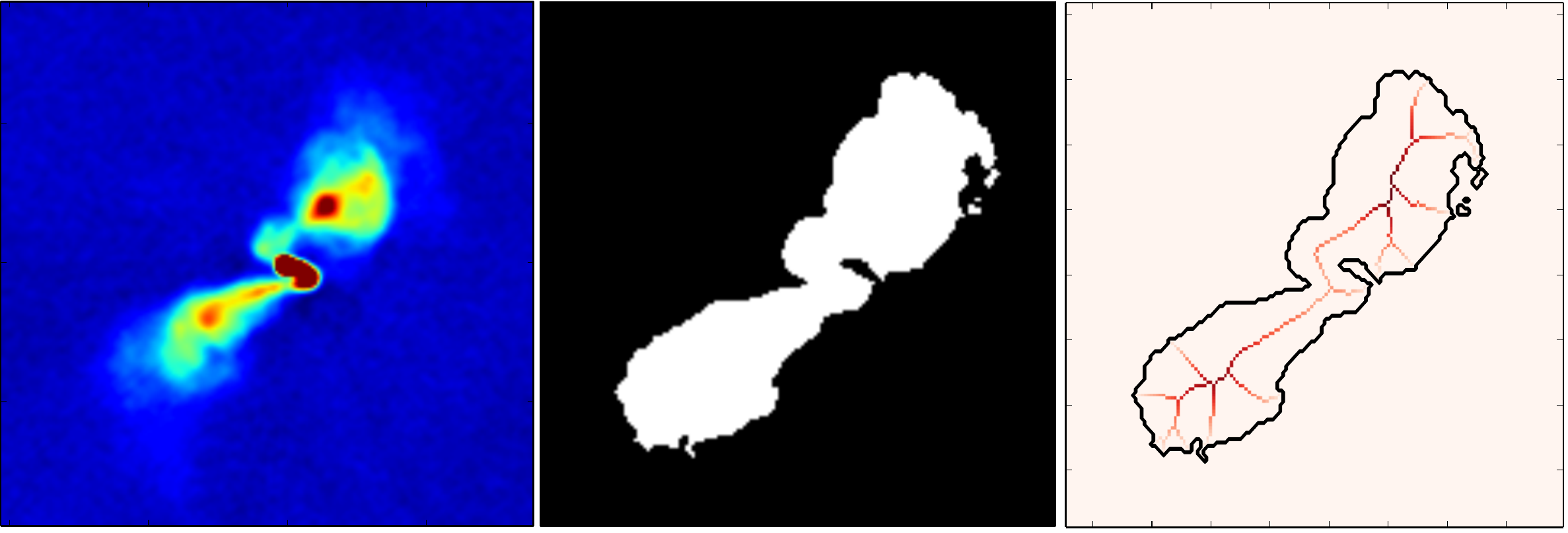}
		
		\caption{\footnotesize ({\bf Left panel}) VLA-COSMOS 10933 as observed in the 3\,GHz VLA-COSMOS radio map. ({\bf Middle panel}) Binary image obtained after setting a threshold of 5$\sigma$. ({\bf Right panel}) Its skeleton (given by the distance transform) is shown in red.     }
		\label{fig:10933_script}
	\end{center}
\end{figure}

Scikit-image, an image-processing toolbox for SciPy, offers a module on ``Skeletonisation algorithms'' which is suitable for our particular purpose. We are using the module ``Medial axis skeletonisation" to derive the skeleton/medial-axis of  islands of emission above the $5\sigma$ level (see Fig.~\ref{fig:10933_script}).  The result is an array of vectors containing information only on the pixels from the skeleton. Each vector is of the form: $skel=[rad,dec,r\_skel,flux]$; where {\it ra} and {\it dec} are the coordinates, $r\_skel$ is the distance from the skeleton to the boundaries (i.e. the local half-width of the island)  and $flux$ is the flux density corresponding to the each pixel in the skeleton (see Fig.~\ref{fig:10933_script}, bottom panel). \\ \\

{\bf IV) Removing compact sources}

Based on the parameter $r\_skel$ we can identify those islands of emission which are resolved, i.e. islands for which its width is larger than the synthesised beam of 0.75 arcsec; given the pixel scale of 0.2 arcsec/pixel  this translates into 3.75 pixels. In order to do so, we leave $r\_skel$ as an input parameter, as it depends on the synthesised beam and pixel scale. In the case of the 3\,GHz VLA-COSMOS map we set $r\_skel=2$, which yields a local width of 4 pixels given that resolved sources will have a local width above 3.75 pixels.  

This allows to get rid of all the  unresolved  sources, however, we still have to deal with {\bf circular} sources which are resolved, i.e. for which $r\_skel>2$.  At this stage, we are taking the position of all the 3\,GHz VLA-COSMOS sources (as specified in the public catalogue) as a prior. We identify all these  sources within the original cut-out of 160$\times$160 arcsec and  discard all the pixels in their skeleton  --  within a given aperture -- for further analyses (e..g. clustering). We leave the radius for this aperture (r$\_$resolved) as an input parameter. In this case, to ensure that all the pixels from the skeleton of resolved point-like sources  are removed from the field, we chose a radius  of 10 pixels, or $\sim$2.5 arcsec ($\sim$3 times the synthesised beam).

{\bf V) Clustering, mean-shift algorithm}

The next question is which islands within the cut-out belong to the same radio source. This is particularly important when dealing with crowded fields and multiple sources, which can also be extended.  Nowadays there are different clustering algorithms on 2D datasets, however, most of them rely on the number of clusters to find as an input. On the other hand, the so-called mean shift algorithm \citep{fukunaga75, cheng95} is a nonparametric clustering technique which does not require prior
knowledge of the number of clusters, and does not constrain the shape of the clusters, which is very convenient for our goal.  

\begin{figure}[!ht]
	\begin{center}
		\includegraphics[width=9cm]{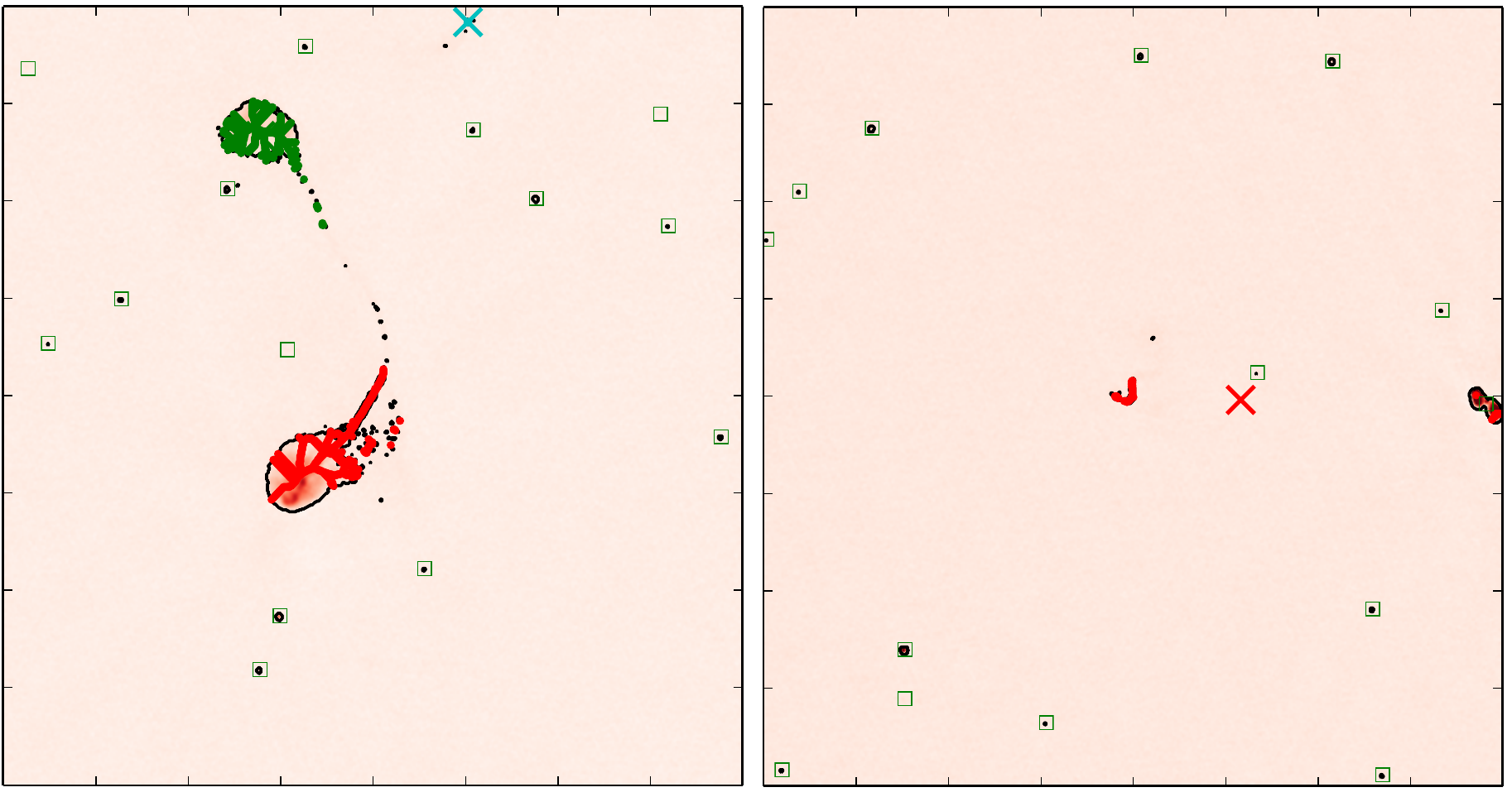}
		
		\caption{\footnotesize ({\bf Left panel}) VLA-COSMOS 10913, with a LAS$\sim$1.5 arcmin is one of the largest source in the sample. Consequently, the bandwidth should be large to recover a proper fit. When using an intermediate value of 0.5, the algorithm splits the source into two clusters (see red and green coloured pixels in the skeleton). ({\bf Right panel}) On the contrary, in the case of VLA-COSMOS 10958 with LAS$\sim$5 arcsec the bandwidth should be small. When using a high value of 0.9, the algorithm associate this source with another nearby extended (but unrelated) source in the field.  Different colours represent the different groups/clusters identified by the algorithm. Squares show the position of the compact radio sources identified with \textsc{blobcat} \citep{hales12}.}
		\label{fig:10913_script}
	\end{center}
\end{figure}

Mean shift clustering is a centroid based algorithm, which works by updating candidates for centroids to be the mean of the points within a given region. These candidates are then filtered in a post-processing stage to eliminate near-duplicates to form the final set of centroids. 
The algorithm automatically sets the number of clusters. However,  it relies on a parameter bandwidth, which dictates the size of the region to search through.
Depending on the used bandwidth  the clustering will be different. 
Then, large values for the  bandwidth tend to find a small amount of cluster and vice versa (see a nice description of the physical meaning of the kernel bandwidth in \href{https://spin.atomicobject.com/2015/05/26/mean-shift-clustering/}{ Matt Nedrich's blog}\footnote{\url{https://spin.atomicobject.com/2015/05/26/mean-shift-clustering/}}).  In other words,  very extended radio sources (e.g. LAS\,$\sim$\,1\,arcmin) would need high values in order to pack all the emission into only one single source; on the contrary, small values for the  bandwidth will suffice to group relatively compact emission (e.g. LAS\,$\sim$\,3\,arcsec) into one source (see Fig.~\ref{fig:10913_script}).

For simplicity reasons we are using a small sample to optimise the bandwidth, which allows to  recover the highest percentage of successful fits. We select a representative sub-sample ($\sim$10 sources) which comprises sources with different angular size. It should be noted that varying the bandwidth value (which goes from 0 to 1) does not strongly affect the final results as the unsuccessful fits usually occur with extreme sources (very extended or compact, see Fig.~\ref{fig:10913_script}) which represents less than 20$\%$ of the sample. We found that a bandwidth of 0.9 yields a 90$\%$ of successful fits (this was verified by eye). We leave the bandwidth as an input parameter, in principle we should  have a prior   based on analytical/statistical predictions (see the ``issues'' paragraph below).

The result from the clustering algorithm is  an array containing the coordinates of those pixel-members of the skeletons, of all the emission islands within the cut-out, along with a label to identify the different clusters found in this step. 
To select those pixels associated with the source of interest we match the label from the nearest pixel  to the central position of the cut-out, which correspond to the position of the host galaxy given in the 3\,GHz VLA-COSMOS source catalogue. 

{\bf VI) Maximising the angular size}

After all the latter steps pixels in the  skeletons of all the islands of emission, across the whole cut-out, would have been grouped into single sources.  Next, we find what is the maximum euclidean distance between two points which belong to the same source. We have to make a small correction to this distance, given that the end points of the skeleton are not the end points of the real source (at the 5$\sigma$ level, see section III). We only need to add the width in pixels of the end points of the skeleton to get the final distance. We  convert the LAS in pixels to arcsec and given the redshift and an assumed cosmology we can convert this to the Largest-projected Linear Size (LLS) of the source.

The information containing the LAS  for all the sources is stored in an ASC II file, while the fits performed in the script (skeletonisation, clustering, maximising the distance between two points) are displayed in the interactive plotting interface of Python.

\begin{figure}[!ht]
  \resizebox{\hsize}{!}
 {\includegraphics{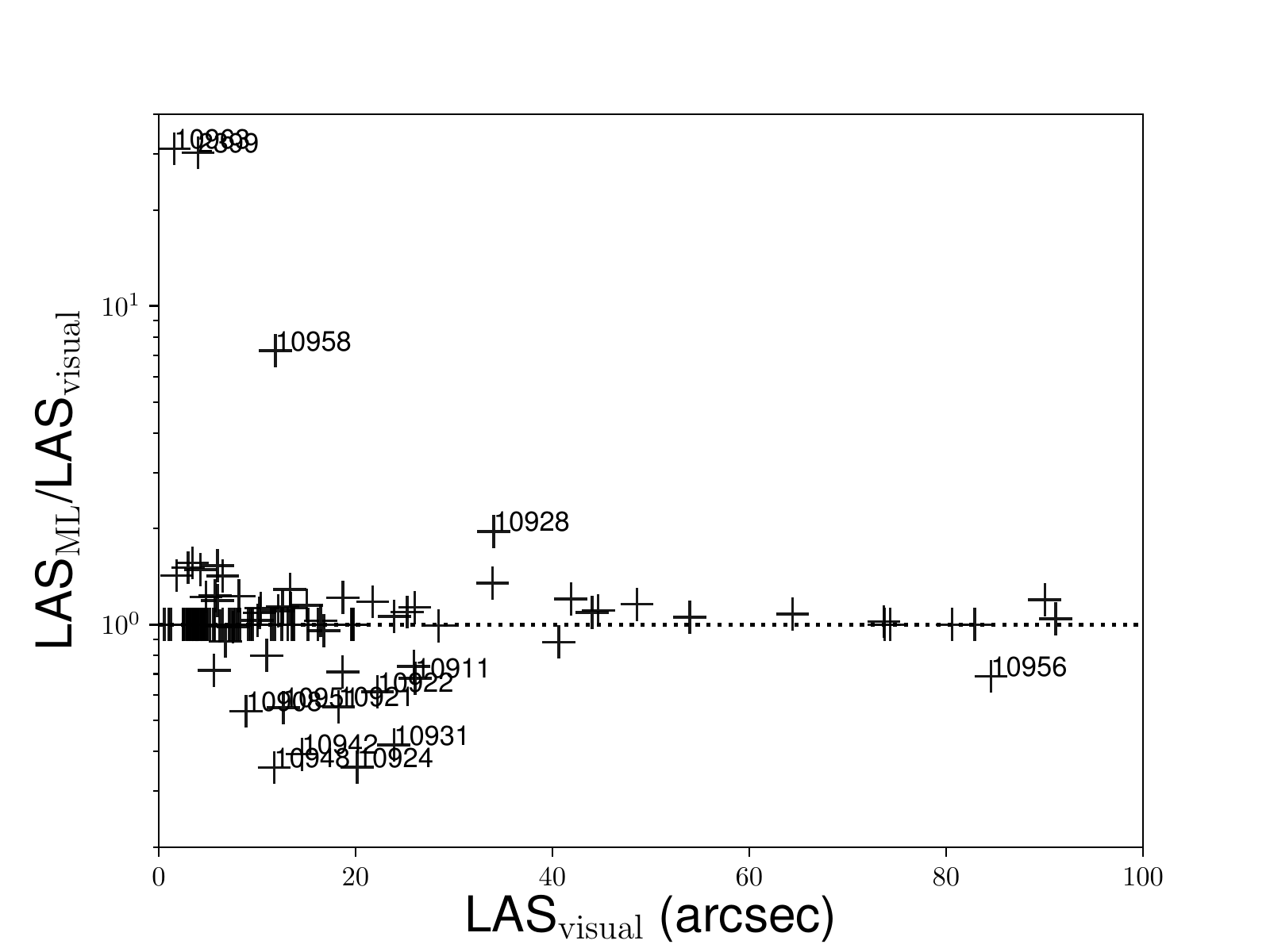}
            }
          
       \caption{Comparison between LAS measured with the semi-automatic ML method described in Appendix~\ref{sec:auto_class} and the parametrised visual method described in Appendix~\ref{sec:measure_FR}. The dotted horizontal line shows the one-to-one relation. 
   }
              \label{fig:las_comp}%
    \end{figure}

\begin{figure}[!ht]
	\begin{center}
		
		\includegraphics[width=9cm]{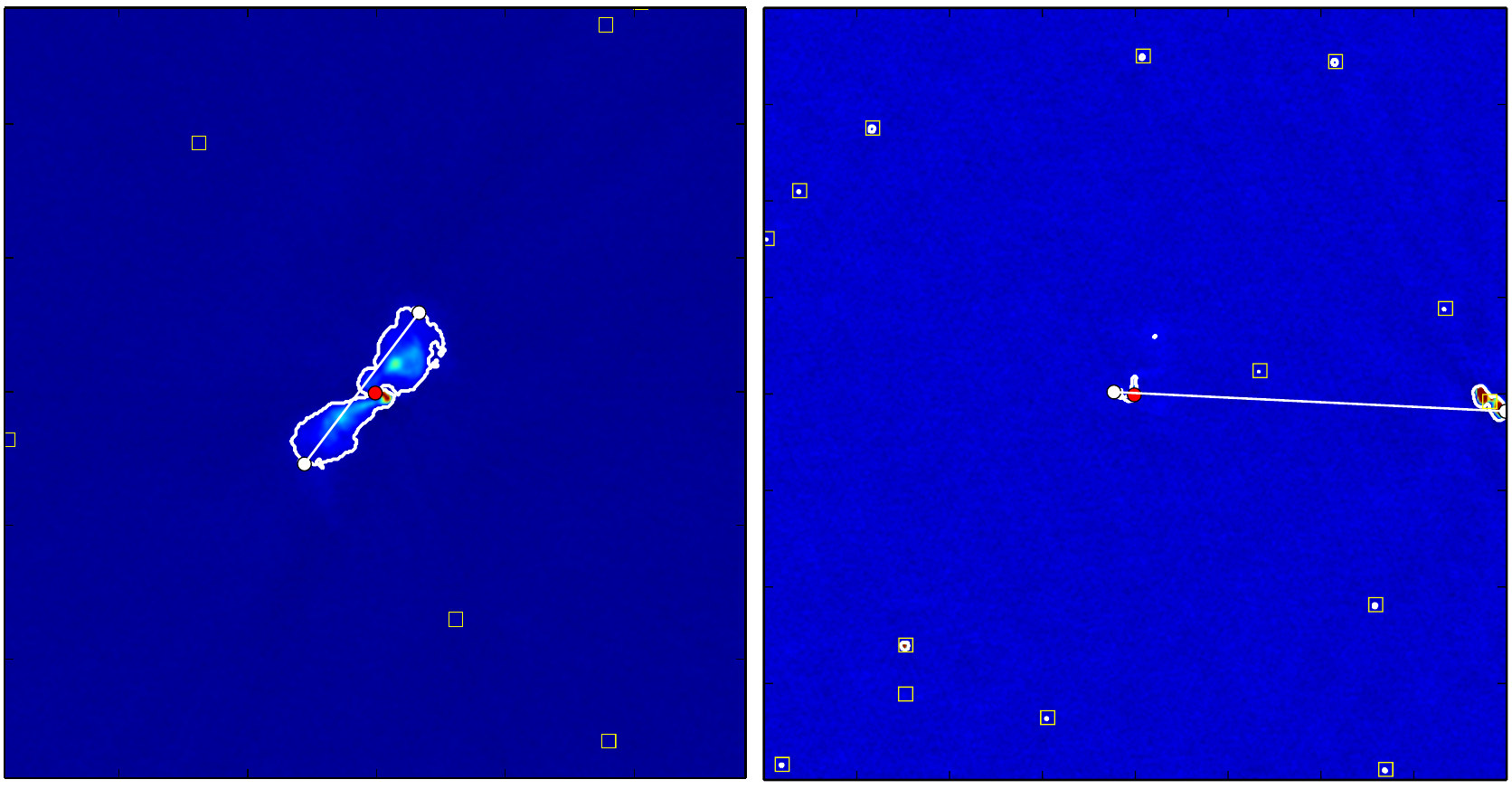}
		\caption{\footnotesize Examples of 3 GHz sources (10933 left, 10958 right) for which the semi-automatic machine learning code did not provide a correct measurement of the largest angular size. The line indicates the LAS measurement.}
		\label{fig:las_fail}
	\end{center}
\end{figure}

In Fig.~\ref{fig:las_comp} we compare the LAS measured from the semi-automatic ML method (LAS$_{\rm ML}$) and the parametrised method described in Appendix~\ref{sec:measure_FR} (LAS$_{\rm visual}$). The ML method overestimates LAS sizes for objects with nearby sources or at the edges of the 3 GHz mosaic where the noise level is higher, and underestimates LAS sizes for diffuse radio sources. Examples where the code fails to provide an accurate LAS are marked with their name in Fig.~\ref{fig:las_comp} and shown in Fig.~\ref{fig:las_fail}.

\clearpage

\begin{sidewaystable*}

\caption{Radio Properties}             
\label{table:data}      
\centering          
\begin{tabular}{l l l l c c c c c c c c c l}     
\hline\hline       
 \\               
\multicolumn{1}{c}{3 GHz}  & & \multicolumn{1}{c}{R.A.} & \multicolumn{1}{c}{Dec.} & \multicolumn{1}{c}{$S_{\rm 3GHz}$} & \multicolumn{1}{c}{$S_{\rm err}$}& \multicolumn{1}{c}{z} & \multicolumn{2}{c}{radio class} & \multicolumn{1}{c}{radio} & \multicolumn{1}{c}{X-ray} & \multicolumn{1}{c}{LAS} & \multicolumn{1}{c}{D} & \multicolumn{1}{c}{1.4GHz ID}\\ 
\hline                    
 \multicolumn{1}{c}{ID} & \multicolumn{1}{c}{COSMOSVLA3} & \multicolumn{1}{c}{(deg)}  & \multicolumn{1}{c}{(deg)} & \multicolumn{1}{c}{(mJy)} & \multicolumn{1}{c}{(mJy)} &  & \multicolumn{1}{c}{3GHz} &  \multicolumn{1}{c}{1.4GHz} &  \multicolumn{1}{c}{excess} &  \multicolumn{1}{c}{group ID}&  \multicolumn{1}{c}{(arcsec)}&  \multicolumn{1}{c}{(kpc)}&  \multicolumn{1}{c}{COSMOSVLADP}\\
 \hline                    
 \multicolumn{1}{c}{(1)} & \multicolumn{1}{c}{(2)}  & \multicolumn{1}{c}{(3)} & \multicolumn{1}{c}{(4)}& \multicolumn{1}{c}{(5)}& \multicolumn{1}{c}{(6)} & \multicolumn{1}{c}{(7)} & \multicolumn{1}{c}{(8)} & \multicolumn{1}{c}{(9)} & \multicolumn{1}{c}{(10)} & \multicolumn{1}{c}{(11)} & \multicolumn{1}{c}{(12)} & \multicolumn{1}{c}{(13)} & \multicolumn{1}{c}{(14)}\\
 \hline 
 \\
26 &J100153.45+021152.5& 150.47275 &    2.1979 &     2.769 &     0.139 & $-$ &FRI &$-$ &$-$ &$-$ &      4.43 &       0.0 &J100153+021152 \\
33 &J100131.14+022924.7& 150.37978 &    2.4902 &     2.970 &     0.148 & 0.349$^{s}$ &FRI/FRII &$-$ &T &$-$ &      4.68 &      23.1 &J100131+022924 \\
38 &J100326.62+020455.4& 150.86095 &    2.0821 &     4.924 &     0.250 & $-$ &FRI/FRII? &$-$ &$-$ &$-$ &      4.34 &     $-$.0 &$-$ \\
44 &J100026.49+024229.7& 150.11040 &    2.7083 &     2.250 &     0.113 & 0.351$^{s}$ &FRII &FRII &T &237 &     13.54 &      67.0 &J100026+024229T \\
64 &J100159.82+023904.8& 150.49928 &    2.6513 &     2.245 &     0.112 & 0.800$^{p}$ &FRI/FRII? &$-$ &T &$-$ &      3.18 &      23.9 &J100159+023904 \\
80 &J095959.61+024608.9& 149.99838 &    2.7691 &     3.021 &     0.151 & 0.166$^{s}$ &FRI &FRII &T &$-$ &     13.11 &      37.3 &J095959+024608T \\
83 &J095957.98+021809.7& 149.99159 &    2.3027 &     3.554 &     0.178 & 1.698$^{p}$ &FRII &FRII &T &$-$ &     12.51 &     105.9 &J095957+021809T \\
89 &J100109.28+021721.7& 150.28871 &    2.2894 &     1.988 &     0.099 & 2.582$^{p}$ &FRII &FRI &T &264 &      4.61 &      36.9 &J100109+021721T \\
112 &J100304.56+024917.4& 150.76901 &    2.8215 &     1.678 &     0.084 & 0.218$^{s}$ &FRI &$-$ &F &$-$ &      4.62 &      16.3 &J100304+024917 \\
115 &J100329.18+025300.6& 150.87160 &    2.8835 &     6.876 &     0.344 & $-$ &FRI/FRII &$-$ &$-$ &$-$ &      8.04 &       0.0 &$-$ \\
123 &J100122.38+020111.8& 150.34329 &    2.0200 &     8.009 &     0.400 & 0.425$^{s}$ &FRI &$-$ &F &$-$ &      7.84 &      43.7 &$-$ \\
137 &J100131.36+022639.1& 150.38069 &    2.4442 &     9.024 &     0.451 & 0.348$^{s}$ &FRII &FRI &T &$-$ &     80.61 &     396.5 &J100131+022639T \\
138 &J095908.93+024813.0& 149.78723 &    2.8036 &    13.975 &     0.699 & 1.114$^{s}$ &FRII &FRI &T &$-$ &     15.18 &     124.3 &J095908+024813T \\
145 &J100047.59+015911.5& 150.19832 &    1.9865 &    10.064 &     0.503 & 0.438$^{s}$ &FRII &$-$ &$-$ &$-$ &     19.68 &     111.6 &$-$ \\
153 &J100057.45+024216.9& 150.23938 &    2.7047 &     0.627 &     0.031 & 1.514$^{p}$ &FRI/FRII &FRII &T &$-$ &      3.71 &      31.4 &J100057+024217T \\
160 &J100007.31+024049.8& 150.03046 &    2.6805 &     2.211 &     0.111 & 1.169$^{p}$ &FRII &FRI &T &$-$ &      5.58 &      46.1 &J100007+024049T \\
164 &J100003.39+020723.0& 150.01416 &    2.1231 &     1.085 &     0.054 & 0.675$^{s}$ &FRI? &$-$ &$-$ &$-$ &      4.00 &      28.2 &J100003+020723 \\
166 &J100124.08+021706.3& 150.35036 &    2.2851 &     1.048 &     0.052 & 1.677$^{p}$ &FRII &$-$ &T &$-$ &      4.36 &      36.9 &J100124+021706 \\
177 &J100058.05+015133.3& 150.24190 &    1.8593 &     5.294 &     0.265 & 1.687$^{s}$ &FRI/FRII &$-$ &$-$ &$-$ &     19.81 &     167.7 &$-$ \\
187 &J095946.30+023602.1& 149.94295 &    2.6006 &     9.294 &     0.465 & 0.344$^{s}$ &FRI &FRI &T &$-$ &      9.50 &      46.4 &J095946+023602T \\
195 &J100002.88+013907.8& 150.01204 &    1.6522 &     0.662 &     0.033 & 1.518$^{s}$ &FRI/FRII &$-$ &T &$-$ &      3.49 &      29.6 &J100002+013907 \\
208 &J095846.38+021602.4& 149.69327 &    2.2673 &     2.898 &     0.145 & 0.905$^{s}$ &FRI/FRII &$-$ &T &$-$ &      5.19 &      40.5 &J095846+021602 \\
213 &J100112.28+014123.7& 150.30116 &    1.6899 &     0.886 &     0.044 & 0.530$^{p}$ &FRI? &$-$ &T &311 &      3.93 &      24.7 &J100112+014123 \\
233 &J100328.30+020650.4& 150.86795 &    2.1140 &     3.392 &     0.170 & $-$ &FRI/FRII? &$-$ &$-$ &$-$ &      5.69 &       0.0 &$-$ \\
236 &J095829.07+020531.1& 149.62114 &    2.0920 &     0.805 &     0.040 & 1.203$^{p}$ &FRI &$-$ &T &$-$ &      3.40 &      28.2 &J095829+020530 \\
237 &J100151.52+022532.3& 150.46469 &    2.4256 &     0.620 &     0.031 & 0.827$^{s}$ &FRI/FRII &$-$ &F &$-$ &      4.94 &      37.5 &$-$ \\
247 &J095901.42+024742.4& 149.75592 &    2.7951 &     3.516 &     0.176 & 0.490$^{s}$ &FRI &$-$ &$-$ &24 &     19.58 &     118.2 &$-$ \\
248 &J100338.97+022622.9& 150.91241 &    2.4397 &    13.837 &     0.692 & $-$ &FRII &$-$ &$-$ &$-$ &      7.34 &       0.0 &$-$ \\
280 &J100218.31+022804.0& 150.57631 &    2.4678 &     0.678 &     0.034 & 0.559$^{s}$ &FRI? &$-$ &T &$-$ &      3.86 &      24.9 &J100218+022804 \\
299 &J095807.61+025501.5& 149.53174 &    2.9171 &     7.043 &     0.352 & 2.15$^{p}$ &FRII &$-$ &$-$ &$-$ &     13.72 &     113.7 &$-$ \\
307 &J100004.93+024628.6& 150.02057 &    2.7746 &     0.593 &     0.030 & 2.091$^{s}$ &FRI/FRII? &$-$ &T &$-$ &     82.91 &     690.4 &J100004+024628 \\
311 &J095818.40+013514.2& 149.57671 &    1.5873 &     0.512 &     0.026 & 2.451$^{p}$ &FRI/FRII &$-$ &T &$-$ &      4.44 &      36.0 &J095818+013514 \\
319 &J100015.46+025231.3& 150.06445 &    2.8754 &     0.909 &     0.045 & 1.041$^{p}$ &FRI &$-$ &T &$-$ &      3.20 &      25.8 &J100015+025231 \\
327 &J095958.79+013756.5& 149.99496 &    1.6324 &     1.092 &     0.055 & 1.278$^{p}$ &FRII &FRII &T &$-$ &      7.64 &      63.9 &J095958+013756T \\
347 &J095751.96+024255.0& 149.46651 &    2.7153 &     0.571 &     0.029 & 0.880$^{s}$ &FRI? &$-$ &T &$-$ &      3.09 &      23.9 &095751+024255 \\
360 &J100201.19+021327.1& 150.50499 &    2.2242 &     2.959 &     0.148 & 0.836$^{s}$ &FRI &FRI &$-$ &$-$ &      8.24 &      62.7 &J100201+021327T \\
386 &J100141.01+015903.6& 150.42090 &    1.9843 &     0.412 &     0.021 & 0.322$^{s}$ &FRI &$-$ &$-$ &$-$ &      3.57 &      16.7 &$-$ \\
404 &J095856.37+024546.7& 149.73488 &    2.7630 &     0.336 &     0.017 & 0.754$^{p}$ &FRI/FRII &$-$ &T &$-$ &      4.01 &      29.5 &J095856+024546 \\

               \hline                  
\end{tabular}
\tablefoot{Basic radio properties and radio classification for the FR objects in our sample. {\bf Column 1 \& 2}: The 3 GHz VLA-COSMOS ID \citep{smolcic17a} and {\bf COSMOSVLA3} name respectively. {\bf Columns 3 \& 4}: Right Ascession and Declination, respectively, of radio position. {\bf Columns 5 \& 6}: Integrated flux density at 3 GHz, and corresponding error; for the multi-component sources the error corresponds to the 5\% of the calibration error, while for the rest is the error given by \textsc{blobcat}. {\bf Column 7}: Redshift and type ('s' for spectroscopic, 'p' for photometric) from \cite{laigle16}. {\bf Columns 8 \& 9} Radio classification performed by visual inspection of the VLA-COSMOS at 3 GHz, as described in Sec.~\ref{sec:measure_FR}, and VLA radio classification at 1.4 GHz given by \cite{schinnerer10}. {\bf Column 10} gives the radio excess as in \cite{delvecchio17}: 'T' for radio excess object, i.e. radio AGN, 'F' for non-radio excess. {\bf Column 11} X-ray group ID \citep[groups from][]{george11}. {\bf Column 12}: Largest angular size, projected, in arcsec, measured by a semi-automatic technique and verified by eye (see Sec.~\ref{sec:auto_class}). {\bf Column 13}: linear projected size in kpc, calculated using $z$ and LAS. {\bf Column 14}: VLA 1.4 GHz ID from \cite{schinnerer10}.
}

\end{sidewaystable*}

\addtocounter{table}{-1}

\begin{sidewaystable*}

\caption{Radio Properties (Continued)}             
\label{table:data}      
\centering          
\begin{tabular}{l l l l c c c c c c c c c l}     
\hline\hline       
 \\               
\multicolumn{1}{c}{3 GHz}  & & \multicolumn{1}{c}{R.A.} & \multicolumn{1}{c}{Dec.} & \multicolumn{1}{c}{$S_{\rm 3GHz}$} & \multicolumn{1}{c}{$S_{\rm err}$}& \multicolumn{1}{c}{z} & \multicolumn{2}{c}{radio class} & \multicolumn{1}{c}{radio} & \multicolumn{1}{c}{X-ray} & \multicolumn{1}{c}{LAS} & \multicolumn{1}{c}{D} & \multicolumn{1}{c}{1.4GHz ID}\\ 
\hline                    
 \multicolumn{1}{c}{ID} & \multicolumn{1}{c}{COSMOSVLA3} & \multicolumn{1}{c}{(deg)}  & \multicolumn{1}{c}{(deg)} & \multicolumn{1}{c}{(mJy)} & \multicolumn{1}{c}{(mJy)} &  & \multicolumn{1}{c}{3GHz} &  \multicolumn{1}{c}{1.4GHz} &  \multicolumn{1}{c}{excess} &  \multicolumn{1}{c}{group ID}&  \multicolumn{1}{c}{(arcsec)}&  \multicolumn{1}{c}{(kpc)}&  \multicolumn{1}{c}{COSMOSVLADP}\\
 \hline                    
 \multicolumn{1}{c}{(1)} & \multicolumn{1}{c}{(2)}  & \multicolumn{1}{c}{(3)} & \multicolumn{1}{c}{(4)}& \multicolumn{1}{c}{(5)}& \multicolumn{1}{c}{(6)} & \multicolumn{1}{c}{(7)} & \multicolumn{1}{c}{(8)} & \multicolumn{1}{c}{(9)} & \multicolumn{1}{c}{(10)} & \multicolumn{1}{c}{(11)} & \multicolumn{1}{c}{(12)} & \multicolumn{1}{c}{(13)} & \multicolumn{1}{c}{(14)}\\
 \hline 
\\
433 &J095901.79+025336.2& 149.75746 &    2.8934 &     1.146 &     0.057 & 0.351$^{s}$ &FRI &$-$ &F &$-$ &     11.77 &      58.3 &$-$ \\
437 &J095823.24+020859.5& 149.59685 &    2.1499 &     2.403 &     0.120 & 0.283$^{s}$ &FRI &FRI &T &$-$ &      9.34 &      40.0 &J095823+020859T \\
503 &J100243.23+015511.2& 150.68015 &    1.9198 &     0.242 &     0.012 & 1.292$^{s}$ &FRI/FRII &FRII &T &$-$ &      3.75 &      31.4 &J100243+015511T \\
516 &J100144.47+021346.0& 150.43533 &    2.2295 &     0.249 &     0.013 & 1.142$^{s}$ &FRI? &$-$ &T &$-$ &      2.96 &      24.4 &J100144+021346 \\
560 &J100256.09+021552.6& 150.73375 &    2.2646 &     0.198 &     0.010 & 1.176$^{p}$ &FRI/FRII? &$-$ &T &$-$ &      2.49 &      20.6 &J100256+021552 \\
566 &J100025.21+024328.4& 150.10507 &    2.7246 &     0.237 &     0.012 & 0.729$^{s}$ &FRI/FRII? &$-$ &T &$-$ &     74.30 &     539.5 &J100025+024328 \\
584 &J095835.90+024954.7& 149.64963 &    2.8319 &     0.307 &     0.016 & 0.705$^{s}$ &FRI &$-$ &T &$-$ &      3.52 &      25.2 &J095835+024955 \\
613 &J100101.33+022159.3& 150.25557 &    2.3665 &     0.344 &     0.017 & 2.910$^{s}$ &FRII &$-$ &$-$ &$-$ &      4.80 &      37.3 &$-$ \\
619 &J095951.42+023126.3& 149.96426 &    2.5240 &     0.184 &     0.009 & 0.687$^{s}$ &FRI &$-$ &T &$-$ &      3.37 &      23.9 &J095951+023126 \\
629 &J095929.23+022844.9& 149.87181 &    2.4791 &     0.580 &     0.029 & 0.735$^{s}$ &FRI/FRII &FRII &T &$-$ &     11.43 &      83.3 &J095929+022844T \\
739 &J095914.66+021509.0& 149.81111 &    2.2525 &     0.167 &     0.009 & 0.380$^{s}$ &FRII? &$-$ &T &246 &      2.57 &      13.4 &J095914+021509 \\
743 &J095932.82+013753.5& 149.88676 &    1.6315 &     0.672 &     0.034 & 1.335$^{p}$ &FRII &FRI &T &$-$ &      6.49 &      54.5 &J095932+013753T \\
746 &J095847.45+023408.9& 149.69771 &    2.5691 &     0.528 &     0.026 & 0.941$^{s}$ &FRI/FRII &$-$ &F &$-$ &      9.12 &      72.0 &$-$ \\
773 &J095815.49+014923.1& 149.56454 &    1.8231 &     0.282 &     0.014 & 1.507$^{s}$ &FRI &$-$ &T &$-$ &      3.99 &      33.8 &J095815+014923T \\
798 &J100128.99+025243.7& 150.37082 &    2.8788 &     0.469 &     0.024 & 0.821$^{p}$ &FRI &$-$ &T &$-$ &      3.84 &      29.1 &J100128+025243 \\
840 &J100147.49+015101.1& 150.44791 &    1.8503 &     0.704 &     0.035 & 0.894$^{s}$ &FRI/FRII &WAT &T &$-$ &     16.16 &     125.7 &J100147+015101T \\
936 &J095929.97+020345.3& 149.87489 &    2.0626 &     0.252 &     0.013 & 0.847$^{s}$ &FRII &FRII &T &$-$ &      4.18 &      31.9 &J095929+020345T \\
942 &J095947.05+024806.7& 149.94608 &    2.8019 &     0.523 &     0.026 & 0.605$^{s}$ &FRI/FRII &FRII &T &$-$ &      7.13 &      47.9 &J095947+024806 \\
976 &J095958.77+020715.0& 149.99490 &    2.1208 &     0.523 &     0.026 & $-$ &FRI &$-$ &$-$ &$-$ &     73.73 &       0.0 &095958+020714 \\
1031 &J100245.96+025647.4& 150.69151 &    2.9465 &     0.406 &     0.021 & $-$ &FRI/FRII &$-$ &$-$ &$-$ &      3.83 &       0.0 &$-$ \\
1290 &J095836.17+013919.0& 149.65074 &    1.6553 &     0.152 &     0.008 & 0.790$^{p}$ &FRI &$-$ &T &$-$ &      3.54 &      26.5 &J095836+013918 \\
1411 &J100115.51+021720.0& 150.31467 &    2.2889 &     0.225 &     0.012 & 1.200$^{p}$ &FRI/FRII? &$-$ &T &$-$ &      2.92 &      24.2 &J100115+021719 \\
2251 &J100304.62+020713.1& 150.76927 &    2.1203 &     0.413 &     0.021 & $-$ &FRI &$-$ &$-$ &239 &      4.41 &     $-$ &100103+023806 \\
2399 &J095945.76+012806.4& 149.94069 &    1.4684 &     0.472 &     0.027 & 1.805$^{p}$ &FRII &$-$ &F &$-$ &      4.00 &      33.8 &$-$ \\
2516 &J100306.82+014518.6& 150.77846 &    1.7552 &     0.160 &     0.009 & 0.966$^{p}$ &FRI/FRII &$-$ &$-$ &259 &      2.46 &      19.5 &J100306+014518 \\
2660 &J095905.54+025144.7& 149.77310 &    2.8624 &     0.142 &     0.008 & 0.901$^{p}$ &FRII &$-$ &T &$-$ &      3.90 &      30.4 &$-$ \\
3065 &J095941.38+024504.7& 149.92244 &    2.7513 &     0.075 &     0.005 & 2.467$^{p}$ &FRI &$-$ &$-$ &$-$ &      1.00 &       8.1 &J095941+024504 \\
3112 &J100331.36+014605.2& 150.88069 &    1.7681 &     0.217 &     0.012 & $-$ &FRII? &$-$ &$-$ &$-$ &      1.21 &       0.0 &$-$ \\
3528 &J100108.67+023022.6& 150.28615 &    2.5063 &     0.080 &     0.005 & 2.217$^{p}$ &FRI/FRII &$-$ &F &$-$ &      2.91 &      24.0 &J100108+023022 \\
3866 &J095950.74+022058.2& 149.96143 &    2.3495 &     0.107 &     0.006 & 0.940$^{s}$ &FRII? &FRI &T &145 &      2.50 &      19.7 &J095950+022058 \\
7087 &J095718.98+021255.1& 149.32912 &    2.2153 &     0.383 &     0.022 & $-$ &FRII? &$-$ &$-$ &$-$ &      0.57 &       0.0 &$-$ \\
10900 &J095908.31+024309.6& 149.78467 &    2.7193 &    35.170 &     1.758 & 1.308$^{s}$ &FRII &WAT &T &$-$ &     28.42 &     238.3 &J095908+024309T \\
10901 &J095758.04+015825.1& 149.49184 &    1.9737 &    18.160 &     0.908 & 2.239$^{s}$ &FRII &$-$ &T &37 &     91.13 &     751.2 &J095758+015825 \\
10902 &J095823.31+022628.4& 149.59712 &    2.4412 &    46.160 &     2.308 & 1.168$^{s}$ &FRII &$-$ &T &$-$ &     73.67 &     608.4 &$-$ \\
10903 &J100208.75+024103.2& 150.53647 &    2.6842 &     6.950 &     0.347 & 1.213$^{s}$ &FRI &$-$ &T &$-$ &     10.35 &      86.0 &J100208+024103 \\
10904 &J100243.20+015943.4& 150.68028 &    1.9958 &    28.420 &     1.421 & 1.206$^{p}$ &FRI/FRII &$-$ &T &$-$ &     33.90 &     281.3 &$-$ \\
10905 &J100229.89+023225.0& 150.62456 &    2.5403 &     3.170 &     0.159 & 0.432$^{s}$ &FRII &FRII &T &379 &     23.92 &     134.6 &J100229+023225T \\
10906 &J100212.06+023135.0& 150.55051 &    2.5264 &     8.651 &     0.433 & 0.948$^{p}$ &FRII &FRI &T &$-$ &     13.37 &     105.7 &J100212+023134T \\
10907 &J100309.43+022714.1& 150.78931 &    2.4539 &     2.814 &     0.141 & 1.211$^{p}$ &FRII &FRII &T &$-$ &      8.14 &      67.6 &J100309+022714T \\
10908 &J100339.13+015546.6& 150.91350 &    1.9296 &    26.230 &     1.311 & 0.922$^{p}$ &FRII &$-$ &F &$-$ &      8.86 &      69.5 &$-$ \\

               \hline                  
\end{tabular}

\end{sidewaystable*}

\addtocounter{table}{-1}

\begin{sidewaystable*}

\caption{Radio Properties (Continued)}             
\label{table:data}      
\centering          
\begin{tabular}{l l l l c c c c c c c c c l}     
\hline\hline       
 \\               
\multicolumn{1}{c}{3 GHz}  & & \multicolumn{1}{c}{R.A.} & \multicolumn{1}{c}{Dec.} & \multicolumn{1}{c}{$S_{\rm 3GHz}$} & \multicolumn{1}{c}{$S_{\rm err}$}& \multicolumn{1}{c}{z} & \multicolumn{2}{c}{radio class} & \multicolumn{1}{c}{radio} & \multicolumn{1}{c}{X-ray} & \multicolumn{1}{c}{LAS} & \multicolumn{1}{c}{D} & \multicolumn{1}{c}{1.4GHz ID}\\ 
\hline                    
 \multicolumn{1}{c}{ID} & \multicolumn{1}{c}{COSMOSVLA3} & \multicolumn{1}{c}{(deg)}  & \multicolumn{1}{c}{(deg)} & \multicolumn{1}{c}{(mJy)} & \multicolumn{1}{c}{(mJy)} &  & \multicolumn{1}{c}{3GHz} &  \multicolumn{1}{c}{1.4GHz} &  \multicolumn{1}{c}{excess} &  \multicolumn{1}{c}{group ID}&  \multicolumn{1}{c}{(arcsec)}&  \multicolumn{1}{c}{(kpc)}&  \multicolumn{1}{c}{COSMOSVLADP}\\
 \hline                    
 \multicolumn{1}{c}{(1)} & \multicolumn{1}{c}{(2)}  & \multicolumn{1}{c}{(3)} & \multicolumn{1}{c}{(4)}& \multicolumn{1}{c}{(5)}& \multicolumn{1}{c}{(6)} & \multicolumn{1}{c}{(7)} & \multicolumn{1}{c}{(8)} & \multicolumn{1}{c}{(9)} & \multicolumn{1}{c}{(10)} & \multicolumn{1}{c}{(11)} & \multicolumn{1}{c}{(12)} & \multicolumn{1}{c}{(13)} & \multicolumn{1}{c}{(14)}\\
 \hline
\\
10909 &J100007.91+024315.3& 150.03293 &    2.7209 &     6.828 &     0.341 & 1.438$^{p}$ &FRII &FRI &T &$-$ &     12.57 &     106.2 &J100007+024315T \\
10910 &J100049.59+014923.7& 150.20662 &    1.8233 &     5.867 &     0.293 & 0.530$^{s}$ &FRI/FRII &WAT &T &$-$ &     40.66 &     255.9 &J100049+014923T \\
10911 &J100114.85+020208.6& 150.31190 &    2.0357 &     3.425 &     0.171 & 0.971$^{s}$ &FRII &FRII &T &$-$ &     26.03 &     207.0 &J100114+020208T \\
10912 &J095802.10+021540.8& 149.50874 &    2.2613 &     1.220 &     0.061 & 0.943$^{s}$ &FRI &$-$ &T &$-$ &     18.69 &     147.5 &J095802+021540 \\
10913 &J100028.28+024103.3& 150.11786 &    2.6843 &    32.090 &     1.605 & 0.349$^{s}$ &FRI &FRI &T &$-$ &     90.02 &     443.9 &J100028+024103T \\
10914 &J100230.19+020913.2& 150.62582 &    2.1537 &     3.327 &     0.166 & 1.437$^{p}$ &FRI/FRII &$-$ &T &$-$ &     18.73 &     158.2 &$-$ \\
10915 &J095959.15+014837.7& 149.99655 &    1.8105 &     3.692 &     0.185 & 2.357$^{p}$ &FRI/FRII &FRII &T &$-$ &      5.95 &      48.6 &J095959+014837T \\
10916 &J100140.13+015129.6& 150.41718 &    1.8583 &     5.438 &     0.272 & 0.459$^{s}$ &FRII &FRI &T &$-$ &     44.65 &     260.2 &J100140+015129T \\
10917 &J100152.22+024535.3& 150.46756 &    2.7598 &     1.523 &     0.076 & 1.446$^{p}$ &FRII &$-$ &T &$-$ &      6.79 &      57.3 &J100152+024535 \\
10918 &J095824.01+024916.1& 149.60008 &    2.8212 &    25.220 &     1.261 & 0.345$^{s}$ &FRII &FRI &T &$-$ &     41.88 &     204.8 &J095824+024916T \\
10919 &J100114.13+015444.1& 150.30887 &    1.9123 &     3.157 &     0.158 & 1.483$^{p}$ &FRII &FRI &T &$-$ &     25.24 &     213.5 &J100114+015444T \\
10920 &J095839.24+013557.7& 149.66354 &    1.5994 &     1.975 &     0.099 & 1.668$^{p}$ &FRII &FRI &T &$-$ &     13.33 &     112.9 &J095839+013557T \\
10921 &J095834.09+022703.3& 149.64206 &    2.4509 &     0.981 &     0.049 & 1.319$^{p}$ &FRI/FRII &FRII &T &$-$ &     18.26 &     153.2 &J095834+022703T \\
10922 &J100343.12+023700.2& 150.92970 &    2.6168 &    24.150 &     1.208 & 1.597$^{p}$ &FRII &$-$ &T &$-$ &     22.23 &     188.3 &$-$ \\
10923 &J100303.67+014735.9& 150.76530 &    1.7933 &    13.170 &     0.659 & 1.203$^{p}$ &FRII &FRI &T &$-$ &     64.40 &     534.3 &J100303+014736T \\
10924 &J095925.79+030100.5& 149.85750 &    3.0168 &     9.353 &     0.468 & $-$ &FRI/FRII &$-$ &F &$-$ &     20.16 &       0.0 &$-$ \\
10925 &J095741.10+015122.4& 149.42126 &    1.8562 &    18.540 &     0.927 & 0.985$^{p}$ &FRII &FRI &T &$-$ &     53.95 &     430.5 &J095741+015122T \\
10926 &J095949.83+015650.3& 149.95770 &    1.9473 &     0.744 &     0.037 & 1.769$^{p}$ &FRI &FRI &T &$-$ &      4.23 &      35.7 &J095949+015650T \\
10927 &J100101.98+020511.4& 150.25827 &    2.0865 &     1.055 &     0.053 & 0.915$^{p}$ &FRI &$-$ &T &$-$ &     18.81 &     147.2 &$-$ \\
10928 &J095822.49+024722.1& 149.59373 &    2.7895 &    11.640 &     0.582 & 0.878$^{s}$ &FRII &$-$ &T &$-$ &     34.03 &     263.2 &$-$ \\
10929 &J100211.45+015458.0& 150.54771 &    1.9161 &     0.298 &     0.015 & 1.716$^{p}$ &FRI &$-$ &T &$-$ &     10.25 &      86.8 &J100211+015458 \\
10930 &J100231.43+015138.1& 150.63098 &    1.8606 &     0.578 &     0.029 & 2.165$^{s}$ &FRII &$-$ &T &$-$ &      6.50 &      53.9 &J100231+015138T \\
10931 &J095828.65+014407.6& 149.61937 &    1.7355 &     0.867 &     0.043 & 0.595$^{s}$ &FRII &WAT &T &$-$ &     23.90 &     159.1 &J095828+014407T \\
10932 &J095945.81+025924.3& 149.94090 &    2.9901 &     0.816 &     0.041 & 1.162$^{p}$ &FRI &$-$ &T &$-$ &      5.74 &      47.4 &$-$ \\
10933 &J100043.19+014607.8& 150.17995 &    1.7689 &    39.300 &     1.965 & 0.346$^{s}$ &FRII &FRII &T &$-$ &     26.02 &     127.6 &J100043+014607T \\
10934 &J100047.58+020958.5& 150.19827 &    2.1663 &     0.380 &     0.019 & 0.669$^{s}$ &FRII &FRII &T &$-$ &     10.04 &      70.4 &J100047+020958T \\
10935 &J095927.25+023729.3& 149.86354 &    2.6248 &     2.382 &     0.119 & 0.954$^{s}$ &FRII &FRI &T &$-$ &     21.73 &     172.0 &J095927+023729T \\
10936 &J100028.23+013508.5& 150.11766 &    1.5857 &    10.050 &     0.502 & 0.839$^{s}$ &FRI/FRII &$-$ &T &347 &     48.59 &     370.6 &$-$ \\
10937 &J095947.83+021023.9& 149.94931 &    2.1733 &     0.230 &     0.012 & 1.128$^{p}$ &FRII &FRI &T &$-$ &      2.96 &      24.3 &J095947+021024T \\
10938 &J100138.54+030105.7& 150.41100 &    3.0327 &     0.849 &     0.042 & 0.913$^{p}$ &FRII &$-$ &F &$-$ &      3.96 &      30.9 &$-$ \\
10939 &J100242.23+013432.1& 150.67598 &    1.5756 &     0.367 &     0.018 & 1.519$^{s}$ &FRI &$-$ &T &161 &      3.42 &      29.0 &J100242+013432 \\
10940 &J095857.37+021315.5& 149.73907 &    2.2210 &     0.204 &     0.010 & 1.024$^{s}$ &FRII &$-$ &T &$-$ &      4.11 &      33.1 &095857+021315 \\
10941 &J100102.80+012925.9& 150.26169 &    1.4905 &     0.549 &     0.027 & 2.244$^{p}$ &FRI &$-$ &T &$-$ &      4.88 &      40.2 &$-$ \\
10942 &J100034.76+014635.7& 150.14484 &    1.7766 &     0.374 &     0.019 & 0.734$^{s}$ &FRII &$-$ &F &$-$ &     14.56 &     105.9 &J100034+014635 \\
10943 &J100104.98+013154.6& 150.27080 &    1.5318 &     0.406 &     0.020 & 1.809$^{p}$ &FRII &$-$ &T &$-$ &      5.96 &      50.3 &$-$ \\
10945 &J100142.19+023049.9& 150.35081 &    2.5139 &     0.205 &     0.010 & 0.689$^{s}$ &FRI &$-$ &T &$-$ &    6.2 &      44.0 &$-$ \\
10947 &J095918.98+014035.9& 149.82909 &    1.6767 &     0.120 &     0.006 & 2.538$^{p}$ &FRII &FRI &T &$-$ &      6.20 &      49.9 &J095919+014036T \\
10948 &J100021.78+015959.9& 150.09076 &    2.0000 &     1.943 &     0.097 & 0.219$^{s}$ &FRII &$-$ &T &73 &     11.72 &      41.5 &J100021+020000 \\
10949 &J100124.06+024936.7& 150.35030 &    2.8269 &     2.718 &     0.136 & 0.826$^{s}$ &FRII &WAT &T &$-$ &     16.78 &     127.4 &J100124+024936T \\
10950 &J095949.01+025516.4& 149.95424 &    2.9212 &     1.439 &     0.072 & 0.126$^{s}$ &FRII &$-$ &T &198 &     10.97 &      24.7 &$-$ \\

               \hline                  
\end{tabular}

\end{sidewaystable*}

\addtocounter{table}{-1}
\begin{sidewaystable*}

\caption{Radio Properties (Continued)}             
\label{table:data}      
\centering          
\begin{tabular}{l l l l c c c c c c c c c l}     
\hline\hline       
 \\               
\multicolumn{1}{c}{3 GHz}  & & \multicolumn{1}{c}{R.A.} & \multicolumn{1}{c}{Dec.} & \multicolumn{1}{c}{$S_{\rm 3GHz}$} & \multicolumn{1}{c}{$S_{\rm err}$}& \multicolumn{1}{c}{z} & \multicolumn{2}{c}{radio class} & \multicolumn{1}{c}{radio} & \multicolumn{1}{c}{X-ray} & \multicolumn{1}{c}{LAS} & \multicolumn{1}{c}{D} & \multicolumn{1}{c}{1.4GHz ID}\\ 
\hline                    
 \multicolumn{1}{c}{ID} & \multicolumn{1}{c}{COSMOSVLA3} & \multicolumn{1}{c}{(deg)}  & \multicolumn{1}{c}{(deg)} & \multicolumn{1}{c}{(mJy)} & \multicolumn{1}{c}{(mJy)} &  & \multicolumn{1}{c}{3GHz} &  \multicolumn{1}{c}{1.4GHz} &  \multicolumn{1}{c}{excess} &  \multicolumn{1}{c}{group ID}&  \multicolumn{1}{c}{(arcsec)}&  \multicolumn{1}{c}{(kpc)}&  \multicolumn{1}{c}{COSMOSVLADP}\\
 \hline                    
 \multicolumn{1}{c}{(1)} & \multicolumn{1}{c}{(2)}  & \multicolumn{1}{c}{(3)} & \multicolumn{1}{c}{(4)}& \multicolumn{1}{c}{(5)}& \multicolumn{1}{c}{(6)} & \multicolumn{1}{c}{(7)} & \multicolumn{1}{c}{(8)} & \multicolumn{1}{c}{(9)} & \multicolumn{1}{c}{(10)} & \multicolumn{1}{c}{(11)} & \multicolumn{1}{c}{(12)} & \multicolumn{1}{c}{(13)} & \multicolumn{1}{c}{(14)}\\
 \hline
 \\
10951 &J095917.74+020927.8& 149.82391 &    2.1577 &     0.890 &     0.045 & 1.693$^{p}$ &FRII &$-$ &T &$-$ &     12.65 &     107.1 &J095917+020927 \\
10952 &J100238.68+022152.2& 150.66118 &    2.3645 &     1.687 &     0.084 & 0.827$^{s}$ &FRII &WAT &T &$-$ &     16.44 &     124.8 &J100238+022152T \\
10953 &J100018.50+023256.2& 150.07710 &    2.5490 &     1.254 &     0.063 & 0.890$^{s}$ &FRII &WAT &T &$-$ &     25.91 &     201.2 &J100018+023256T \\
10955 &J100307.46+023655.8& 150.78116 &    2.6155 &     1.058 &     0.053 & 0.370$^{p}$ &FRI &WAT &T &$-$ &      7.53 &      38.5 &J100307+023655 \\
10956 &J100027.43+022123.3& 150.11432 &    2.3565 &     4.379 &     0.219 & 0.220$^{s}$ &FRII &$-$ &T &$-$ &     84.56 &     300.6 &$-$ \\
10957 &J100015.55+020731.4& 150.06483 &    2.1254 &     0.514 &     0.026 & 0.661$^{s}$ &FRI &$-$ &T &$-$ &      5.62 &      39.2 &J100015+020731 \\
10958 &J100136.46+022640.7& 150.40193 &    2.4450 &     0.889 &     0.044 & 0.123$^{s}$ &FRI/FRII &$-$ &T &149 &     11.85 &      26.2 &$-$ \\
10959 &J100245.40+024516.1& 150.68919 &    2.7545 &     8.576 &     0.429 & 0.986$^{p}$ &FRII &$-$ &T &$-$ &     44.02 &     351.4 &$-$ \\
10962 &J100251.18+024248.0& 150.71330 &    2.7134 &    80.250 &     4.013 & 0.974$^{p}$ &FRII &$-$ &T &$-$ &     15.03 &     119.6 &$-$ \\
10963 &J100008.92+024010.4& 150.03725 &    2.6697 &     0.152 &     0.008 & 1.599$^{s}$ &FRI? &$-$ &F &$-$ &      1.56 &      13.2 &J100008+024010 \\
10964 &J100211.28+013706.8& 150.54750 &    1.6186 &     0.092 &     0.005 & 1.592$^{s}$ &FRI? &$-$ &F &$-$ &      1.80 &      15.3 &J100211+013706 \\
10966 &J100106.73+013320.3& 150.27806 &    1.5557 &     0.861 &     0.043 & 0.361$^{p}$ &FRII &$-$ &T &$-$ &     12.12 &      61.1 &J100106+013320T \\

\hline                  
\end{tabular}

\end{sidewaystable*}

%
%

\begin{table}
\caption{Host properties of FR objects}             
\label{table:hostprop}      
\centering          
\begin{tabular}{| l r c c l l |}     
\hline               
\multicolumn{1}{c}{3 GHz}  & SFR$_{\rm IR+UV}$ & log$_{10}$& SED& \multicolumn{1}{c}{counterpart} &
\multicolumn{1}{c}{Optical}\\                
 \multicolumn{1}{c}{ID} & (M$_{\odot}$/yr) & ($M^{*}$/M$_{\odot}$) &AGN& \multicolumn{1}{c}{ID} & \multicolumn{1}{c}{class}\\                  
 \multicolumn{1}{c}{(1)} & \multicolumn{1}{c}{(2)}  & \multicolumn{1}{c}{(3)}& \multicolumn{1}{c}{(4)}& \multicolumn{1}{c}{(5)} & \multicolumn{1}{c}{(6)}\\
 \hline  \hline   
26  &   $-$ &   $-$  & $-$  &  $-$  & D$^{B}$ \\
33  &    2.54 &   11.30  &F  & 786830$^{C15}$ & D$^{B}$ \\
38  &   $-$ &   $-$  & $-$  &  $-$ &  $-$ \\
44  &    2.63 &   11.21  &F  & 923481$^{C15}$ & E \\
64  &    3.71 &   10.77  &F  & 890793$^{C15}$ & E \\
80  &    1.47 &   11.13  &F  & 1688919$^{I}$ & $-$ \\
83  &   37.18 &   11.22  &F  & 660985$^{C15}$ & $-$ \\
89  &   12.36 &   10.93  &F  & 651098$^{C15}$ & $-$ \\
112  &   29.88 &   $-$  &F  & 307378$^{IR}$ & $-$ \\
115  &   $-$ &   $-$  & $-$  &  $-$ &  $-$ \\
123  &  430.55 &   $-$  &F  & 126603$^{IR}$ & E \\
137  &    1.95 &   11.19  &F  & 1198880$^{I}$ & E \\
138  &   29.39 &   11.12  &F  & 987678$^{C15}$ & D$^{B}$ \\
145  &   $-$ &   $-$  & $-$  &  $-$ & E \\
153  &   50.98 &   11.37  &F  & 925744$^{C15}$ & $-$ \\
160  &    7.20 &   10.58  &F  & 909080$^{C15}$ & D \\
164  &   $-$ &   $-$  & $-$  &  $-$ & D$^{B}$ \\
166  &  142.30 &   11.60  &F  & 650786$^{C15}$ & D \\
177  &    1.97 &   11.09  & $-$  & 374455$^{C15}$ &  $-$ \\
187  &    2.48 &   11.05  &F  & 858706$^{C15}$ & D$^{B}$ \\
195  & 1038.50 &   10.98  &F  & 244448$^{C15}$ & IRR \\
208  &    6.02 &   10.69  &F  & 639085$^{C15}$ & D$^{B}$ \\
213  &    2.80 &   11.24  &F  & 268559$^{C15}$ & E \\
233  &   $-$ &   $-$  & $-$  &  $-$ &  $-$ \\
236  &   16.18 &   11.45  &F  & 522790$^{C15}$ & D \\
237  &   16.18 &   $-$  &F  & 218129$^{IR}$ &  $-$ \\
247  &   -0.41 &   11.75  & $-$  & 975876$^{C15}$ &  $-$ \\
248  &   $-$ &   $-$  & $-$  &  $-$ &  $-$ \\
280  &    3.00 &   11.15  &F  & 771790$^{C15}$ & D$^{B}$ \\
299  &   2.40 & 11.34   & $-$  &  1066821$^{C15}$ &  $-$ \\
307  &  296.58 &   11.27  &F  & 968549$^{C15}$ & $-$ \\
311  &  367.53 &   11.17  &F  & 202645$^{C15}$ & $-$ \\
319  &   22.37 &   11.54  &F  & 1037768$^{C15}$ & D \\
327  &   10.54 &   11.01  &F  & 231725$^{C15}$ & $-$ \\
347  &   11.66 &   11.18  &F  & 933872$^{C15}$ &  $-$ \\
360  &   $-$ &   $-$  & $-$  &  $-$ & E \\
386  &   $-$ &   $-$  & $-$  &  $-$ & E \\
404  &    3.06 &   11.18  &T  & 962971$^{C15}$ & E \\
433  &   13.56 &   $-$  &F  & 323662$^{IR}$ & $-$ \\
437  &    0.46 &   11.17  &F  & 556464$^{C15}$ & E \\
503  &   10.50 &   10.82  &F  & 412242$^{C15}$ & $-$ \\
516  &   35.29 &   11.53  &F  & 614988$^{C15}$ & E \\
560  &   22.75 &   10.79  &F  & 635990$^{C15}$ & D \\
566  &   19.57 &   11.64  &F  & 1671795$^{I}$ & E \\
584  &    6.45 &   11.24  &F  & 1008200$^{C15}$ & E \\
613  &  139.02 &   10.11  &T  & 703742$^{C15}$ &  $-$ \\
619  &   13.32 &   11.36  &T  & 808541$^{C15}$ & D \\

  \hline                  
\end{tabular}
\tablefoot{Host-galaxy properties. {\bf Column 1}: 3 GHz ID; {\bf Columns 2 \& 3}: SFR in M$_{\odot}$/yr and stellar mass $M^{*}$ in M$_{\odot}$ from the fit to the IR+UV SED  \cite{delvecchio17}, respectively;  {\bf Column 4}: AGN based on the SED fit as in \cite{delvecchio17}: "T" for AGN, "F" no AGN; {\bf Column 5}: counterpart ID from \cite{smolcic17b}; C15 stands for COSMOS2015; IR is for IRAC ID; I is for i-band ID; {\bf Column 5}: Gives the optical morphology from \cite{schinnerer10}: "E" for early-type elliptical, "D" for disk galaxy, "IRR" for irregular/peculiar galaxy, "B" for bulge dominated disk. 
}

\end{table}

\addtocounter{table}{-1}

\begin{table}
\caption{Host properties of FR objects (continued)}             
\label{table:data2}      
\centering          
\begin{tabular}{|  l r c c l l  |}     
\hline          
\multicolumn{1}{c}{3 GHz}  & SFR$_{\rm IR+UV}$ & log$_{10}$& SED& \multicolumn{1}{c}{counterpart} &
\multicolumn{1}{c}{Optical}\\                  
 \multicolumn{1}{c}{ID} & (M$_{\odot}$/yr) & ($M^{*}$/M$_{\odot}$) &AGN& \multicolumn{1}{c}{ID} & \multicolumn{1}{c}{class}\\
                   
 \multicolumn{1}{c}{(1)} & \multicolumn{1}{c}{(2)}  & \multicolumn{1}{c}{(3)}& \multicolumn{1}{c}{(4)}& \multicolumn{1}{c}{(5)} & \multicolumn{1}{c}{(6)}\\
 \hline  \hline 
629  &   11.26 &   11.21  &F  & 779565$^{C15}$ & E \\
739  &    2.31 &   11.32  &F  & 628153$^{C15}$ & D$^{B}$ \\
743  &   45.47 &   11.38  &F  & 231975$^{C15}$ & D \\
746  &   12.80 &   $-$  &F  & 250473$^{IR}$ &  $-$ \\
773  &  169.46 &   10.67  &T  & 638834$^{I}$ & $-$ \\
\hline                  
\end{tabular}

\end{table}

\addtocounter{table}{-1}

\begin{table}
\caption{Host properties of FR objects (continued)}             
\label{table:data2}      
\centering          
\begin{tabular}{ |  l r c c l l  | }     
\hline             
\multicolumn{1}{c}{3 GHz}  & SFR$_{\rm IR+UV}$ & log$_{10}$& SED& \multicolumn{1}{c}{counterpart} &
\multicolumn{1}{c}{Optical}\\                  
 \multicolumn{1}{c}{ID} & (M$_{\odot}$/yr) & ($M^{*}$/M$_{\odot}$) &AGN& \multicolumn{1}{c}{ID} & \multicolumn{1}{c}{class}\\
 \multicolumn{1}{c}{(1)} & \multicolumn{1}{c}{(2)}  & \multicolumn{1}{c}{(3)}& \multicolumn{1}{c}{(4)}& \multicolumn{1}{c}{(5)} & \multicolumn{1}{c}{(6)}\\
 \hline \hline
10932  &   30.81 &   11.61  &F  & 2011437$^{I}$ & $-$ \\
10933  &    3.93 &   11.53  &F  & 305535$^{C15}$ & $-$ \\
10934  &    5.71 &   11.21  &F  & 570506$^{C15}$ & E \\
10935  &   36.20 &   11.33  &F  & 873867$^{C15}$ & E \\
10936  &    4.76 &   11.01  &F  & 202465$^{C15}$ & D$^{B}$ \\
\hline                  
\end{tabular}

\end{table}

%
\begin{table}[!ht]
\caption{Eddington ratios for FRs and COM AGN}             
\label{tab:edd_qjet}      
\centering                          
\begin{tabular}{| l  l l l l l l | }       
\hline
ID  &    radio      &              $\lambda_{r}$ &           $\lambda_{rk}$  &     log$_{10}($      &   $L_{\rm X} $/   &    log$_{10}$( \\
&             class                          &             &                     &                               $L_{\rm jet}$ /                      &          $L_{\rm radio}$              &  $M_{\rm BH} /$ \\
&                        &                 &                     &                                 erg~s$^{-1})$                       &                &  $M_{\odot})$ \\
\hline\hline                 
2 & {\small COM AGN} &-0.93 &-0.4 &45.86 & 1.78 &8.27 \\
3 & {\small COM AGN} &-1.00 &-0.2 &45.35 & 1.80 &7.51 \\
6 & {\small COM AGN} &-2.77 &-1.0 &45.67 & 0.69 &8.53 \\
7 & {\small COM AGN} &-1.26 &-1.1 &45.44 & 2.59 &8.92 \\
8 & {\small COM AGN} &-0.68 &-0.6 &45.28 & 2.85 &8.38 \\
\hline 
\end{tabular}

\end{table}

\addtocounter{table}{-1}

\begin{table}[!ht]
\caption{Eddington ratios for FRs and COM AGN (continued)}             
\label{tab:edd_qjet}      
\centering                          
\begin{tabular}{| l l l l l l l | }       
\hline
ID  &    radio     &           $\lambda_{r}$ &           $\lambda_{rk}$  &     log$_{10}($      &   $L_{\rm X} $/   &    log$_{10}$( \\
&             class                   &             &                     &                               $L_{\rm jet}$ /                      &          $L_{\rm radio}$              &  $M_{\rm BH} /$ \\
&                       &                  &                     &                                 erg~s$^{-1})$                       &                &  $M_{\odot})$ \\
\hline\hline    
307 & {\small FRI/FRII} &-2.45 &-1.3 &45.33 & 1.47 &8.57 \\
333 & {\small COM AGN} &-2.09 &-1.9 &43.83 & 3.45 &8.19 \\
350 & {\small COM AGN} &-2.06 &-1.4 &45.03 & 2.08 &8.42 \\
352 & {\small COM AGN} &-3.10 &-2.5 &43.82 & 2.72 &8.40 \\
358 & {\small COM AGN} &-0.24 &-0.2 &44.53 & 4.06 &8.24 \\
\hline 
\end{tabular}

\end{table}

\addtocounter{table}{-1}

\begin{table}[!ht]
\caption{Eddington ratios for FRs and COM AGN (continued)}             
\label{tab:edd_qjet}      
\centering                          
\begin{tabular}{| l l l l l l l | }       
\hline
ID  &    radio        &           $\lambda_{r}$ &           $\lambda_{rk}$  &     log$_{10}($      &   $L_{\rm X} $/   &    log$_{10}$( \\
&             class                           &             &                     &                               $L_{\rm jet}$ /                      &          $L_{\rm radio}$              &  $M_{\rm BH} /$ \\
&                       &                       &                     &                                 erg~s$^{-1})$                       &                &  $M_{\odot})$ \\
\hline\hline    
1061 & {\small COM AGN} &-2.66 &-2.5 &43.72 & 3.48 &8.59 \\
1076 & {\small COM AGN} &-1.21 &-1.0 &44.30 & 3.16 &7.69 \\
1082 & {\small COM AGN} &-3.56 &-3.2 &43.22 & 3.25 &8.59 \\
1136 & {\small COM AGN} &-2.88 &-2.4 &43.85 & 2.79 &8.29 \\
1139 & {\small COM AGN} &-2.34 &-1.8 &44.49 & 2.51 &8.36 \\
\hline 
\end{tabular}

\end{table}

\addtocounter{table}{-1}

\begin{table}[!ht]
\caption{Eddington ratios for FRs and COM AGN (continued)}             
\label{tab:edd_qjet}      
\centering                          
\begin{tabular}{| l l l l l l l | }       
\hline
ID  &    radio      &           $\lambda_{r}$ &           $\lambda_{rk}$  &     log$_{10}($      &   $L_{\rm X} $/   &    log$_{10}$( \\
&             class                          &             &                     &                               $L_{\rm jet}$ /                      &          $L_{\rm radio}$              &  $M_{\rm BH} /$ \\
&                       &                      &                     &                                 erg~s$^{-1})$                       &                &  $M_{\odot})$ \\
\hline\hline    
2236 & {\small COM AGN} &-1.94 &-1.8 &43.88 & 3.65 &8.32 \\
2244 & {\small COM AGN} &-0.51 &-0.5 &44.31 & 4.07 &8.14 \\
2267 & {\small COM AGN} &-2.36 &-2.1 &43.53 & 3.42 &7.95 \\
2287 & {\small COM AGN} &-2.80 &-2.6 &42.68 & 3.74 &7.59 \\
2291 & {\small COM AGN} &-2.15 &-2.1 &43.47 & 3.95 &8.28 \\
\hline 
\end{tabular}

\end{table}

\addtocounter{table}{-1}

\begin{table}[!ht]
\caption{Eddington ratios for FRs and COM AGN (continued)}             
\label{tab:edd_qjet}      
\centering                          
\begin{tabular}{| l l l l l l l | }       
\hline
ID  &    radio       &           $\lambda_{r}$ &           $\lambda_{rk}$  &     log$_{10}($      &   $L_{\rm X} $/   &    log$_{10}$( \\
&             class                          &             &                     &                               $L_{\rm jet}$ /                      &          $L_{\rm radio}$              &  $M_{\rm BH} /$ \\
&                       &                  &                     &                                 erg~s$^{-1})$                       &                &  $M_{\odot})$ \\
\hline\hline    
4417 & {\small COM AGN} &-2.13 &-2.0 &43.49 & 3.75 &8.02 \\
4492 & {\small COM AGN} &-2.71 &-2.5 &43.16 & 3.69 &8.09 \\
4599 & {\small COM AGN} &-1.67 &-1.6 &43.90 & 3.85 &8.34 \\
4710 & {\small COM AGN} &-1.85 &-1.6 &44.05 & 3.22 &8.01 \\
4761 & {\small COM AGN} &-2.47 &-2.2 &44.38 & 3.03 &8.90 \\
\hline 
\end{tabular}

\end{table}

\addtocounter{table}{-1}

\begin{table}[!ht]
\caption{Eddington ratios for FRs and COM AGN (continued)}             
\label{tab:edd_qjet}      
\centering                          
\begin{tabular}{| l l l l l l l | }       
\hline
ID  &    radio        &           $\lambda_{r}$ &           $\lambda_{rk}$  &     log$_{10}($      &   $L_{\rm X} $/   &    log$_{10}$( \\
&             class                               &             &                     &                               $L_{\rm jet}$ /                      &          $L_{\rm radio}$              &  $M_{\rm BH} /$ \\
&                       &                          &                     &                                 erg~s$^{-1})$                       &                &  $M_{\odot})$ \\
\hline\hline    
10902 & {\small FRII} &-1.11 & 0.0 &46.03 & 1.12 &7.91 \\
10903 & {\small FRI} &-1.82 &-1.0 &45.35 & 1.75 &8.31 \\
10905 & {\small FRII} &-2.29 &-2.1 &44.43 & 2.98 &8.76 \\
10906 & {\small FRII} &-2.18 &-1.0 &45.49 & 1.35 &8.41 \\
10909 & {\small FRII} &-2.73 &-1.4 &45.63 & 1.13 &8.93 \\
10910 & {\small FRI/FRII} &-3.57 &-1.7 &44.97 & 0.91 &8.60 \\
10911 & {\small FRII} &-1.68 &-1.4 &45.20 & 2.49 &8.79 \\
10912 & {\small FRI} &-2.64 &-2.2 &44.44 & 2.65 &8.76 \\
10913 & {\small FRI} &-4.28 &-1.9 &45.19 & 0.26 &8.98 \\
10914 & {\small FRI/FRII} &-2.52 &-1.2 &45.33 & 1.27 &8.42 \\
10918 & {\small FRII} &-3.78 &-1.9 &45.04 & 0.90 &8.89 \\
10919 & {\small FRII} &-0.73 &-0.4 &45.63 & 2.27 &8.30 \\
10920 & {\small FRII} &-1.31 &-0.6 &45.52 & 1.77 &8.06 \\
10923 & {\small FRII} &-2.31 &-1.0 &45.98 & 1.04 &8.93 \\
10930 & {\small FRII} &-2.16 &-1.5 &45.20 & 2.03 &8.71 \\
10933 & {\small FRII} &-3.73 &-1.7 &45.19 & 0.68 &8.83 \\
10936 & {\small FRI/FRII} &-2.08 &-0.8 &45.59 & 1.21 &8.31 \\
10940 & {\small FRII} & 0.64 & 0.6 &44.25 & 4.66 &7.69 \\
10948 & {\small FRII} &-4.12 &-3.0 &43.66 & 2.08 &8.61 \\
10949 & {\small FRII} &-2.45 &-1.8 &44.94 & 2.11 &8.70 \\
10958 & {\small FRI/FRII} &-4.12 &-2.7 &43.25 & 1.88 &7.63 \\
10962 & {\small FRII} &-2.40 &-0.1 &46.23 & 0.03 &8.25 \\
10964 & {\small FRI} &-1.22 &-1.2 &43.92 & 7.31 &9.42 \\
\hline 
\end{tabular}

\end{table}

\clearpage

   \begin{figure*}[!ht]
    \resizebox{\hsize}{!}
            {\includegraphics{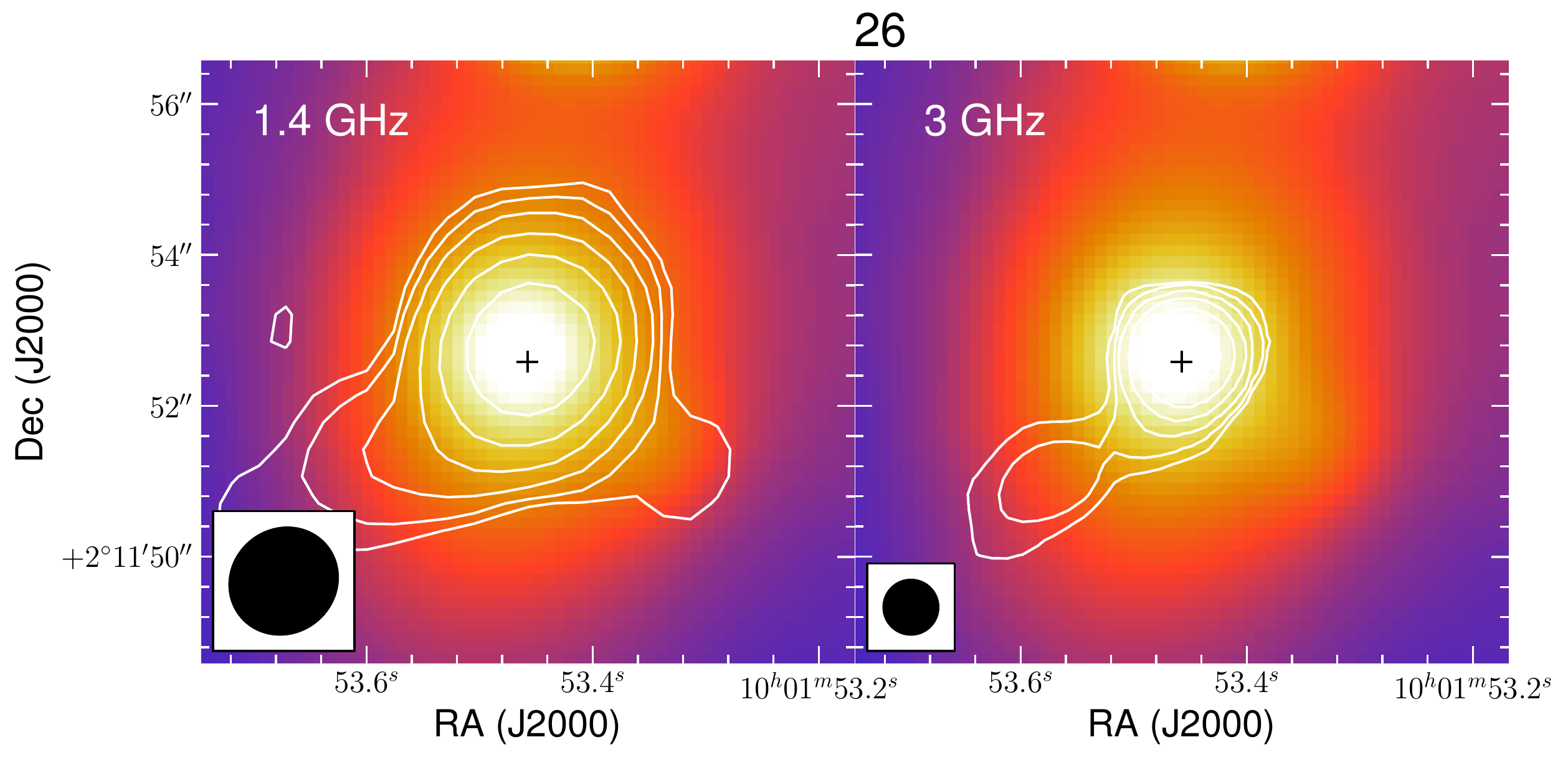}
            \includegraphics{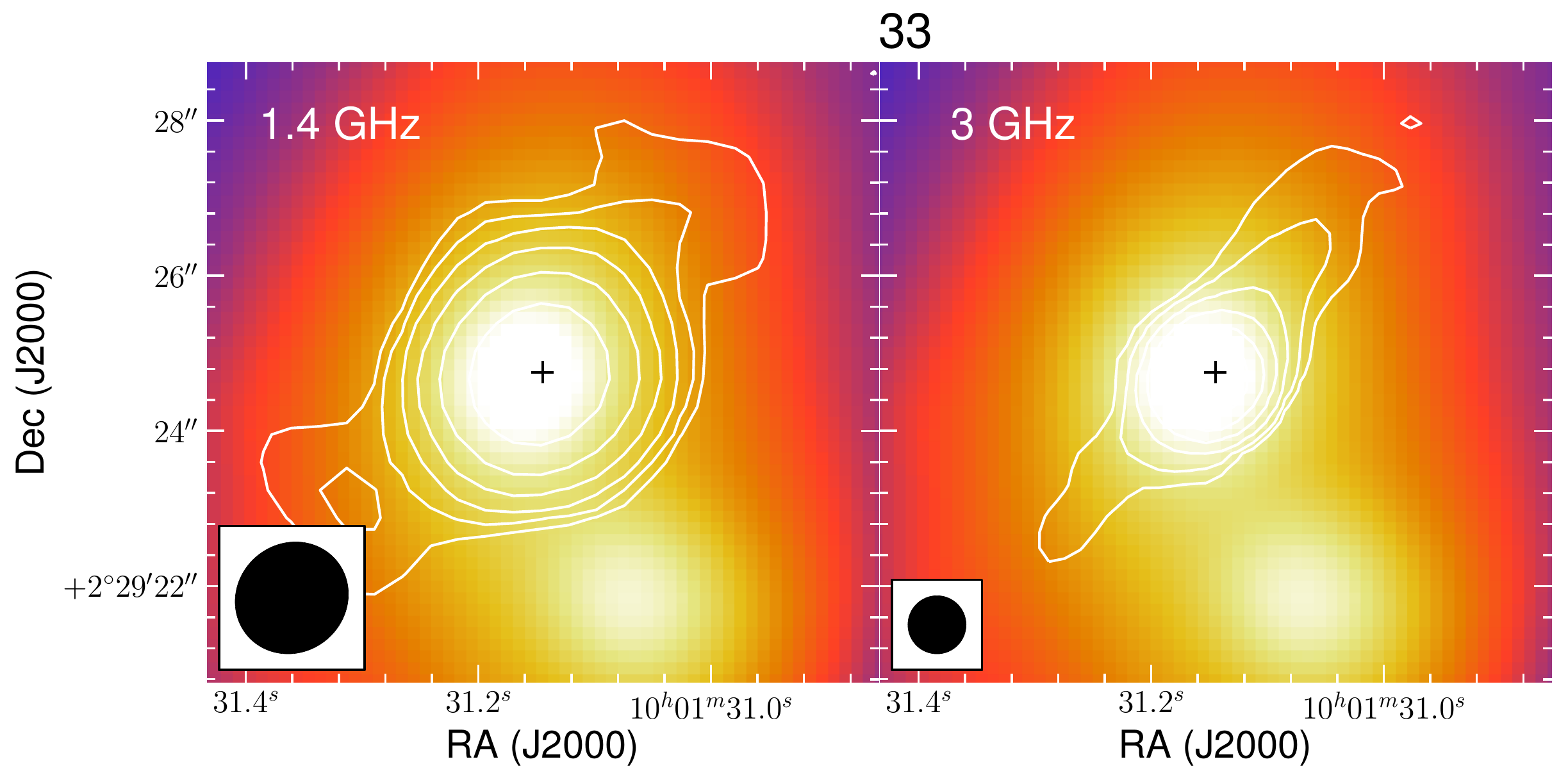}
            \includegraphics{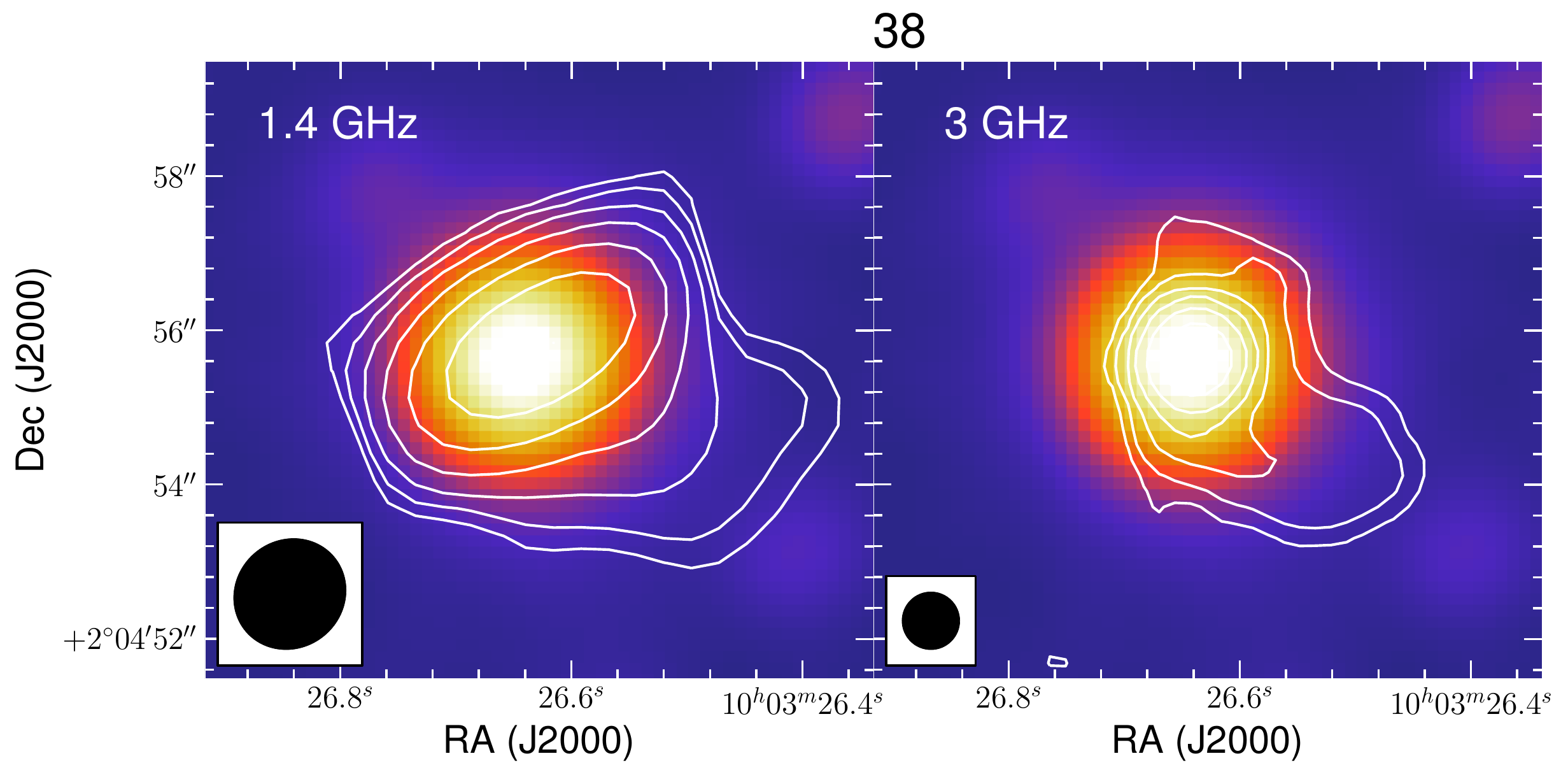}
            }
            \\ \\ 
               \resizebox{\hsize}{!}
            {\includegraphics{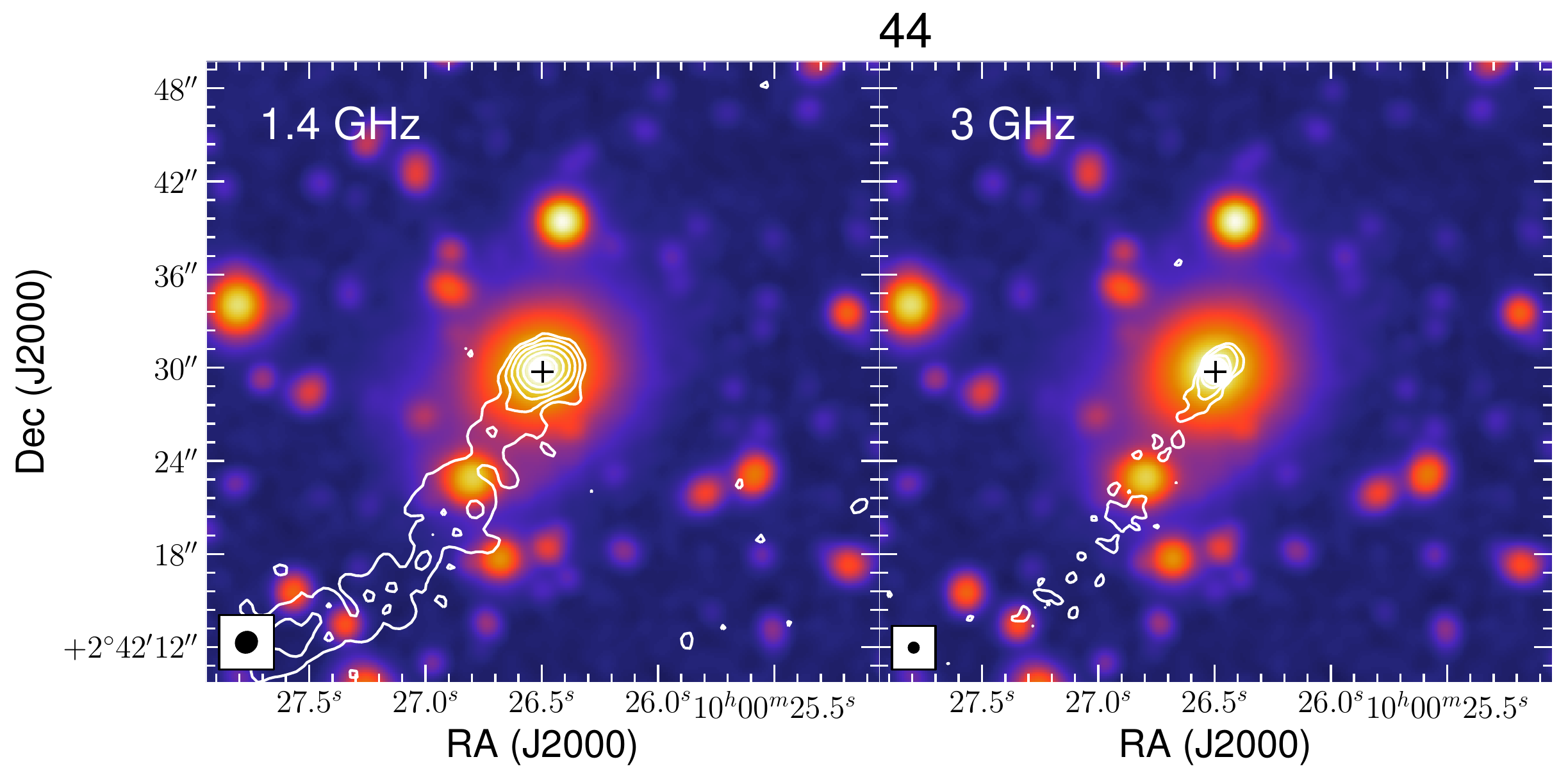}
            \includegraphics{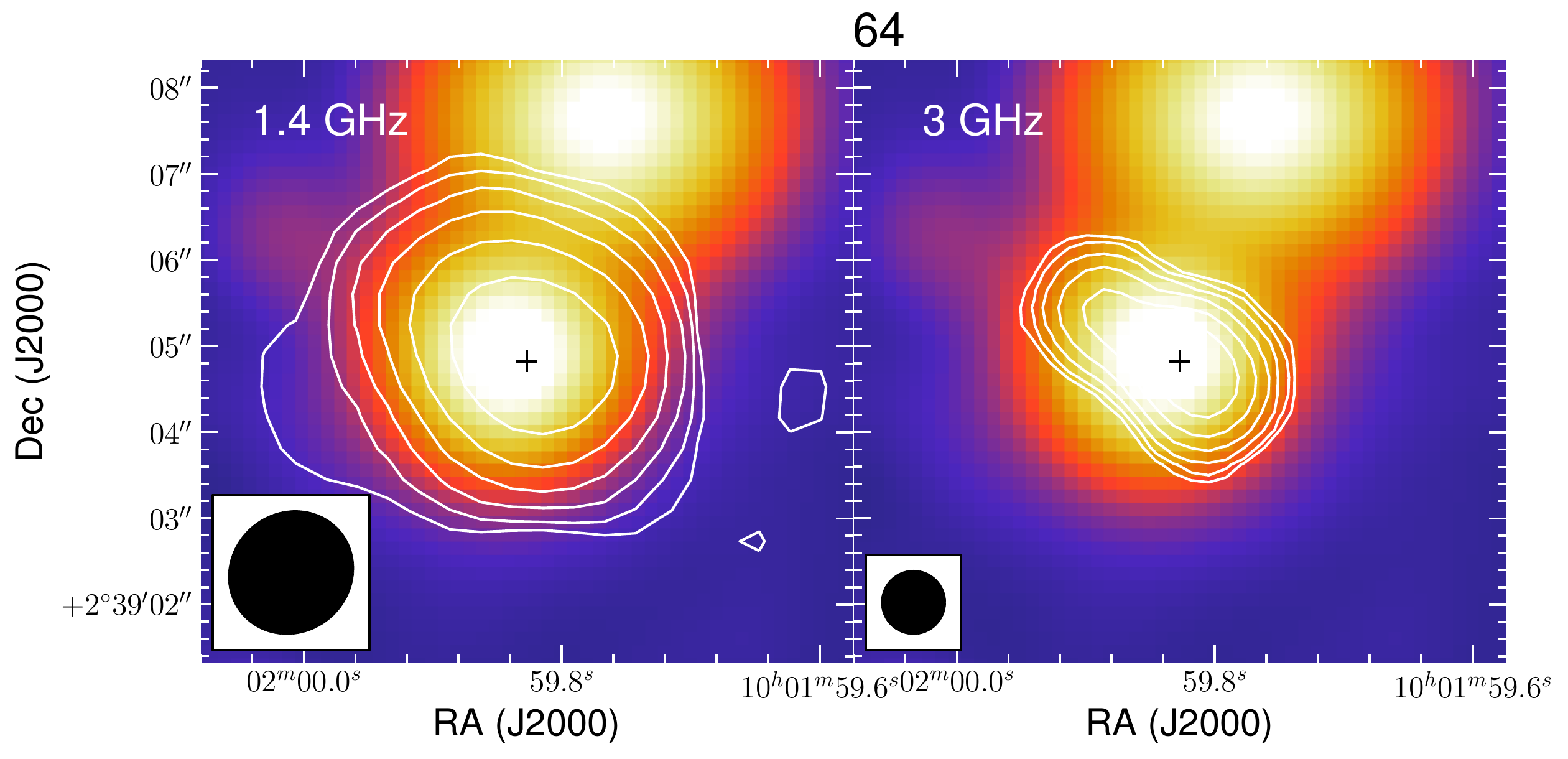}
            \includegraphics{JVLA80-ultrajvlavla.pdf}            
            }
             \\ \\ 
               \resizebox{\hsize}{!}
            {\includegraphics{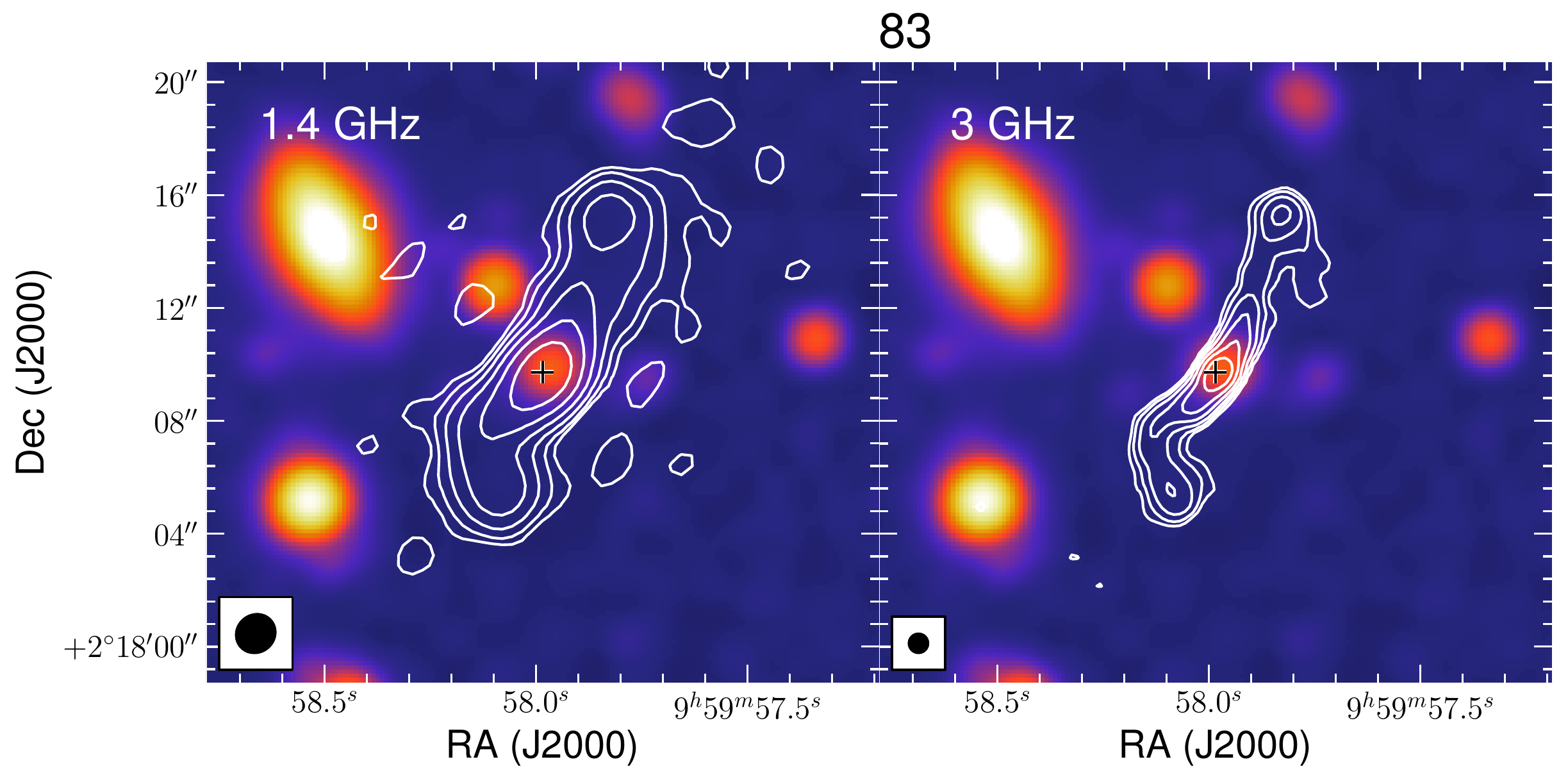}
            \includegraphics{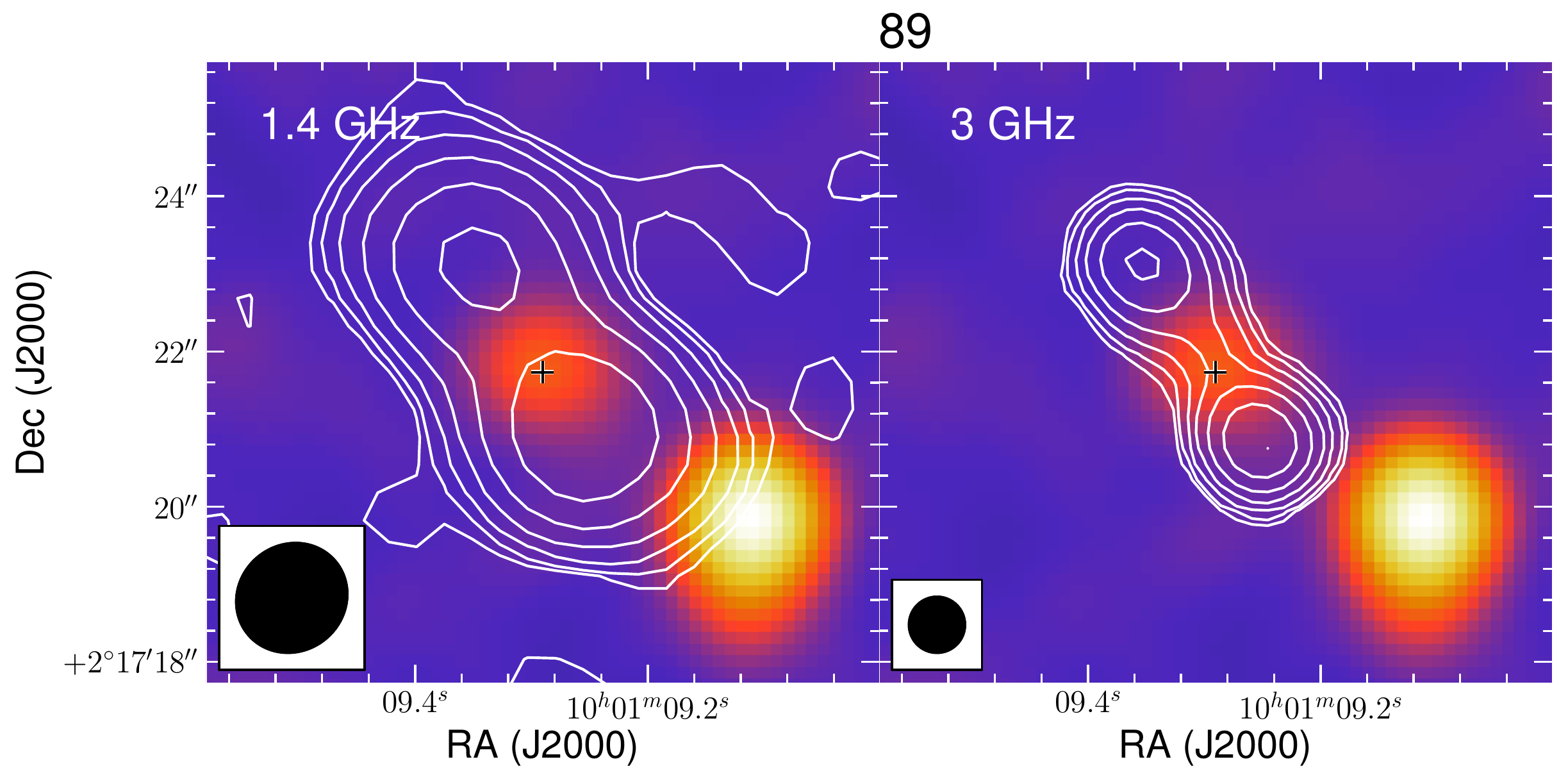}
            \includegraphics{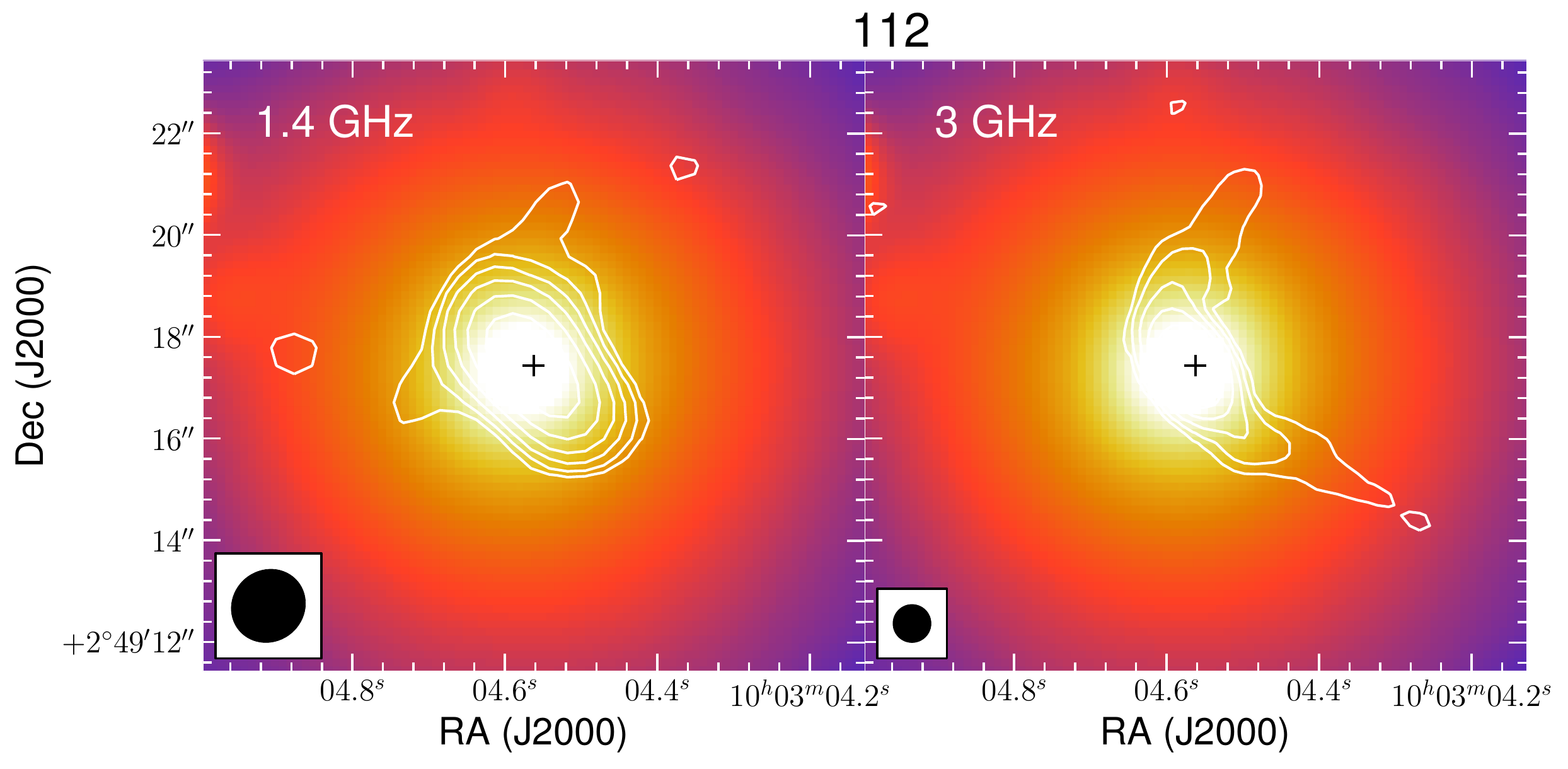}
            }
  \\ \\
   \resizebox{\hsize}{!}
{\includegraphics{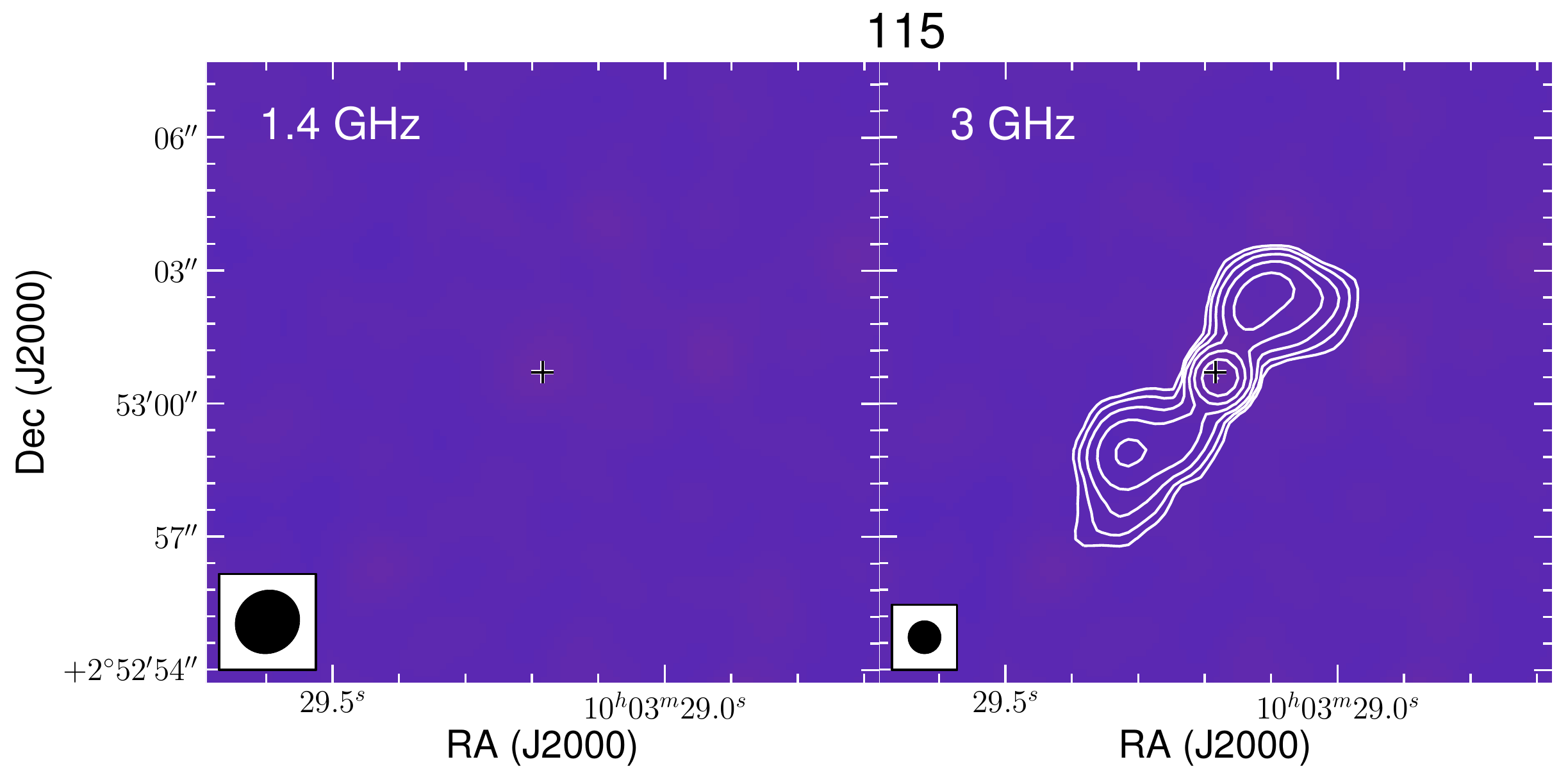}
 \includegraphics{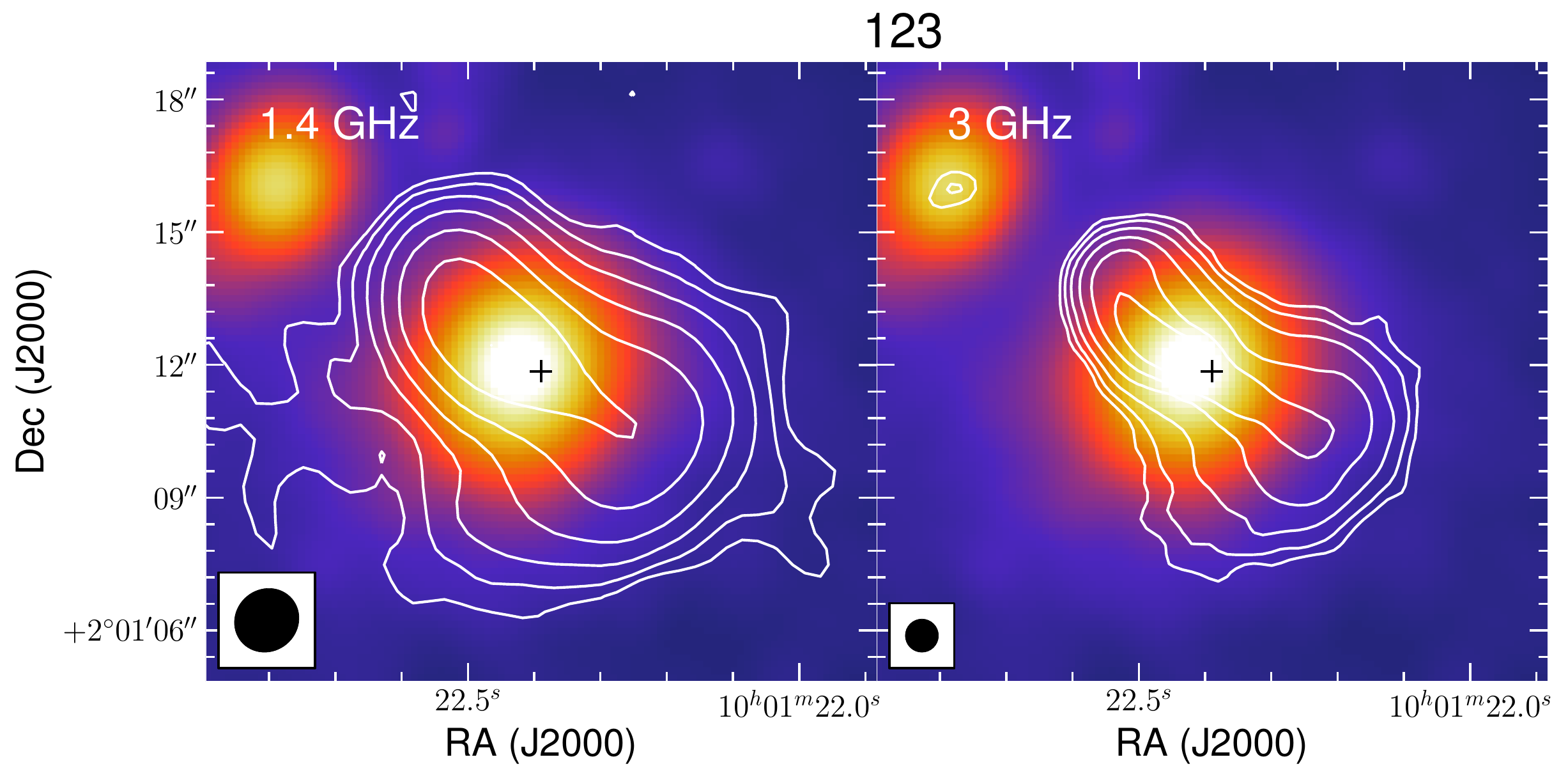}
 \includegraphics{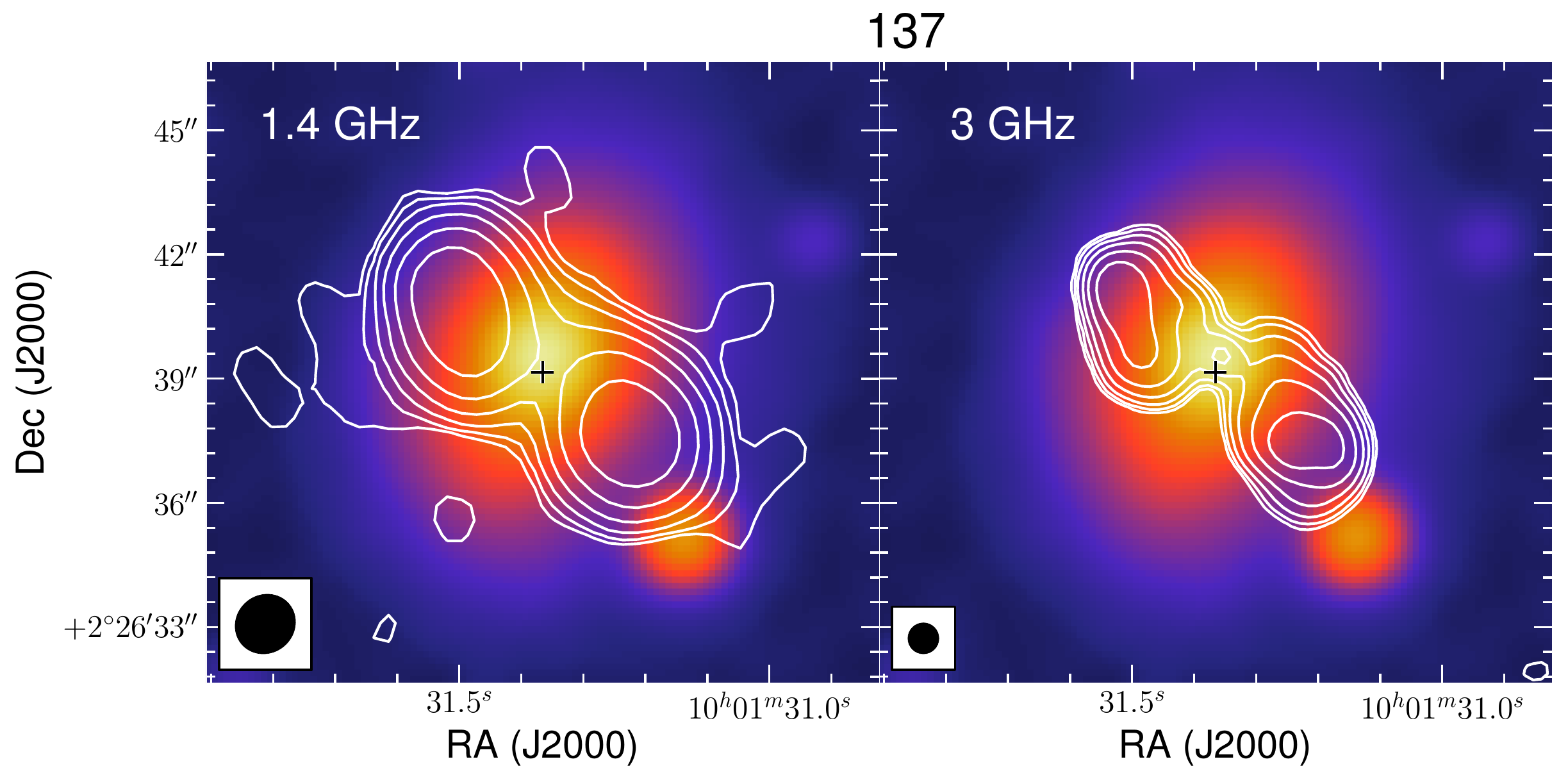}
            }
            \\ \\ 
  \resizebox{\hsize}{!}
 {\includegraphics{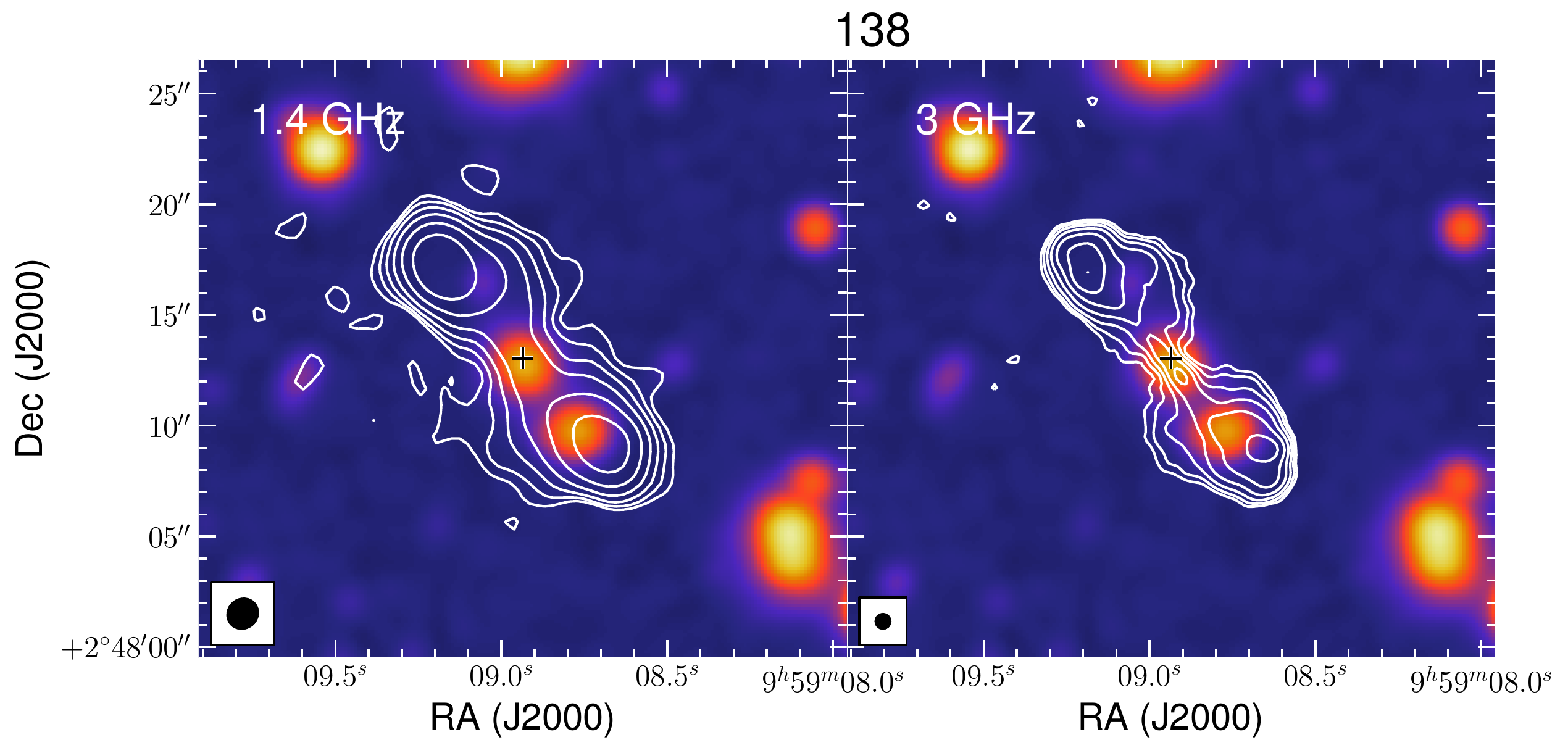}
    \includegraphics{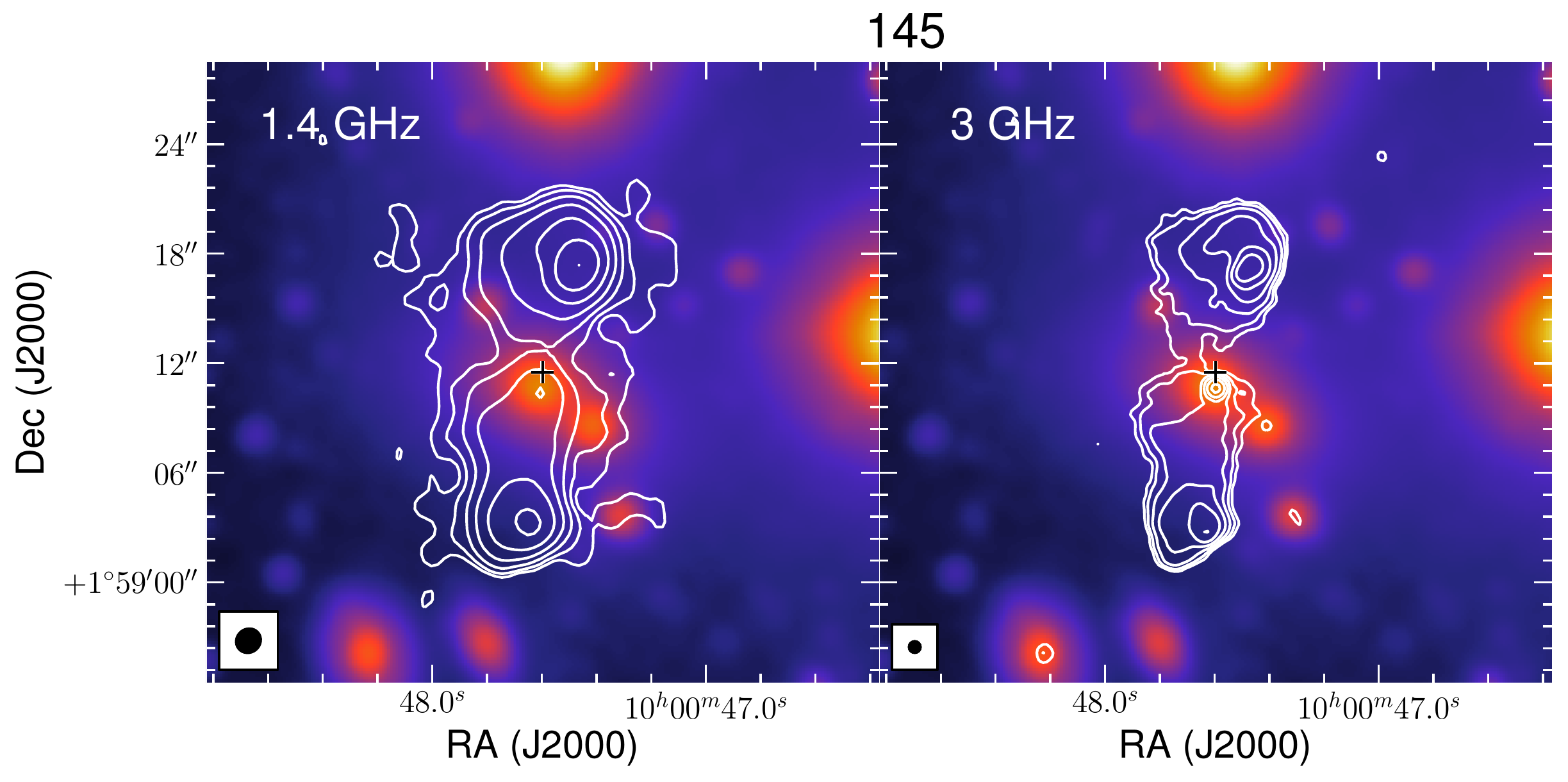}
    \includegraphics{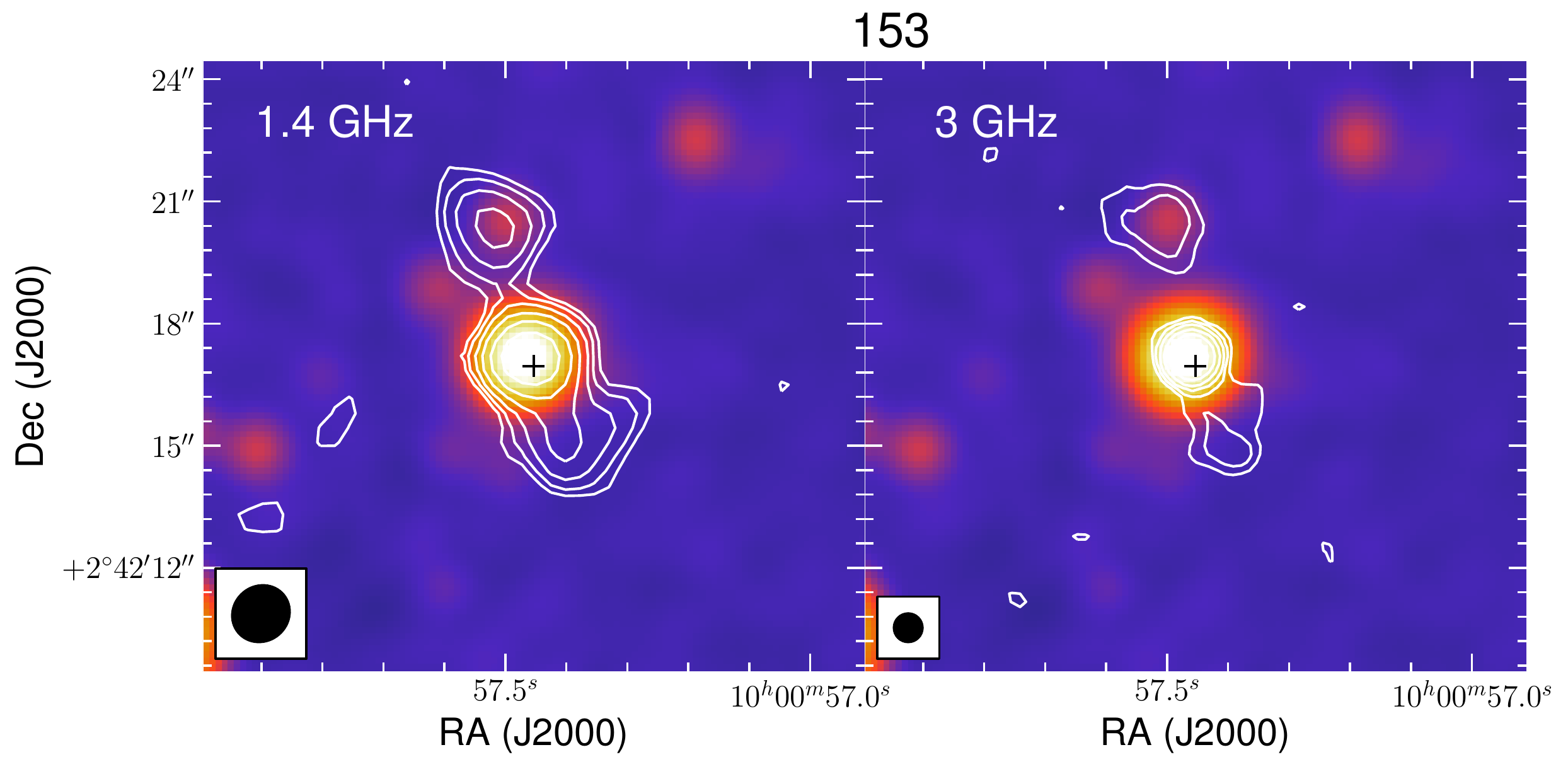}
            }
             \\ \\ 
      \resizebox{\hsize}{!}
       {\includegraphics{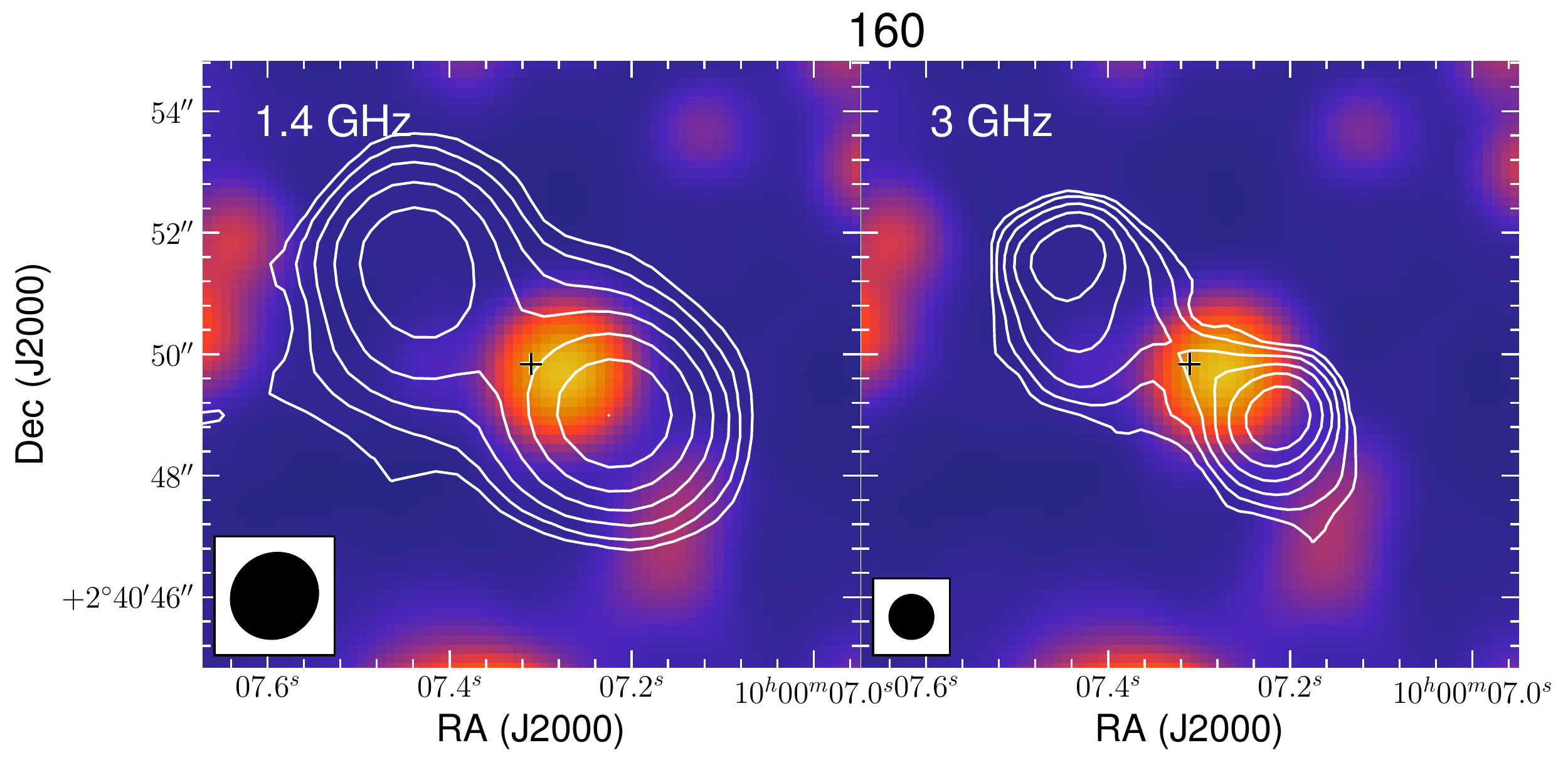}
        \includegraphics{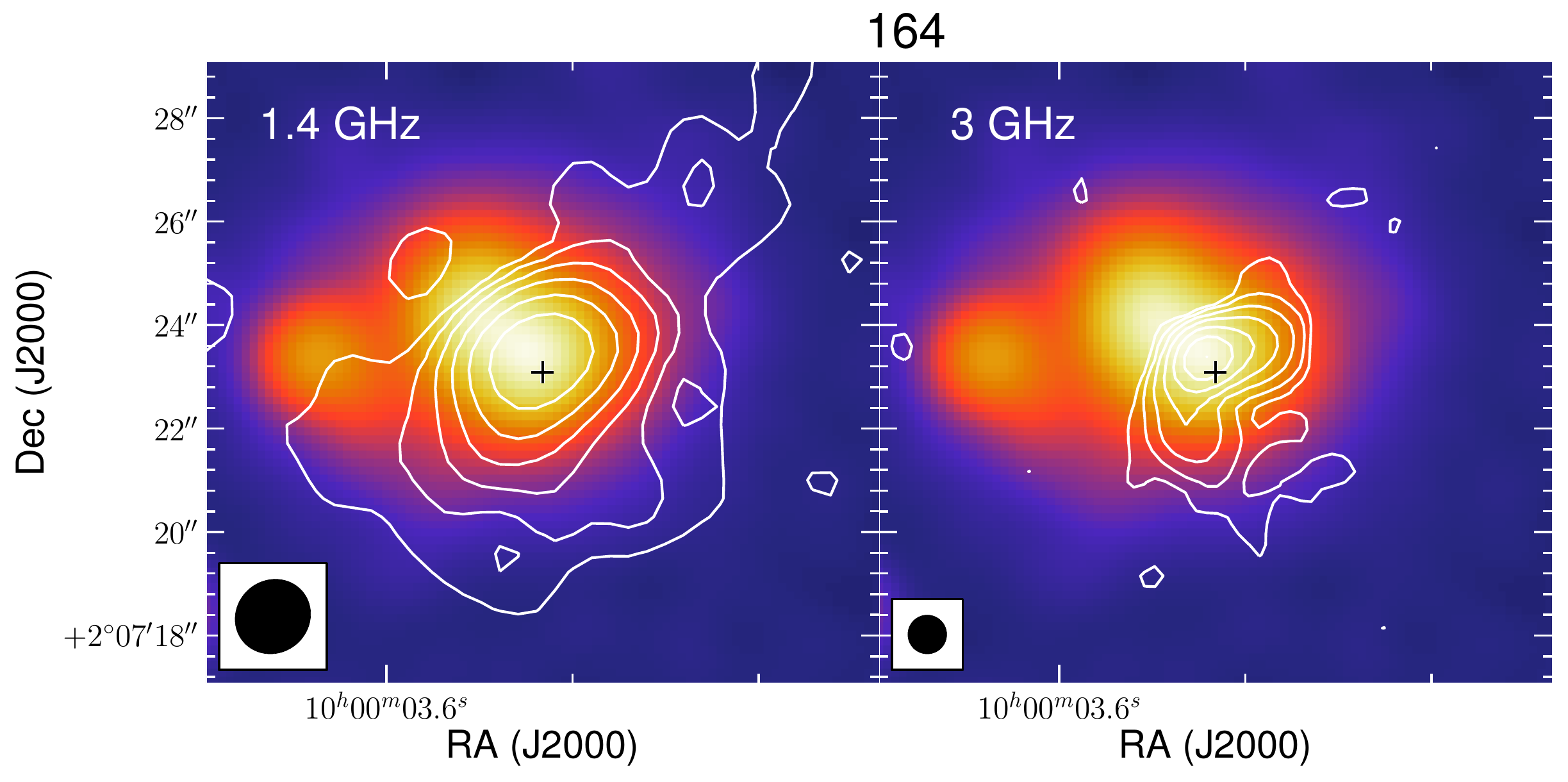}
       \includegraphics{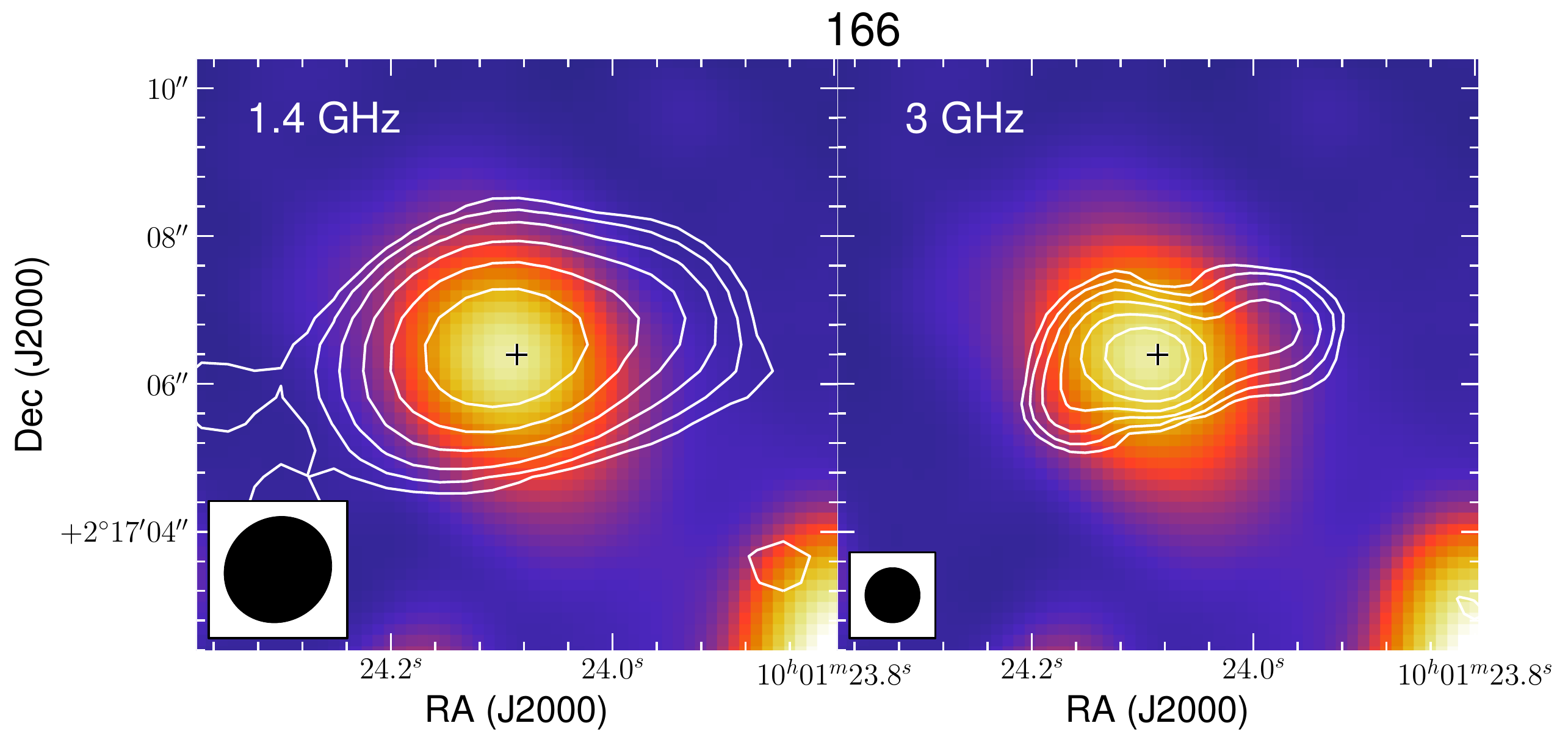}
            }
   \caption{Set of overlays at 1.4 GHz (left) and 3 GHz (right) of VLA-COSMOS for the FR objects in our sample, shown as white contours. The colour scale is the UltraVISTA stacked mosaic. Data for these sources are shown in Table~\ref{table:data}. Objects without 1.4 GHz or UltraVISTA maps lie in masked regions or outside the data coverage.
   }
              \label{fig:maps2}%
    \end{figure*}
\addtocounter{figure}{-1}
\begin{figure*}[!ht]
 \resizebox{\hsize}{!}
{\includegraphics{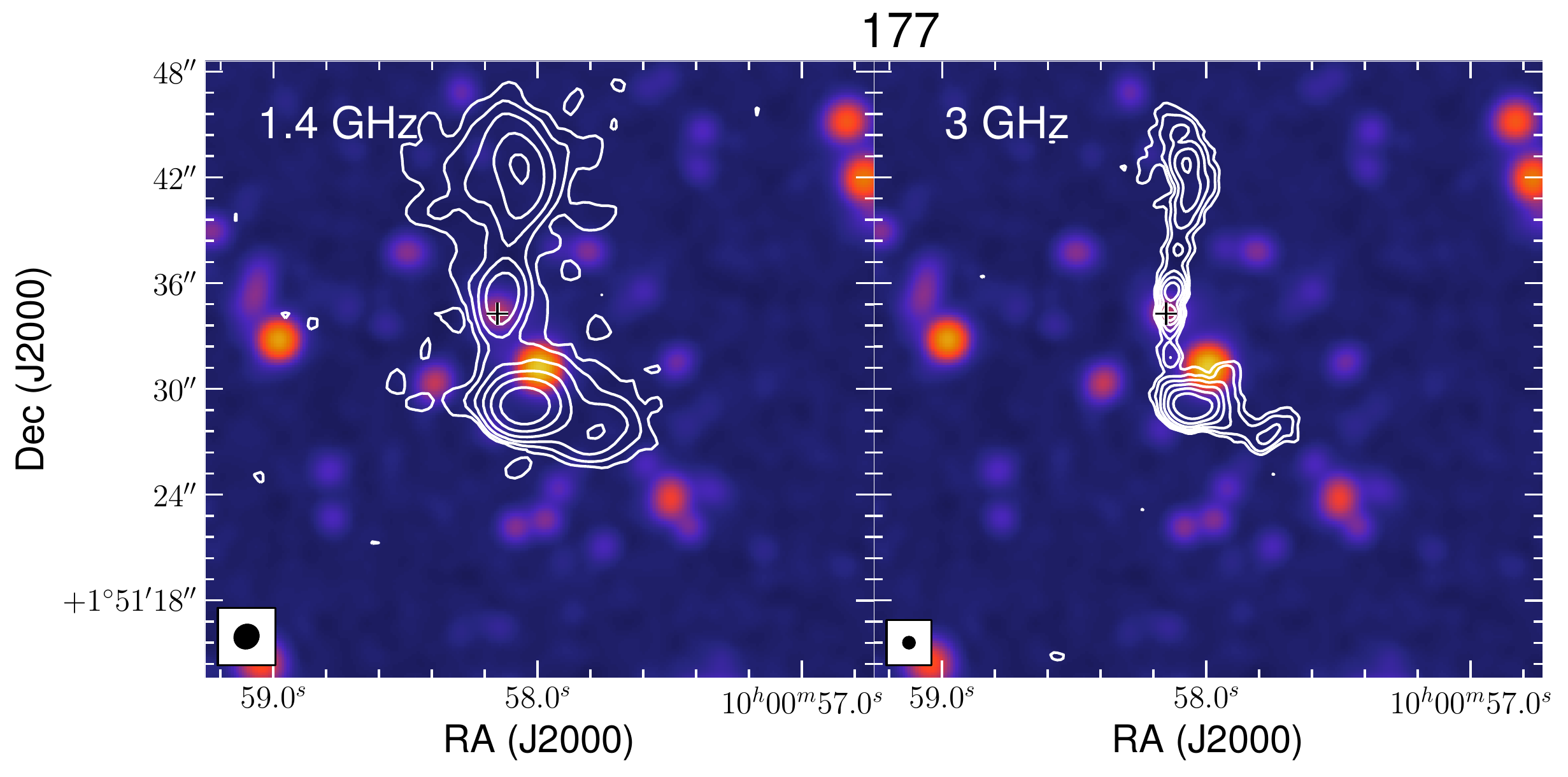}
 \includegraphics{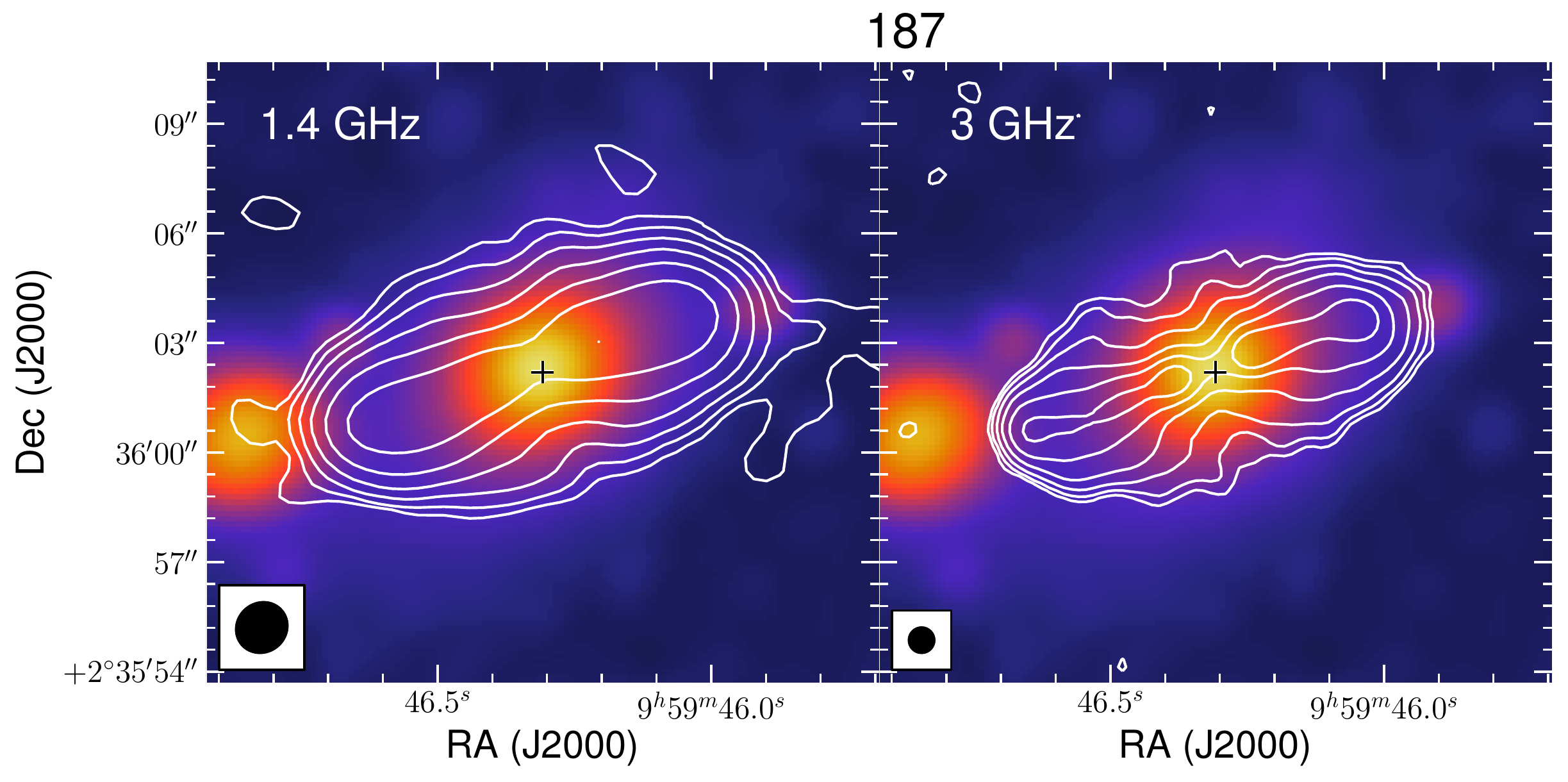}
 \includegraphics{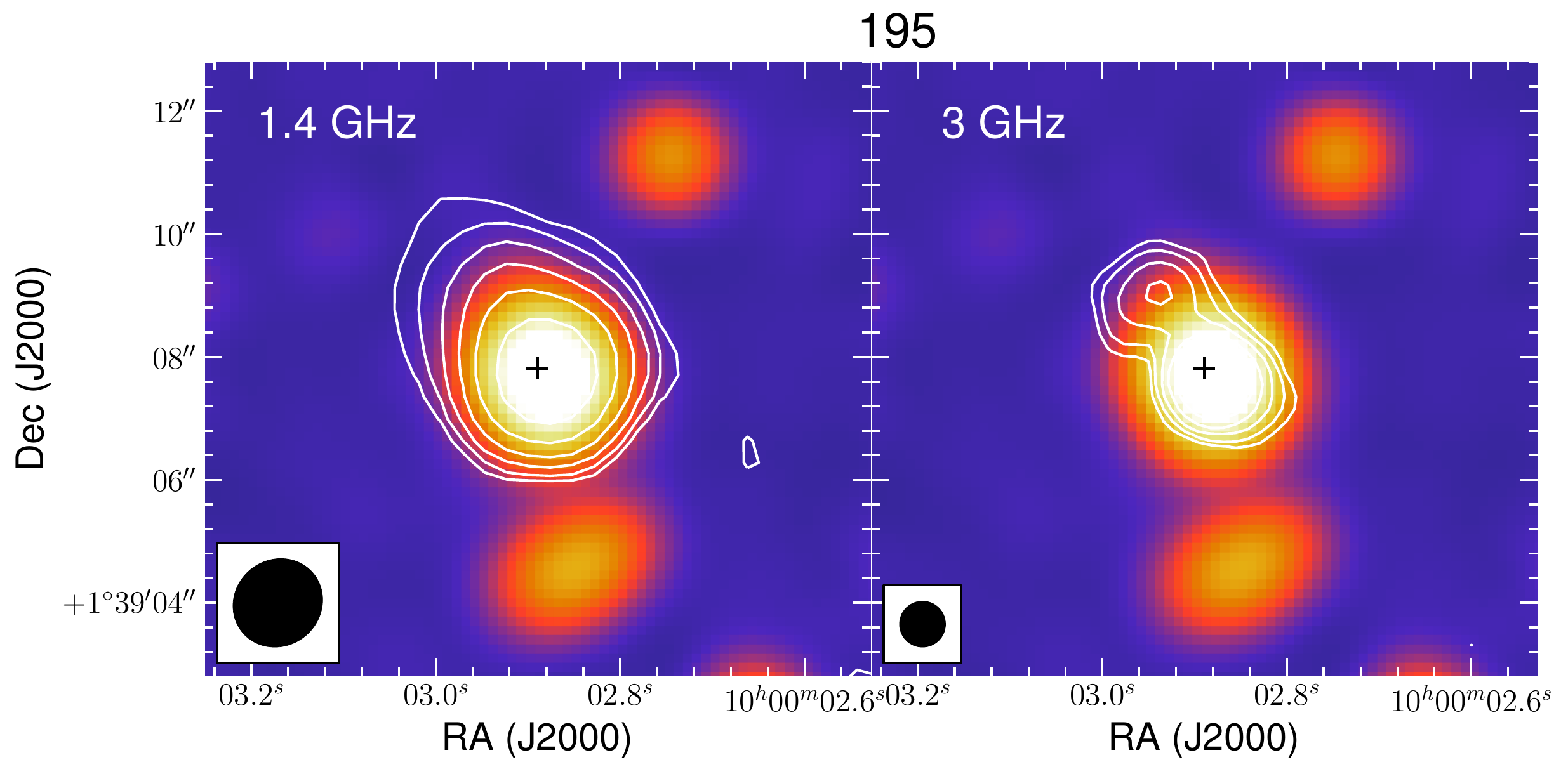}
            }
            \\ \\ 
  \resizebox{\hsize}{!}
 {\includegraphics{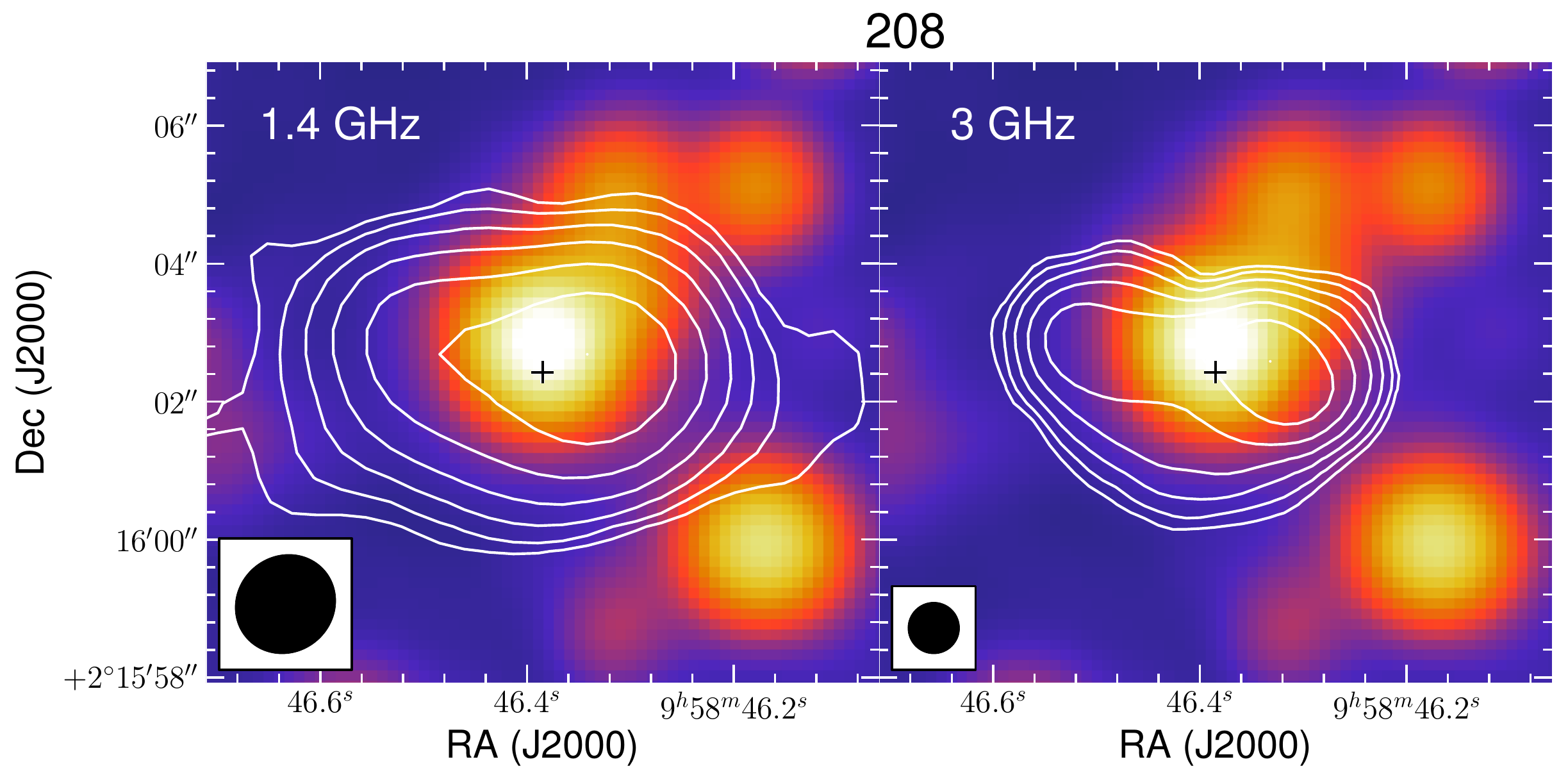}
    \includegraphics{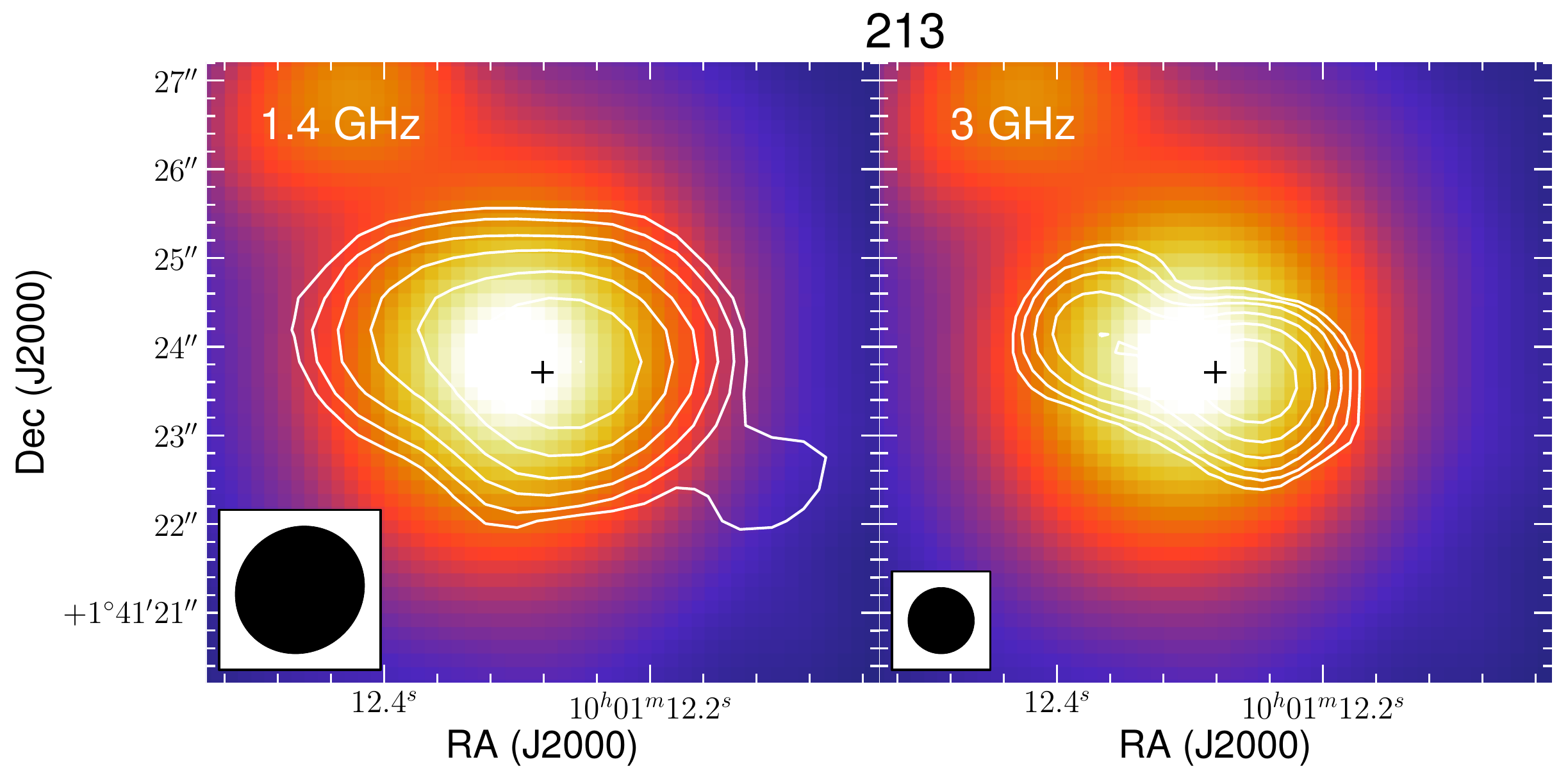}
    \includegraphics{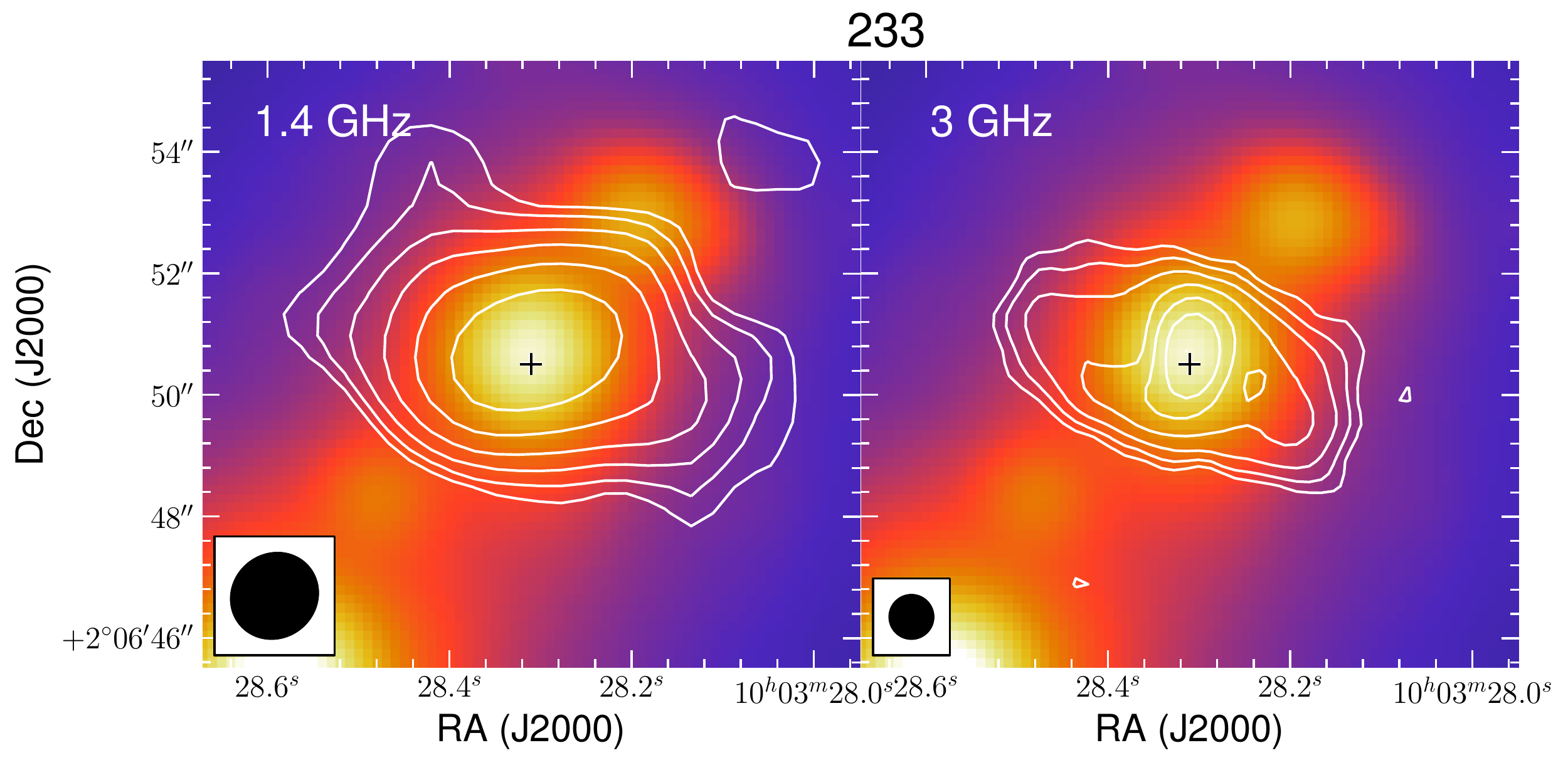}
            }
             \\ \\ 
      \resizebox{\hsize}{!}
       {\includegraphics{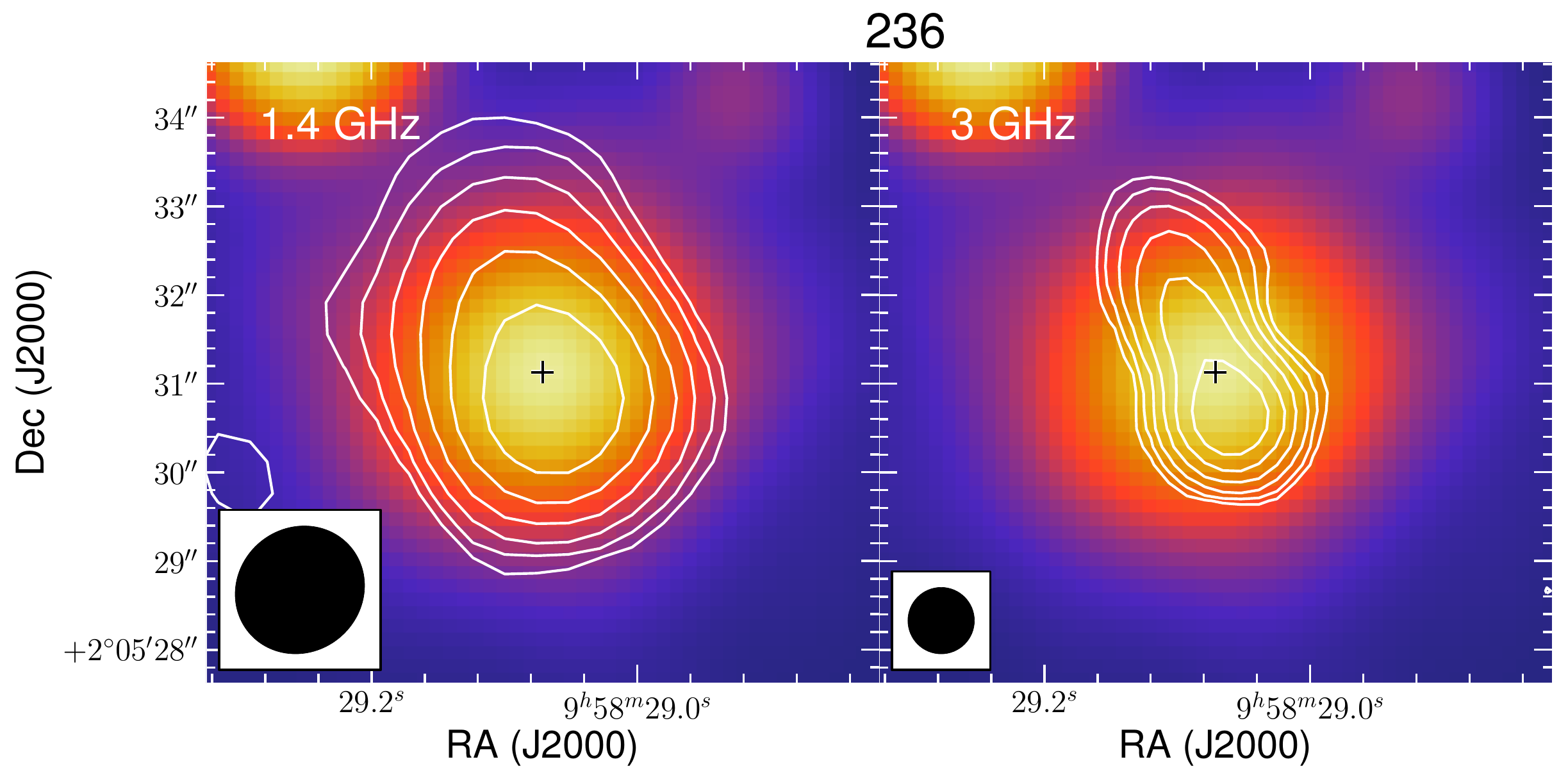}
        \includegraphics{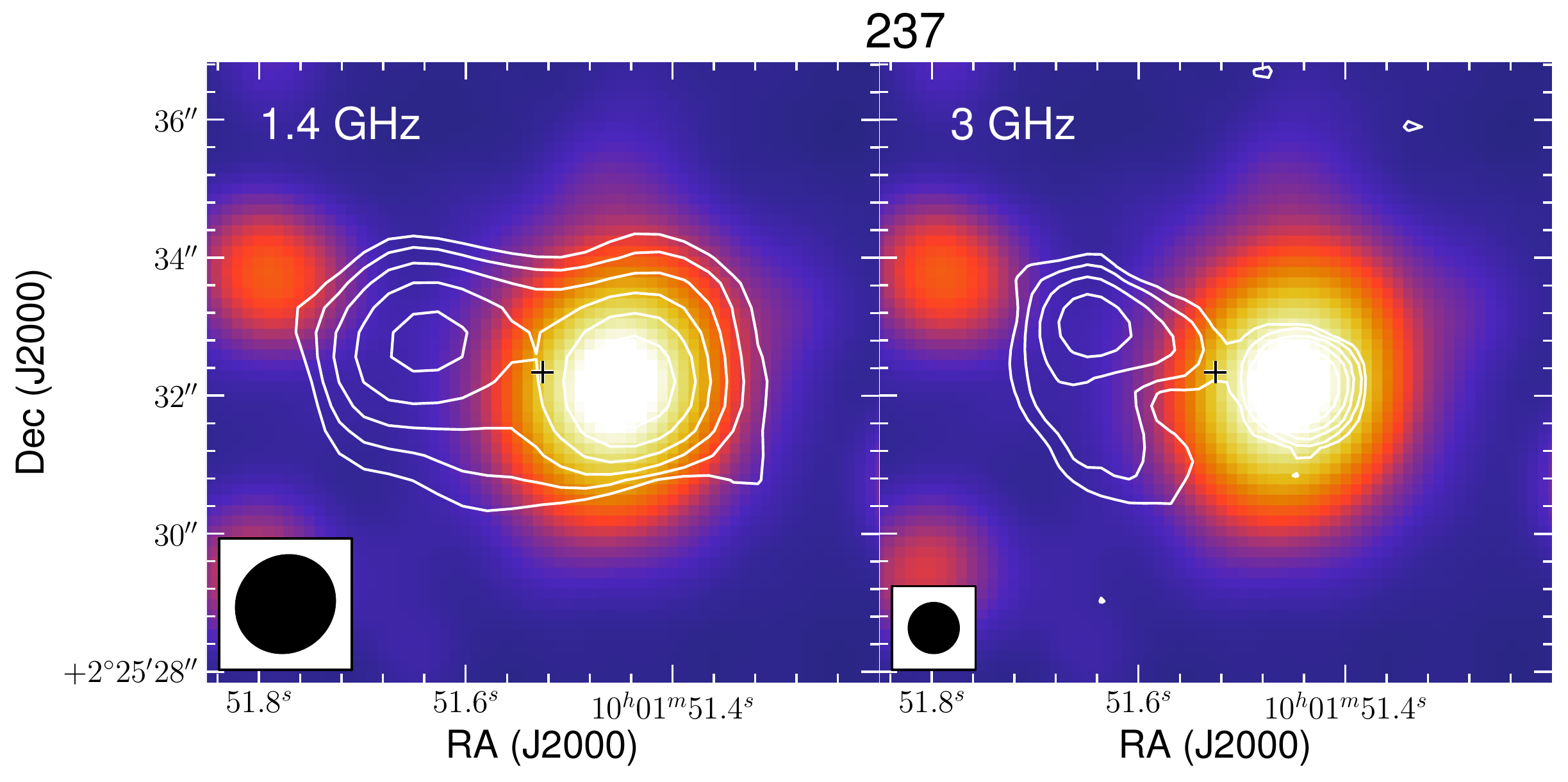}
       \includegraphics{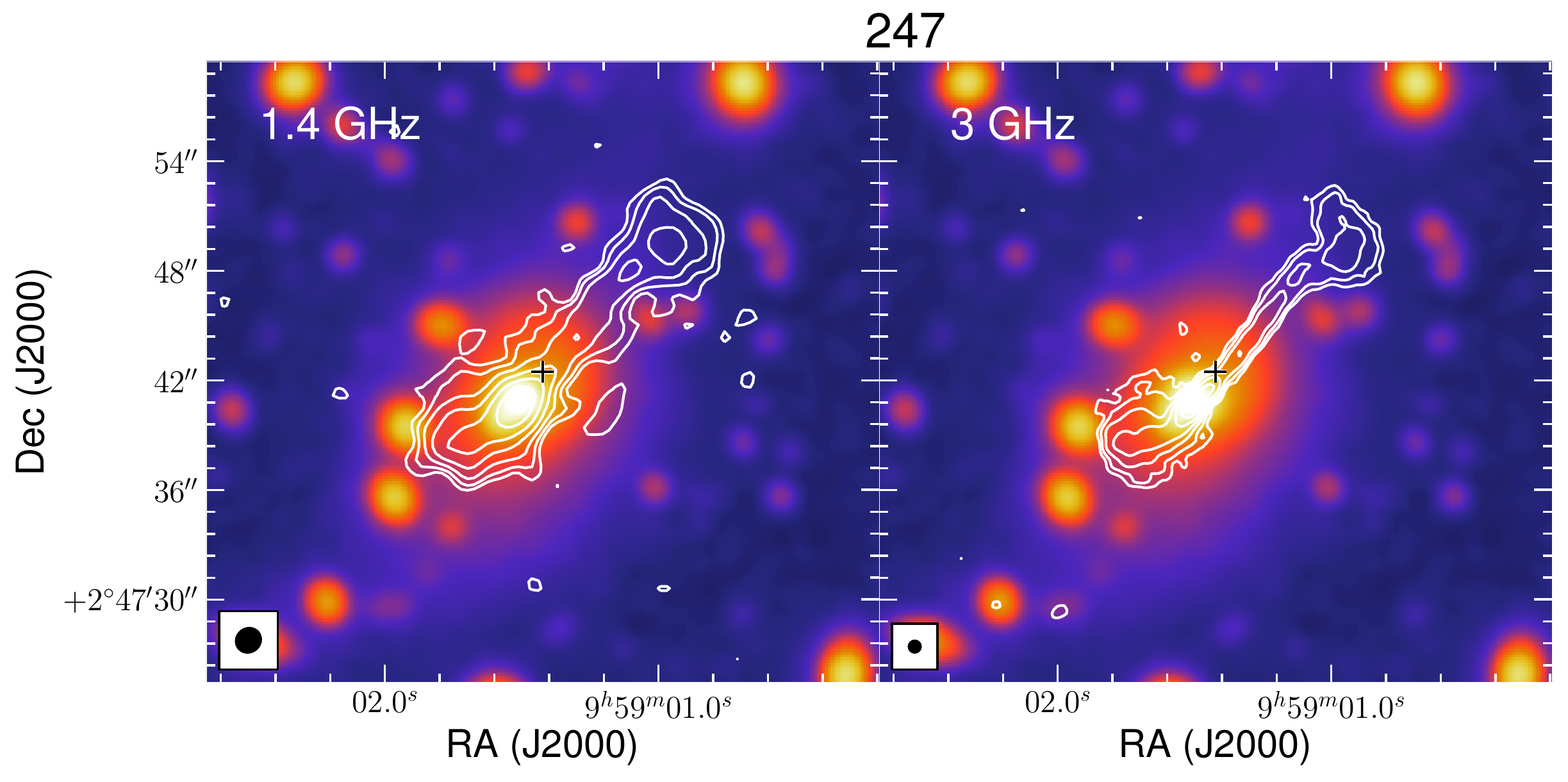}
            }
  \\ \\
 \resizebox{\hsize}{!}
{\includegraphics{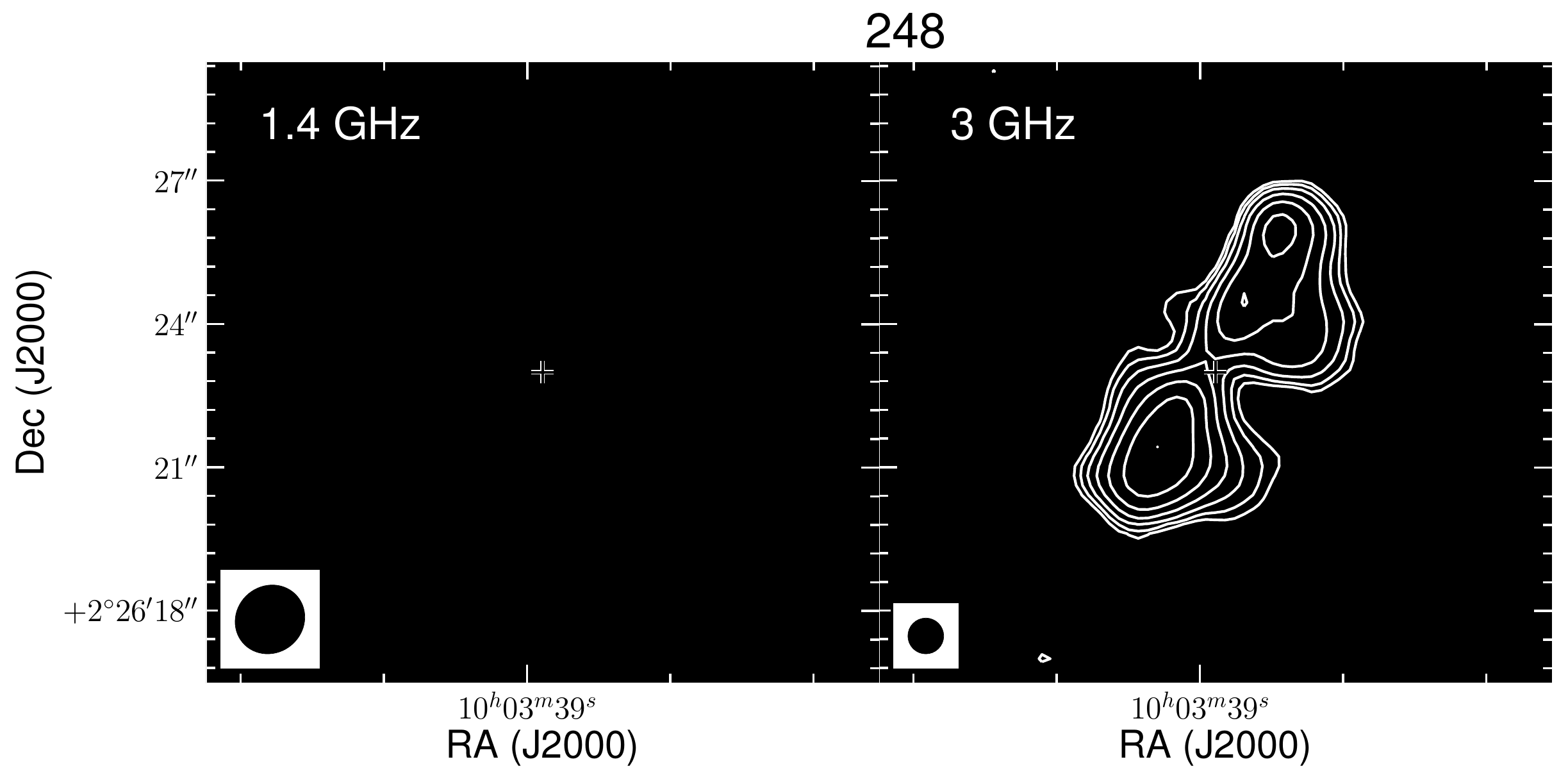}
 \includegraphics{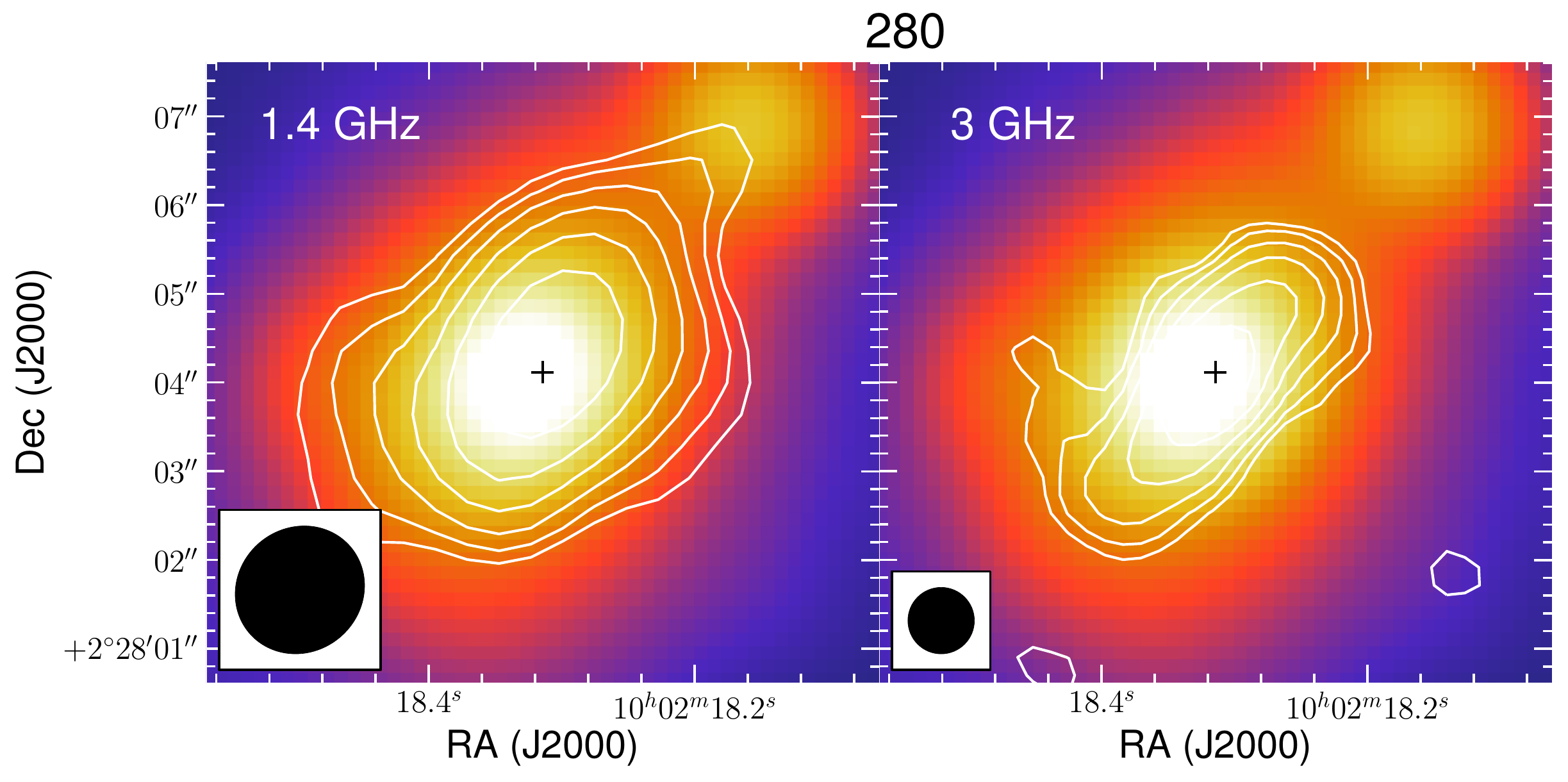}
 \includegraphics{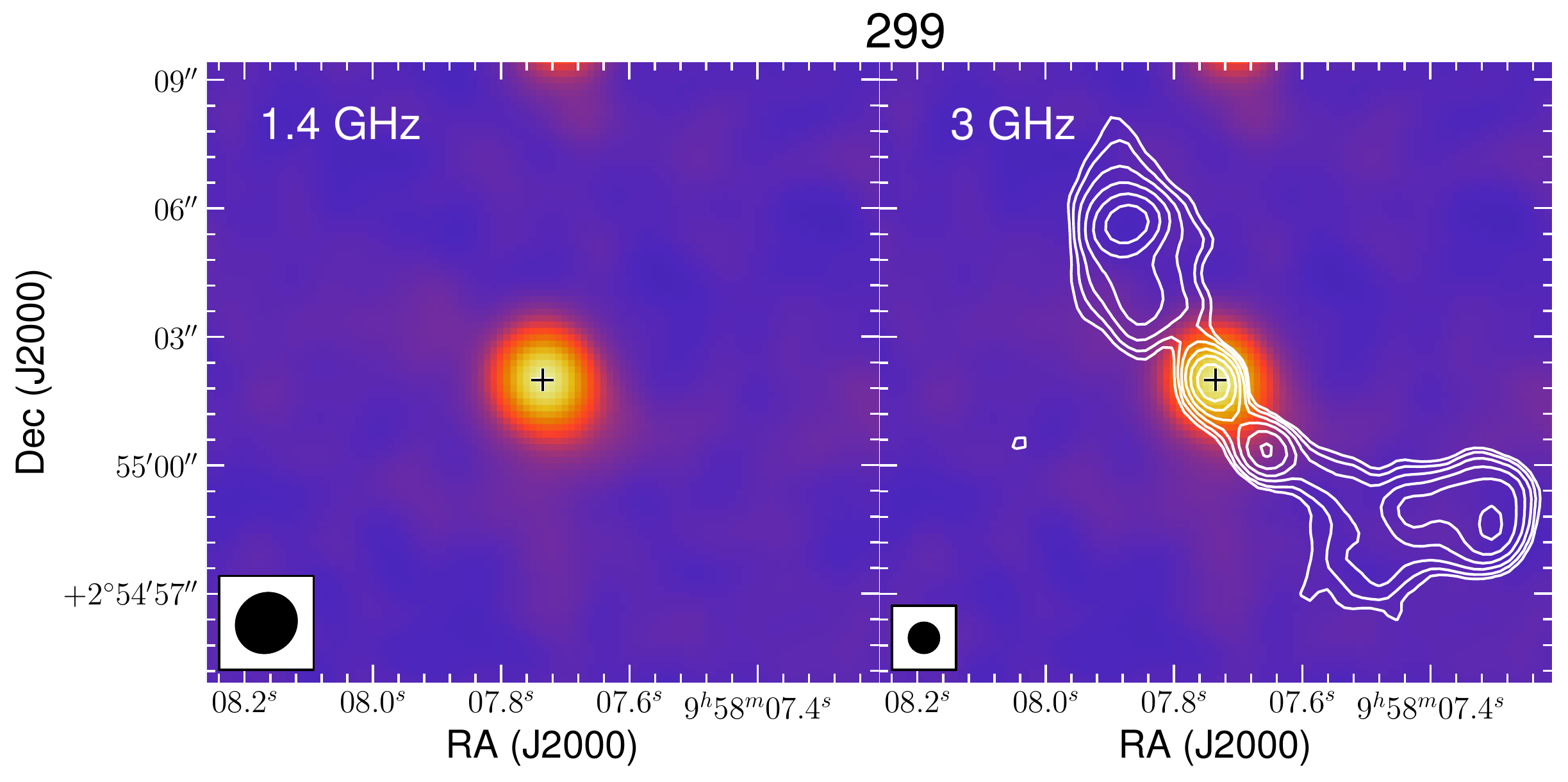}
            }
            \\ \\ 
  \resizebox{\hsize}{!}
 {\includegraphics{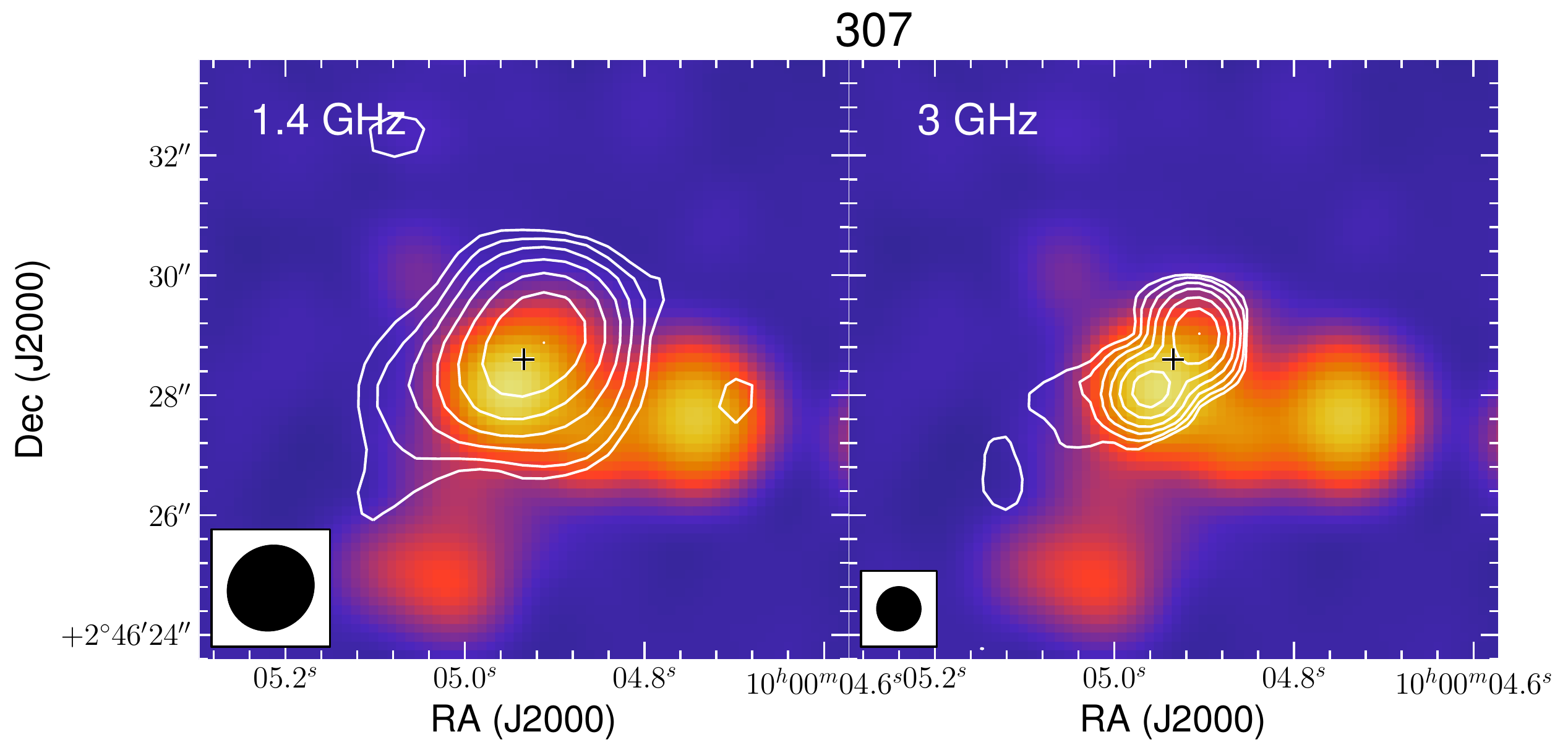}
    \includegraphics{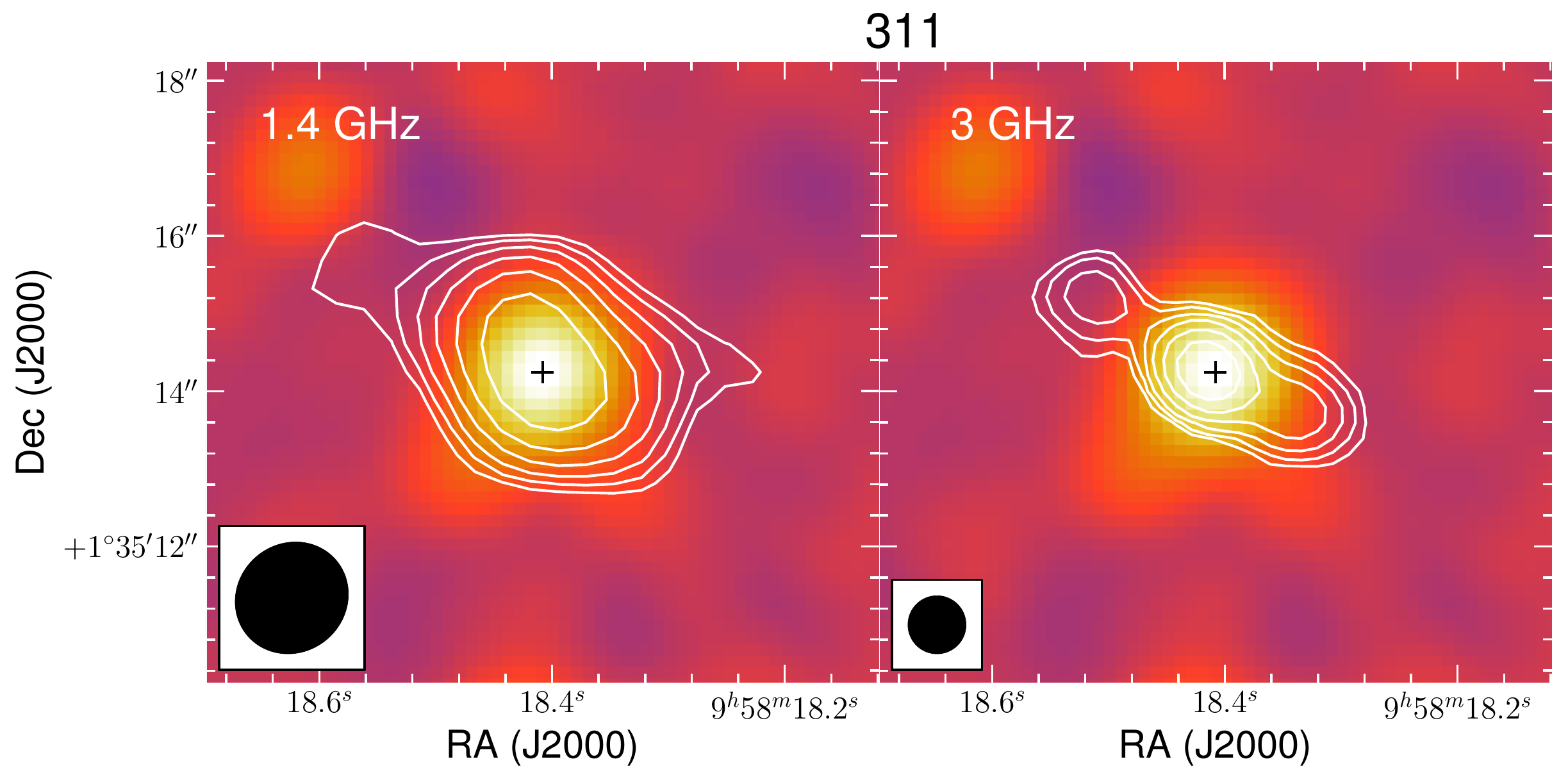}
    \includegraphics{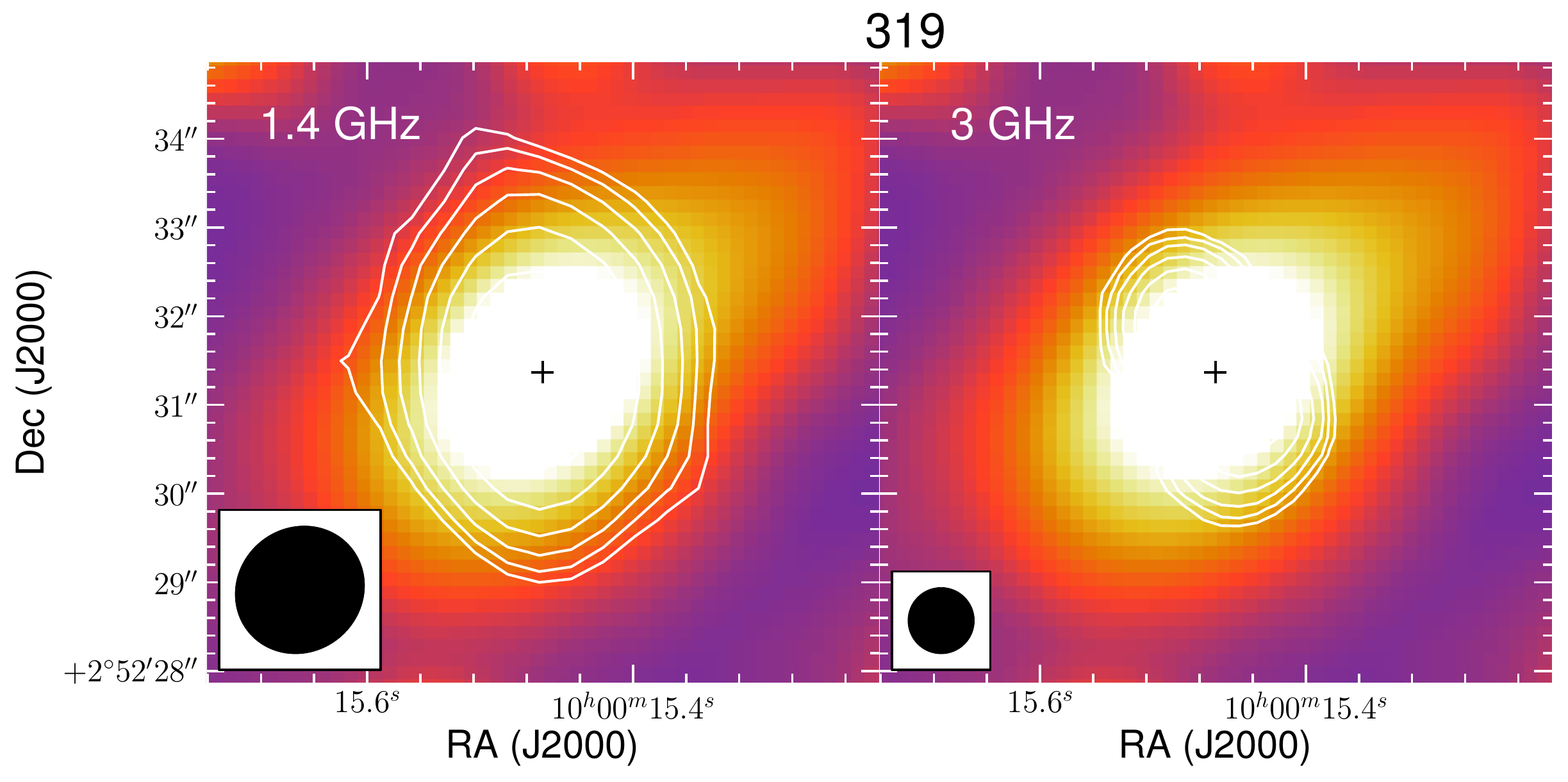}
            }
             \\ \\ 
      \resizebox{\hsize}{!}
       {\includegraphics{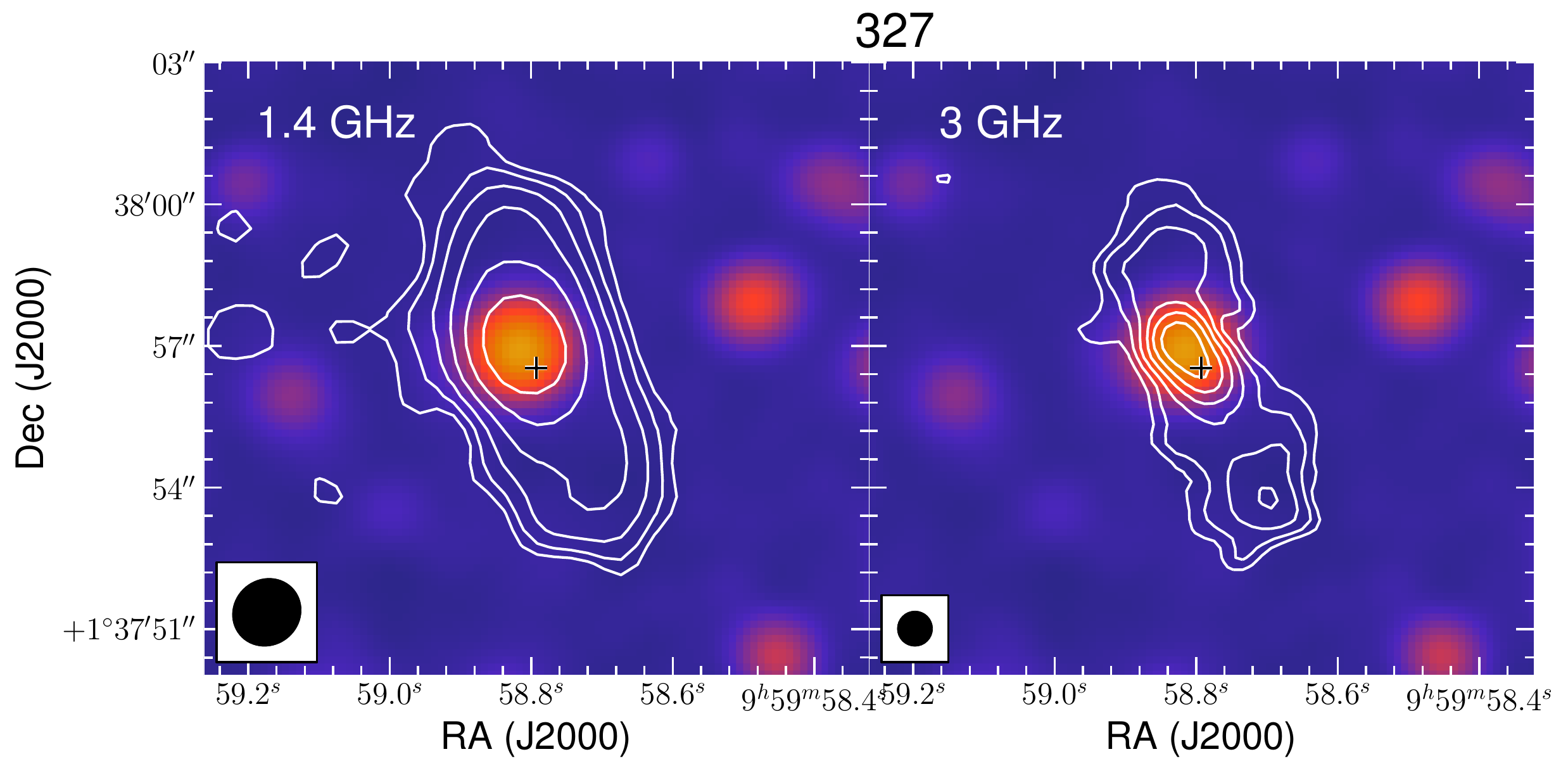}
        \includegraphics{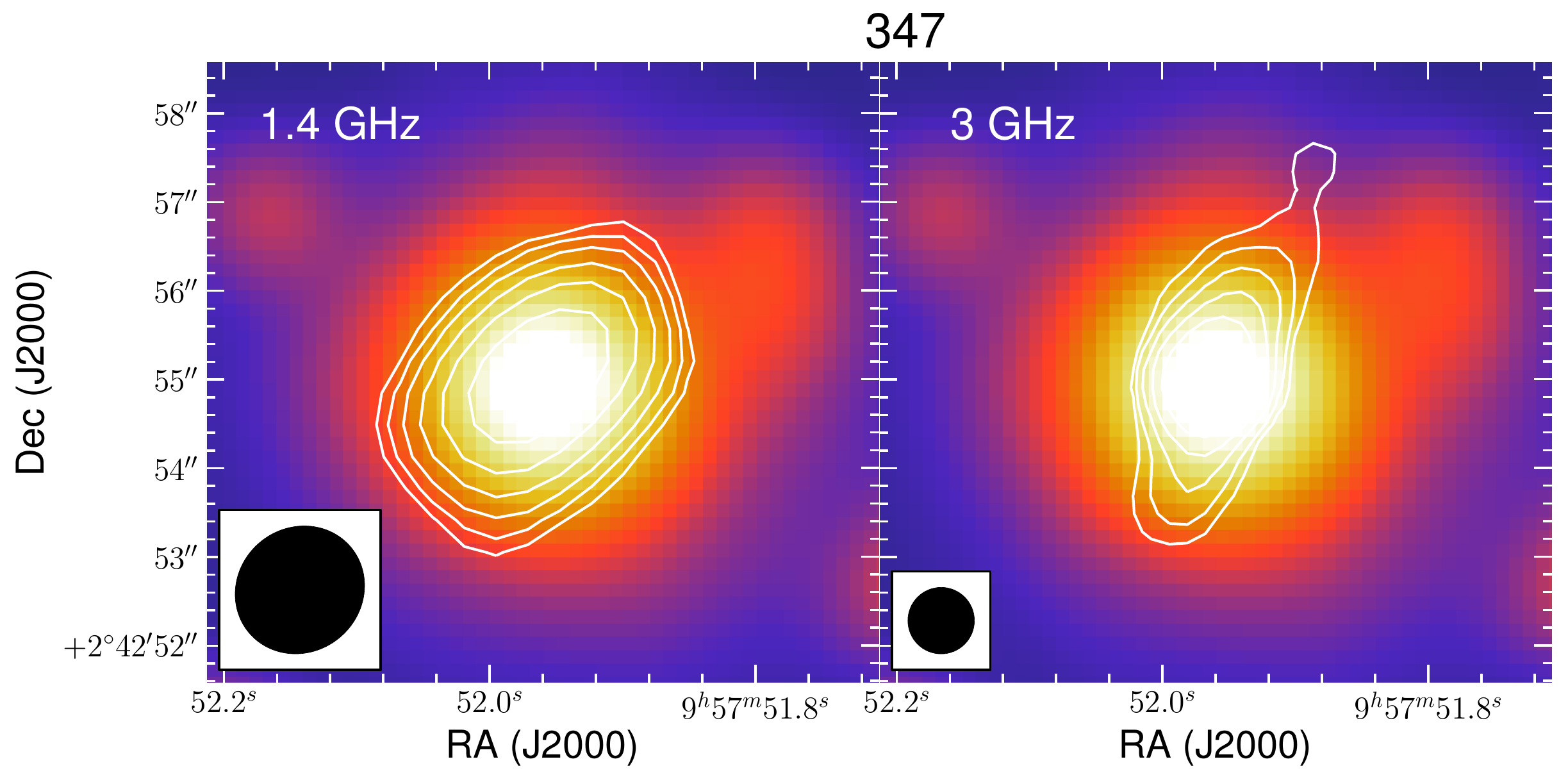}
       \includegraphics{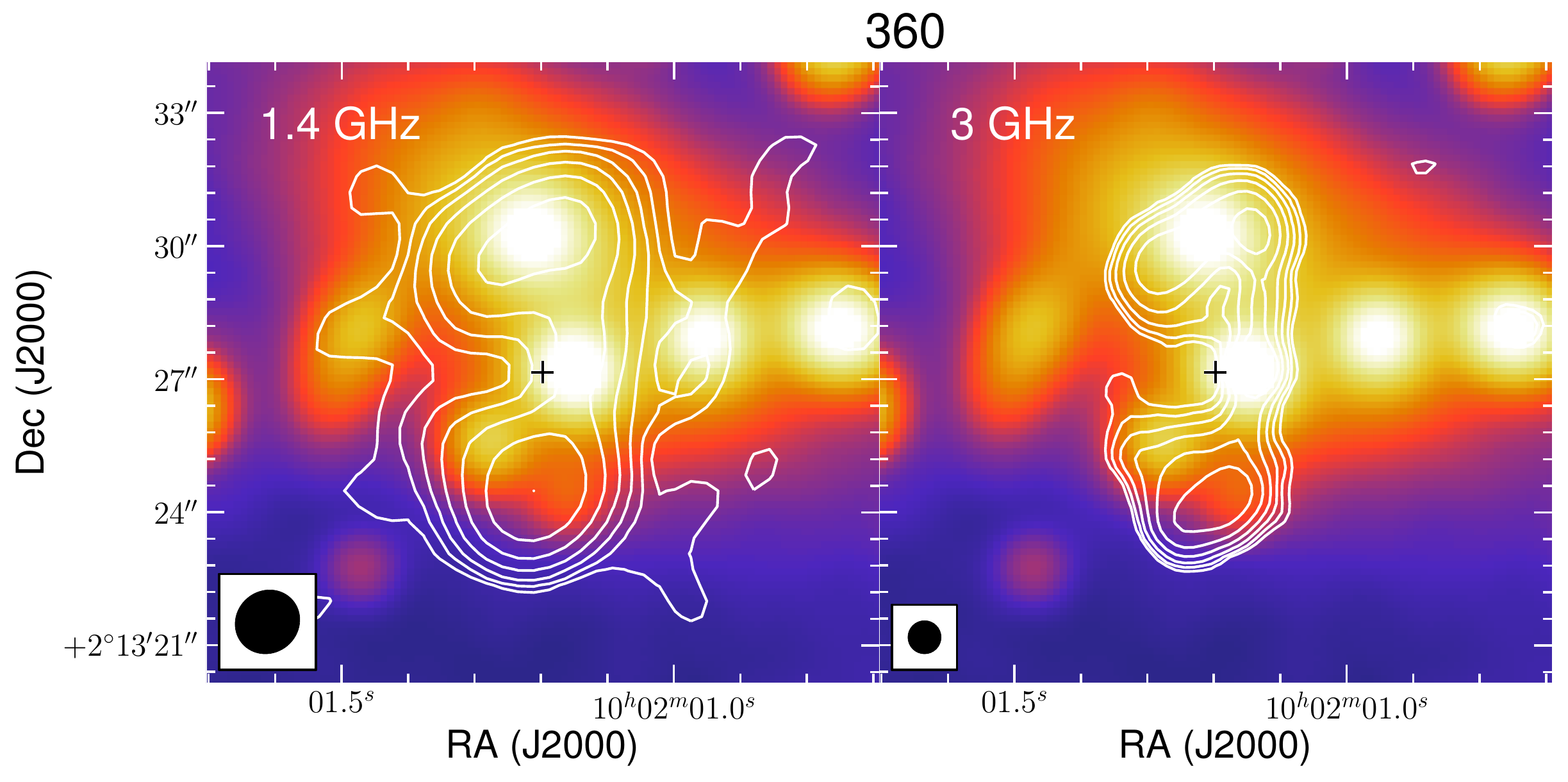}
            }
            \\ \\ 
             \resizebox{\hsize}{!}
{\includegraphics{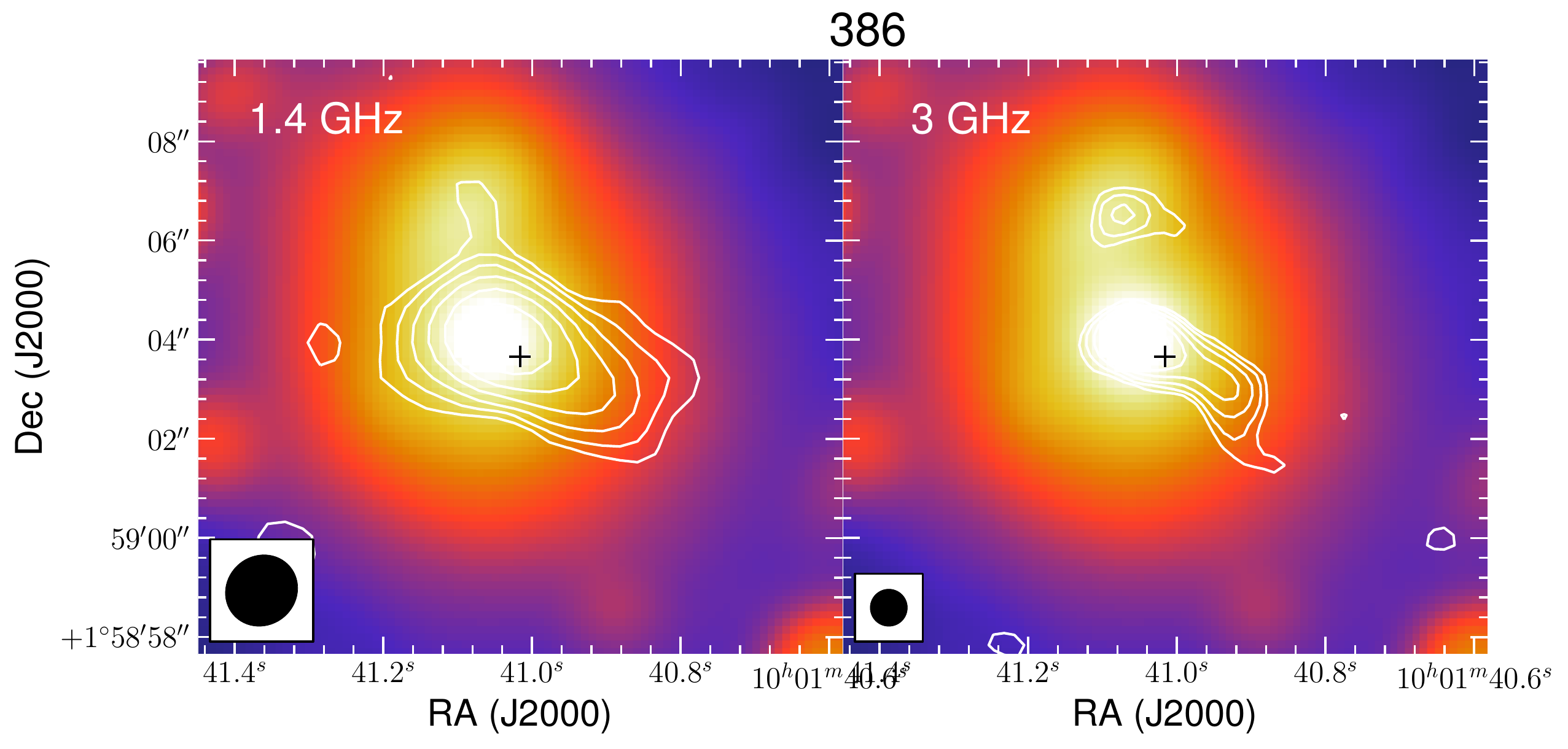}
 \includegraphics{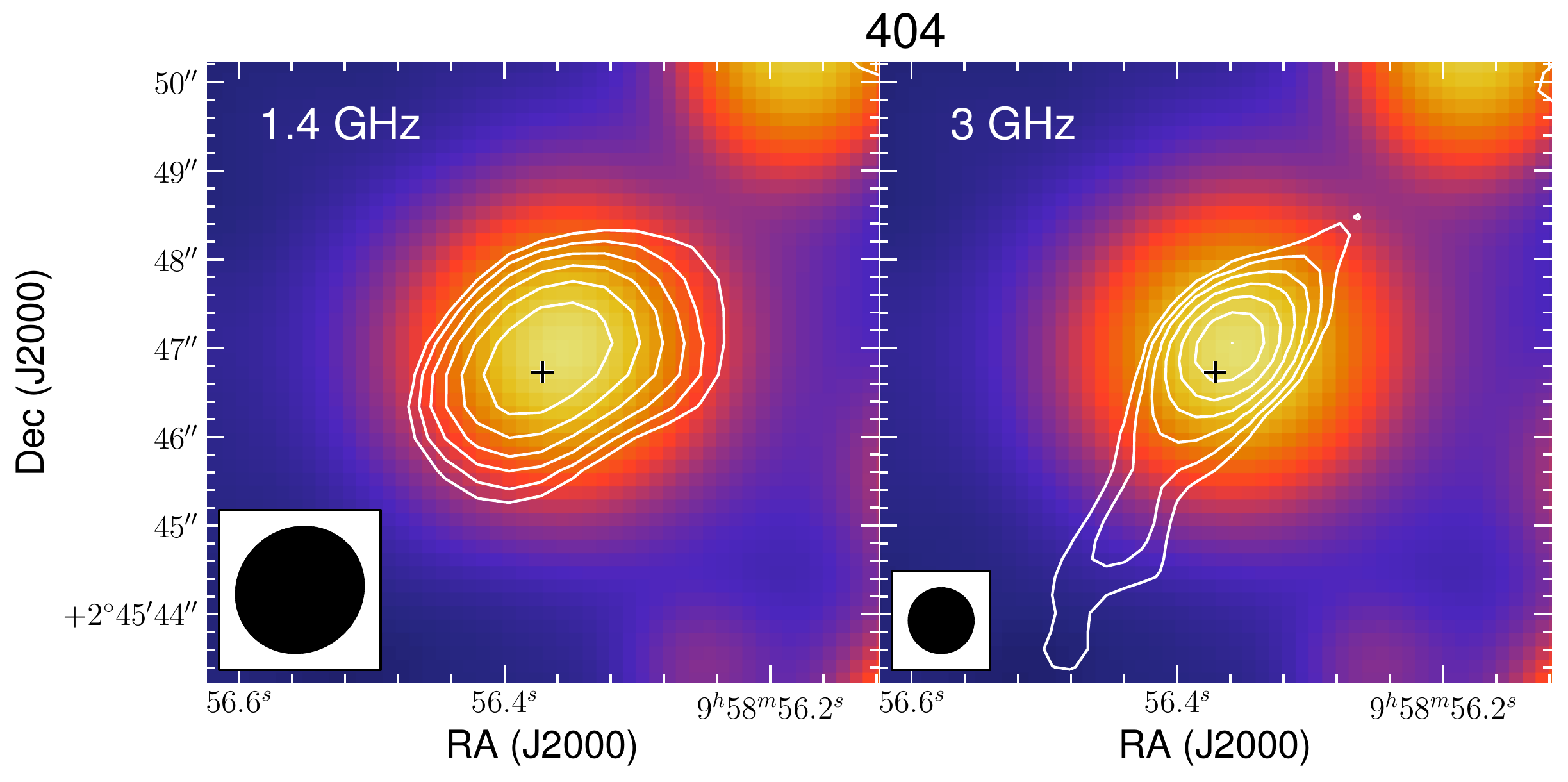}
 \includegraphics{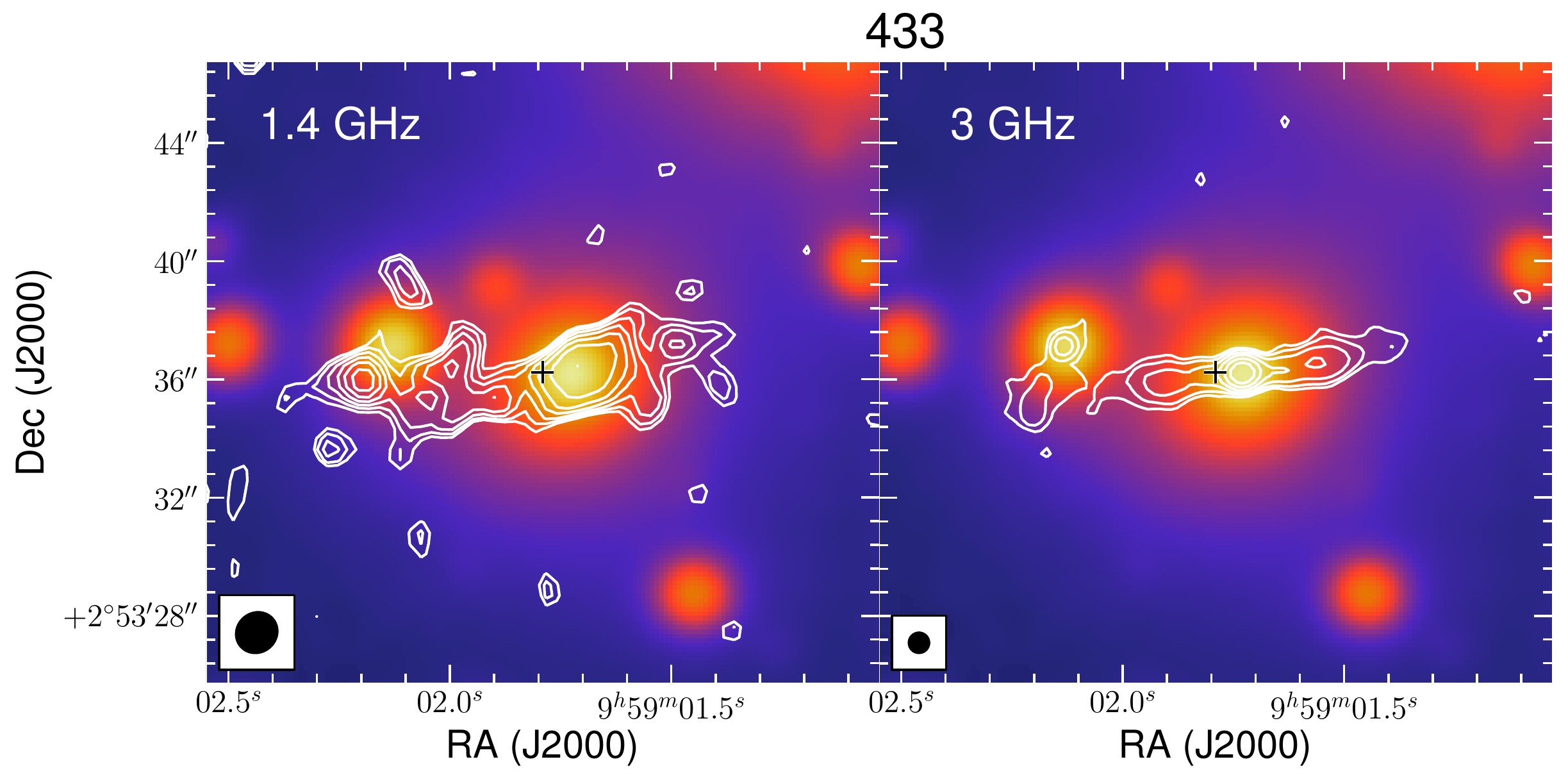}
            }

   \caption{(continued)
   }
              \label{fig:maps2}%
    \end{figure*}
\addtocounter{figure}{-1}
\begin{figure*}[!ht]
 
  \resizebox{\hsize}{!}
 {\includegraphics{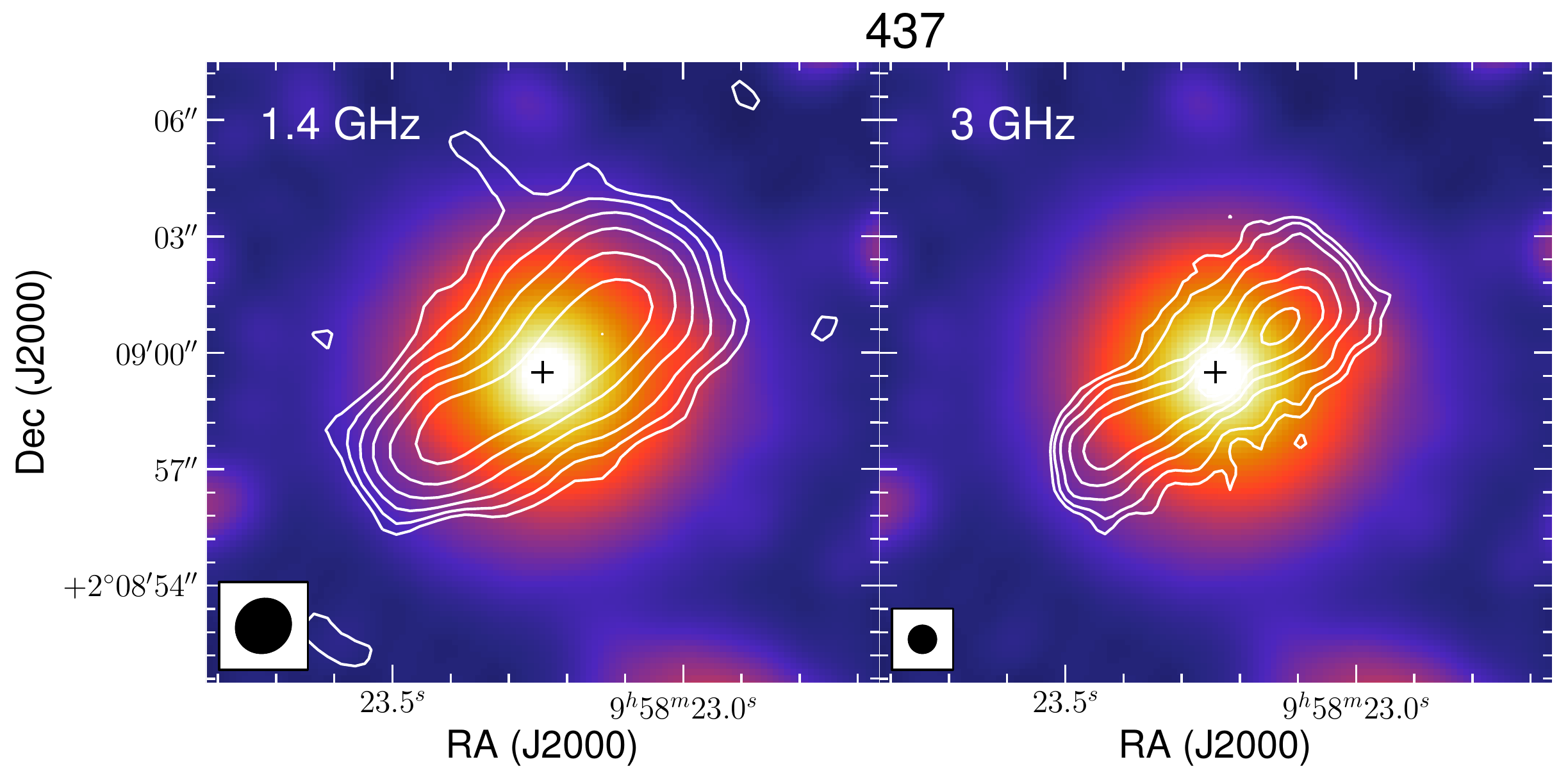}
    \includegraphics{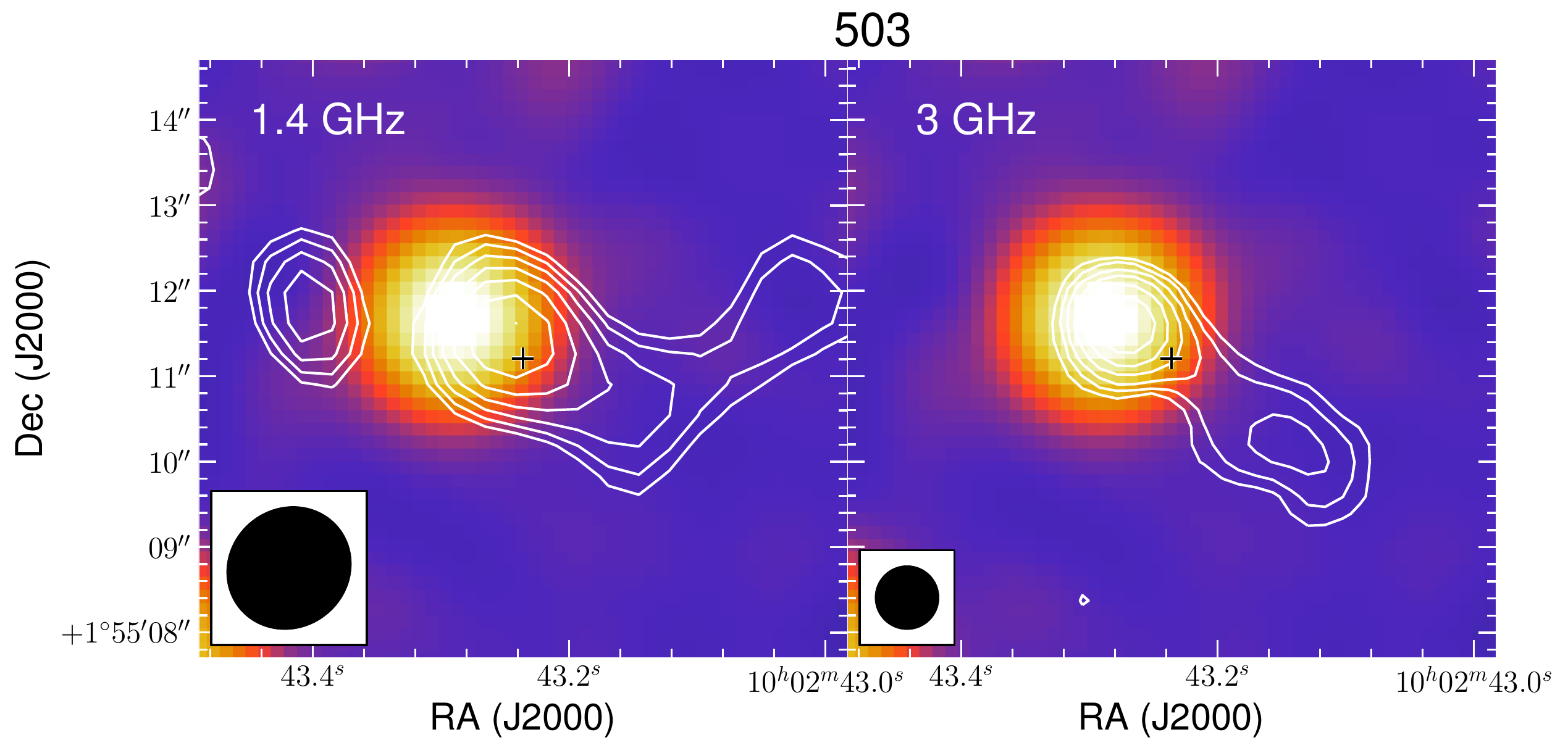}
    \includegraphics{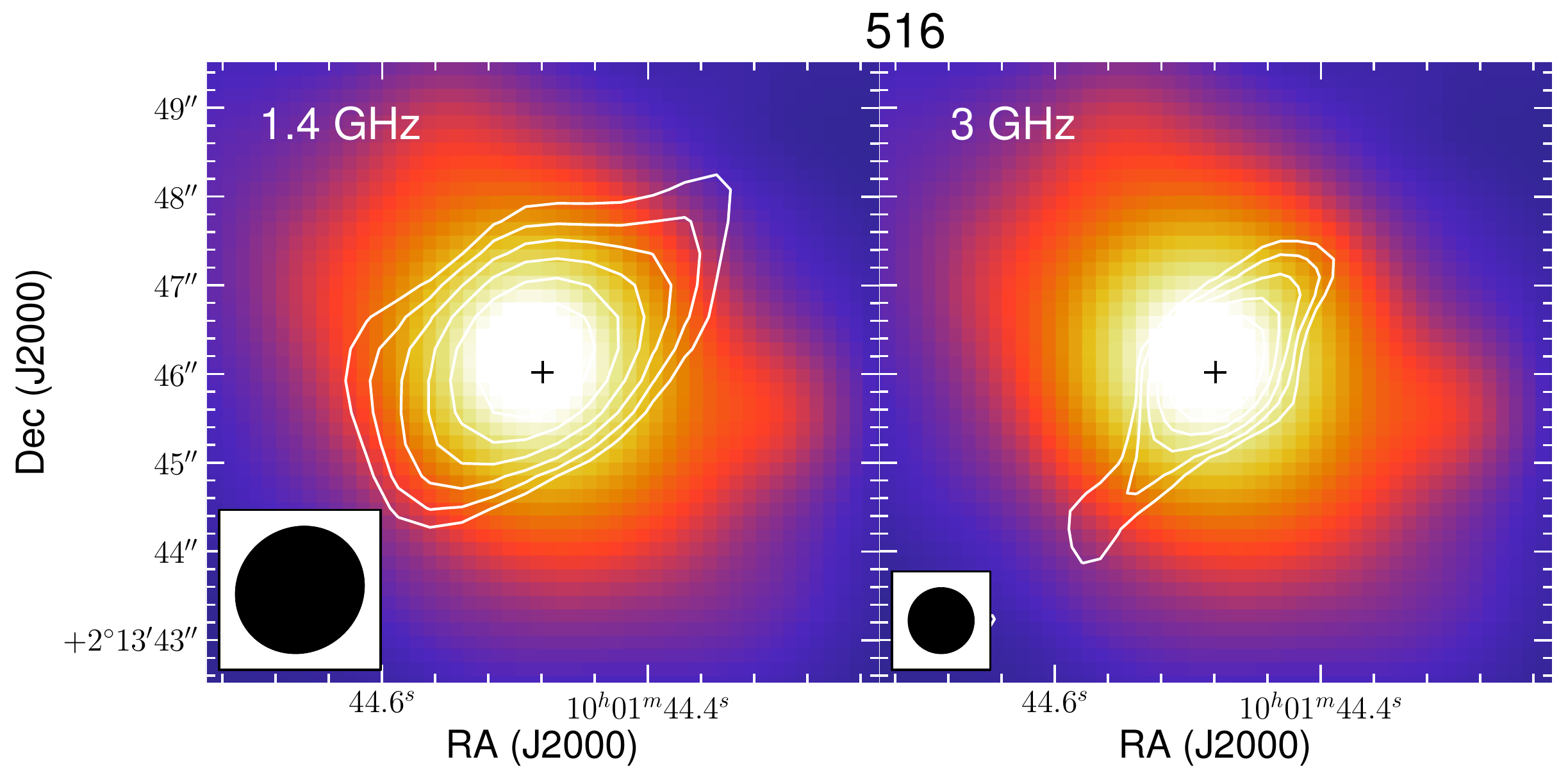}
            }
             \\ \\ 
      \resizebox{\hsize}{!}
       {\includegraphics{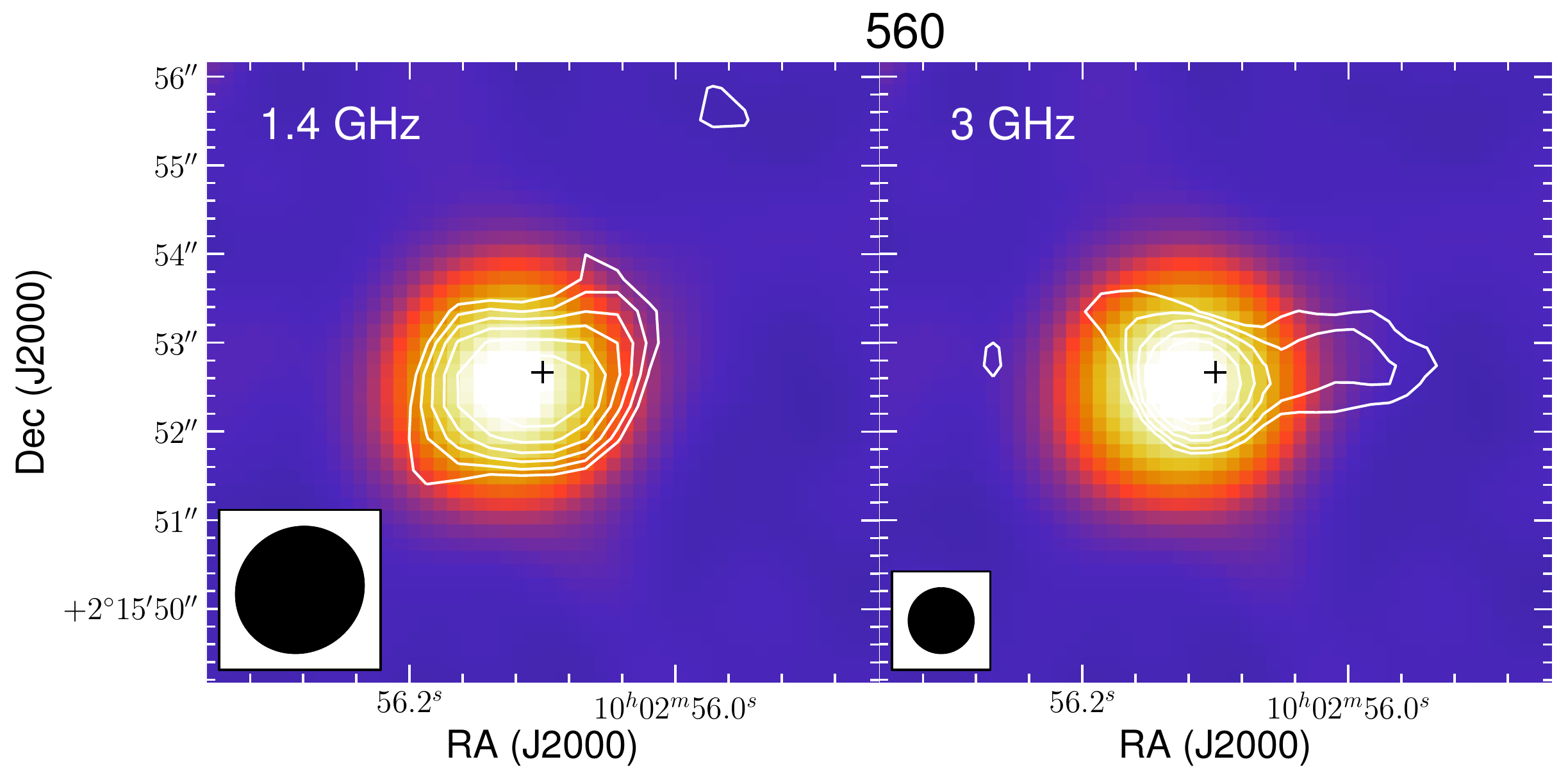}
        \includegraphics{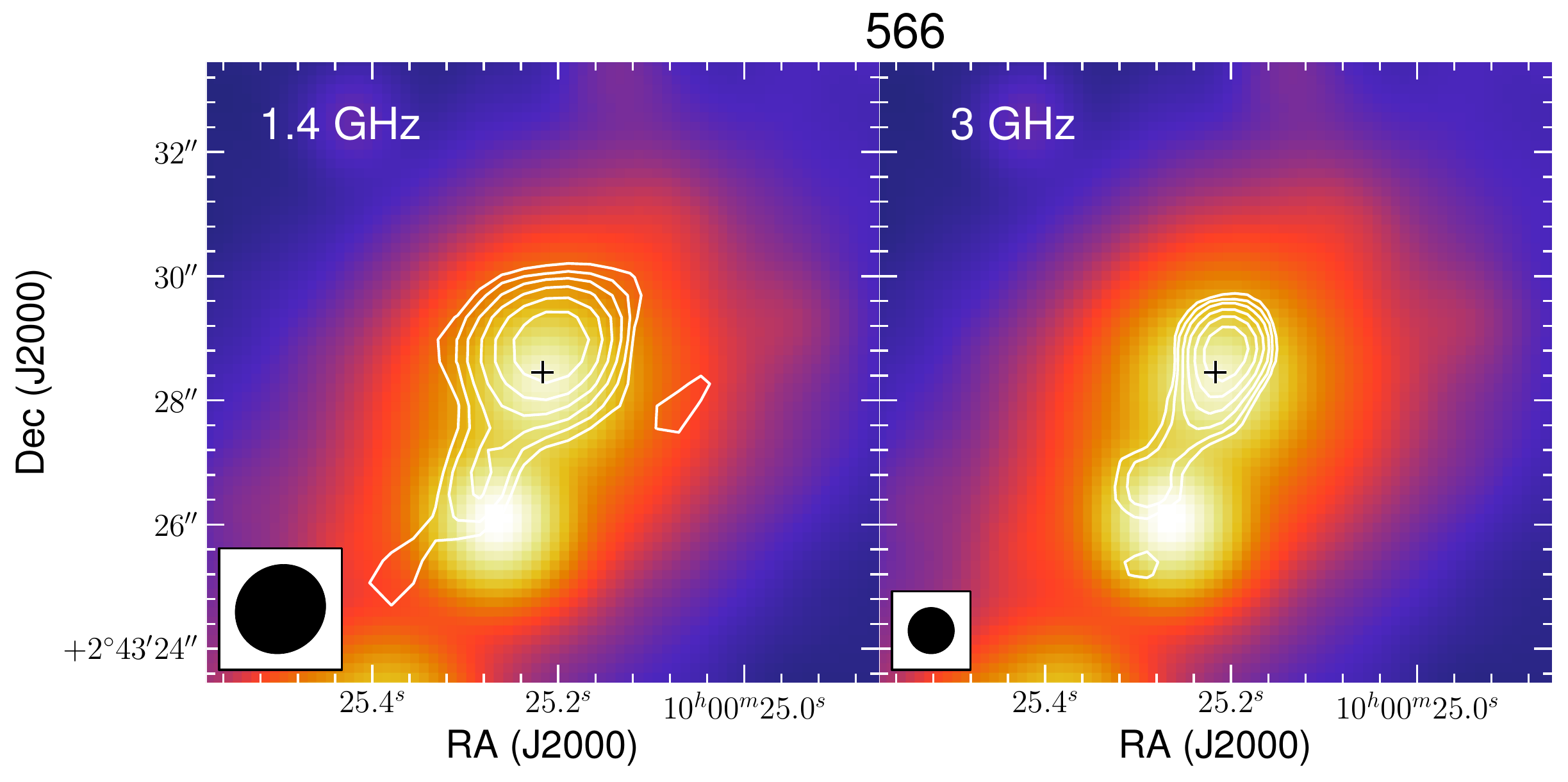}
       \includegraphics{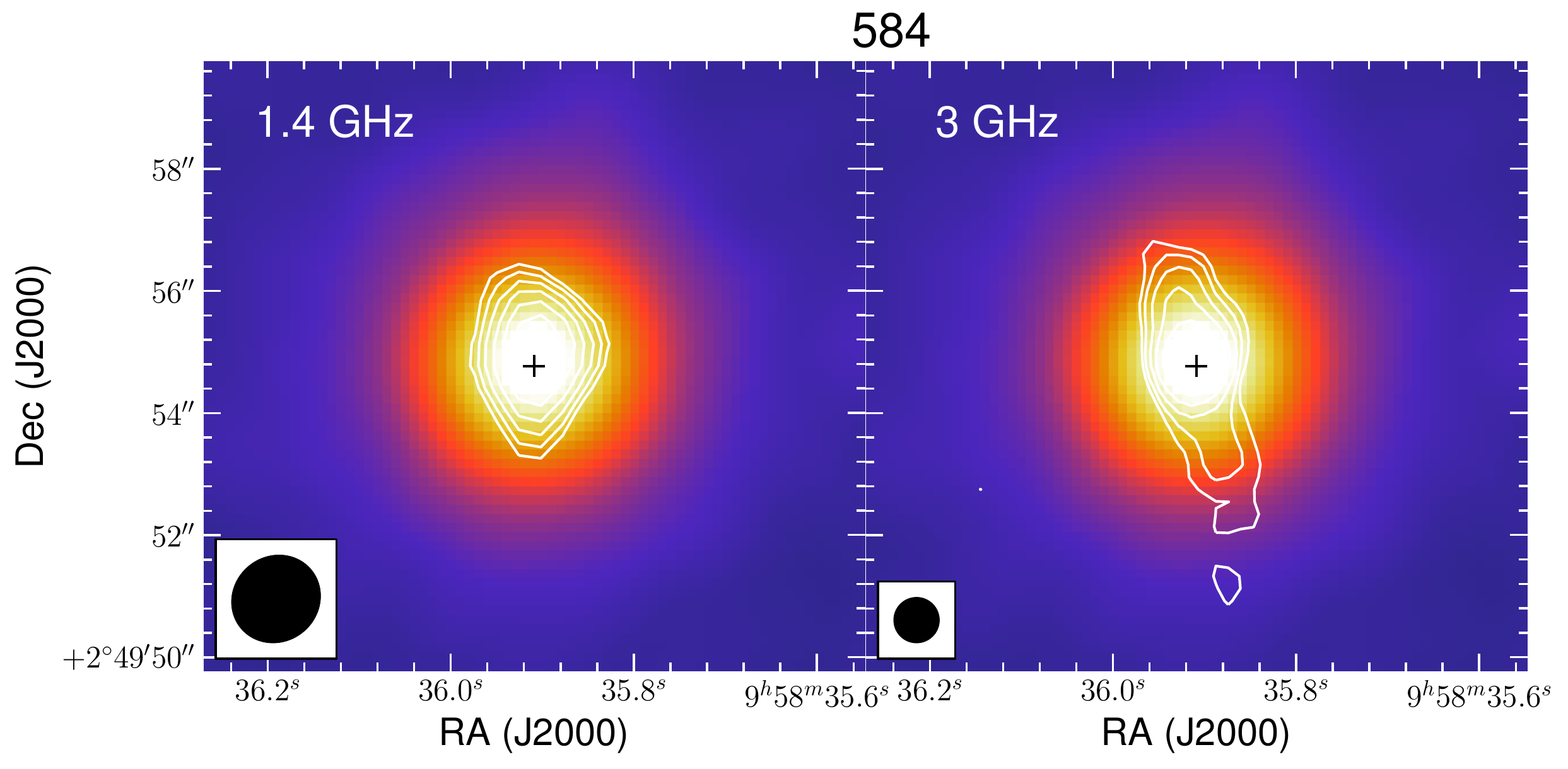}
            }
 \\ \\
  \resizebox{\hsize}{!}
{\includegraphics{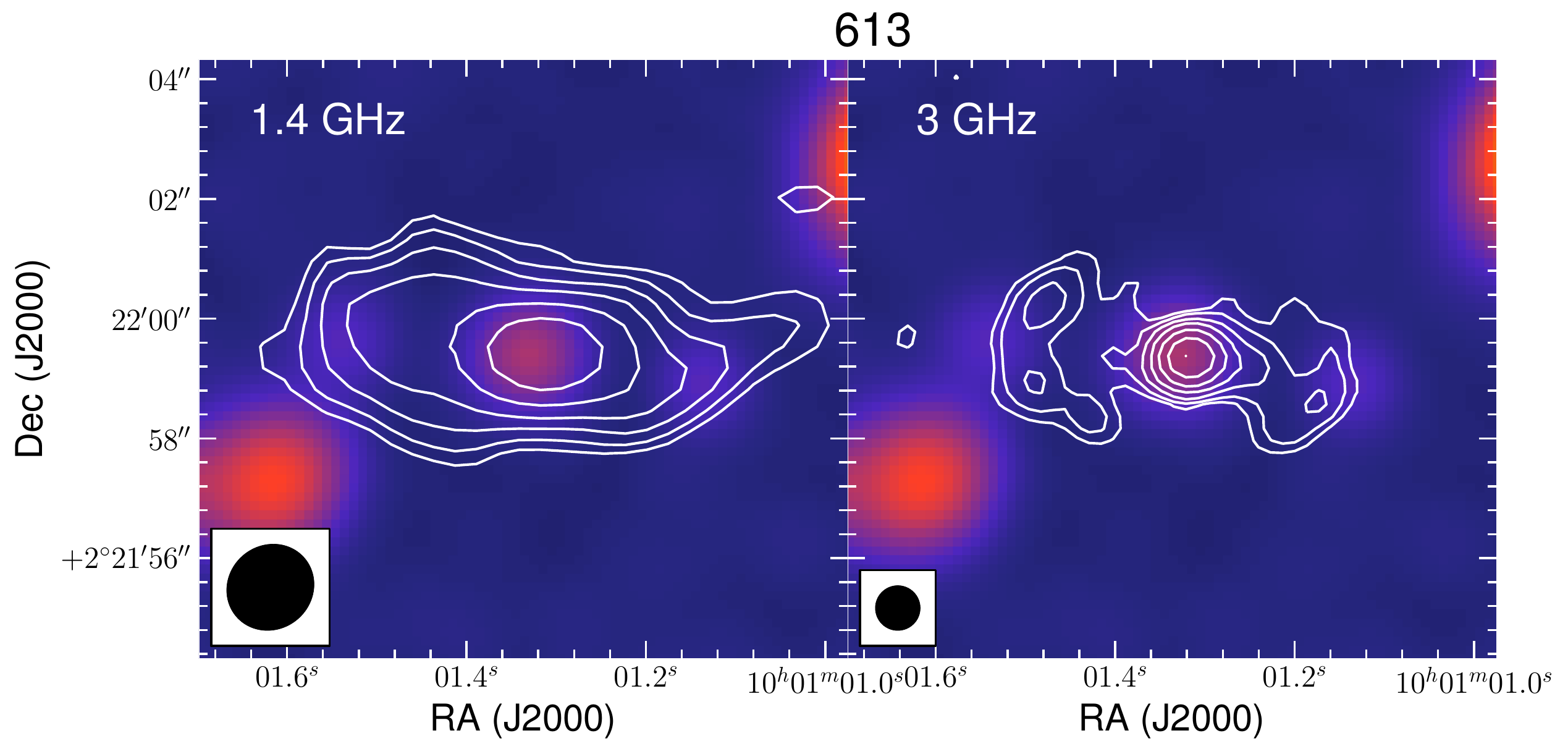}
 \includegraphics{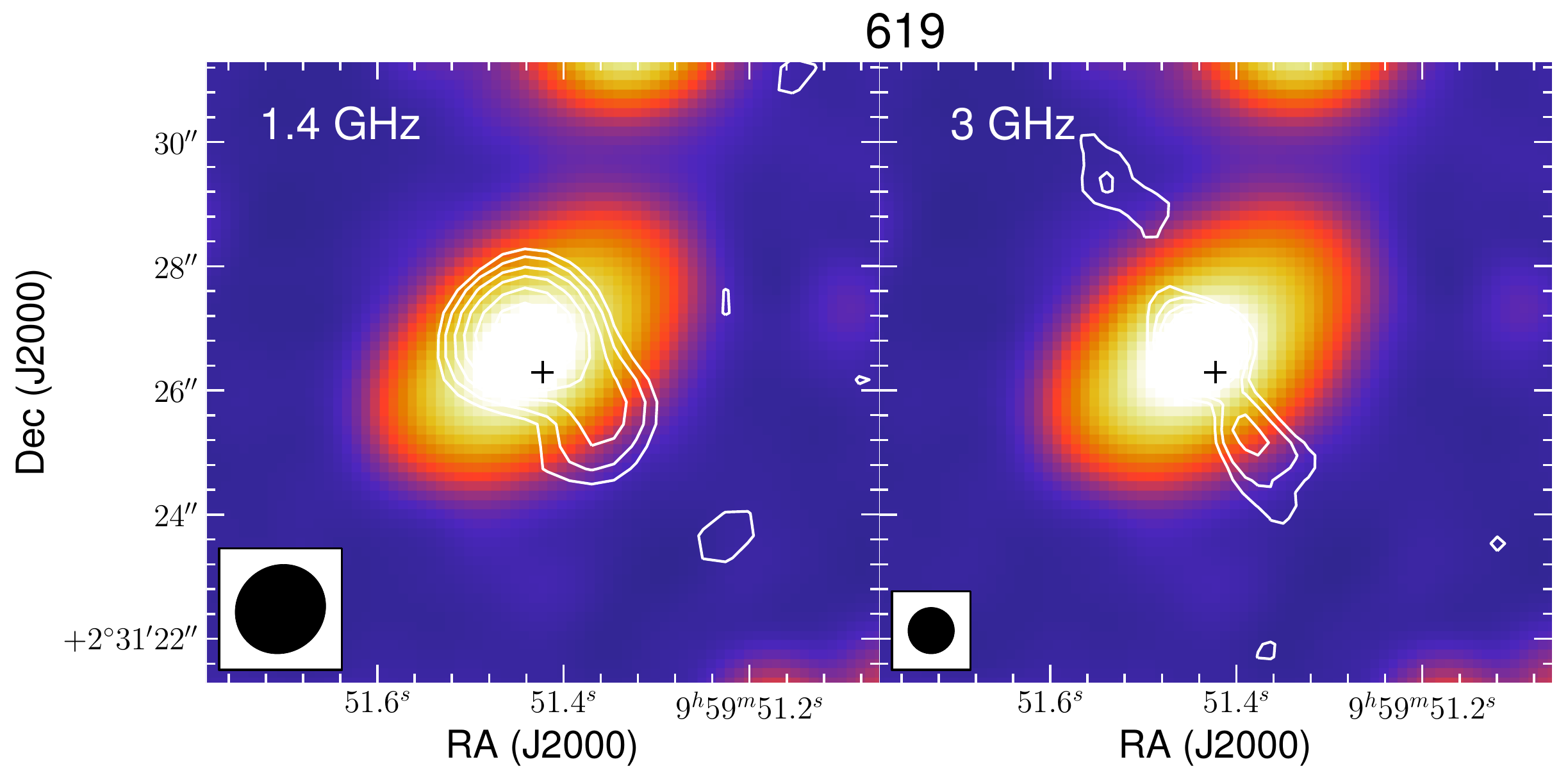}
 \includegraphics{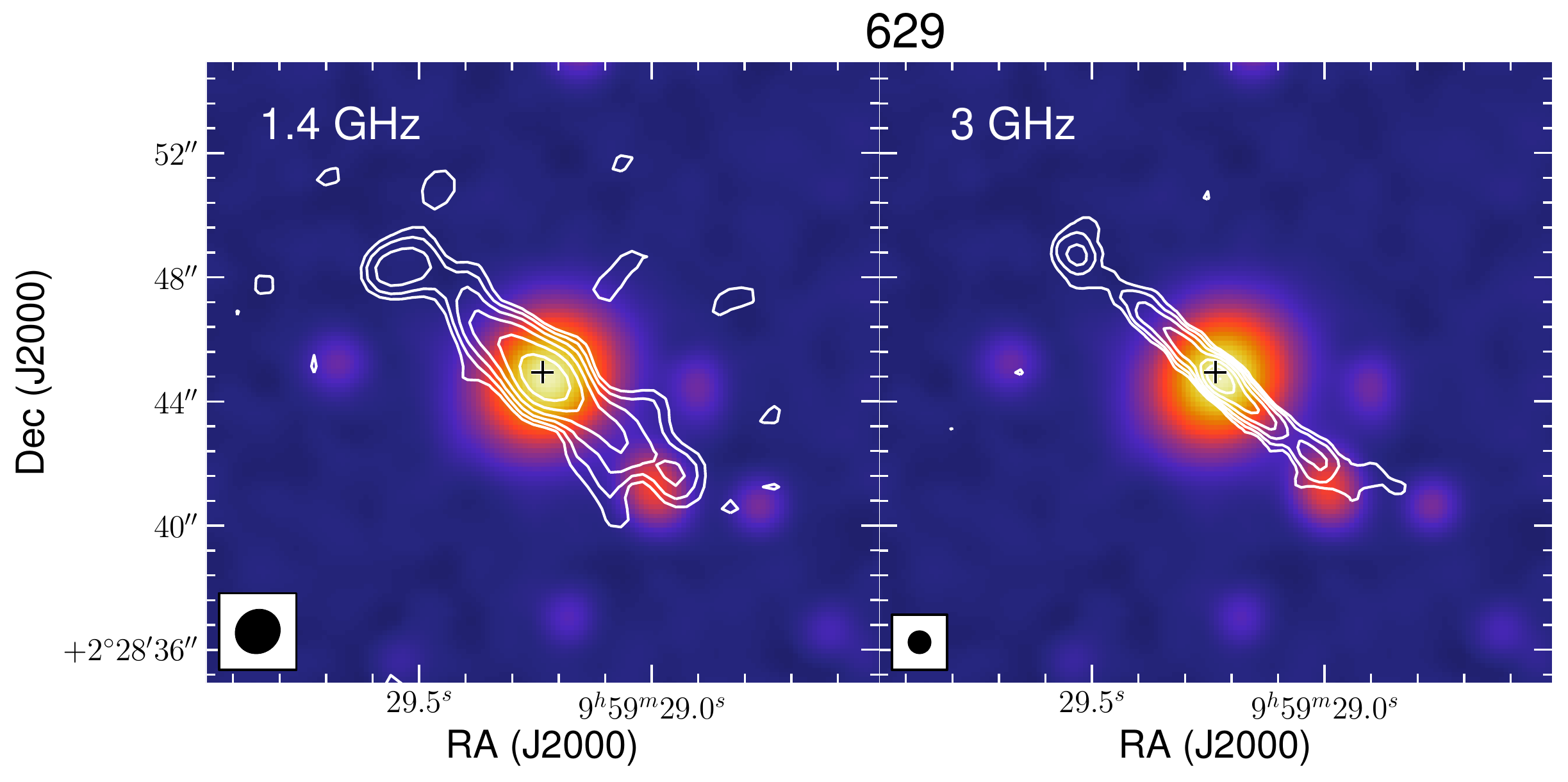}
            }
            \\ \\ 
  \resizebox{\hsize}{!}
 {\includegraphics{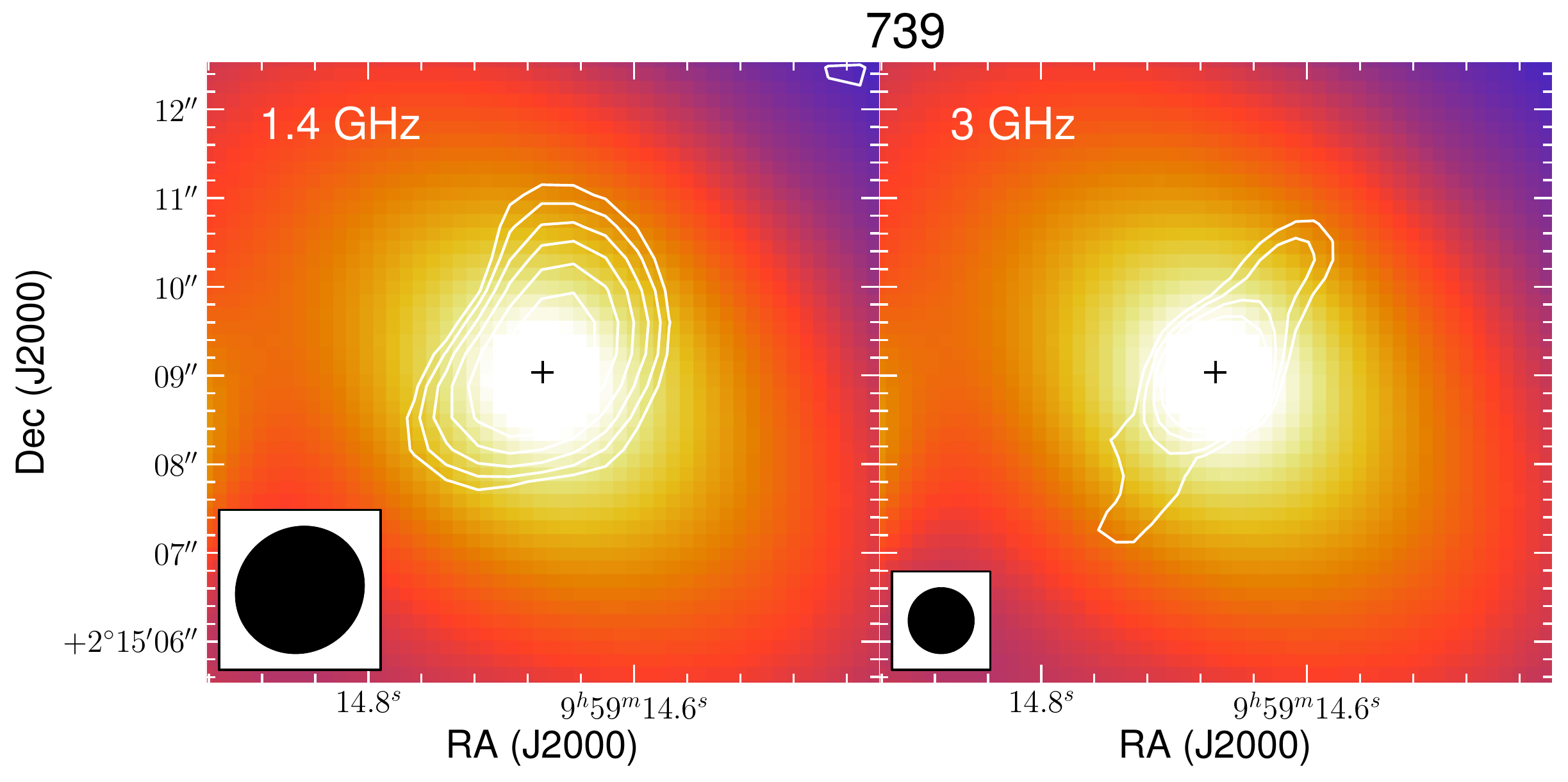}
    \includegraphics{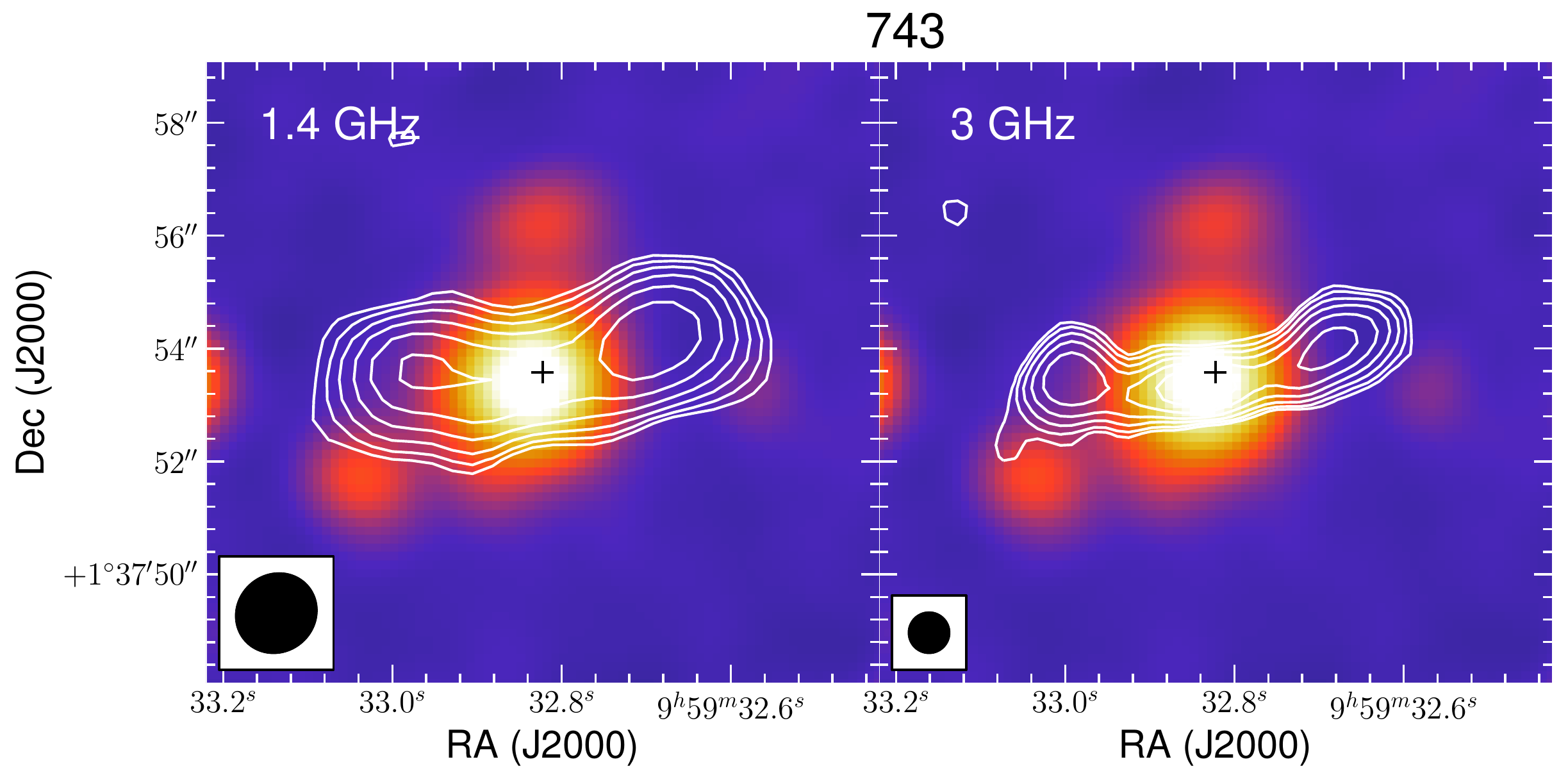}
    \includegraphics{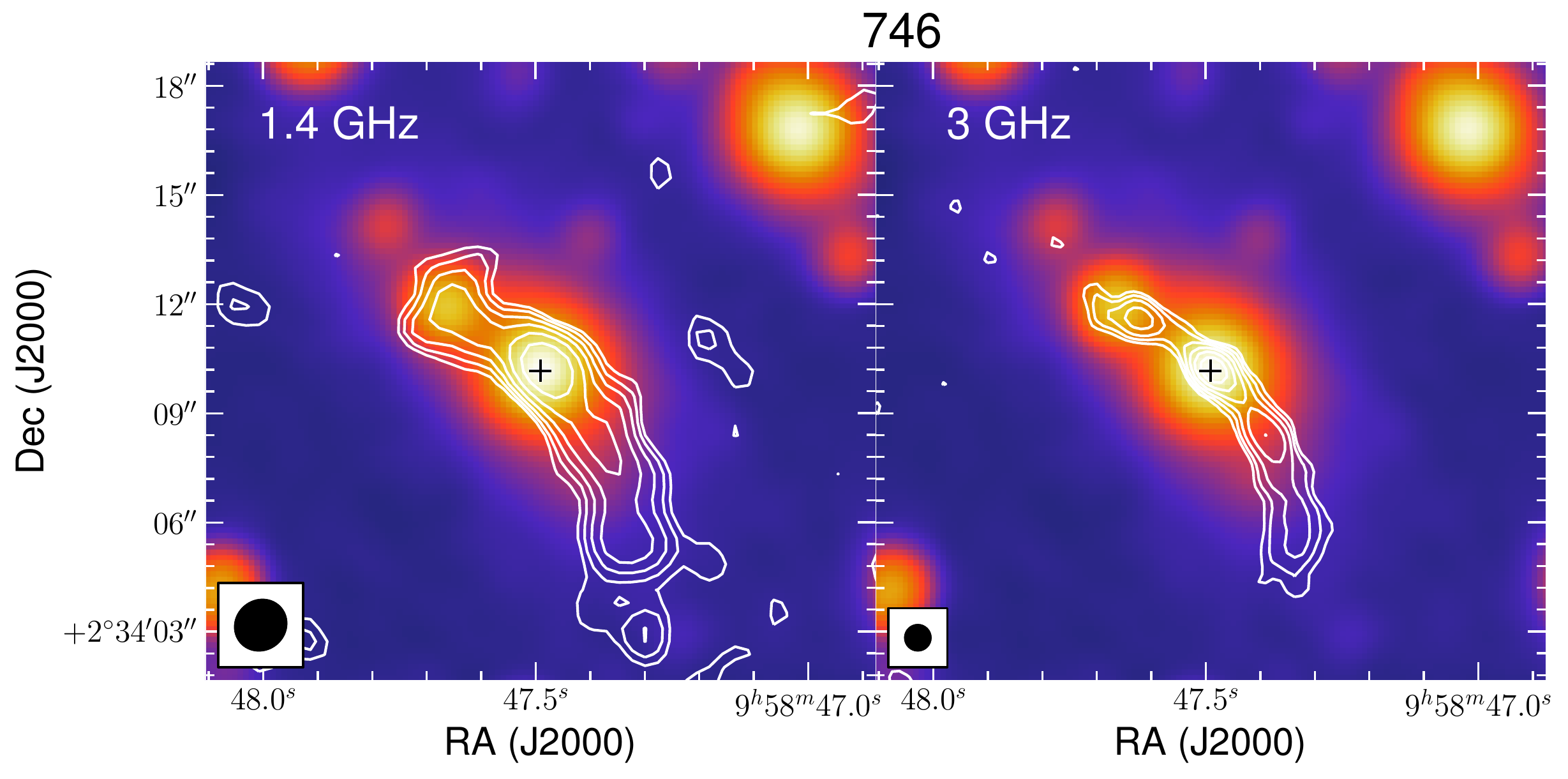}
            }
             \\ \\ 
      \resizebox{\hsize}{!}
       {\includegraphics{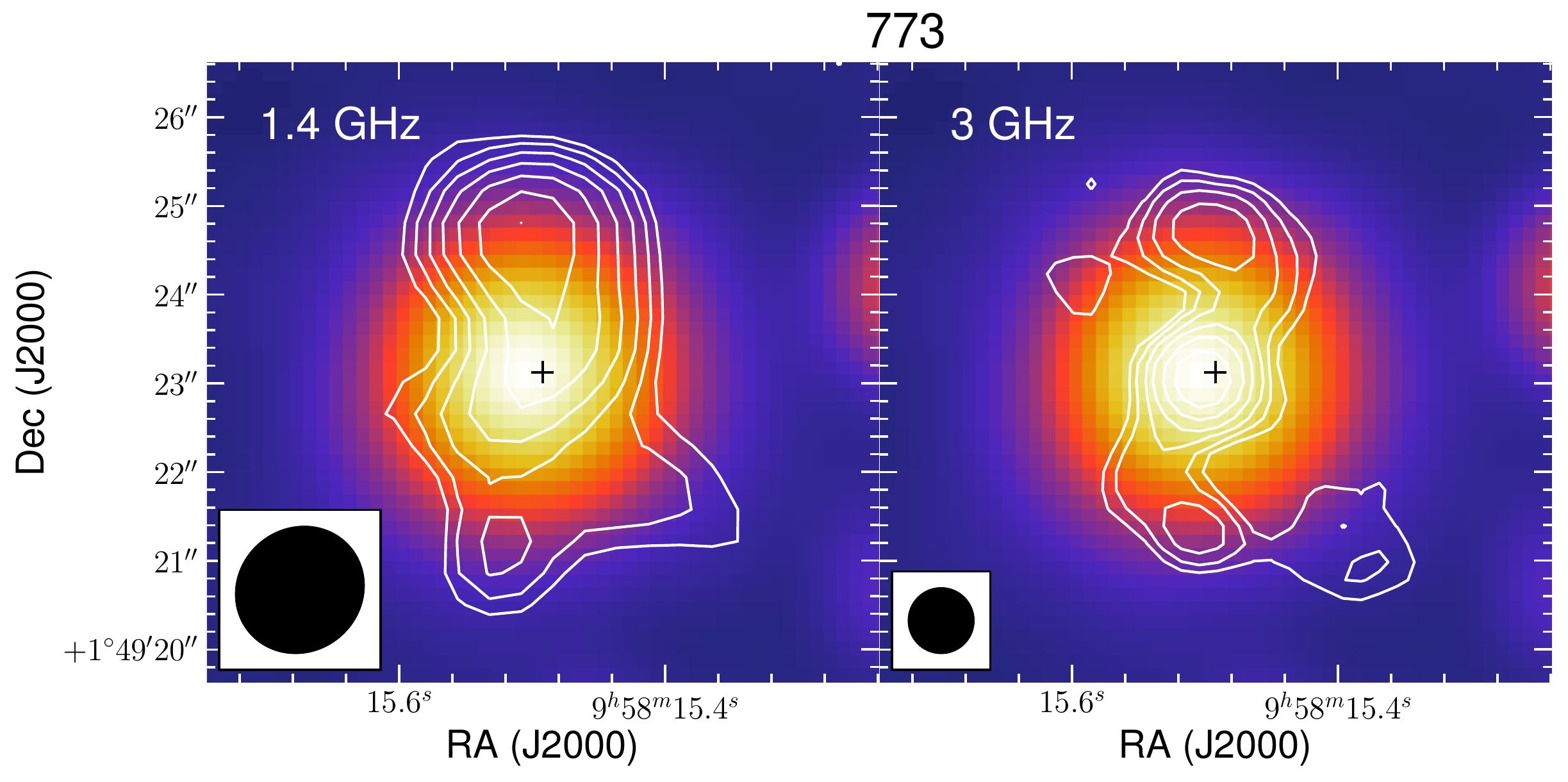}
        \includegraphics{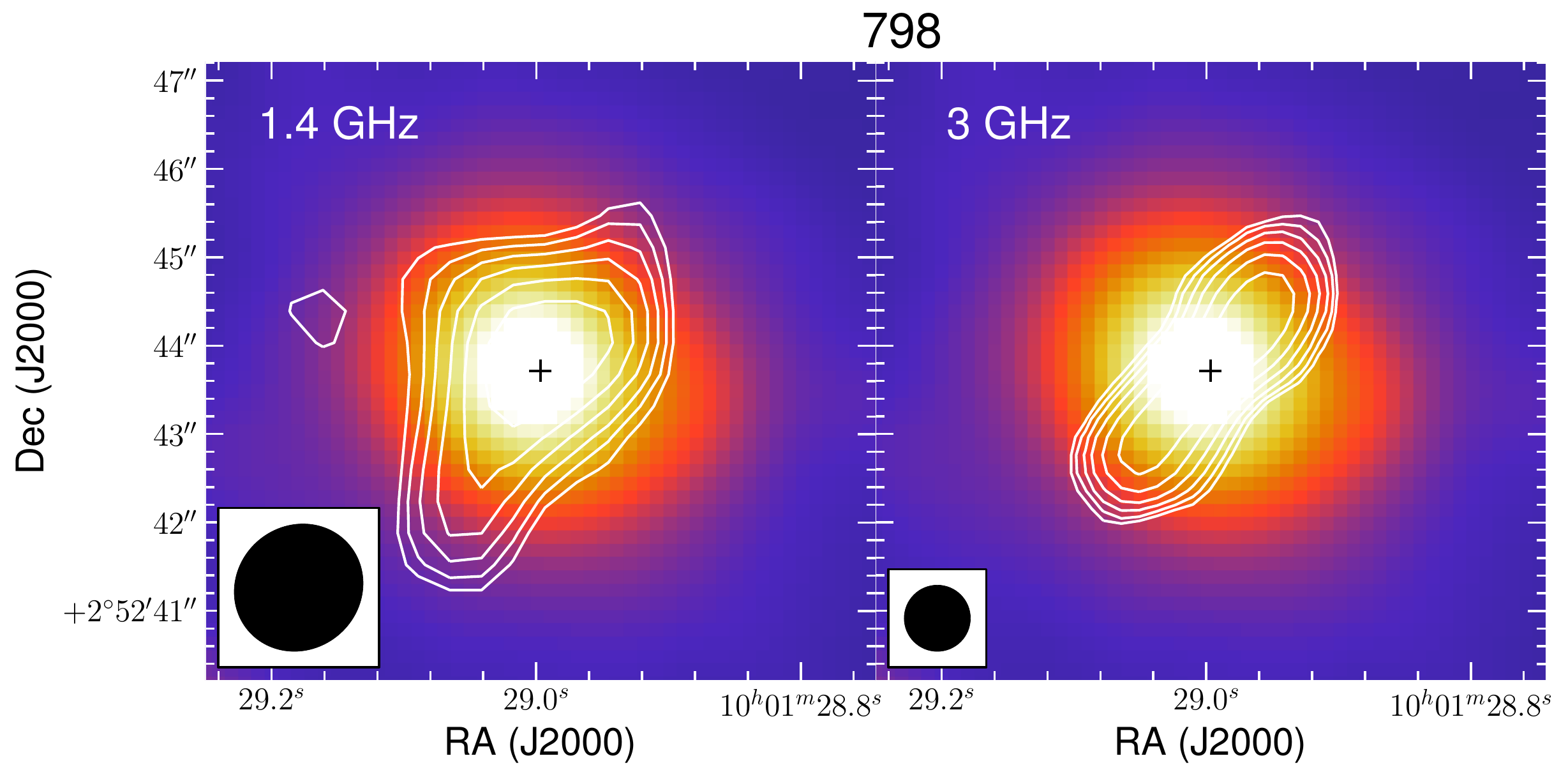}
       \includegraphics{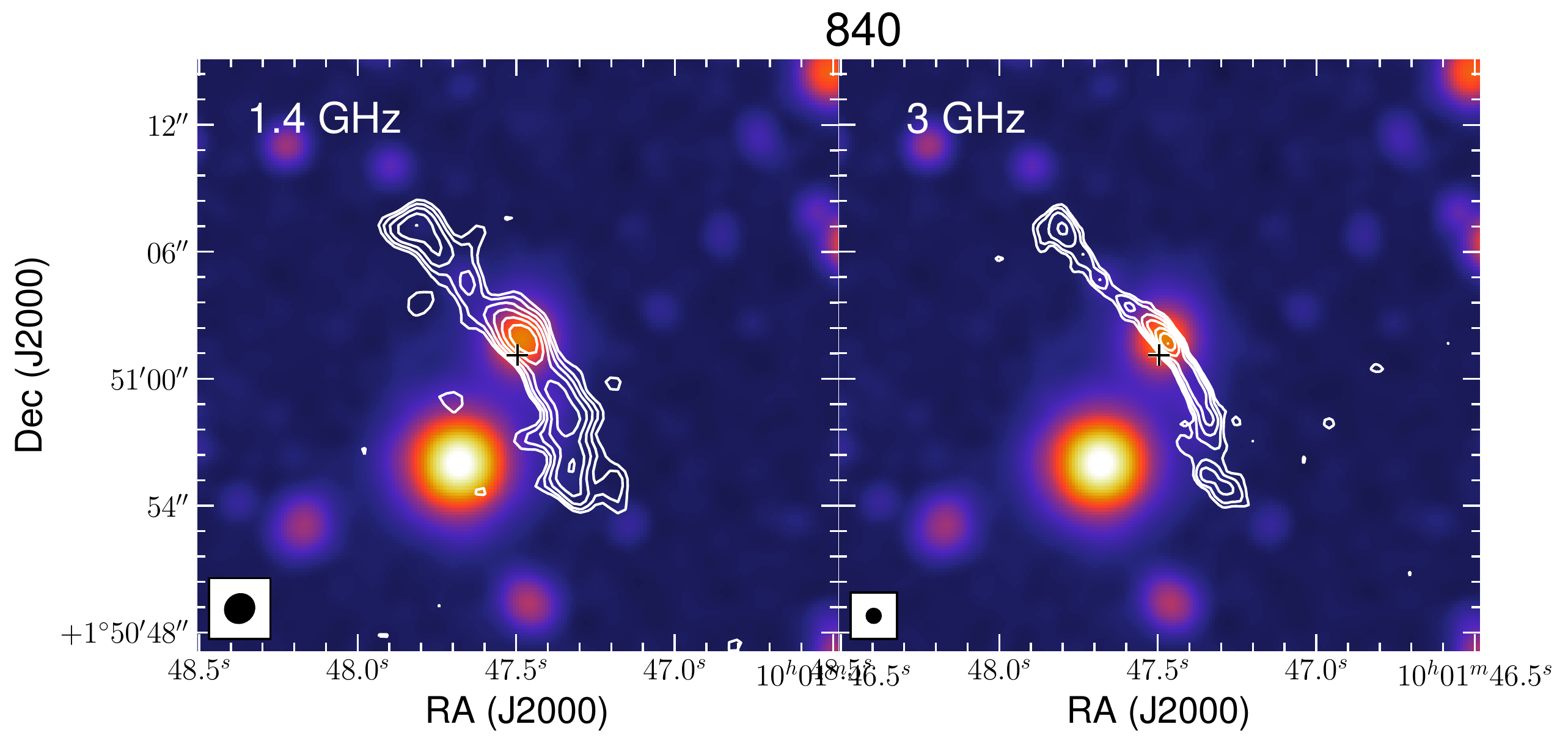}
            }
  \\ \\
 \resizebox{\hsize}{!}
{\includegraphics{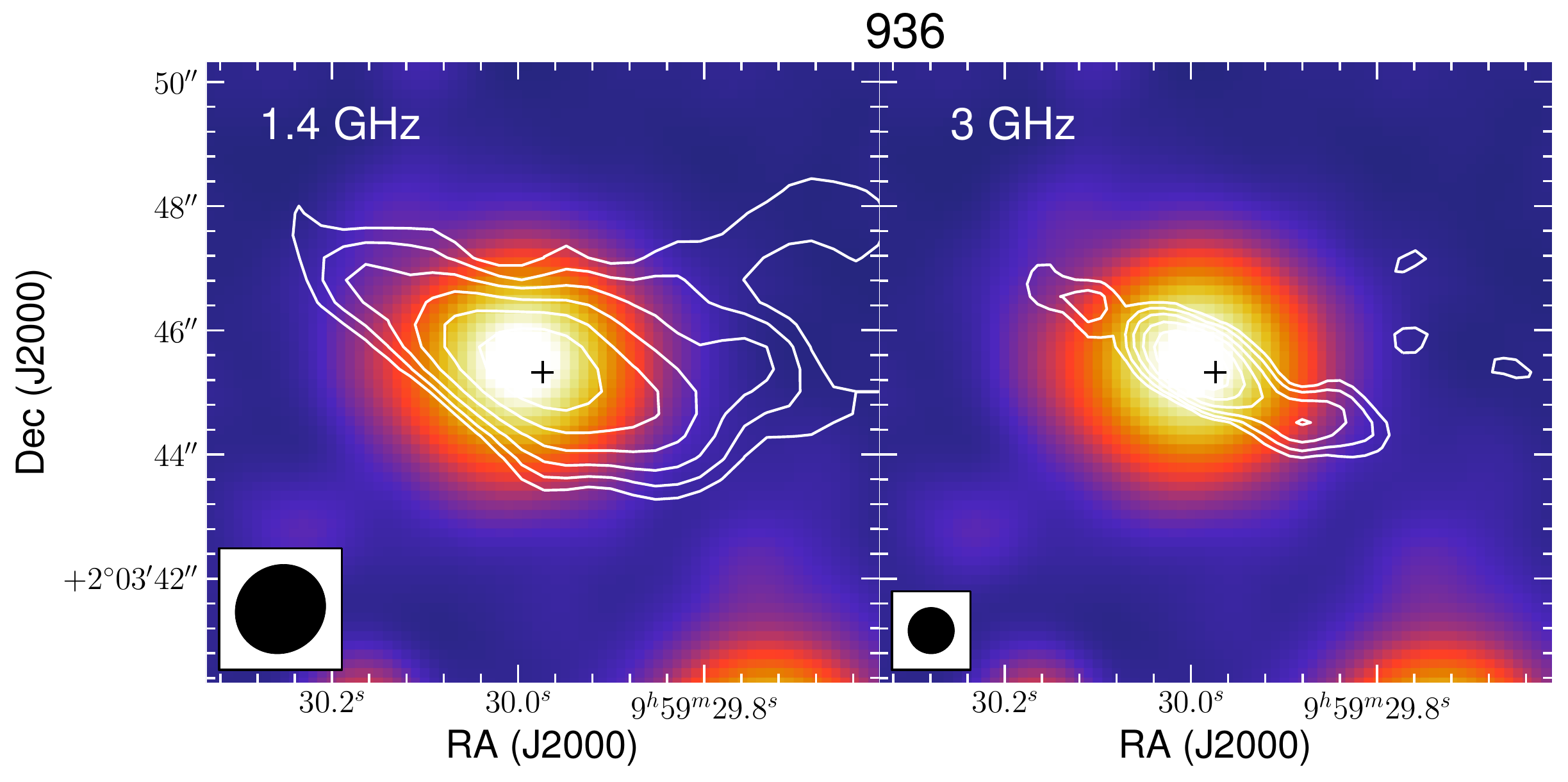}
 \includegraphics{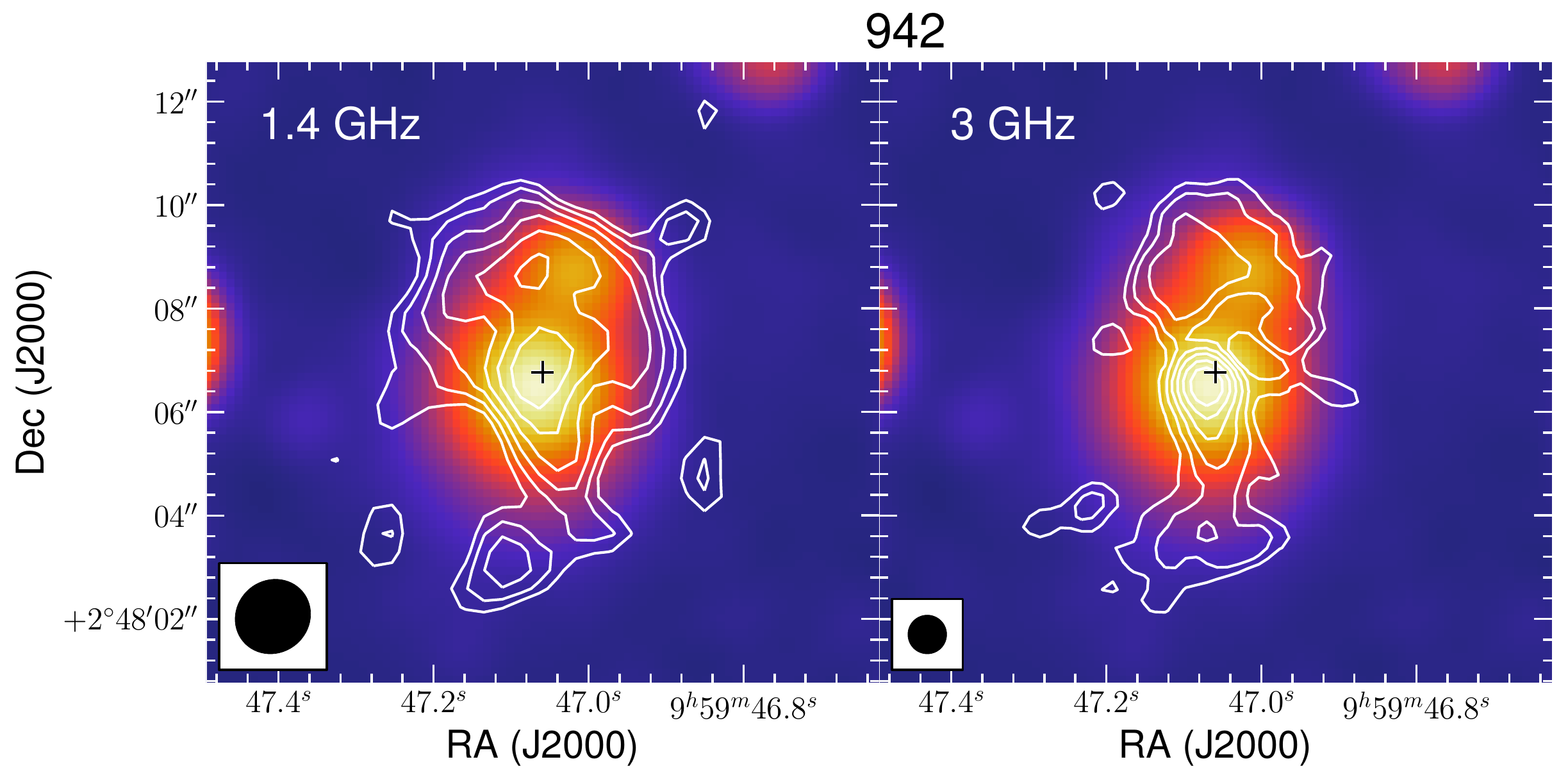}
 \includegraphics{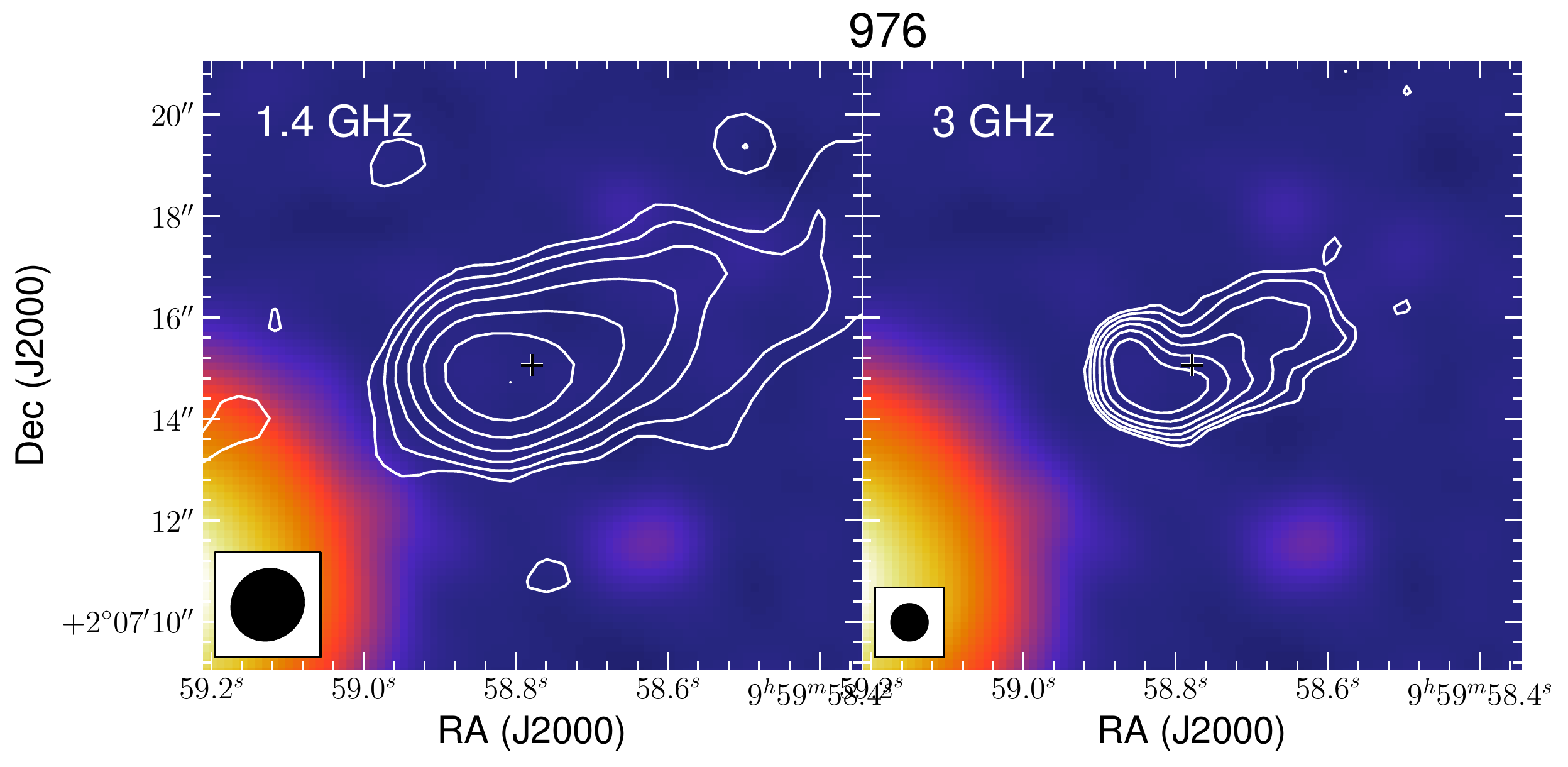}
            }
            \\ \\ 
  \resizebox{\hsize}{!}
 {\includegraphics{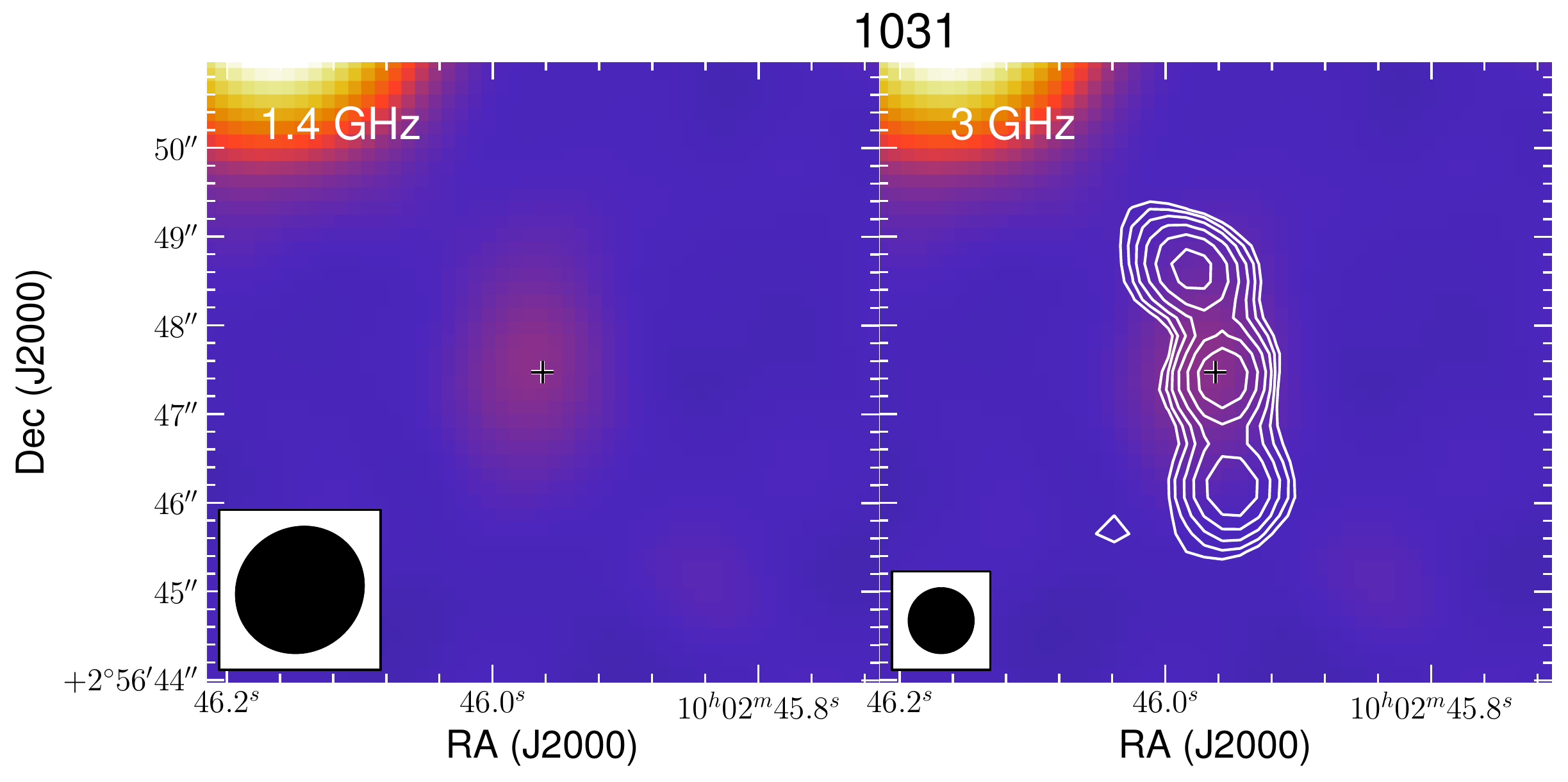}
    \includegraphics{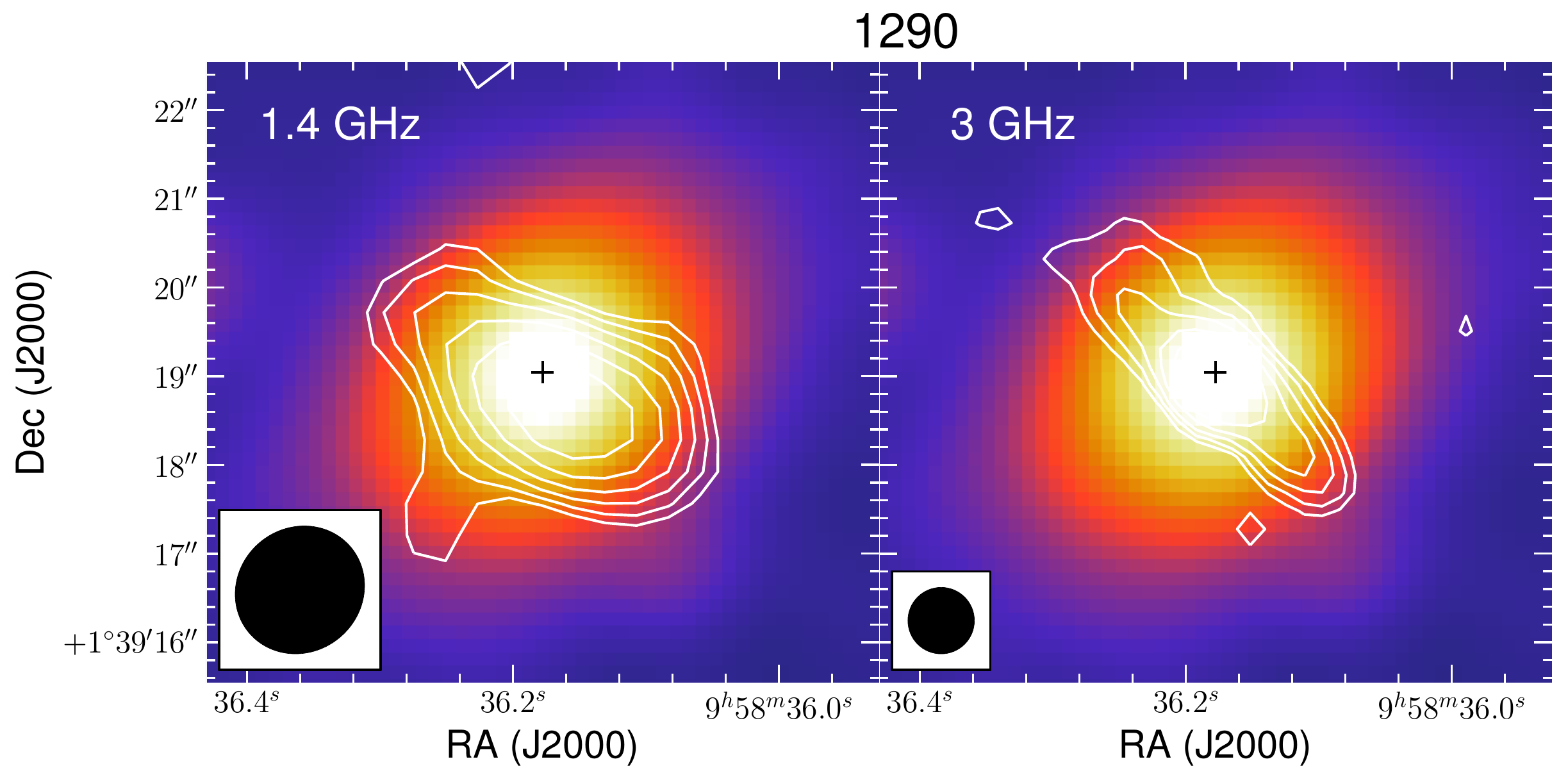}
    \includegraphics{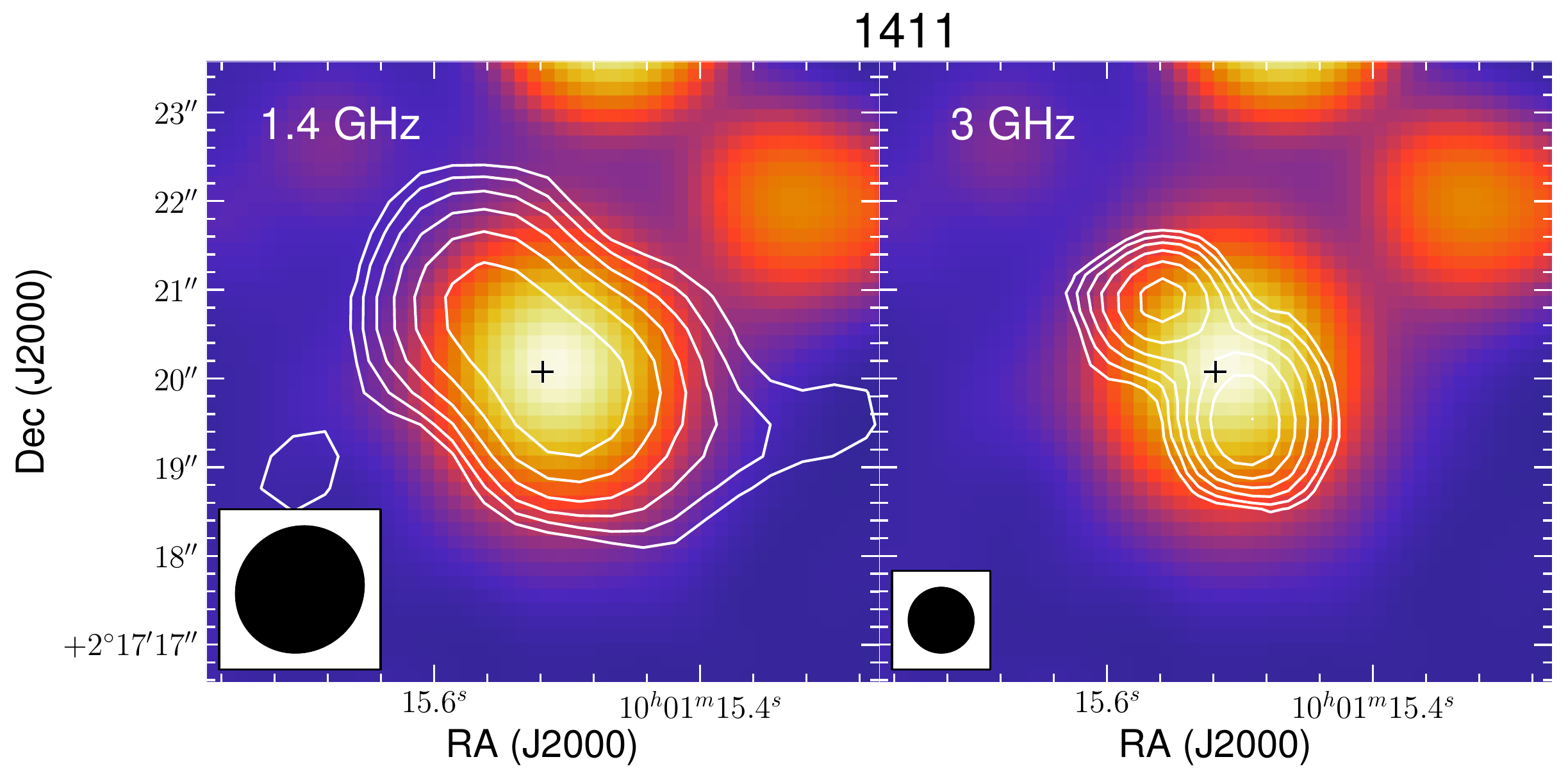}
            }
           
   \caption{(continued)
   }
              \label{fig:maps2}%
    \end{figure*}
\addtocounter{figure}{-1}
\begin{figure*}[!ht]
  \resizebox{\hsize}{!}
       {\includegraphics{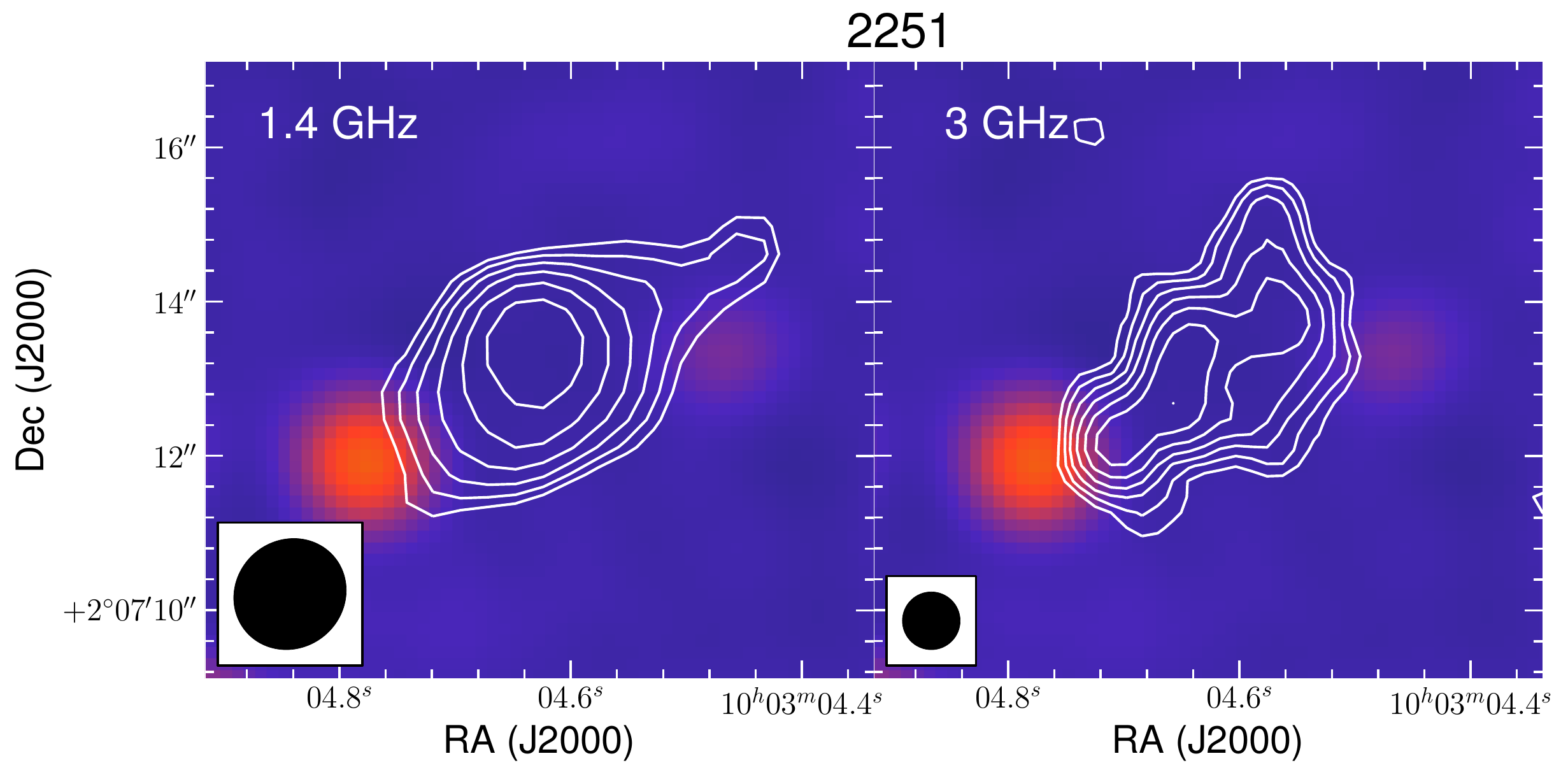}
        \includegraphics{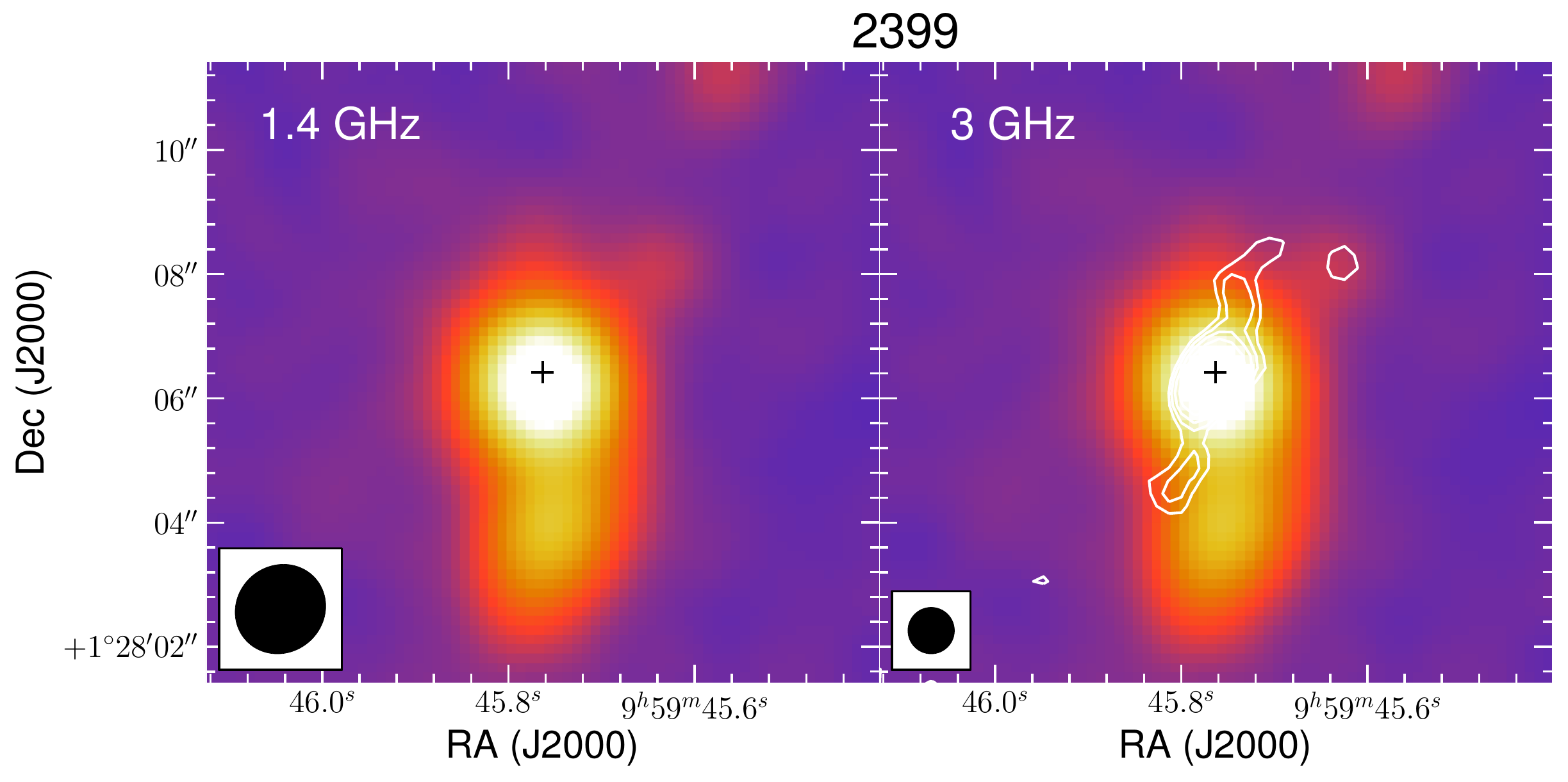}
       \includegraphics{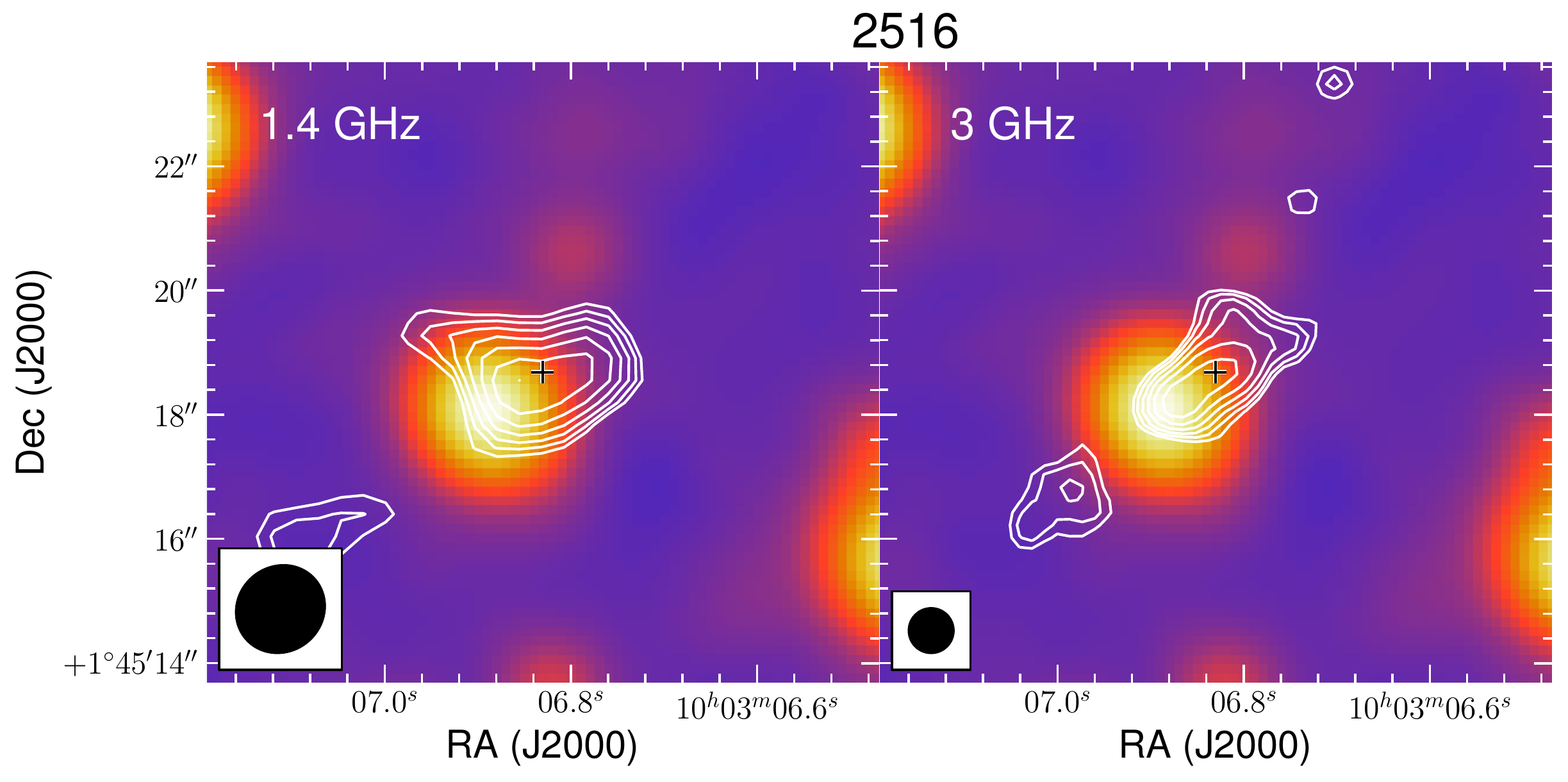}
            }
            \\ \\
 \resizebox{\hsize}{!}
{\includegraphics{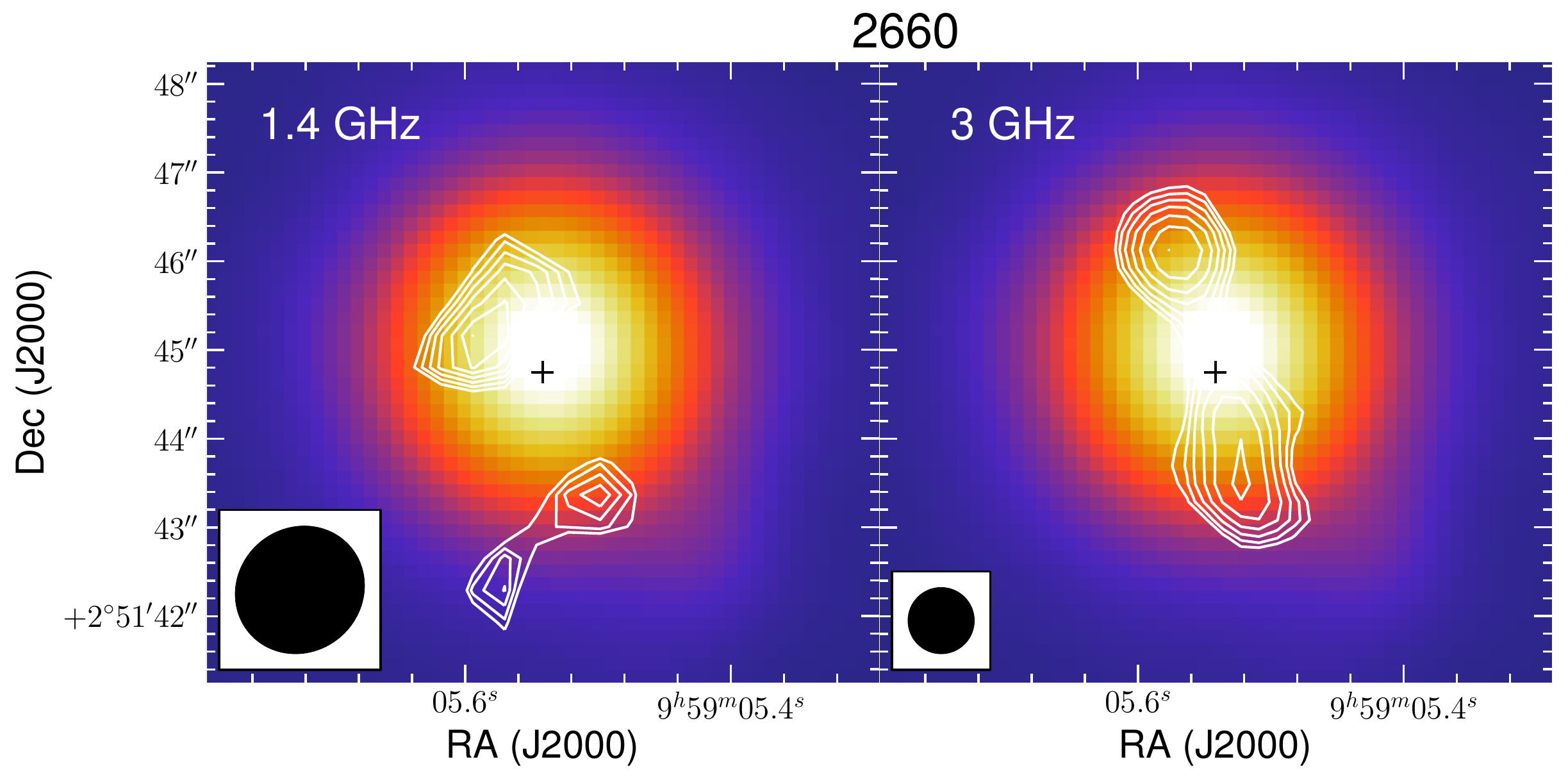}
 \includegraphics{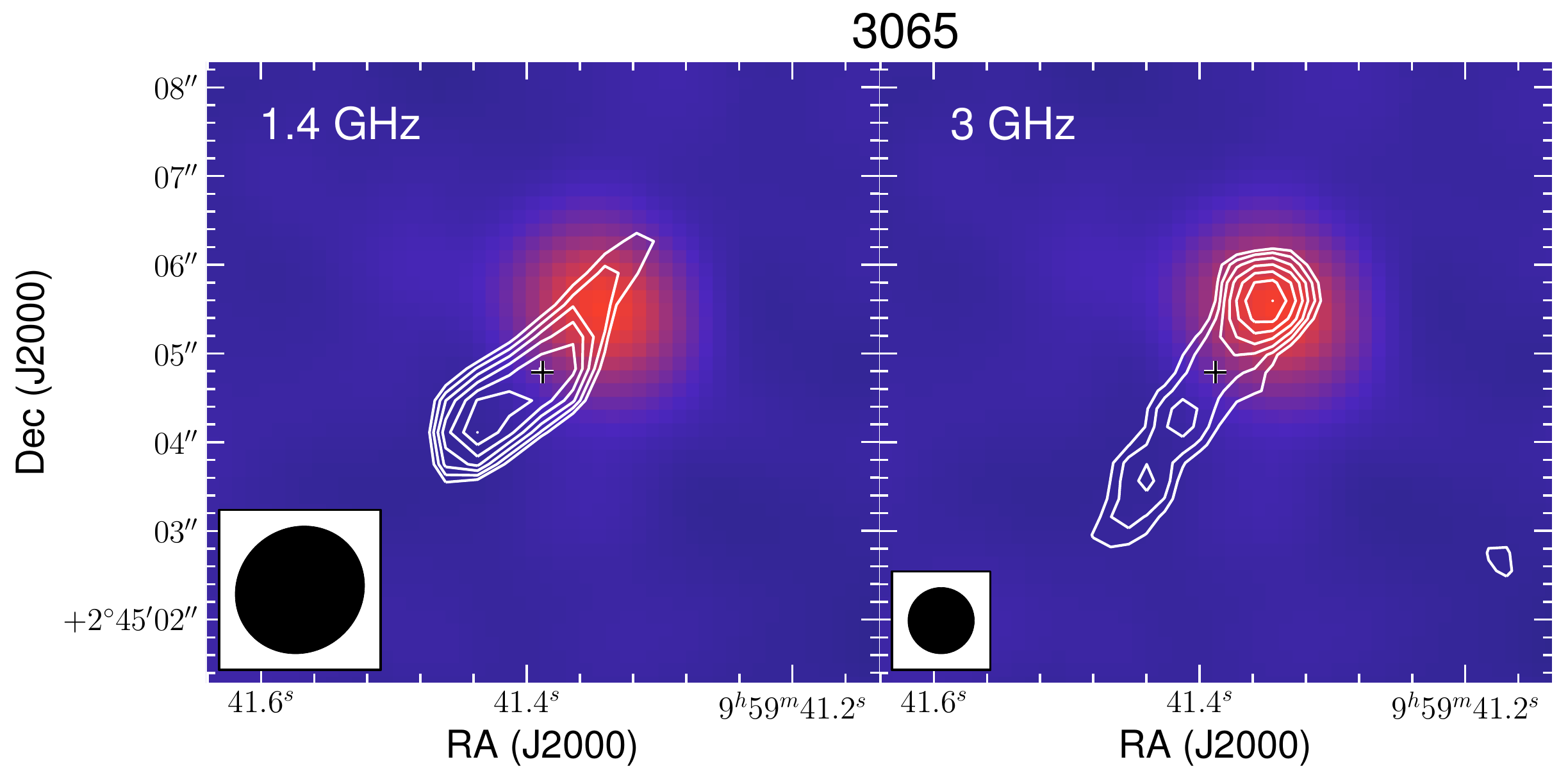}
 \includegraphics{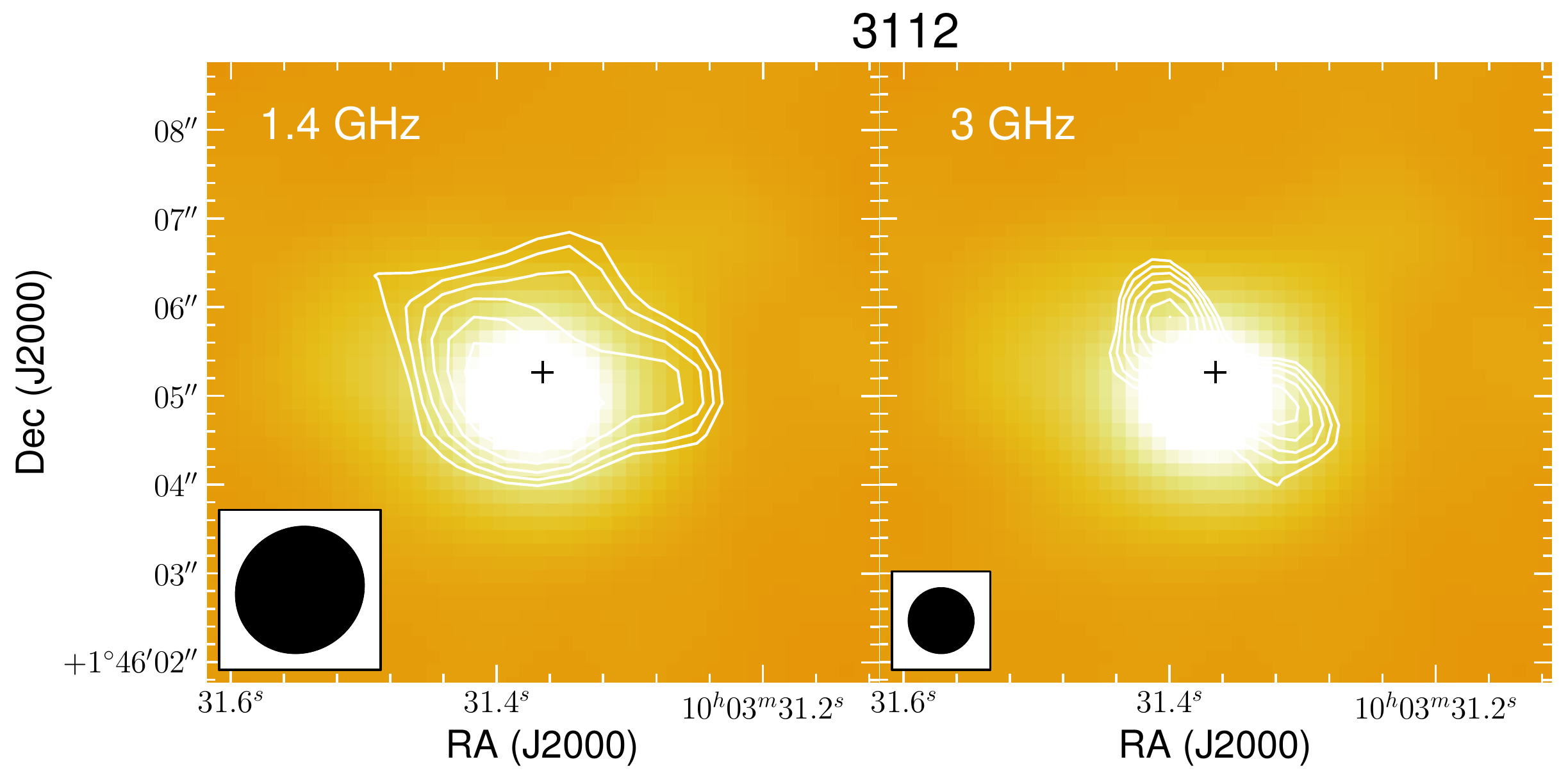}
            }
            \\ \\ 
  \resizebox{\hsize}{!}
 {
    \includegraphics{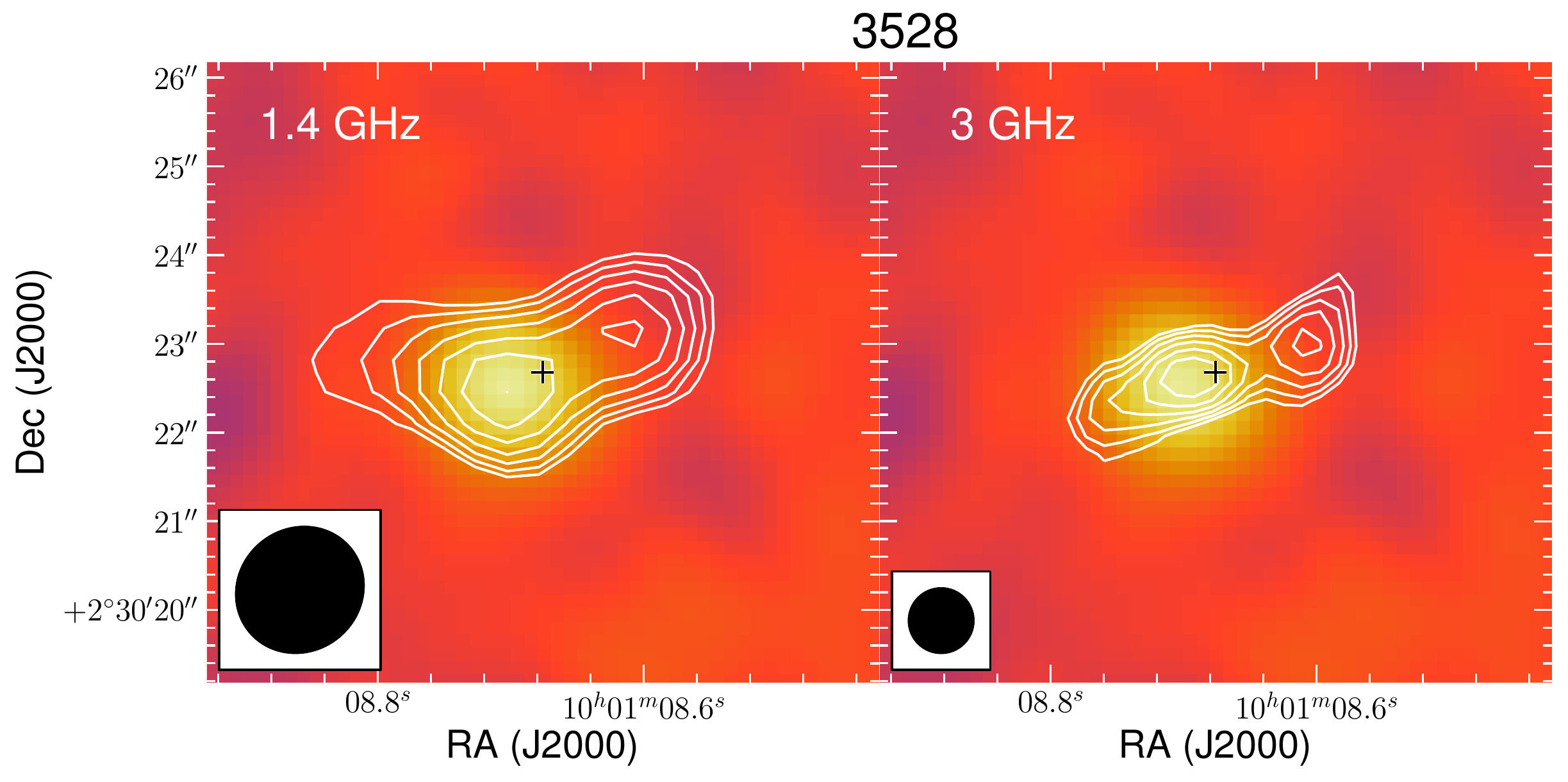}
    \includegraphics{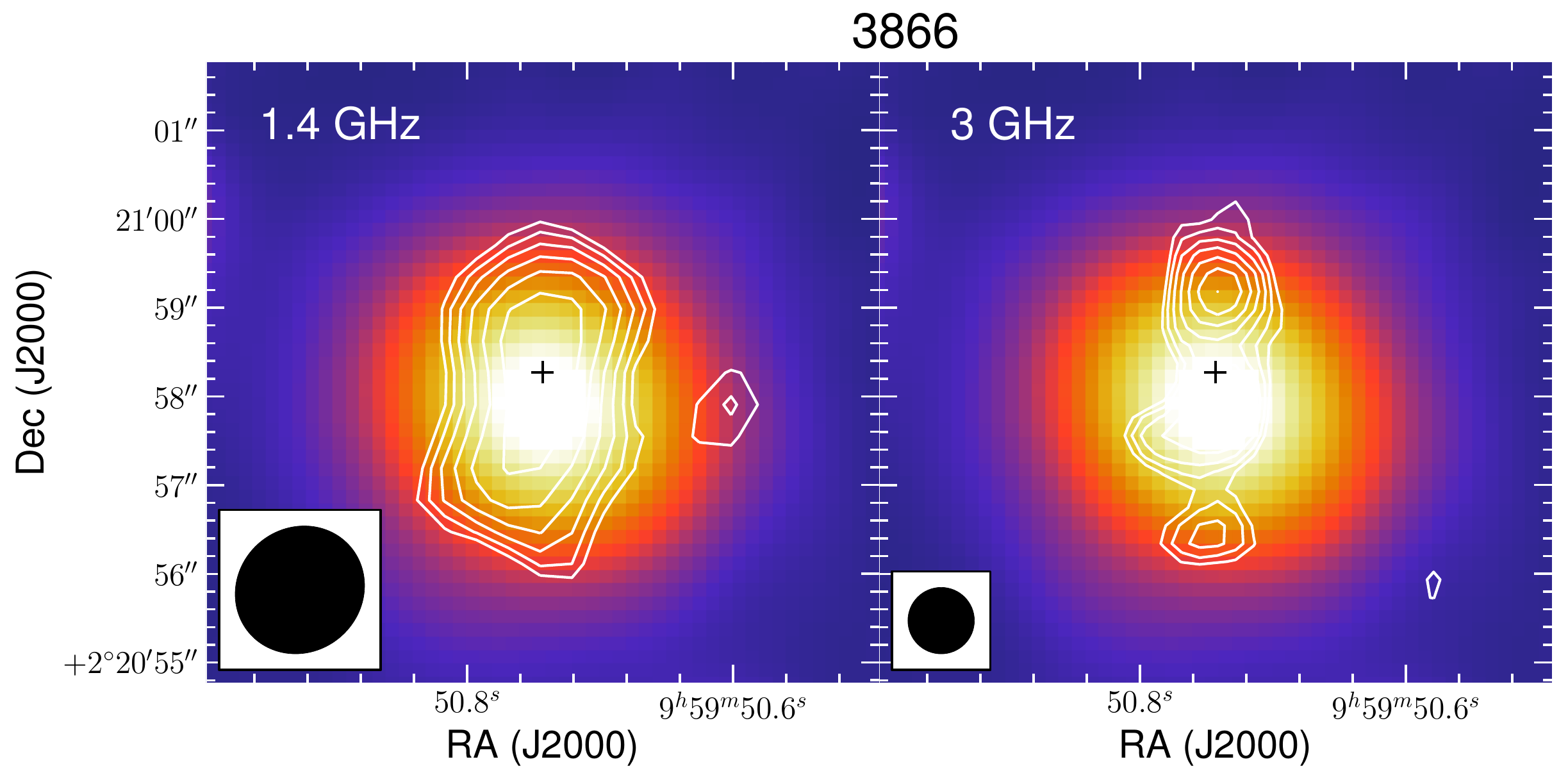}
            }
             \\ \\ 
      \resizebox{\hsize}{!}
       {\includegraphics{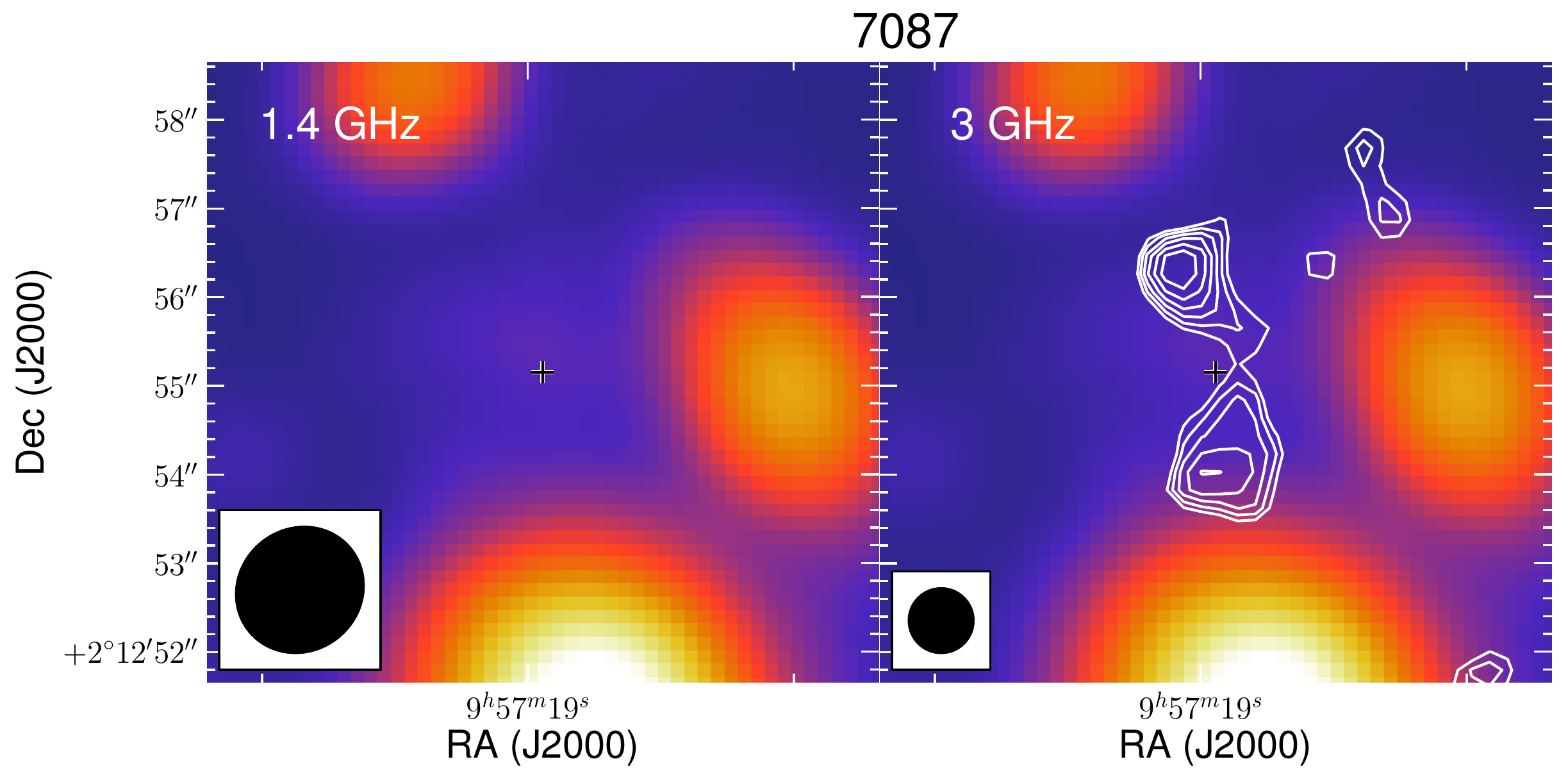}
        \includegraphics{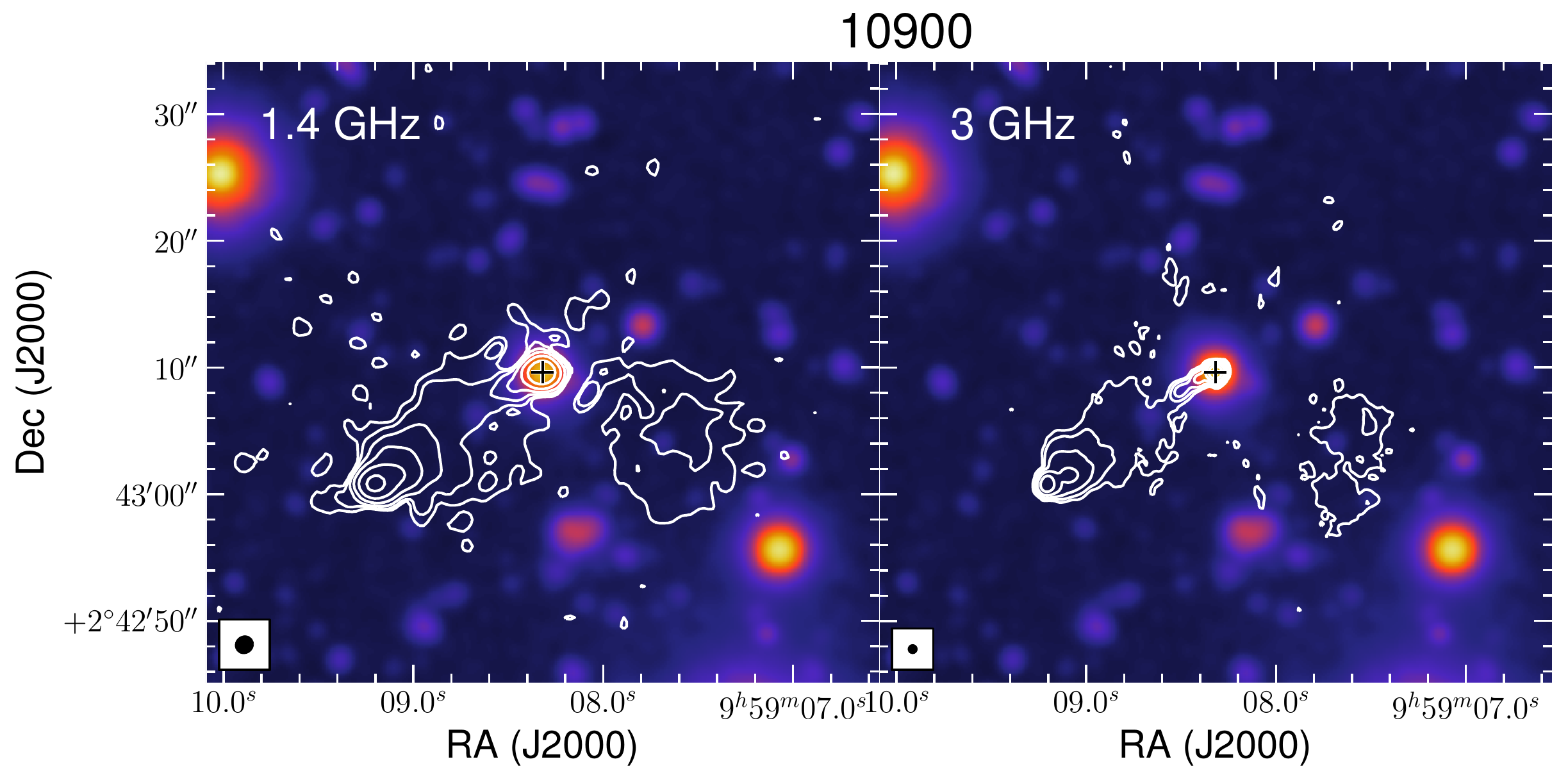}
       \includegraphics{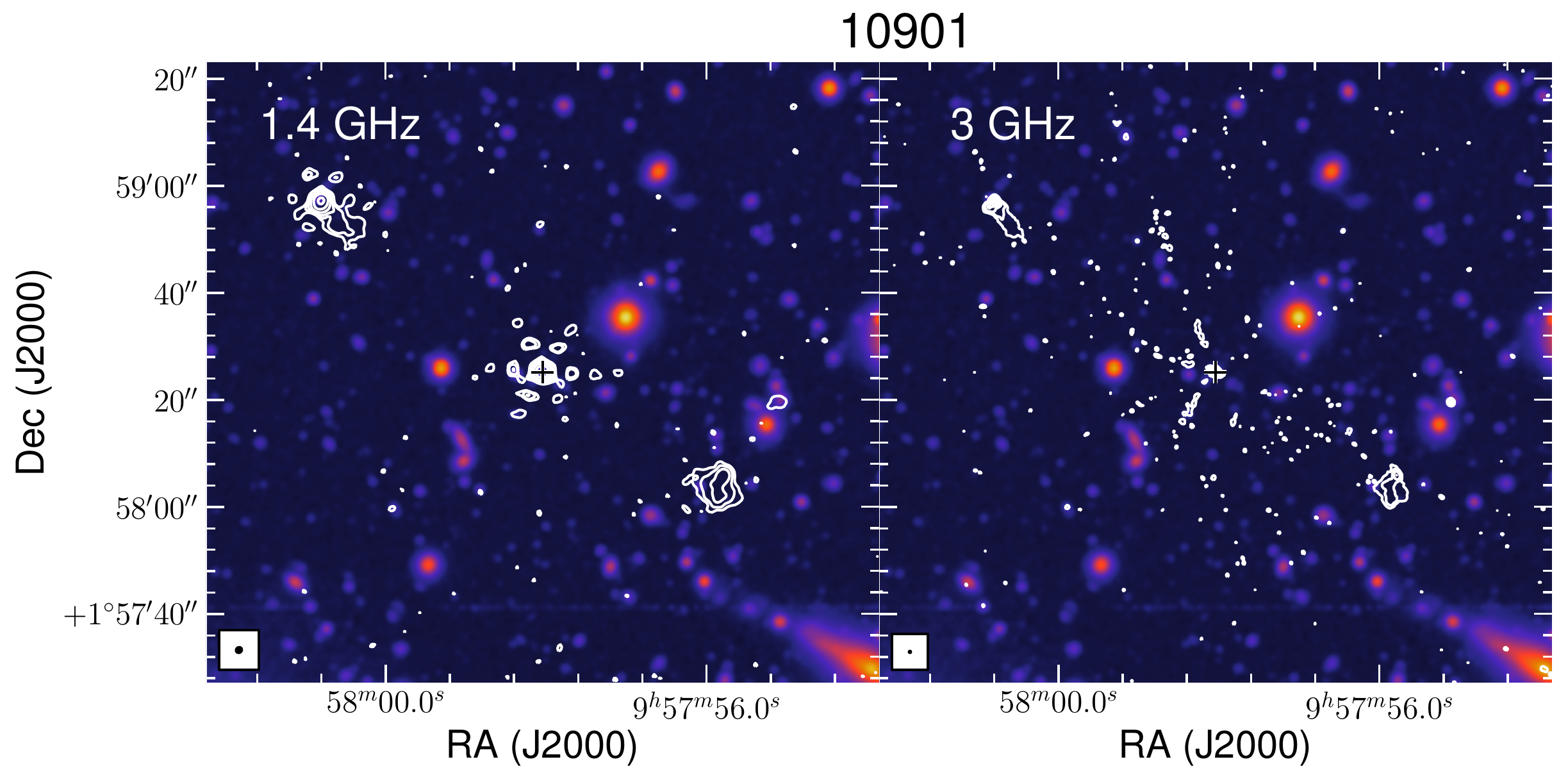}
            }
\\ \\
 \resizebox{\hsize}{!}
{\includegraphics{JVLA10902-ultrajvlavla.pdf}
 \includegraphics{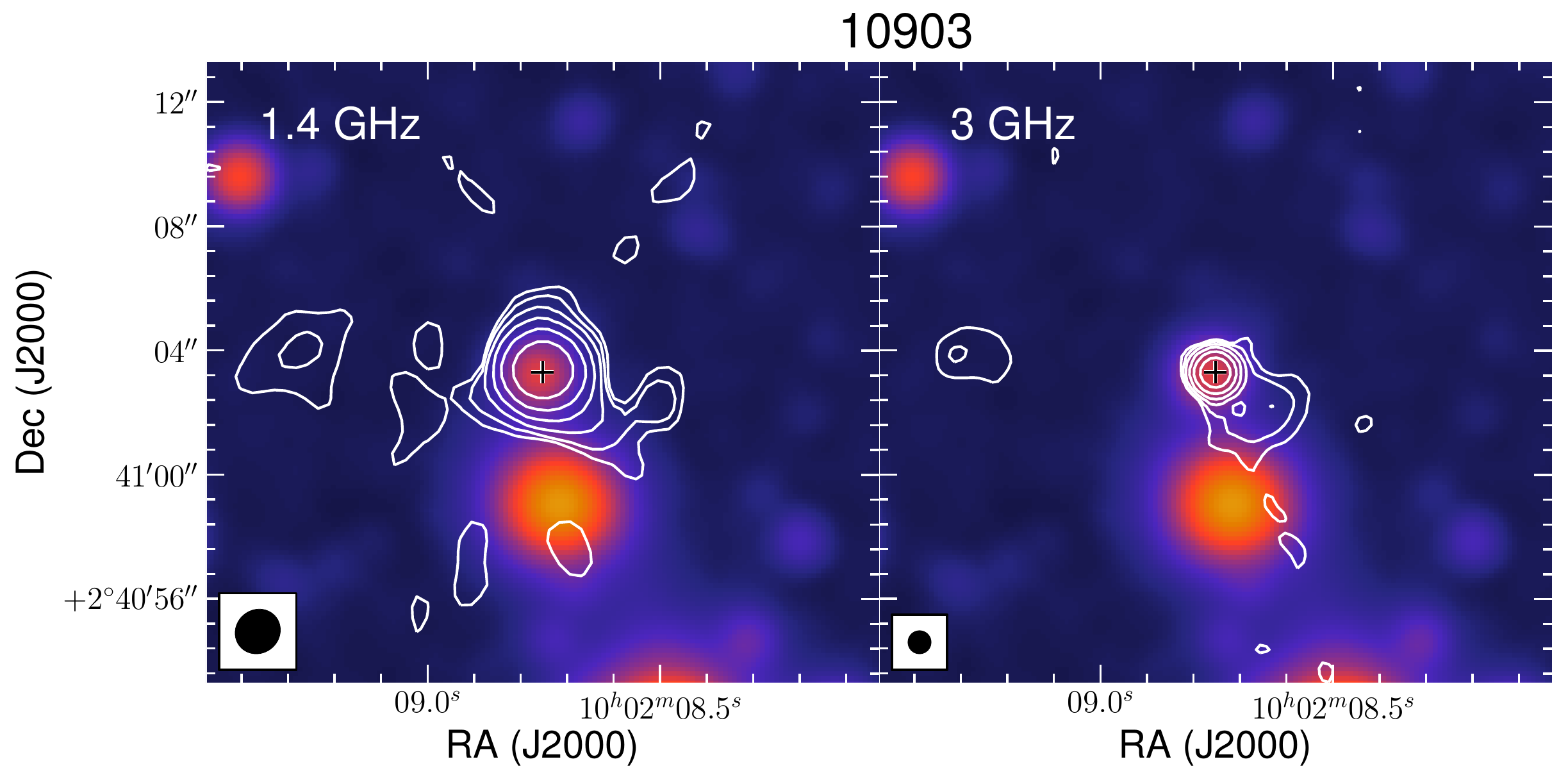}
 \includegraphics{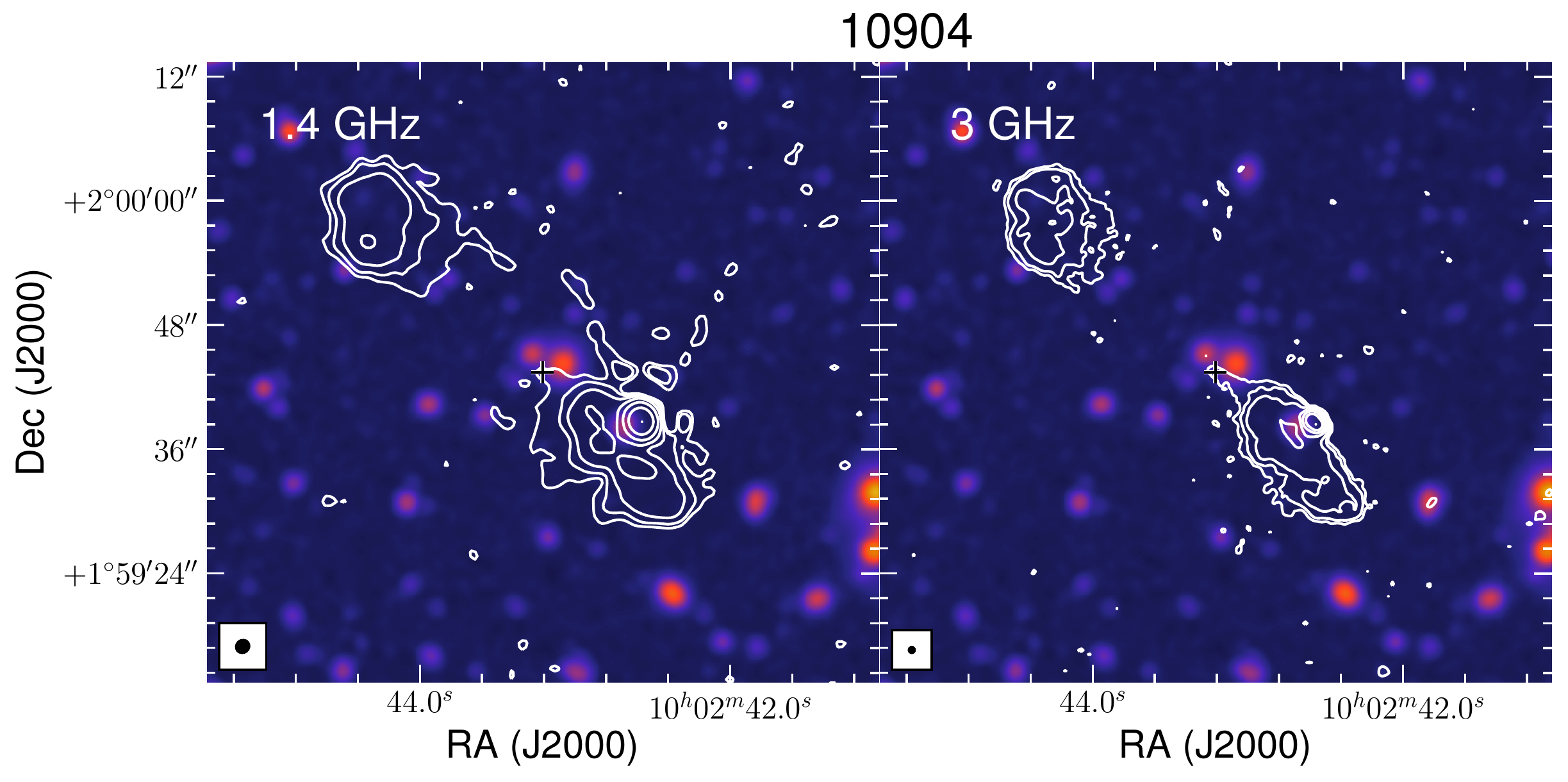}
            }
            \\ \\ 
  \resizebox{\hsize}{!}
 {\includegraphics{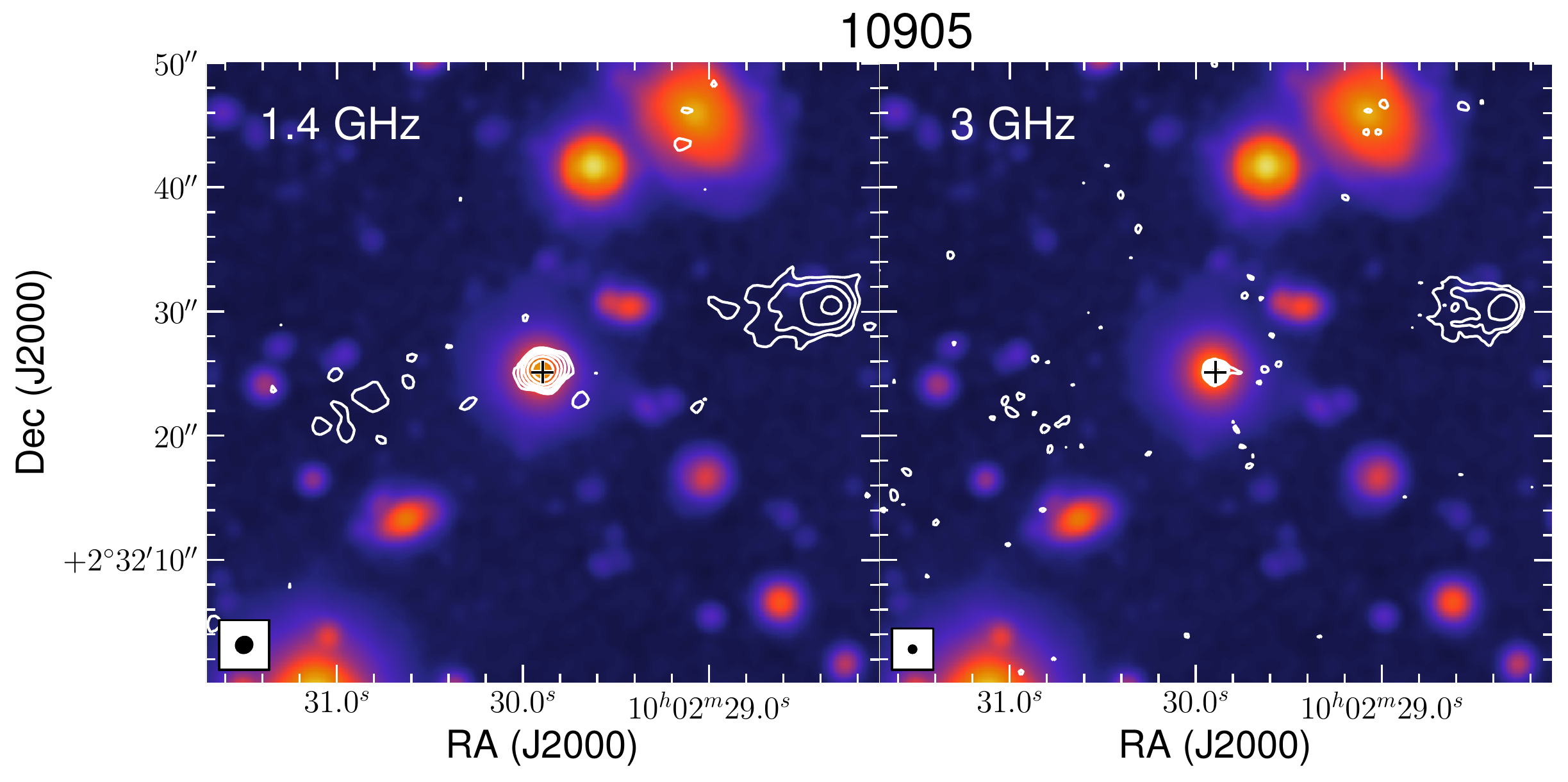}
    \includegraphics{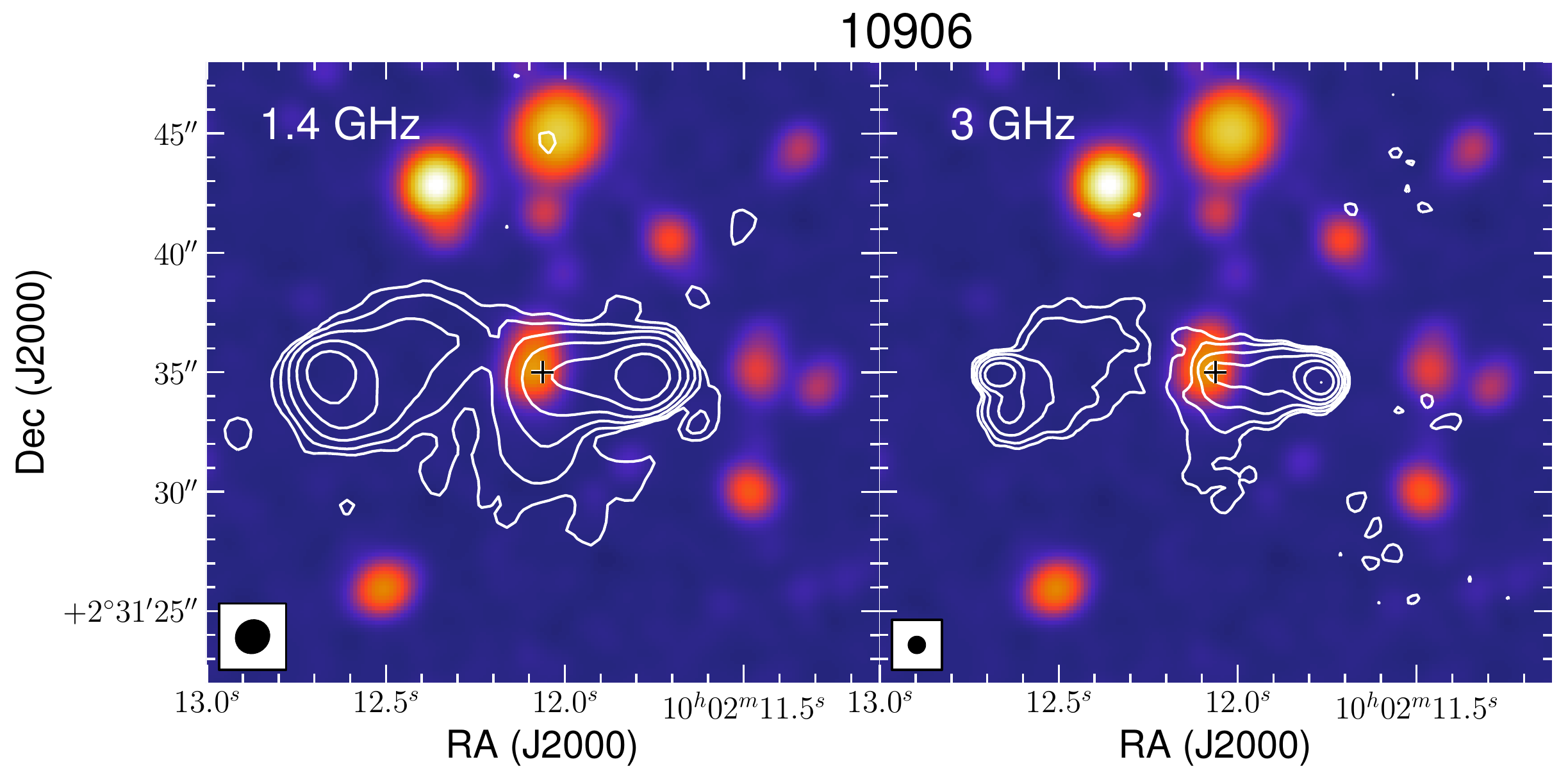}
    \includegraphics{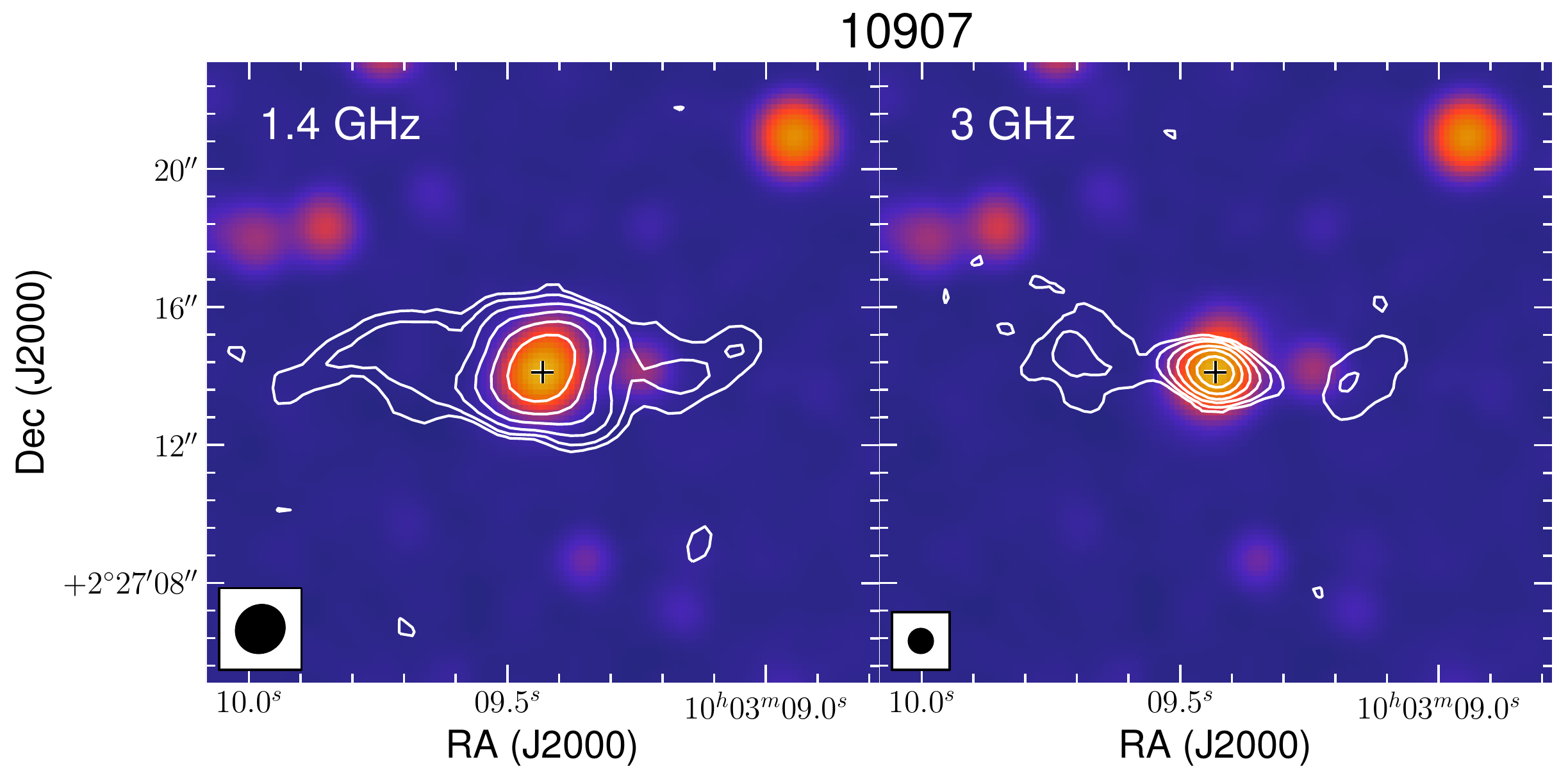}
            }
             \\ \\ 
      \resizebox{\hsize}{!}
       {\includegraphics{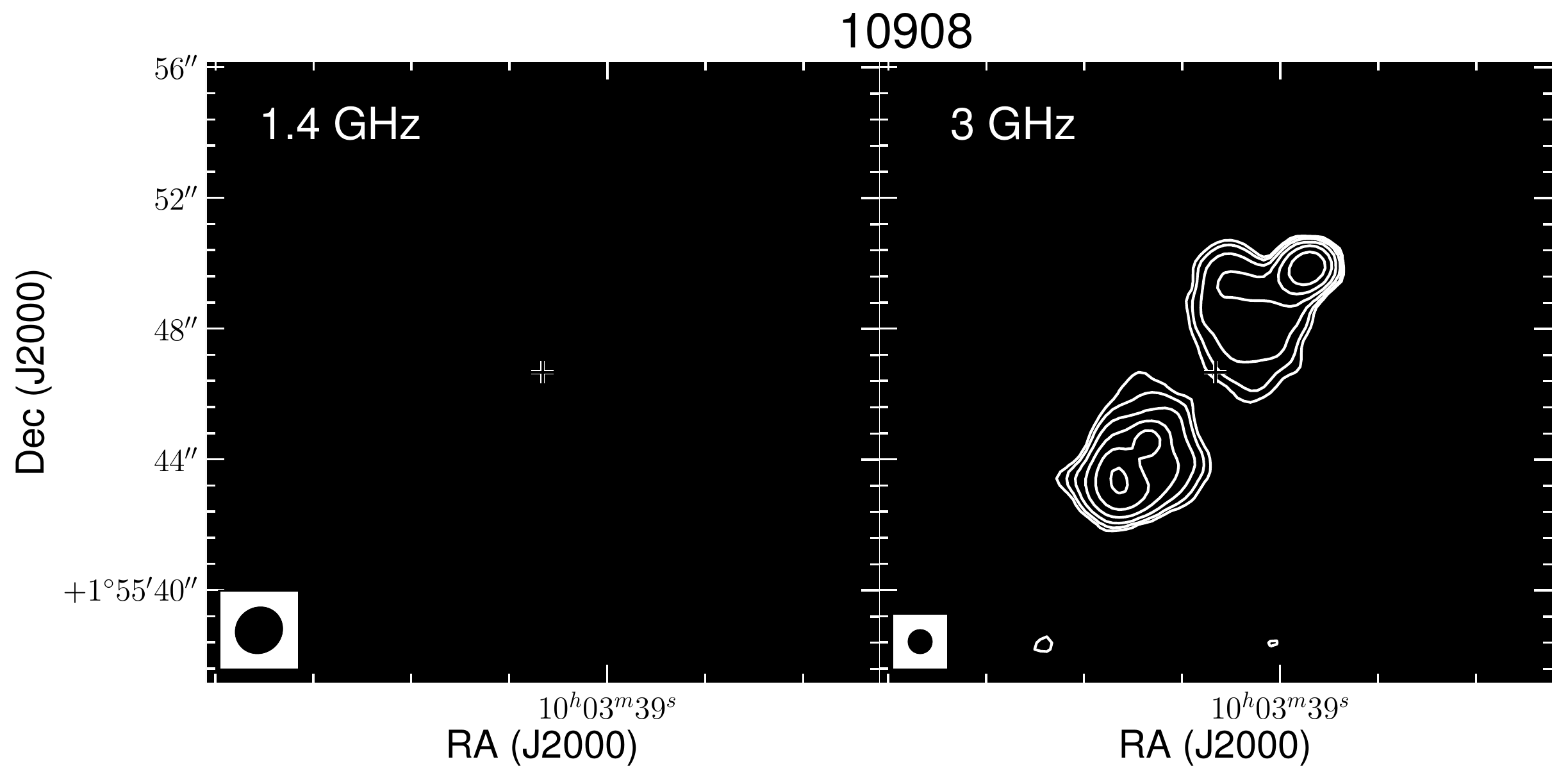}
        \includegraphics{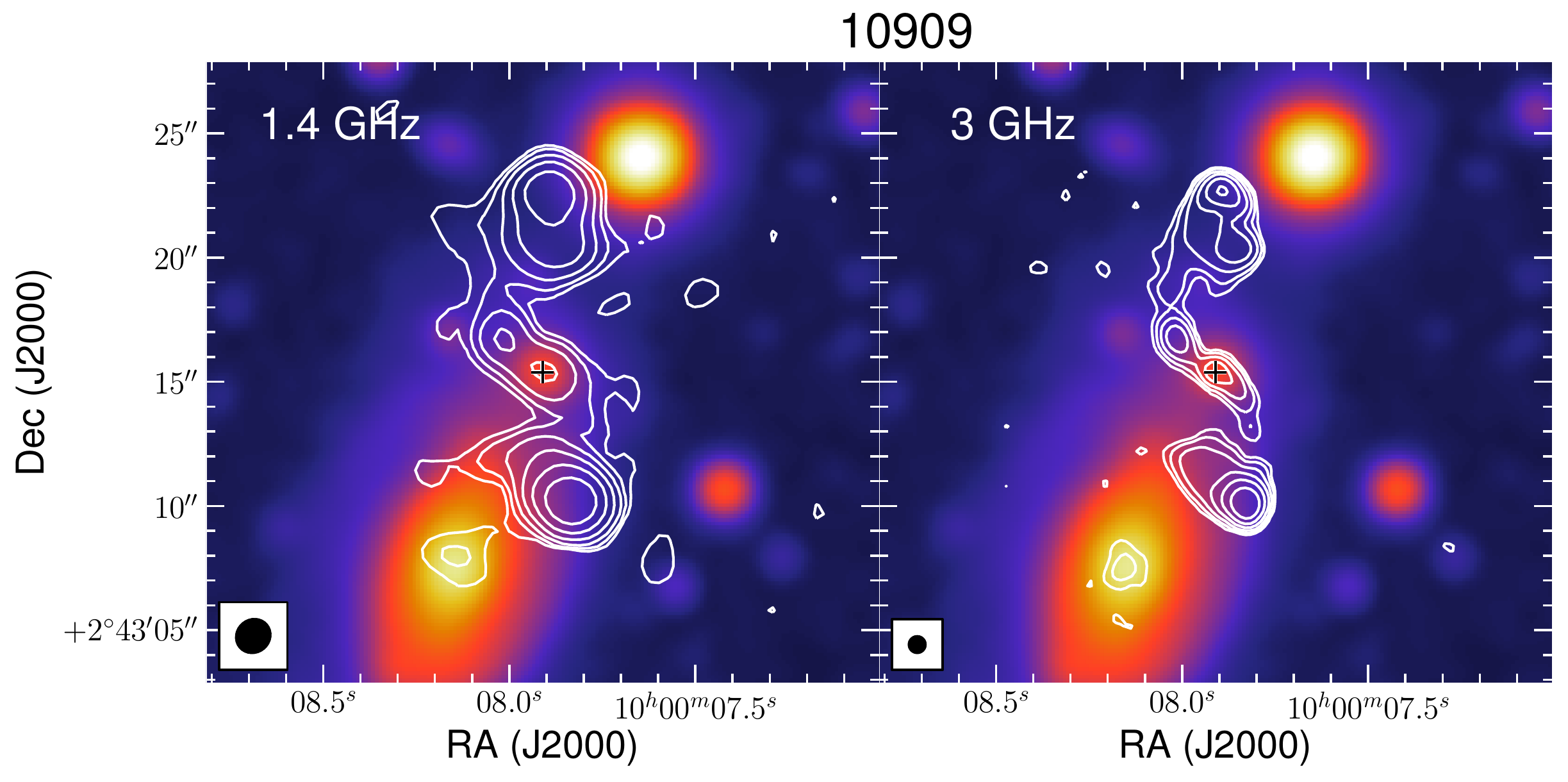}
       \includegraphics{JVLA10910-ultrajvlavla.pdf}
            }
   \caption{(continued)
   }
              \label{fig:maps2}%
    \end{figure*}
\addtocounter{figure}{-1}
\begin{figure*}[!ht]
 \resizebox{\hsize}{!}
{\includegraphics{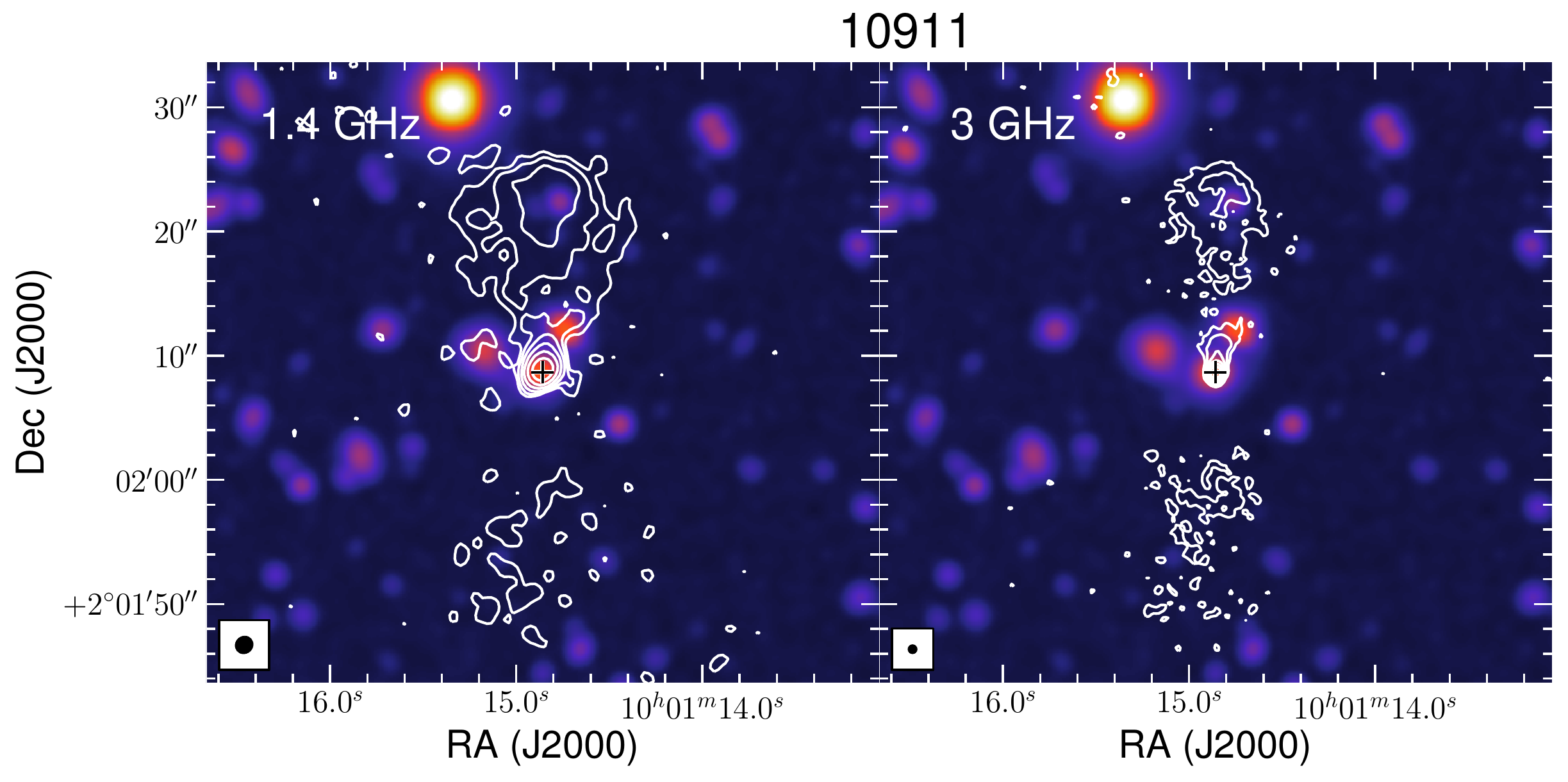}
 \includegraphics{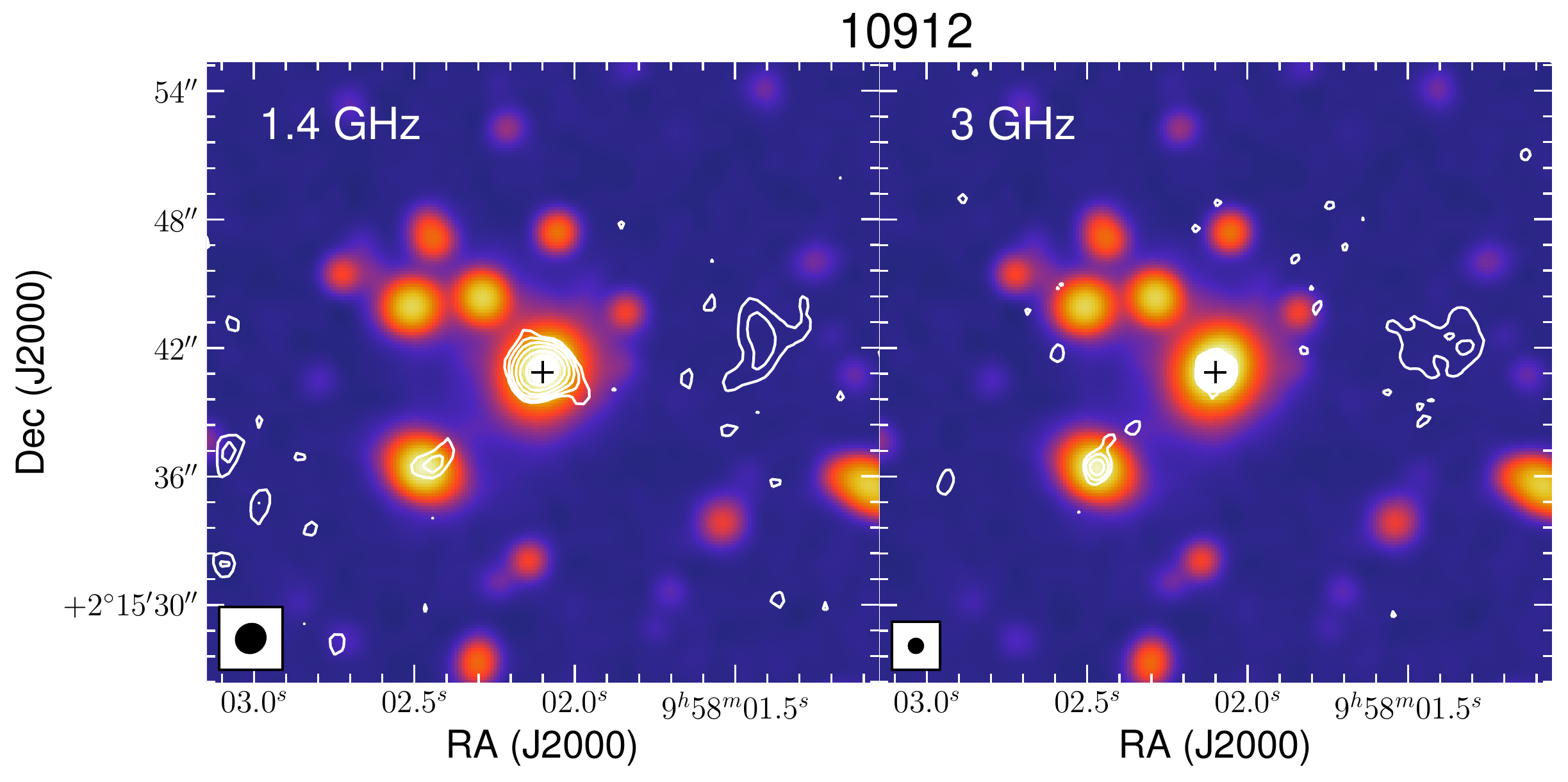}
 \includegraphics{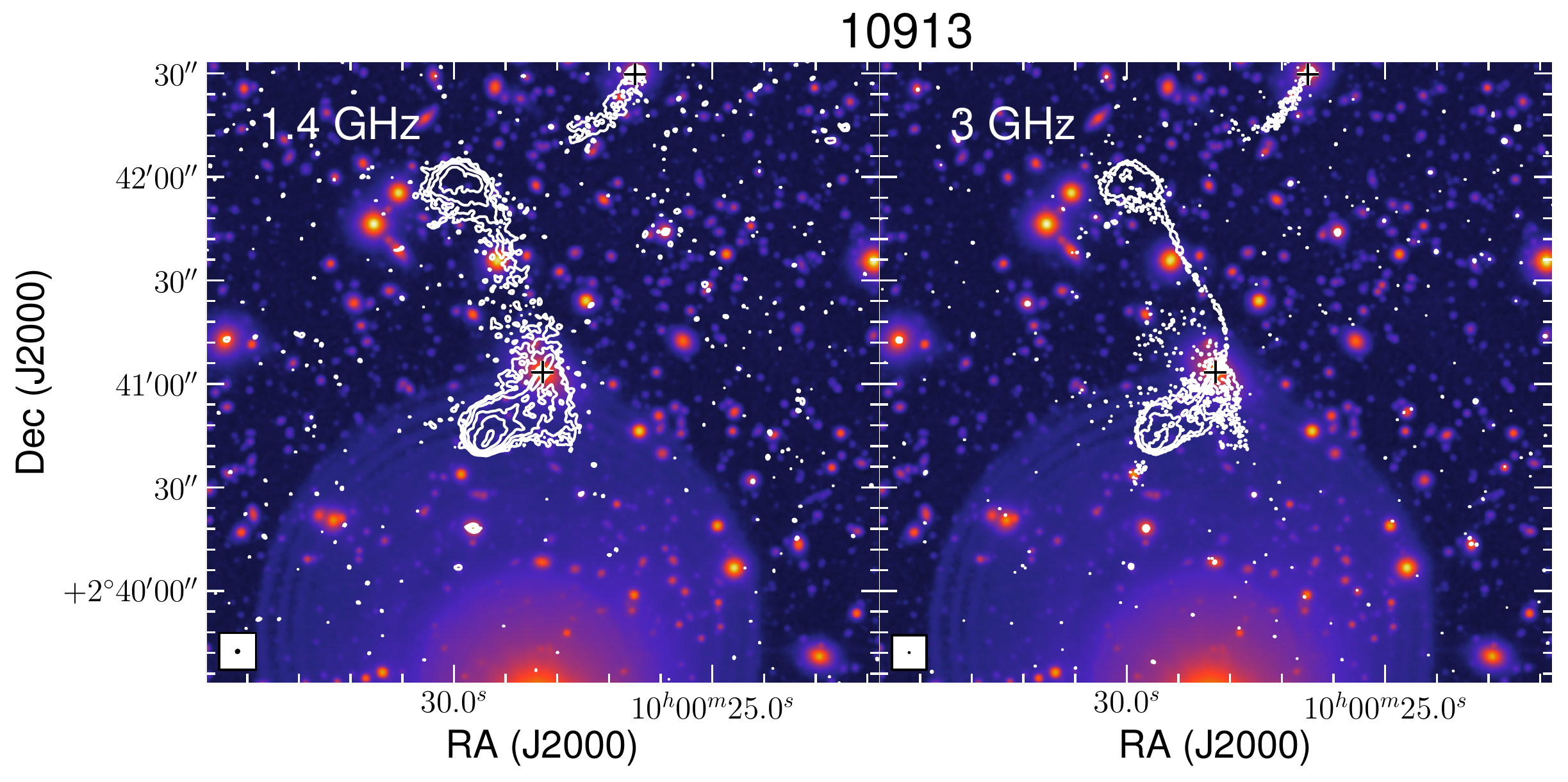}
            }
            \\ \\ 
  \resizebox{\hsize}{!}
 {\includegraphics{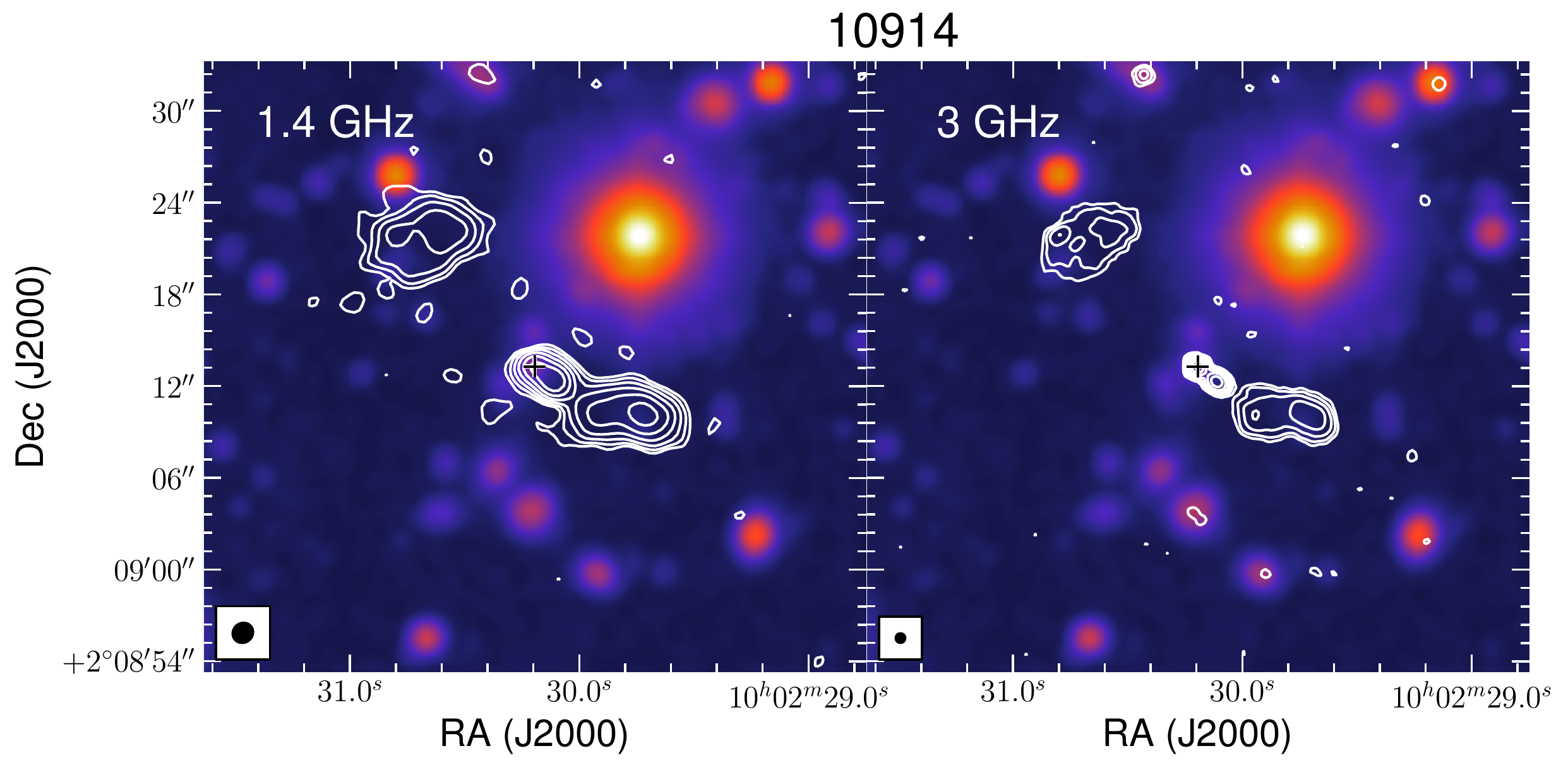}
    \includegraphics{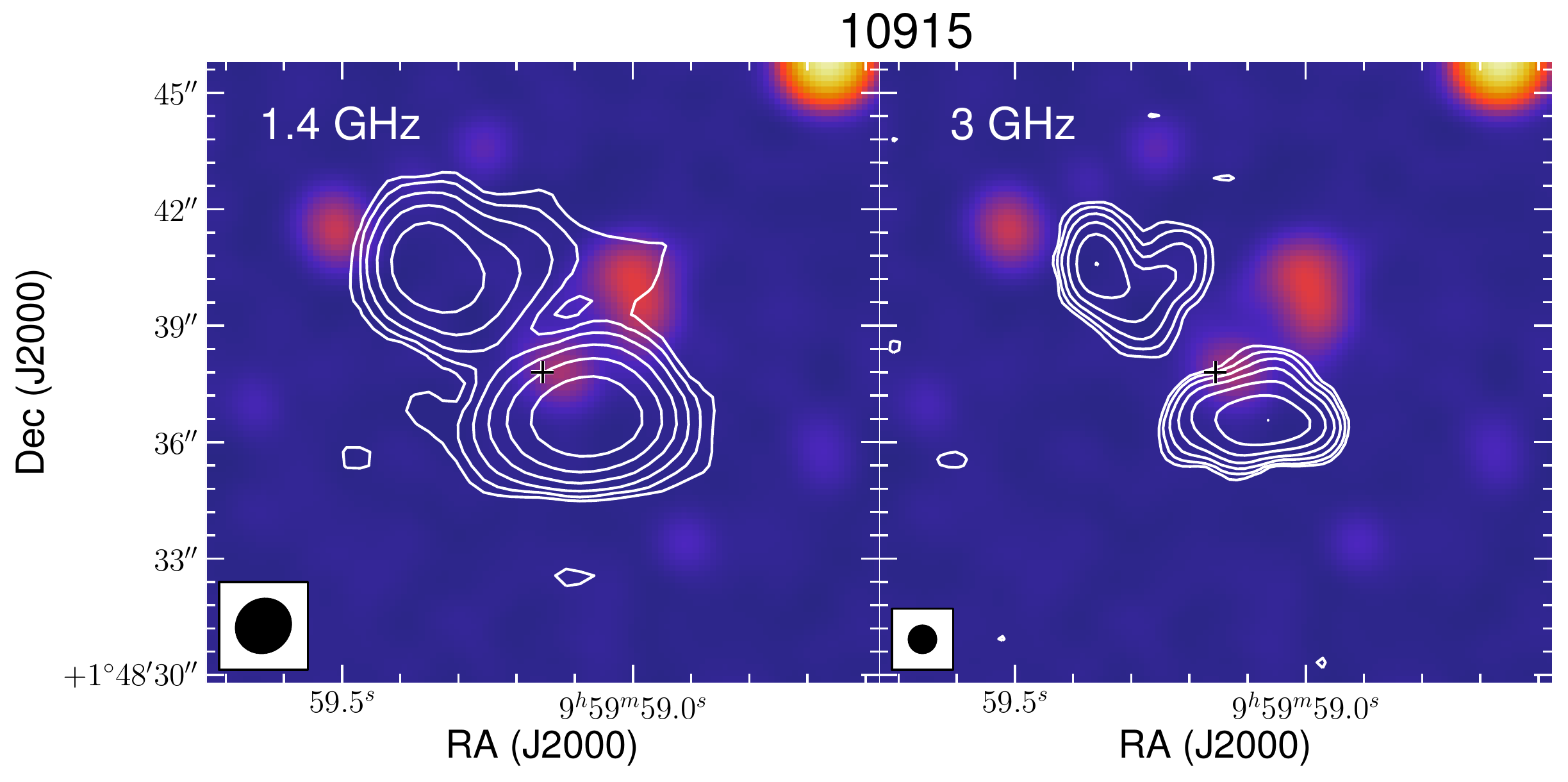}
    \includegraphics{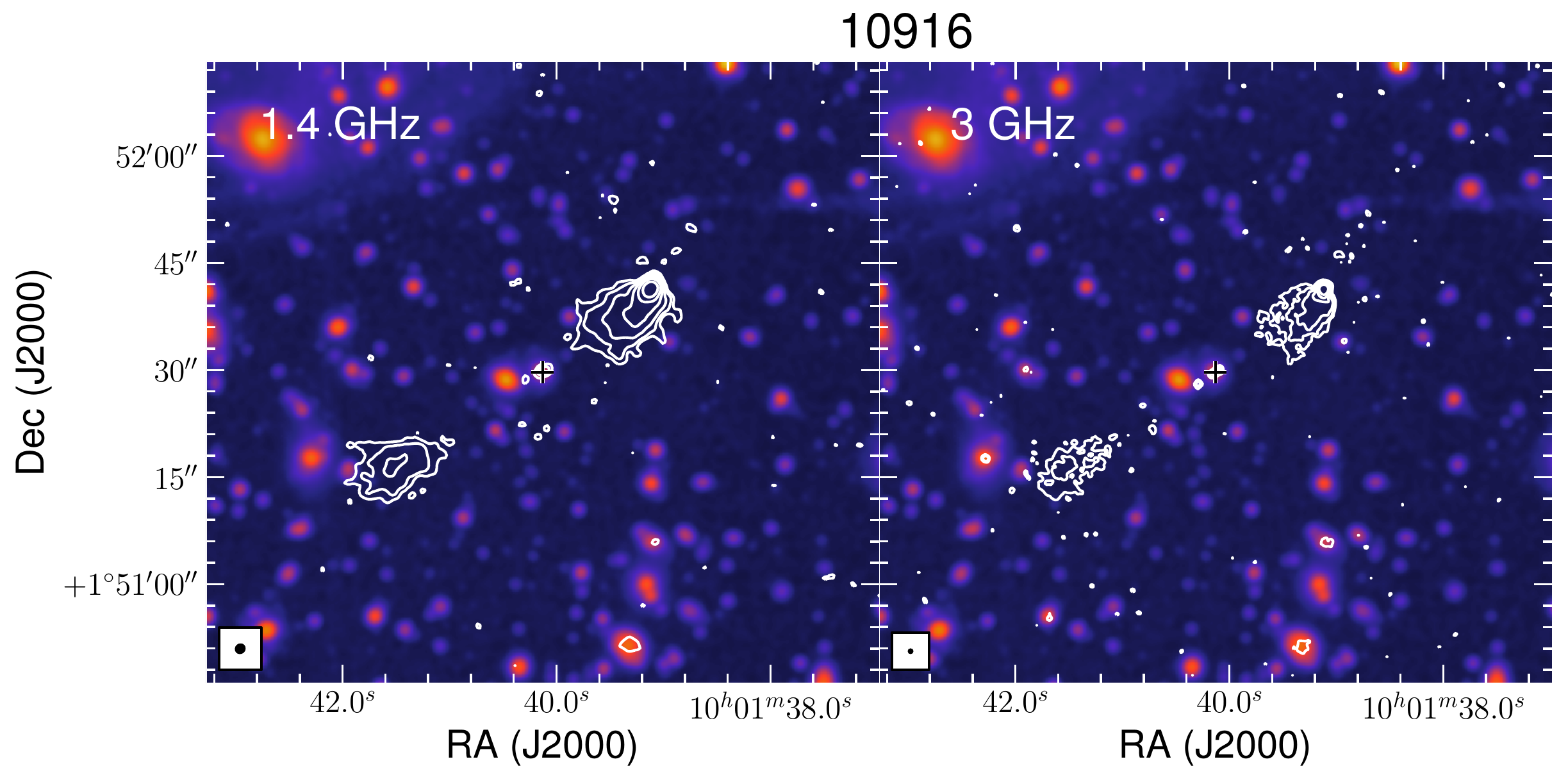}
            }
             \\ \\ 
      \resizebox{\hsize}{!}
       {\includegraphics{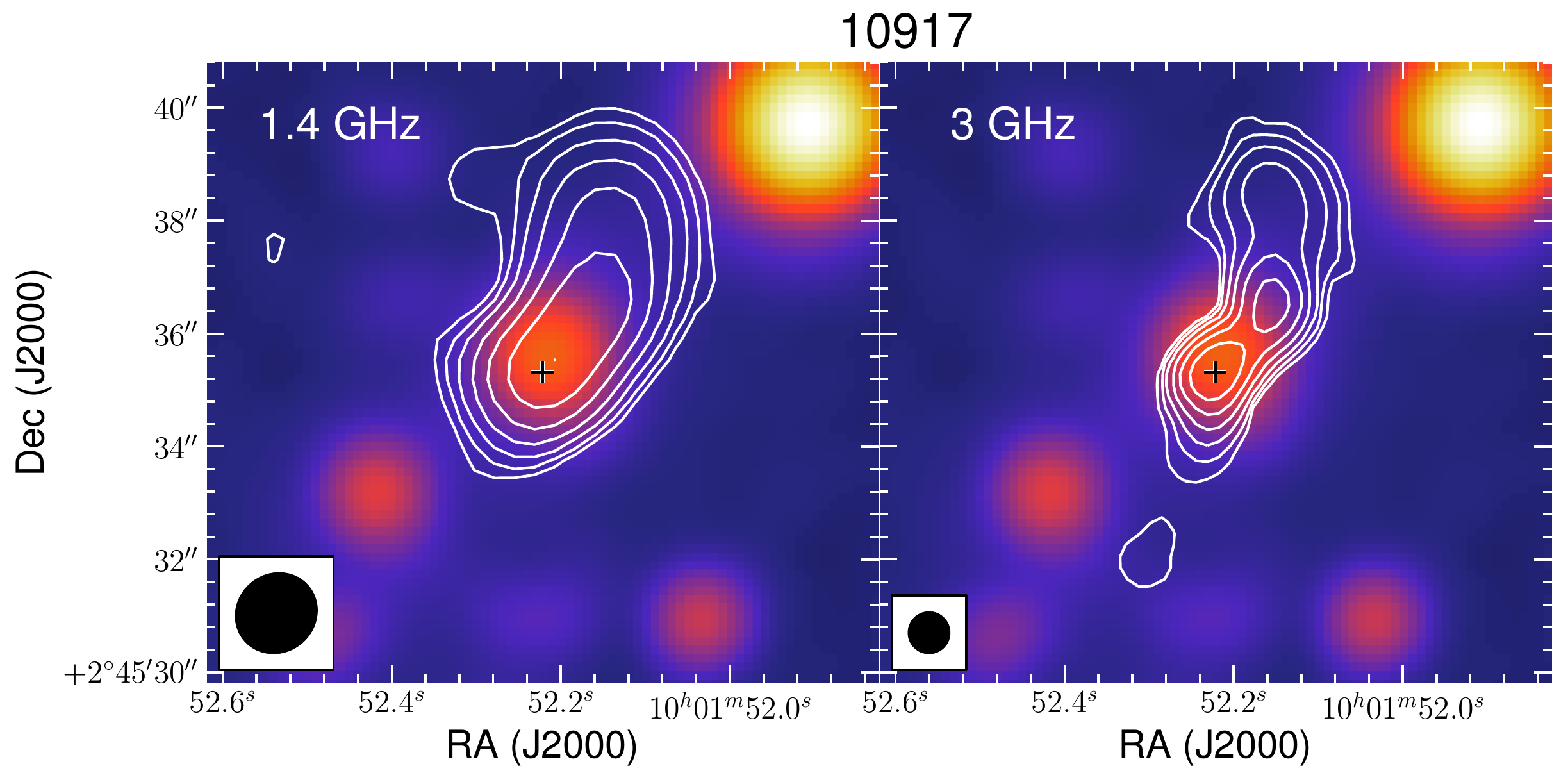}
        \includegraphics{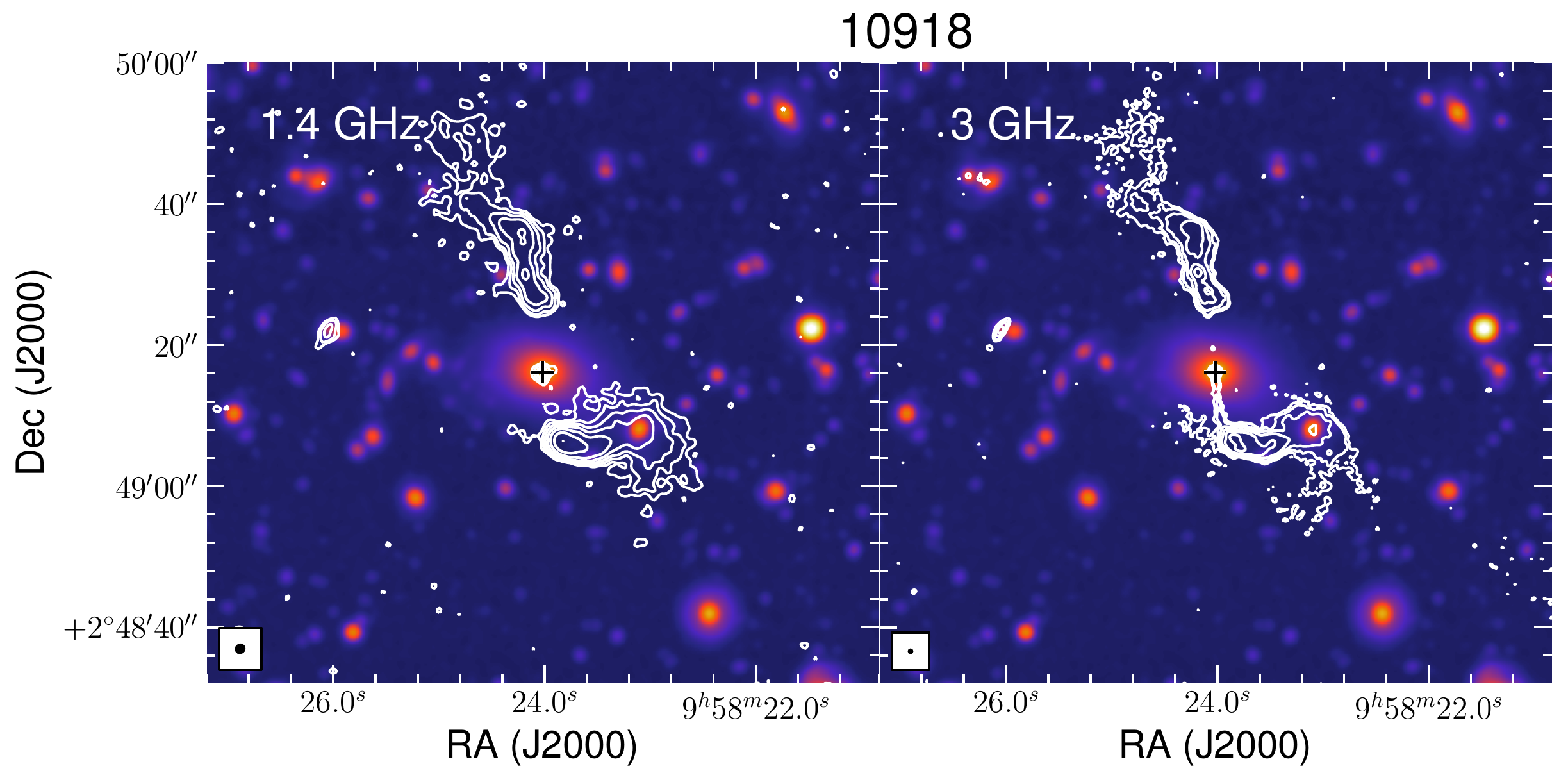}
       \includegraphics{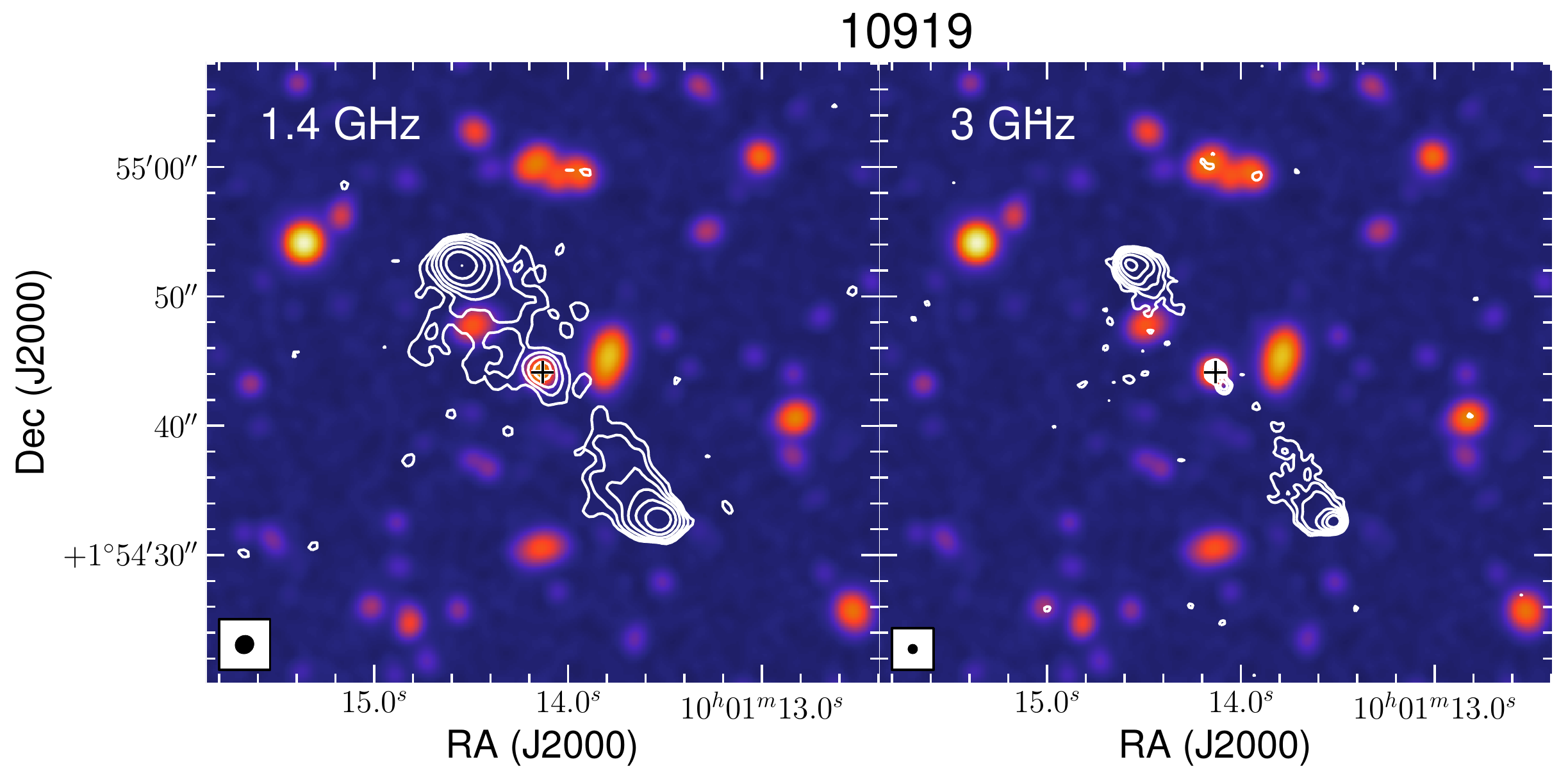}
            }
 \\ \\
  \resizebox{\hsize}{!}
 {\includegraphics{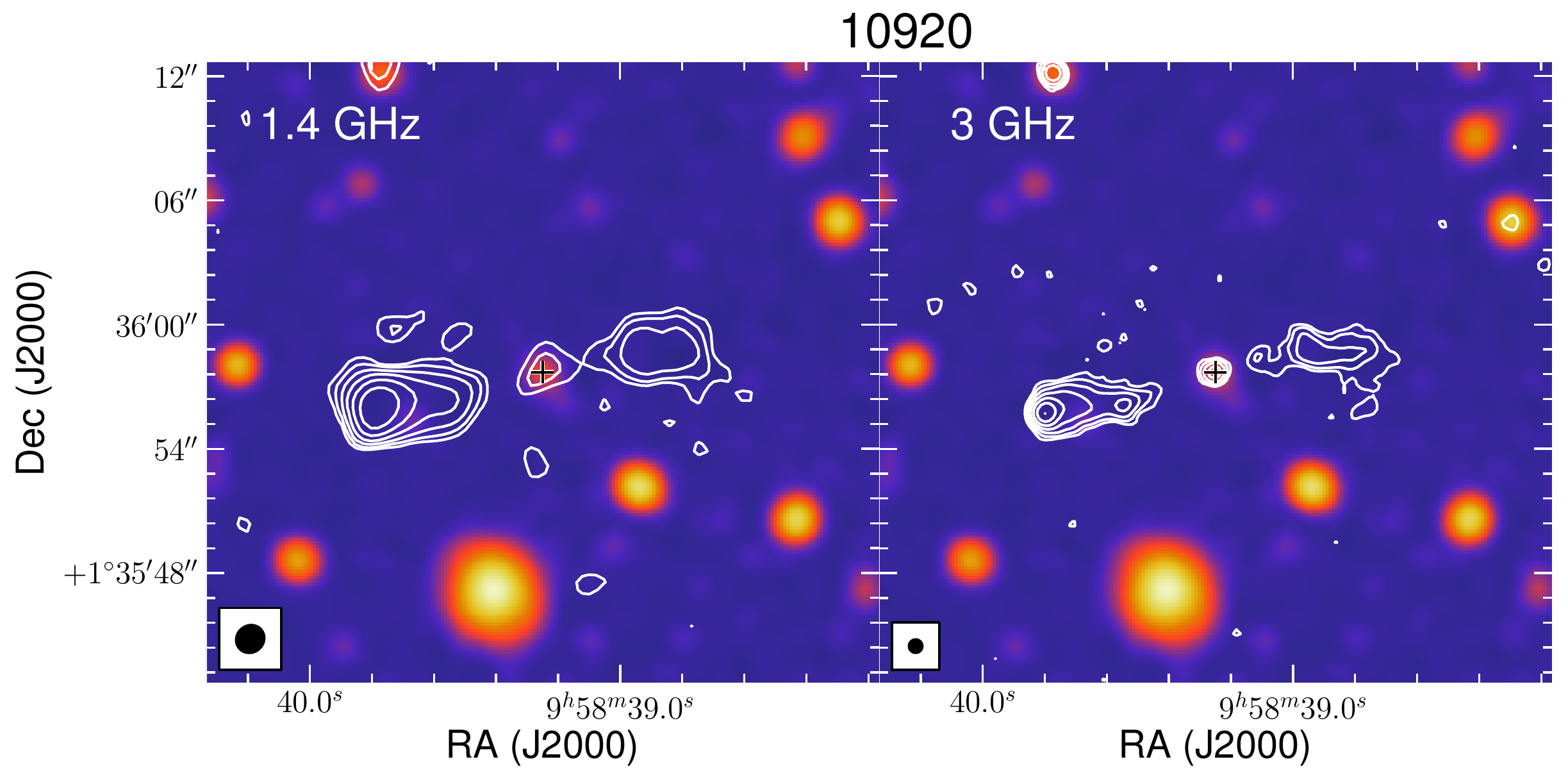}
 \includegraphics{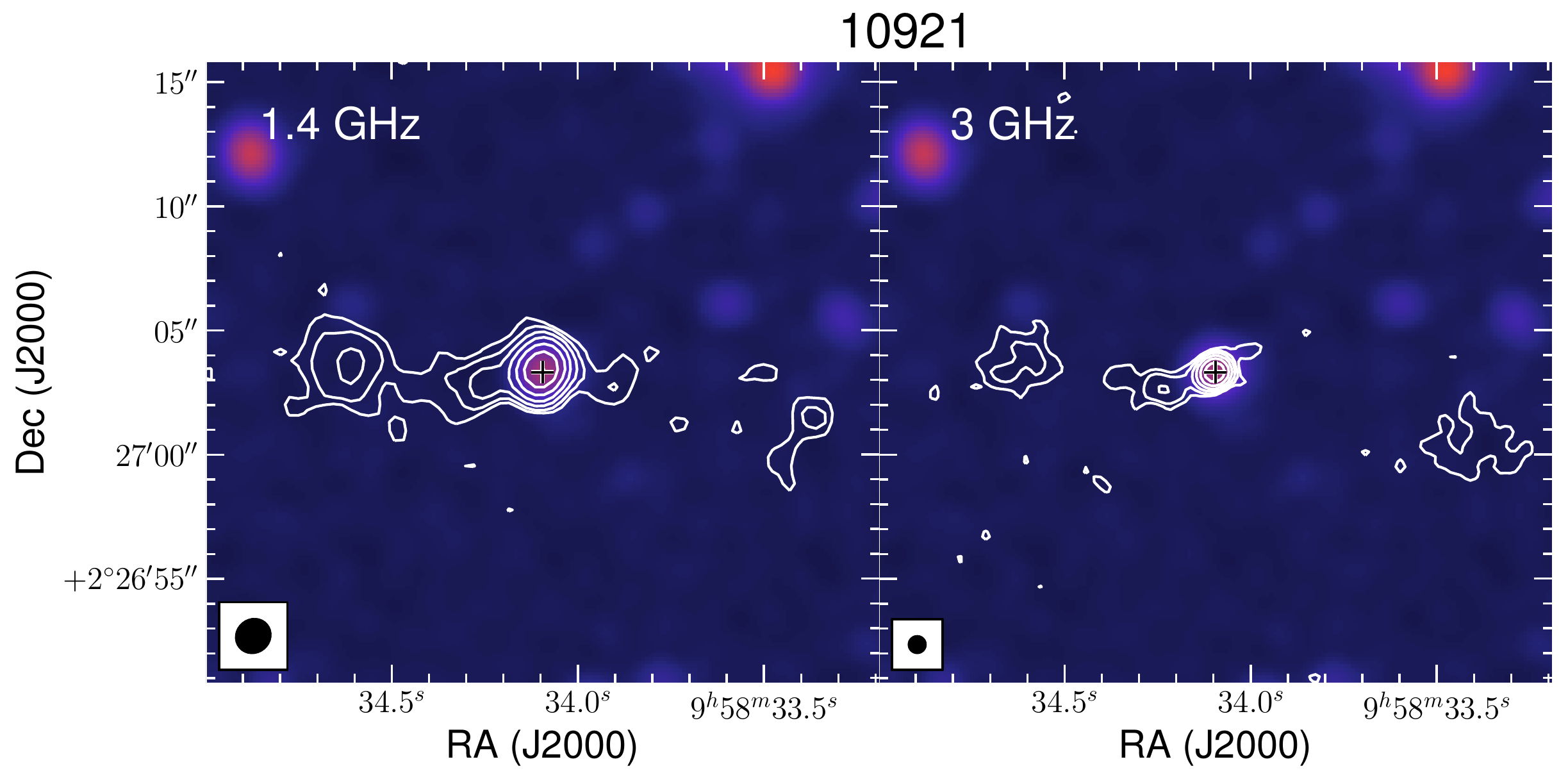}
 \includegraphics{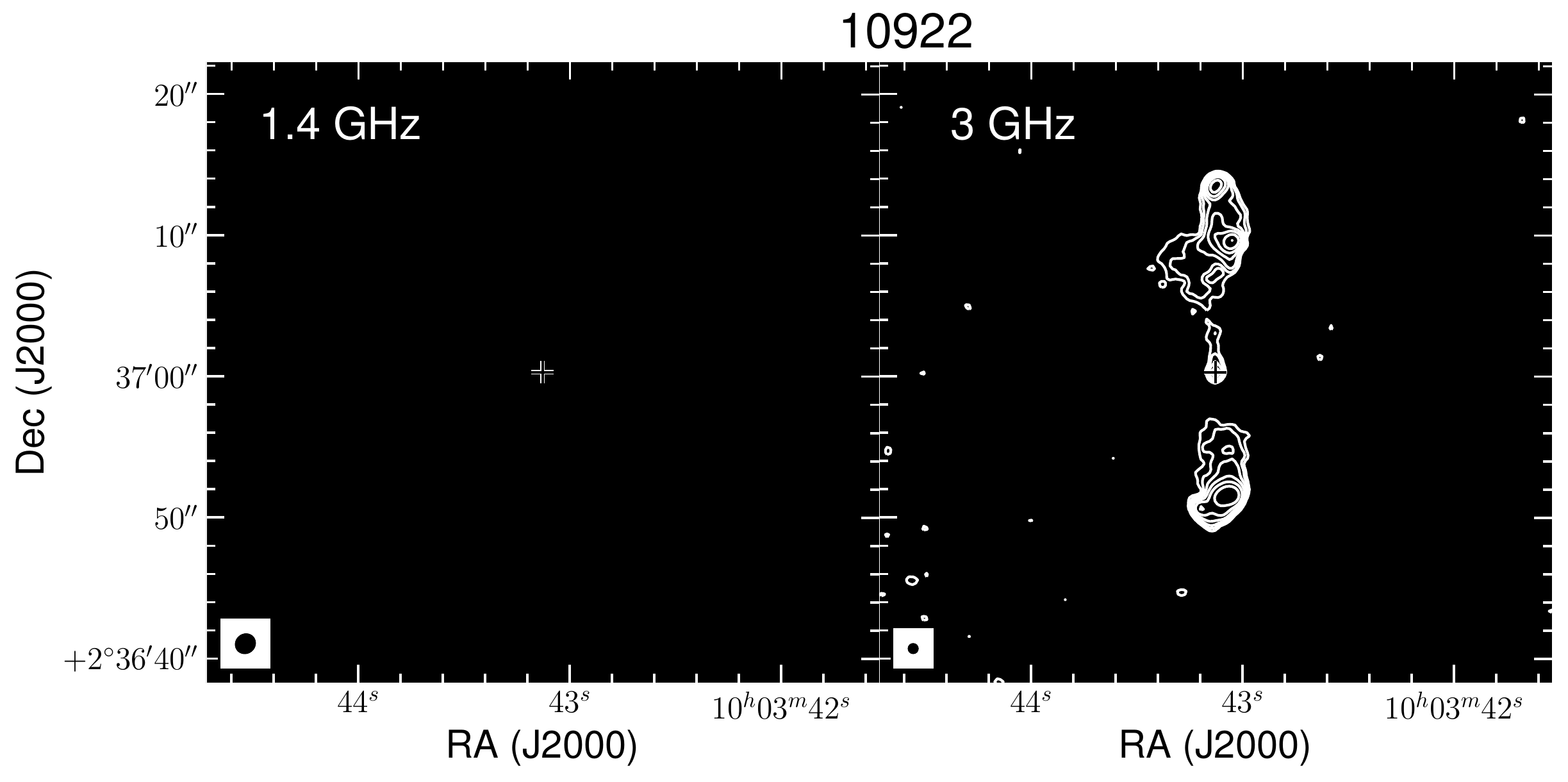}
            }
             \\ \\
      \resizebox{\hsize}{!}
       {        \includegraphics{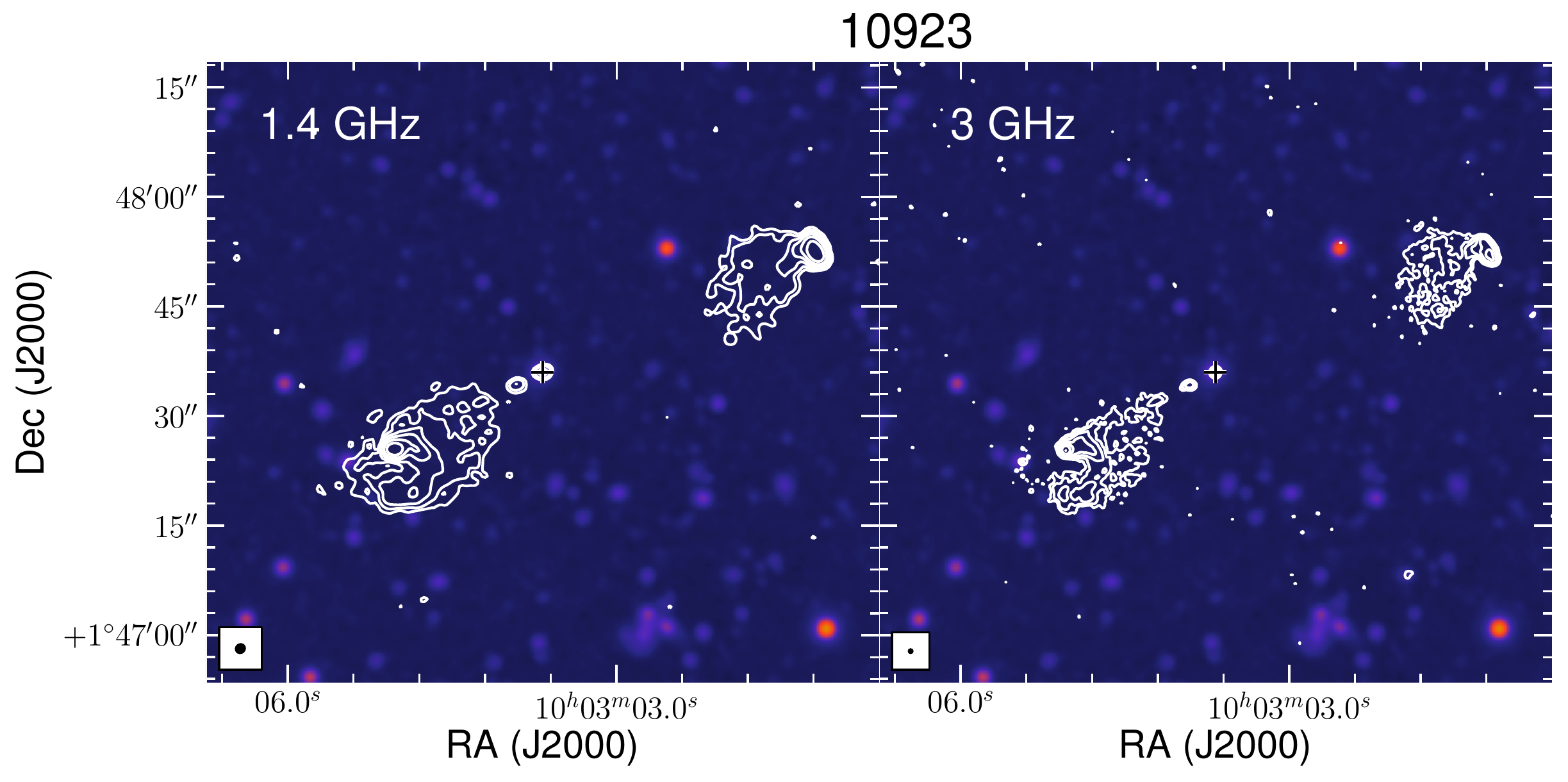}
        \includegraphics{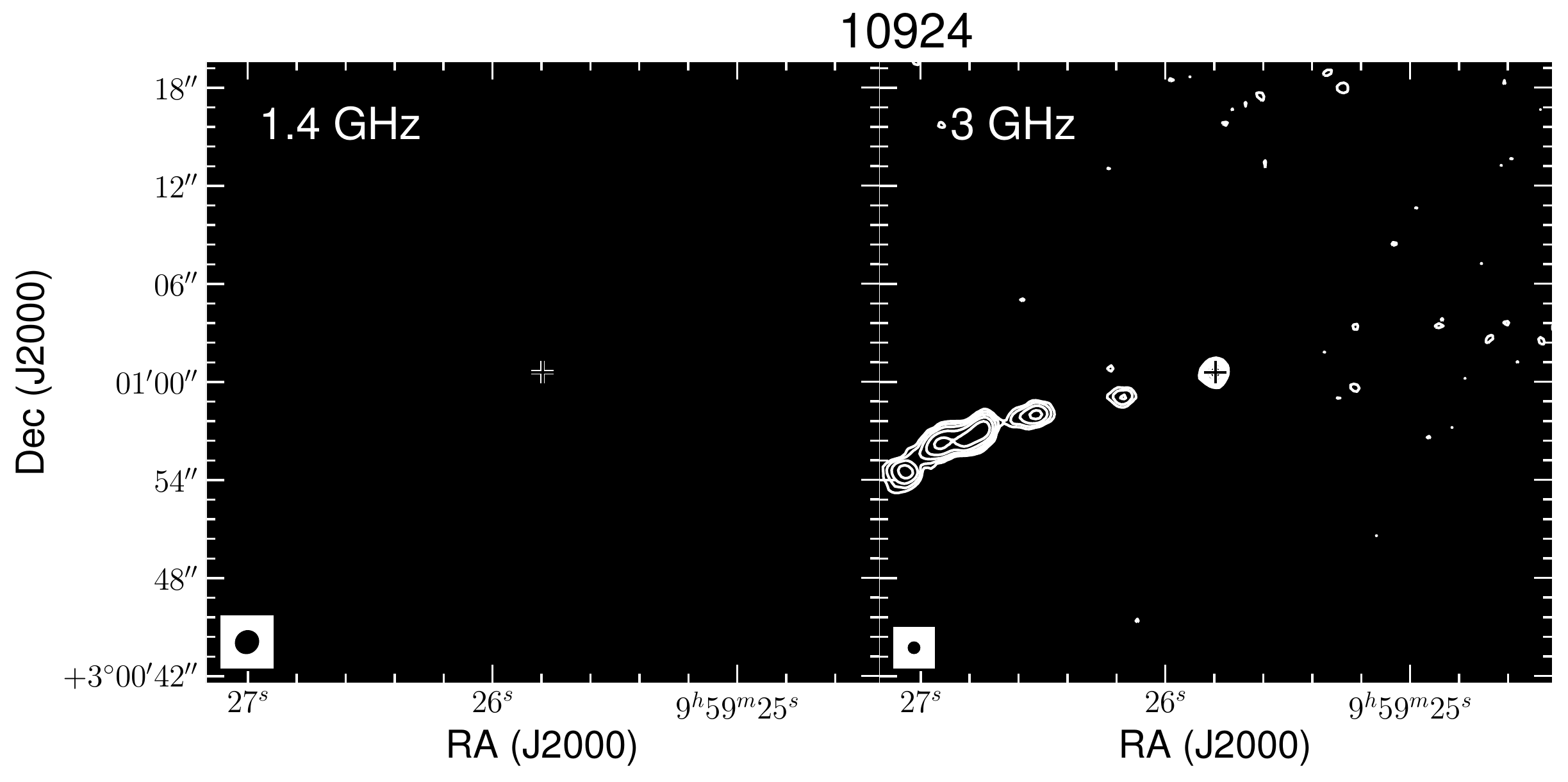}
 \includegraphics{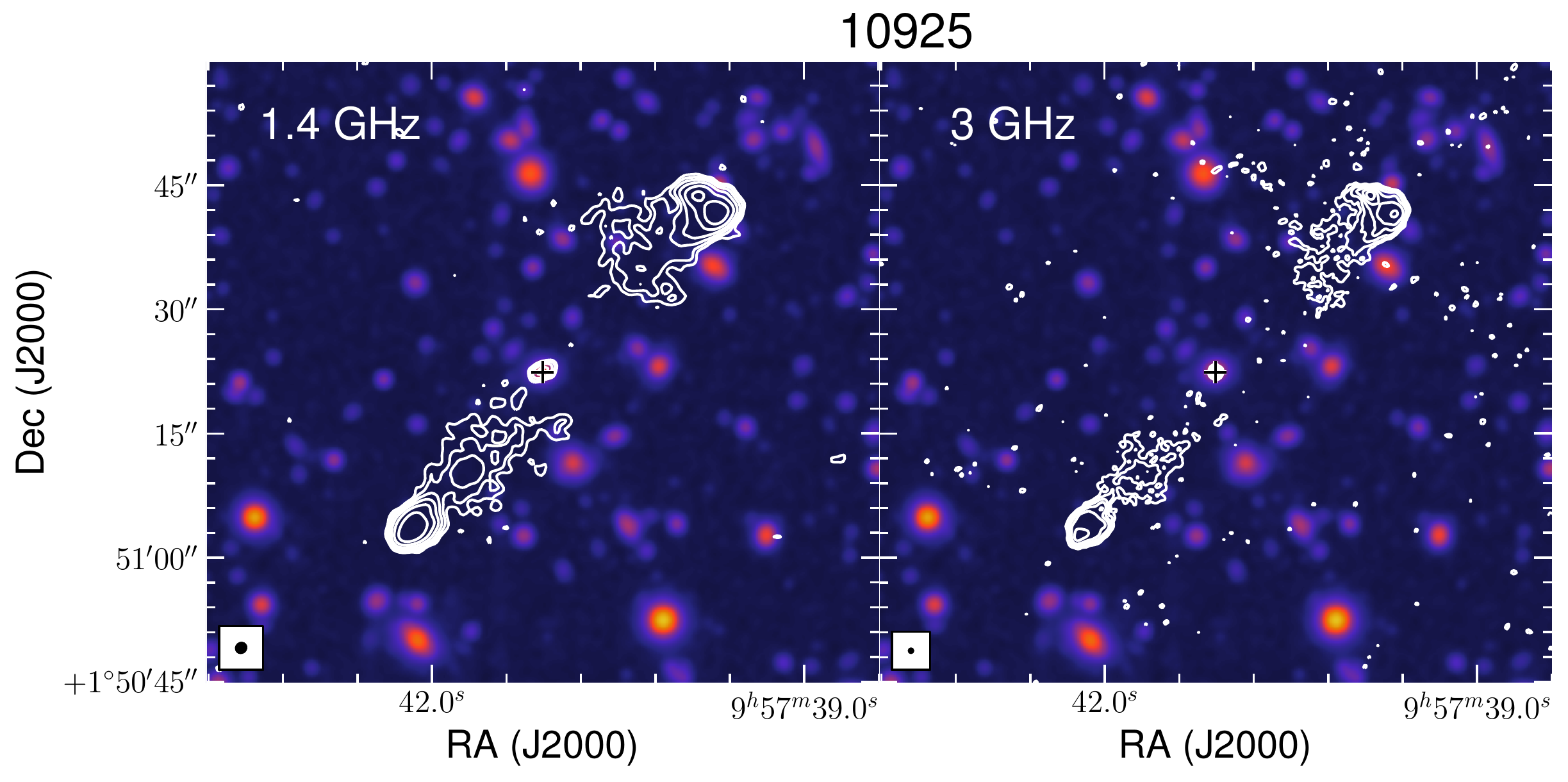}
            }
 \\ \\
 \resizebox{\hsize}{!}
{\includegraphics{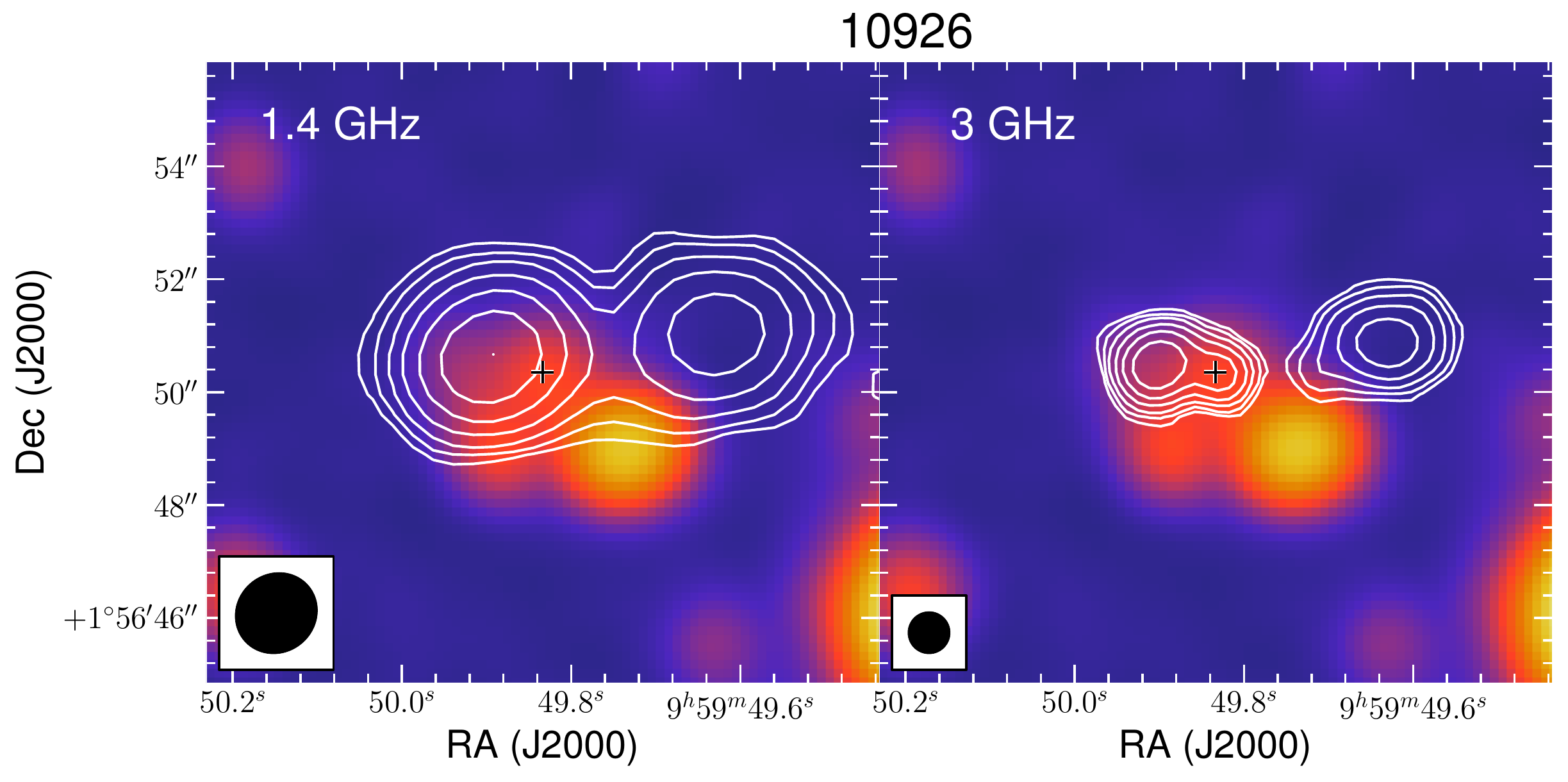}
    \includegraphics{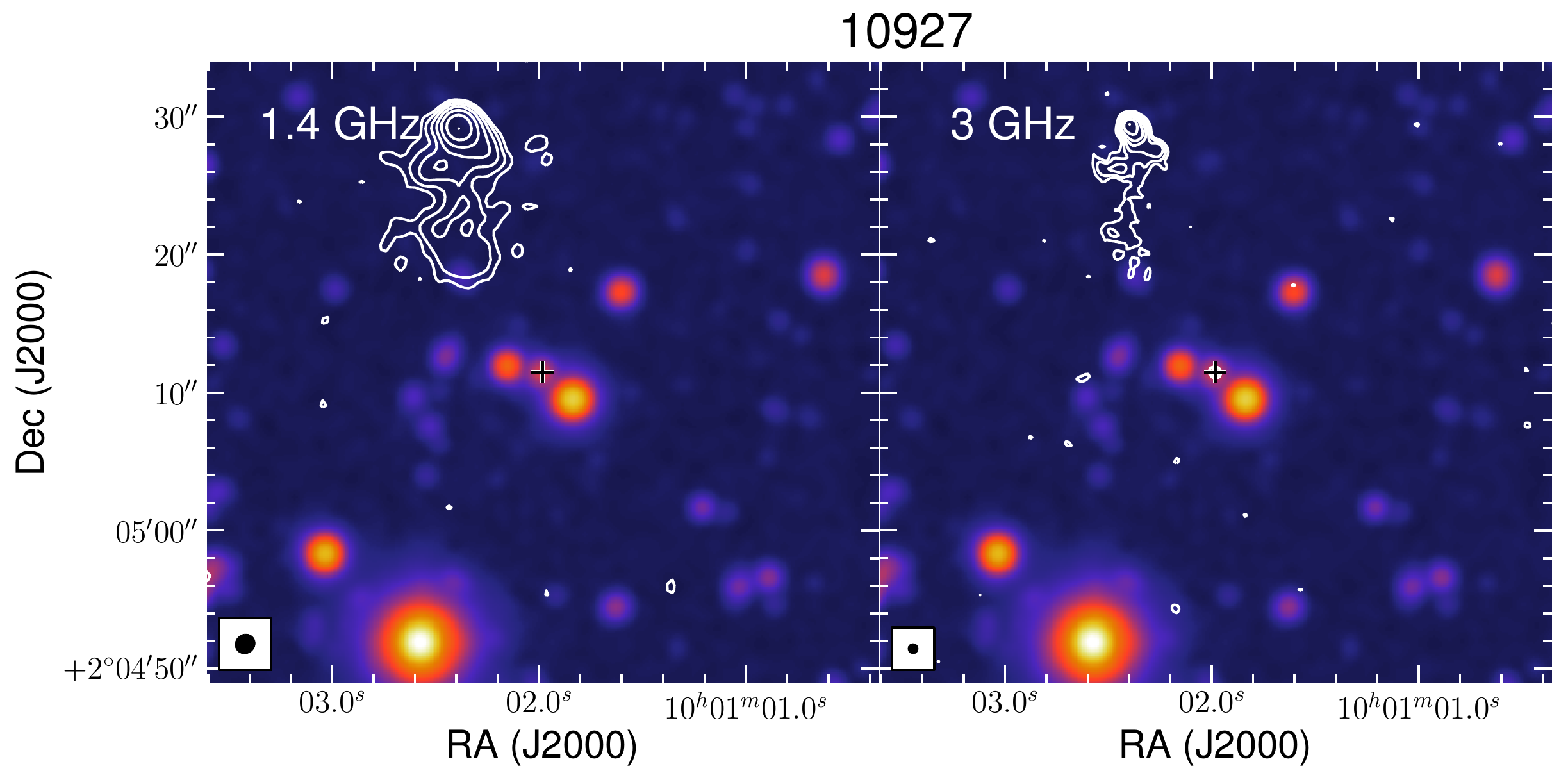}
    \includegraphics{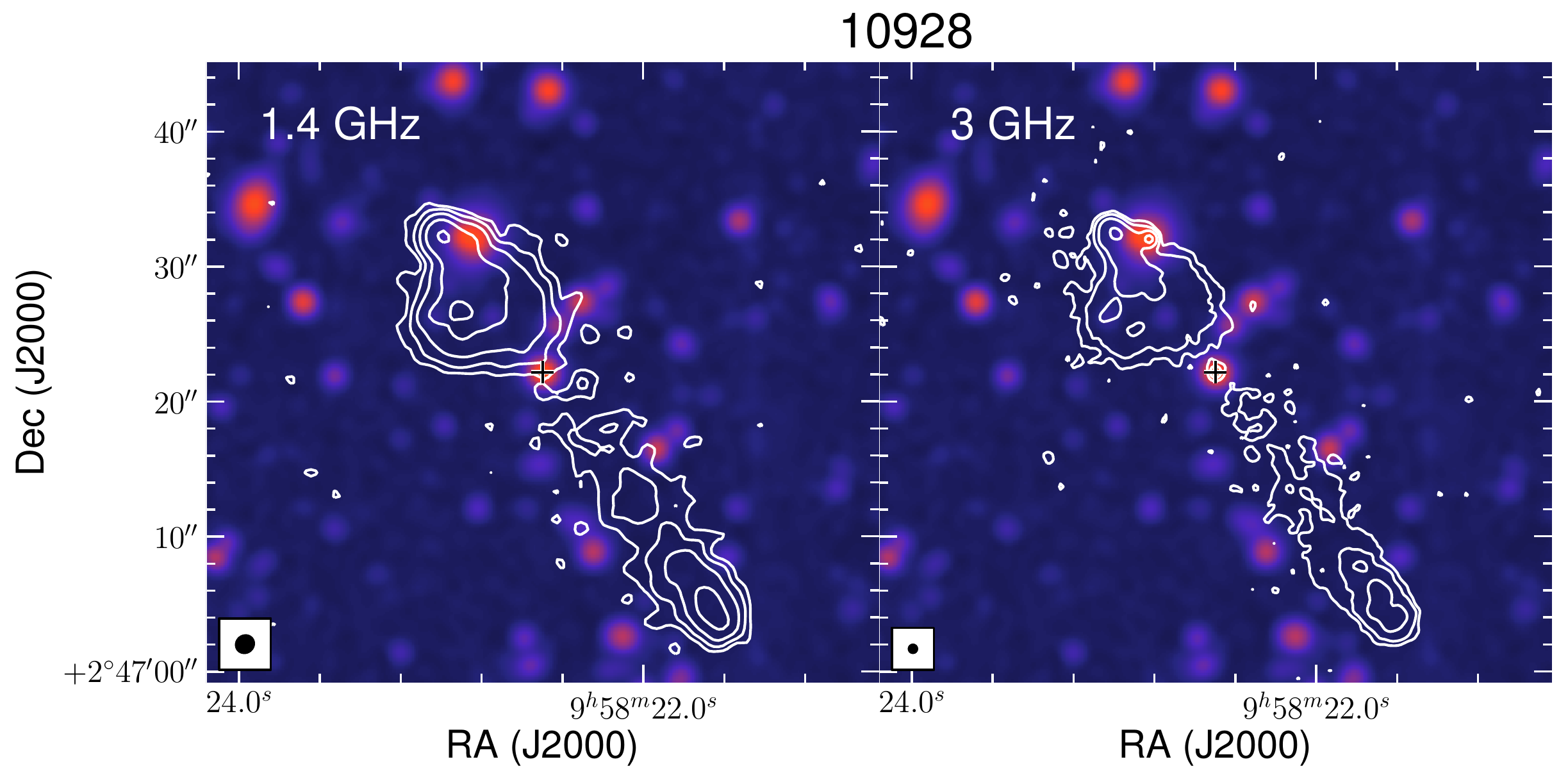}
            }
            \\ \\
  \resizebox{\hsize}{!}
 { \includegraphics{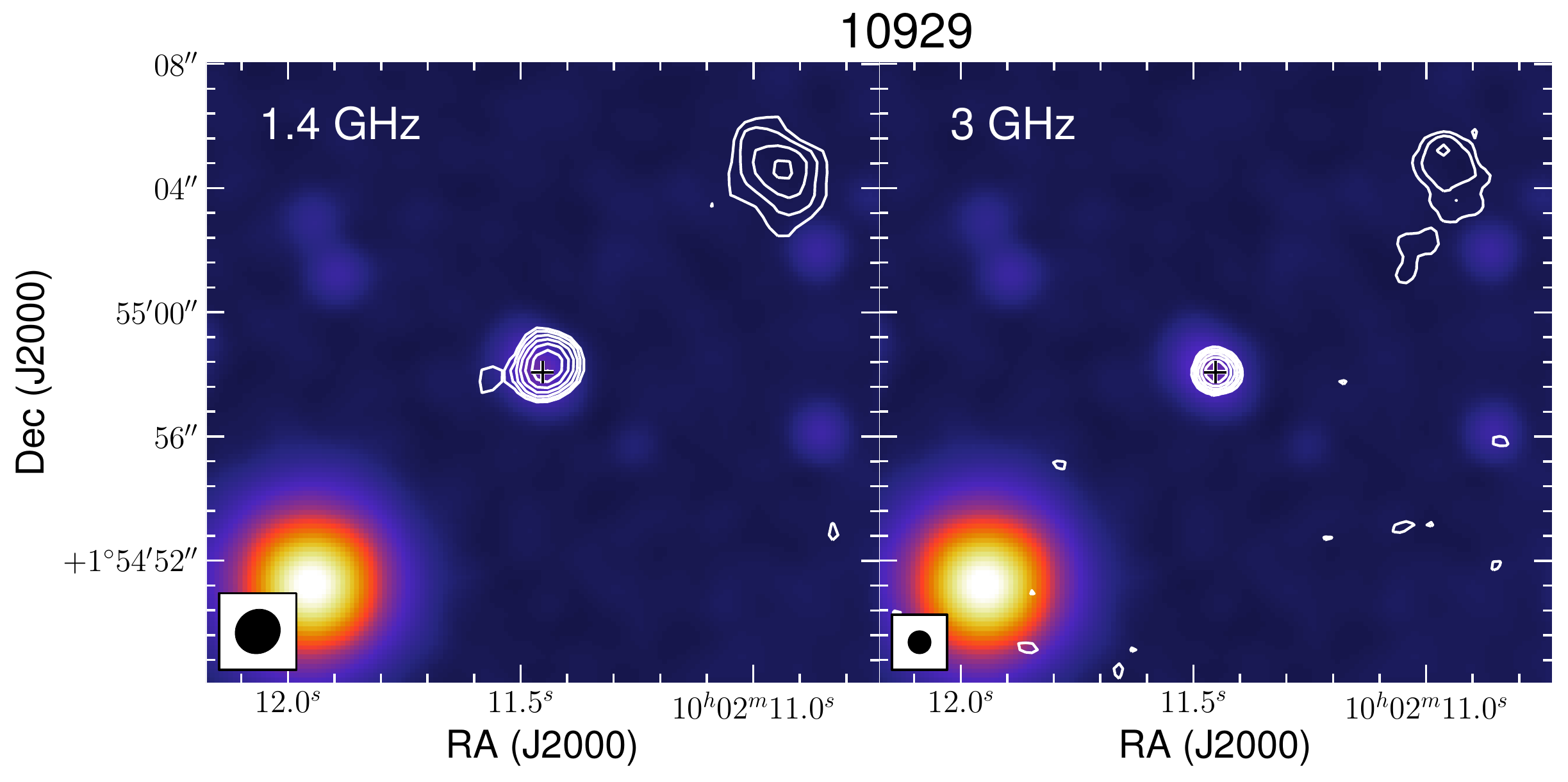}
 \includegraphics{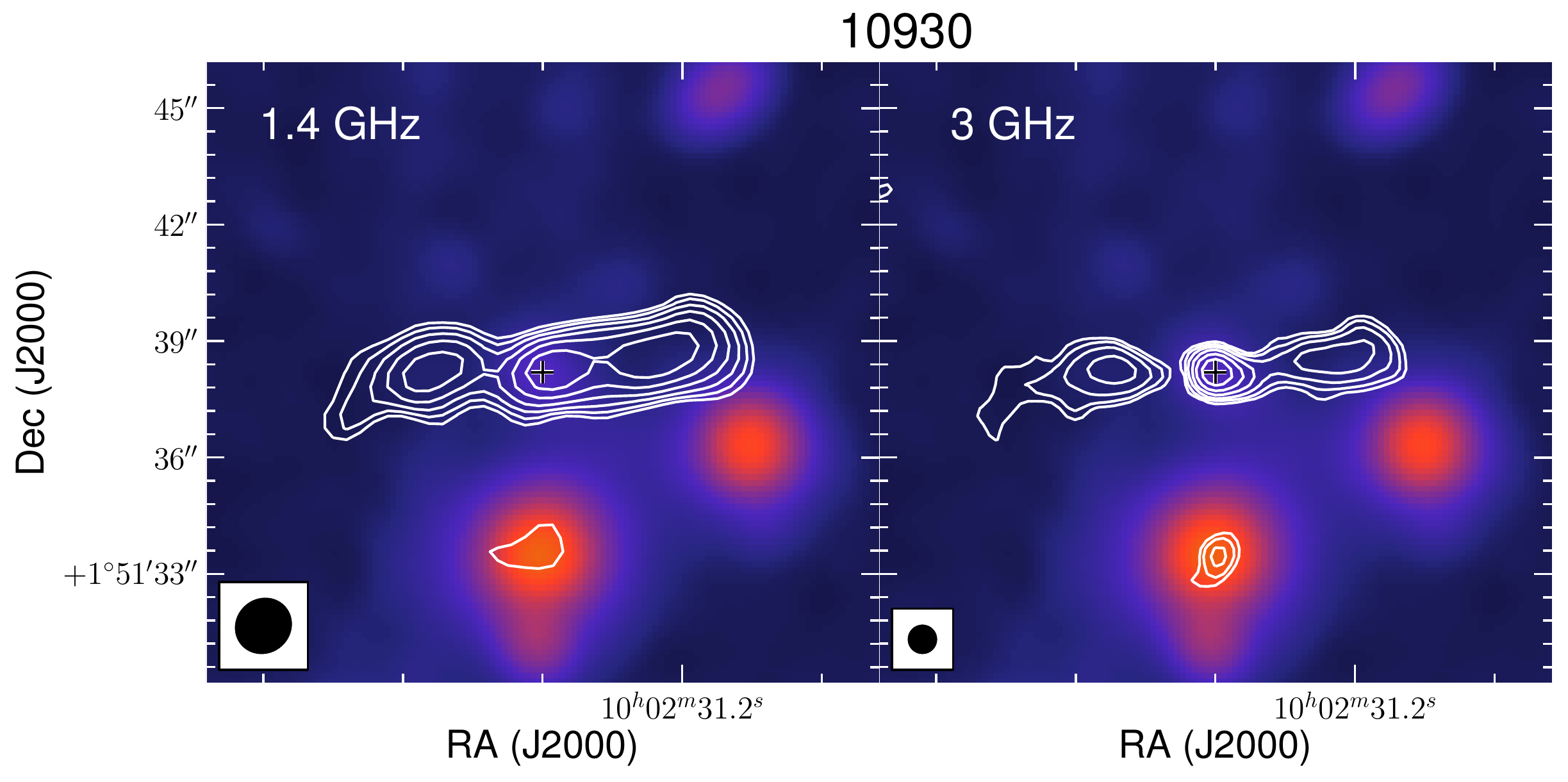}
        \includegraphics{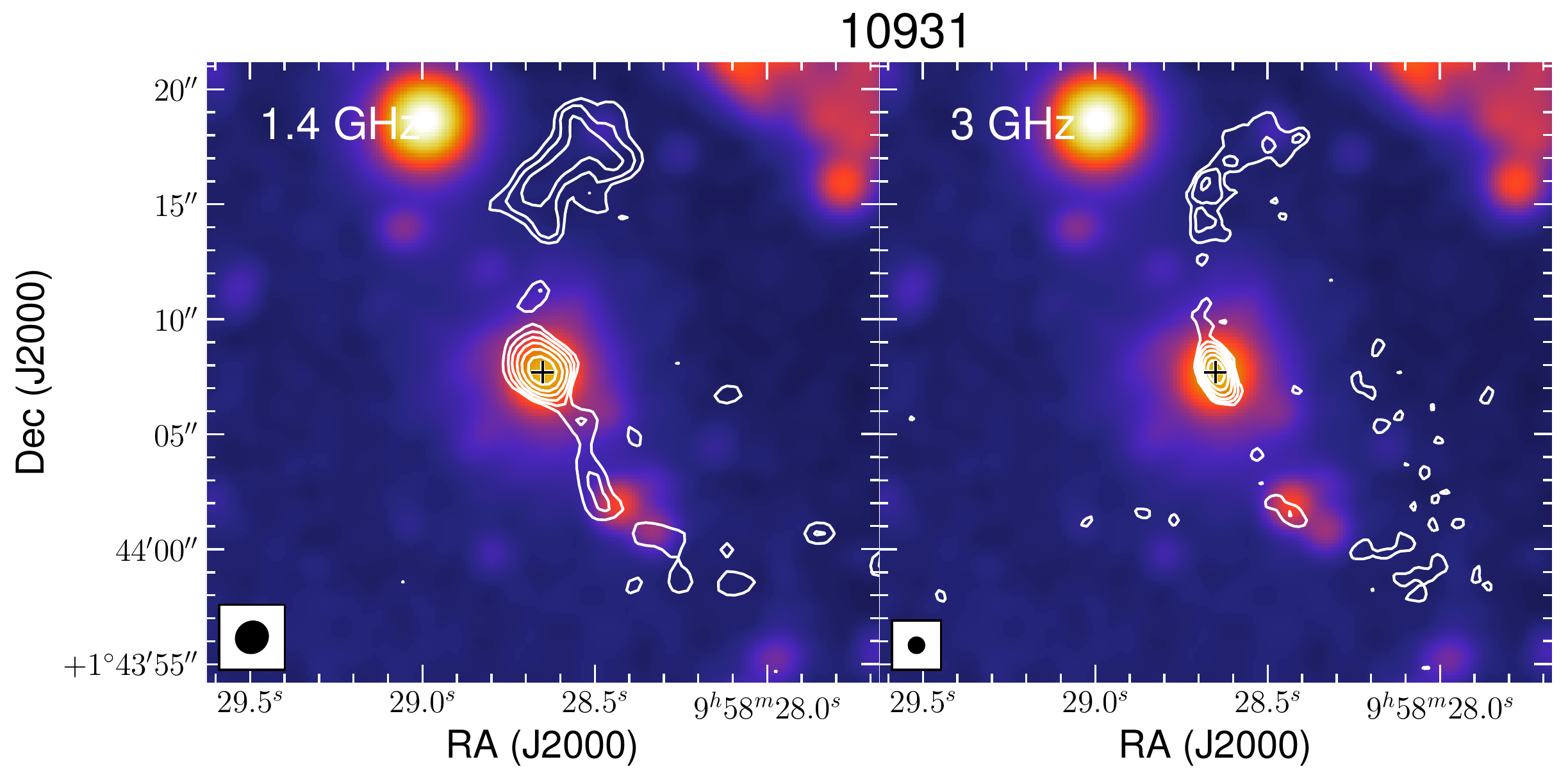}
            }
           
   \caption{(continued)
   }
              \label{fig:maps2}%
    \end{figure*}
\addtocounter{figure}{-1}
\begin{figure*}[!ht]
  \resizebox{\hsize}{!}
       {\includegraphics{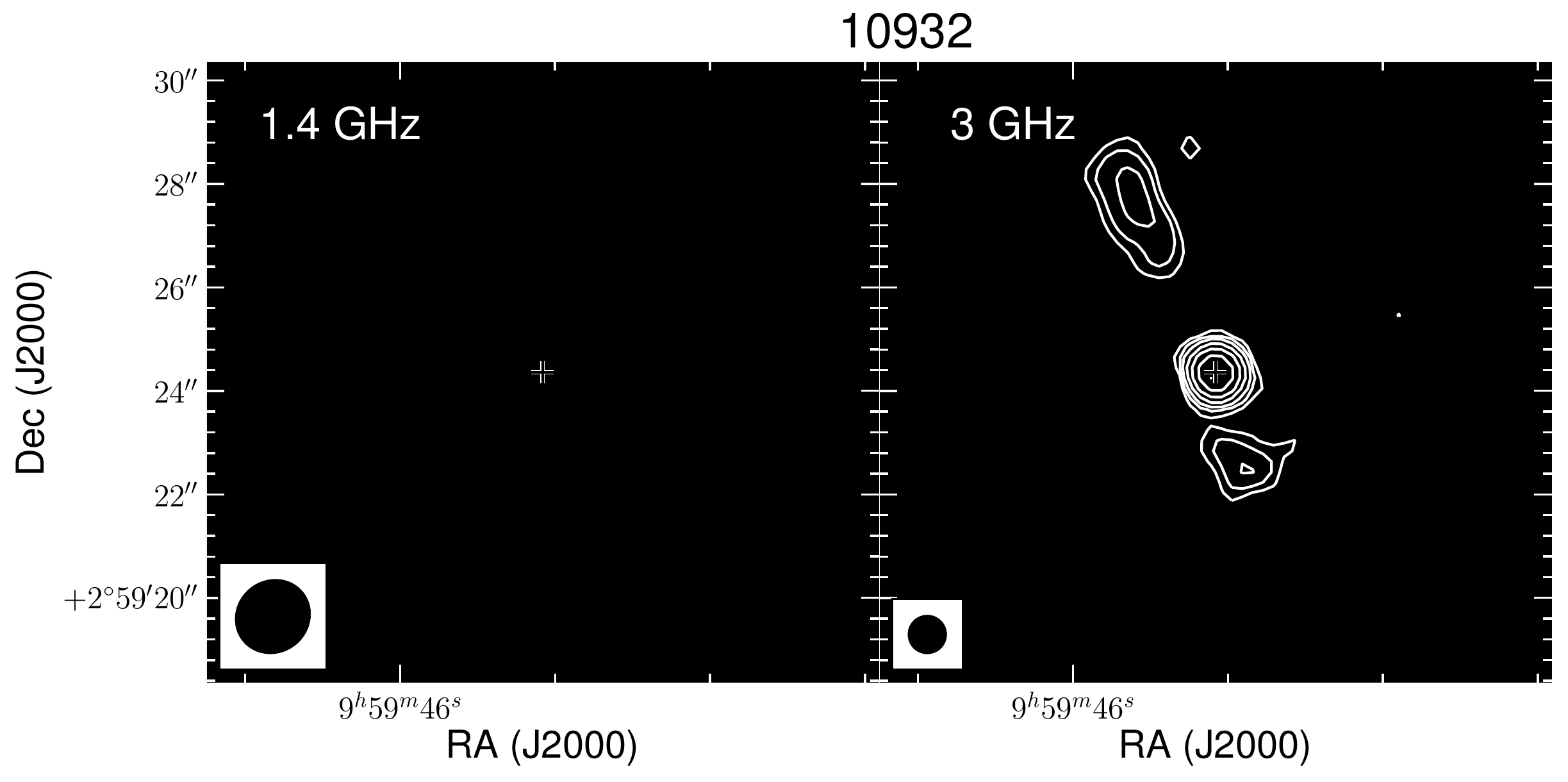}
 \includegraphics{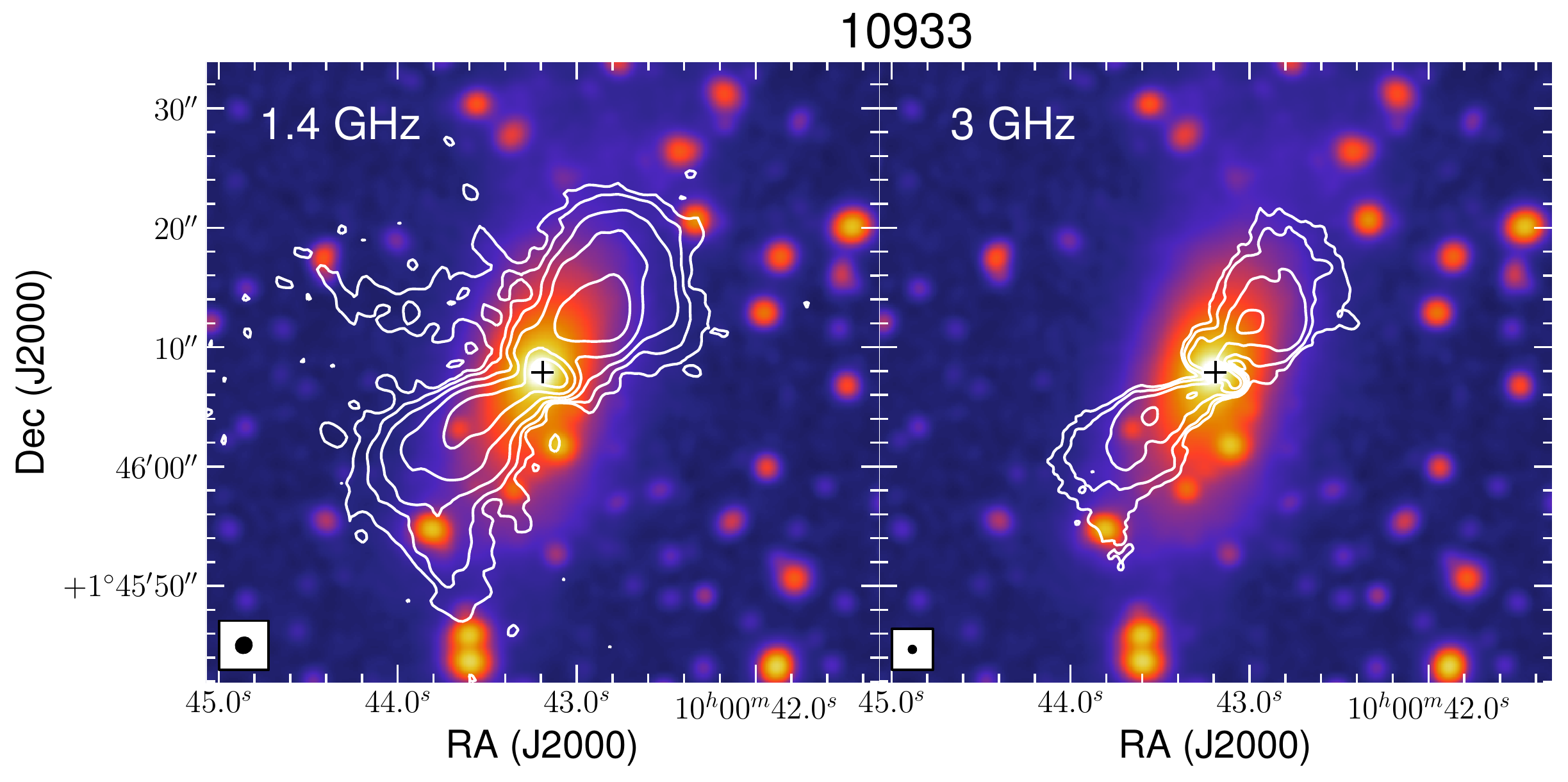}
       \includegraphics{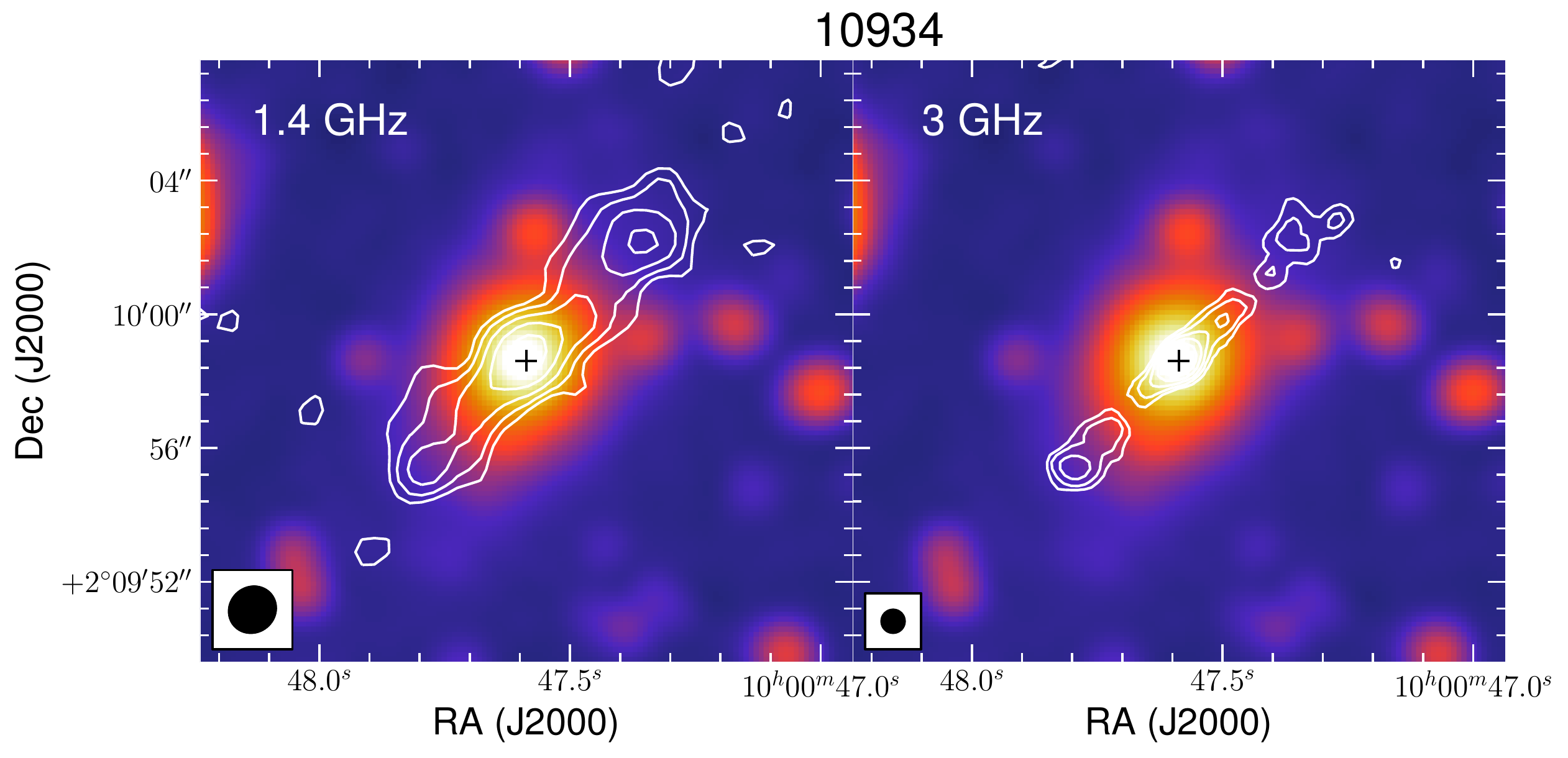}
            }
            \\ \\
    \resizebox{\hsize}{!}
       {\includegraphics{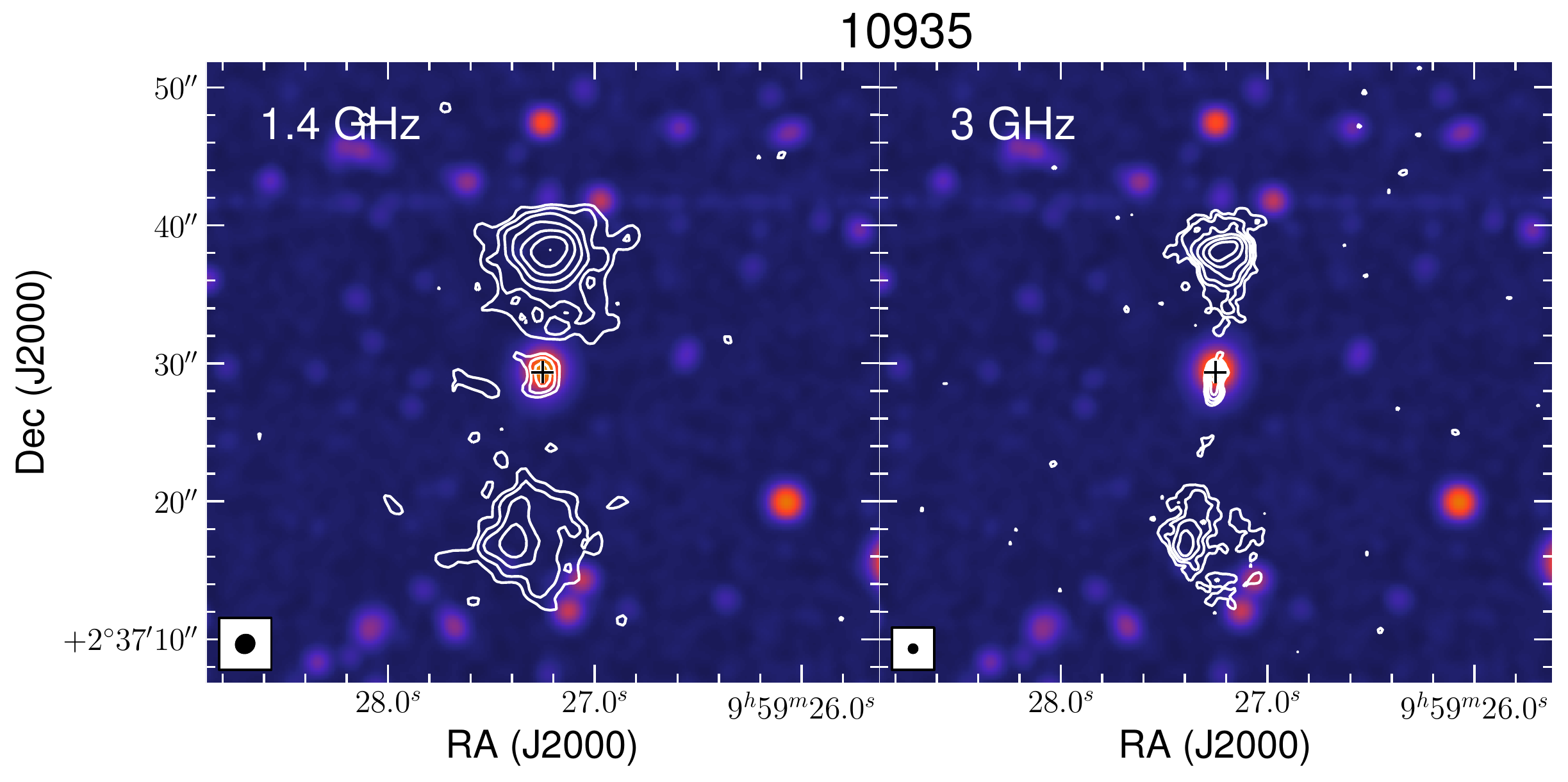}
       \includegraphics{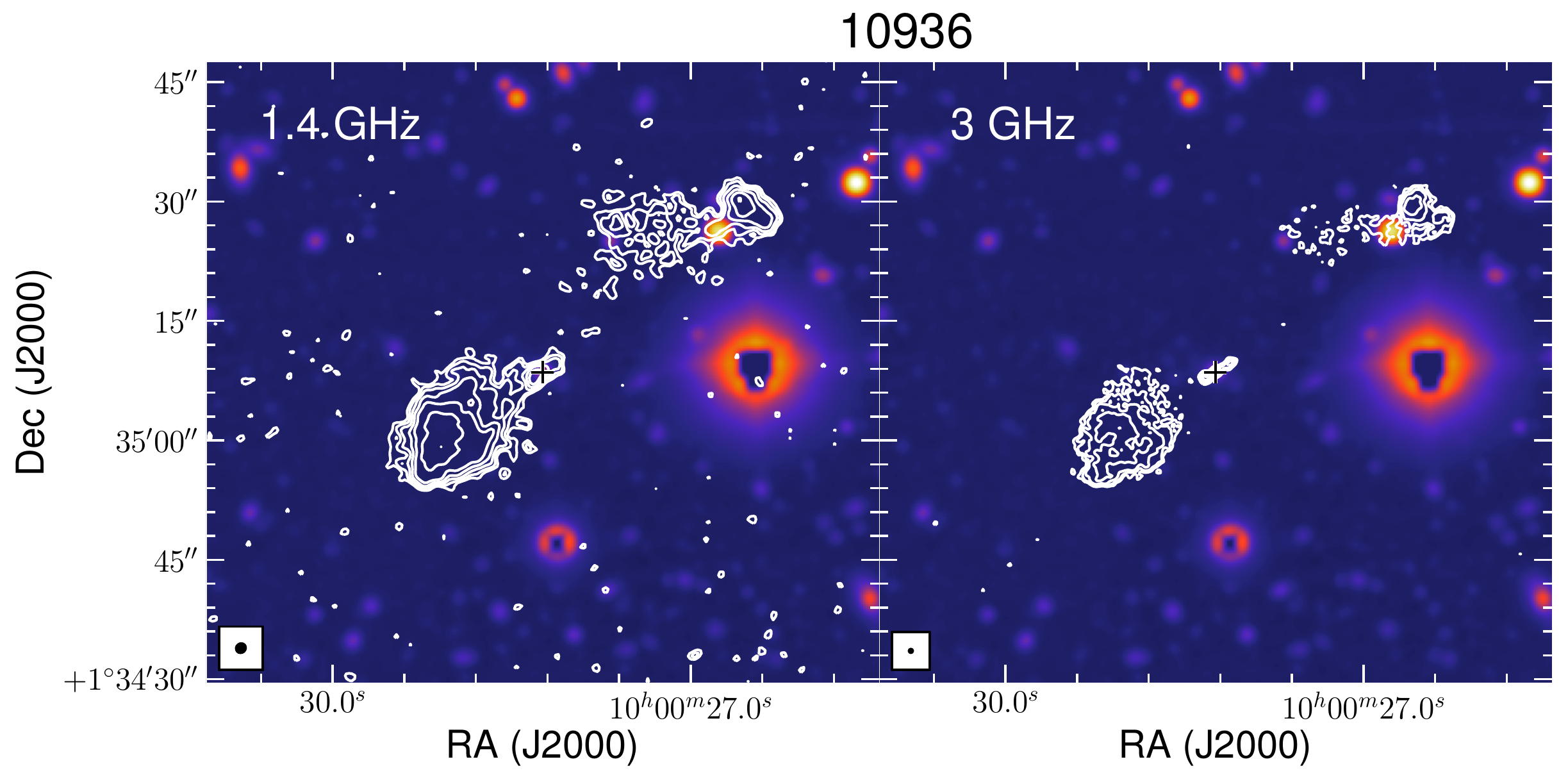}
        \includegraphics{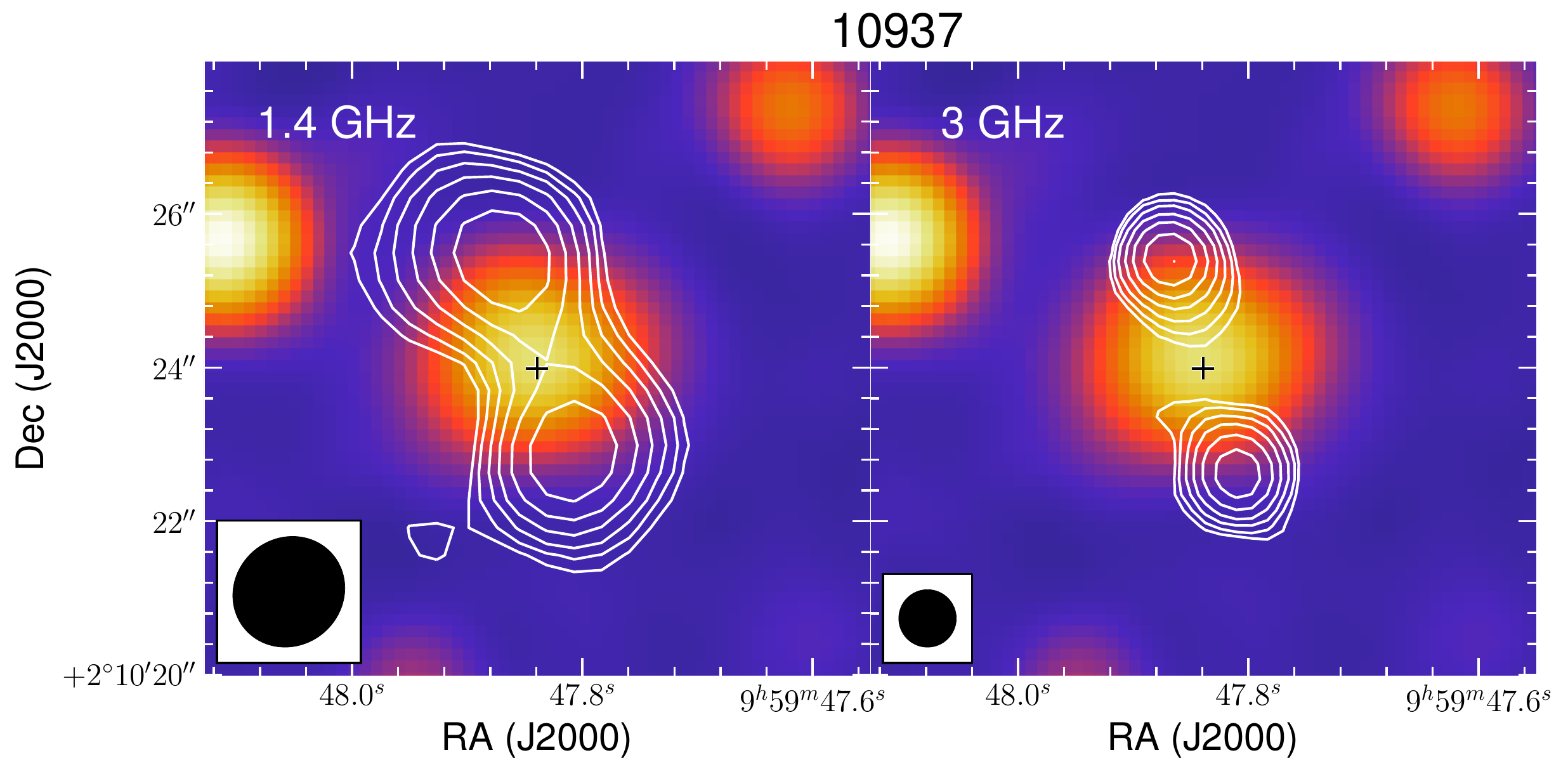}
            }
            \\ \\
 \resizebox{\hsize}{!}
{\includegraphics{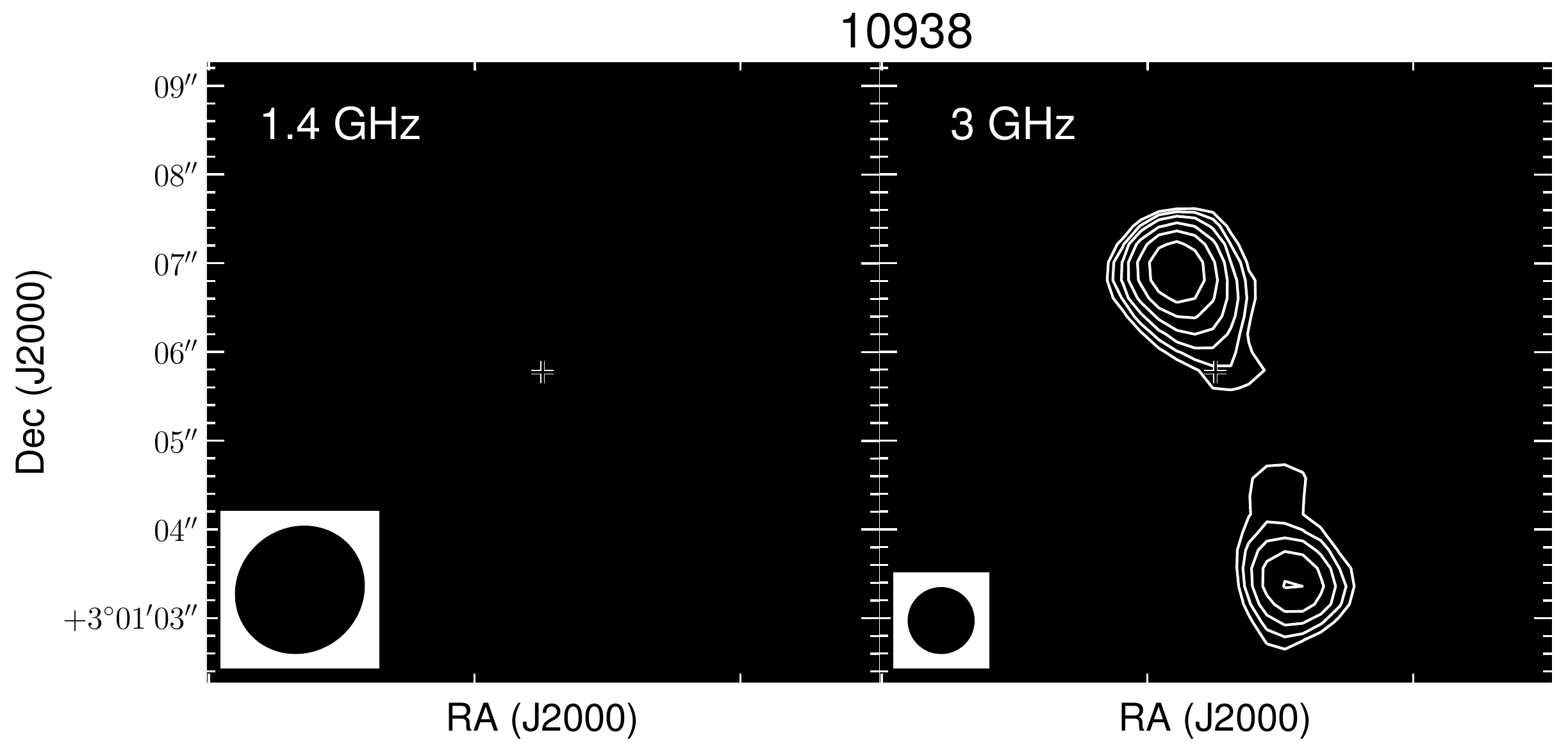}
 \includegraphics{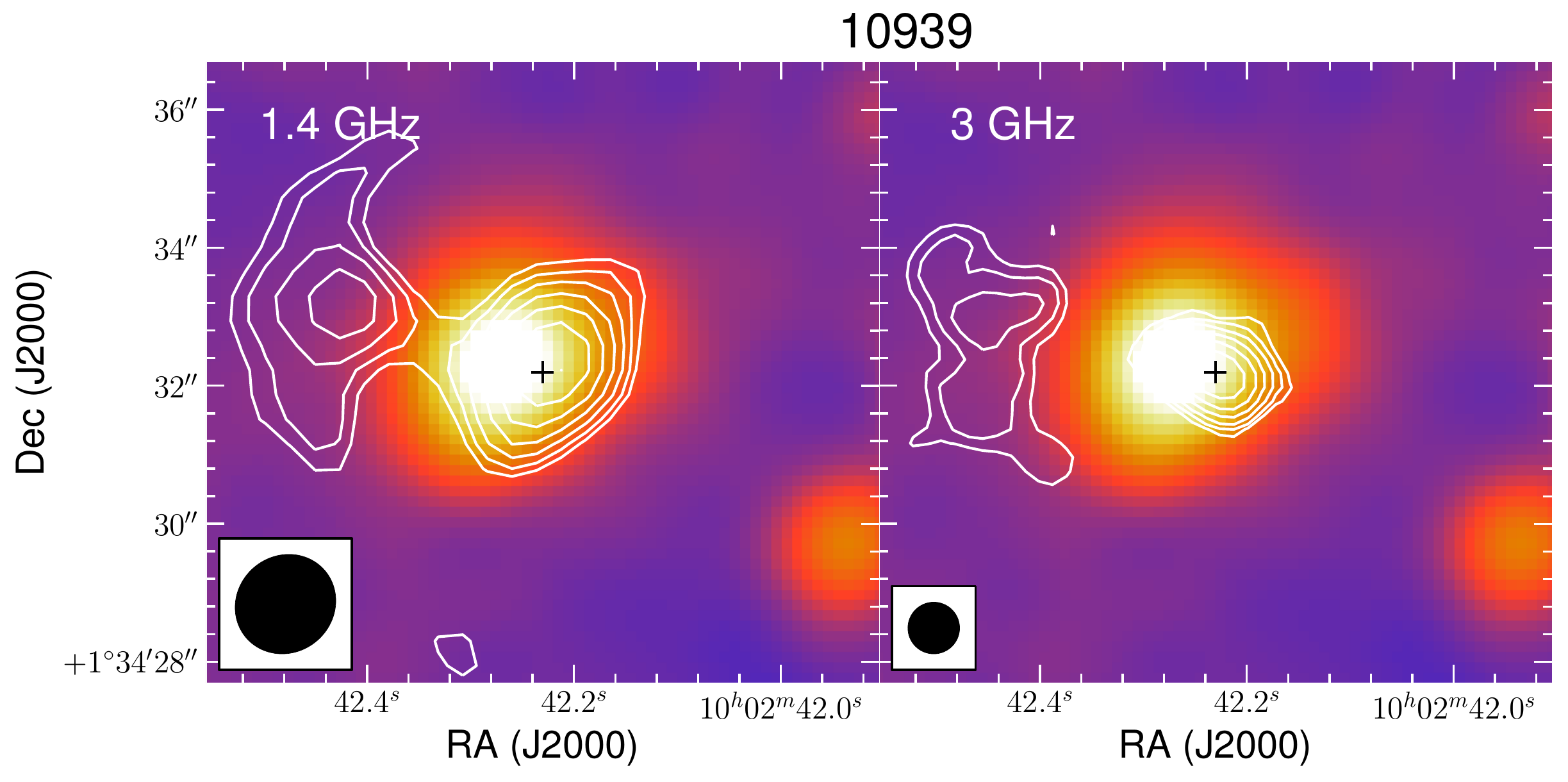}
 \includegraphics{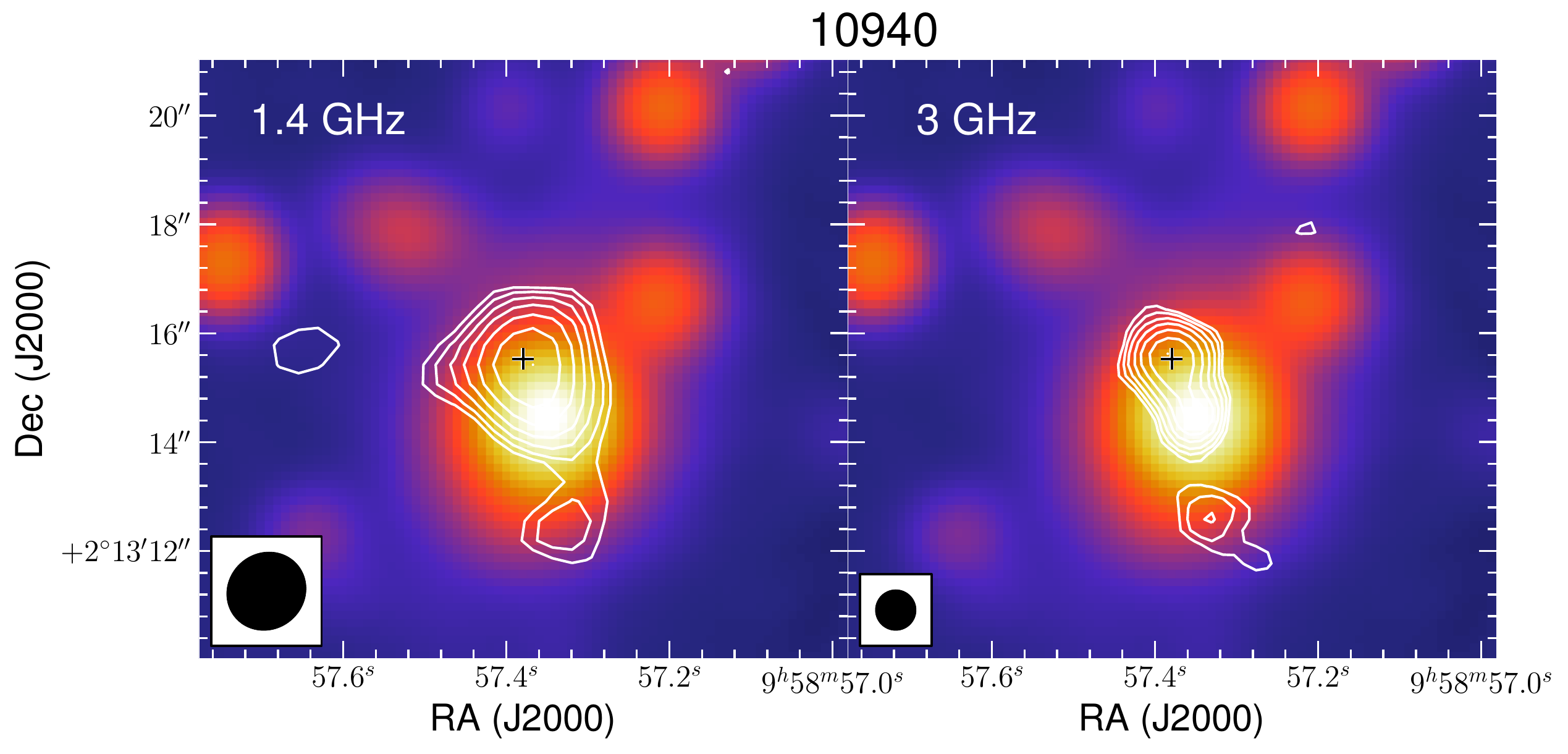}
            }
            \\ \\
  \resizebox{\hsize}{!}
 { \includegraphics{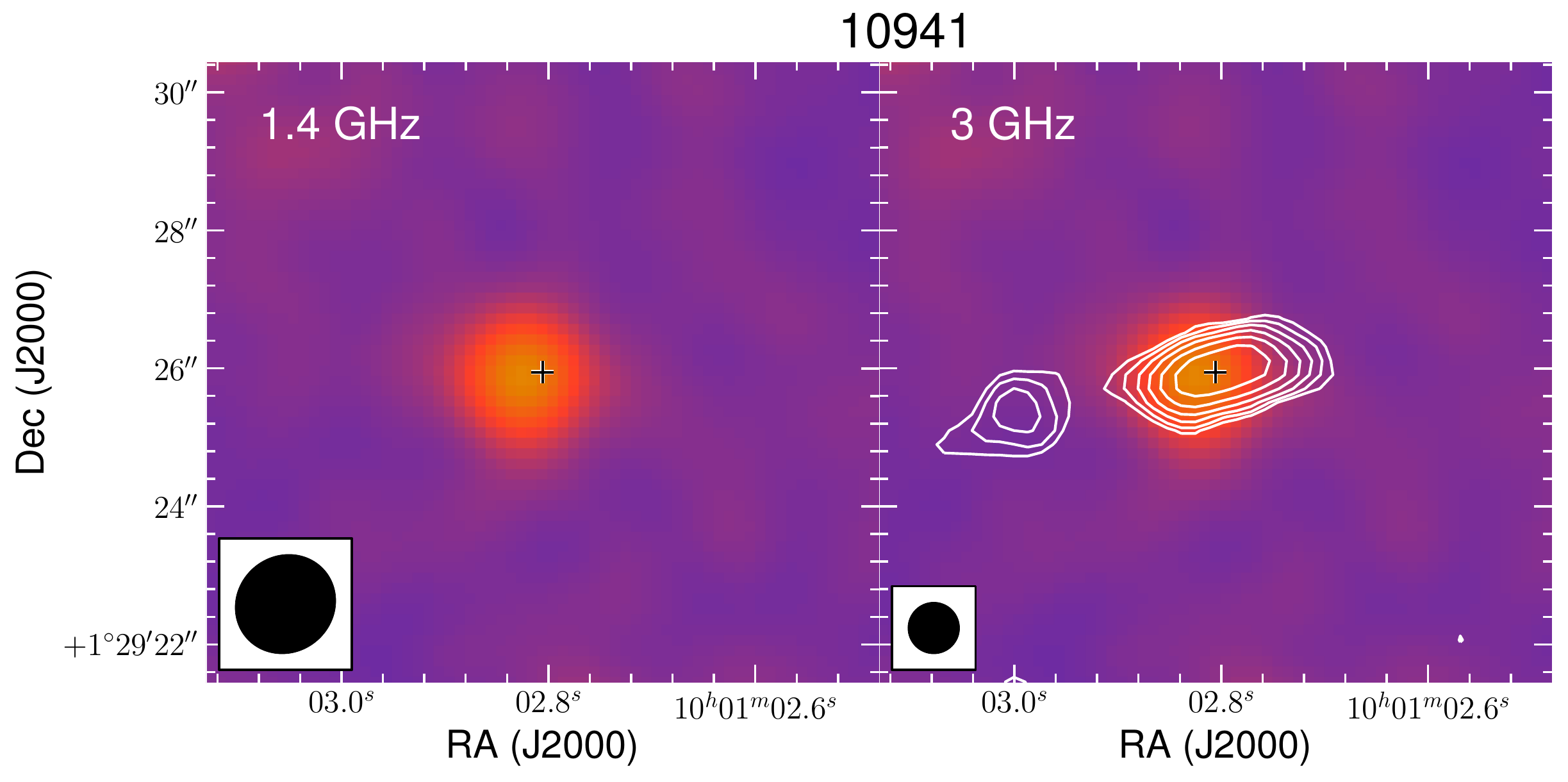}
    \includegraphics{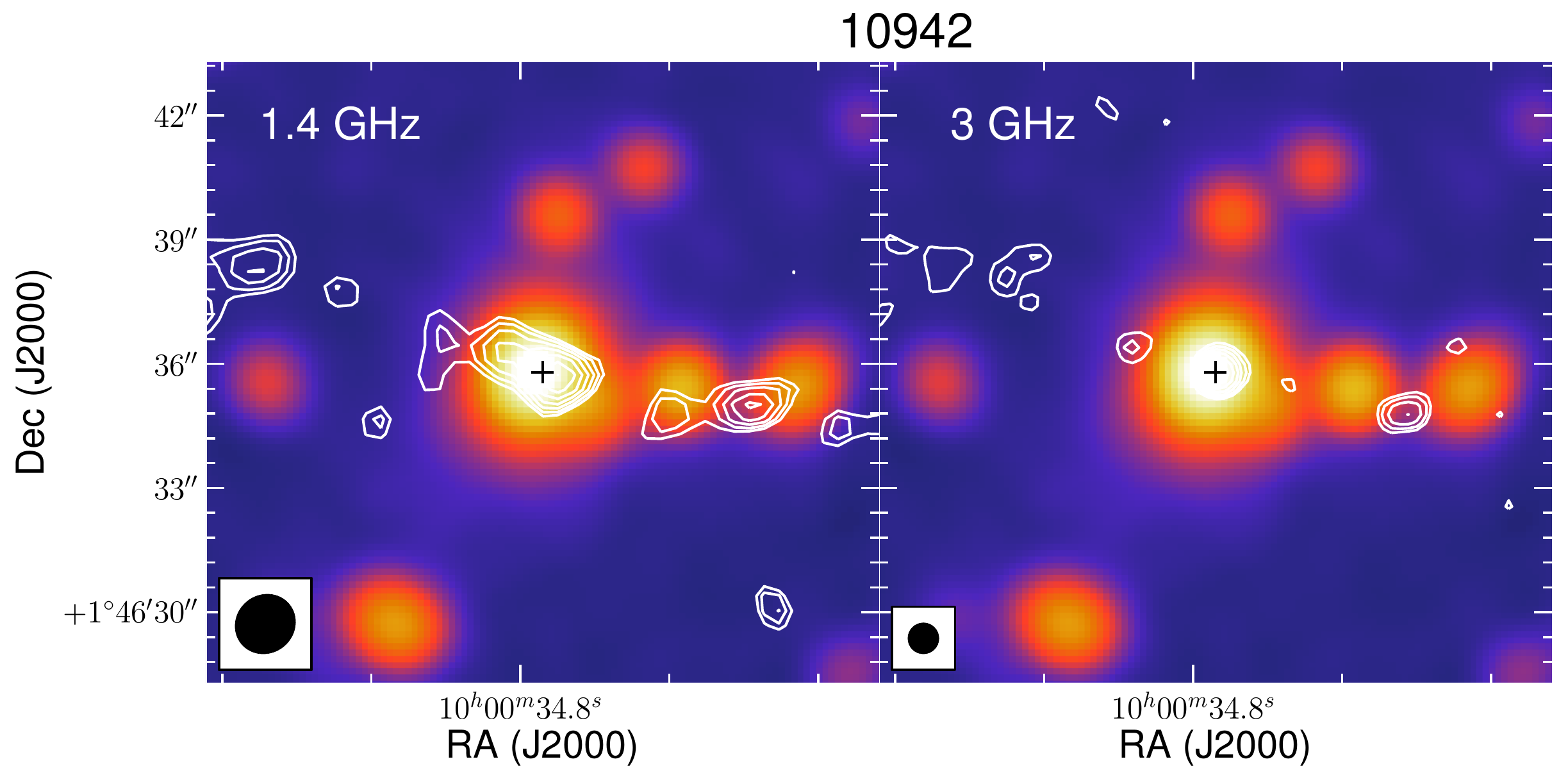}
 \includegraphics{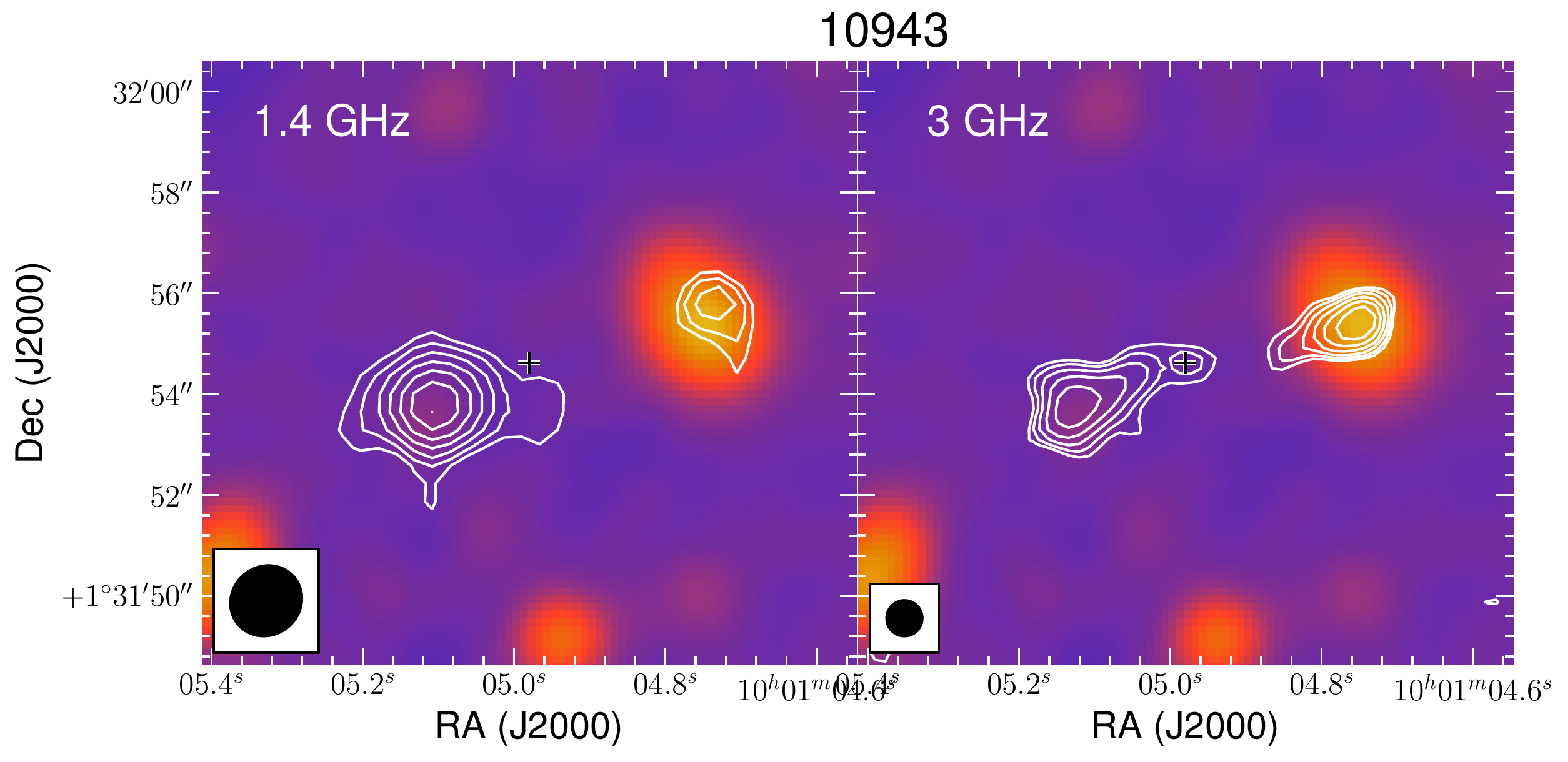}
            }
             \\ \\ 
      \resizebox{\hsize}{!}
       {\includegraphics{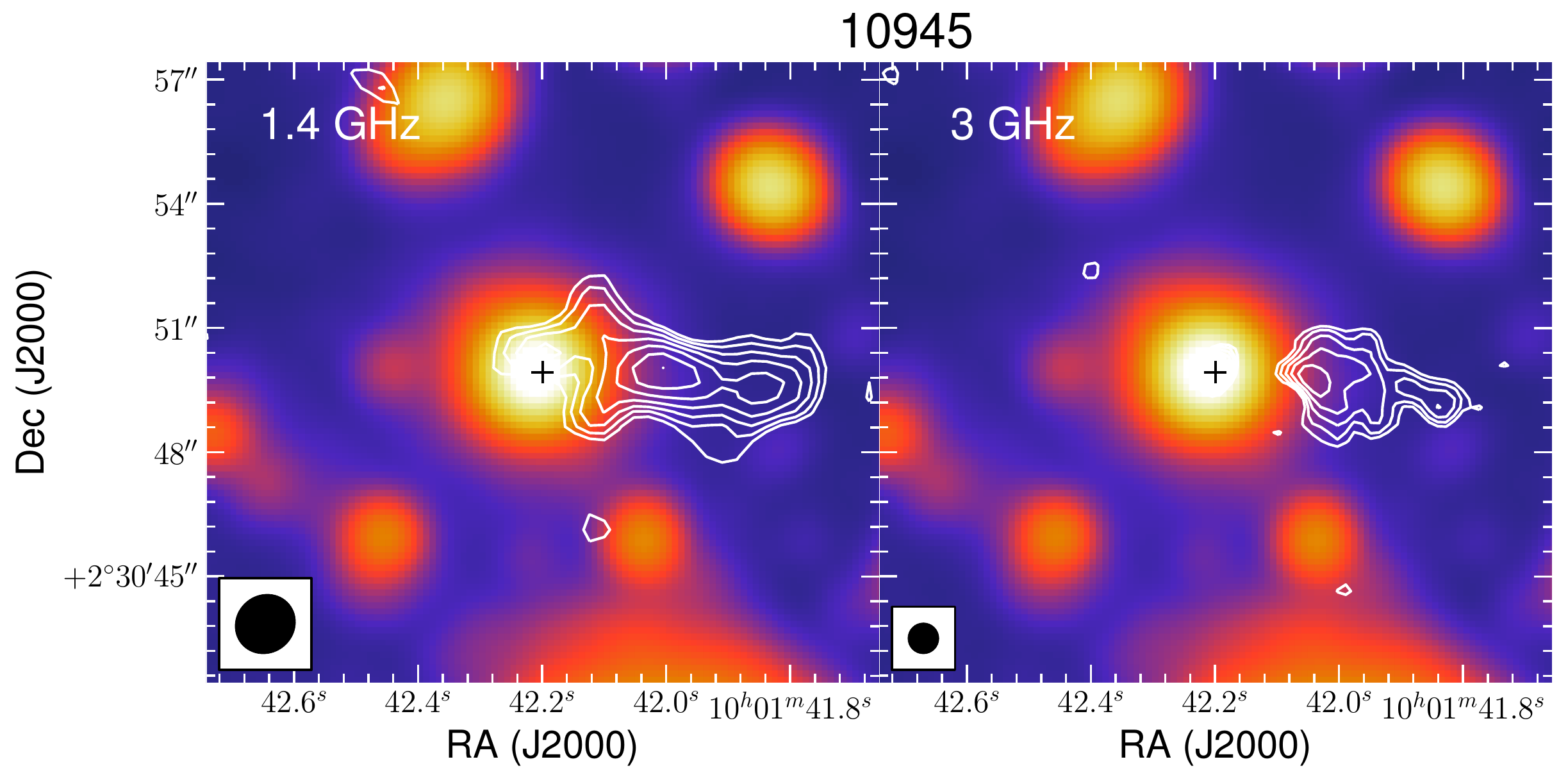}
    \includegraphics{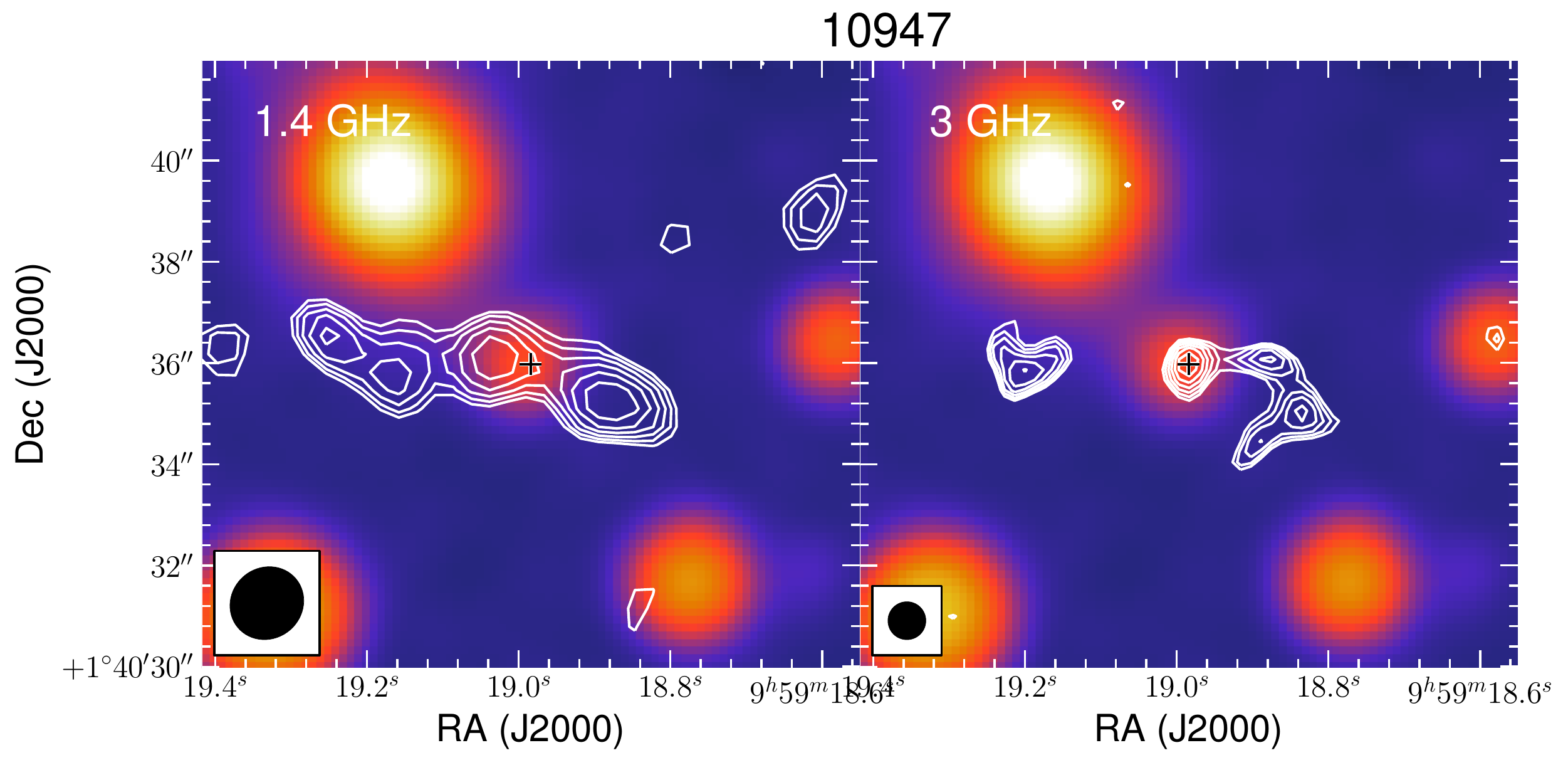}
    \includegraphics{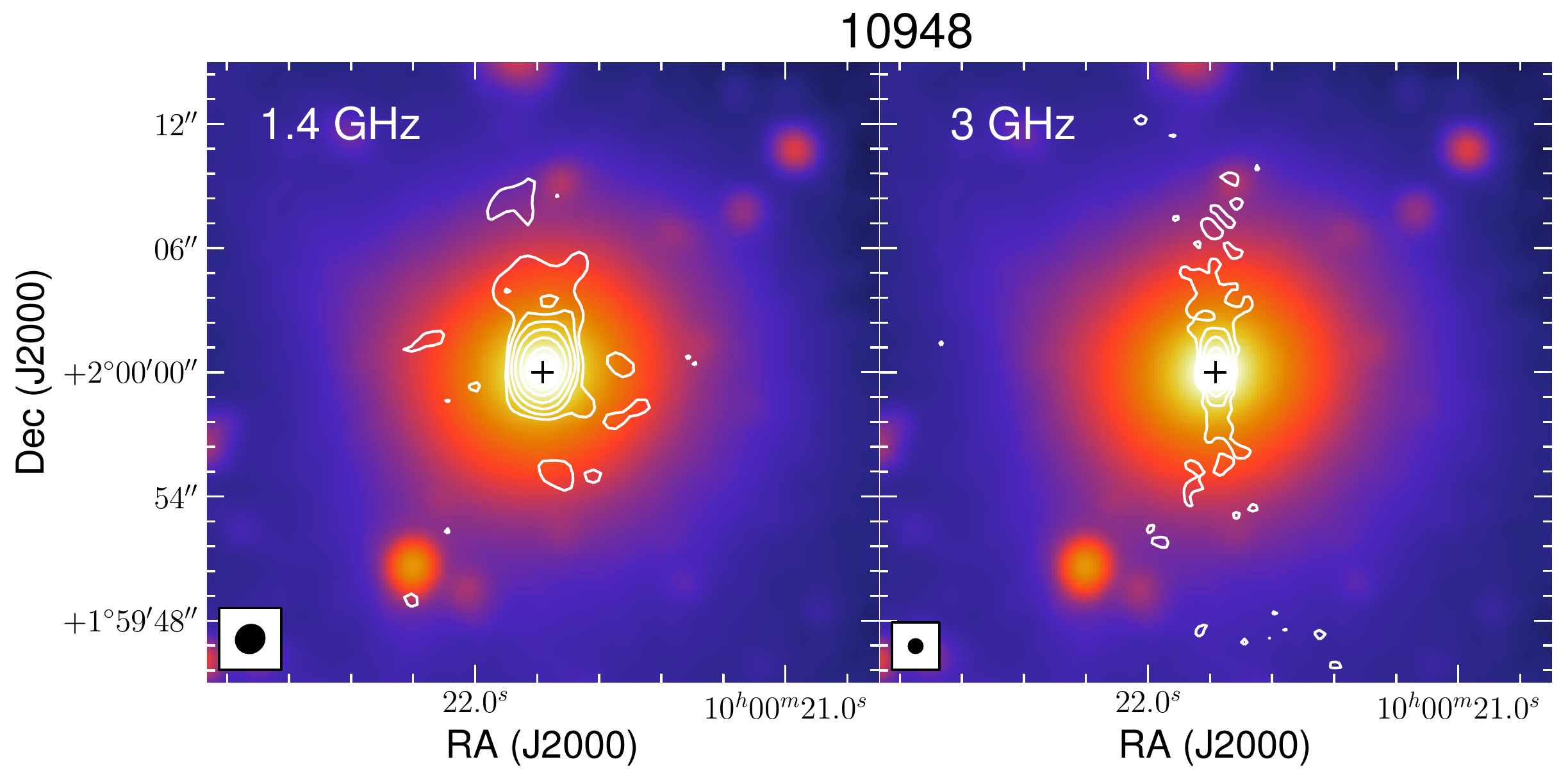}
            }
\\ \\
 \resizebox{\hsize}{!}
{ \includegraphics{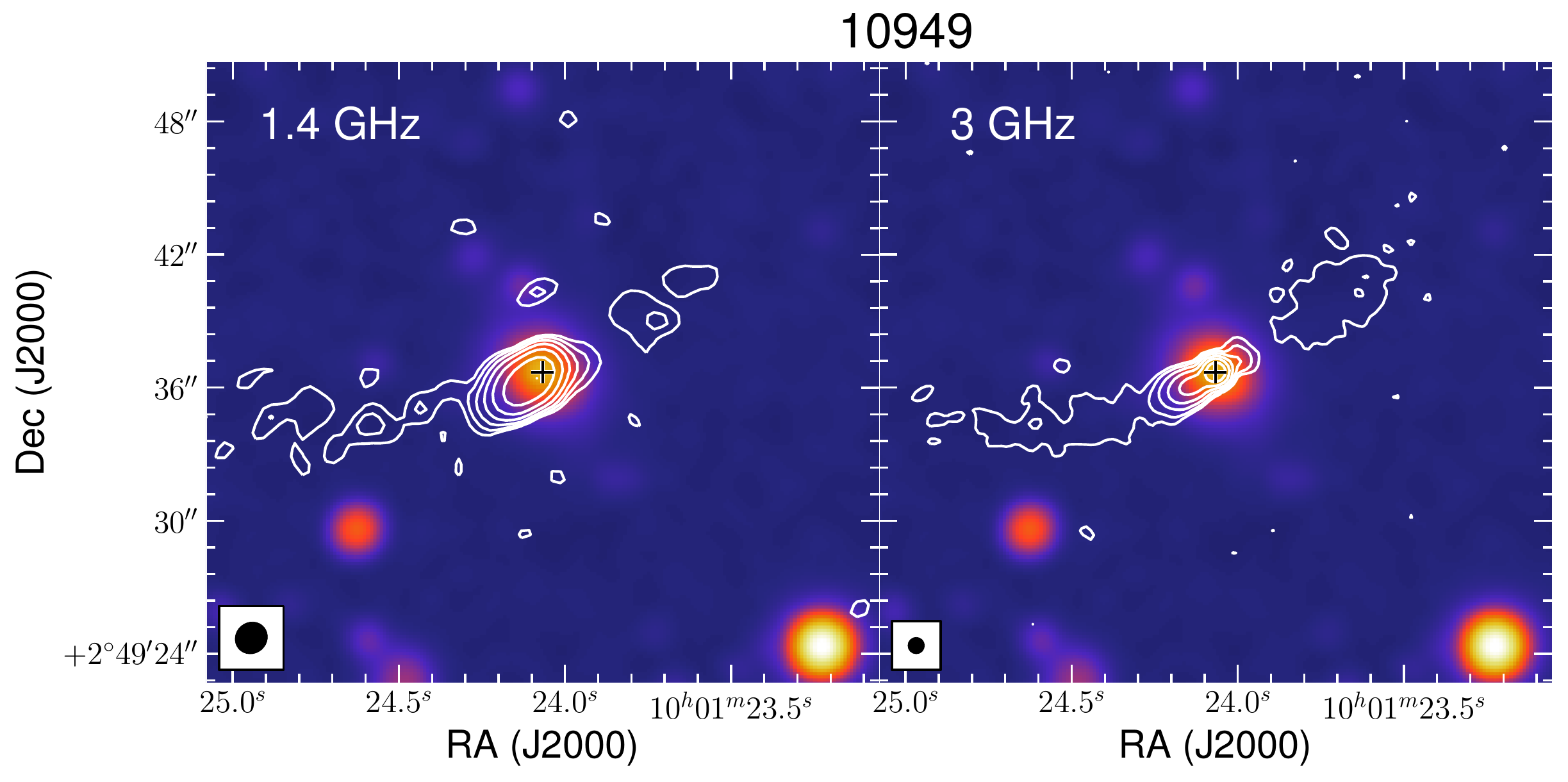}
\includegraphics{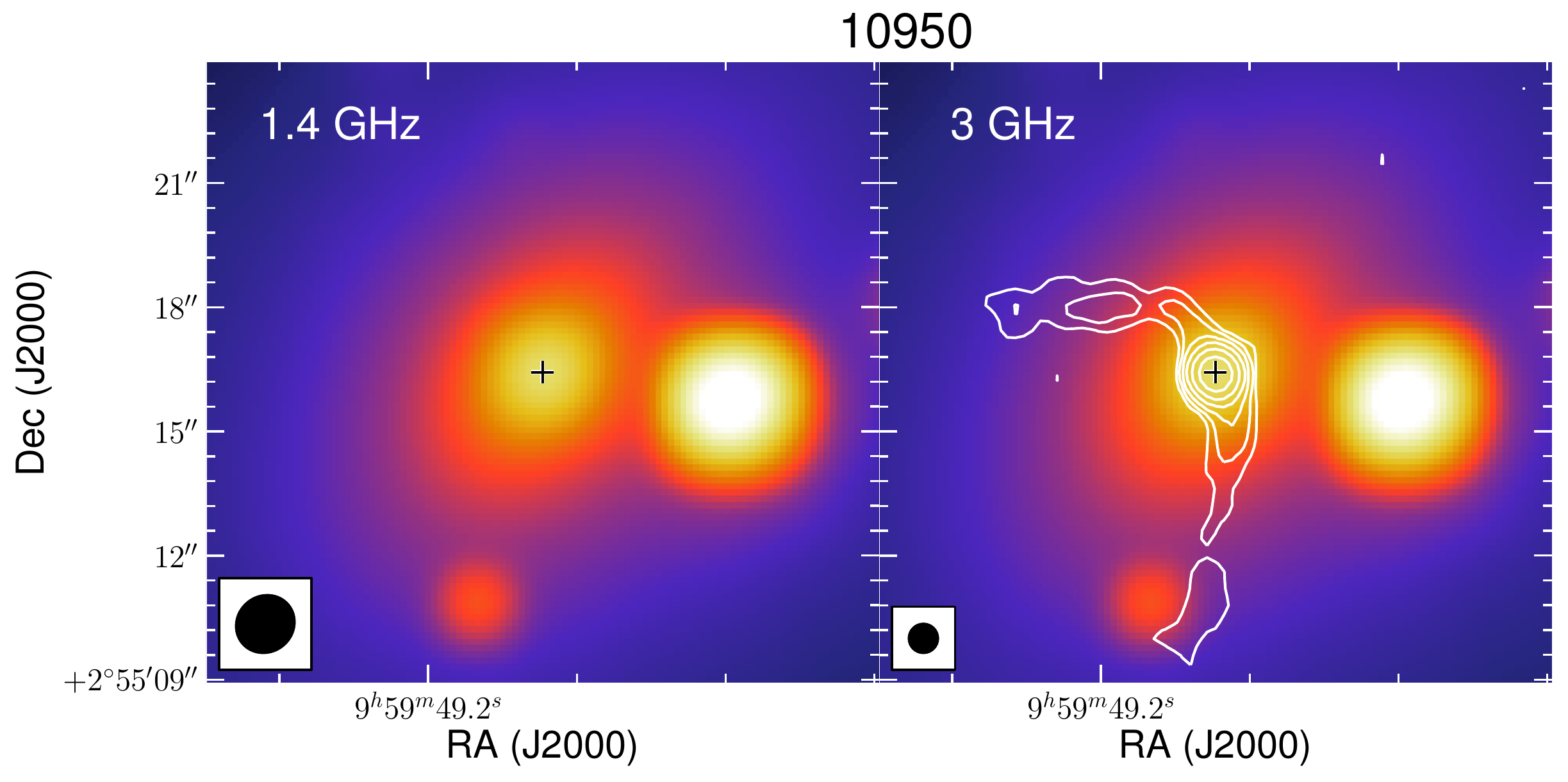}
 \includegraphics{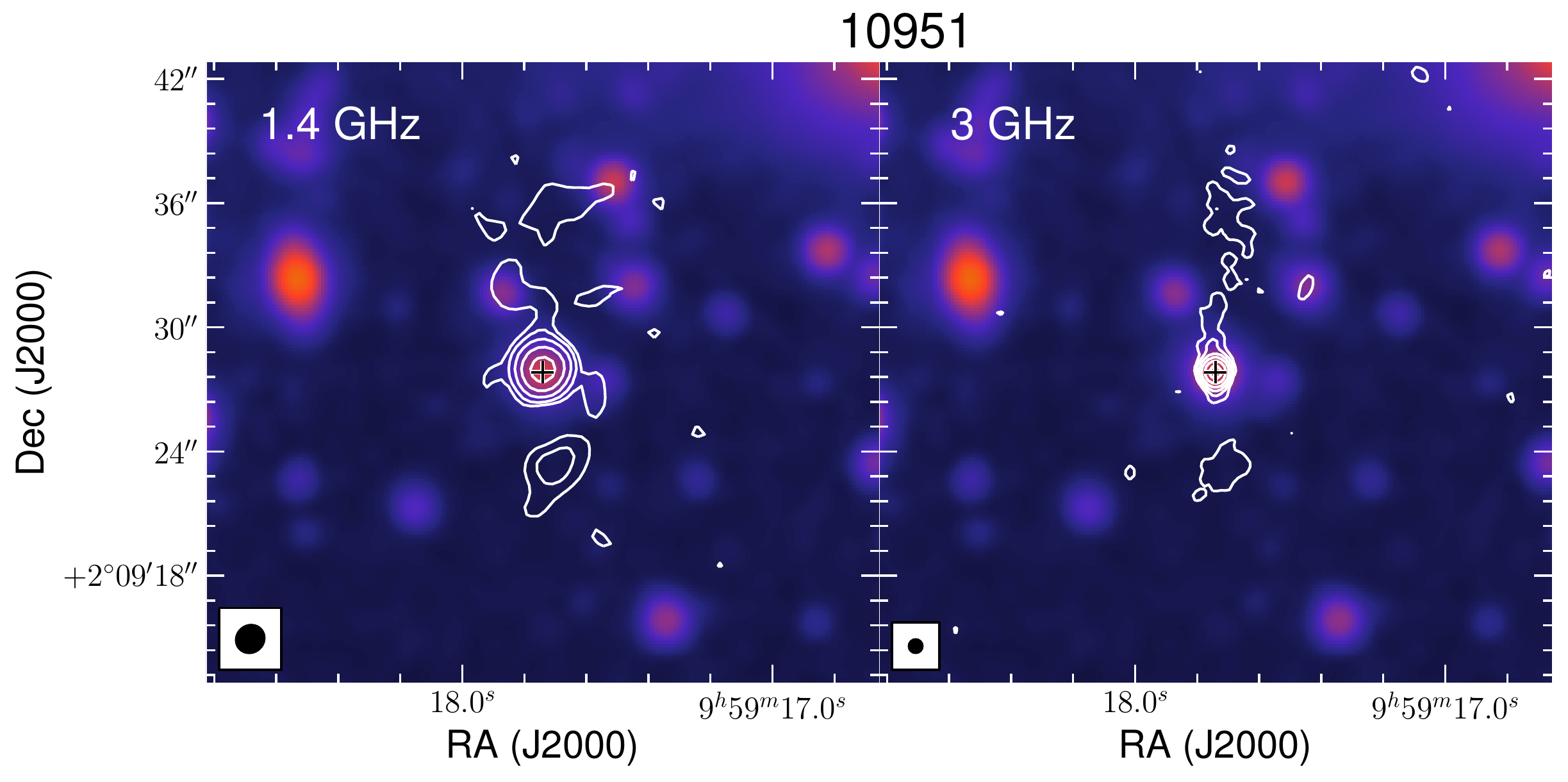}
            }
            \\ \\
              \resizebox{\hsize}{!}
       {\includegraphics{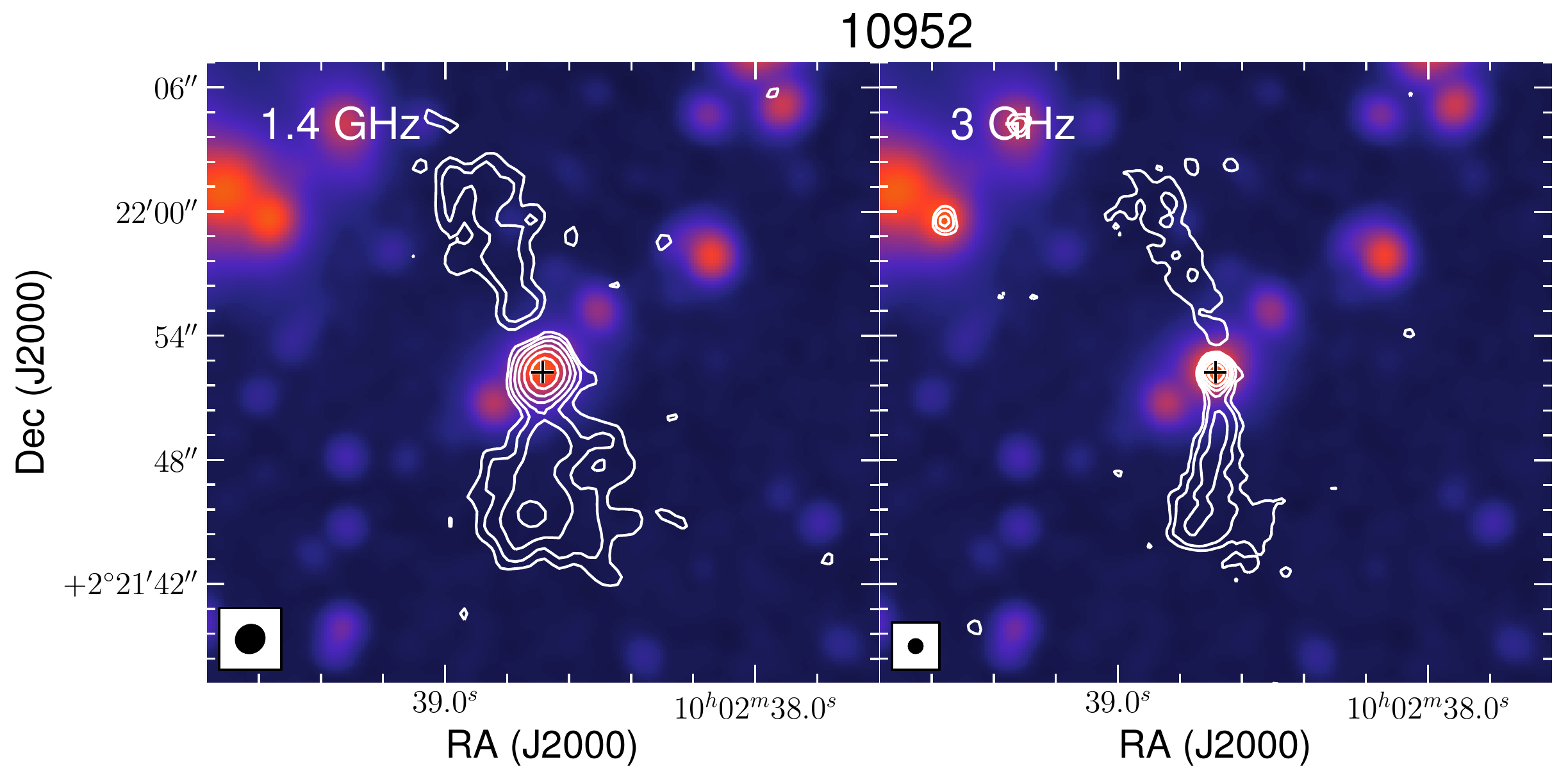}
        \includegraphics{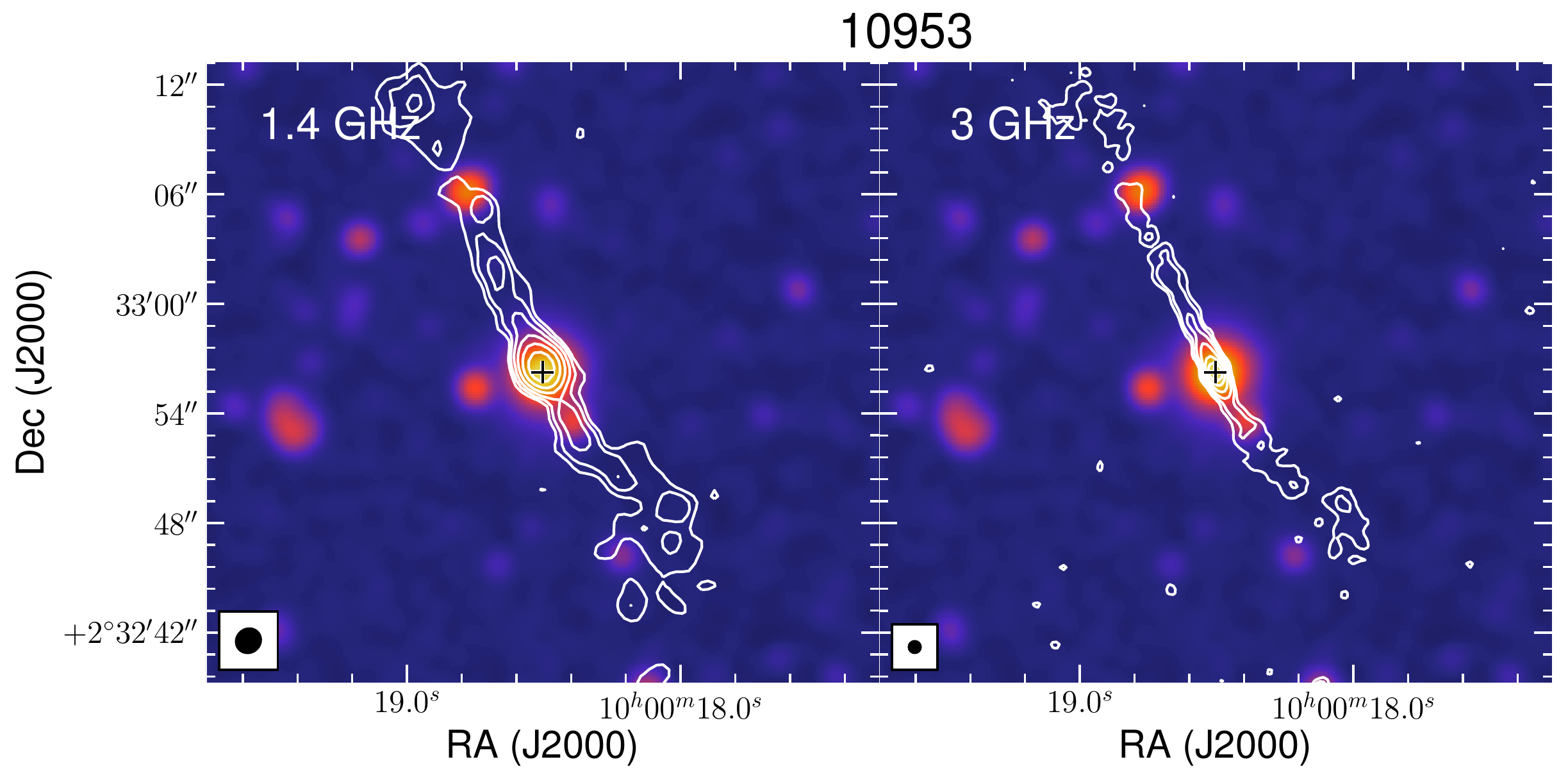}
       \includegraphics{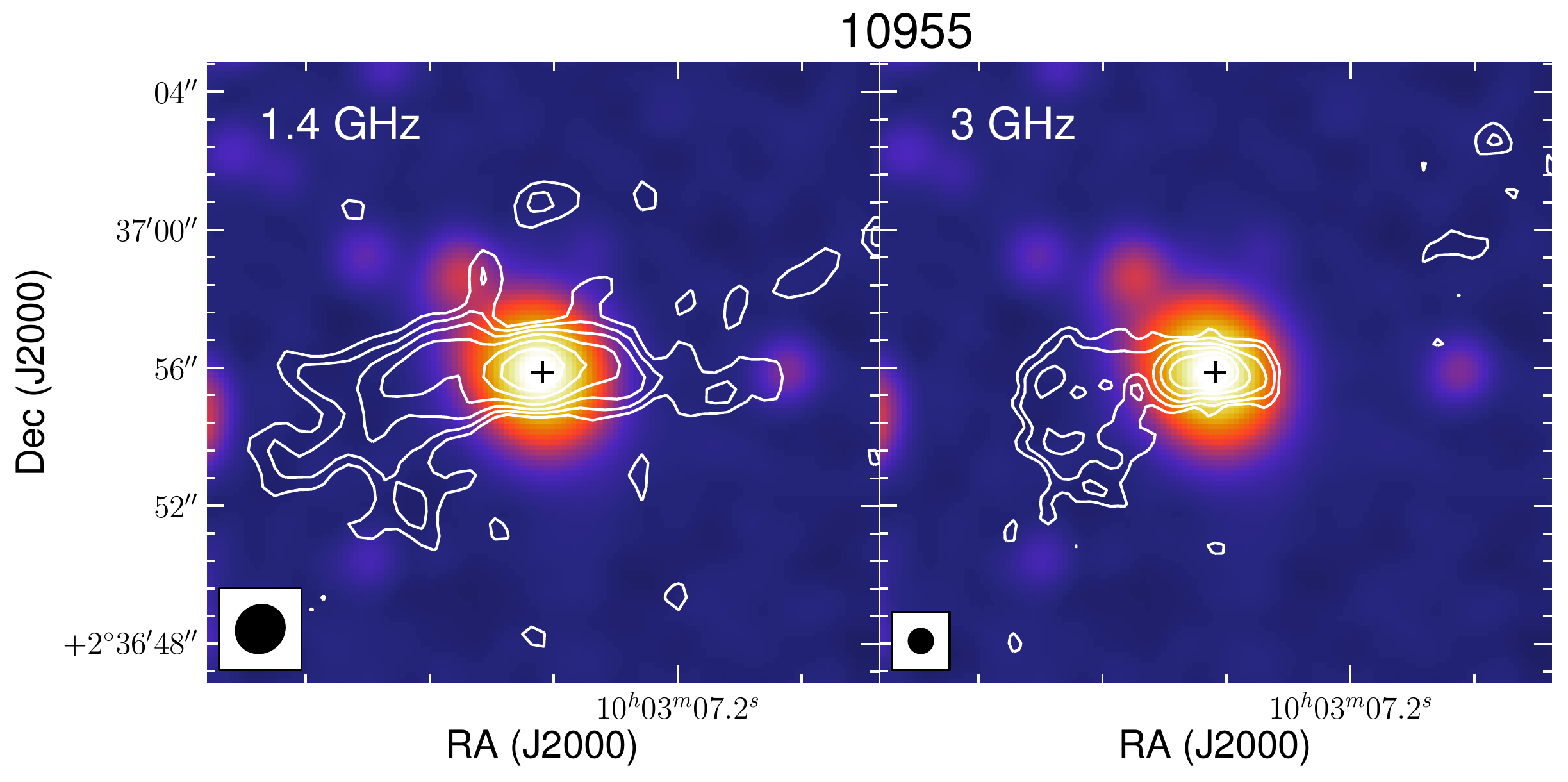}
            }

   \caption{(continued)
   }
              \label{fig:maps2}%
    \end{figure*}
\addtocounter{figure}{-1}
\begin{figure*}[!ht]
   \resizebox{\hsize}{!}
{            \includegraphics{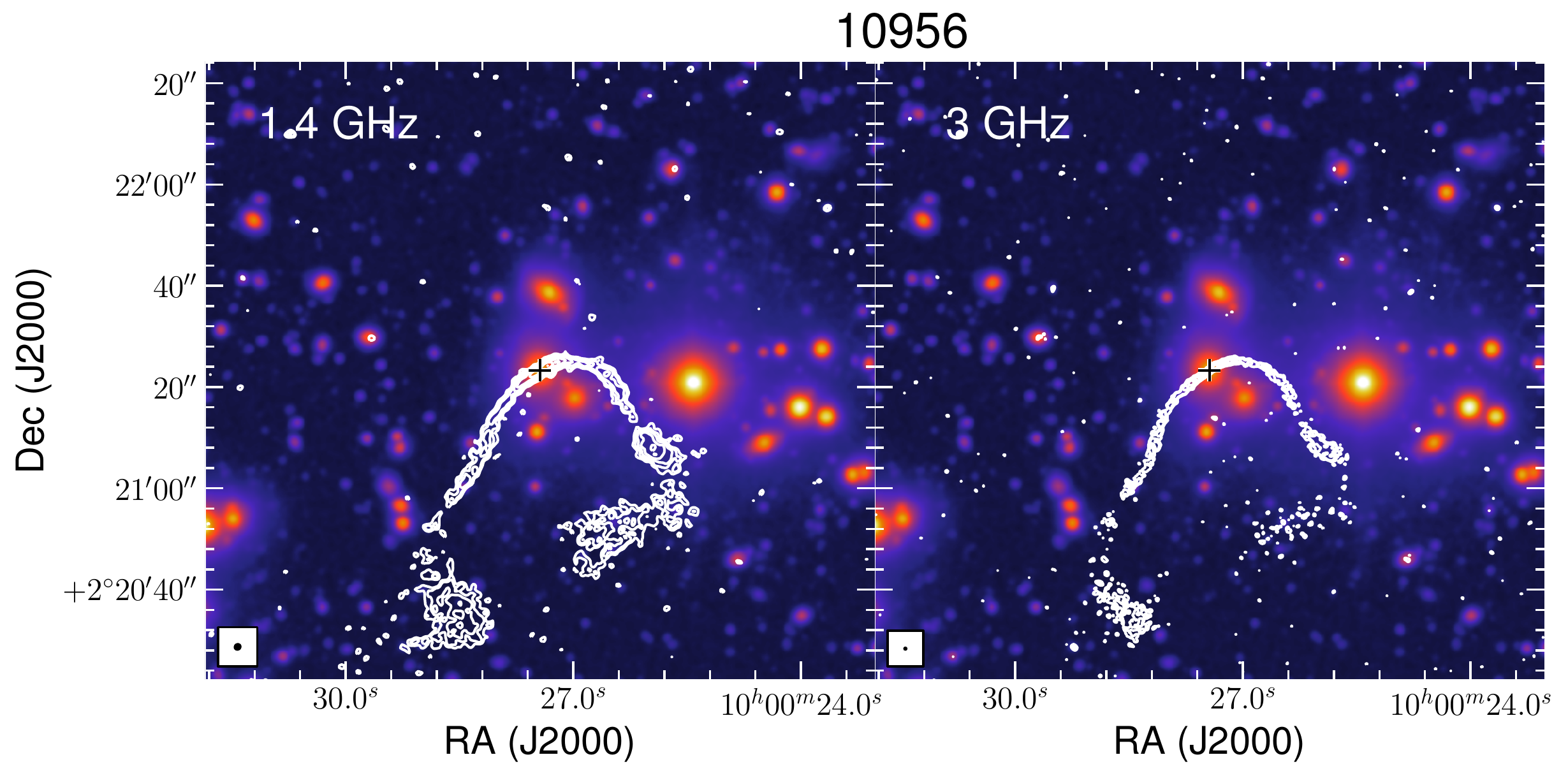}
\includegraphics{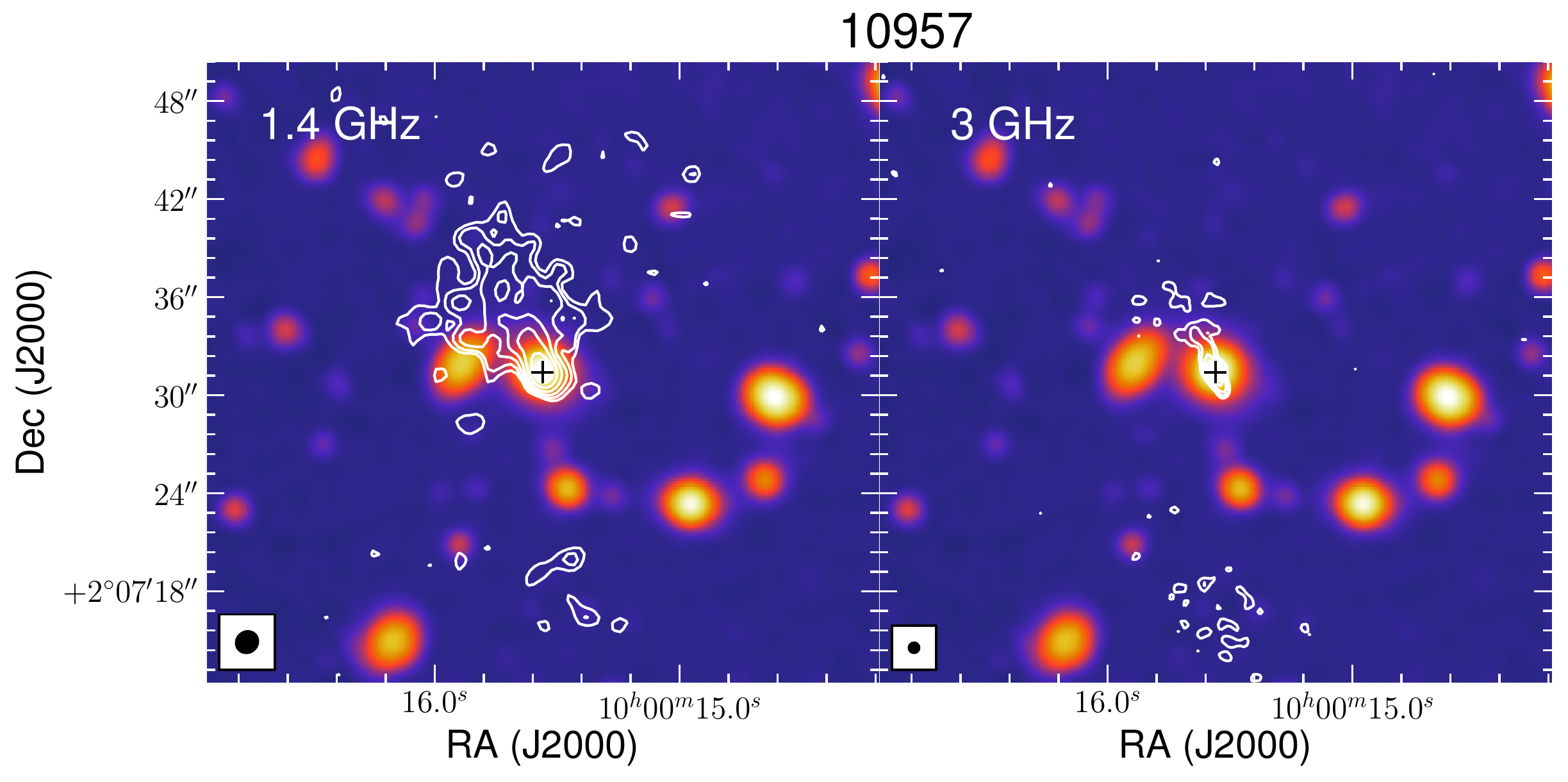}
    \includegraphics{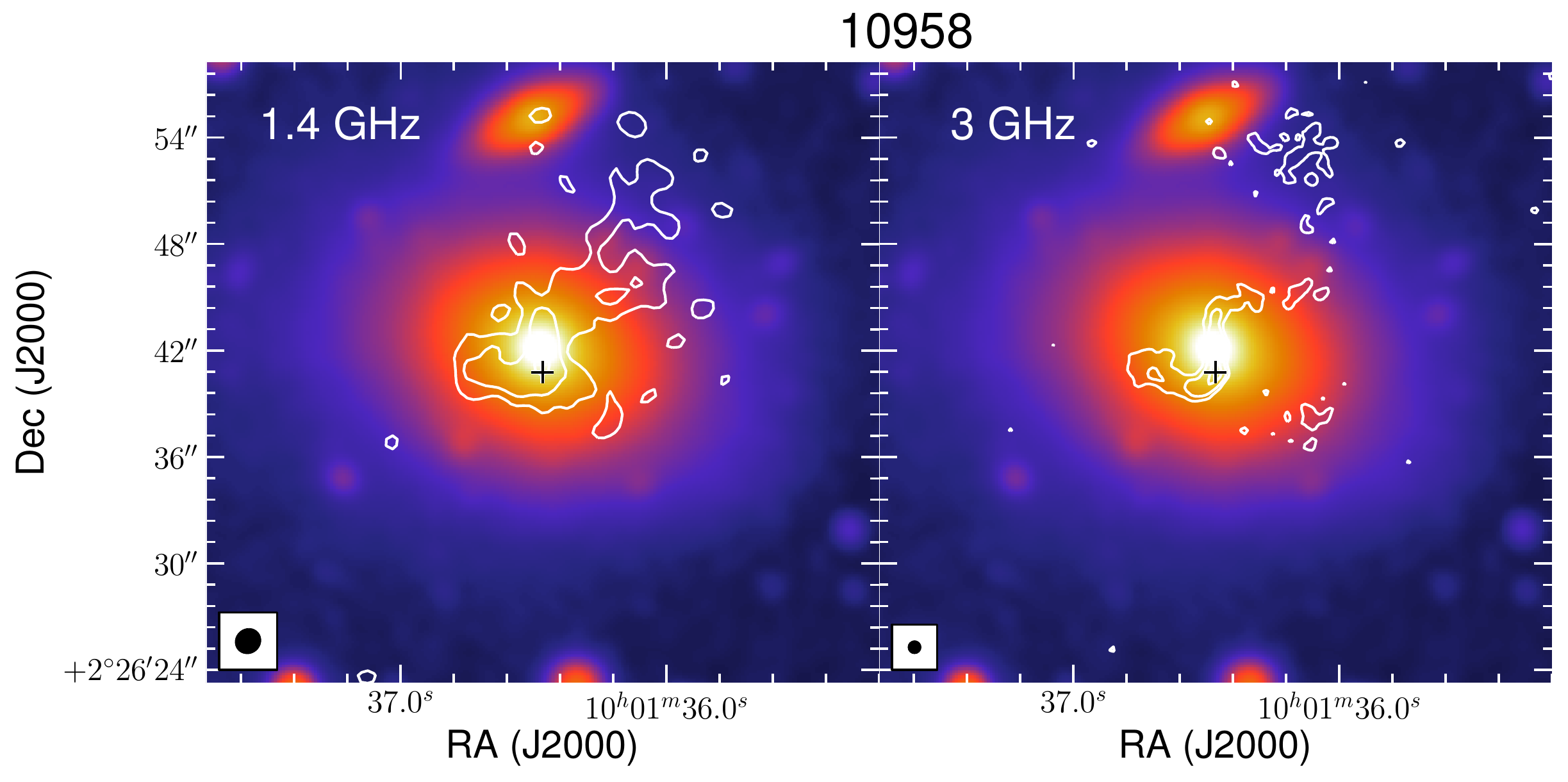}
 }
            \\ \\
  \resizebox{\hsize}{!}
 {\includegraphics{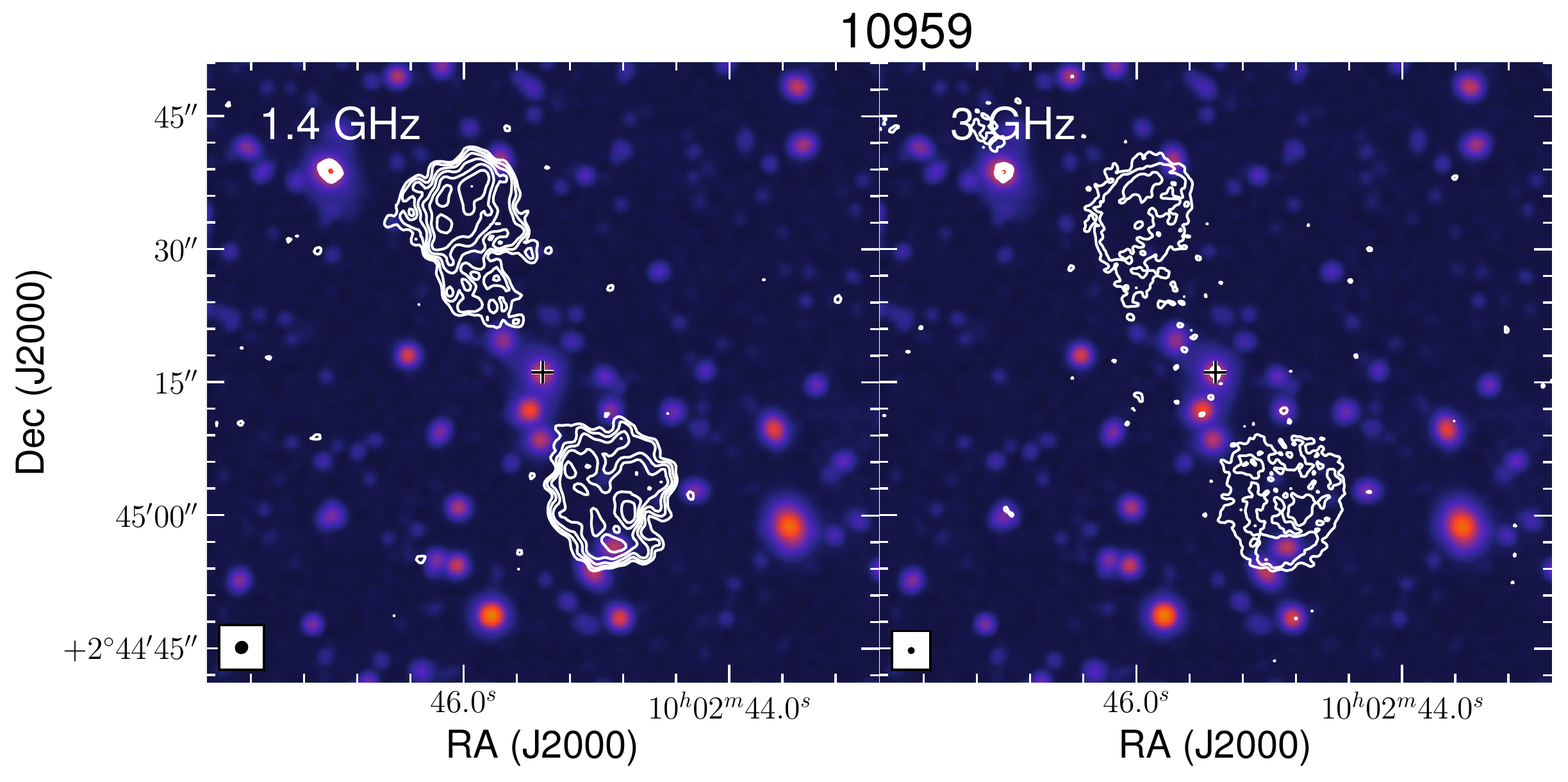}
        \includegraphics{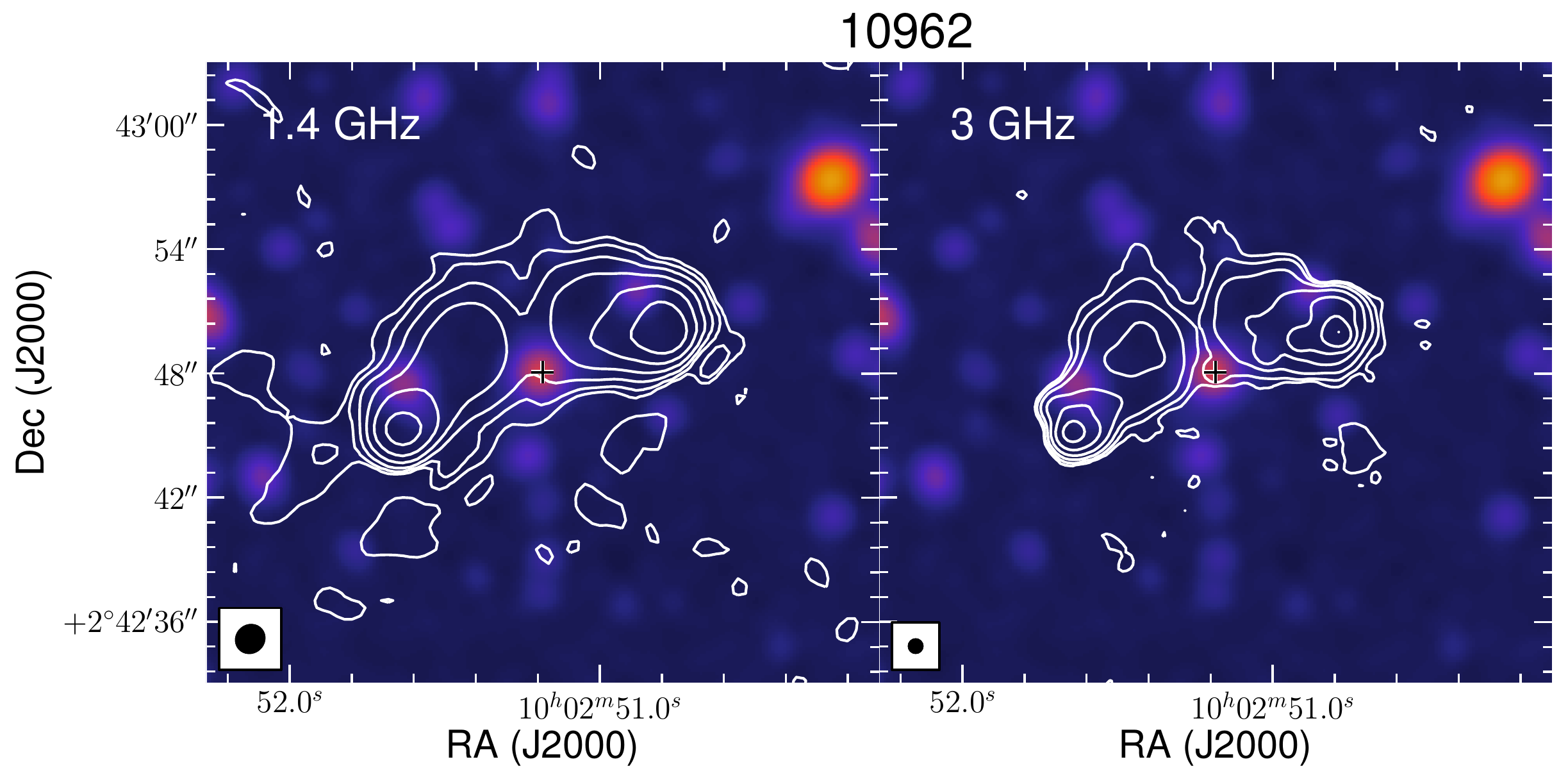}
       \includegraphics{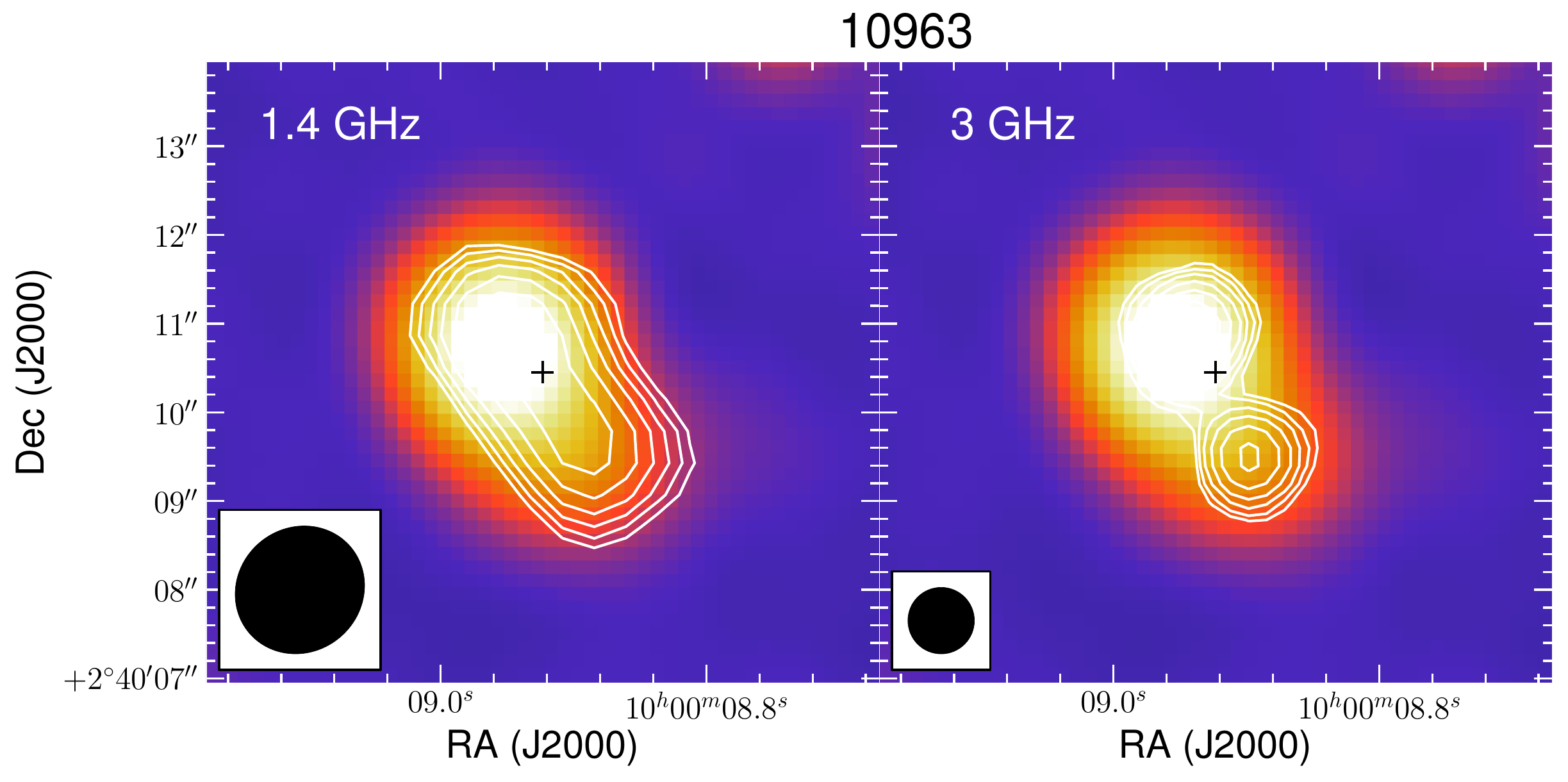} 
            }
             \\ \\ 
      \resizebox{\hsize}{!}
       { \includegraphics{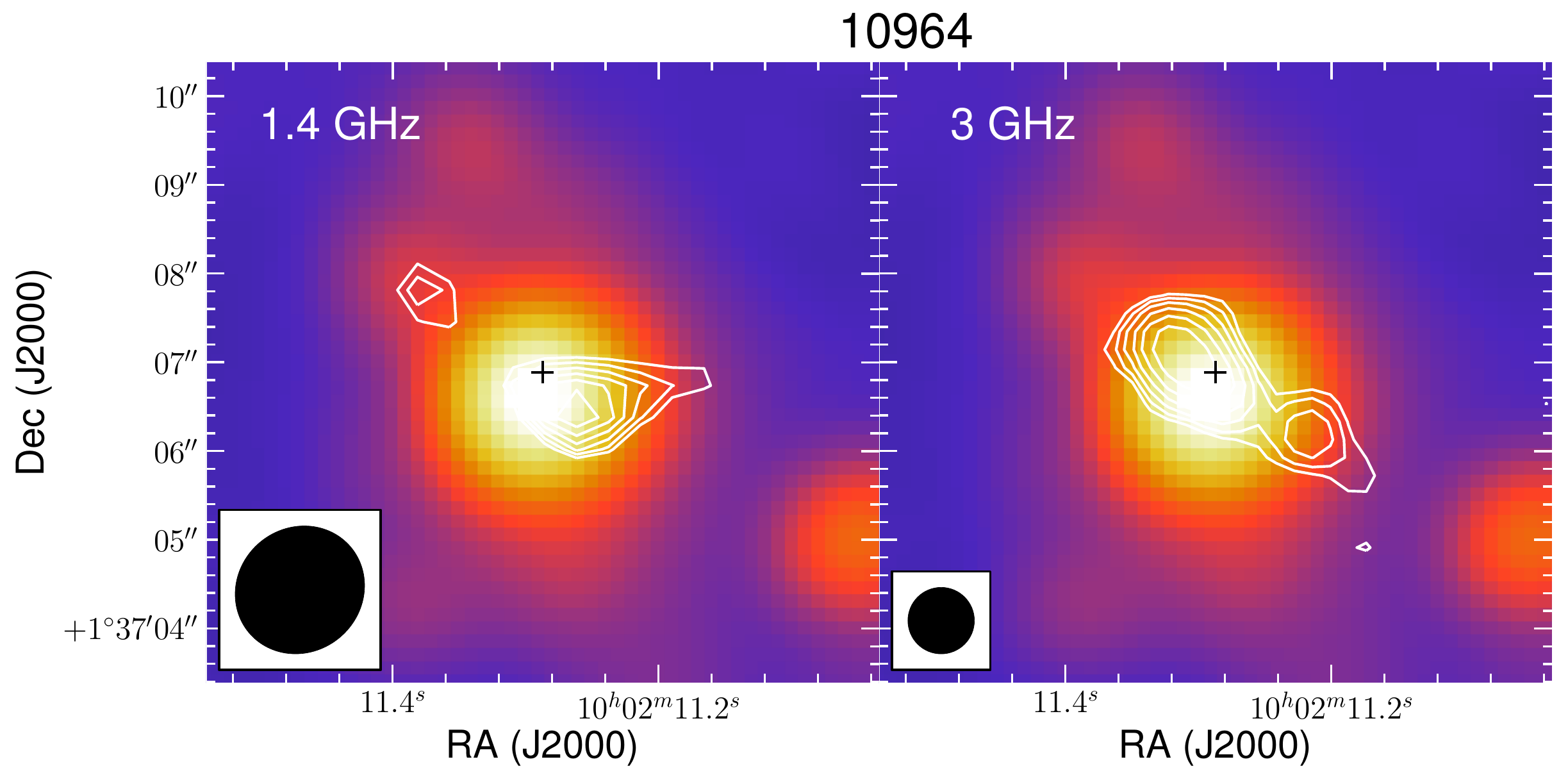}
       \includegraphics{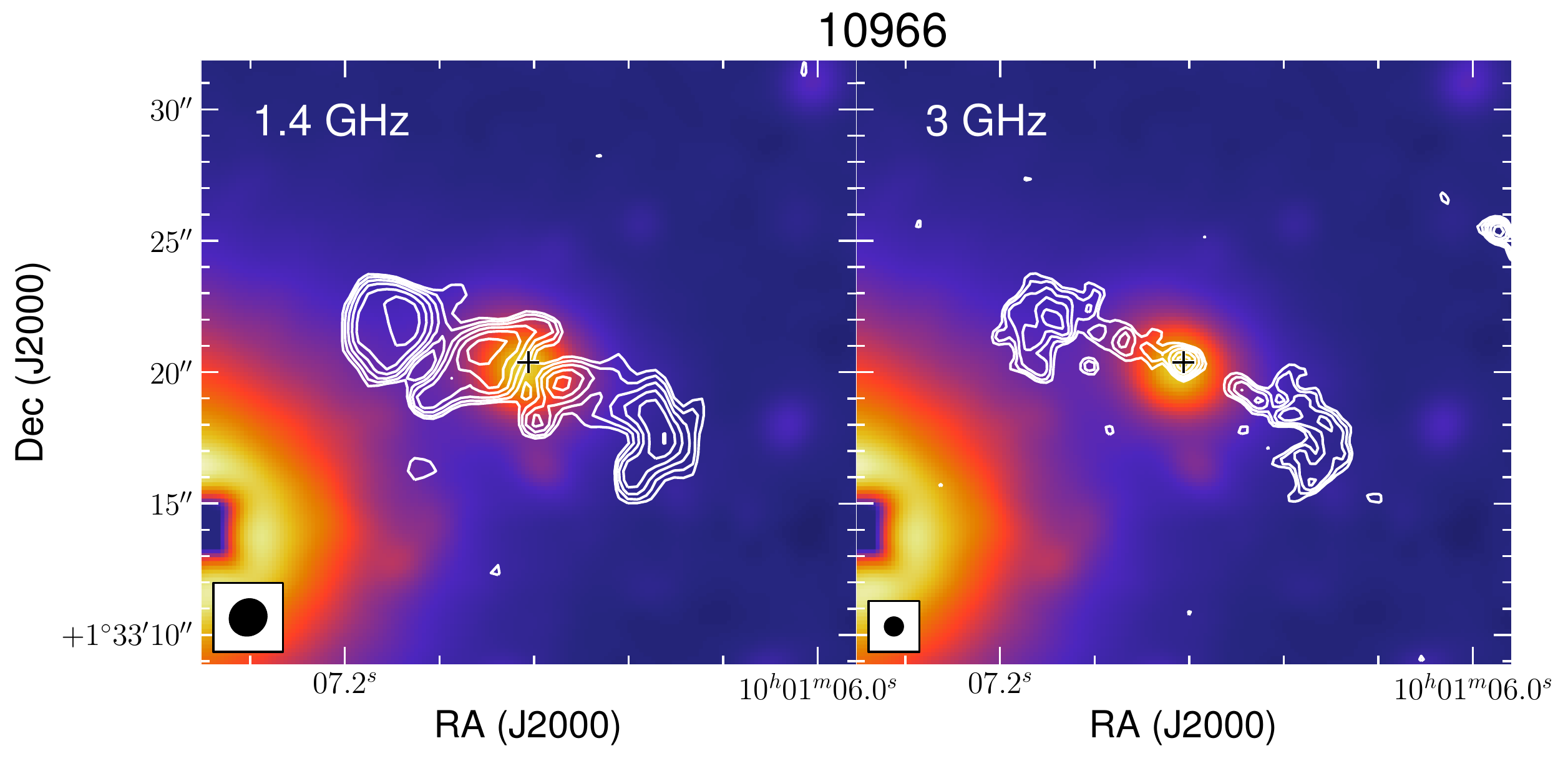}
            }

   \caption{(continued)
   }
              \label{fig:maps2}%
    \end{figure*}
%
%
\clearpage
\begin{table}[!ht]
\caption{Intrinsic radio sizes of COM AGN}             
\label{tab:com_D}      
\centering                          
\begin{tabular}{|   l l   |  l l   |   ll   |  ll   |}       
\hline
ID  &   $D$ & ID  &   $D$ & ID  &   $D$& ID  &   $D$\\
&    (kpc)        &              &          (kpc)   & &          (kpc) & &          (kpc)          \\
\hline\hline    
1 &  0.0 &105 &  0.5 &219 &  0.6 &345 &  0.5 \\
2 &  0.0 &106 &  1.4 &224 &  0.6 &346 &  1.1 \\
3 &  0.0 &108 &  0.0 &226 &  0.8 &350 &  7.8 \\
5 &  0.9 &109 &  0.7 &227 &  0.0 &352 &  0.9 \\
6 &  3.0 &114 &  0.7 &229 &  4.8 &353 &  1.7 \\
7 &  4.4 &119 &  0.7 &231 &  1.3 &356 &  1.4 \\
\hline 
\end{tabular}
\tablefoot{Intrinsic sizes are estimated after using \textsc{pyBDSF} \citep{mohan15} on the 3 GHz mosaic \citep{jimenez19}.}
\end{table}

\addtocounter{table}{-1}

\begin{table}[!ht]
\caption{Intrinsic radio sizes of COM AGN (continued)}             
\label{tab:com_D}      
\centering                          
\begin{tabular}{|   l l   |  l l   |   ll   |  ll   |}       
\hline
ID  &   $D$ & ID  &   $D$ & ID  &   $D$& ID  &   $D$\\
&    (kpc)        &              &          (kpc)   & &          (kpc) & &          (kpc)          \\
\hline\hline   
456 &  1.9 &603 &  2.4 &723 &  1.4 &862 &  0.0 \\
457 &  0.0 &605 &  1.8 &724 &  2.3 &863 &  2.2 \\
458 &  0.0 &608 &  1.0 &725 &  2.2 &869 &  0.0 \\
460 &  3.4 &611 &  1.6 &729 &  0.0 &872 &  0.0 \\
461 &  2.3 &612 &  1.0 &730 &  2.2 &876 &  0.0 \\
468 & 10.2 &614 &  0.0 &734 &  2.3 &877 &  2.0 \\
\hline 
\end{tabular}

\end{table}

\addtocounter{table}{-1}

\begin{table}[!ht]
\caption{Intrinsic radio sizes of COM AGN (continued)}             
\label{tab:com_D}      
\centering                          
\begin{tabular}{|   l l   |  l l   |   ll   |  ll   |}       
\hline
ID  &   $D$ & ID  &   $D$ & ID  &   $D$& ID  &   $D$\\
&    (kpc)        &              &          (kpc)   & &          (kpc) & &          (kpc)          \\
\hline\hline   
1020 &  0.0 &1216 &  8.1 &1408 &  3.4 &1607 &  2.0 \\
1024 &  0.0 &1222 &  2.2 &1409 &  2.9 &1608 &  0.0 \\
1026 &  2.7 &1224 &  0.0 &1413 &  3.2 &1620 &  0.0 \\
1029 &  0.0 &1225 &  5.1 &1415 &  1.8 &1631 &  3.3 \\
1030 &  1.8 &1228 &  2.4 &1421 &  1.6 &1639 &  0.0 \\
1035 &  3.3 &1230 &  2.2 &1423 &  0.0 &1643 &  3.1 \\
\hline 
\end{tabular}

\end{table}

\addtocounter{table}{-1}

\begin{table}[!ht]
\caption{Intrinsic radio sizes of COM AGN (continued)}             
  
\centering                          
\begin{tabular}{|   l l   |  l l   |   ll   |  ll   |}       
\hline
ID  &   $D$ & ID  &   $D$ & ID  &   $D$& ID  &   $D$\\
&    (kpc)        &              &          (kpc)   & &          (kpc) & &          (kpc)          \\
\hline\hline   
1829 &  9.1 &2052 &  0.0 &2297 &  3.2 &2581 &  2.4 \\
1830 &  2.7 &2053 &  0.0 &2307 &  2.8 &2585 &  3.4 \\
1836 &  1.5 &2054 &  0.0 &2312 &  0.0 &2594 &  0.0 \\
1838 &  2.4 &2060 &  0.0 &2320 &  0.0 &2596 &  4.5 \\
1845 &  4.9 &2067 &  1.2 &2321 &  0.0 &2602 &  0.0 \\
1847 &  1.5 &2073 &  2.9 &2329 &  2.6 &2604 &  0.0 \\
\hline 
\end{tabular}

\end{table}

\addtocounter{table}{-1}

\begin{table}[!ht]
\caption{Intrinsic radio sizes of COM AGN (continued)}             
  
\centering                          
\begin{tabular}{|   l l   |  l l   |   ll   |  ll   |}       
\hline
ID  &   $D$ & ID  &   $D$ & ID  &   $D$& ID  &   $D$\\
&    (kpc)        &              &          (kpc)   & &          (kpc) & &          (kpc)          \\
\hline\hline   
2853 &  0.0 &3151 &  0.0 &3439 &  9.6 &3844 &  0.0 \\
2866 &  4.9 &3161 &  0.0 &3444 &  7.5 &3850 &  3.2 \\
2867 &  1.0 &3163 &  0.0 &3445 &  0.0 &3853 &  2.4 \\
2876 &  0.0 &3165 &  4.7 &3449 &  4.2 &3862 &  3.4 \\
2877 &  0.0 &3172 &  2.7 &3461 &  0.0 &3869 &  4.4 \\
2883 &  0.0 &3182 &  5.5 &3471 &  2.5 &3871 &  3.8 \\
\hline 
\end{tabular}

\end{table}

\addtocounter{table}{-1}

\begin{table}[!ht]
\caption{Intrinsic radio sizes of COM AGN (continued)}             
   
\centering                          
\begin{tabular}{|   l l   |  l l   |   ll   |  ll   |}       
\hline
ID  &   $D$ & ID  &   $D$ & ID  &   $D$& ID  &   $D$\\
&    (kpc)        &              &          (kpc)   & &          (kpc) & &          (kpc)          \\
\hline\hline   
4254 &  0.0 &4674 &  8.8 &5125 &  4.4 &5756 &  6.0 \\
4260 &  4.0 &4675 &  3.1 &5136 &  0.0 &5761 &  0.0 \\
4277 &  0.0 &4680 &  0.0 &5142 &  3.6 &5763 &  0.0 \\
4278 &  0.0 &4683 &  0.0 &5164 &  4.0 &5767 &  4.1 \\
4284 &  7.6 &4689 &  4.5 &5165 &  0.0 &5775 &  3.9 \\
4289 &  3.1 &4701 &  5.4 &5196 &  0.0 &5779 &  0.0 \\
\hline 
\end{tabular}

\end{table}

\addtocounter{table}{-1}

\begin{table}[!ht]
\caption{Intrinsic radio sizes of COM AGN (continued)}             
    
\centering                          
\begin{tabular}{|   l l   |  l l   |   ll   |  ll   |}       
\hline
ID  &   $D$ & ID  &   $D$ & ID  &   $D$& ID  &   $D$\\
&    (kpc)        &              &          (kpc)   & &          (kpc) & &          (kpc)          \\
\hline\hline  
6231 &  7.8 &6860 &  0.0 &7757 &  0.0 &8338 &  0.0 \\
6232 &  0.0 &6880 &  0.0 &7779 &  0.0 &8341 &  0.0 \\
6240 &  2.9 &6906 &  3.9 &7781 &  9.5 &8355 &  7.1 \\
6263 &  6.8 &6930 &  7.5 &7787 &  3.9 &8356 &  0.0 \\
6283 &  3.0 &6933 & 12.6 &7788 &  8.5 &8361 &  0.0 \\
6286 &  0.0 &6935 &  3.2 &7802 &  0.0 &8365 &  0.0 \\
\hline 
\end{tabular}

\end{table}
\addtocounter{table}{-1}

\begin{table}[!ht]
\caption{Intrinsic radio sizes of COM AGN (continued)}             
  
\centering                          
\begin{tabular}{|   l l   |  l l   |   ll   |  }       
\hline
ID  &   $D$ & ID  &   $D$ & ID  &   $D$\\
&    (kpc)        &              &          (kpc)   & &          (kpc)      \\
\hline\hline 
9029 &  0.0 &9851 &  6.5 &10629 &  0.0 \\
9059 &  2.8 &9853 &  0.0 &10631 &  0.0 \\
9063 &  0.0 &9890 &  0.0 &10633 &  0.0 \\
9069 &  0.0 &9899 &  0.0 &10637 &  0.0 \\
9074 &  7.7 &9909 &  6.2 &10696 &  0.0 \\
9103 &  0.0 &9925 &  0.0 &10698 &  0.0 \\
9106 &  6.0 &9933 &  0.0 &10734 &  0.0 \\
9113 &  0.0 &9938 & 10.0 &10748 &  0.0 \\
9120 &  6.3 &9947 &  0.0 &10754 &  0.0 \\
9123 & 15.5 &9965 &  0.0 &10765 &  0.0 \\
9124 &  4.2 &9985 &  0.0 &10781 &  5.2 \\
9135 &  8.8 &10010 &  2.4 &10811 &  0.0 \\
9138 &  0.0 &10026 &  0.0 &10814 &  0.0 \\
9141 &  0.0 &10048 &  4.7 &10824 &  0.0 \\
9144 &  0.0 &10054 &  0.0 &10843 &  0.0 \\
9151 &  5.0 &10074 &  0.0 &10857 &  0.0 \\
9164 &  8.5 &10075 &  0.0 &10896 &  0.0 \\
9201 & 10.6 &10077 &  3.7 &10946 &  6.4 \\
\hline 
\end{tabular}

\end{table}

\end{document}